\documentclass{aa}
\usepackage[varg]{txfonts}
\usepackage{graphicx}
\usepackage{amsmath}
\usepackage{amssymb}
\usepackage{footmisc}
\usepackage{lscape}
\usepackage{color}
\usepackage{longtable}
\usepackage{natbib,twoopt}

\begin{document}

\title{Gamma-ray burst optical light-curve zoo: comparison with X-ray observations}

\author{E. Zaninoni\inst{1,}\inst{2}, M. G. Bernardini\inst{1}, R. Margutti\inst{3}, S. Oates\inst{4}, and G. Chincarini\inst{1,}\inst{5}}

\institute{INAF - Osservatorio Astronomico di Brera, via Bianchi 46, Merate 23807, Italy
  \and University of Padova, Physics \& Astronomy Dept. Galileo Galilei, vicolo dell'Osservatorio, 3- I-35131 Padova, Italy
  \and Harvard-Smithsonian Center for Astrophysics, 60 Garden Street, Cambridge, MA02138
  \and Mullard Space Science Laboratory, University College London, Holmbury St. Mary, Dorking, Surrey RH5 6NT, UK
  \and Univerisit\`a Milano Bicocca, Dip. Fisica G. Occhialini, P.zza della Scienza 3, Milano 20126, Italy}

\offprints{elena.zaninoni@brera.inaf.it}

\date{Received DD Mmmm YYYY / Accepted DD Mmmm YYYY}

\abstract{\textit{Aims}. We present a comprehensive analysis of the optical and X-ray light curves (LCs) and spectral energy distributions (SEDs) of a large sample of gamma-ray burst (GRB) afterglows to investigate the relationship between the optical and X-ray emission after the prompt phase. We consider all data available in the literature, which where obtained with different instruments.\\
\textit{Methods}. We collected the optical data from the literature and determined the shapes of the optical LCs. Then, using previously presented X-ray data we modeled the optical/X-ray SEDs. We studied the SED parameter distributions and compared the optical and X-ray LC slopes and shapes.\\
\textit{Results}. The optical and X-ray spectra become softer as a function of time while the gas-to-dust ratios of GRBs are higher than the values calculated for the Milky Way and the Large and Magellanic Clouds. For 20\% of the GRBs the difference between the optical and X-ray slopes is consistent with 0 or $1/4$ within the uncertainties (we did it not consider the steep decay phase), while in the remaining 80\% the optical and X-ray afterglows show significantly different temporal behaviors. Interestingly, we find an indication that the onset of the forward shock in the optical LCs (initial peaks or shallow phases) could be linked to the presence of the X-ray flares. Indeed, when X-ray flares are present during the steep decay, the optical LC initial peak or end plateau occurs during the steep decay; if instead the X-ray flares are absent or occur during the plateau, the optical initial peak or end plateau takes place during the X-ray plateau.\\
\textit{Conclusions}. The forward-shock model cannot explain all features of the optical (e.g. bumps, late re-brightenings) and X-ray (e.g. flares, plateaus) LCs. However, the synchrotron model is a viable mechanism for GRBs at late times. In particular, we found a relationship between the presence of the X-ray flares and the shape of the optical LC that indicates a link between the prompt emission and the optical afterglow.
}


\titlerunning{The gamma-ray burst optical light-curve zoo: comparison with the X-ray observations}
\authorrunning{E. Zaninoni et al.}
\maketitle 
%%%%%%%%%%%%%%%%%%%%%%%%%%%
\section{Introduction}

Gamma-ray bursts (GRBs) are the most powerful sources of electromagnetic radiation in the Universe, with an isotropic luminosity that can reach values of $10^{54}$ erg s$^{-1}$. The \textit{Swift} satellite \citep{2004ApJ...611.1005G}, launched in November 2004, opened a new era for the study and understanding of GRB phenomena; thanks to the rapid response of the instruments with a small field of view, it was discovered that both the X-ray light-curve (LC) (e.g. \citealt{2006ApJ...642..389N}, \citealt{2006ApJ...642..354Z}) and the optical LC (e.g. \citealt{2009ApJ...690..163R}) have a complex shape. The rapid computation of the GRB position by the \textit{Swift} Burst Alert Telescope (BAT; \citealt{2005SSRv..120..143B}), refined with an accuracy of few arcseconds by the \textit{Swift} X-ray Telescope (XRT, \citealt{2005SSRv..120..165B}), and  the instantaneous dissemination to the community via the GCN\footnote{Gamma-ray Coordinates Network (GCN), http://gcn.gsfc.nasa.gov/gcn3\_archive.html} allows a growing number of robotic telescopes to promptly repoint to the source. Some examples are the Robotic Optical Transient Search Experiment (ROTSE-III, \citealt{2003PASP..115..132A}), the Rapid Eye Mount telescope (REM, \citealt{2001AN....322..275Z,2003Msngr.113...40C}), the Gamma-Ray Burst Optical/Near-Infrared detector (GROND, \citealt{2008PASP..120..405G}), Liverpool (LT) and Faulkes telescopes \citep{2006NCimB.121.1303G}, T\'elescopes \`a Action Rapide pour les Objets Transitoires (TAROT, \citealt{2008AN....329..275K}), and others. 

Some generic features have been previously found in optical LCs. Optical and X-ray LCs are different at early times in the majority of cases (\citealt{2008ApJ...686.1209M,2009ApJ...702..489R,2009MNRAS.395..490O,2011MNRAS.412..561O}). In particular, \citet{2009MNRAS.395..490O,2011MNRAS.412..561O} noted that the optical LCs can decay or rise before 500 s after the trigger in the observer frame and do not show the steep decay as the X-ray LCs; after 2000 s the optical and X-ray LCs have similar slopes. \citet{2008MNRAS.387..497P,2011MNRAS.414.3537P} divided the optical LCs according to their initial behavior (peaky or shallow). Peaks were associated to impulsive ejecta releases, while plateau phases were assumed to belong to the energy released by a long-lived central engine. Chromatic and achromatic breaks have been found in the optical and X-ray LCs \citep{2008ApJ...686.1209M,2009ApJ...702..489R,2009MNRAS.395..490O,2011MNRAS.412..561O,2011MNRAS.414.3537P}. Moreover, the brighter optical LCs decay faster \citep{2009MNRAS.395..490O,2011MNRAS.412..561O,2012MNRAS.426L..86O}. When optical and X-ray LCs do not share the same temporal decay, X-ray LCs have been found to decay faster \citep{2009MNRAS.395..490O,2011MNRAS.412..561O,2011MNRAS.414.3537P}. For only a few GRBs with shallow X-ray decay phases we find a corresponding shallow decay in the optical \citep{2009ApJ...702..489R,2012ApJ...758...27L}. Flares can occasionally appear in optical LCs and likely linked to the long term central engine activity \citep{2012ApJ...758...27L}.

Previous works mainly concentrated on data obtained by a single telescope (e.g. \citealt{2008ApJ...686.1209M,2009AJ....137.4100K,2009ApJ...693.1484C,2009MNRAS.395..490O,2011MNRAS.412..561O,2009ApJ...702..489R}) and only a few authors compared the data from different instruments (e.g. \citealt{2009ApJ...701..824N, 2010ApJ...720.1513K,2011ApJ...734...96K,2012ApJ...758...27L,2012arXiv1210.5142L}). For example, \citet{2010ApJ...720.1513K,2011ApJ...734...96K} focused on the classification of the optical LCs and the host galaxy extinction. \citet{2012ApJ...758...27L} and \citet{2012arXiv1210.5142L} concentrated on the optical LC shapes and particular features, as bumps, plateaus, late rebrightenings. Other works studied the dust extinction of the GRB host galaxies (e.g. \citealt{2012A&A...537A..15S,2010MNRAS.401.2773S,2011A&A...532A.143Z}) or the circumburst density profiles around GRB progenitors \citep{2011A&A...526A..23S}. 

In this paper we analyze a large sample of 68 GRBs with optical and X-ray observations and known redshift, detected between December 2004 and December 2010. Our starting sample includes 165 GRBs with known redshift presented by \citealt{2013MNRAS.428..729M} (hereafter M13). We collected the optical data from the literature and obtained well-sampled optical LCs for 68 GRBs from different telescopes and instruments. To compare the optical and X-ray observations, we used the X-ray data extracted and analyzed in M13. We focused on the relationship between the optical and X-ray emission, comparing their rest-frame temporal and spectroscopic properties and their energetics. In particular, we investigated the forward-shock model and the synchrotron emission in the GRB afterglow. In Sect.~\ref{sample} we detail the sample selection criteria, the data selection and reduction, the procedure followed for fitting the optical LCs and of the spectral energy distributions (SEDs). The results of our analysis are presented in Sect.~\ref{risultati} and are discussed in Sect.~\ref{discussione}. The main conclusions are drawn in Sect.~\ref{conclusioni}. We adopt standard values of the cosmological parameters: $H_{0}=70$ km s$^{-1}$ Mpc$^{-1}$, $\Omega_{\rm M}$=0.27, and $\Omega_{\Lambda}$=0.73. For the temporal and spectral energy index, $\alpha$ and $\beta$, we used the convention $F_{\nu}(t,\nu)\propto t^{-\alpha}\nu^{-\beta}$. Errors are given at 1$\sigma$ confidence level unless otherwise stated. 

%%%%%%%%%%%%%%%%%%%%%%%%%%%%
\section{Sample selection and data analysis}\label{sample}

We considered the 165 GRBs with known and secure redshift\footnote{From \cite{2013MNRAS.428..729M} we used only optical spectroscopic redshifts and photometric redshifts for which we are able to exclude sources of degeneracy. We list the redshifts and luminosity distances of the GRBs of our sample in table5c.dat at CDS.} observed by \textit{Swift}/XRT between December 2004 and December 2010, presented in M13. Among these GRBs, we selected those with optical observations and with optical data available in the literature. We used only the data from refereed papers and with more than five data points per filter. In this way we obtained a subsample of 68 long GRBs (Table~\ref{elenco}, \textit{Online Material}). This criterion automatically excluded short GRBs. The selection in spectroscopic redshift from the optical afterglow introduces a bias against highly absorbed optical afterglows \citep{fynbo09,perley09,greiner11}. In fact, GRBs with optical spectroscopy have a substantially lower X-ray excess absorption and a substantially smaller fraction of dark bursts \citep{fynbo09}. On the other hand, our final aim is to compare X-ray and optical rest frame properties, and this can be carried out only with bright and well-sampled optical LCs. Within these constraints we collected a large number of data from more than one hundred telescopes with different instruments and filters (Table~\ref{elenco}, \textit{Online Material}). We analyzed the energetics and luminosities of these GRBs and calculated the SED in the optical/X-ray frequency range. 

%************************************************************
\subsection{Optical data}

Magnitudes were converted into flux densities following standard practice (see Appendix~\ref{appendice1} for details). For this analysis, we used only LCs that had more than five data points per filter and excluded upper limits. This is the best compromise between statistics (in the sense that we do not discard too many GRBs) and reliability (robust fit and energy measurement). All collected data will be available online\footnote{The data will be available on the web site www.elenazaninoni.com}.

For each filter we fitted the optical LCs with the same fit functions as for the X-ray data in M13. We chose these functional forms because they represent the optical LC shapes well and it facilitates comparing optical with X-ray data. We used optical data not corrected for reddening and these fit functions:
\begin{enumerate}
\item Single power-law:
\begin{eqnarray}
F_\nu(t)=N\ t^{-\alpha}.\label{plaw}
\end{eqnarray}
\item One or more smoothed broken power-laws:
\begin{eqnarray}
F_\nu(t)=\sum_i N_{i}\left( \left( \frac{t}{t_{\rm b,\it i}}\right)^{-\frac{\alpha_{1,i}}{s_{i}}}+ \left(\frac{t}{t_{\rm b,\it i}}\right)^{-\frac{\alpha_{2,i}}{s_{i}}} \right)^{s_{i}}.
\end{eqnarray}
\item Sum of power-law and smoothed broken power-law:
\begin{eqnarray}
F_\nu(t)=N_1\,t^{-\alpha_1}+N_2\left( \left( \frac{t}{t_{\rm b}}\right)^{-\frac{\alpha_2}{s}}+ \left(\frac{t}{t_{\rm b}}\right)^{-\frac{\alpha_3}{s}} \right)^{s},\label{bro4}
\end{eqnarray}
\end{enumerate}
where $\alpha$ is the power-law decay index, $t_{\rm b}$ the break time, $s$ the smoothness parameter (always fixed to -0.3, -0.5 or -0.8) and $N$ the normalization. The best-fit parameters were determined using the IDL Levenberg-Marquard least-squares fit routine (MPFIT) supplied by \citet{2009ASPC..411..251M}\footnote{http://www.physics.wisc.edu/$\sim$craigm/idl/fitting.html}. The best-fitting function was chosen considering the $\chi^2$ statistics. The best-fitting parameters are reported in Table~\ref{table1_a} (\textit{Online Material}\footnote{The complete and machine-readable form of the table is provided at CDS (table1c.dat).}). The best-fit of the optical LCs and their residuals are shown in Figures~\ref{confronto1}-\ref{confronto9} (\textit{Online Material}).
 
%************************************************************
\subsection{X-ray data}

The X-ray spectra were extracted using the method presented in M13 (see also references therein). We fitted them with \textit{Xspec} and the function \texttt{tbabs*ztbabs*pow}, which considers the hydrogen column density absorption of the Milky Way ($N_{\rm H,MW}$) and of the host galaxy ($N_{\rm H,host}$). The $N_{\rm H,MW}$ was calculated with the \texttt{nh} tool, which uses the weighted average value from the \citet{2005A&A...440..775K} map.  The output data obtained from the X-ray spectrum are $N_{\rm H,X}$ and the X-ray photon index\footnote{The spectral index $\beta$ is related to the photon index $\Gamma$ by $\Gamma=\beta+1$.\label{fn:photonindex}} ($\Gamma_{\rm X}$) (Table~\ref{table4_a}, \textit{Online Material}\footnote{The complete and machine-readable form of the table is provided at CDS (table4c.dat).}). 
 
 %************************************************************
\subsection{Optical/X-ray SEDs}\label{SED}

For each GRB, we created optical/X-ray SEDs at one or more epochs (Table~\ref{ftest}, \textit{Online Material}). The time intervals for the SEDs were chosen taking into account the shape of the X-ray and optical shapes: \textit{a)} they belong to a determined phase of the X-ray LC that is steep decay, plateau or normal decay to avoid the X-ray LC breaks. In this way we obtained a SED both at early times (where the afterglow emission could be influenced by the prompt emission) and at late times (where the afterglow emission is very unlikely to be contaminated by the prompt emission); \textit{b)} sometimes the SEDs were constructed during X-ray and optical flares. For the optical data, we did not extrapolate the optical LC, so that if for a given filter no data were available, the filter was excluded from the SED. For each filter with data in this time range, we calculated the flux density by integrating the optical LC over the considered time interval.
%Then, for each GRB, we chose significant time intervals of the LCs to investigate their optical/X-ray SED. We did not perform any extrapolation of the optical LC, so that if, for a given filter, no data are available, the filter is not included in the SED. For each filter with data in this time range, we calculated the flux density integrating the optical LC over the considered interval time, to be used for the optical/X-ray SED (Sec. \ref{SED}). We selected time intervals: $a)$ before and after an optical LC break; $b)$ in correspondence to a plateau in the X-ray LC, an X-ray flare or optical fluctuation; $c)$ at early times, in the few cases with contemporaneous optical and X-ray observations; these cases could be influenced by the prompt emission; $d)$ at late times, where the afterglow emission is very unlikely to be contaminated by the prompt emission.

We fitted the optical/X-ray SED accounting for absorption in the optical and X-ray ranges both locally (i.e., in the GRB host galaxy) and arising from the Milky Way (MW). For the optical band we used the extinction laws given by \cite{1992ApJ...395..130P} (Eq.~20 and Table~4 therein) for the MW and the Large Magellanic Cloud (LMC) and Small Magellanic Cloud (SMC). For the X-ray data, we considered the model for the photoelectric cross section per HI-atom units for a given metallicity presented by \cite{1983ApJ...270..119M}, assuming solar metallicity.

We first considered the case that the X-ray and the optical bands lie in the same spectral segment, hence the SED-fitting function is a combination of the absorption laws presented above and a single power-law:
\begin{eqnarray}
f_\nu(\nu) =f_0\ \nu_{\rm obs}^{-\beta_{\rm op,X}},
\end{eqnarray}
where $\nu_{\rm obs}$ is the observed frequency, $\beta_{\rm op,X}$ the spectral index\footref{fn:photonindex} and $f_0$ the normalization. From the input parameters, the Galactic hydrogen column density  ($N_{\rm H,MW}$), the Galactic reddening ($E(B-V)_{\rm MW}$\footnote{The $E(B-V)$ values were taken from NASA/IPAC Extragalactic Database (NED) website (http://ned.ipac.caltech.edu/forms/calculator.html), which uses the \cite{1998ApJ...500..525S} maps.}) and the redshift ($z$), we obtained the host galaxy hydrogen column density $\mathbf{N_{\rm H,op,X}}$, reddening $E(B-V)_{\rm host}$, and the spectral index $\beta_{\rm op,X}$.

Then, we examined the hypothesis that the cooling frequency is between the optical and the X-ray bands and fit the data using the absorption laws plus a broken-power law:
\begin{eqnarray}
f_\nu(\nu)&=&F_0(\nu_{\rm obs}^{-\beta_{\rm op}}step(\nu_{\rm obs,BR}-\nu_{\rm obs})+\nonumber\\
         &   &+\nu_{\rm obs}^{-\beta_{\rm X}}\nu_{\rm obs,BR}^{\beta_{\rm X}-\beta_{\rm op}}step(\nu_{\rm obs}-\nu_{\rm obs,BR})),
\end{eqnarray}
where $step$ is the step function, $\nu_{\rm obs,BR}$ the observer frame break frequency between the optical and X-ray band, $\beta_{\rm op}$ the optical spectral index, and $\beta_{\rm X}$ the X-ray spectral index. The fit was performed in two ways: with $\beta_{\rm op}$ let free to vary or fixed as $\beta_{\rm op}=\beta_{\rm X}-0.5$, as predicted theorically by \citet{1998ApJ...497L..17S} and empirically by \citet{2011A&A...532A.143Z}.  Letting $\beta_{\rm op}$ free to vary did not lead to reliable results. Therefore the best-fit functions of the optical/X-ray SEDs may either be a single power-law or a broken power-law with $\beta_{\rm op}=\beta_{\rm X}-0.5$. To determine if a broken or a single power-law was required, we used an F-test probability $<$5\% as threshold. The results of this selection are presented in Table~\ref{ftest} (\textit{Online Material}) and Figures~\ref{sed1}-\ref{sed8}. The fit parameters are listed in Table~\ref{table2_a} (\textit{Online Material}\footnote{The complete and machine-readable forms of the tables are provided at CDS (table2c.dat, table3c.dat).}). In Table~\ref{table6_a} (\textit{Online Material}\footnote{The complete and machine-readable form of the table is provided at CDS (table6c.dat).}) we list the optical data used for the SEDs.

%%%%%%%%%%%%%%%%%%%%%%%%%%%%
\section{Results}\label{risultati}

%----------------
\begin{figure}[!]
\centering
     \includegraphics[width=0.8 \hsize,clip]{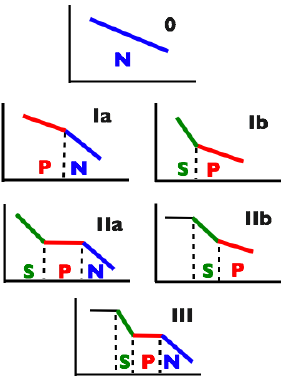}
     \caption{\small{Cartoon representing the X-ray LCs types. For the X-ray LC shapes we used the code presented in M13. Following the prescription of \citet{2012A&A...539A...3B} and M13, we denoted the different parts of the LCs as \textit{a)} steep decay (\textit{S, green}): first segment of type $Ib$ and $IIa$ LCs; the second segment of type $IIb$ and $III$ LCs; \textit{b)} plateau (\textit{P, red}): the first segment of type $Ia$ LCs; the second segment of type $Ib$ and $IIa$ LCs; the third segment of type $IIb$ and $III$ LCs; \textit{c)} normal decay (\textit{N, blue}): type $0$ LCs; the second segment of type $Ia$ LCs; the third segment of type $IIa$ LCs; the forth segment of type $III$ LCs.}}
     	\label{cart}
\end{figure}
%----------------

%************************************************************
\subsection{Spectral parameter distributions}

We considered the parameters obtained from fitting the optical/X-ray SEDs ($\beta$, $N_{\rm H}$, $E(B-V)$, $\nu_{\rm BR}$) with a single or broken power-law, as selected in Sec.~\ref{SED} (Table~\ref{ftest}, \textit{Online Material}), and with a p-value\footnote{The p-value is a number between zero and one and it is the probability of obtaining a test statistic at least as extreme as the one that was actually observed, assuming that the null hypothesis is true.} higher than 0.05. We eliminated the results with errors larger than the data themselves and set to zero negative data that where consistent with 0 within the uncertainties. In total, 78\% of our fits have p-value>0.05 and 33 GRBs have more than one SED with a p-value>0.05. For the steep-decay SEDs, we obtained a good fit (p-value$>0.05$) with a single power-law for 9/13 SEDs and with a broken power-law for 1/13. We where unable to fit three SEDs. Accordingly, we did not consider parameters obtained with a broken power-law for the steep decay (i.e. $\nu_{\rm rest,BR}$, $\beta_{\rm op}$, $\beta_{\rm X}$, $N_{\rm H,BR}$) in the distribution.

\begin{figure*}[!]
\centering
    \includegraphics[width=0.45 \hsize,clip]{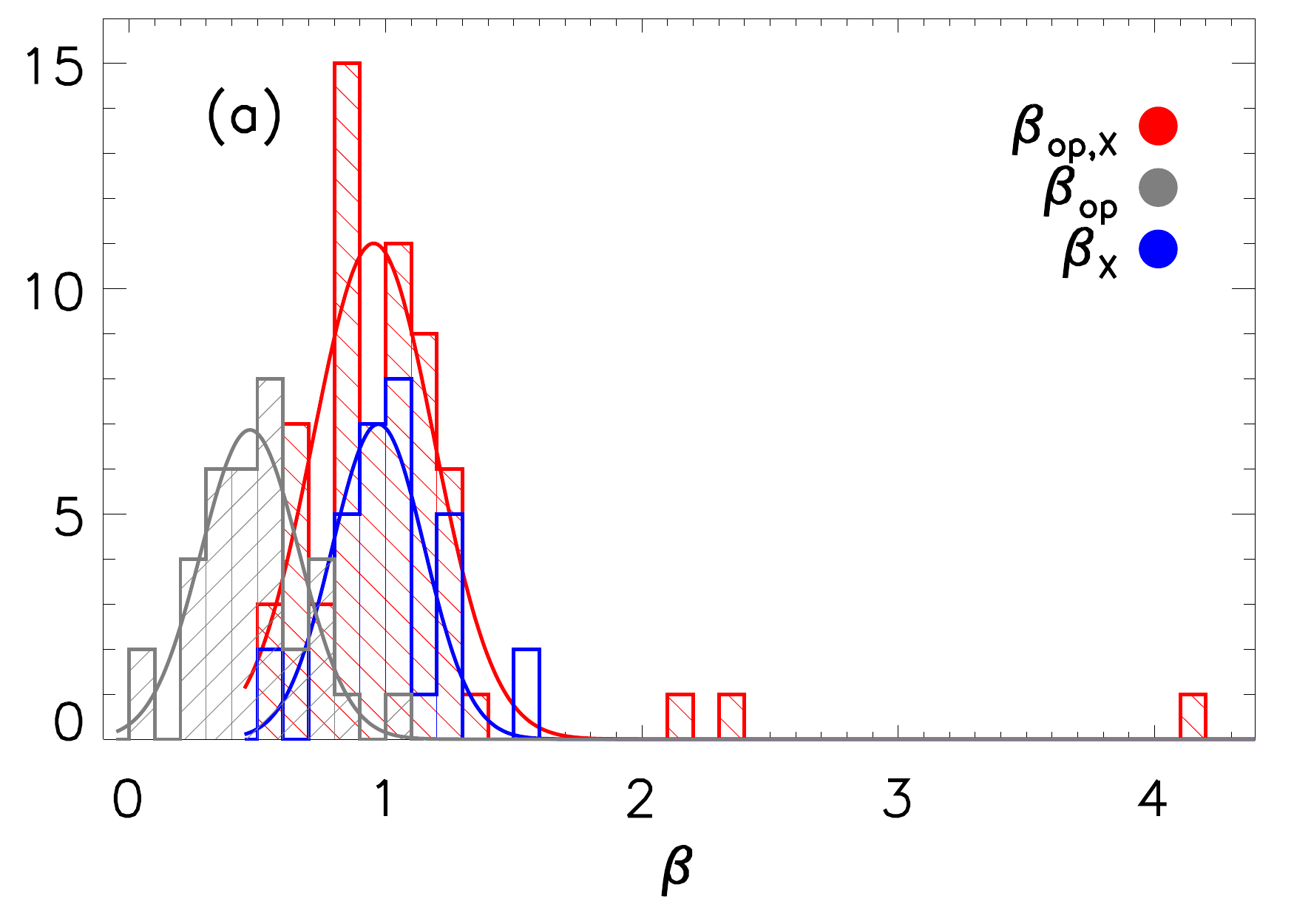}
      \includegraphics[width=0.45 \hsize,clip]{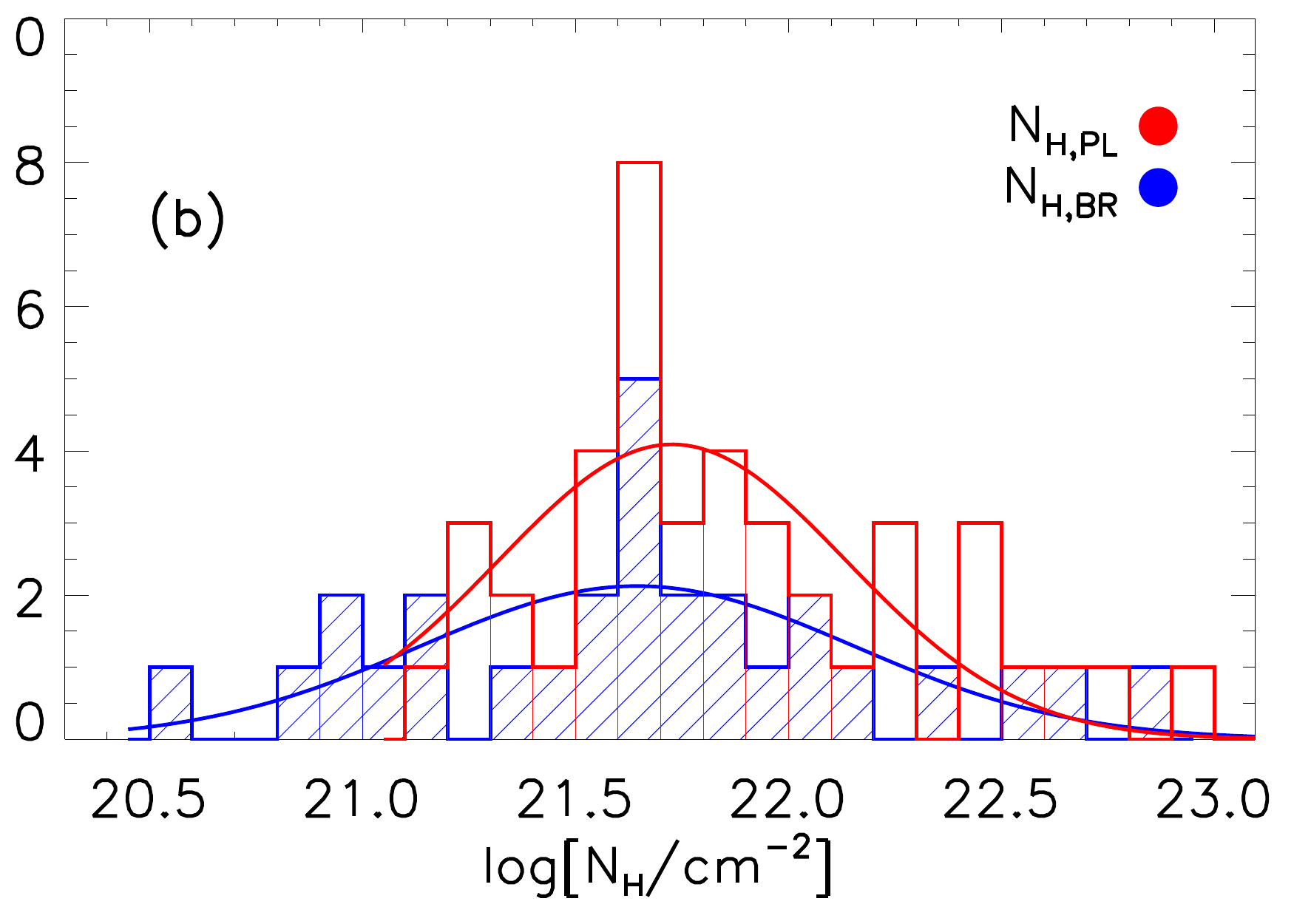}\\      
         \includegraphics[width=0.45 \hsize,clip]{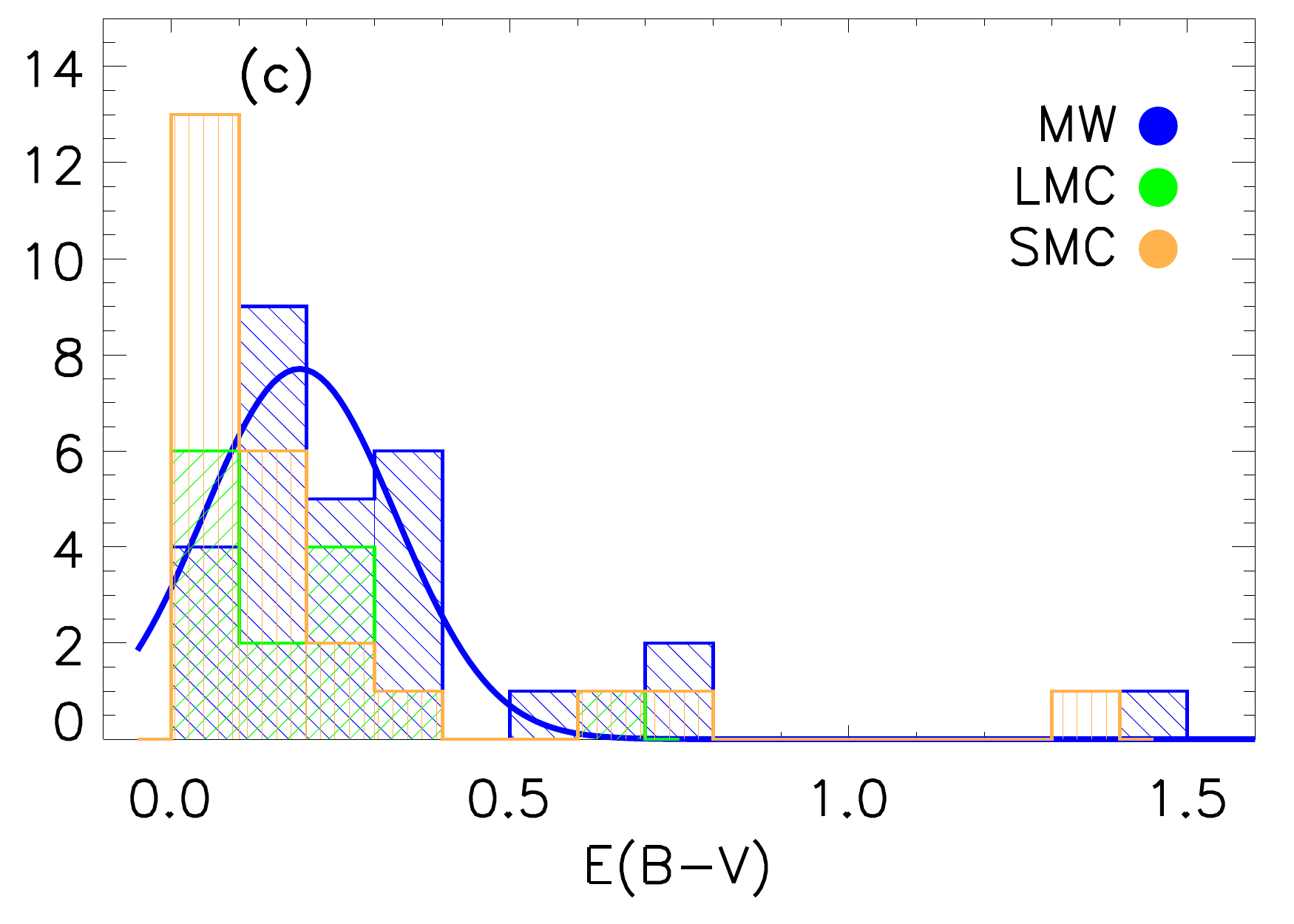}
            \includegraphics[width=0.45 \hsize,clip]{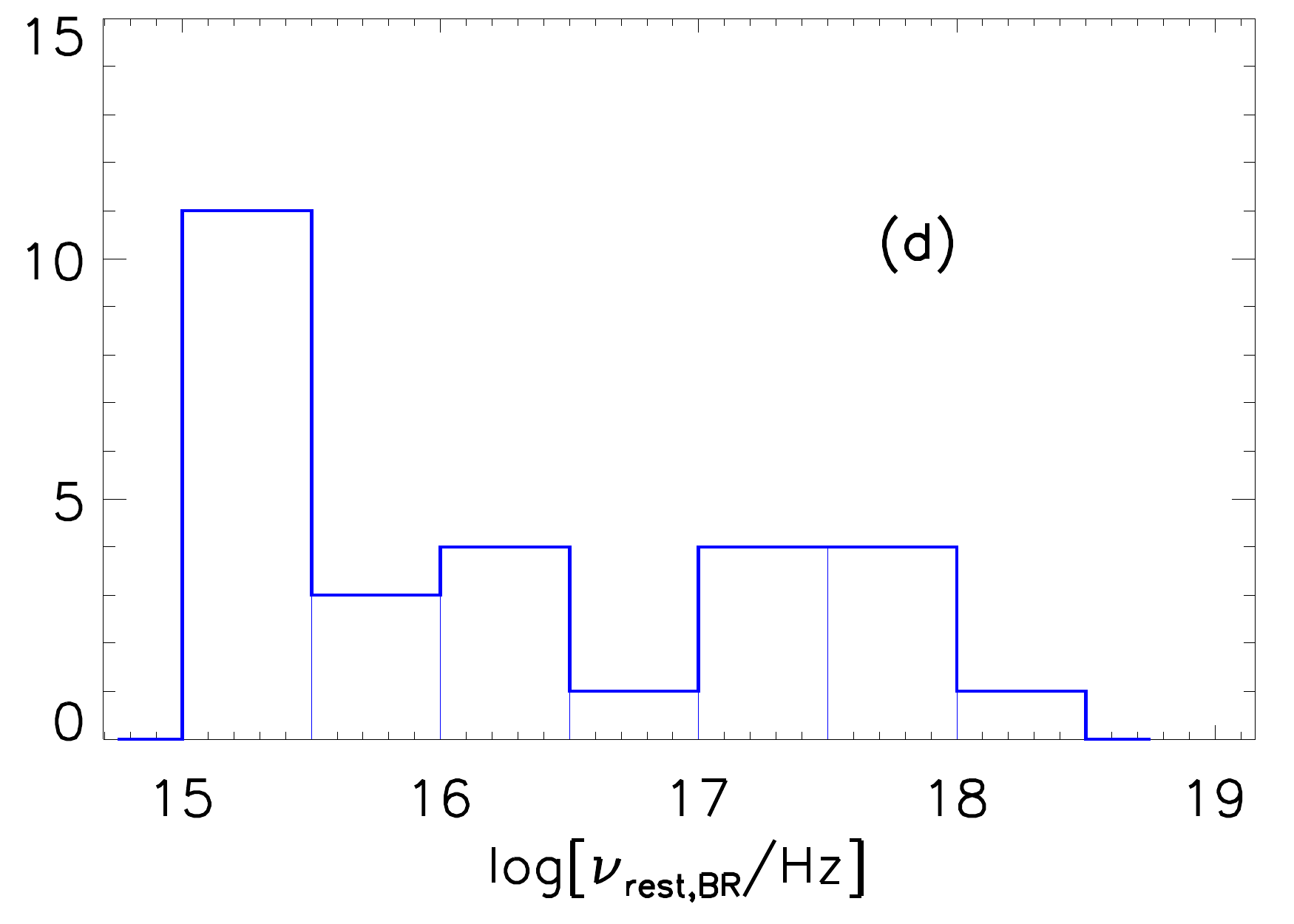}
      \caption{\small{Parameter distributions. The color-coding separates different SED best-fitting functions. \textit{Top panels}. \textit{Blue}: results obtained by fitting the SEDs with a broken power-law and the relative Gaussian fit (\textit{solid line}). \textit{Red}: results obtained by fitting the SEDs with a power-law and the relative Gaussian fit (\textit{solid line}). \textit{a)} The spectral indices ($\beta$) calculated fitting the SED with a single power-law ($\beta_{\rm op,X}$) and with a broken power-law ($\beta_{\rm op}$, \textit{gray}, and $\beta_{\rm X}$).  \textit{b)} The hydrogen column density ($N_{\rm H}$). \textit{Bottom panels.} \textit{c)} The optical extinction ($E(B-V)$) distributions separated according to the different extinction laws: MW (\textit{blue}), LMC (\textit{green}), and SMC (\textit{orange}). \textit{d)} The rest frame break frequency ($\nu_{\rm rest,BR}$) calculated by fitting the SEDs with a broken-power law.}}
     \label{parametri1}
\end{figure*}

\begin{figure*}[!]
\centering         
    \includegraphics[width=0.42 \hsize,clip]{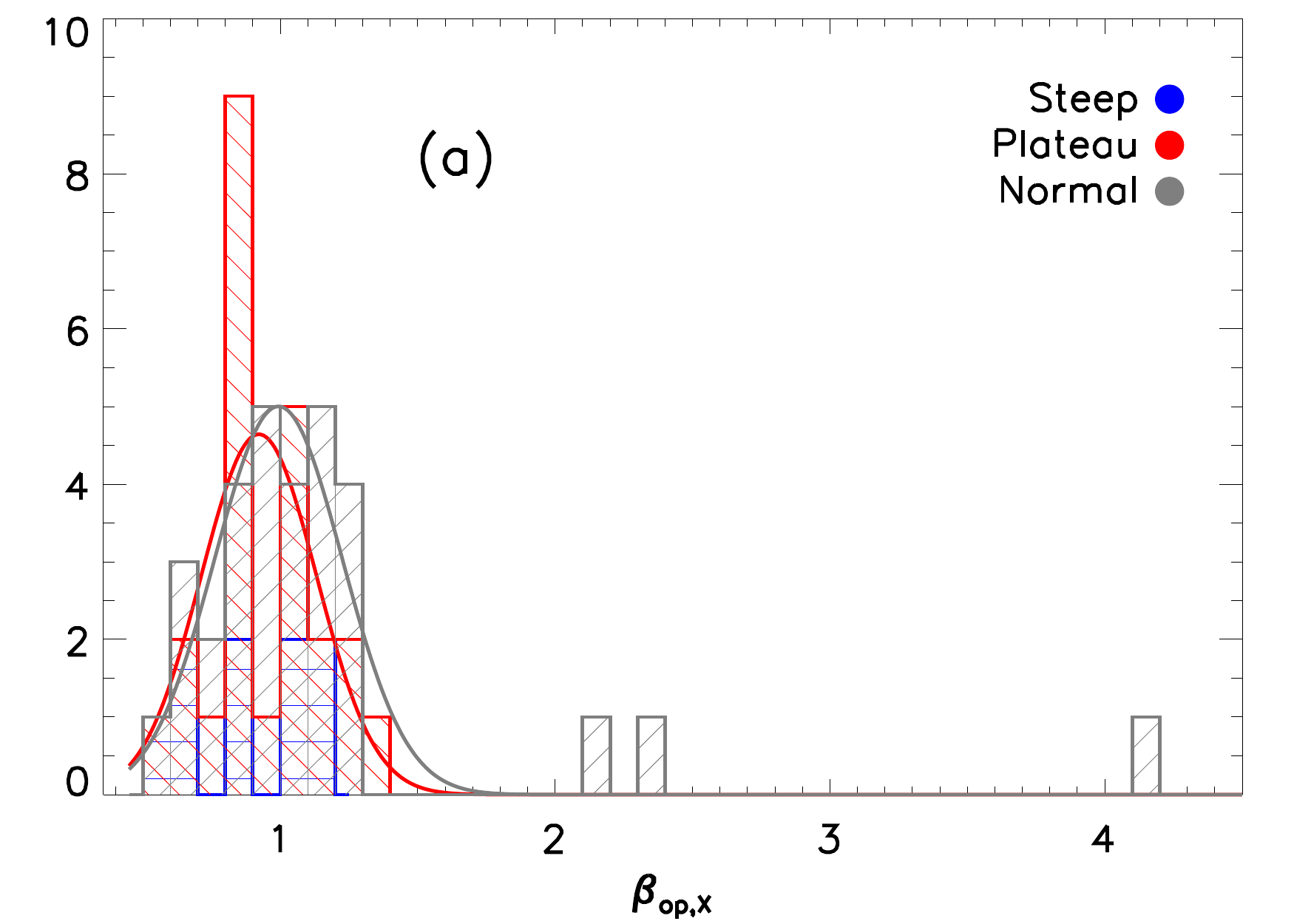}
        \includegraphics[width=0.42 \hsize,clip]{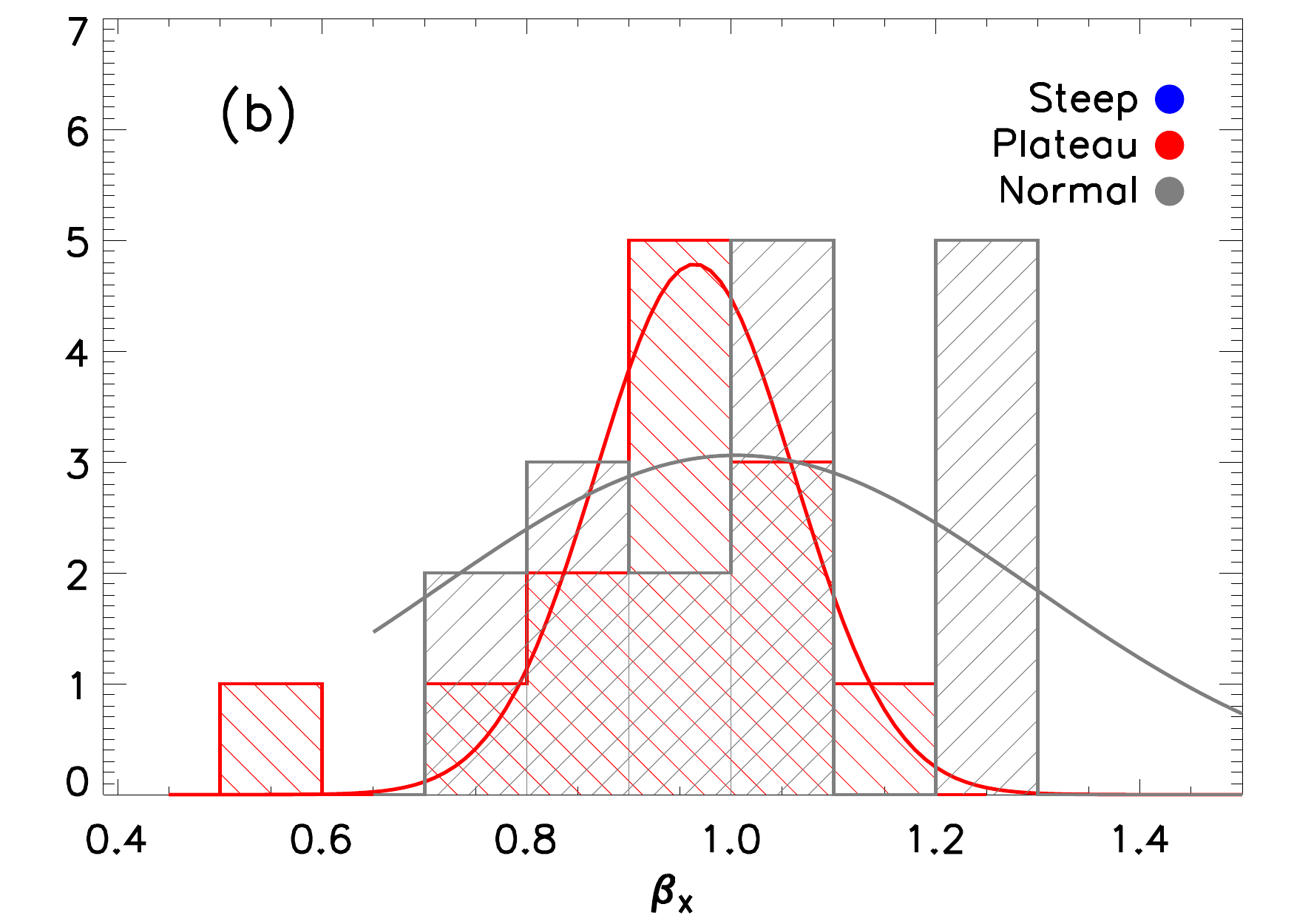}\\
            \includegraphics[width=0.42 \hsize,clip]{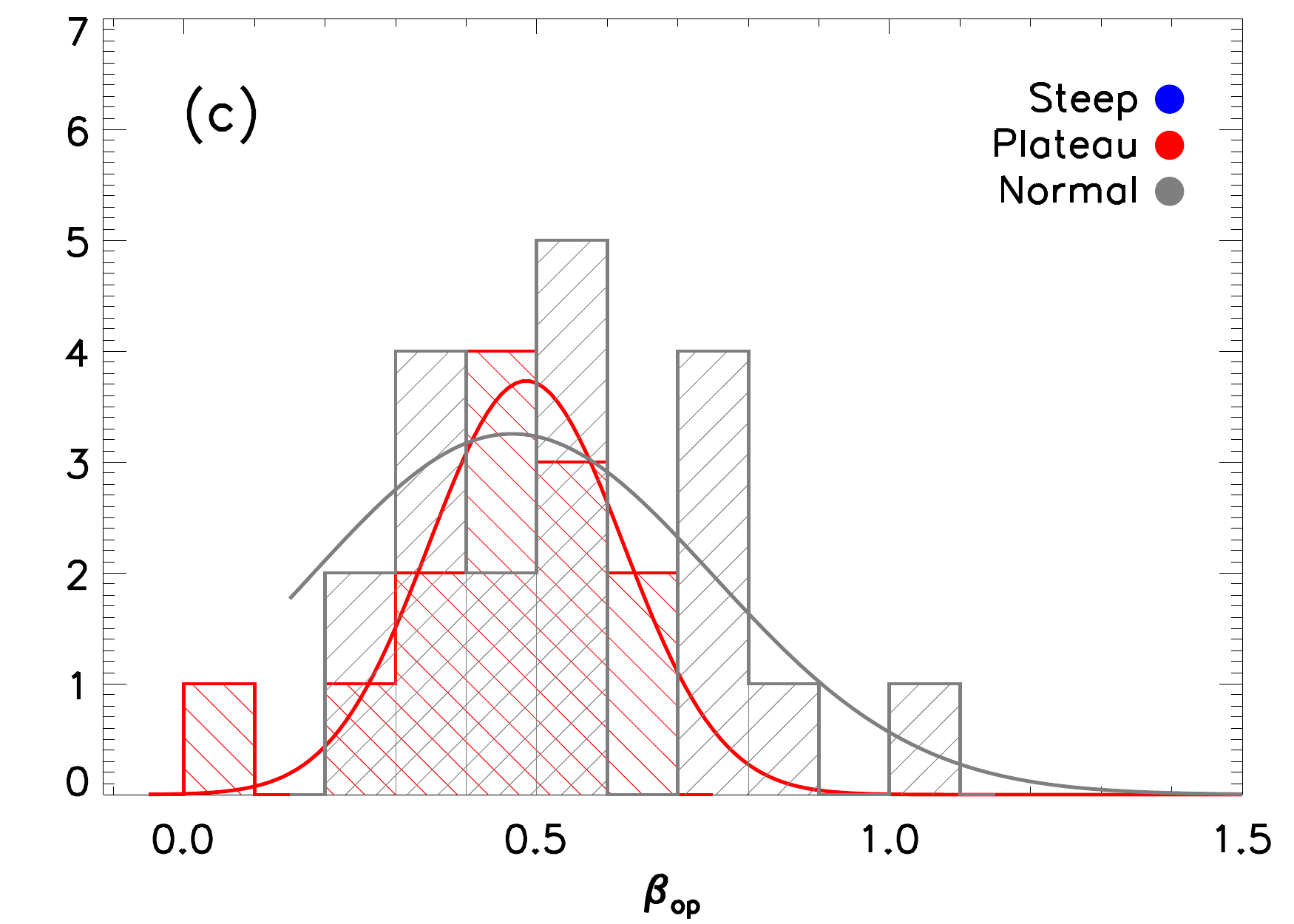}
      \includegraphics[width=0.42 \hsize,clip]{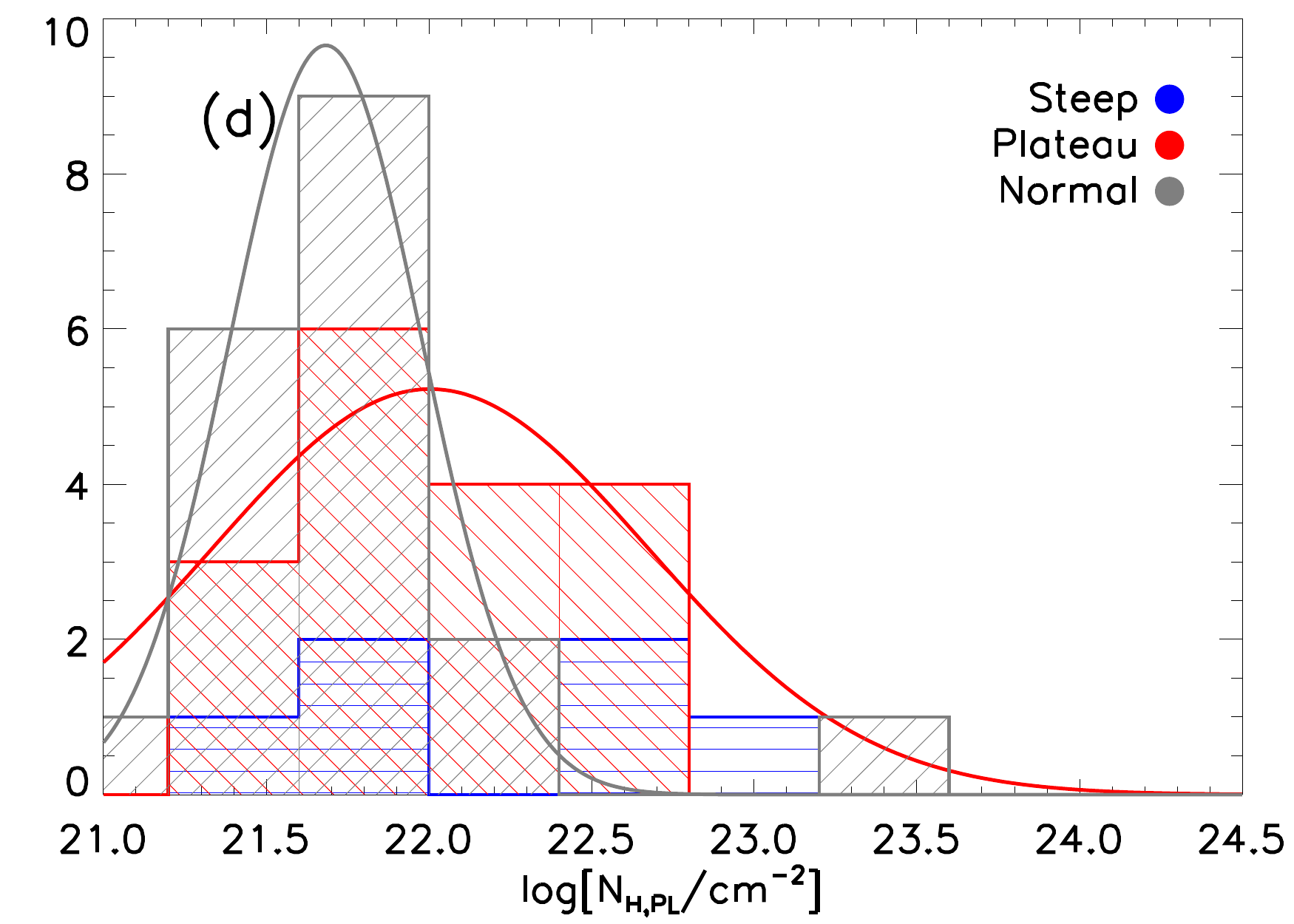}\\
            \includegraphics[width=0.42 \hsize,clip]{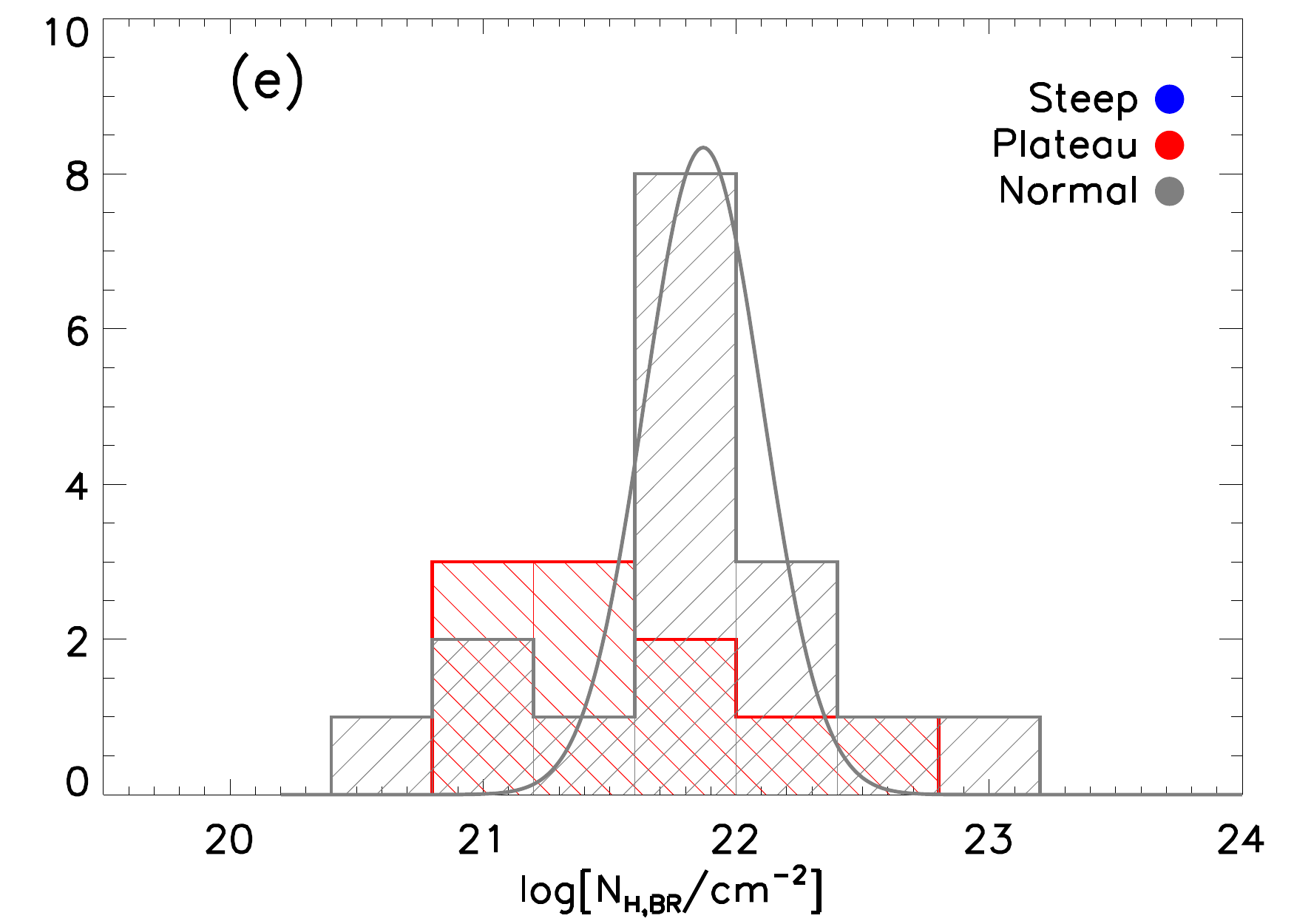}
                              \includegraphics[width=0.42 \hsize,clip]{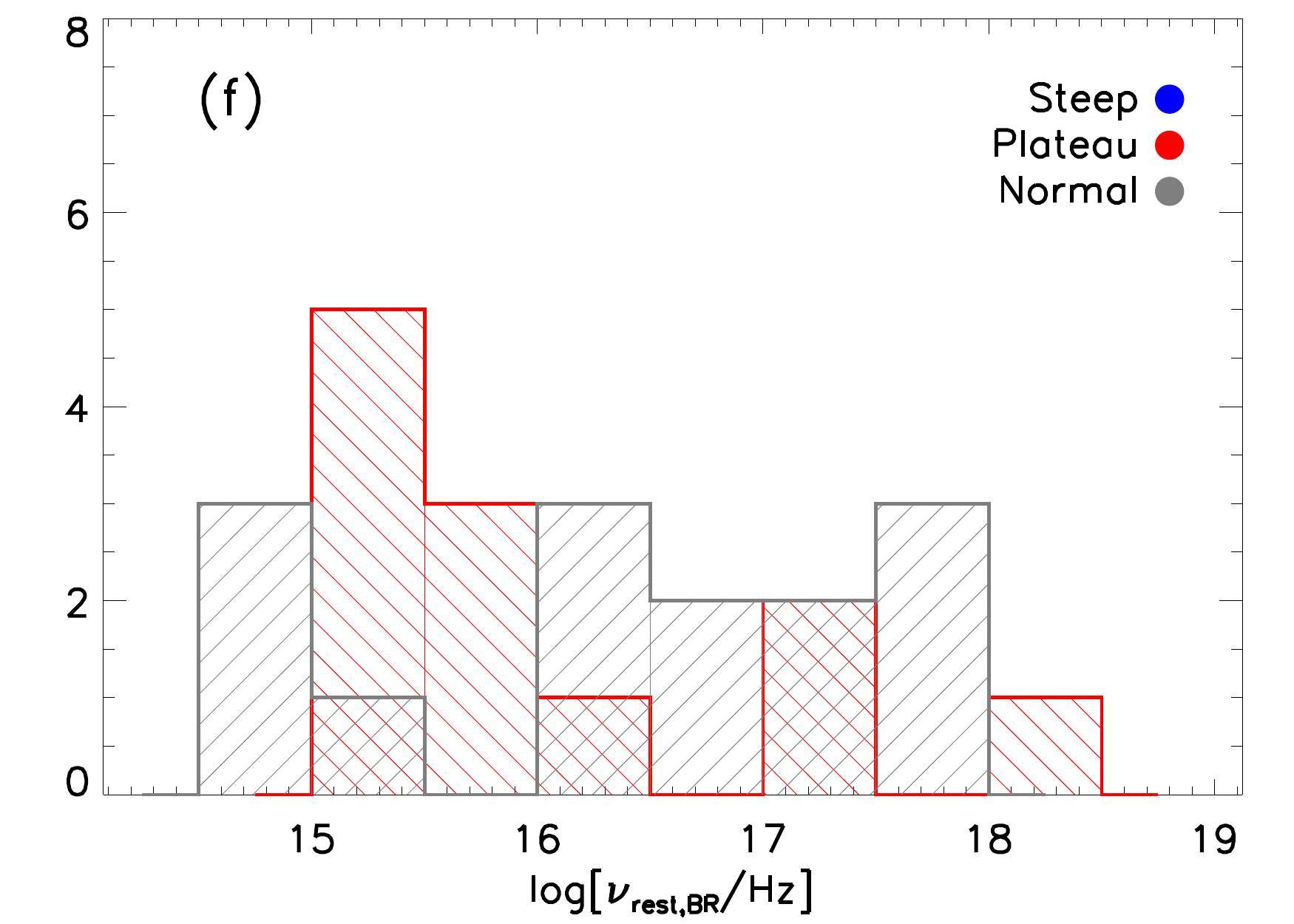}
     \caption{\small{Parameter distributions considering the X-ray LC part of the SED. \textit{Blue lines}: steep-decay phase. \textit{Red lines}: plateau. \textit{Gray lines}: normal decay phase. \textit{a)} $\beta_{\rm op,X}$: the spectral slopes calculated using a power-law as fitting function. \textit{b)} $\beta_{\rm X}$ and \textit{c)} $\beta_{\rm op}$: the broken power law spectral slopes for the X-ray and optical data, respectively.  \textit{d)} $N_{\rm H,PL}$ and \textit{e)} $N_{\rm H,BR}$: the hydrogen column densities obtained using as SED fitting function a single power-law and a broken power-law, respectively. \textit{f)} $\nu_{\rm rest,BR}$ : the rest frame break frequency calculated fitting the SEDs with a broken-power law.}}
     \label{parametri2}
\end{figure*}

For every distribution of the best-fitting values, we calculated the mean ($m$), the standard deviation ($SD$), and the median ($M$). When possible, we fitted the distributions with a Gaussian function obtaining the mean ($\mu$) and the standard deviation ($\sigma$). All results are listed in Table~\ref{tab_par}. In Figure~\ref{parametri1} (\textit{top panels}) we show the parameter distributions differentiating between the data obtained by fitting the SEDs with a single-power law (\textit{red}, PL) or a broken-power law (\textit{blue}, BR) and in Figure~\ref{parametri2} the distributions distiguishing the SED parameters extracted during the X-ray steep-decay phase (\textit{blue}, S), the plateau (\textit{red}, P) or the normal decay phase (\textit{gray}, N) ( see Figure~\ref{cart}). We defined the X-ray LCs shapes as in M13 (Figure~\ref{cart}): 0 if there are no breaks, Ia or Ib if there is a break, IIa or  IIb if there are two breaks, and III if there are three breaks. The differentiation between model Ia and Ib  depends on the smoothness parameter $s < 0$ and $s > 0$. Type IIa is the canonical shape (e.g. \citealt{2006ApJ...642..389N}, \citealt{2006ApJ...642..354Z}), while type IIb starts with a shallow phase followed by a steep decay and then a normal decay.  

We present in the bottom left panel of Figure~\ref{parametri1} the distributions of host $E(B-V)$ for the MW, SMC and LMC for the SEDs for which we were able to differentiate between the extinction laws used.

%----------------
%----------------     
%----------------
%----------------
\begin{table*}
\caption{\small{Characteristic quantities describing the parameter distributions (number of elements (\#), mean ($m$), median ($M$), standard deviation ($SD$)), and best-fitting values from a Gaussian fit (mean ($\mu$), standard deviation ($\sigma$)). The subscript PL indicates the values obtained by fitting the SED with a single power-law, while BR indicates a broken power-law. S stands for steep decay, P for plateau, and N for normal decay. The values of $E(B-V)$ depend on the extinction model law used to fit the SED: Milky Way (MW), Large Magellanic Cloud (LMC) and Small Magellanic Cloud (SMC).}}
\centering
\begin{tabular}{lllllll}
\hline\hline
Name &  \# &$\mu$&$\sigma$&$m$&$SD$& $M$ \\
\hline
  $\beta_{\rm op,X}$  &     65    &        0.95$\pm$0.01     &         0.24$\pm$0.02      &                  1.03      &        0.50     &         0.95\\
  $\beta_{\rm X}$      &      34  &         0.97$\pm$0.02    &          0.18$\pm$0.02     &                   0.99     &         0.22    &          0.95\\
  $\beta_{\rm op}$     &     34    &        0.47$\pm$0.02     &         0.19$\pm$0.02     &                 0.48      &        0.20     &         0.48\\
 $\log(N_{\rm H,BR}/ \rm cm^{-2})$ &     28    &         21.6$\pm$0.11     &         0.51$\pm$0.11     &                 21.7      &        0.57     &          21.7\\
  $\log(N_{\rm H,PL}/ \rm cm^{-2})$ &     43    &         21.7$\pm$0.06     &         0.41$\pm$0.06      &                  21.9      &        0.48     &          21.7\\
 $E(B-V)_{\rm MW}$ &     30  &         0.19$\pm$0.02& 0.14$\pm$0.02&11.1&59.1&0.22\\
$E(B-V)_{\rm LMC}$ &    14   &        -            &     - &         0.20       &    0.16    &       0.18\\
$E(B-V)_{\rm SMC}$  &   25   &       -&-&0.20&0.29&0.10\\  
$A_{\rm V,MW}$        &    30    &       0.56$\pm$0.10&0.42$\pm$0.11     &       34.2      &      182.       &    0.69\\
$A_{\rm V,LMC}$        &    14 &          -&-&     0.63  &   0.51       &    0.57\\
$A_{\rm V,SMC}$         &  25&           -&  -&0.59& 0.87&  0.29\\
$\log(\nu_{\rm rest,BR}/\rm Hz)$&    28     &        -     &      -      &       16.20&     0.98 &   16.1\\
\hline
$\log[(N_{\rm H}/\rm cm^{-2})/(\it A_{\rm{V}}/\rm mag)]_{\rm MW}$&20&21.9$\pm$0.05&0.04$\pm$0.05&22.0&0.56&21.8\\
$\log[(N_{\rm H}/\rm cm^{-2})/(\it  A_{\rm{V}}/\rm mag)]_{\rm LMC}$&12&22.6$\pm$0.08&0.32$\pm$0.11&22.4&0.95&22.6\\
$\log[(N_{\rm H}/\rm cm^{-2})/(\it  A_{\rm{V}}/\rm mag)]_{\rm SMC}$&20&21.8$\pm$0.16&0.87$\pm$0.18&22.4&2.05&22.2\\
%$\log[(N_{\rm H}/\rm cm^{-2})/(\it E(B-V)/\rm mag)]_{\rm MW}$&20&22.4$\pm$0.05&0.04$\pm$0.05&22.4&0.56&22.3\\
%$\log[(N_{\rm H}/\rm cm^{-2})/(\it E(B-V)/\rm mag)]_{\rm LMC}$&12&23.1$\pm$0.08&0.32$\pm$0.11&22.9&0.95&23.1\\
%$\log[(N_{\rm H}/\rm cm^{-2})/(\it E(B-V)/\rm mag)]_{\rm SMC}$&20&22.3$\pm$0.14&0.79$\pm$0.15&22.9&2.05&22.7\\
\hline
 $\beta_{\rm op,X}^{\rm S}$    &      9       &       - & -   &   0.85    &       0.22     &        0.85 \\ 
 $\beta_{\rm op,X}^{\rm P}$    &     24       &      0.92$\pm$0.03         &    0.21$\pm$0.04        &             0.95       &      0.27       &     0.95\\
 $\beta_{\rm op,X}^{\rm N}$    &     31       &      0.99$\pm$0.03         &    0.23$\pm$0.03        &           2.35       &       1.08       &       2.35\\
 $\beta_{\rm X}^{\rm P}$       &     13       &      0.96$\pm$0.02         &   0.10$\pm$0.02        &          0.85       &      0.22       &       0.85\\
 $\beta_{\rm X}^{\rm N}$       &     19       &      1.00$\pm$0.08         &    0.29$\pm$0.09        &        1.15       &      0.27       &       1.15\\
 $\beta_{\rm op}^{\rm P}$     &      13      &       0.49$\pm$0.04        &     0.14$\pm$0.04       &         0.35      &       0.22      &       0.35\\
 $\beta_{\rm op}^{\rm N}$     &      19      &       0.47$\pm$0.08        &     0.29$\pm$0.09       &           0.65      &       0.27      &       0.65\\
 $\log(N_{\rm H,PL}^{\rm S}/\rm cm^{-2})$   &       6      &       -        &      -       &    22.2      &       0.63      &        22.2\\
 $\log(N_{\rm H,PL}^{\rm P}/\rm cm^{-2})$   &    17&22.00$\pm$0.14&0.67$\pm$0.23&22.0&0.52&22.2\\
 $\log(N_{\rm H,PL}^{\rm N}/\rm cm^{-2})$   &     19&21.70$\pm$0.04&0.30$\pm$0.04&22.2&0.86&22.2\\
 $\log(N_{\rm H,BR}^{\rm P}/\rm cm^{-2})$  &        10& 21.00$\pm$1.20&0.93$\pm$0.870&21.8&0.632&21.8\\
 $\log(N_{\rm H,BR}^{\rm N}/\rm cm^{-2})$  &        17&21.90$\pm$0.05&0.23$\pm$0.04& 21.8&0.86&21.8\\
 $\log(\nu_{\rm rest,BR}^{\rm P}/\rm Hz)$ &        14    &  -& -&          16.80    &         1.08    &          16.80\\
 $\log(\nu_{\rm rest,BR}^{\rm N}/\rm Hz)$ &       16     &          -     &       -       &             16.20     &        1.08    &   16.20\\
\hline
$\log(L_{\rm R,500s}/(\rm erg\ s^{-1}))$&64&45.90$\pm$0.06&0.83$\pm$0.06&46.00&3.72&45.90\\
$\log(L_{\rm R,1hr}/(\rm erg\ s^{-1}))$&57&45.40$\pm$0.06&0.73$\pm$0.06&45.90&4.77&45.40\\
$\log(L_{\rm R,11hr}/(\rm erg\ s^{-1}))$&40&44.50$\pm$0.08&0.74$\pm$0.08&45.70&7.44&44.50\\
$\log(L_{\rm R,1day}/(\rm erg\ s^{-1}))$&32&44.20$\pm$0.11&0.83$\pm$0.13&44.20&0.68&44.20\\
\hline
$\log(F_{\rm R,920-1200s}/(\rm erg\ cm^{-2}\ s^{-1}))$ &46&-11.41$\pm$0.07&0.34$\pm$0.07&-11.31&0.80&-11.37\\
$\log(F_{\rm 1keV,920-1200s}/(\rm erg\ cm^{-2}\ s^{-1}))$&52&-12.54$\pm$0.06&0.49$\pm$0.06&-12.40&0.72&-12.45\\
\hline
 \end{tabular}
\label{tab_par}
\end{table*}
%----------------------------------------------------------------

%************************************************************
\subsubsection{Spectral index}
The mean spectral slope computed by fitting a single power-law is $\mu(\beta_{\rm op,X})=0.95\pm0.01$. This value is consistent with the spectral slope $\mu(\beta_{\rm X})=0.97\pm0.02$ obtained using a broken power-law; in fact the fit is largely weighted over the numerous X-ray data. The mean spectral slope of the optical part of the SED is $\mu(\beta_{\rm op})=0.47\pm0.02$, computed by fixing $\beta_{\rm X}-\beta_{\rm op}=0.5$, hence it is simply a rigid shift of the distribution. 

\begin{figure}[!]
\centering
   \includegraphics[width=1. \hsize,clip]{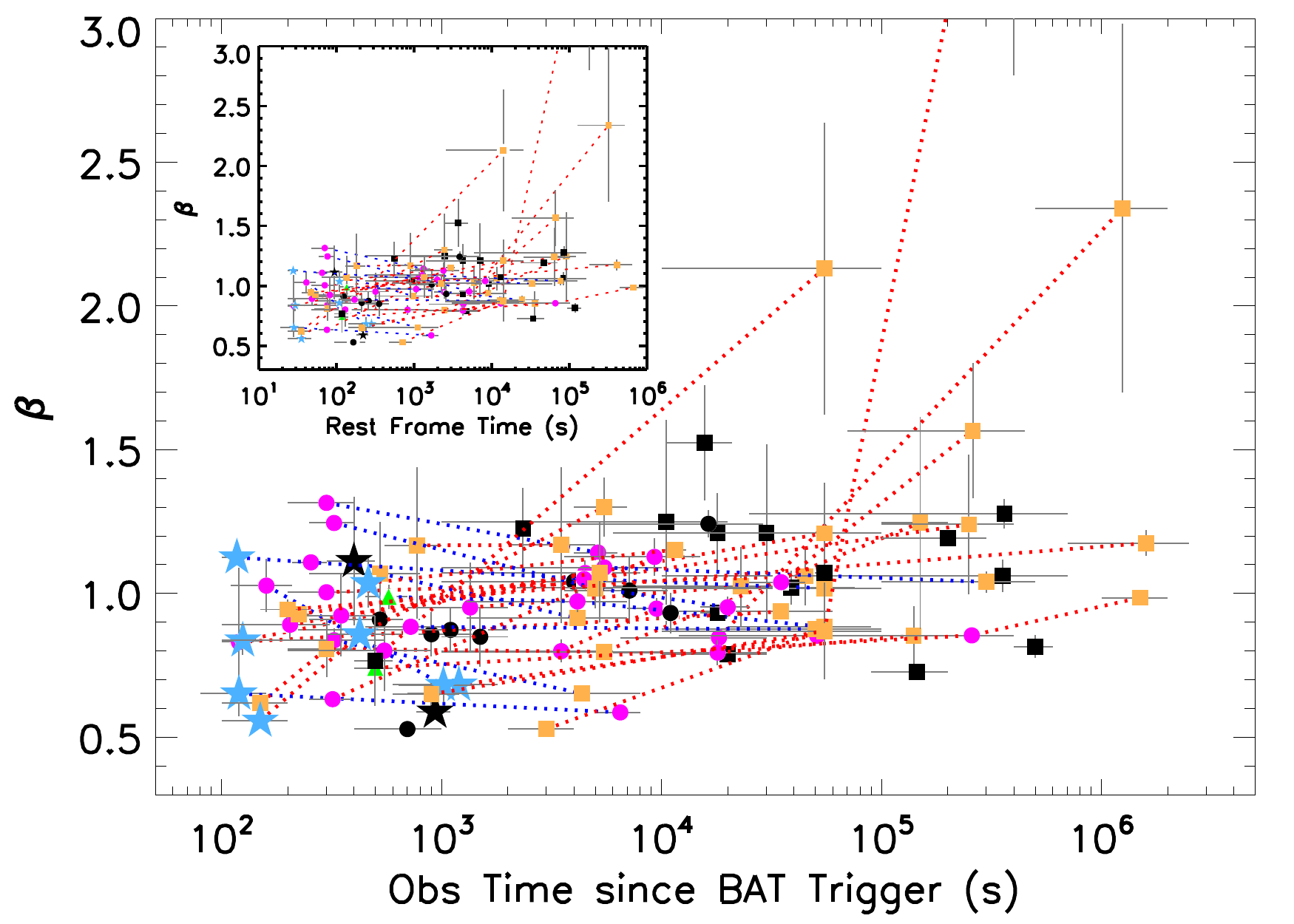}
     \caption{\small{Evolution of $\beta$ with time for individual GRBs. For every GRB we considered the ``correct'' spectral index as selected in Table~\ref{ftest}, hence $\beta$ can be $\beta_{\rm op,X}$ or $\beta_{\rm X}$ depending on the chosen SED fitting function, a single power-law or a broken power-law. \textit{Blue dotted lines}: the initial spectral slope is steeper than the final spectral slope. \textit{Red dotted lines}: the initial spectral slope is flatter than the final spectral slope. \textit{Light blue stars}: steep decay data. \textit{Magenta dots}: plateau data. \textit{Orange squares}: normal decay data. \textit{Black}: only one SED is available for these GRBs and precisely during the steep decay (\textit{stars}), the plateau (\textit{dots}), and normal decay (\textit{squares}). \textit{Inset}: the same as the principal plot, but in the rest frame.}} 
    \label{beta_evoluzione}
\end{figure}

The distributions of $\beta$ computed over the three different parts of the X-ray LCs (steep decay phase, plateau, normal decay phase) have the following mean values (Figure~\ref{parametri2}, \textit{a, b, c}): \textit{a)} $m(\beta_{\rm op,X}^{\rm S})=0.85$ (with $SD=0.22$\footnote{There are too few data to fit a Gaussian over the distribution.\label{fn:nodata}}), $\mu(\beta_{\rm op,X}^{\rm P})=0.92\pm0.03$, $\mu(\beta_{\rm op,X}^{\rm N})=0.99\pm0.03$. \textit{b)} $\mu(\beta_{\rm X}^{\rm P})=0.96\pm0.02$, $\mu(\beta_{\rm X}^{\rm N})=1.00\pm0.08$. \textit{c)} $\mu(\beta_{\rm op}^{\rm P})=0.49\pm0.04$, $\mu(\beta_{\rm op}^{\rm N})=0.47\pm0.08$. From these distributions we note that the mean spectral index during the plateau is lower than during the normal decay phase, even though they are  consistent within 2$\sigma$; in addition, the normal decay spectral index distribution is broader than during the plateau. Therefore we tested the evolution of $\beta$ for each GRB (Figure~\ref{beta_evoluzione}), with $\beta=\beta_{\rm op,X}$ or $\beta=\beta_{\rm X}$ depending of the fitting function used for each single SED (Table~\ref{ftest}, \textit{Online Material}). In most cases the spectrum becomes softer (22 GRBs, \textit{red lines}), and only for ten GRBs it becomes harder (\textit{blue lines}). For 26 GRBs we have only one valid SED fit (\textit{black dots}). If we examine these relationships in the rest frame (\textit{inset}), in particular only the plateau and normal decay data (\textit{magenta dots} and \textit{orange squares}) because we have only few data for the steep-decay phase (\textit{light blue stars}) and the unclassified phase (\textit{green triangles}), then the relationships do not change.

%----------------
%----------------

%************************************************************
\subsubsection{Hydrogen column density and optical extinction}\label{NHEBV} 

The intrinsic hydrogen equivalent column density determines the X-ray absorption and measures the quantity of gas contained in the GRB host galaxy. The origin of  this absorption is still debated, but is most likely due to absorption by the intergalactic medium, intervening absorbers or He in the HII region hosting the GRB (e.g. M13, \citealt{2010MNRAS.402.2429C,2012MNRAS.421.1697C,2011ApJ...734...26B,2012arXiv1212.4492W}).

\begin{figure}[!]
   \includegraphics[width=1. \hsize,clip]{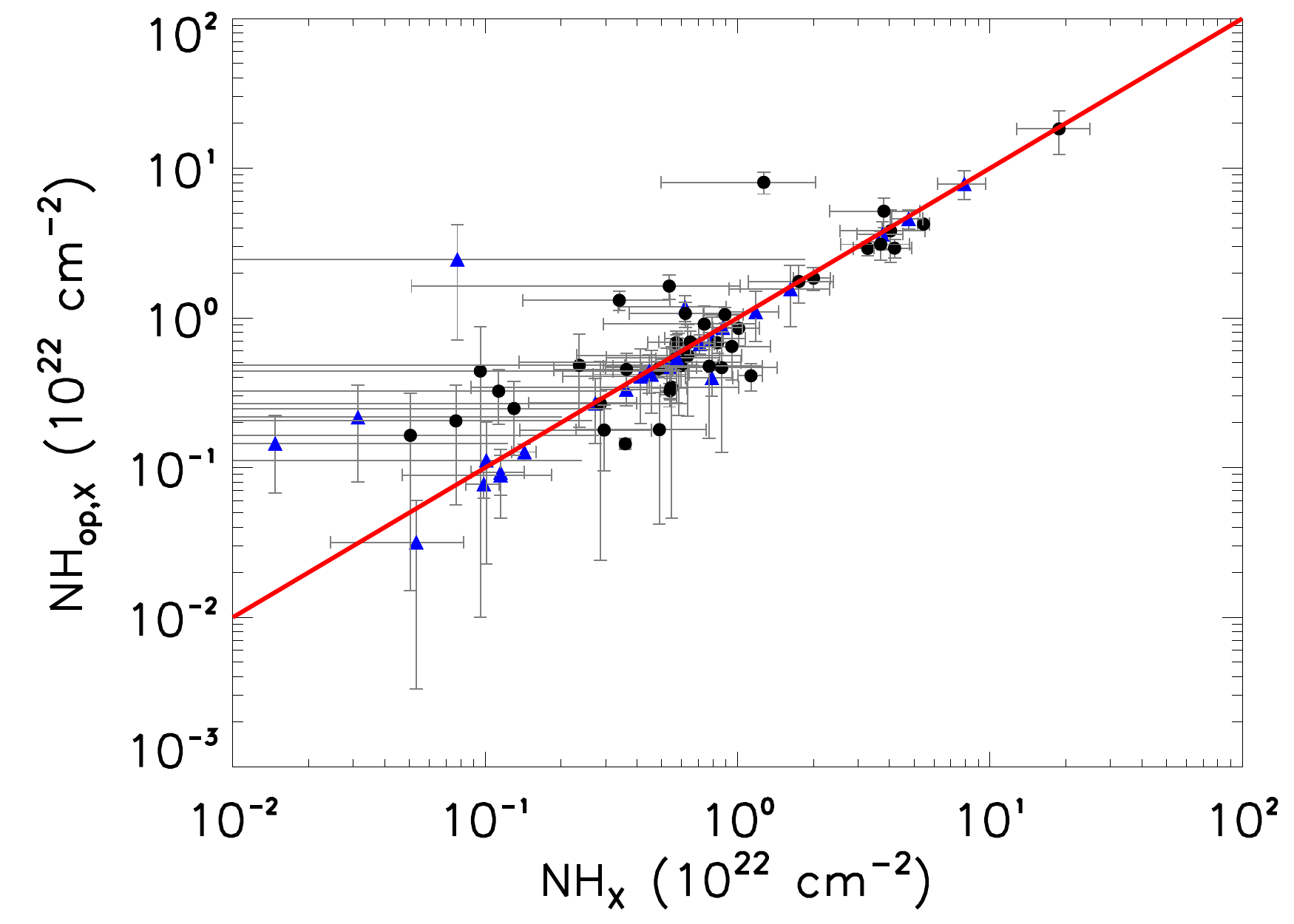}
     \caption{\small{Comparison between the $N_{\rm H}$ calculated from the X-ray spectrum ($N_{\rm H,X}$) and the optical/X-ray SED ($N_{\rm H,op,X}$). \textit{Blue triangles} stand for the broken power-law fit function and \textit{black dots} for the simple power-law. \textit{Red line}: $N_{\rm H,X}=N_{\rm H,op,X}$.}}
     	\label{nhcomp}
\end{figure}

We calculated the intrinsic hydrogen equivalent column density after subtracting of the MW contribution, both by fitting the X-ray spectrum alone (N$_{\rm H,X}$) and by a joined fit of optical and X-ray SED ($N_{\rm H,op,X}$). $N_{\rm H,op,X}$ was computed following the model presented by \cite{1983ApJ...270..119M}, which takes into account the photoelectric cross section per HI-atom units and for solar metallicity (Sec.~\ref{SED}). The $N_{\rm H}$ values found with the two methods are consistent, as shown in Figure~\ref{nhcomp}, even the low values of $N_{\rm H}$ ($<10^{21}$ cm$^{-2}$) are consistent within two sigma.

We therefore restricted our analysis to the intrinsic hydrogen equivalent column densities derived by the optical/X-ray SEDs ($N_{\rm H,op,X}\equiv N_{\rm H}$). The distributions of the $\mathbf{N}_{\bf{\rm H,op,X}}$ of the host galaxies derived from the single and broken power-law fits are consistent: $\mu(\log(N_{\rm H,PL}/\rm cm^{-2}))=21.60\pm0.11$ and $\mu(\log(N_{\rm H,BR}/\rm cm^{-2}))=21.70\pm0.06$ (Figure~\ref{parametri1}\textit{b}). As shown by the distributions of $N_{\rm H}$ calculated in the different parts of the LCs (Figure~\ref{parametri2}\textit{d,e}), this parameter does not evolve with time because it has a similar mean value, within error, for the steep decay phase, the plateau and, normal decay phase. The values we found for the $N_{\rm H}$ are consistent with those of M13 within 1$\sigma$.

In Figure~\ref{parametri1}\textit{c} we show the reddening distributions ($E(B-V)$), differentiating between the best-fitting extinction laws (MW, LMC, SMC). The mean values are $\mu(E(B-V)_{\rm MW})=0.19\pm0.02$ mag, $m(E(B-V)_{\rm LMC})=0.20$ mag ($SD$=0.16), $m(E(B-V)_{\rm SMC})=0.20$ mag, which corresponds to the host galaxy visual extinction ($A_{\rm V}$) $\mu(A_{\rm V,MW})=0.56\pm0.10$ mag, $m(A_{\rm V,LMC})=0.63$ mag, $m(A_{\rm V,SMC})=0.59$ mag\footnote{$A_{\rm V}=R_{\rm V}\ E(B-V)$ with $R_{\rm V}^{\rm MW}=3.08$, $R_{\rm V}^{\rm LMC}=3.16$ and $R_{\rm V}^{\rm SMC}=2.93$ \citep{1992ApJ...395..130P}.}. The mean $A_{\rm V,SMC}$ is agrees with the value presented by \citet{2011A&A...532A.143Z}. 

\begin{figure}[!]
    \includegraphics[width=1. \hsize,clip]{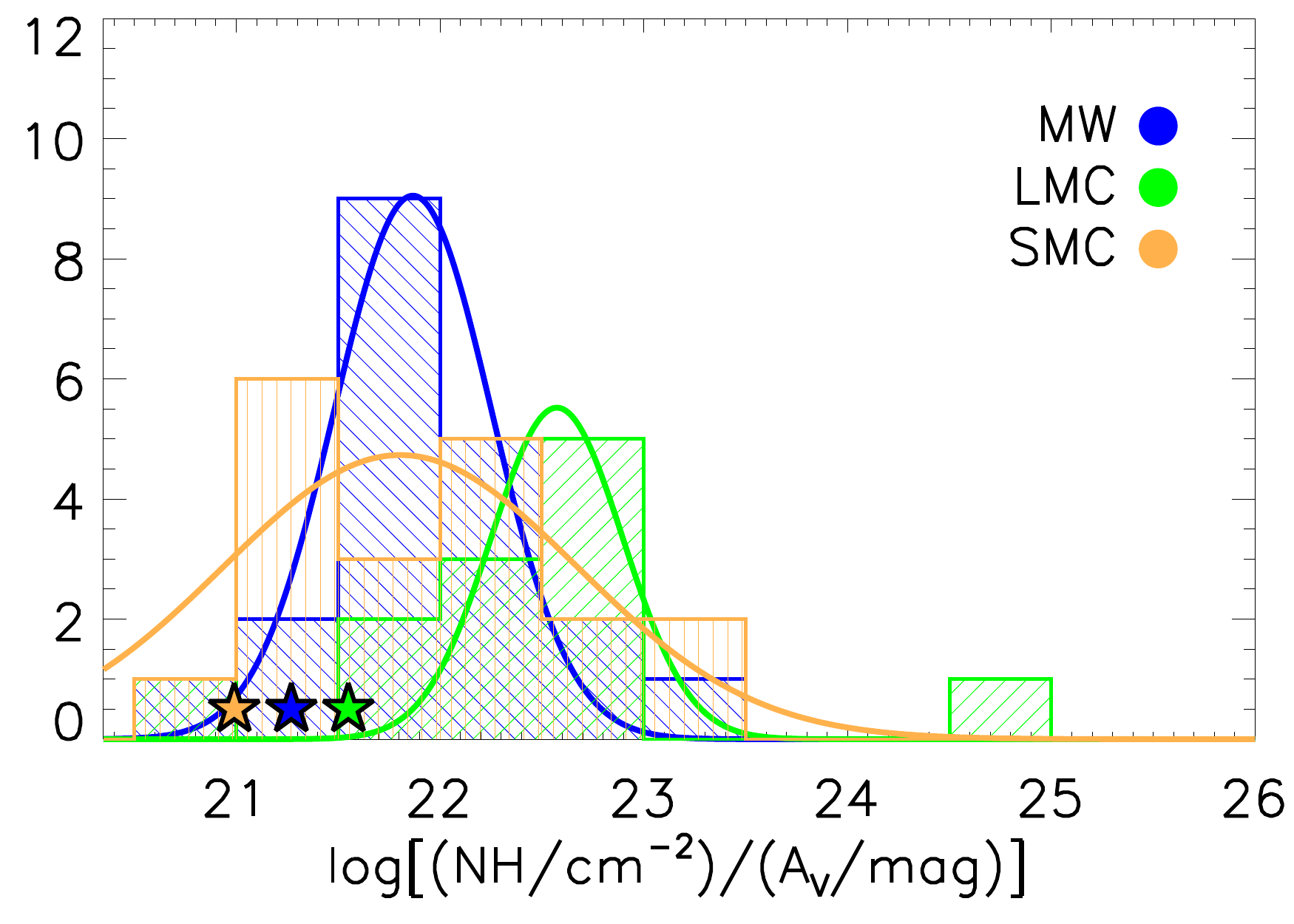}
     \caption{\small{Distribution of $\log((N_{\rm H}/\rm cm^{-2})/(A_{\rm{V}}/\rm mag))$ considering the three different extinction laws used: MW (\textit{blue}), LMC (\textit{green}), and SMC (\textit{orange}). \textit{Stars}: reference values of the ratios $N_{\rm H}/A_{\rm V}$ from the literature. \textit{Blue star}: $\log((N_{\rm H}/\rm cm^{-2})/(\it A_{\rm V}/\rm mag))_{\rm MW}= 21.27$ \citep{1978ApJ...224..132B}. \textit{Green star}: $\log((N_{\rm H}/\rm cm^{-2})/(\it A_{\rm V})/\rm mag))_{\rm LMC}=21.55$ (assuming sub solar abundances). \textit{Orange star}: $\log((N_{\rm H}/\rm cm^{-2})/(\it A_{\rm V}/\rm mag))_{\rm SMC}=20.99$ (assuming sub solar abundances).}}
     	\label{nhcomp2}
\end{figure}

We studied the properties of the GRB host galaxy environment through the gas-to-dust ratio ($N_{\rm H}/A_{\rm V}$, Figure~\ref{nhcomp2}). We considered $\bf{N_{\rm H,op,X}}$ and obtained for different extinction laws $\mu(\log(N_{\rm H}/\rm cm^{-2})/(\it A_{\rm{V}}/\rm mag))_{\rm MW})=21.90\pm0.05$ (\textit{blue}), $\mu(\log((N_{\rm H}/\rm cm^{-2})$ $/(\it A_{\rm V}/\rm mag))_{\rm LMC})=22.60\pm0.08$ (\textit{green}) and $\mu(\log((N_{\rm H}/\rm cm^{-2})/(\it A_{\rm V}/\rm mag))_{\rm SMC})=21.80\pm0.16$ (\textit{orange}). We compared these results with the $N_{\rm H}/\it A_{\rm V}$ values available in the literature for the MW, LMC and SMC: $\log((N_{\rm H}/\rm cm^{-2})/(\it A_{\rm V}/\rm mag))_{\rm MW}= 21.27$ (Figure~\ref{nhcomp2}, \textit{blue star}; \citealt{1978ApJ...224..132B}), $\log((N_{\rm H}/\rm cm^{-2})/(\it A_{\rm V}/\rm mag))_{\rm LMC}=21.84$ \citep{2001ApJ...548..296W} and $\log((N_{\rm H}/\rm cm^{-2})/(A_{\rm V}/\rm mag))_{\rm SMC}=22.19$ \citep{1989A&A...215..219M}. To compare the Magellanic Clouds data of the $N_{\rm H}$ from the literature with our results, calculated assuming solar abundances, we converted the values from the literature assuming a metallicity $Z=0.26\ Z_{\odot}$ for the LMC and $Z=0.14\ Z_{\odot}$ for the SMC (\citealt{2003ARA&A..41..241D} and references therein). We obtained $\log((N_{\rm H}/\rm cm^{-2})/(\it A_{\rm V}/\rm mag))_{\rm LMC}=21.55$ and $\log((N_{\rm H}/\rm cm^{-2})/(\it A_{\rm V}/\rm mag))_{\rm SMC}=20.99$ (Figure~\ref{nhcomp2}, \textit{green} and \textit{orange stars}, respectively).

Our analysis shows that the gas-to-dust ratios of GRBs are higher than the values calculated for the MW, the LMC, and SMC assuming sub solar abundances (e.g. \citealt{2010MNRAS.401.2773S,2012A&A...537A..15S}). We caution, however, that our distributions characterize GRBs that are not heavily absorbed in the X-rays and in the optical band, because our sample is redshift-selected.

%----------------
%----------------
%----------------
%************************************************************
\subsubsection{Break frequency} 

The rest frame break frequency distribution has a peak around $\log(\nu_{\rm rest,BR}/\rm Hz)\sim16$ (Figure~\ref{parametri1}\textit{d}). The values are spread between the optical and the X-ray band frequencies and it is not possible to fit a Gaussian function to these data. 

\begin{figure*}[!]
\centering         
            \includegraphics[width=0.45 \hsize,clip]{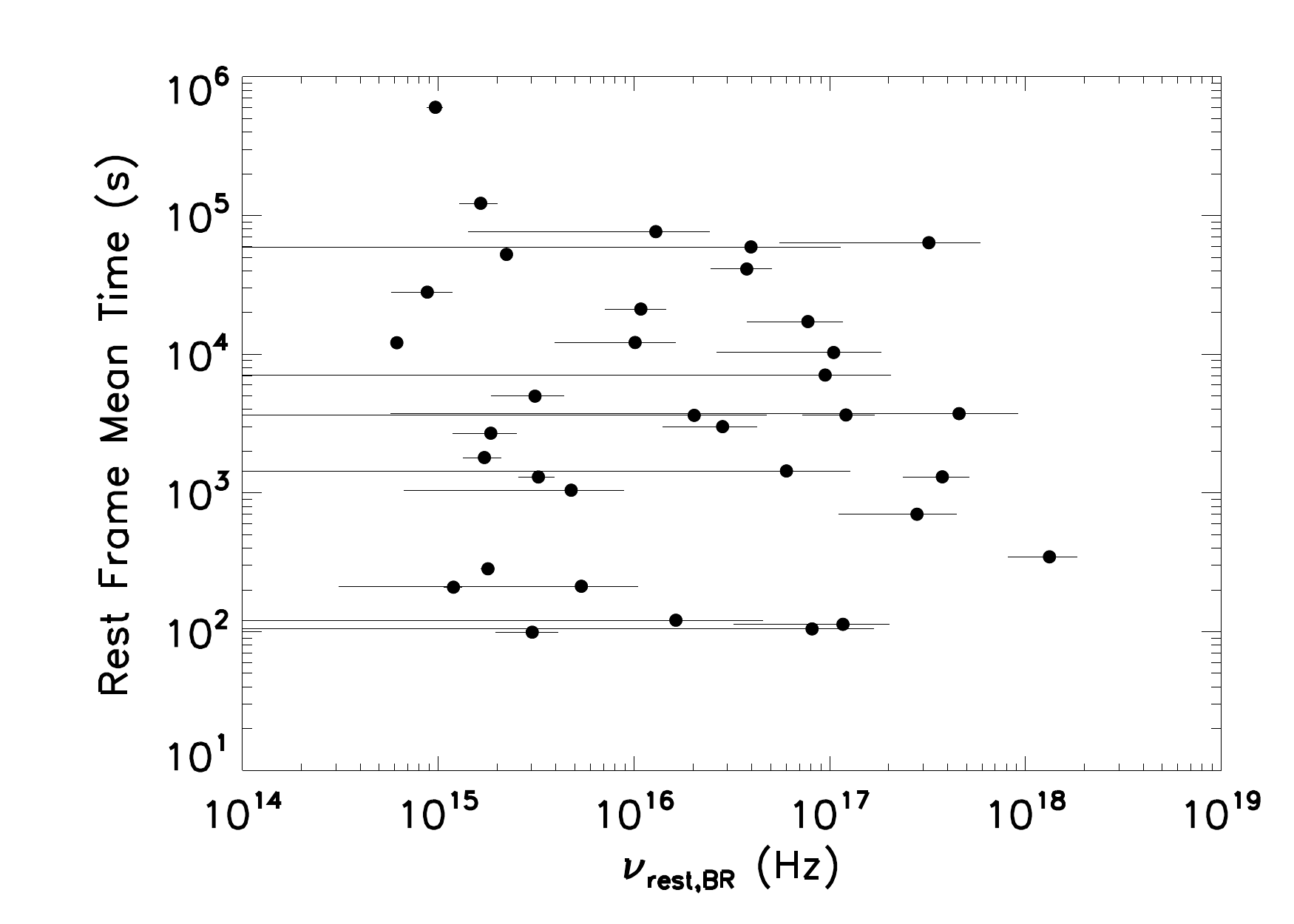}
    \includegraphics[width=0.45 \hsize,clip]{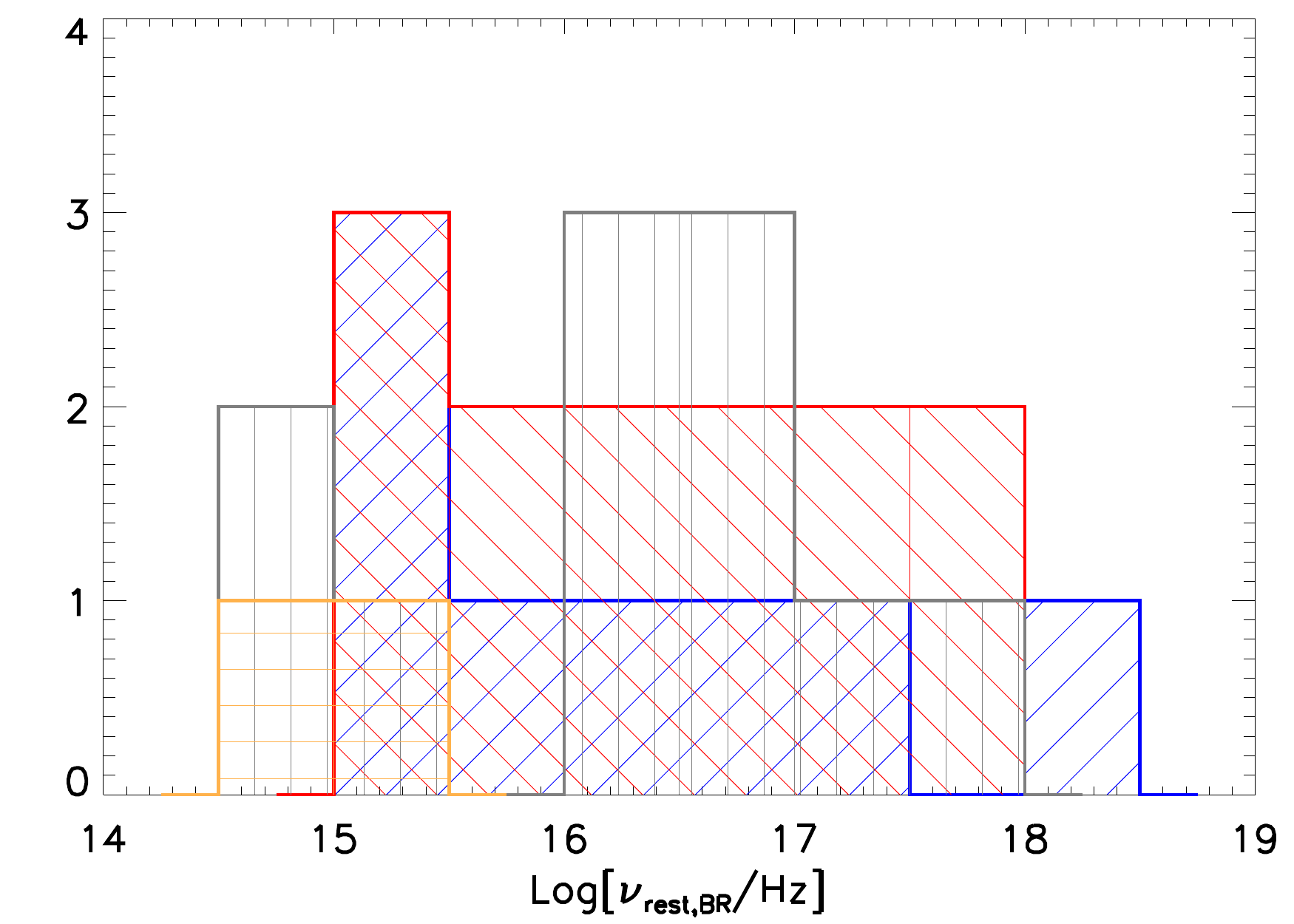}
     \caption{\small{Break frequency. \textit{Left}: break frequency ($\nu_{\rm rest,BR}$) vs. the mean time ($t_{\rm rest,m}$) of the interval in which the SED is calculated. \textit{Right}: the distribution of the break frequencies. \textit{Blue}: $t_{\rm rest,m}<500$ s. \textit{Red}: $500<t_{\rm rest,m}<10^4$ s. \textit{Gray}: $10^4<t_{\rm rest,m}<10^5$ s. \textit{Orange}: $t_{\rm rest,m}>10^5$ s. The time intervals have been arbitrarily chosen.}}
     \label{parametri4}
\end{figure*}

Because most of our data where taken at late times, they probably correspond to a slow cooling regime for a homogeneous medium, when the break-frequency evolves as $t^{-1/2}$ \citep{1998ApJ...497L..17S}, moving from the X-ray toward the optical frequencies. Since we cannot follow these changes for a single burst, we tested whether we could find any correlation between the mean time at which we measure the break frequency and the break frequency itself. If GRBs had a similar behavior, the time and the break frequency would be correlated. Figure~\ref{parametri4} shows that the peak at low frequencies is spread over a long time interval, and there is no evidence of a correlation between time and frequency (Figure~\ref{parametri4}, \textit{left}). This may be due above all to the dependence of the the break frequency on other parameters, in part to the uncertainties in its measurement, and also because we considered data from different GRBs. 
%----------------
%----------------
%************************************************************
\subsection{Luminosity and energetics}

\begin{figure}[!]
\centering
  \resizebox{\hsize}{!}{\includegraphics{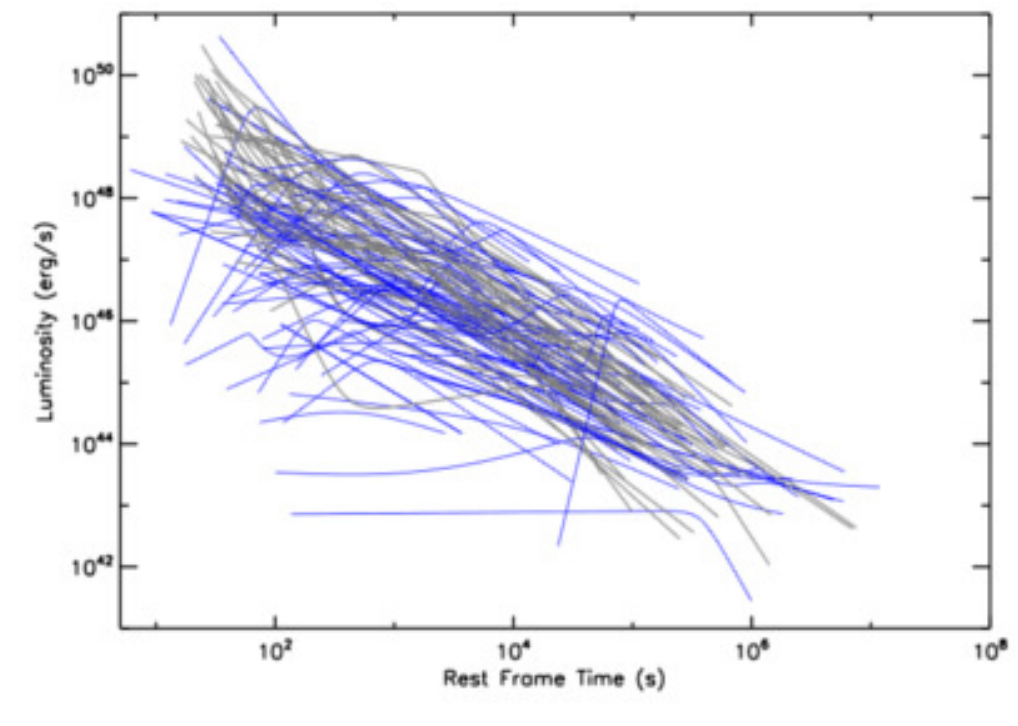}}
     \caption{\small{X-ray (1 keV, \textit{gray}) and optical ($R$ band, \textit{blue}) LCs in the rest frame.}}
      \label{op_X_lum}
\end{figure}

In Figure~\ref{op_X_lum} we plot the X-ray (1 keV, \textit{gray lines}) and optical ($R$ band) \textit{blue lines}) rest-frame LCs of the GRBs in our sample\footnote{The X-ray and optical data are k-corrected. Optical data are not corrected for Galactic and host galaxy absorption; X-ray data are corrected for Galactic and host galaxy absorption. Therefore, the optical luminosity derived is a lower limit of the real value. However, since the GRBs considered have smaller absorption (see Section~\ref{NHEBV}) the real luminosity is about a factor 2 higher than the one considered.}. The optical and X-ray LCs have similar luminosities; our redshift-selected sample favors bright optical GRBs.

\begin{figure}[!]
\centering
\resizebox{\hsize}{!}{\includegraphics{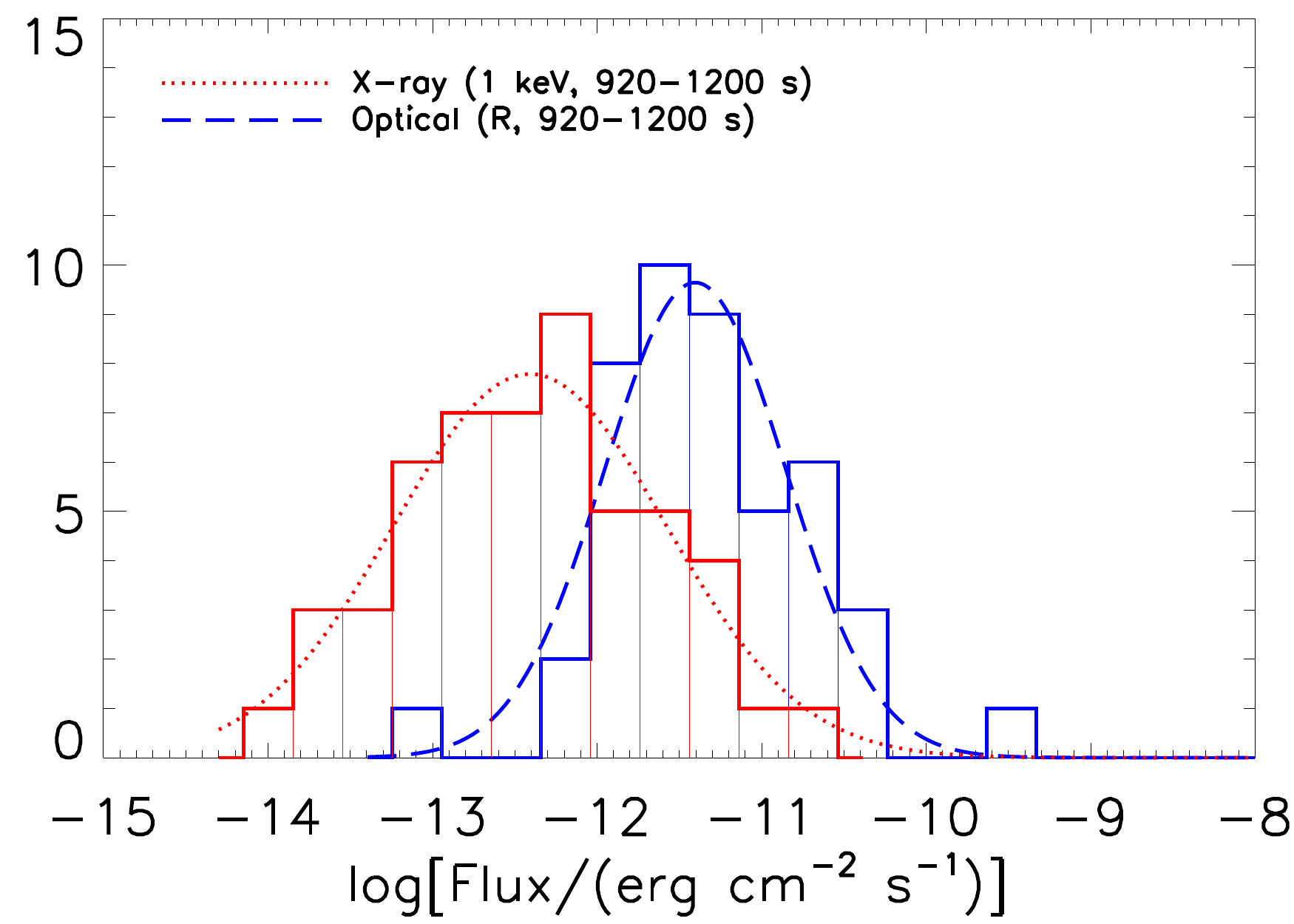}}
\caption{\small{Energetics. Distribution of the X-ray (1 keV, \textit{red, solid line}) and optical (R band, \textit{blue, solid line}) flux calculated in a common rest frame time interval (920-1200 s) for our sample and their distributions (\textit{red, dotted line} and \textit{blue, dashed line}, respectively).}}
\label{energia}
\end{figure}

For the GRBs in our sample, we compared the optical ($R$ band) and X-ray (at 1 keV) flux (Figure~\ref{energia}) in a common rest frame time interval (920-1200 s): the X-ray emission ($\log(\mu(F_{\rm X})/(\rm erg\ cm^{-2}\ s^{-1}))=-12.54$, $\sigma=0.49$) is on average one order of magnitude fainter than the optical ($\log(\mu(F_{\rm op})/(\rm erg\ cm^{-2}\ s^{-1}))=-11.41$, $\sigma=0.34$).
%The X-ray energy integrated over the 0.3-30 keV energy band (M13) is two orders of magnitude greater than the optical energy, because of the amplitude of the energy band. Then, we computed the energy  (Figure~\ref{energia}, Table~\ref{tab_par}) in a common rest frame time interval, 920-1200 s, for the 1 keV X-ray energy band\footnote{To compare homogenous quantities, we calculated the X-ray energy emitted in a band large as the optical $R$ band.} (\textit{red dotted line}) and the $R$ optical band\footref{fn:rband} (\textit{blue dashed line}): the X-ray energy is one order less than the optical energy. These energies increase of one order if we do not consider a common time interval, but the total time interval of the observations.

%----------------
%----------------
%
%\begin{figure}
%\centering
% \resizebox{\hsize}{!}{\includegraphics{FIGURE/energy_OK-eps-converted-to.pdf}}
%\caption{\small{Energetics. Distribution of the X-ray total energy (1 keV) for the sample presented in M13 (\textit{orange, solid line}), of the X-ray energy (1 keV) calculated in a common rest frame time interval (920-1200 s) for our sample (\textit{red, dotted line}), of the X-ray total energy (0.3-30 keV) for the sample presented in M13 (\textit{gray, dash-triple-dotted line}), and of the X-ray energy (0.3-30 keV) calculated in a common rest frame time interval (920-1200 s) for our sample (\textit{black, dash-dotted line}). Distribution of the total optical energy in a common rest frame band ($R$) (\textit{light blue, dashed line}) and calculated in a common rest frame time interval (\textit{blue, long-dashed line}).}}
%\label{energia}
%\end{figure}
%----------------

The optical LCs can show an early-time rise or a quasi-constant phase (optical plateau), followed by a decay. \citet{2011MNRAS.414.3537P} claimed that there are some correlations involving the energies and luminosities calculated at the peak of the early-time rise or at the end of the optical plateau, which are predicted by theoretical models. We verified these relations in the observer and the rest frames considering the GRBs with an optical LC with an initial peak and with an initial optical plateau (Table~\ref{tab_divisione_ottico}).

\begin{figure}[!]
\centering
\includegraphics[width=0.90 \hsize,clip]{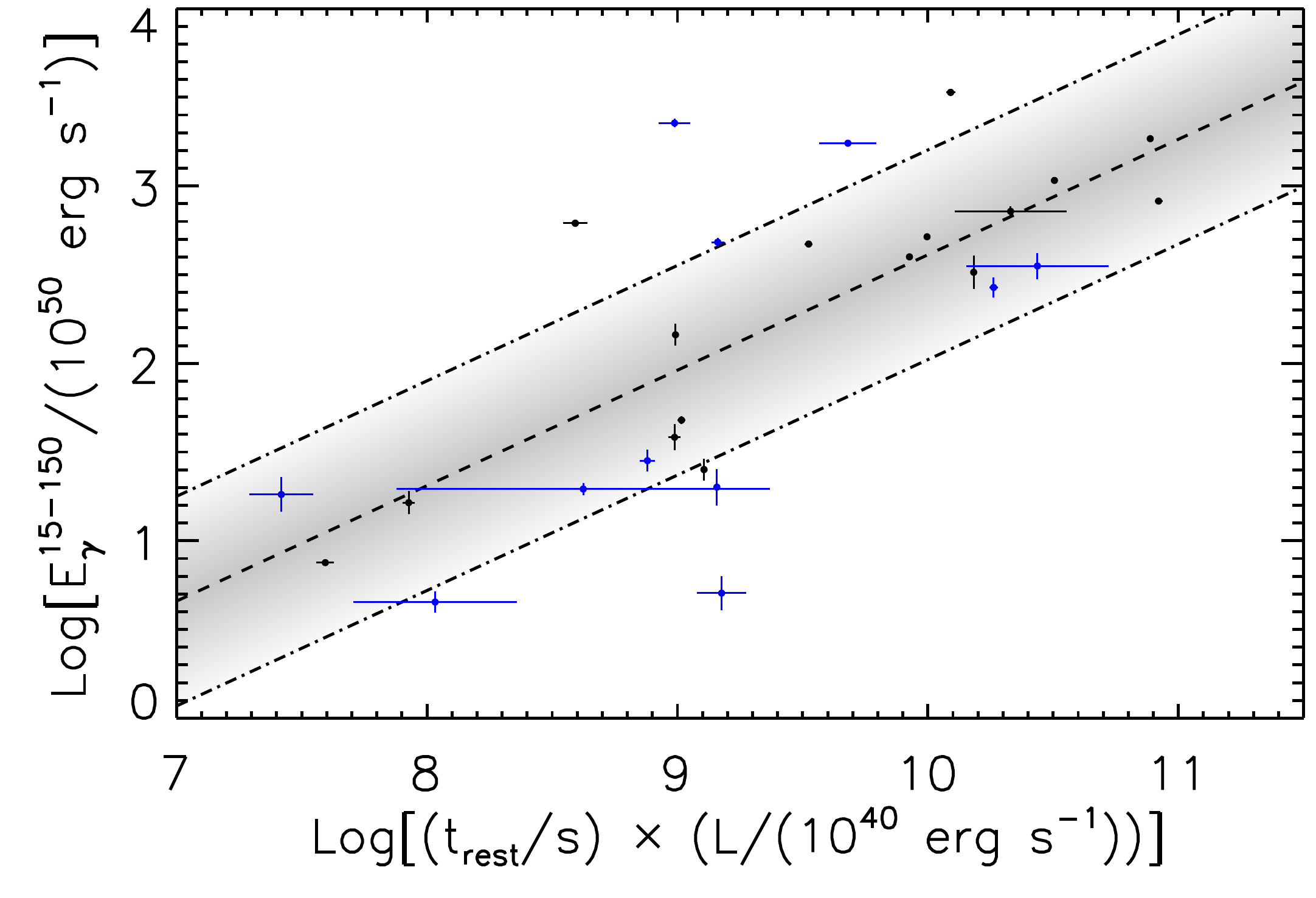}\\
\includegraphics[width=0.90 \hsize,clip]{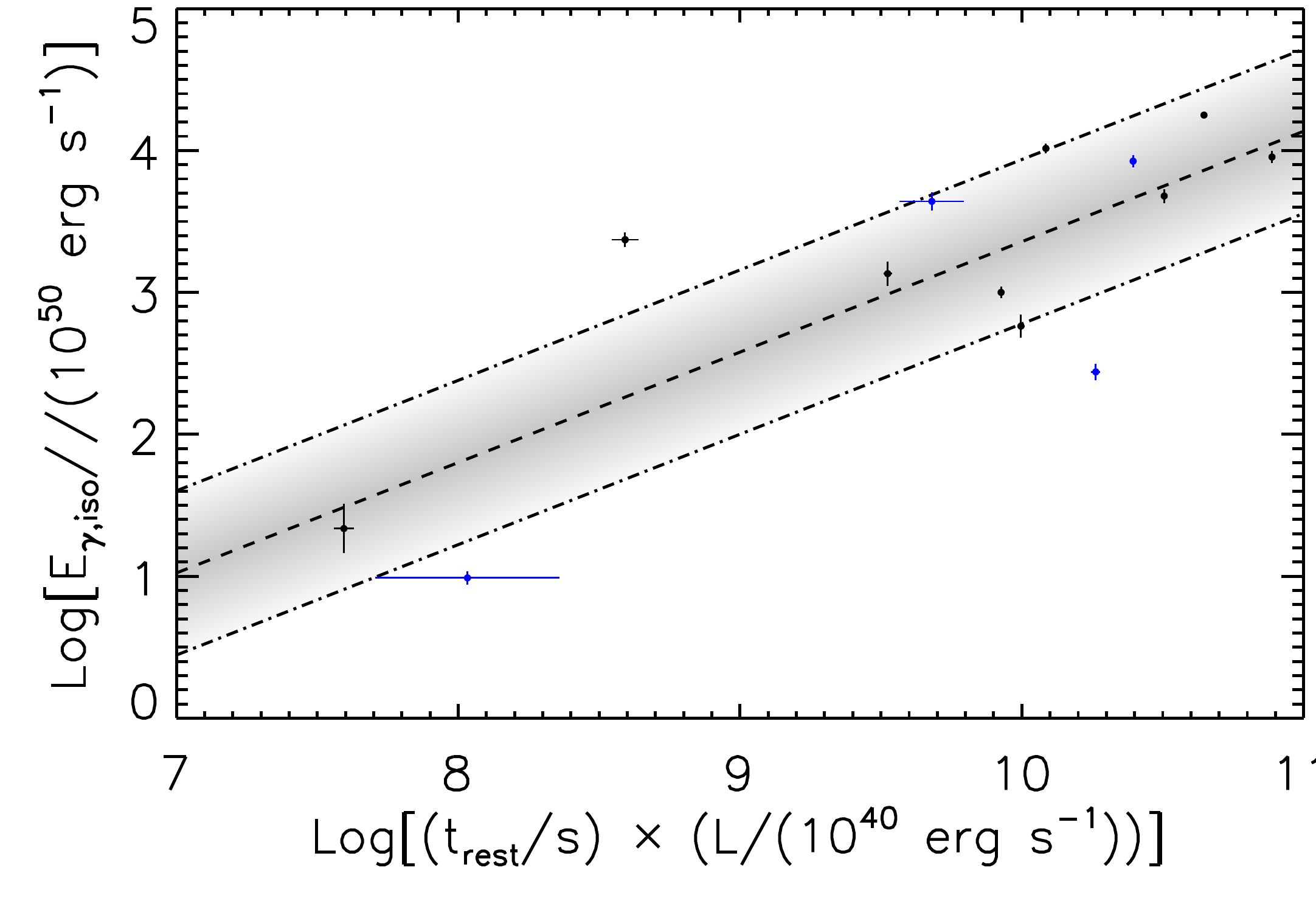}
\caption{\small{Relations between the optical energy (time$\times$luminosity) of the optical plateau end (\textit{blue dots}) and of the peak (\textit{black dots}) and the 15-150 keV BAT energy (\textit{top}) and the isotropic prompt emission energy (\textit{bottom}). \textit{Dashed line}: best-fitting power-law model obtained accounting for the sample variance \citep{2005physics..11182D}. \textit{Gray area}: the 68\% confidence region around the best fit. The results of the fit are listed in Table~\ref{opt_corr}.}}
\label{CORR3}
\end{figure}

\begin{figure*}[!]
\centering
\includegraphics[width=0.45 \hsize,clip]{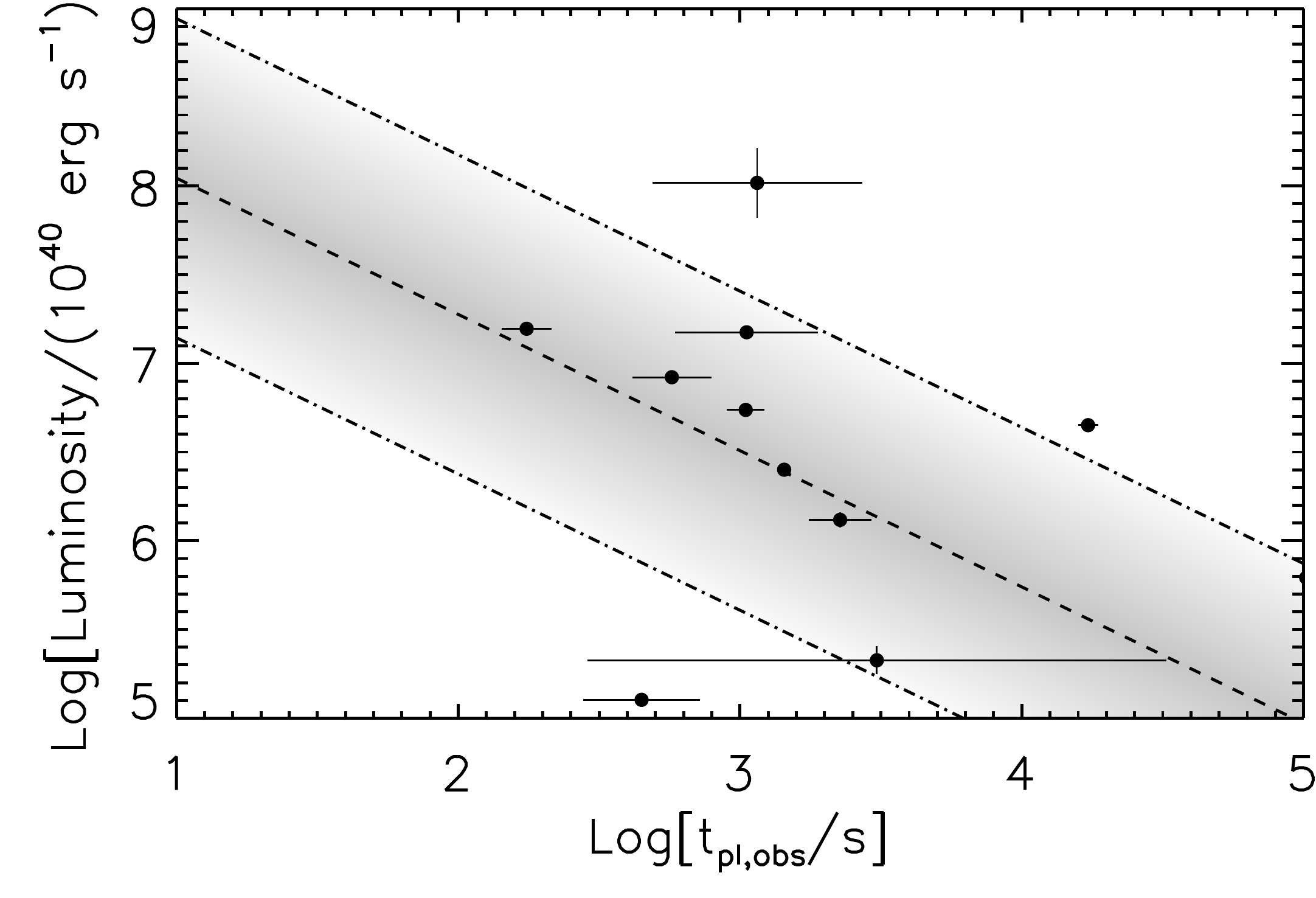}
\includegraphics[width=0.45 \hsize,clip]{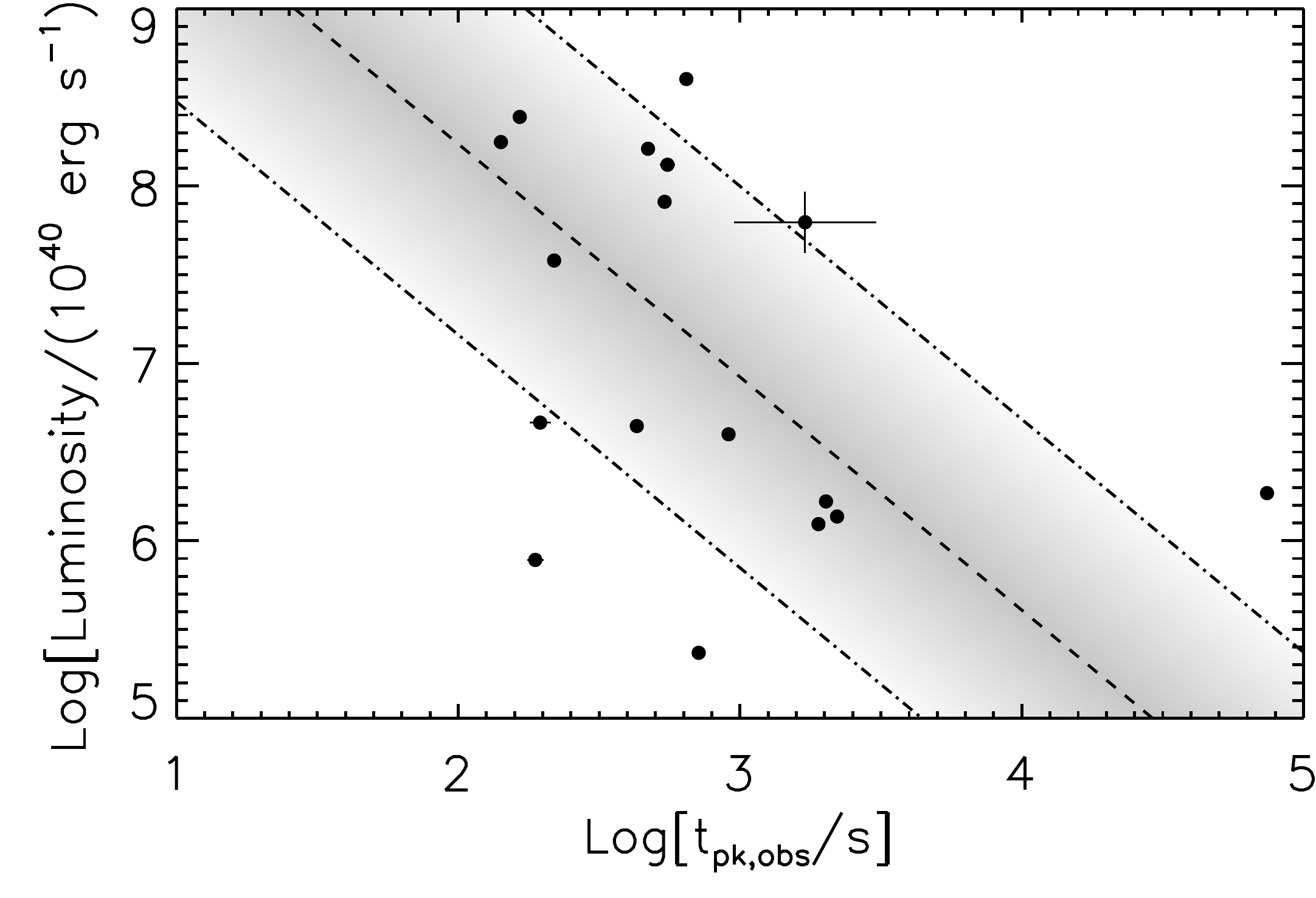}\\
\includegraphics[width=0.45 \hsize,clip]{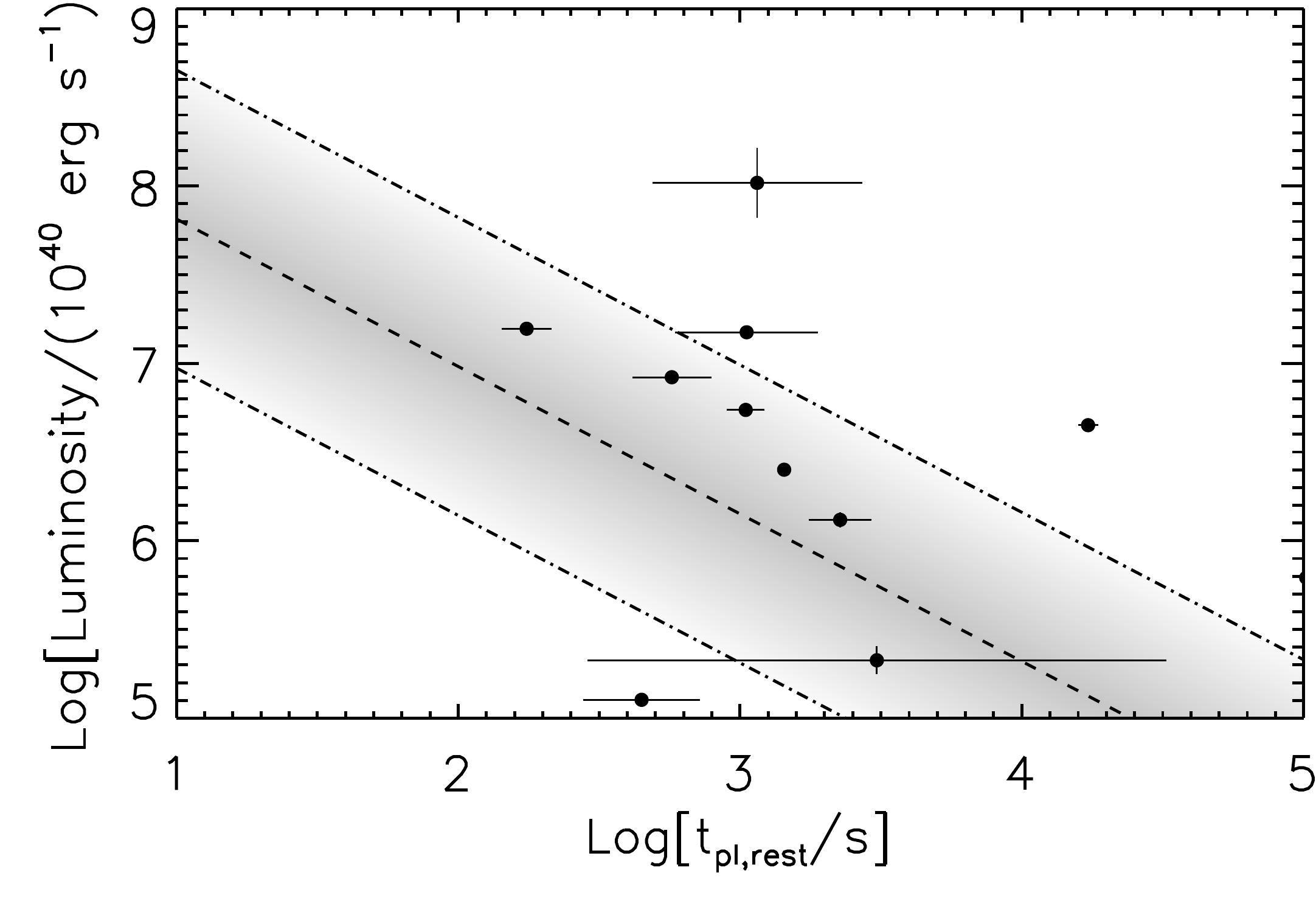}
\includegraphics[width=0.45 \hsize,clip]{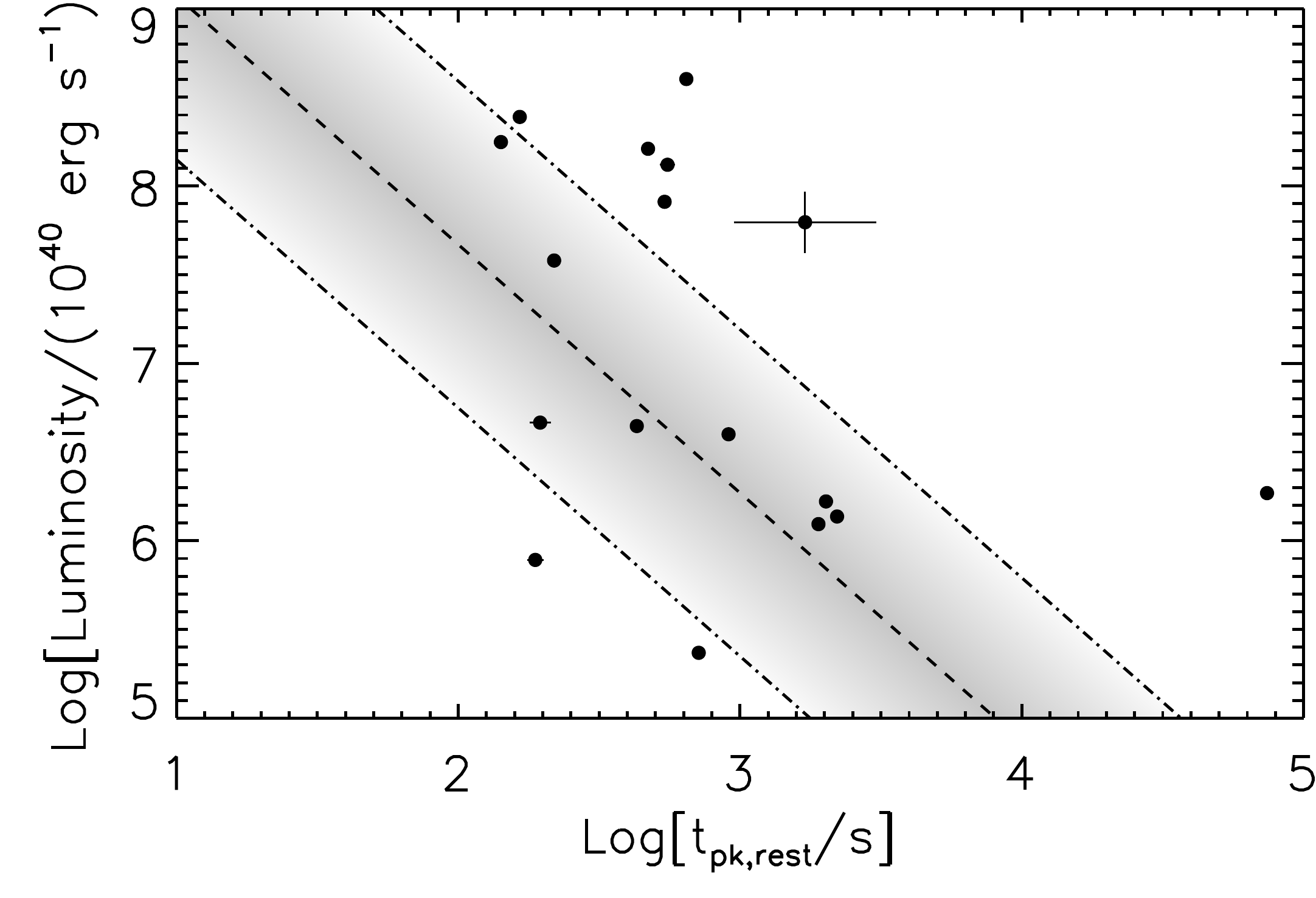}
\caption{\small{Relations between the optical luminosity of the end of the plateau (\textit{left}) and of the peak (\textit{right}) and the relative observer (\textit{top}) and rest (\textit{bottom}) frame time. \textit{Dashed line}: best-fitting power-law model obtained accounting for the sample variance \citep{2005physics..11182D}. \textit{Gray area}: the 68\% confidence region around the best fit. The results of the fit are listed in Table~\ref{opt_corr}.}}
\label{CORR2}
\end{figure*}

\begin{figure}[!]
\centering
\includegraphics[width=0.90 \hsize,clip]{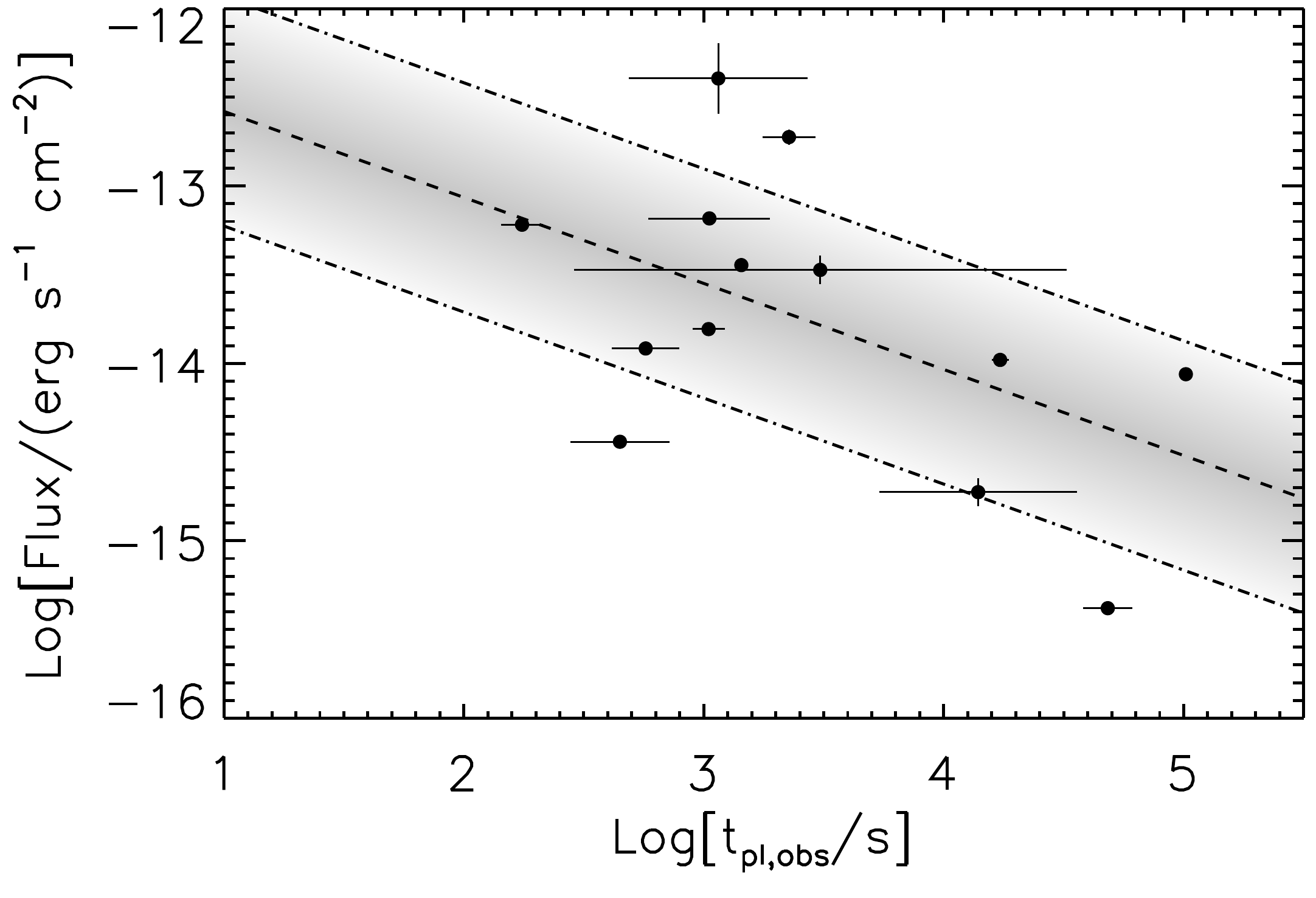}\\
\includegraphics[width=0.90 \hsize,clip]{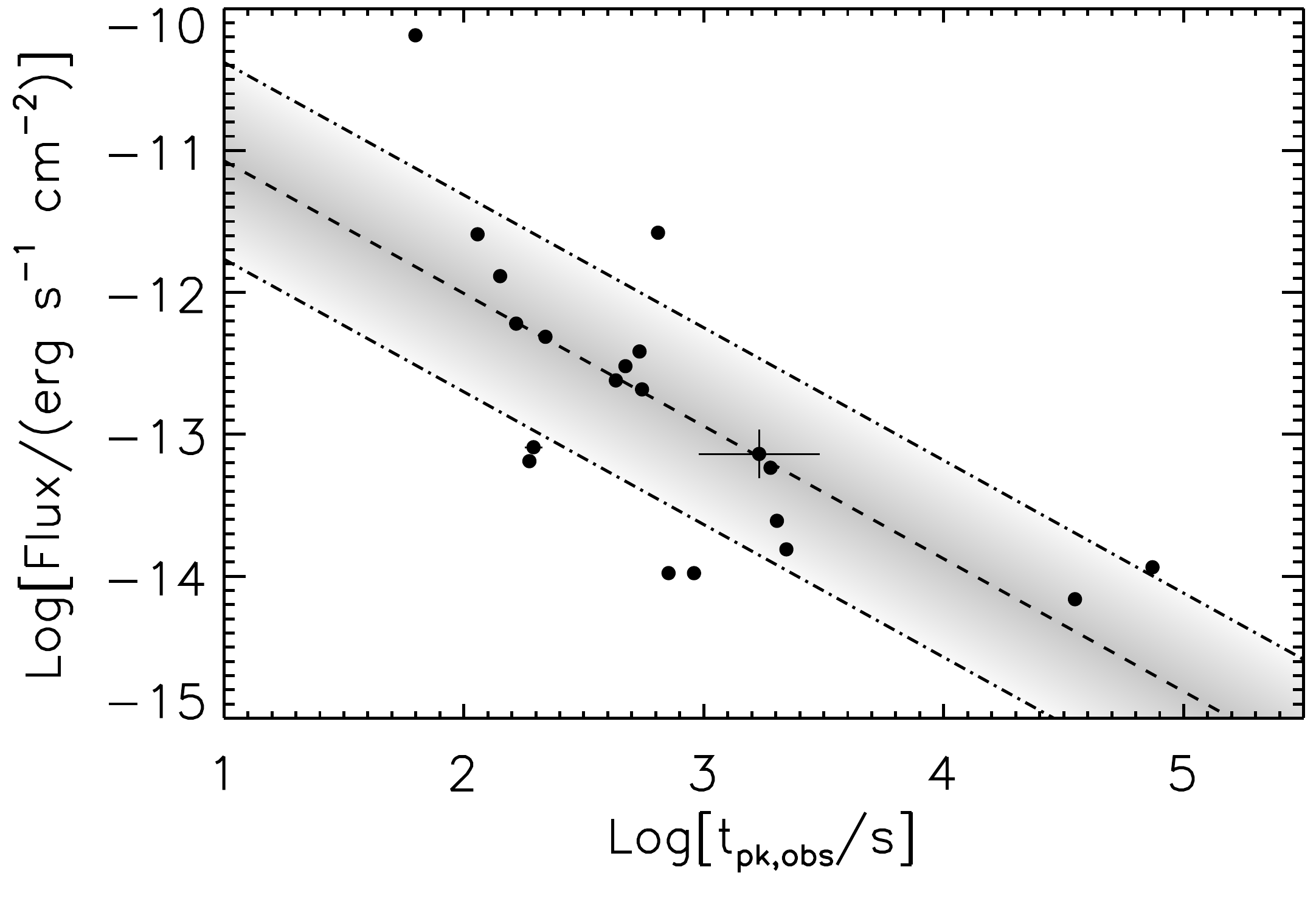}
\caption{\small{Relations between the optical flux and the observer time of the end of the plateau (\textit{top}) and the peak (\textit{bottom}). \textit{Dashed line}: best-fitting power-law model obtained accounting for the sample variance \citep{2005physics..11182D}. \textit{Gray area}: the 68\% confidence region around the best fit. The results of the fit are listed in Table~\ref{opt_corr}.}}
\label{CORR1}
\end{figure}

To compute the relations between two parameters, we used the best-fitting procedure, which accounts for the sample variance \citep{2005physics..11182D}. All results are listed in Table~\ref{opt_corr} and presented in Figures~\ref{CORR3}, \ref{CORR2} and, \ref{CORR1}.

We confirm the correlation between the optical energy ($L\times t_{\rm rest}$ with $L=L_{\rm end},L_{\rm pk}$) and the isotropic gamma-ray energy\footnote{We used the values presented in M13.\label{fn:repeat}} ($E_{\rm \gamma,iso}$, \citealt{2008MNRAS.391..577A}) and the optical energy and the energy calculated in the BAT energy band\footref{fn:repeat} ($E_{\rm \gamma}^{\rm 15-150}$) (Figure~\ref{CORR3}). However, for all these correlations the data show very broad distributions. 

There is a weak indication that the optical plateau end luminosities and the relative observer and rest frame times are correlated (Figure~\ref{CORR2}, \textit{left}). The same occurs for the peak luminosities and the relative observer and rest frame times (Figure~\ref{CORR2}, \textit{right}). Since there are few elements in our sample, the correlation is not reliable.

In the observer frame, the peak flux correlates with the peak time (Figure~\ref{CORR1}, \textit{bottom}), but the optical plateau end fluxes and their times are not related (Figure~\ref{CORR1}, \textit{top}). We had only few data for this as well: 19 GRBs for the peak relation and 14 for the plateau, and a few discrepant cases.

%----------------------------------------------------------------
\begin{table*}
\caption{\small{Two-parameter correlations involving optical luminosities and fluxes. From left to right: $X$ and $Y$ parameters to be correlated (the best-fitting law reads $\log(Y)=q+m\log(X)$); best-fitting parameters as obtained accounting for the sample variance \citep{2005physics..11182D}: slope ($m$), normalization ($q$), intrinsic scatter ($\sigma$); errors are given at 95\% c.l. The last three columns list the value of the Spearman rank ($\rho$), Kendall coefficient $K$, and $R$-index $r$ statistics.}}
\centering
\begin{tabular}{llllllll}
\hline\hline
$X$&$Y$&$m$&$q$&$\sigma$&$\rho$&$K$&$R$\\
\hline
$L\times t_{\rm rest}$&$E_{\rm \gamma}^{\rm 15-150}$&0.65$\pm$0.03&-0.39$\pm$2.52&0.59$\pm$0.01&0.64&0.47&0.72\\
$L\times t_{\rm rest}$&$E_{\rm \gamma,iso}$&0.78$\pm$0.05&-4.42$\pm$4.99&0.58$\pm$0.03&0.72&-0.49&0.80\\
$t_{\rm rest,PL}$         &$L_{\rm end,PL}$ &-0.83$\pm$0.14&8.64$\pm$1.36&0.84$\pm$0.06&-0.66&-0.51&-0.65\\
$t_{\rm obs,PL}$         &$L_{\rm end,PL}$&-0.77$\pm$0.18&8.81$\pm$2.35&0.90$\pm$0.07&-0.55&-0.41&-0.55\\
$t_{\rm rest,PK}$        &$L_{\rm end,PK}$&-1.40$\pm$0.13&10.47$\pm$0.81&0.92$\pm$0.04&-0.79&-0.64&-0.77\\
$t_{\rm obs,PK}$         &$L_{\rm end,PK}$&-1.32$\pm$0.20&10.86$\pm$0.75&1.08$\pm$0.06&-0.66&-0.48&-0.67\\
$t_{\rm obs,PL}$         &$F_{\rm PL}$      &-0.48$\pm$0.10&-12.09$\pm$1.19&0.65$\pm$0.03&-0.39&-0.26&-0.53\\
$t_{\rm obs,PK}$         &$F_{\rm PK}$     &-0.93$\pm$0.08&-10.14$\pm$0.72&0.69$\pm$0.02&-0.78&-0.62&-0.71\\
\hline
\end{tabular}
\label{opt_corr}
\end{table*}
%----------------------------------------------------------------

%----------------
%----------------
%----------------
%----------------

\begin{figure}[!]
\centering
\includegraphics[width=0.80\hsize]{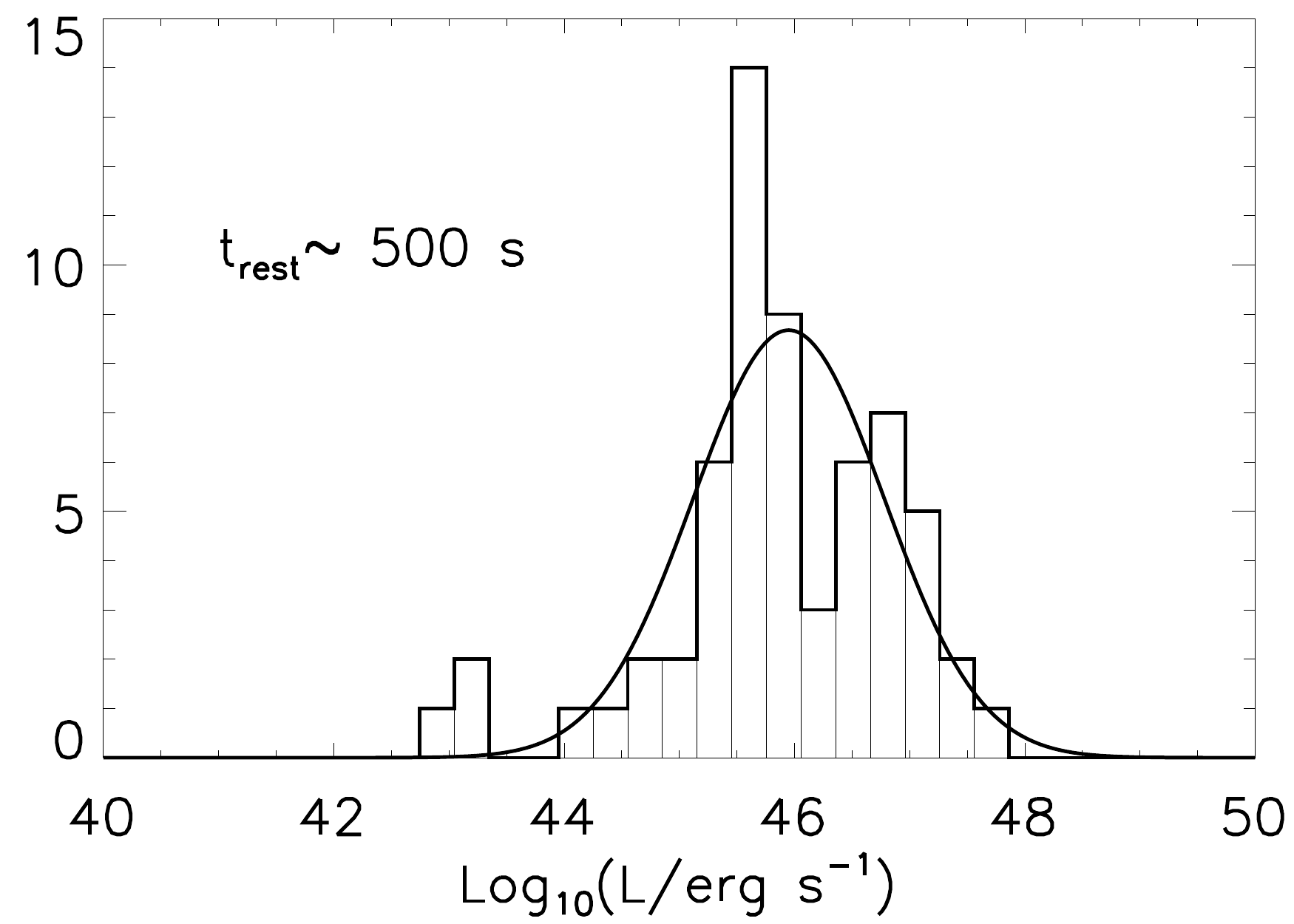}
\includegraphics[width=0.80\hsize]{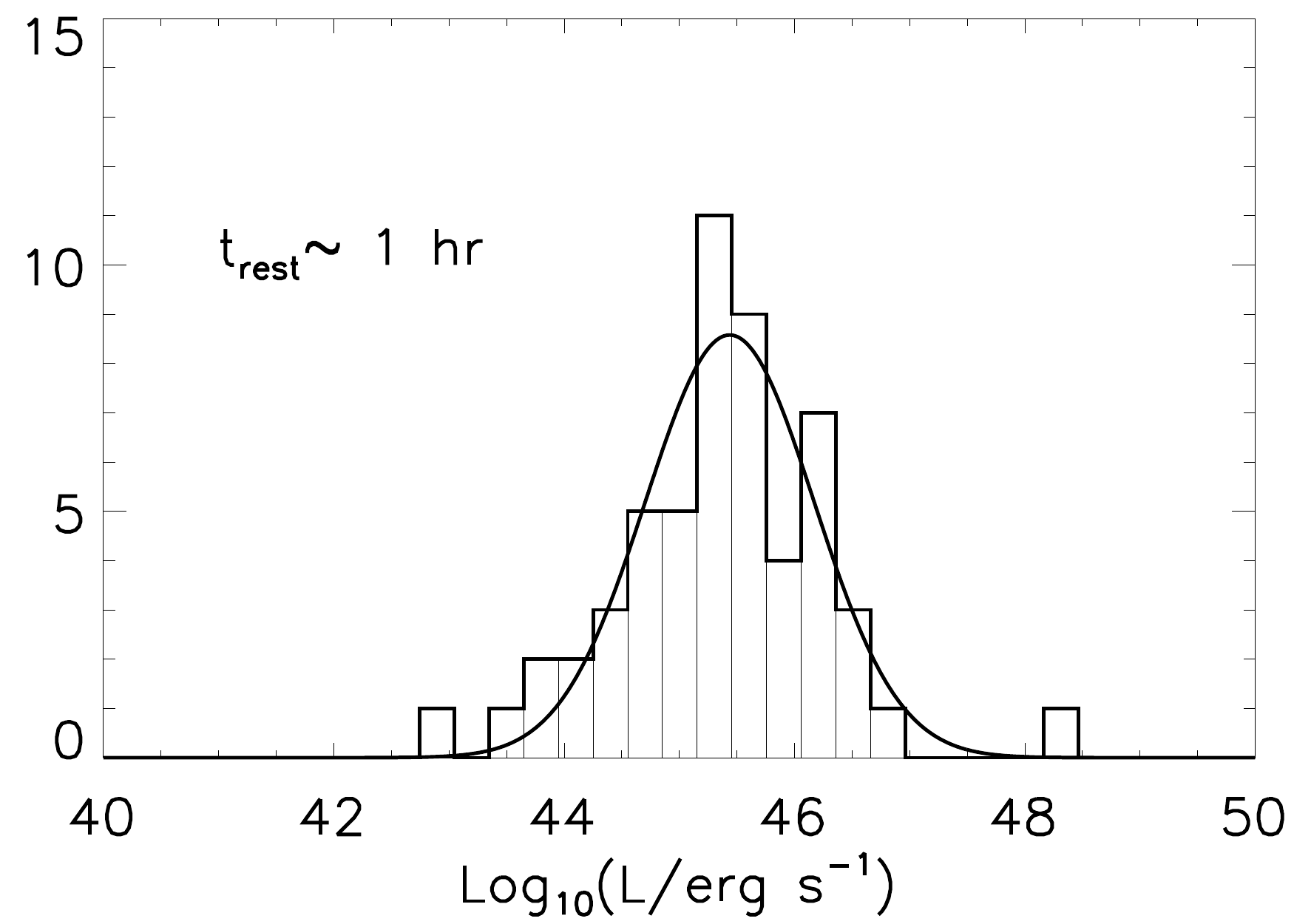}
\includegraphics[width=0.80\hsize]{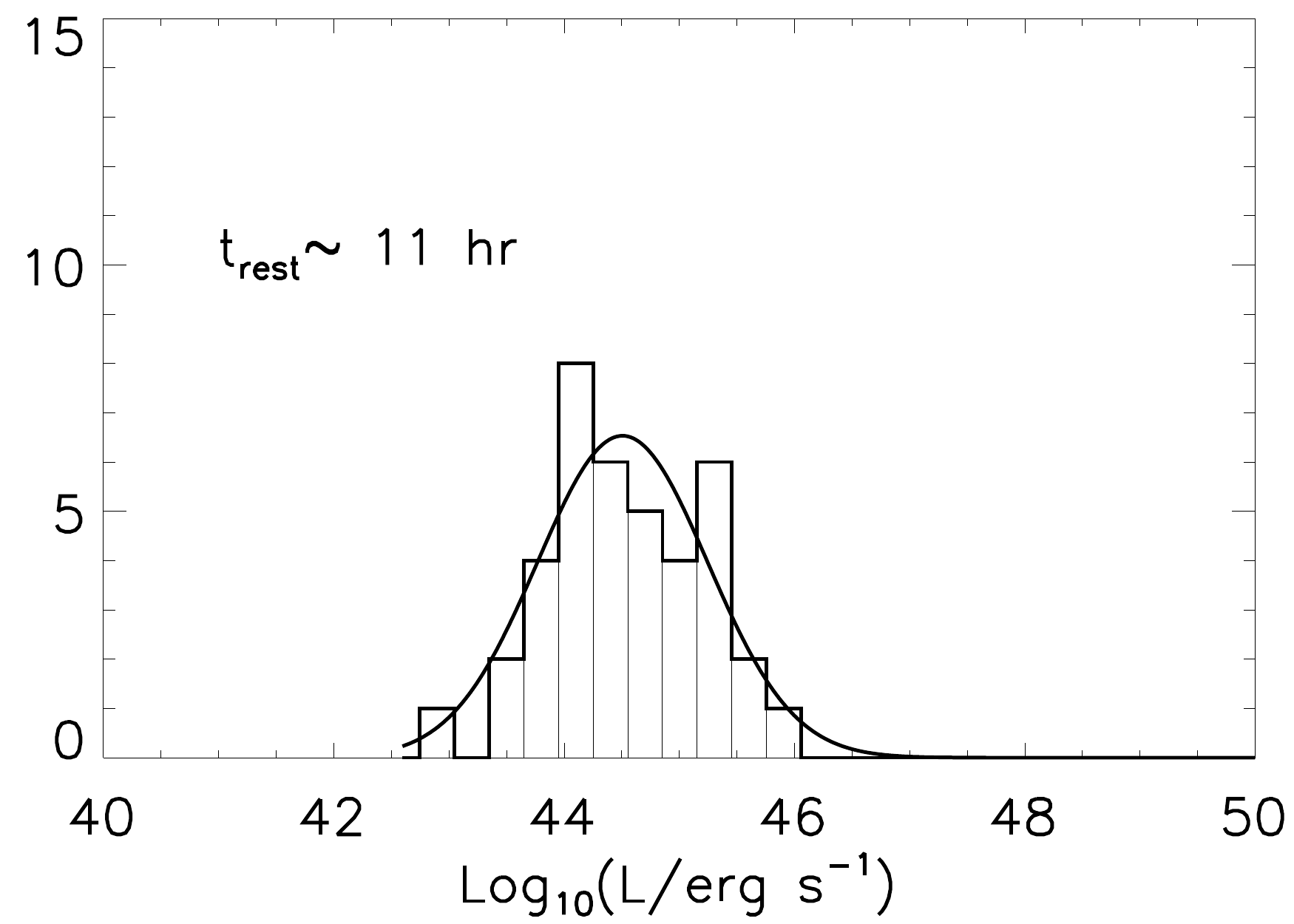}
\includegraphics[width=0.80\hsize]{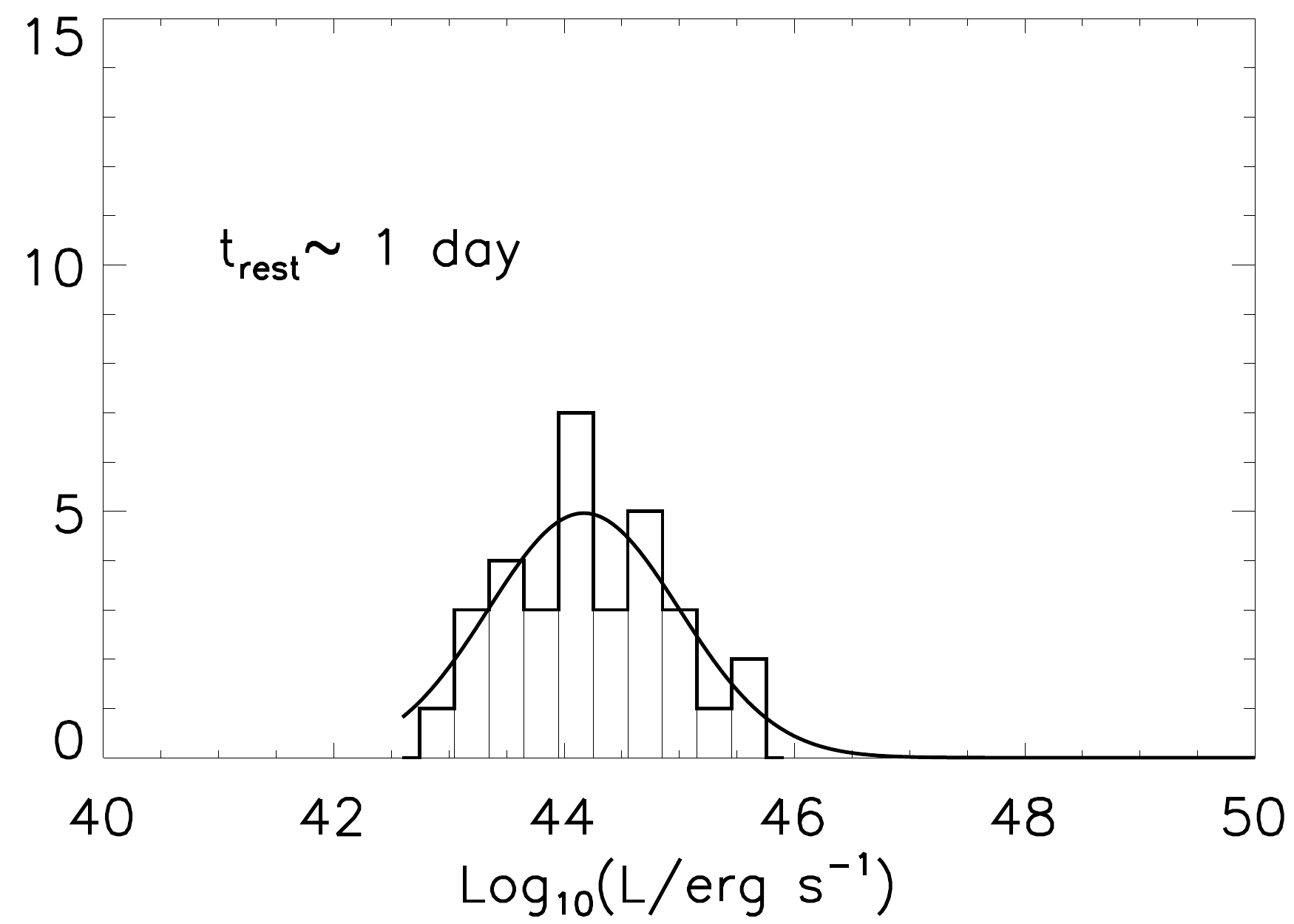}

\caption{\small{Distribution of the optical $R$ luminosity calculated for four different rest frame times: 500 s, 1 hr, 11 hr, and 1day. The black solid line corresponds to the Gaussian fit of the data. The results are listed in Table~\ref{tab_par}.}}
\label{luminosita}
\end{figure}

We also measured the optical luminosity distributions at four different rest frame times: 500 s, 1 hr, 11 hr, and 1 day. The 11-hr time frame is commonly chosen because it is reasonable to assume that the cooling frequency has not passed the optical band yet \citep{2001ApJ...547..922F,2001ApJ...560L.167P}, in addition, 11 hr and 1 day are approximately the times at which several authors found a bimodal distribution\footnote{Possibly caused by a bimodality in the optical luminosity function or by the absorption of gray dust in a fraction of bursts \citep{2008MNRAS.383.1049N}.} of the luminosities (\citealt{2006MNRAS.369L..37L}, \citealt{2006A&A...451..821N}, \citealt{2006ApJ...641..993K}, \citealt{2008MNRAS.383.1049N}). We found no bimodal distribution in our data, as asserted in recent studies \citep{2008ApJ...686.1209M,2009MNRAS.395..490O,2011MNRAS.412..561O,2010ApJ...720.1513K,2011ApJ...734...96K}. The mean luminosity simply decreases with time (Figure~\ref{luminosita}): $\mu(\log(L_{\rm 500s}/\rm erg\ s^{-1}))=45.90\pm0.06$ (64 GRBs), $\mu(\log(L_{\rm 1hr}/\rm erg\ s^{-1}))=45.40\pm0.06$ (57 GRBs), $\mu(\log(L_{\rm 11hr}/\rm erg\ s^{-1}))=44.50\pm0.07$ (40 GRBs) and $\mu(\log(L_{\rm 1day}/\rm erg\ s^{-1}))=44.20\pm0.09$ (32 GRBs) (Table~\ref{tab_par}).  

%----------------
%----------------
%************************************************************
\subsection{The optical LCs}\label{optical}

The optical LCs show different shapes and features (Table~\ref{tab_divisione_ottico}; See also \citealt{2012ApJ...758...27L}, their Figure 2). For each GRB, we selected the optical LC observed with the filter with the widest temporal coverage and the most reliable fit (Table~\ref{table1_a}, \textit{Online Material}\footnote{The complete and machine-readable form of the table is provided at CDS (table1c.dat).}).

In general, they have a rising or constant part, which can occur at any time. This occurs for 53 GRBs in our sample. Only ten GRBs show a simple power-law trend and five GRBs have an LC with an initial decay followed by an almost constant optical flux. The optical LCs with a single power-law decay have an initial time $>$600 s in the observer frame (GRB 050824\footnote{In this case we consider the overall trend, since it is difficult to fit the LC with a more complicated fit function. For a detailed study of this LC see \citet{sollerman07}.}, GRB 050908, GRB 060502A) or they are poorly sampled (GRB 050904, GRB 051111, GRB 080721, and GRB 091018). GRB 050401 and GRB 050922C show weak variability in their optical LCs, even though their best-fit function is a simple power-law. GRB 060912A has a well-sampled optical LC, fitted with a single power-law. Of the five GRBs with an initial decay followed by almost constant flux, GRB 060908 and GRB 090424 show a shallow phase at late times ($\gtrsim 10^6$), which may be due to the host galaxy. For GRB 061126, GRB 070529, and GRB 090102 the LC break occurs at $\sim10^3-10^4$ s (observer frame). For these five GRBs the initial time of the optical observations is $\sim$100 s. Even though these GRBs do not show variable LCs, we do not know what happened before the observations; they could have a very variable LC similar to GRB 080319B \citep{2008Natur.455..183R}.  

There are 14 GRBs with an optical LC with an early peak (i.e., an initial rise followed by a decay) and 14 with a quasi-constant phase (i.e., \textit{optical plateaus,} an initial quasi-constant phase followed by a decay). The optical plateaus and rises in the LCs are interpreted as the onset of the forward shock emission, when the blast wave decelerates: the peaked LCs correspond to an impulsive ejecta release, where all ejecta have the same Lorentz factor after the burst phase; the plateaus are caused by the energy injection in the forward-shock due to an extended ejecta release, a wide distribution of the ejecta initial Lorentz factors or both (e.g. \citealt{2011MNRAS.414.3537P,2009MNRAS.395..490O}), or the onset of the afterglow for the wind medium (e.g. \citealt{1999ApJ...520L..29C,2012MNRAS.420..483G}).

Sixteen GRBs show a late-time re-brightnening (i.e., at late times the LC displays a rise phase followed by a decay phase with the same slope as before). The re-brightening may be related to the jet structure, and seems to agree with the on-axis two-component jet model, with the re-brightening corresponding to the emergence of the slow component  (e.g. \citealt{2004NewA....9..435J}; \citealt{2008Natur.455..183R,2012arXiv1210.5142L}).

Three GRBs show a series of initial large bumps (i.e., more than one peak). GRB 060904B shows two bumps during the X-ray plateau and a shallow decay starting roughly at the beginning of the X-ray normal decay. The two optical peaks are not correlated with the high-energy emission and the subsequent optical bump is assumed to trace the onset of the forward shock \citep{2009ApJ...702..489R}. The optical LC of GRB 060906 has two bumps that coincide with the X-ray plateau. These bumps could be associated to a change of the circumburst density \citep{2002A&A...396L...5L,2009ApJ...693.1484C}. The GRB 080928 optical LC was modeled by multiple energy injections into the forward shock, and not with the central engine, since the fluctuations occur on a long timescale \citep{2011AA...529A.142R}. The first peak is assumed to be the onset of the afterglow, while the following two bumps are produced by the central engine activity \citep{2011AA...529A.142R}.

Five optical LCs show small bumps (i.e., weak fluctuations over the power-law decay). The optical bumps could be related to the erratic late-time central engine activity \citep{2012ApJ...758...27L}.

The optical LCs have a complex behavior, and during a well-defined X-ray LC phase the optical LC can rise and then decay, or vice versa. Specifically\footnote{The percentage refers to single LC parts, not to the total number of GRBs.}:
\begin{itemize}
\item Steep decay: 42\% of the optical LCs rise, 32\% decay, 16\% are constant, and 10\% have a complex behavior (rise-decay, bumps).
\item Plateau: 16\% of the optical LCs rise, 47\% decay, 8\% are constant, 26\% rise and decay and 4\% have one or more bumps.
\item Normal decay: 77\% of the optical LCs decay and 23\% rise and decay or have a more complex behavior.
\end{itemize}

The complexity of the LCs decreases as a function of time.  %Furthermore, the percentage of optical LCs that show a simple power-law decay is 32\% during the X-ray steep decay (not necessary with the same slope), 47\% during the X-ray plateau, and 77\% during the X-ray normal decay.
%----------------------------------------------------------------
\begin{table}[!]
\caption{\small{Subdivision of the GRBs in our sample according to the optical LC features.}}
\centering
\begin{tabular}{p{1.1cm}p{1.1cm}p{1.1cm}p{1.1cm}p{1.1cm}}
\hline\hline
 \multicolumn{5}{l}{\textbf{Initial peak}}\\
050730\tablefootmark{a} & 050820A\tablefootmark{a}& 060418 &060605 &060607A\\ 
060614\tablefootmark{a} &061007 &061121& 070318 &070419A\\ 
070802\tablefootmark{a} &071025 &071031& 071112C &080603A \\
080710\tablefootmark{a} &080810 &081008 &081203A& \\
\hline
 \multicolumn{5}{l}{\textbf{Initial shallow phase}}\\
050408 &050416A& 060124& 060210& 060526\\ 
060729 &070125  & 070208 &070411 &071010A\\
080330 &090313 &090426 &090618 & \\
\hline
 \multicolumn{5}{l}{\textbf{Late-time re-brightening}}\\
050820A &060206 &060526 &060927 &061121 \\
071003 &071010A &080310 &080607 &080913 \\
081008 &081029 &091127 &100418A &100901A\\
\hline
 \multicolumn{5}{l}{\textbf{Series of initial large bumps}}\\
060904B &060906 &080928& & \\
\hline
 \multicolumn{5}{l}{\textbf{Small bumps}}\\
050401\tablefootmark{b} & 060607A &071025 & 071031 & 090313\tablefootmark{b}\\ 
\hline
\end{tabular}
\tablefoot{
\tablefoottext{a}{At late times.}
\tablefoottext{b}{For the $R_{C}$ observations.}
}
\label{tab_divisione_ottico}
\end{table}
%----------------------------------------------------------------
\subsection{Comparison between the optical and X-ray LCs}\label{comparison}

\begin{figure}[!]
\centering
\includegraphics[width=0.90 \hsize,clip]{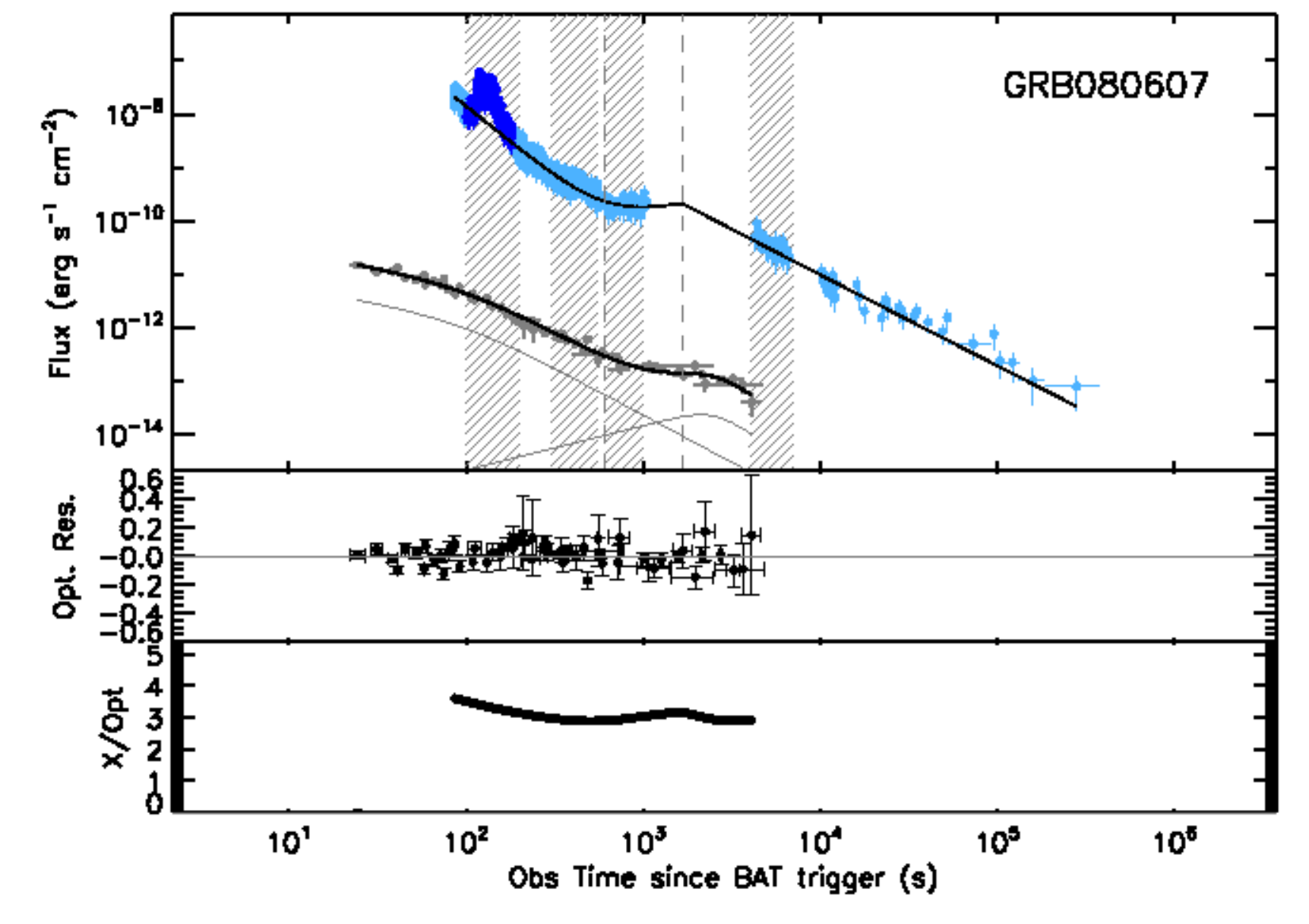}
\includegraphics[width=0.90 \hsize,clip]{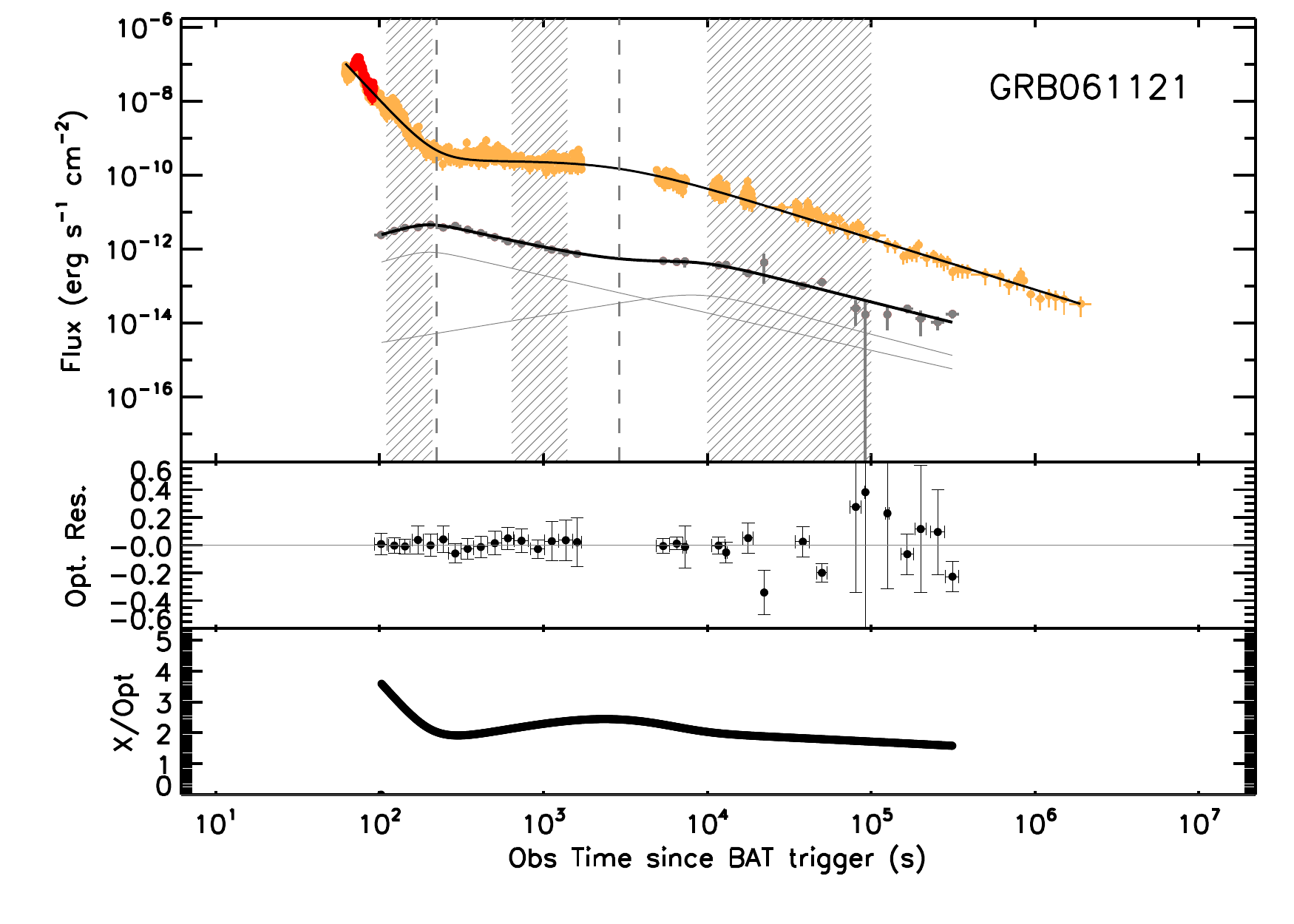}
\includegraphics[width=0.90 \hsize,clip]{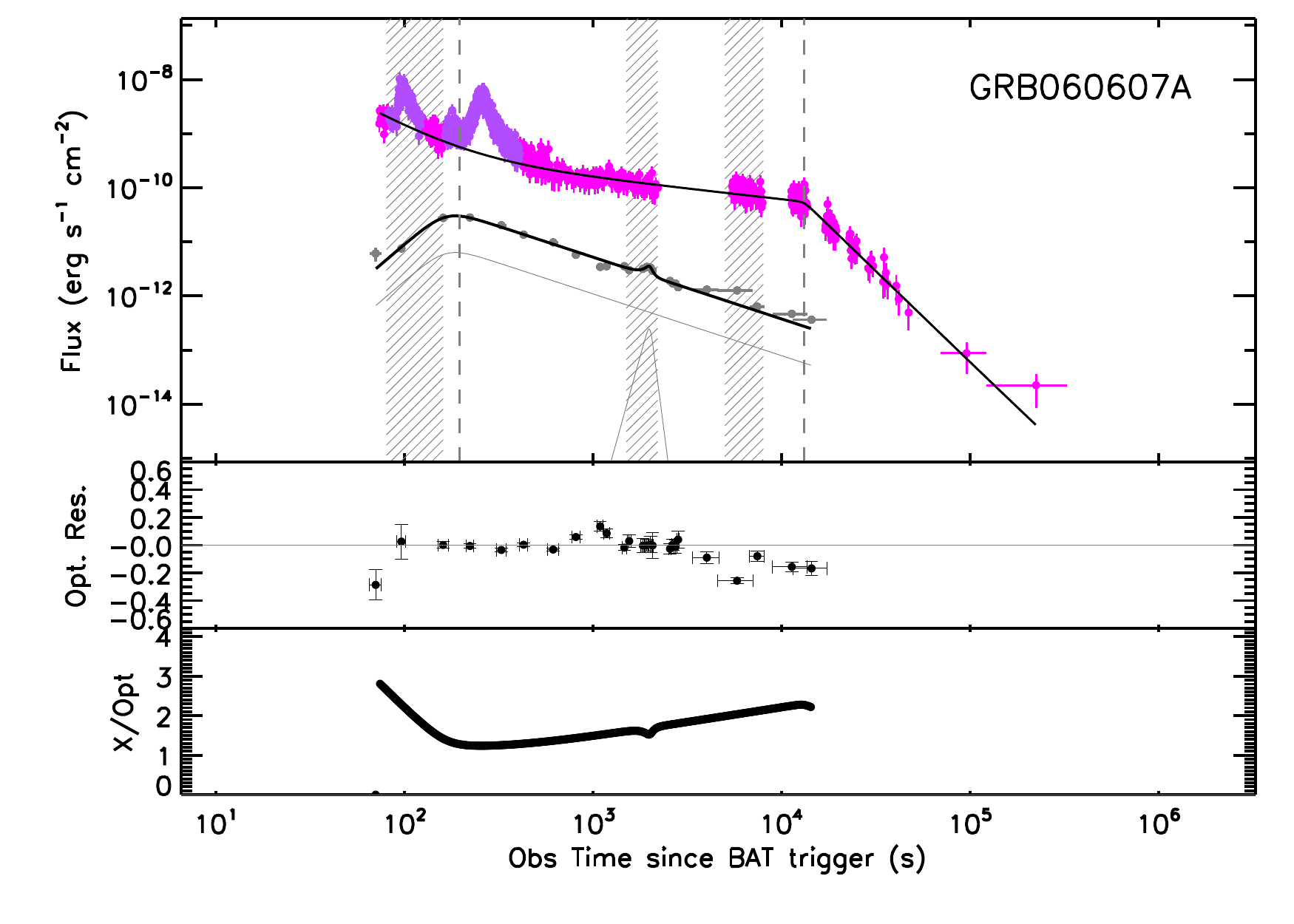}

\caption{\small{Comparison between optical and X-ray LCs, examples of the GRBs in each group. \textit{Group A}: GRB 080607 
(\textit{blue/light blue}). \textit{Group B}: GRB 061121 (\textit{red/orange}). \textit{Group C}: GRB 060607A (\textit{purple/magenta}). For each panel: \textit{Top}. \textit{Colored points}: X-ray data. The data in light color and bright color represent the continuum and the flaring portions, respectively, as calculated by M13. \textit{Gray dashed lines}: X-ray break times. \textit{Gray points}: optical data. \textit{Black solid line}: fit to the data. \textit{Gray solid lines}: components of the fit function used to fit the optical data. \textit{Hashed gray boxes}: SED time intervals. \textit{Middle}. Ratio between the optical data and their fit function. \textit{Bottom.} Ratio between the fit to the X-ray continuum and the optical LC. See Figures \ref{confronto1}-\ref{confronto9} (\textit{Online Material}) for the other GRBs of our sample.}}
\label{classificazione}
\end{figure}

For every GRB we compared the optical LC slopes with the contemporaneous X-ray LC slopes. For both the optical and the X-ray LCs, we considered only the continuum part of the LC, excluding small bumps and flares (see M13). As in the previous section, for each GRB, we selected the optical LC observed with the filter with the widest temporal coverage and the most reliable fit (Table~\ref{table1_a}, \textit{Online Material}\footnote{The complete and machine-readable form of the table is provided at CDS (table1c.dat).}); the X-ray LC parameters are those derived in M13. From the synchrotron spectrum \citep{1998ApJ...497L..17S}, if $\nu_{\rm c}<\nu_{\rm op}<\nu_{\rm X}$, with $\nu_{\rm c}$ the cooling frequency, the difference between the contemporaneous optical and X-ray slopes is $\Delta\alpha=0$. If $\nu_{\rm op}<\nu_{\rm c}<\nu_{\rm X}$, for the slow cooling regime, $\Delta\alpha=\pm1/4$\footnote{This is valid in the slow cooling regime for the constant interstellar medium (ISM) and the wind case: for $\nu_{\rm m}<\nu<\nu_{\rm c}$, $\alpha_1=3(\rm p-1)/4$, and for $\nu>\nu_{\rm c}$, $\alpha_2=(3p-2)/4$ in the ISM case, so $\alpha_1-\alpha_2=(3p-3-3p+2)/4=-1/4$; for the wind case, for $\nu_{\rm m}<\nu<\nu_{\rm c}$, $\alpha_1=(3p-1)/4$, and for $\nu>\nu_{\rm c}$, $\alpha_2=(3p-2)/4$, so $\alpha_1-\alpha_2=(3p-1-3p+2)/4=1/4$.}. We subdivided our sample into three groups, depending on whether the pairs of the optical/X-ray slopes follow the relation $\Delta\alpha=0,1/4$, or not at all within 1$\sigma$\footnote{A similar method was used by \citet{2011MNRAS.414.3537P} to classify coupled and decoupled LCs.} (Table~\ref{gruppi_tab} and Figure~\ref{classificazione}):    
\begin{itemize}
\item Group A: all pairs of slopes of the same GRB satisfy the relation $\Delta\alpha=0,\pm1/4$ (13 GRBs).
\item Group B: some slopes of the same GRB satisfy the relation $\Delta\alpha=0,\pm1/4$ (27 GRBs).
\item Group C: no slopes of the same GRB satisfy the relation $\Delta\alpha=0,\pm1/4$ (28 GRBs).
\end{itemize}

Some X-ray LCs show an initial steep decay; this is generally not present in the optical LCs, which display a rise, a plateau, or a normal decay. The X-ray steep decay is well explained as the decay of the prompt emission, and its slope value is particularly sensitive to the chosen zero time of the power-law decay, $t_{0}\sim 0$ (e.g. the BAT trigger time) or $t_{0}=t_{90}$ (for details see M13). For this reason, the steep-decay phase was not considered in our classification.

\begin{figure}[!]
\centering
   \resizebox{\hsize}{!}{\includegraphics{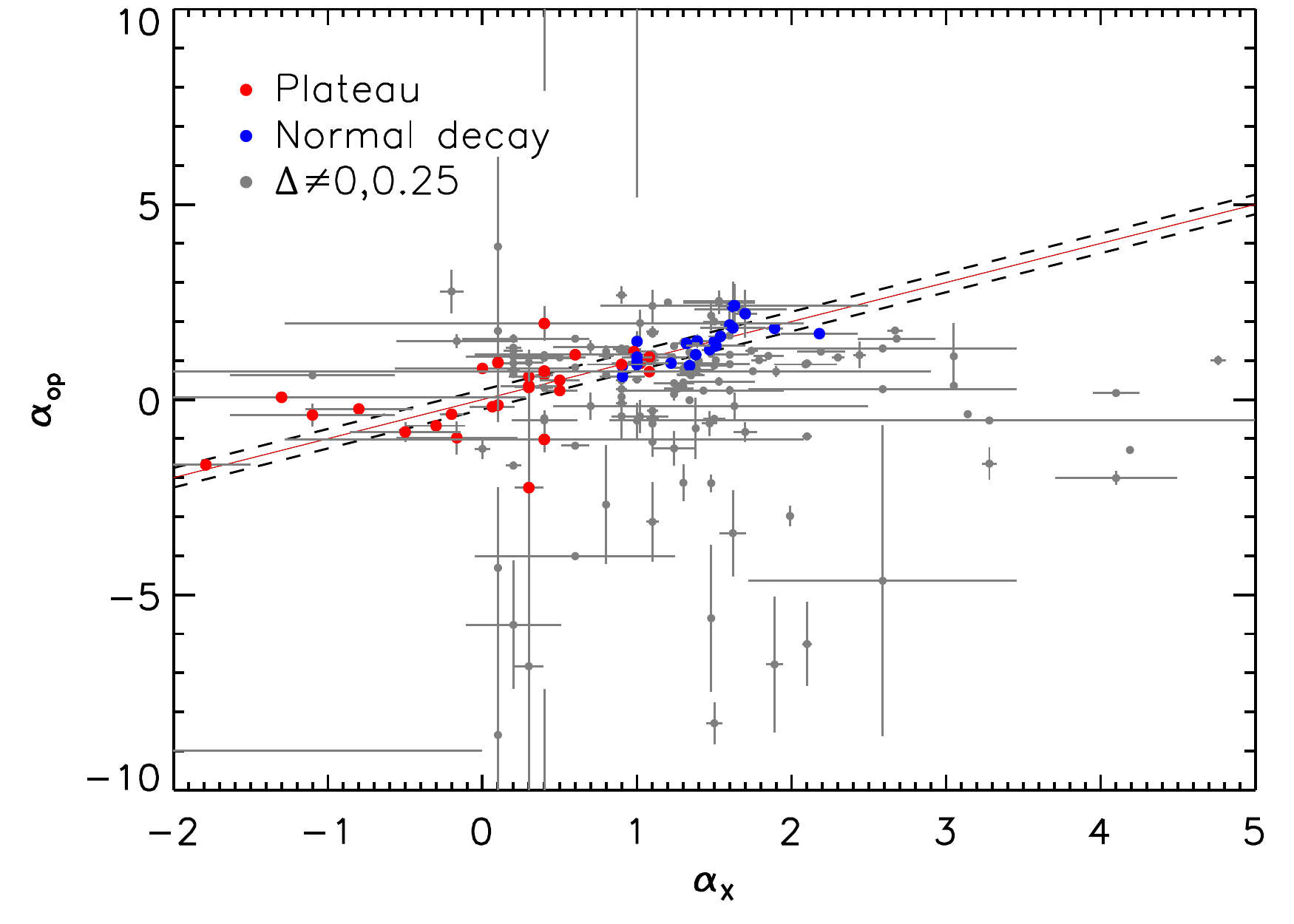}}
     \caption{\small{Comparison between the X-ray LC slope ($\alpha_{X}$) and the optical one ($\alpha_{op}$). \textit{Red (blue) dots}: data for the plateau (normal decay) phase that agree with the $\Delta\alpha=0,1/4$ relation whitin the 1$\sigma$ errors. \textit{Gray dots}: the data that do not follow the $\Delta\alpha=0,1/4$ relation. \textit{Red solid line}: $\Delta\alpha=0$. \textit{Gray dashed lines}: $\Delta\alpha=\pm1/4$.}}
      \label{slope_ok}
\end{figure}

We plot $\alpha_{\rm X}$ vs. $\alpha_{\rm op}$ for every GRB in Figure~\ref{slope_ok} (e.g. \citealt{2007ApJ...668L..95U}). About half of the $\alpha_{\rm X}$ vs. $\alpha_{\rm op}$ couples refer to the X-ray LC normal decay phase and the other half to the plateau. We noted that the GRBs in Group A and B have more complex LCs than the Group C GRBs, indeed most of the X-ray and optical LCs of Group A and B GRBs have Type IIa or III shapes. Therefore, when GRBs have LCs that are well sampled and have a good time coverage, hence with more complicated shapes, the X-ray and the optical LCs show a similar trend. When we have fewer data, we cannot compare some parts of the LCs and perhaps the observed slope is different from the real behavior of the LC.

X-ray flares do not influence the relation between the X-ray and optical LCs, because in Group A there are 5/14 GRBs with flares (36\%), in Group B there are 8/26 (31\%), and in Group C are 8/28 (29\%). This agrees with the percentage found in other samples \citep{2010MNRAS.406.2113C, 2013MNRAS.428..729M}.

%----------------
%----------------
%----------------------------------------------------------------
\begin{table}[!]
\caption{\small{List of GRBs in the three groups. See Appendix~\ref{gruppi_descrizione} for more details.}}
\centering
\begin{tabular}{p{1.cm}p{1.cm}p{1.cm}p{1.cm}p{1.cm}}
\hline\hline
\multicolumn{5}{l}{\textbf{GRB} }\\
\hline
\multicolumn{5}{l}{\textbf{Group A} (13 GRBs)}\\
050416A&050824&  050904 & 060912A& 061007\\
080310& 080319B &080330& 080603A& 080607\\
 080913& 100418A & 100901A& &\\
\hline
\multicolumn{5}{l}{\textbf{Group B} (27 GRBs)}\\
050319B&  050401  &050408 &050525A & 051109A  \\
060124 & 060502A& 060512 &060526&  060614 \\
060904B &060908 &061121 & 060729& 070208 \\
071003& 071031& 071010A & 080413B& 080710\\
080928 & 081008& 081203A & 090313& 090618\\
091018& 090926A & & &\\
\hline
\multicolumn{5}{l}{\textbf{Group C} (28 GRBs)}\\
050730  &050820A& 050908& 050922C& 051111 \\
060206 &060210&  060418&  060605&  060607A \\
060906 &060927 & 061126&  070125& 070318\\
070411 & 070419A&  070529 & 070802 & 071025\\
071031 &071112C& 080721&  080810 & 081029 \\
090102&  090424& 090426 &091127& \\
\hline
\end{tabular}
\label{gruppi_tab}
\end{table}
%----------------------------------------------------------------

%************************************************************
%%%%%%%%%%%%%%%%%%%%%%%%%%%%
\section{Discussion}\label{discussione}
We presented the analysis of a large and homogeneous data set, useful for studying the GRB rest frame properties and for comparing the optical and X-ray emission. 

The comparison between the X-ray and the optical LCs and SEDs enables us to investigate the nature of their emission mechanism and to verify if they have the same origin. For the internal-external shock model \citep{1998ApJ...497L..17S}, the forward shock propagating into the external medium gives rise to the X-ray and optical emission. If the optical and the X-ray LCs have similar shapes and slopes, they could be caused by synchrotron emission and probable are produced by the forward shock (e.g. \citealt{2006ApJ...642..354Z}). Indeed, the X-ray emission is mainly influenced by the central engine activity: the steep decay is thought to be the tail of the prompt emission \citep{2000ApJ...541L..51K} or it is direct emission from the central engine \citep{2009MNRAS.395..955B}. The plateau reflects the effect of energy injection into the forward shock (e.g. \citealt{2006ApJ...642..354Z}).

The optical LCs show various features: initial peaks or constant phases, which are probably caused by the onset of the forward shock (e.g. \citealt{2011MNRAS.414.3537P}), late-time re-brightenings that may depend on the structure of the jet (e.g. \citealt{2008Natur.455..183R}), and small bumps linked to the central engine activity \citep{2012ApJ...758...27L}. 

Thanks to our sample of LCs and SEDs, we were able to discuss the similarities and differences of the optical and X-ray emission by comparing their LCs (Sect.~\ref{curvediluce}). In Sect.~\ref{closure} we considered the forward-shock model and the closure relations (e.g. \citealt{1998ApJ...497L..17S,2006ApJ...642..354Z}), and in Sect.~\ref{radio_sect} we presented the radio/optical/X-ray SED of GRB 071003, which is well fitted by the synchrotron spectrum. Finally, we investigated the role of the optical emission in the three-parameter correlation between $E_{\rm X,iso}$-$E_{\rm \gamma,iso}$-$E_{\rm pk}$ (M13, \citealt{2012MNRAS.425.1199B}).
%************************************************************
\subsection{LC phases}\label{curvediluce}

%**********************************
\subsubsection{Steep-decay phase, the plateau, and the X-ray flares}

The steep-decay phase of the X-ray LCs is either the high-latitude emission after the end of the prompt emission \citep{2000ApJ...541L..51K}, or it may be part of the prompt emission itself, as proposed by \citet{2009MNRAS.395..955B} and \citet{2008MNRAS.388.1729K}, since the X-ray flux is smoothly connected to the $\gamma$-ray emission \citep{2005Natur.436..985T,2006A&A...449...89G}  and is characterized by a strong hard-to-soft spectral evolution \citep{2007ApJ...668..400B}. In this phase most of the optical LCs rise (42\%) or have a complex behavior, with bumps, peaks, and plateaus (26\%), which makes their slopes different from the X-ray steep-decay slopes. The remaining 32\% decay during this phase, but of these there are only two cases whose optical LC slopes are similar to the X-ray slopes (e.g. GRB 080607, GRB 100901A). In some cases, there are X-ray flares superimposed on the steep decay, which are linked to the central engine activity (e.g. \citealt{2006ApJ...642..354Z,2010MNRAS.406.2113C}).
%The steep decay phase of the X-ray LCs is the tail of the prompt $\gamma$-ray emission \citep{2000ApJ...541L..51K} or the prompt emission itself \citep{2009MNRAS.395..955B}

The X-ray plateau is interpreted as an injection of energy into the forward shock (e.g. \citealt{2006ApJ...642..354Z}) and agrees with this prediction since there is no significant spectral evolution \citep{2012A&A...539A...3B}. The source of the energy injection could be the power emitted by a spinning-down newly-born magnetar \citep{1998A&A...333L..87D,2001ApJ...552L..35Z,2009ApJ...702.1171C} that refreshes the forward shock \citep{2011A&A...526A.121D} or the fall-back and accretion of the stellar envelope on the central black hole \citep{2008MNRAS.388.1729K}. During this phase the optical LCs behave in different ways and $\sim$46\% of them rise or show peaks or bumps.

\begin{figure*}[!]
 \centering
\includegraphics[width=0.3 \hsize,clip]{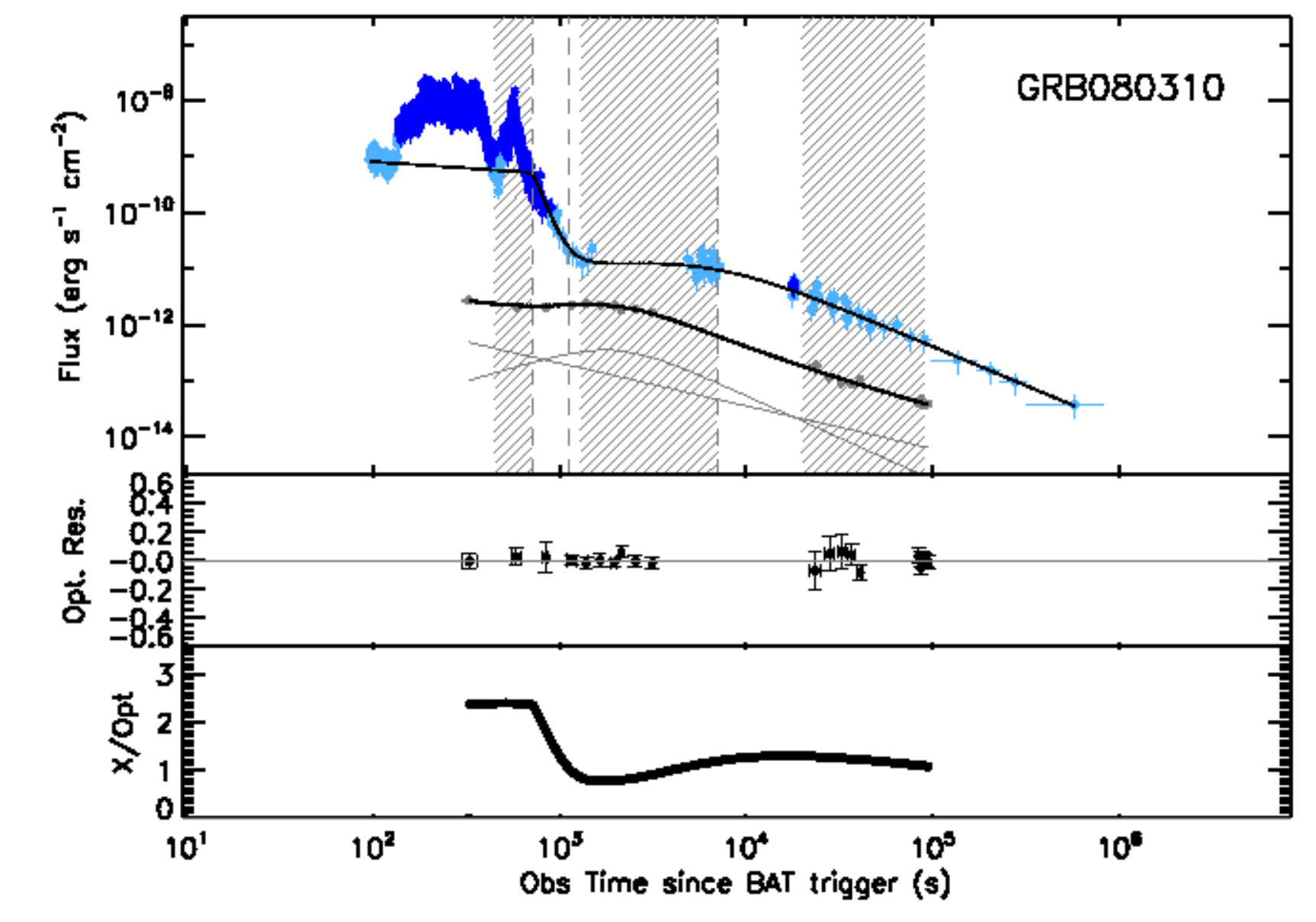}
\includegraphics[width=0.3 \hsize,clip]{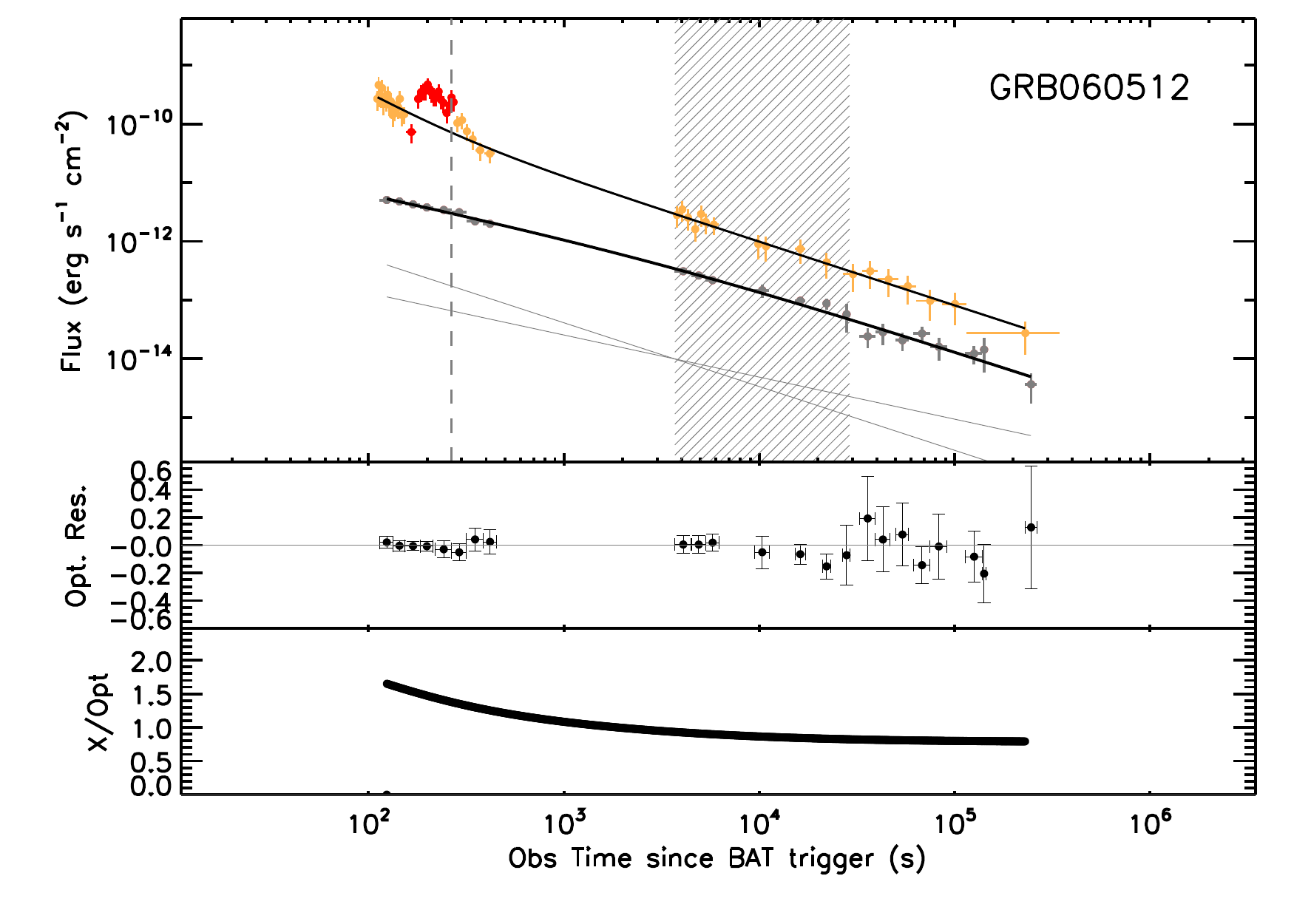}
\includegraphics[width=0.3 \hsize,clip]{FIGURE/LC/061121_OP1X-eps-converted-to.pdf}\\ 
\includegraphics[width=0.3 \hsize,clip]{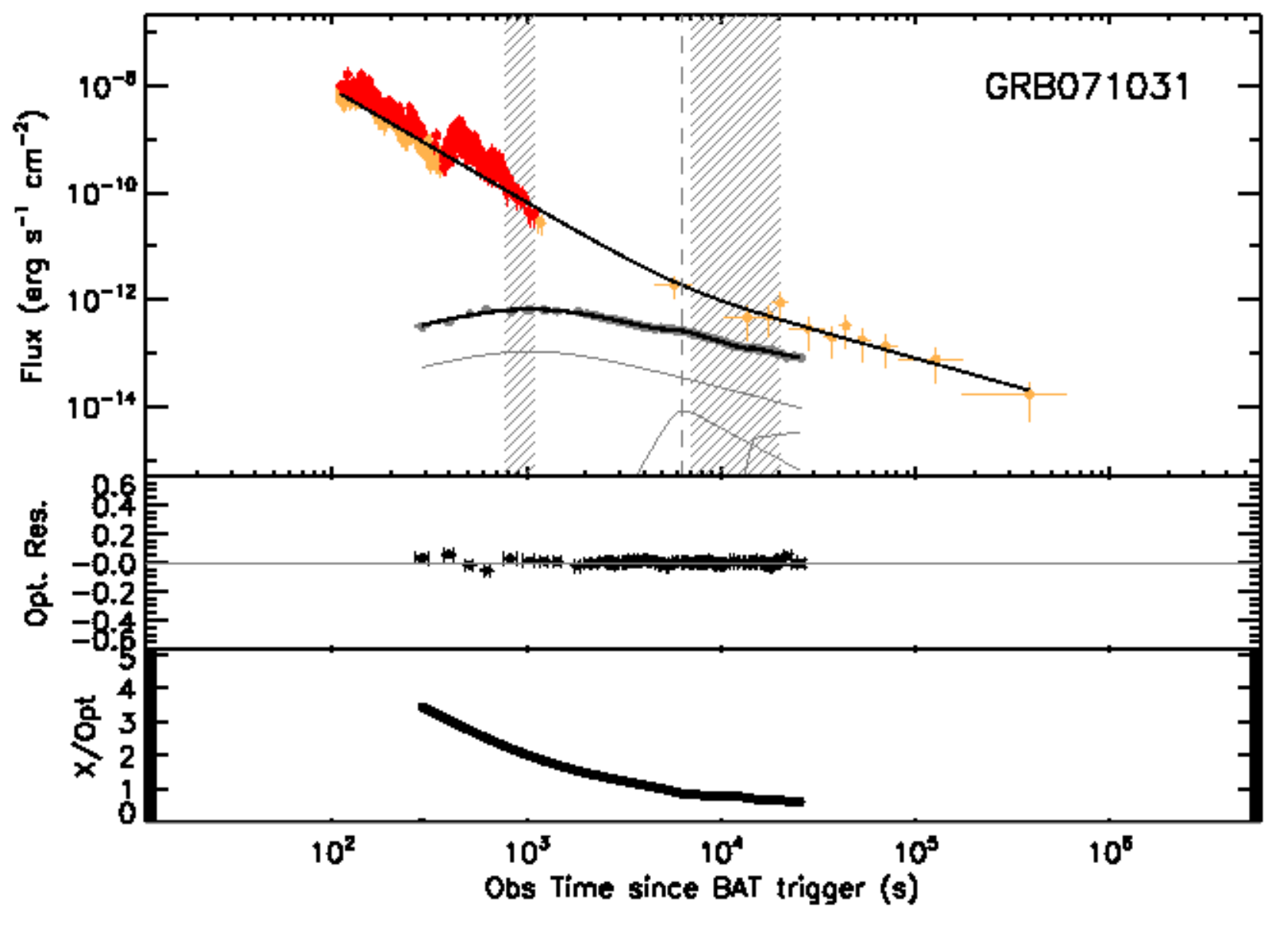}
\includegraphics[width=0.3 \hsize,clip]{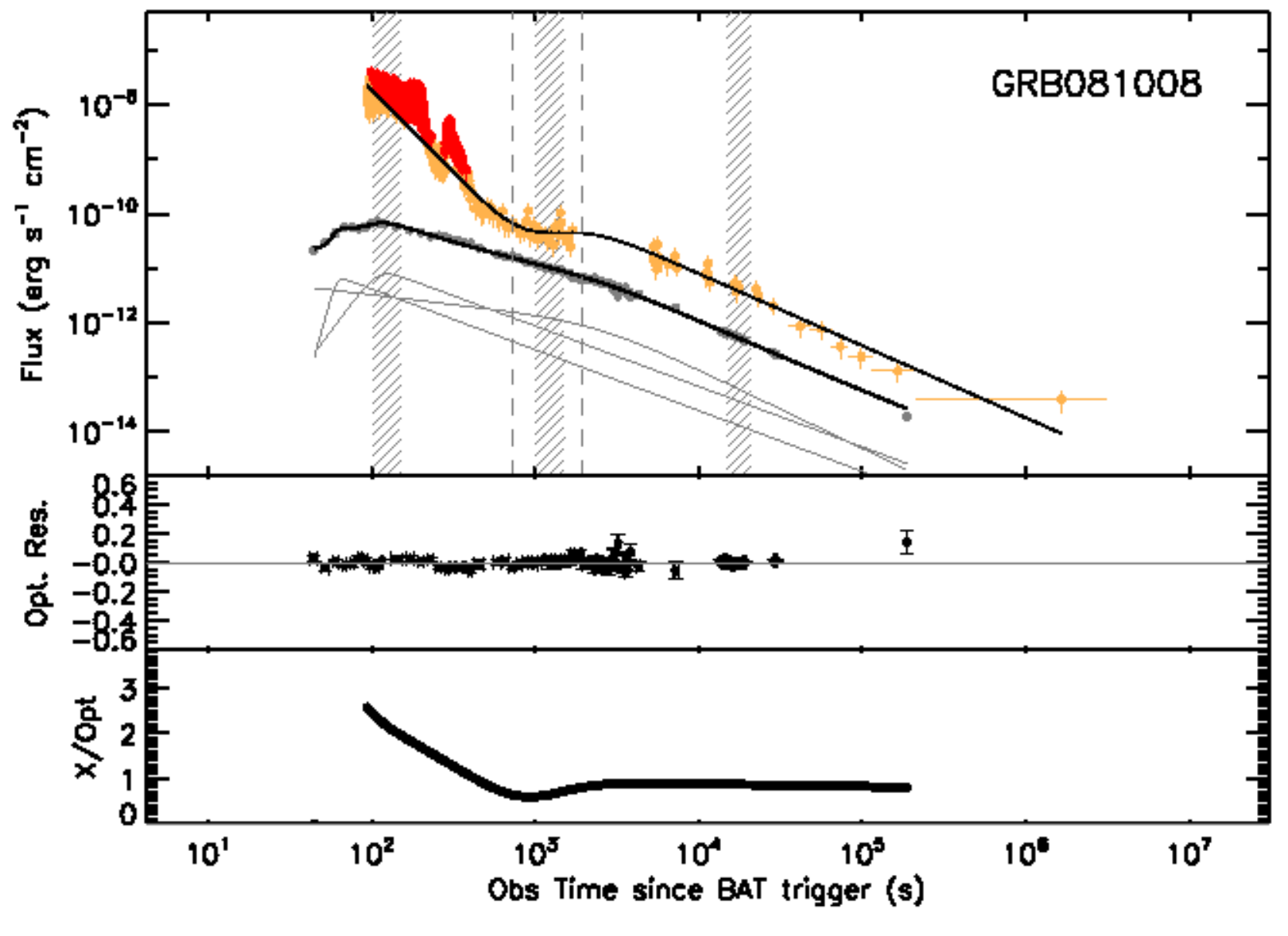}
\includegraphics[width=0.3 \hsize,clip]{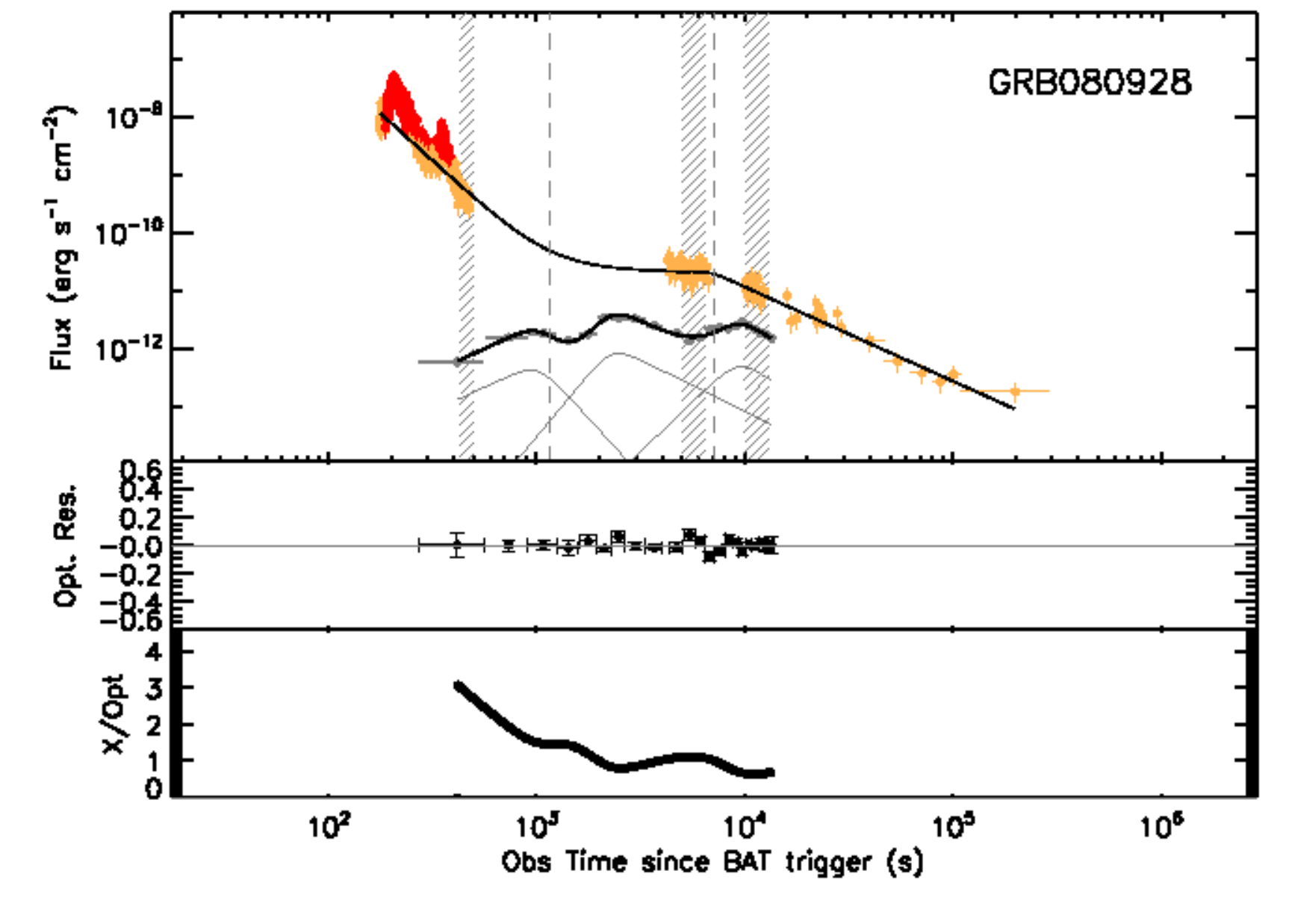}\\
\includegraphics[width=0.3 \hsize,clip]{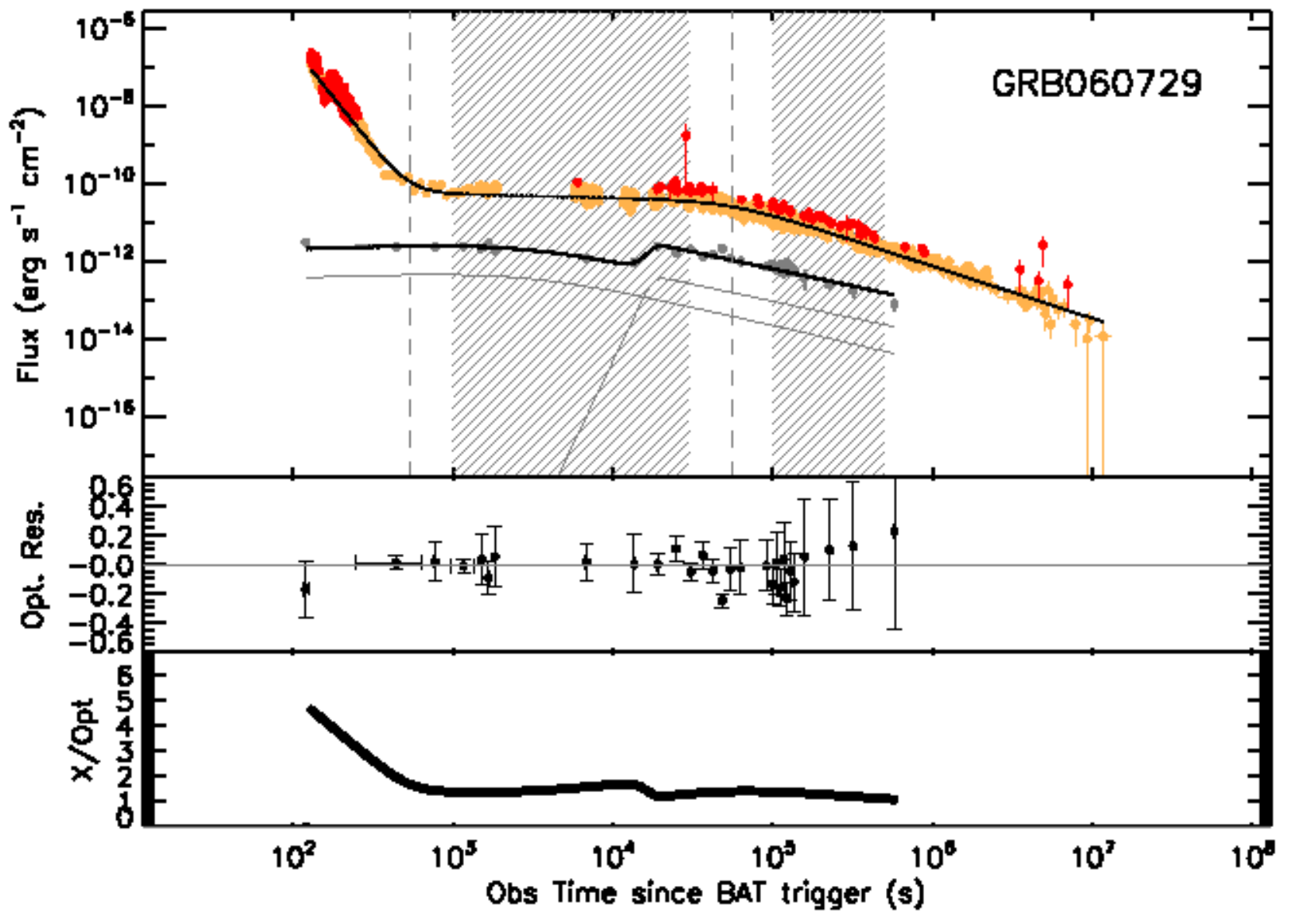}
\includegraphics[width=0.3 \hsize,clip]{FIGURE/LC/080607_OP1X-eps-converted-to.pdf} 
\includegraphics[width=0.3 \hsize,clip]{FIGURE/LC/060607A_OP1X-eps-converted-to.pdf}\\
\includegraphics[width=0.3 \hsize,clip]{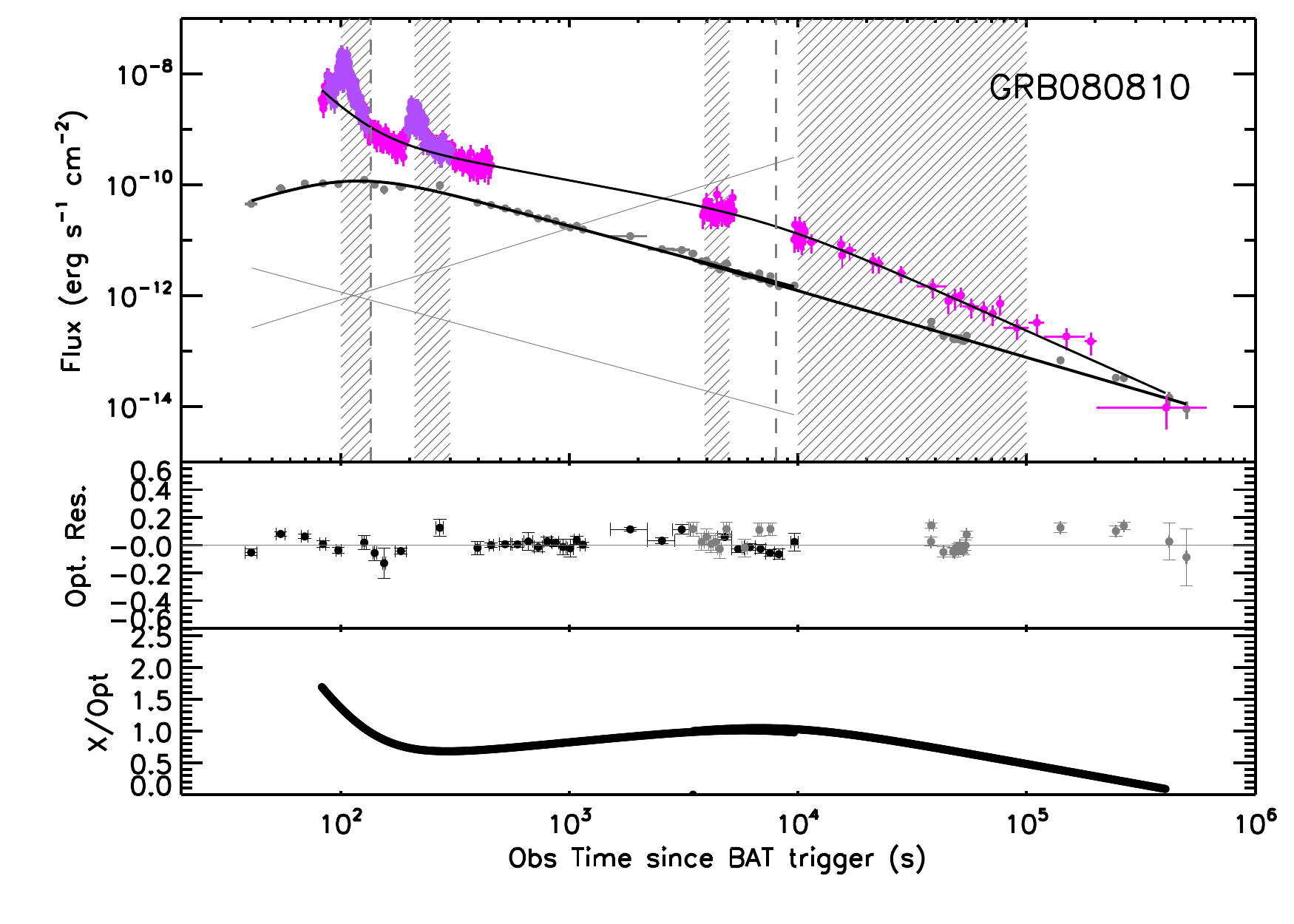} 
\caption{\small{GRBs with X-ray flares during the steep-decay phase and the optical peak during or at the end of the X-ray steep decay. Color code as in Figure~\ref{classificazione}.}}\label{ref1}
\end{figure*}

\begin{figure*}[!]
\centering
\includegraphics[width=0.3 \hsize,clip]{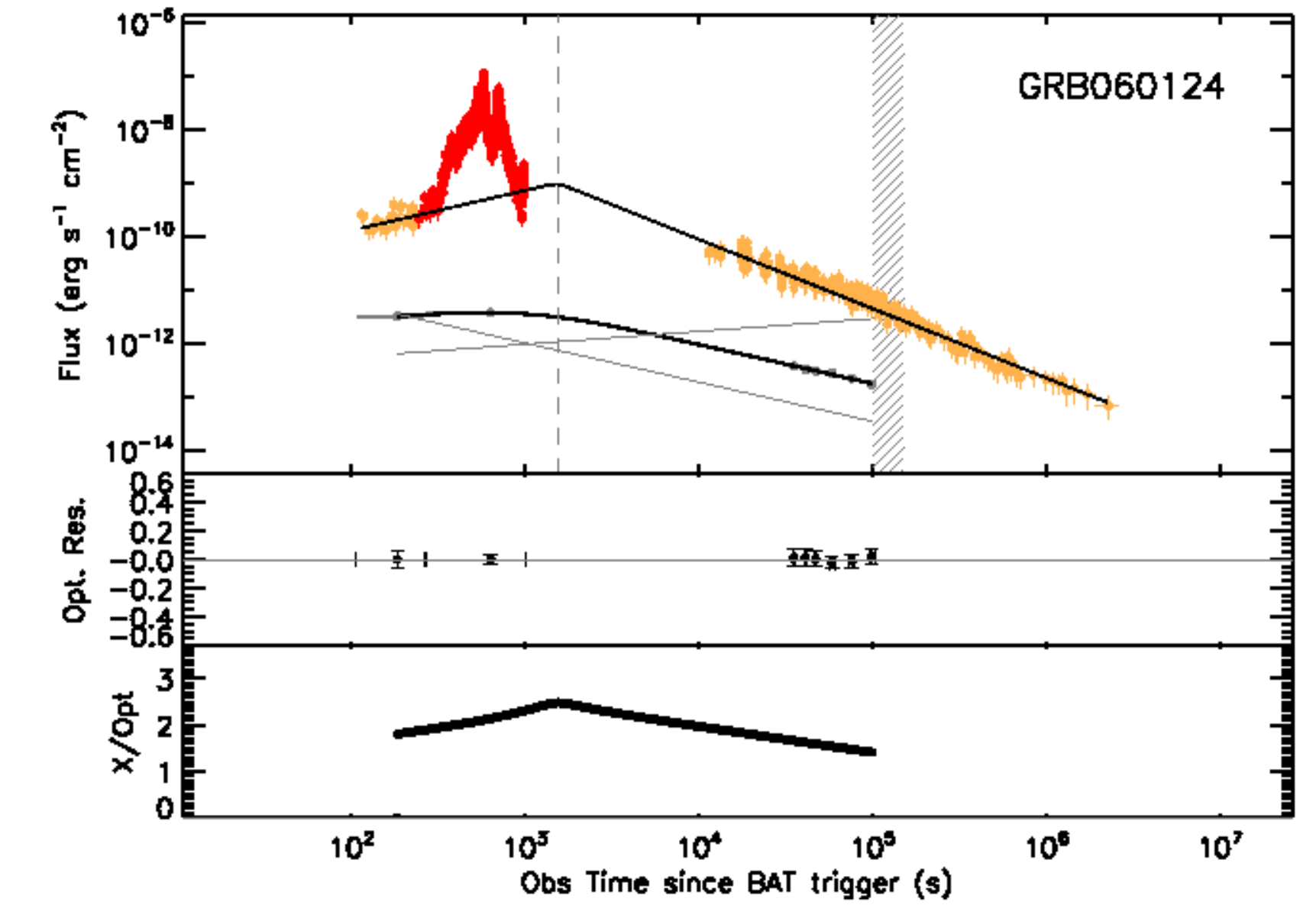}
\includegraphics[width=0.3 \hsize,clip]{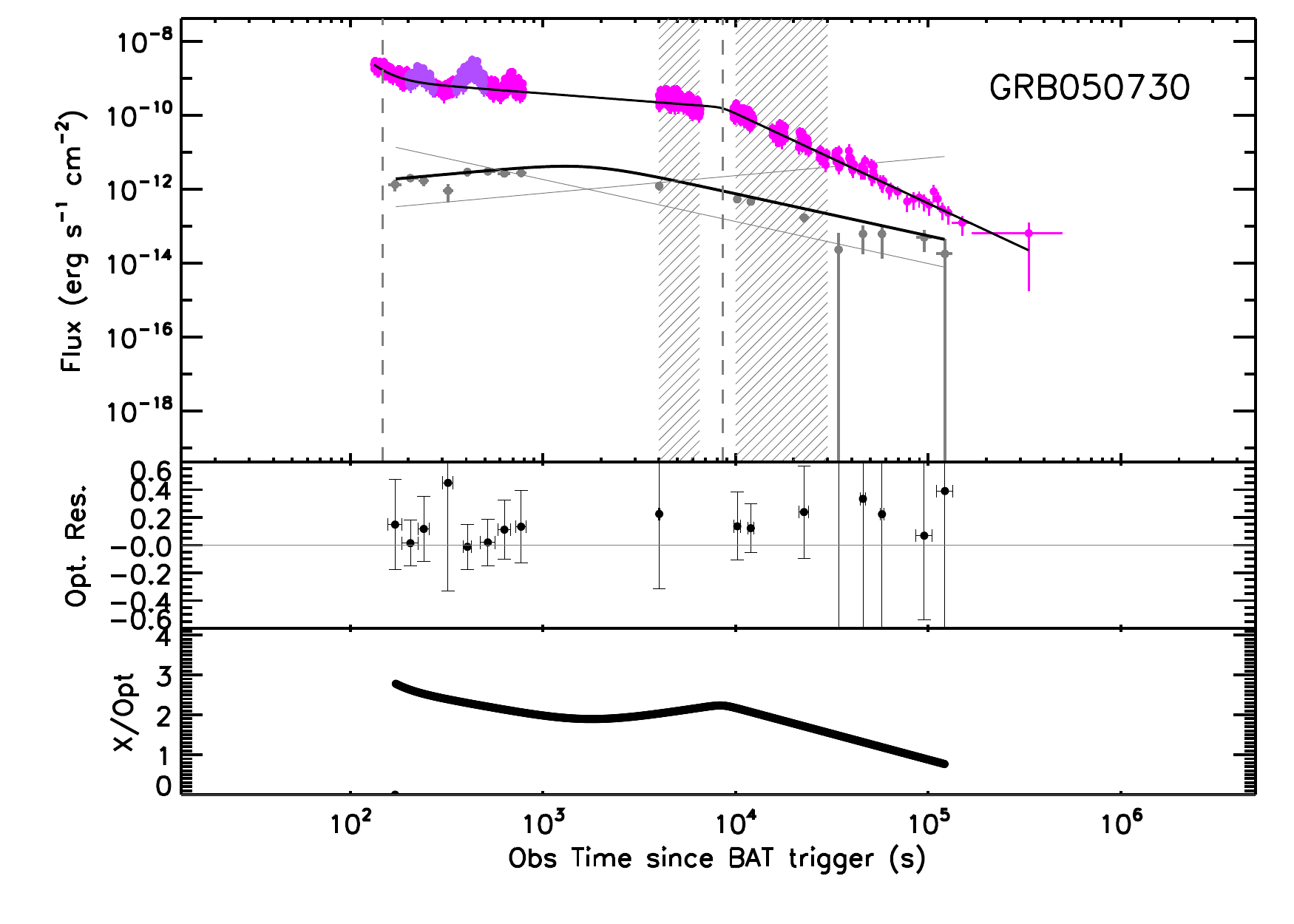}
\includegraphics[width=0.3 \hsize,clip]{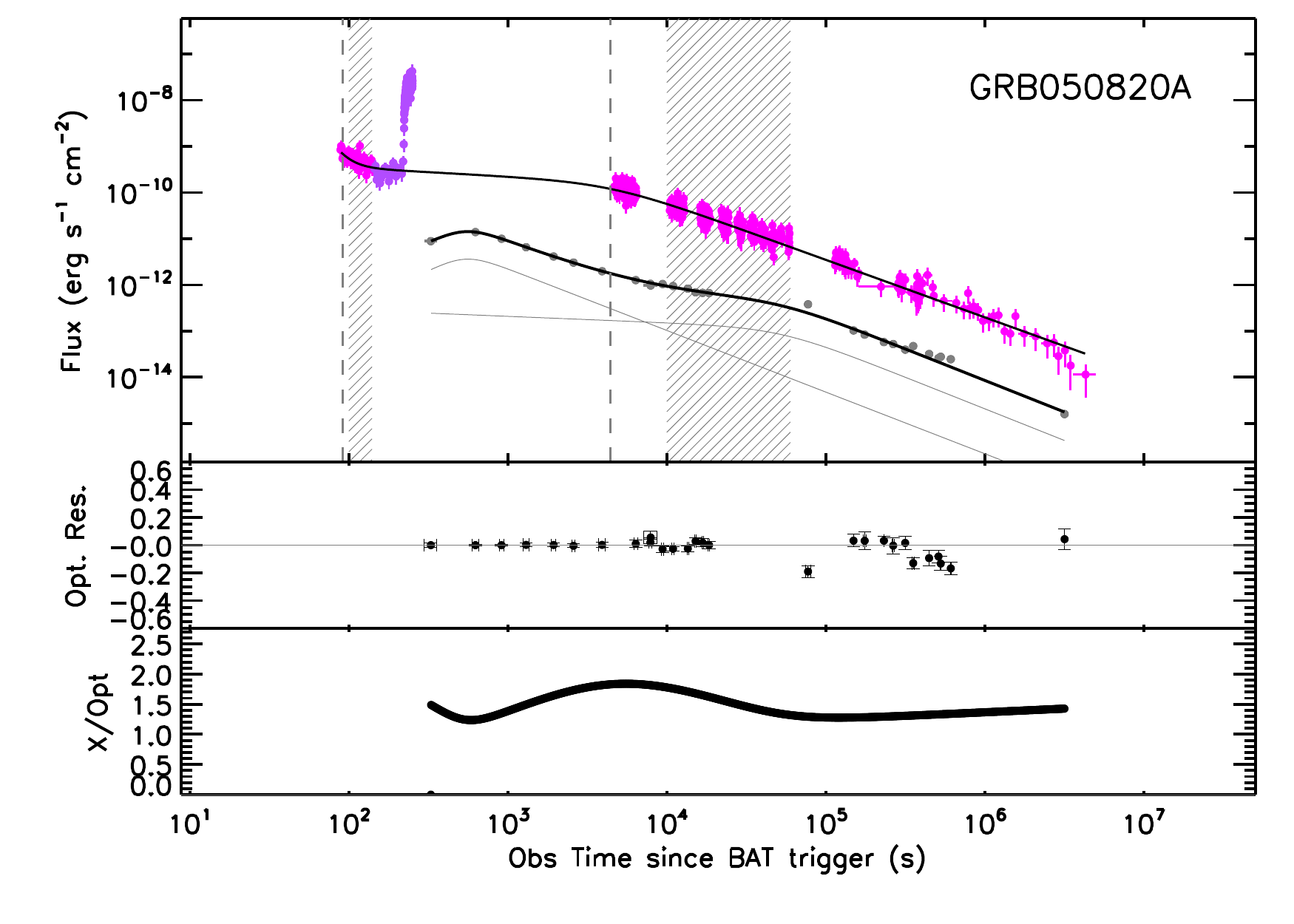}\\ 
\includegraphics[width=0.3 \hsize,clip]{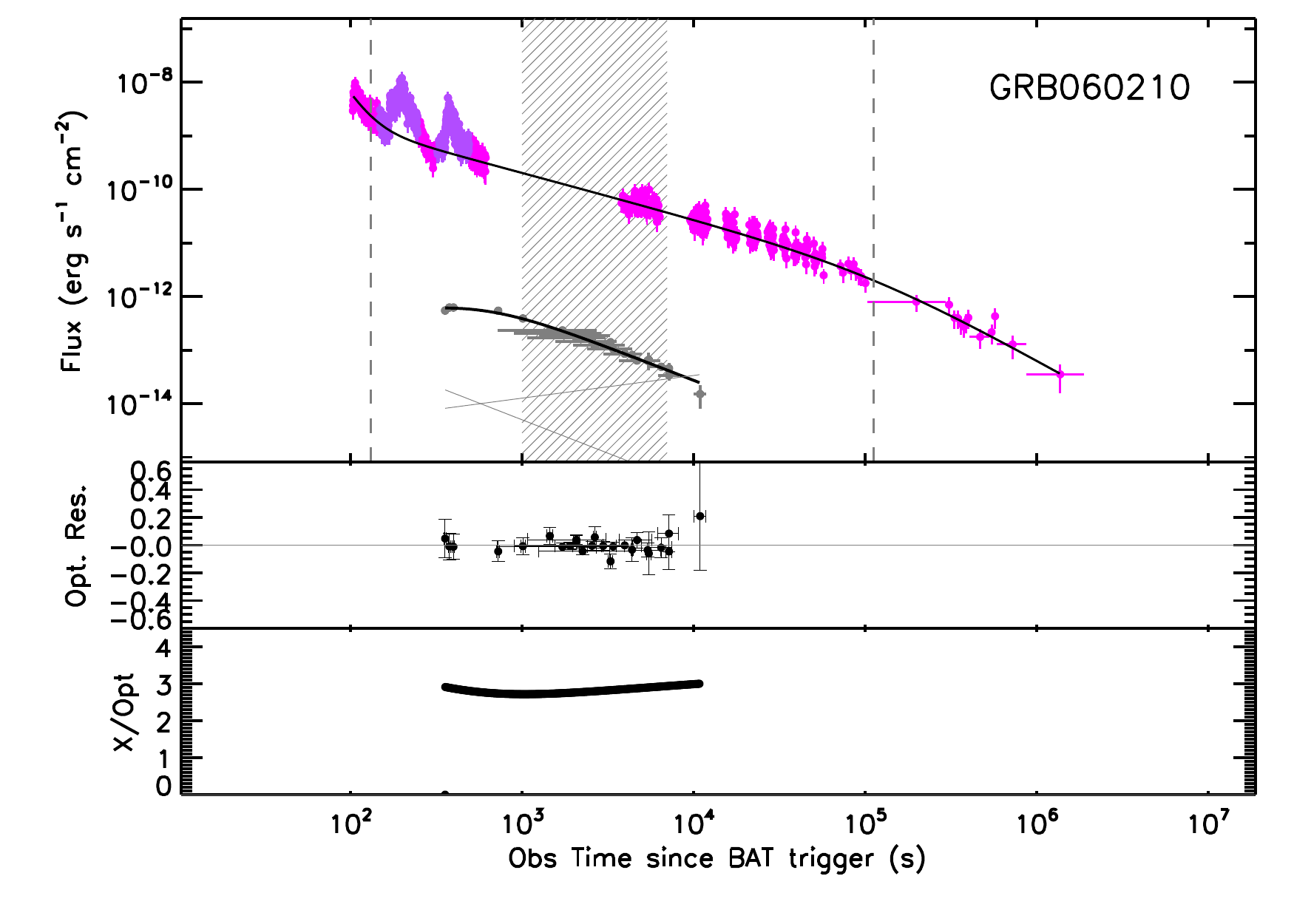}
\includegraphics[width=0.3 \hsize,clip]{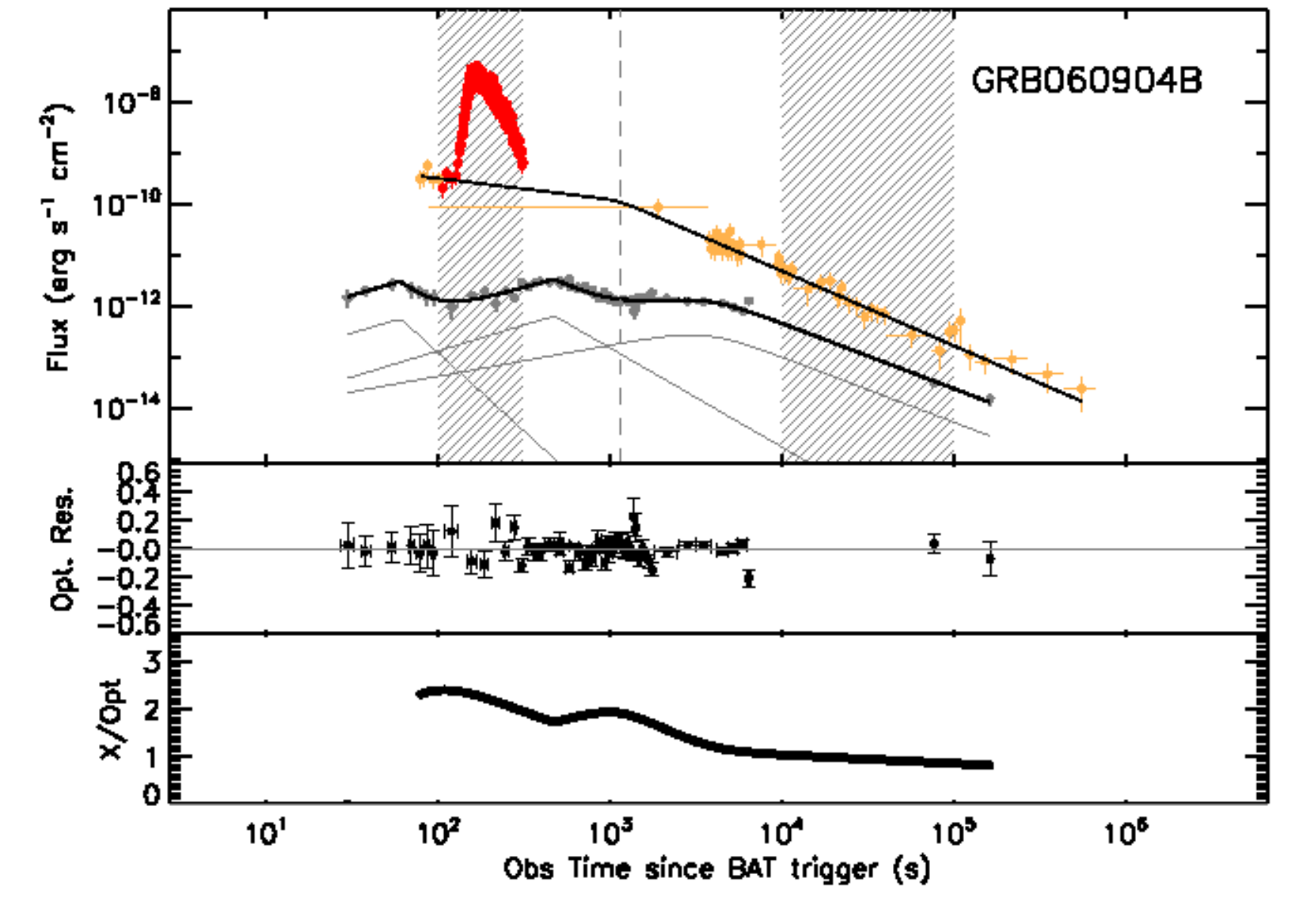}
\caption{\small{GRBs with X-ray flares during the plateau and the optical peak during or at the end of the X-ray plateau. Color code as in Figure~\ref{classificazione}.}}\label{ref2}
\end{figure*}

\begin{figure*}[!]
\centering
\includegraphics[width=0.3 \hsize,clip]{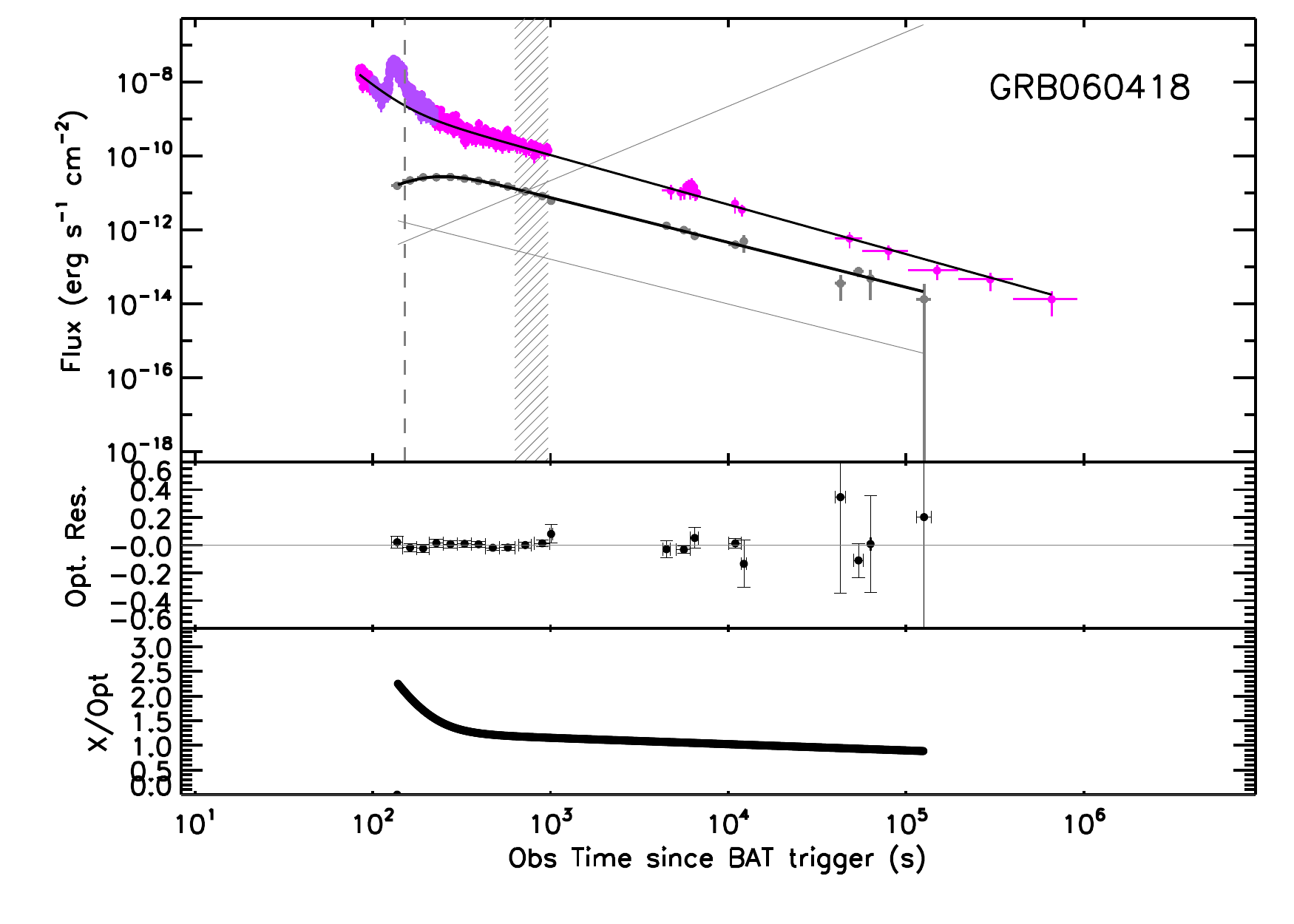}
\includegraphics[width=0.3 \hsize,clip]{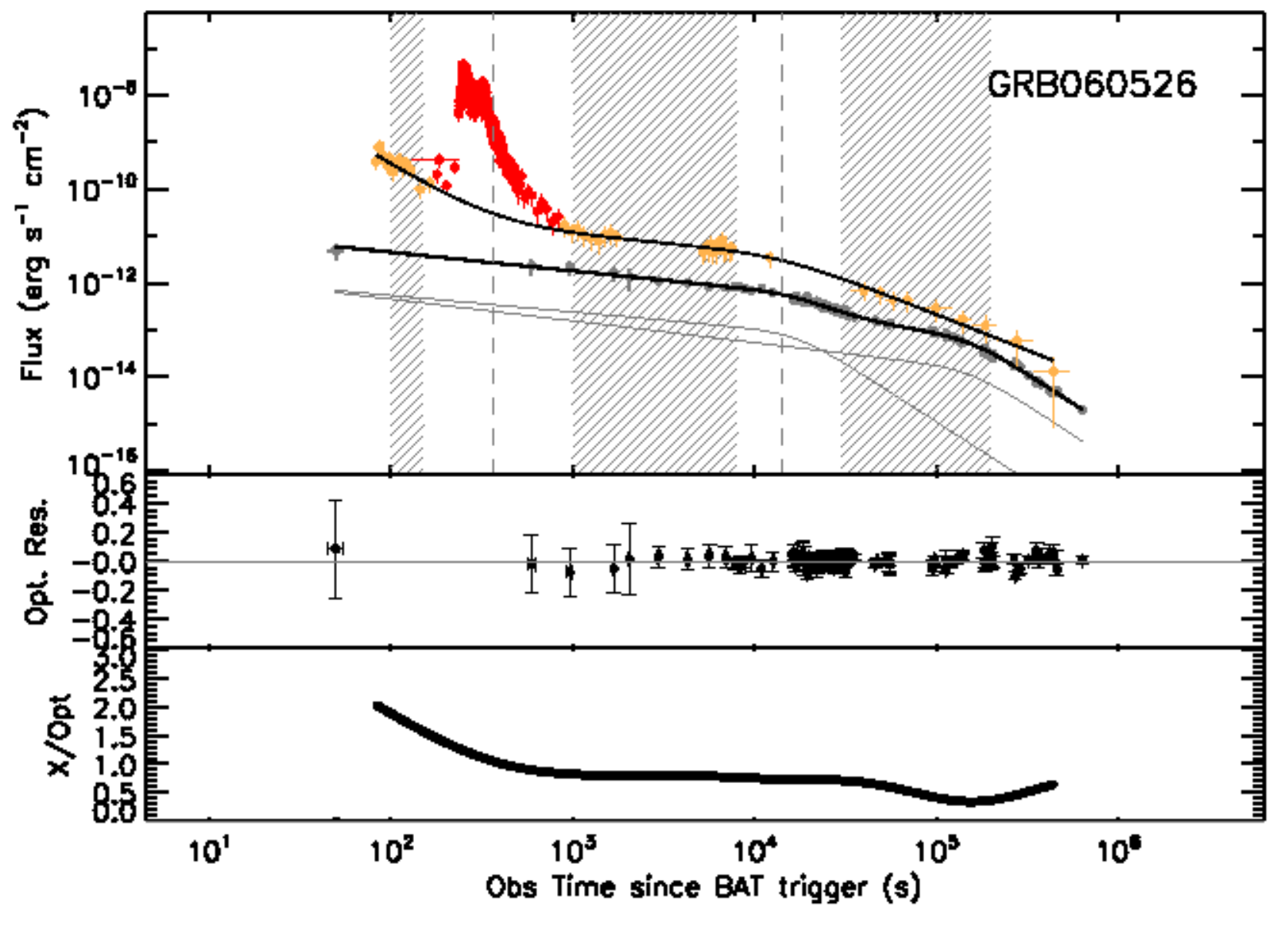}\\
\includegraphics[width=0.3 \hsize,clip]{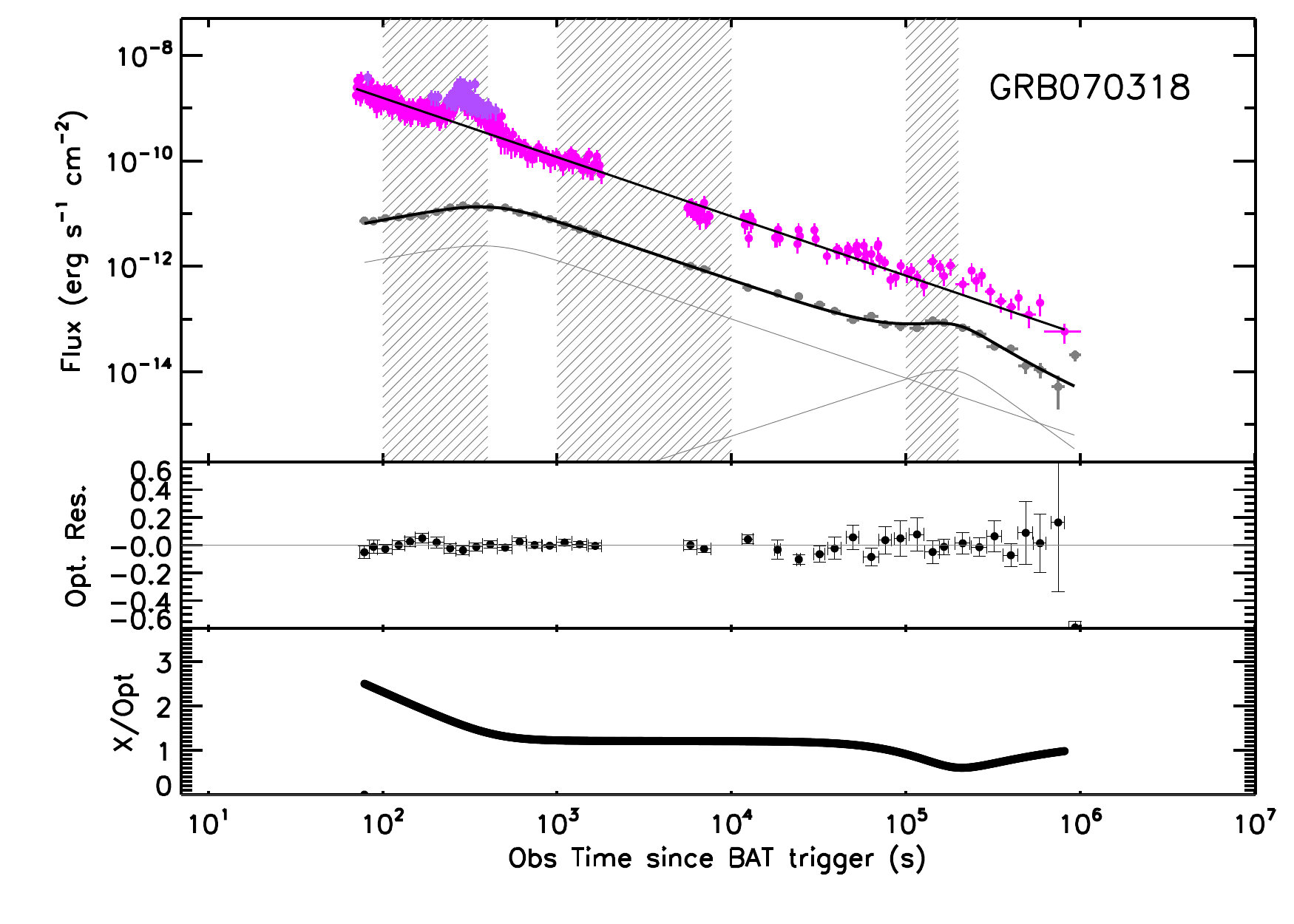}
\includegraphics[width=0.3 \hsize,clip]{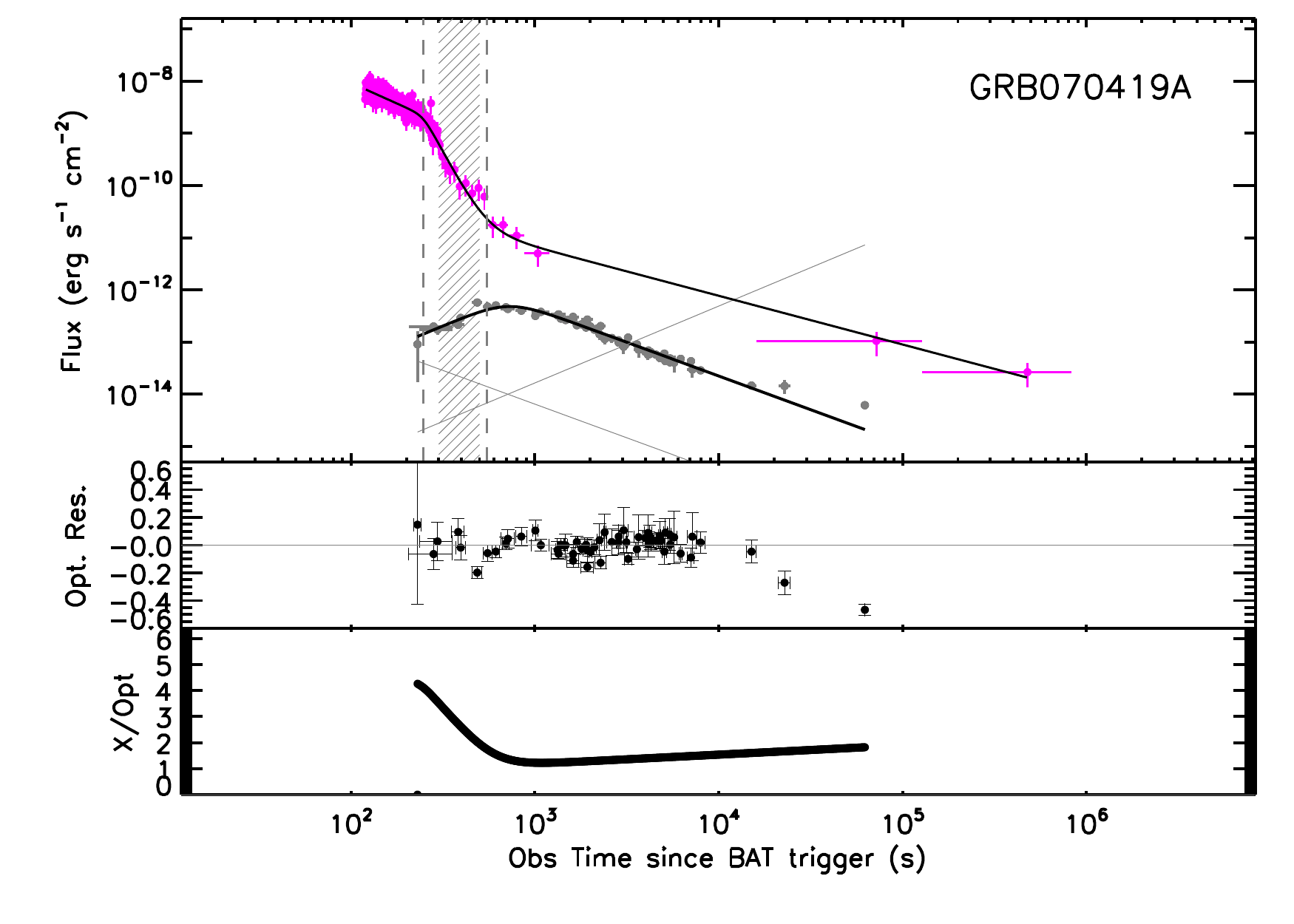}
\caption{\small{Particular cases. Color code as in Figure~\ref{classificazione}.}}\label{ref3}
\end{figure*}

From the comparison of the X-ray and optical LCs, we noted that there is a relation between the occurrence of the X-ray flares and the peak time of the optical LCs (Figures~\ref{ref1}-\ref{ref2}-\ref{ref3}). That is, when the flares are observed during the X-ray steep decay phase, the optical peak occurs early and before the beginning of the X-ray plateau, while if there are no flares or late-time flares, the optical peak occurs during the X-ray plateau.

The peak of the optical LC occurs during or at the end of the steep-decay phase if there are X-ray flares in this phase, as in GRB 080310, GRB 060512, GRB 061121, GRB 071031, GRB 081008, GRB 080928, GRB 060729, GRB 080607, GRB 060607A, and GRB 080810.

If the X-ray flares occur during the X-ray plateau, the optical LC peak or the end of the optical plateau occurs during the X-ray plateau: GRB 060124, GRB 050730, GRB 050820A, GRB 060210, GRB0 60904B, and GRB 060512. 
 
In some cases it is difficult to evaluate this relation between the X-ray flares and the optical LCs. For GRB 060418 the computation of the X-ray break time is influenced by the presence of a very bright X-ray flare (e.g. \citealt{2010MNRAS.406.2149M} for details). Accordingly, if we consider a break time $\sim$200 s, the peak of the optical LC would be synchronous with the end of the steep-decay phase, and in this case it would be part of the group of GRBs with the break of the optical LC occurring during or at end of the steep decay. The GRB 060526 \textit{v} filter data show a variability that corresponds with the X-ray flare, even if there is no true break. GRB 070318 has a Type 0 X-ray LC with a superimposed flare that temporally corresponds to the optical peak. At late time it shows another optical re-brightening that corresponds to a weak X-ray flux variation. GRB 070419A belongs to the 17/437 Type $IIb$ GRBs with complete LCs (M13). The end of the X-ray steep decay corresponds to the peak of the optical LC even though there are no flares during the steep decay. Unlike, the X-ray LC is not well sampled after the steep decay phase.

If there are no flares, the optical break occurs during the X-ray plateau. GRB 100418A and GRB 100901A simply follow the trend of the X-ray data. The LC of GRB 060614 is  similar to that of GRB 100418A and GRB 100901A, but there are no data during the X-ray steep decay. For GRB 050319, GRB 081203A, GRB 060605, GRB 060906, GRB 070802, and GRB 071025 some data are lacking during the plateau and the steep decay is not well sampled or is of short duration. GRB 071112C and  GRB 080413B have Type Ia X-ray LC (M13) and so there is no steep decay phase. GRB 081029 has no X-ray data before the optical peak. For GRB 091127 we have only late-time data ($t_{\rm start,X}\sim5\times10^3$ s). The optical peak of GRB 051109A corresponds to the end time of the X-ray plateau, but there are no X-ray data during the plateau. GRB 050408, whose observations started 2000 s after the trigger, has a Type $0$ LC and shows an optical break at about $2\times10^4$ s, even though the X-ray LC does not show flux variations.  

For GRB 050904, GRB 061007, GRB 080603A, GRB 050525A, GRB 080710, GRB 090313, GRB 090926A, GRB 070125, GRB 060729, GRB 080330, GRB 071003, GRB 071010A, GRB 070208, GRB 070411, and GRB 090426 the X-ray data are very poor.

The relation between the X-flares and the optical peaks and plateaus is displayed also in GRB 070110, GRB 080319A, GRB 081126, GRB 090812, and GRB 100906A studied by  \citet{2012ApJ...758...27L} and \citet{2012arXiv1210.5142L}.

%----------------
%----------------
%----------------
%----------------
%**********************************
\subsubsection{Normal decay phase}

The X-ray normal decay is present in most of the X-ray LCs of our sample (63/68). Seventy-seven per cent of the optical LCs decay during the X-ray normal phase, but only 62\% have a similar slope. In addition, only a few cases follow the closure relations.   

Twenty-three per cent of the optical LCs contemporaneous to the X-ray normal decay have a variable shape with bumps and late-time re-brightenings. Small bumps could represent late central engine activity (e.g. \citealt{2012ApJ...758...27L}) or a change in the circumburst density \citep{2002A&A...396L...5L}, while the late-time rebrightenings could be a result of the jet structure (e.g. \citealt{2012arXiv1210.5142L}).

The normal decay phase of the X-ray LC and the contemporaneous optical emission can be considered as afterglow emission, even though the central engine activity is revealed by the late-time X-ray flares (e.g. \citealt{2011A&A...526A..27B}) and optical bumps and re-brightenings.

%************************************************************
\subsection{Closure relations}\label{closure}

The standard fireball model predicts a link between the characteristic quantities of the spectrum and the LC of a GRB afterglow \citep{1998ApJ...497L..17S}. We compared the temporal decay and spectral indices derived from our analysis of the X-ray and optical LCs in the plateau and normal decay phases with these predictions (see also \citealt{2006ApJ...642..354Z}, their Table 2). We restricted our analysis to the slow-cooling regime, with either a constant ISM or a wind medium, including the possibility of energy injection\footnote{In particular, we considered the following cases:
\textit{a)} $\alpha_{\rm op}-\alpha_{\rm X}=0$: if there is no energy injection, $\alpha=(3\beta-1)/2$; if there is energy injection, $\alpha=(q-2+(2+q)\beta)/2$.
\textit{b)} $\alpha_{\rm op}-\alpha_{\rm X}=-1/4$ corresponds to a constant density medium in the slow cooling regime with no energy injection, so $\alpha_{\rm op}=3\beta_1/2$ and $\alpha_{\rm X}=(3\beta_2-1)/2$.
\textit{c)} $\alpha_{\rm op}-\alpha_{\rm X}=+1/4$ corresponds to a wind medium in the slow cooling regime, so $\alpha_{\rm op}=(3\beta_1+1)/2$ and $\alpha_X=(3\beta_2-1)/2$.
\textit{d)} $\alpha_{\rm op}-\alpha_{\rm X}=(q-2)/4$ corresponds to a constant density medium in the slow cooling regime with energy injection, so $\alpha_{\rm op}=(q-1)+(\beta_1/2)(2+q)$ and $\alpha_{\rm X}=((q-2)+\beta(2+q))/2$.
\textit{e)} $\alpha_{\rm op}-\alpha_{\rm X}=(2-q)/4$ corresponds to a wind medium in the slow cooling regime with energy injection, so $\alpha_{\rm op}=(q+\beta_2(2+q)\beta)/2$ and $\alpha_{\rm X}=((q-2)+\beta(2+q))/2$.
In the energy injection case, we considered three values of $q$, 0, 0.3, 0.5, which depend on the strength of the injection.}. 

Overall, we find that 4\% of the GRBs of our sample are consistent with the closure relations within 2 $\sigma$ during the plateau phase (GRBs in group B) and 12\% during the normal decay (4\% in group A and 8\% in group B). We define as $\alpha$ the value of the LC slope calculated fitting the LCs and $\tilde{\alpha}$ the LC slope expected following the closure relations, determined using the spectral indices computed with the SEDs. For Group A (defined in Sect.~\ref{comparison}), the closure relations are valid in some GRBs only for the normal decay phase:
\begin{itemize}
\item GRB 080607: its optical/X-ray SED, extracted during the normal decay phase, is fitted with a broken power-law function and follows the relation $\alpha_{\rm op}-\alpha_{\rm X}\sim1/4$ because $\alpha_{\rm op}=2.20\pm0.61$ and $\alpha_{\rm X}=1.70\pm0.07$ (constant-density medium in the slow cooling regime without energy injection). Therefore $\tilde{\alpha}_{\rm op}=(3\beta_1+1)/2$ and $\tilde{\alpha}_{\rm X}=(3\beta_2-1)/2$. From the spectrum $\beta_{\rm op}=0.80\pm0.10$ and $\beta_{\rm X}=1.30\pm0.10$, then the expected slopes are $\tilde{\alpha}_{\rm op}=1.70\pm0.15$ and $\tilde{\alpha}_{\rm X}=1.45\pm0.15$. 
\item GRB 050416A: during the normal decay, the best fit of the SED is a broken power-law. $\alpha_{\rm op}=0.86\pm0.15$ and $\alpha_{\rm X}=0.90\pm0.04$ and $\alpha_{\rm op}-\alpha_{\rm X}\sim\pm1/4$ within 2$\sigma$. This agrees with a slow cooling regime without energy injection in a constant density medium: $\tilde{\alpha}_{\rm op}=3\beta_{\rm op}/2=0.78\pm0.23$ and $\tilde{\alpha}_{\rm X}=(3\beta-1)/2=1.03\pm0.21$. 
\item GRB 100418A: the best-fitting function of the SED is a single power-law. $\alpha_{\rm op}=1.38\pm0.07$, $\alpha_{\rm X}=1.51\pm0.19$ and $\beta_{\rm op,X}=1.19\pm0.02$. If there is no energy injection, ISM, or wind medium, for $\nu_{\rm c}<\nu_{\rm op}<\nu_{\rm X}$ the expected slopes are $\tilde{\alpha}_{\rm op}=\tilde{\alpha}_{\rm X}=(3\beta_{\rm op,X}-1)/2=1.39\pm0.03$ in agreement, within errors, with our data.  
\end{itemize}

For Group B, for the normal decay:
\begin{itemize}
\item GRB 071003: the X-ray LC has a simple power-law shape, whereas the optical LC has an initial shallow decay followed by a normal decay. For this last part, $\alpha_{\rm op}=1.91\pm0.36$ and $\alpha_{\rm X}=1.60\pm0.06$. The best fit of the SED is a simple power-law ($\beta_{\rm op,X}=1.28\pm0.05$), so the expected $\tilde{\alpha}=1.42\pm0.08$ for no energy injection, ISM, or wind medium, and $\nu_{\rm c}<\nu_{\rm op}<\nu_{\rm X}$. This agrees with the X-ray and the optical slopes within 2$\sigma$.
\item GRB 080413B: for the normal decay phase, $\alpha_{\rm op}=1.84\pm0.08$ and $\alpha_{\rm X}=1.62\pm0.21$. The best fit of the SED is a broken power-law ($\nu_{\rm op}<\nu_{\rm c}<\nu_{\rm X}$). The expected X-ray slope is $\tilde{\alpha}_{\rm X}=1.36\pm0.36$ for the case without energy injection and $\tilde{\alpha}_{\rm op}=1.11\pm0.36$ for the slow cooling regime and  $\tilde{\alpha}_{\rm op}=1.61\pm0.36$ for the fast cooling regime. Therefore the normal decay observations agrees with an ISM medium, no energy injection and slow (2$\sigma$) or fast (1$\sigma$) cooling regime. 
\item GRB 080913: the X-ray LC is a simple power-law with $\alpha_{\rm X}=1.00\pm0.06$. The optical LC shows a late-time re-brightning, but at the beginning, between 600-1200 s after the trigger, $\alpha_{\rm op}=0.98\pm0.04$. The best fit of the SED is a simple power-law and follows the closure relation for the ISM or wind medium in the slow cooling regime and $\nu_{\rm m}<\nu_{\rm op}<\nu_{\rm X}<\nu_{\rm c}$; in fact $\tilde{\alpha}_{\rm op}=\tilde{\alpha}_{\rm X}=0.98\pm0.06$ in the ISM case and $\tilde{\alpha}_{\rm op}=\tilde{\alpha}_{\rm X}=1.06\pm0.05$ in the wind case.
\item GRB080928: for the normal decay, $\alpha_{\rm op}=2.38\pm0.65$ and $\alpha_{\rm X}=1.62\pm0.09$. The best-fit function is a power-law, hence there is one spectral index, $\beta_{\rm op,X}=1.15\pm0.02$. The expected slope both for the X-ray and the optical LC is $\tilde{\alpha}=1.69\pm0.02$ for the case of slow cooling, energy injection, and wind medium, and $\nu_{\rm m}<\nu_{\rm op}<\nu_{\rm X}<\nu_{\rm c}$.  $\tilde{\alpha}$ is consistent with $\alpha_{\rm X}$ (1$\sigma$) and $\alpha_{\rm op}$ (2$\sigma$).
\item GRB 081008: for the X-ray normal decay $\alpha_{\rm X}=1.32\pm0.08$ and the corresponding optical slope is $\alpha_{\rm op}=1.44\pm0.13$. The spectral index is $\beta_{\rm op,X}=0.93\pm0.01$. Without energy injection, ISM, and $\nu_{\rm m}<\nu_{\rm op}<\nu_{\rm X}<\nu_{\rm c}$, the expected slope is $\tilde{\alpha}=1.39\pm0.01$, which agrees with the optical and X-ray slopes of this GRB.
\end{itemize}

For Group B, for the plateau:
\begin{itemize}
\item GRB 060502A: $\alpha_{\rm op}=0.50\pm0.05$ and $\alpha_{\rm X}=0.45\pm0.13$. The best fit of the SED is a broken power-law, hence $\beta_{\rm op}=0.51\pm0.12$ and $\beta_{\rm X}=1.01\pm0.12$. For the case of energy injection, ISM, and slow cooling regime, we obtain $\tilde{\alpha}_{\rm op}=0.14\pm0.15$ and $\tilde{\alpha}_{\rm X}=0.51\pm0.16$. These values agree with the optical (3$\sigma$) and the X-ray slope (1$\sigma$).
\item GRB 060512: $\alpha_{\rm op}=1.09\pm0.15$ and $\alpha_{\rm X}=1.08\pm0.09$. The best fit of the SED is a single power-law ($\beta_{\rm op,X}=1.24\pm0.05$). Therefore we obtain $\tilde{\alpha}=1.05\pm0.06$, which is consistent with the optical and X-ray slopes. This result was obtained considering energy injection, ISM, slow cooling regime, and $\nu_{\rm m}<\nu_{\rm op}<\nu_{\rm X}<\nu_{\rm c}$.
\item GRB 071031: $\alpha_{\rm op}=0.93\pm0.03$ and $\alpha_{\rm X}=1.02\pm0.18$. The best fit of the SED is a single power-law with $\beta_{\rm op,X}=0.99\pm0.01$. The expected LC slope is $\tilde{\alpha}=0.99\pm0.02$ for the case without energy injection and $\nu_{\rm c}<\nu_{\rm op}<\nu_{\rm X}$. This value agrees with the optical LC slope (2$\sigma$) and the X-ray slope (1$\sigma$).
\end{itemize}

We found no GRB that followed the closure relations for all phases of the LC, indeed, all GRBs of our sample are inconsistent with the closure relations at least in one of the phases of their LC.

The closure relations correspond to the broadband spectrum and LC of synchrotron radiation from a power-law distribution of electrons in an adiabatically expanding relativistic shock, as expected in the standard afterglow theory. The inconsistency with the data implies that this model, at least in its  simplest formulation, is unlikely to produce the observed afterglow emission. Since optical and X-ray observations are well fitted with a synchrotron spectrum, a more complex hydrodynamic evolution of the outflow must be considered, or alternatively, a direct influence from central engine.

%************************************************************
\subsection{Broadband SEDs}\label{radio_sect}
From the GRB sample with radio data presented by \citet{2012ApJ...746..156C} we selected two GRBs (GRB 071003 and GRB 090313) whith radio observations contemporaneous with the optical/X-ray data presented here to test the broadband behavior of the SEDs. 

\begin{figure}[!]
\centering
   \resizebox{\hsize}{!}{\includegraphics{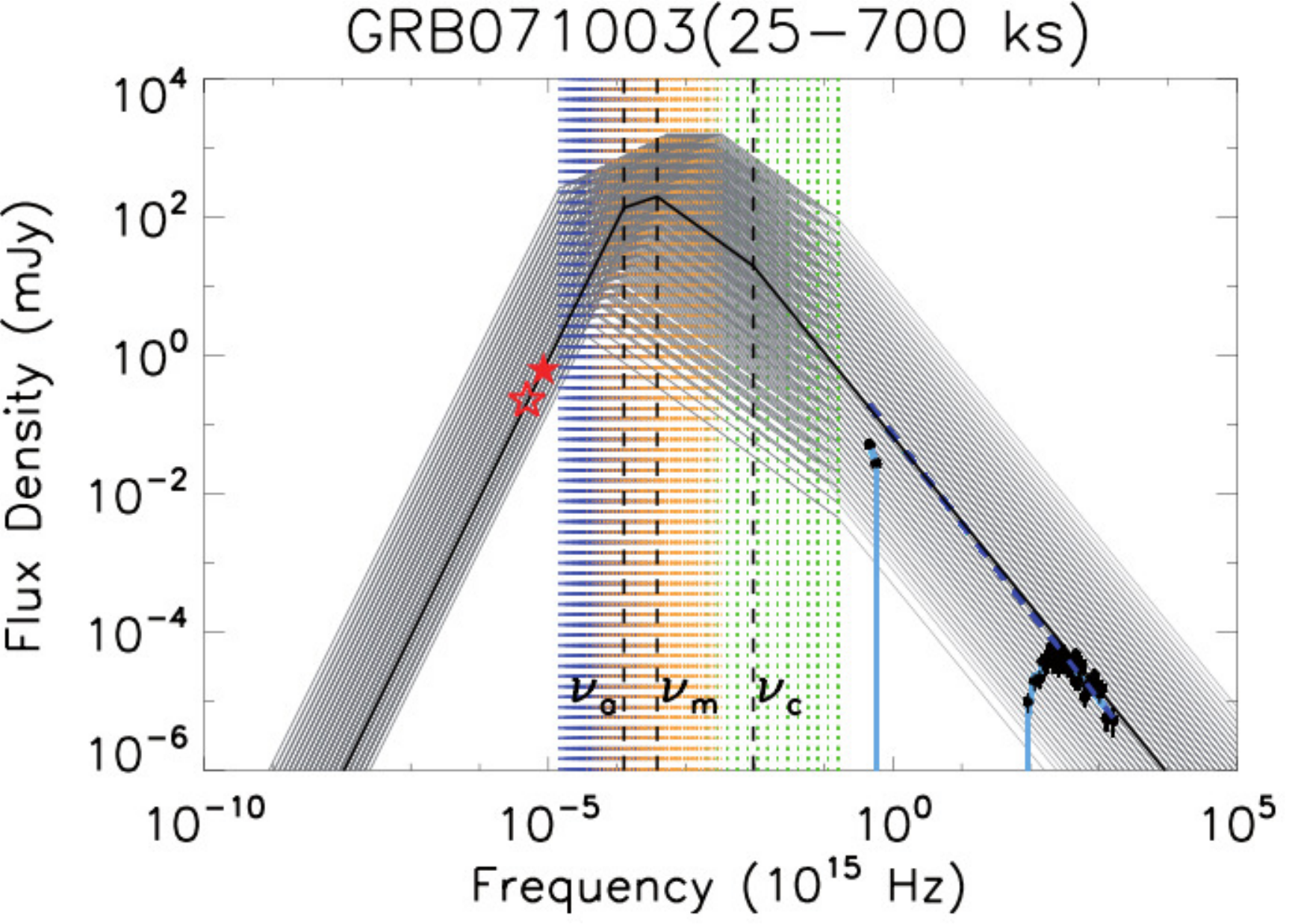}}
      \resizebox{\hsize}{!}{\includegraphics{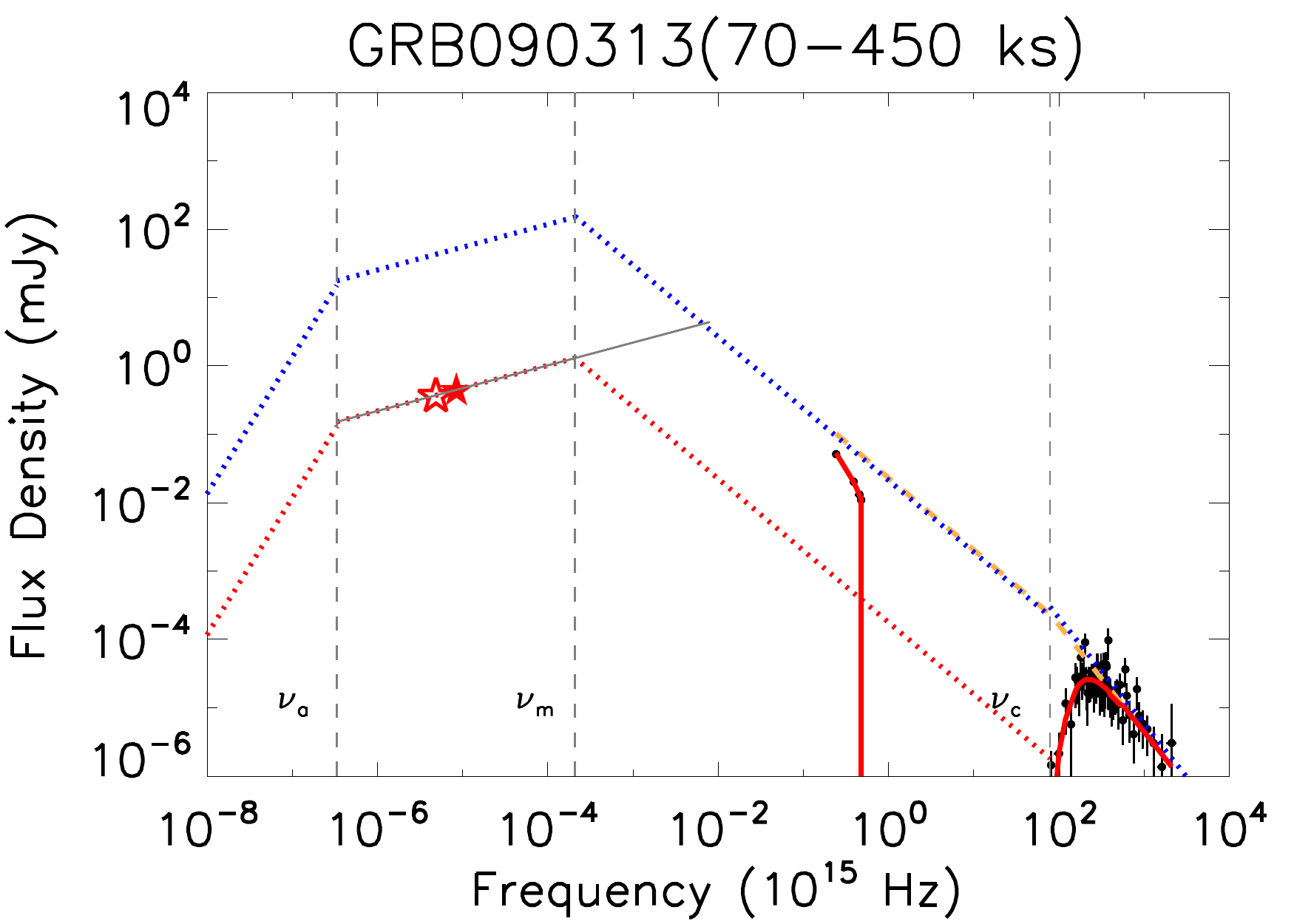}}
     \caption{\small{Radio/optical/X-ray SEDs. \textit{Filled (empty) red star}: radio data (upper limit). \textit{Black dots}: optical and X-ray data.  \textit{Upper panel}: GRB 071003. \textit{Light blue solid line}: optical/X-ray SED. \textit{Blue dashed line}: power-law that fits the data. \textit{Black solid line}: radio/optical/X-ray SED fit function. \textit{Black dashed line}: the absorption frequency ($\nu_a$), the synchrotron frequency ($\nu_m$), and the cooling frequency ($\nu_c$). \textit{Gray lines}:  radio/optical/X-ray SED tests with relative absorption frequencies (\textit{blue dashed lines}), synchrotron frequencies (\textit{orange dashed lines}), and cooling frequencies (\textit{green dashed lines}). \textit{Lower panel}: GRB 090313. \textit{Red solid line}: the optical/X-ray SED fit function. \textit{Orange dashed line}: broken power-law that fits the data. \textit{Blue dotted line}: radio/optical/X-ray SED calculated considering $\nu_{BR}=\nu_{c}$. \textit{Red dotted line}: radio/optical/X-ray SED calculated using the same characteristic frequencies as the red dotted line, but considering the normalization for the radio data. \textit{Gray dashed line}: the absorption frequency ($\nu_a$), the synchrotron frequency ($\nu_m$), and the cooling frequency ($\nu_c$).}}
      \label{radio}
\end{figure}

For GRB 071003 we calculated a set of illustrative radio/optical/X-ray SEDs starting from different values of the cooling frequency ($\nu_{\rm c}$) and the maximum flux ($F_{\rm \nu,max}$), as presented by \citet{1998ApJ...497L..17S} for the slow cooling regime\footnote{For an adiabatic evolution, the equations that describe the emitted radiation are \citep{1998ApJ...497L..17S,2000ApJ...534L.163G}: $F_{\nu}=(\nu_{\rm a}/\nu_{\rm m})^{1/3}(\nu/\nu_{\rm a})^{2}F_{\rm \nu,max}$ for $\nu<\nu_{\rm a}$; $F_{\nu}=(\nu/\nu_{\rm m})^{1/3}F_{\rm \nu,max}$ for $\nu_{\rm m}>\nu$; $F_{\nu}=(\nu/\nu_{\rm m})^{-(p-1)/2}F_{\rm \nu,max}$ for $\nu_{\rm c}>\nu>\nu_{\rm m}$; $F_{\nu}=(\nu_{\rm c}/\nu_{\rm m})^{-(p-1)/2}(\nu/\nu_{\rm c})^{-p/2}F_{\rm \nu,max}$ for $\nu>\nu_{\rm c}$. The absorption frequency is $\nu_{\rm a}=0.247(4.24\times10^9)((p+2)/(3p+2))^{3/5}((p-1)^{8/5}/(p-2))\epsilon_{\rm e}^{-1}\epsilon_{\rm B}^{1/5}E_{52}^{1/5}n_{1}^{3/5}$ Hz; the synchrotron frequency $\nu_{\rm m}=(5.7\times10^{14})\epsilon_{\rm B}^{1/2}\epsilon_{\rm e}^{2}E_{52}^{1/2}t_{\rm d}^{-3/2}$ Hz; the cooling frequency $\nu_{\rm c}=(2.7\times10^{12})\epsilon_{\rm B}^{-3/2}E_{52}^{-1/2}n_{1}^{-1}t_{\rm d}^{-1/2}$ Hz; the maximum flux, which is the flux at $\nu_{\rm m}$, $F_{\rm \nu,max}=(1.1\times10^{5})\epsilon_{\rm B}^{1/2}E_{52}n_{1}^{1/2}D_{28}^{-2}$ $\mu$Jy. $t_{\rm d}$ is the time in days, $E_{52}=E/10^{52}$ ergs, the density $n_1$ in cm$^{-3}$, the distance $D_{28}=D/10^{28}$ cm.}.  The optical/X-ray SED was fitted with a simple power-law, so the cooling frequency is probably below the optical band. In this way we found a possible set of data that is consistent with the radio/optical/X-ray SED is: $\nu_{\rm a}=1.2\times10^{11}$ Hz, $\nu_{\rm m}=3.08\times10^{11}$ Hz, $\nu_{\rm c}=9.45\times10^{12}$ Hz and $F_{\rm \nu,max}=198$ mJy (Figure~\ref{radio}). The total energy, calculated in the frequency interval $\nu=10^{7}-10^{19}$ Hz, is $2.72\times10^{52}$ erg. For the different parts of the spectrum, the energy is $E_{10^7\rm Hz-\nu_{\rm a}}=1.25\times10^{50}$ erg, $E_{\nu_{\rm a}-\nu_{\rm m}}=9.56\times10^{50}$ erg, $E_{\nu_{\rm m}-\nu_{\rm c}}=8.59\times10^{51}$ erg, and $E_{\nu_{\rm c}-10^{19}Hz}=1.76\times10^{52}$ erg.

This example shows that the synchrotron model can adequately fit the GRB spectral properties at late time.

For GRB 090313 we know the cooling frequency, which is the break frequency calculated by fitting the optical/X-ray SED with a broken power-law. At this time the data are unlikely to be influenced by the host galaxy, and we cannot reconstruct the radio/optical/X-ray SED, as shown in Figure~\ref{radio}.  

%----------------
%----------------

%************************************************************
\subsection{Three-parameter correlation}\label{3parametri}

In M13 we found a three-parameter correlation, which we discussed in \citet{2012MNRAS.425.1199B}. It involves the isotropic energy emitted in the rest-frame 1-10$^4$ keV energy band during the prompt emission (E$_{\gamma,iso}$), the peak of the prompt emission energy spectrum (E$_{pk}$) and the X-ray energy emitted in the 0.3-30 keV observed energy band (E$_{X,iso}$), which is integrated over the observed duration of each LC. The X-ray energy was calculated in a specific energy band, and we did not extrapolate the spectrum to lower energies because we did not know the behavior of the spectrum at those energies. In this work we have presented the GRB spectra covering the energy band from the infrared to X-rays and calculated the models that fit the data better. Using these models, for every GRB we calculated the ratio between the X-ray energy band (0.3-30 keV) and the total energy emitted from the IR to X-rays: $\sim$70\% of the total energy is radiated in the X-rays. The mean value of the IR-optical-UV energy\footnote{We considered the 0.5-6 keV band.} is $\lesssim$10\%.

The correlation is not quantitatively modified by including the optical energy. However, since the origin of the correlation is still unknown, it is not clear if including the optical energy is conceptually necessary or not.
  
%%%%%%%%%%%%%%%%%%%%%%%%%%%%
\section{Conclusions}\label{conclusioni}

We performed a systematic analysis of a large sample of well-monitored optical LCs (68 GRBs) obtained with different instruments and compared them with the X-ray emission. Since the GRBs of this sample have known redshift, we considered their rest frame properties. From this large sample of observations: 

\begin{itemize}
\item we fitted the optical LCs collected from different instruments and the optical/X-ray SEDs at different times for every GRB;
\item we studied the distribution of the parameters computed from the SEDs, which yield information on the medium and the spectrum. We found that \textit{a)} there is a slight \textit{softening} of the optical/X-ray spectrum with time, \textit{b)} the gas-to-dust ratios ($NH/A_{\rm V}$) of GRBs are higher than the values calculated for the MW, the LMC, and SMC assuming subsolar abundances (e.g. \citealt{2010MNRAS.401.2773S,2012A&A...537A..15S}), and \textit{c)} the break frequencies are spread between the optical and X-ray bands;
\item for a given rest frame time range (920-12000 s), the optical flux in the $R$ band is $\sim$2 orders of magnitude brighter than the X-ray flux in the 1 keV band;
\item there is a correlation between the energy of the peak of the optical LCs and E$_{\gamma,iso}$ and E$_{\gamma}^{15-150}$, confirming the result of \citet{2011MNRAS.414.3537P}. The optical plateau end luminosity and rest frame time are correlated, and similarly, the peak luminosity and the peak times are correlated (e.g. \citealt{2011MNRAS.414.3537P});
\item the optical LCs have a complex shape, with initial plateaus and peaks, bumps, and late-time re-brightenings. With time, the complexity of the LCs decreases and more and more of them decay;
\item only for 13 GRBs of our sample (Group A), all LC segments follow $\alpha_{\rm op}-\alpha_{\rm X}=0,\pm1/4$\footnote{We did not consider the steep decay, since during this phase the emission from the two bands clearly comes from different mechanisms or emitting regions.}, in the other cases the X-ray and optical LCs show a different behavior;
\item the onset of the forward shock observed in the optical LCs could be linked to the presence of the X-ray flares. Indeed when there are X-ray flares during the steep decay, the optical LC peak or the end of the initial plateau occurs during or just at the end of the X-ray steep-decay phase, while if there are no flares or the flares take place during the X-ray plateau, the optical peak or plateau end occurs during the X-ray plateau;
\item the forward shock model cannot explain all features of the optical and X-ray LCs, such as bumps, flares, re-brightenings, steep decays, and plateaus, either at early or late time;
\item the contribution of the optical energy to the three-parameter correlation (M13, \citealt{2012MNRAS.425.1199B}) is low, less than 10\% of the total energy emitted from the IR and X-rays. Therefore the correlation is not quantitatively modified by including the optical energy. However, since the origin of the correlation is still unknown, it is not clear whether including the optical energy is conceptually necessary or not.    
\end{itemize}

From this study it clearly emerges that the complex shapes of the optical and X-ray LCs cannot be explained simply by the forward shock, although we can confirm that the synchrotron is a viable emission mechanism for GRBs at late times. Moreover, we showed the importance of a systematic analysis of the GRB multi-wavelength observations. To improve the knowledge of the physics of GRBs and their origin, very fast detectors and multi-wavelength observations are needed. In particular, we need an optical follow-up starting the latest a few seconds after the trigger to collect homogeneous spectroscopic and photometric data (e.g. \citealt{2010arXiv1005.1569C}).

%%%%%%%%%%%%%%%%%%%%%%%%%%%%
\appendix
%************************************************************
\section{Optical data conversion factors}\label{appendice1}

In the literature data are generally given in different photometric systems and based on the Vega or AB magnitude convention. The Vega system is defined as the system for which the magnitude of the Vega star is zero: $m-0=-2.5\log f_{\lambda}+2.5\log f_{\rm \lambda,Vega}$, with $f_{\lambda}$ and $f_{\rm \lambda,Vega}$ in erg cm$^{-2}$ s$^{-1}$ \AA$^{-1}$. In this case, the conversion formula used is $f_{\lambda}=f_{\lambda,Vega}10^{-0.4\ m}$. The relation between flux density in frequency and wavelength units is $f_{\nu}=\lambda^{2} f_{\lambda}/c$, with $c$ the speed of light and $\lambda$ the central wavelength of the considered filter. For the AB system the reference spectrum is flat in units of frequency density (erg cm$^{-2}$ s$^{-1}$ Hz$^{-1}$) and the system is defined by $m^{\rm Vega}_{\rm V} \equiv m^{\rm AB}_{\rm V}\equiv0$ in the visual band; this occurs at 3631 Jy. So $m-m_{\rm AB,0}=-2.5\log f_{\nu}-2.5 \log(3631 \times 10^{-23})=-2.5\log f_{\nu}-(48.585\pm0.005)$, with $f_{\nu}$ in erg cm$^{-2}$ s$^{-1}$ Hz$^{-1}$. The conversion formula is $f_{\nu}=10^{-0.4(m+48.585)}$. In various articles, the photometric system is not always specified, which leads to mistakes in converting from magnitude to flux. For UVOT data the calibration has been provided by the UVOT team, and we used their conversion factors \citep{2008MNRAS.383..627P}. Some errors might occur when the photometric system is not specified, especially for the Gunn system or the standard Johnson system; for example, for the $R$ filter, the Gunn conversion factor is twice as much the standard system conversion factor. On the other hand, the IR filters used in literature are very similar. In this case, the choice of the photometric system does not noticeably influence the following analysis, and we preferred to choose only standard systems listed in Table~\ref{conversion_factors} (\textit{Online Material}).
%************************************************************
\section{Description of the GRBs in the three groups}\label{gruppi_descrizione}

\textbf{Group A.} All pairs of optical/X-ray slopes of the same GRB satisfy the relation $\alpha=0,\pm1/4$. The X-ray plateau and normal decay phase follows the optical slopes for seven GRBs: GRB 080607, GRB 100901A, GRB 050416A, GRB 080310, GRB 080319B, GRB 100418A, and GRB 050824.

For six GRBs we can compare the optical and X-ray LCs only during the X-ray plateau or X-ray normal decay, because of the poor data. GRB 060912A, GRB 061007, and GRB 080913 have a Type 0 complete X-ray LC. GRB 060912A has a simple power-law optical LC. The GRB 061007 optical LC rises at early time, where there are no X-ray observations, and then follows the normal decay of the X-ray LC. Finally, the GRB 080913 optical LC traces the X-ray LC during the first and the last part for its optical LC. Unfortunately, there are no optical observations between 10$^4$ and 10$^5$ s. For GRB 080603A there are no observations before 10$^4$ s; GRB 050904 ($\alpha_{\rm op}-\alpha_{\rm X}=0.25$) has a Type $Ib$ X-ray LC, so there is only the normal decay; for GRB 080330 there are no observations after the plateau phase.

\textbf{Group B.} Some pairs of optical/X-ray slopes of the same GRB satisfy the relation $\alpha=0,\pm1/4$. This group can be subdivided into two classes according to the LC shape: six GRBs have a similar shape for the X-ray and optical LCs, while 21 GRBs have different shapes.

GRBs with similar optical/X-ray LC shapes are GRB 051109, GRB 060526, GRB 071010A, GRB 060124, GRB 070208, and GRB 090618, where the optical and X-ray LC coincide during the X-ray plateau. In particular: the 060526 optical LC traces the X-ray LC at late time, even if the X-ray LC has few data points; the 090618 optical LC shows the presence of a supernova (SN) at late time \citep{2010ApJ...708L.112D,2011MNRAS.413..669C}, so at late time optical and X-ray LCs do not coincide.

We divided GRBs with different optical-X-ray LC shapes into four subgroups: optical LC more complex than X-ray LC (five GRBs), X-ray LC more complex than optical LC (five GRBs), optical bumps (two GRBs), peculiar cases (nine GRBs). In the first subgroup are GRB 050319B (coincidence during the X-ray plateau phase), GRB 061121, and GRB 050408 (coincidence during the X-ray normal decay phase); the GRB 080413B optical LC resembles the canonical X-ray shape and the first and last phases decay as the X-ray LC and the central part shows a plateau. The GRB 071031 normal decay has the same slope for the optical and X-ray LCs; we cannot conclude about the emission mechanism during this phase because the SEDs are too uncertain. There are two bumps superimposed on this segment. In the second subgroup are GRB 050401 and GRB 060502A (correspondence during the X-ray plateau phase), GRB 060908, GRB 060614, and GRB 091018 (correspondence during the X-ray normal decay phase). In the third group (GRB 060904B, GRB 080928) are the GRBs that show optical bumps that coincide with the X-ray steep and plateau phase; the optical and X-ray LC coincide during the normal decay phase (after the optical bumps). In the fourth subgroup are GRB 050525A\footnote{In this case there is coincidence because the X-ray LC plateau slope has a relatively large error: $\alpha_{\rm X,PL}=0.600\pm0.648$ and $\alpha_{\rm op}=1.150\pm0.036$.}, GRB 060512, GRB 080710, and GRB 090313, whose LCs coincide during the plateau phase; GRB 090926A, GRB 071003, GRB 081008, and GRB 081203A, whose LCs agree during the normal decay. The slopes of GRB 060729 optical LC are similar to the slopes of the X-ray LC during the plateau and the normal decay, but there is an optical re-brightening coincident with the end of the X-ray plateau.

\textbf{Group C.} No pairs of optical/X-ray slopes of the same GRB satisfy the relation $\alpha=0,\pm1/4$. GRB 050730, GRB 060210, GRB 060605, GRB 060607A, GRB 070802, and GRB 080810 have type $IIa$ X-ray LC and $Ia$ optical LC (four with X-ray flares); 070529 has type $IIa$ X-ray LC and $Ib$ optical LC. Nine GRBs have a simple power-law X-ray LC (six truncated\footnote{As defined in M13, complete X-ray LCs correspond to GRBs re-pointed by XRT at t$_{\rm rep}<$300 s for which we were able to follow the fading of the XRT flux down to a factor $\sim$5-10 from the background limit (or, equivalently, $t_{\rm end}\ge 4\times 10^{5}$ s). The LCs do not follow this criterion are classified as ``truncated''.\label{fn:completezza}}): GRB 051111, GRB 061126, GRB 070125, GRB 070411, GRB 090102, GRB 090426, GRB 091127, GRB 070318, and GRB 060206. Eight GRBs have a Type $Ia$ X-ray LC (four complete\footref{fn:completezza}): GRB 050908, GRB 060927, GRB 080721, GRB 081029, GRB 090424, GRB 050820A, GRB 050922C, and GRB 071112C. Only two GRBs have Type $Ib$ X-ray LC (GRB 060418, GRB 071025) and one GRB has Type $IIb$ X-ray LC (GRB 070419B). GRB 060906 is observed in the optical band only during the X-ray plateau phase and it is not a regular shape, but there are optical bumps.

%%%%%%%%%%%%%%%%%%%%%%%%%%%%
\begin{acknowledgements}
We thank the anonymous referee for the helpful comments that have improved this paper. EZ thanks Daniele Malesani for the useful discussions, suggestions and support during the preparation of the paper; Paolo D'avanzo and Andrea Melandri for the useful advices; Thomas Kr\"uler for sharing the data of GRB 070208 and Fang Yuang for the data of GRB 081008; Craig B. Markwardt for the help with the MPFIT routine. This research has made use of the XRT Data Analysis Software (XRTDAS) developed under the responsibility of the ASI Science Data Center (ASDC), Italy. This work was supported by ASI grant Swift I/011/07/0 and in part by I/004/11/0, by Ministero degli Affari Esteri and the University of Milano - Bicocca.
\end{acknowledgements}
%%%%%%%%%%%%%%%%%%%%%%%%%%%%
\bibliographystyle{aa} % style aa.bst
\bibliography{biblio.bib} % your references Yourfile.bib
%%%%%%%%%%%%%%%%%%%%%%%%%%%%
\Online

%%----------------------------------------------------------------
%\begin{appendix}
\appendix
\onecolumn

\addtocounter{section}{2} 

\section{Tables and figures}
\begin{figure}
\includegraphics[width=0.45 \hsize,clip]{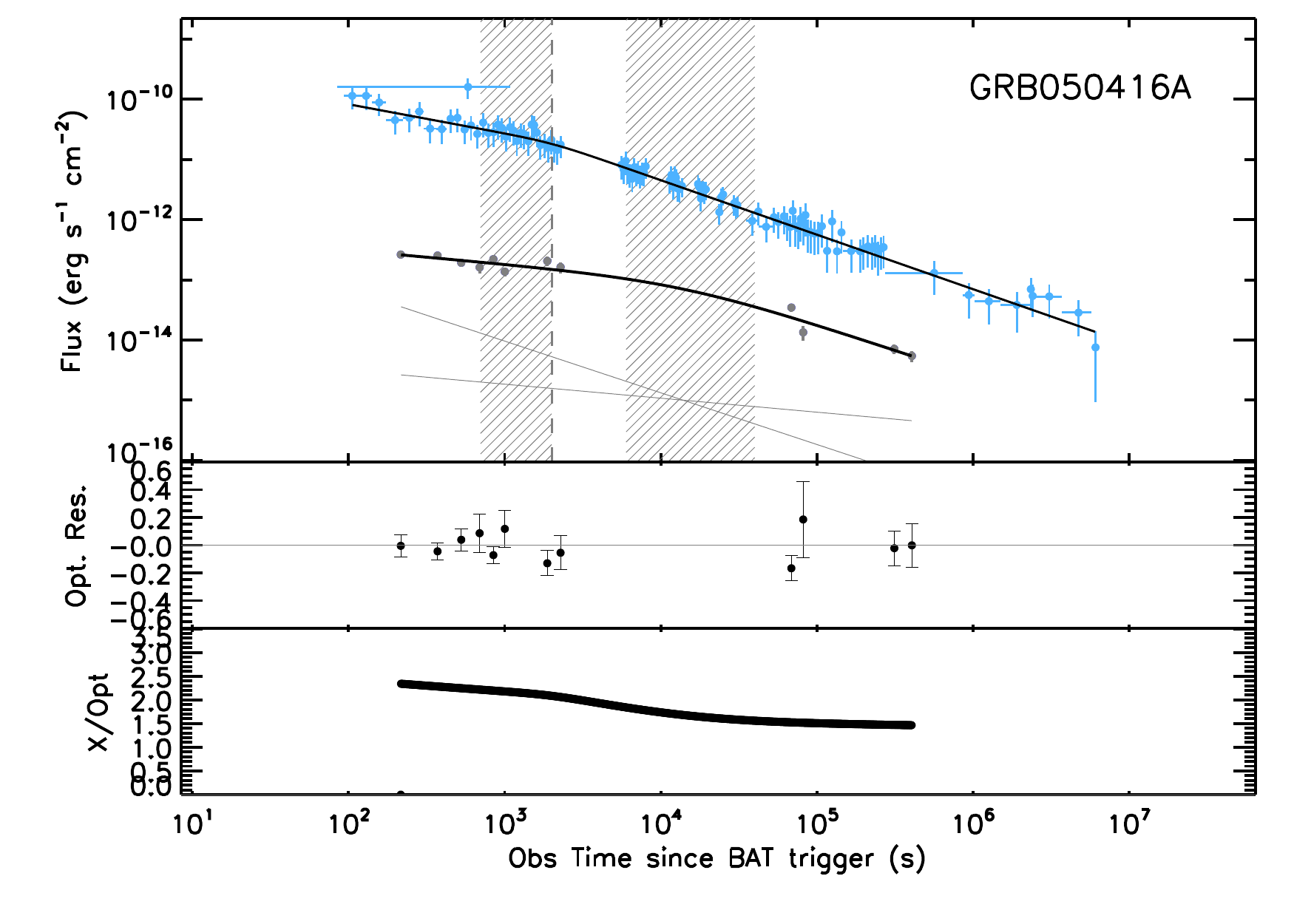}
\includegraphics[width=0.45 \hsize,clip]{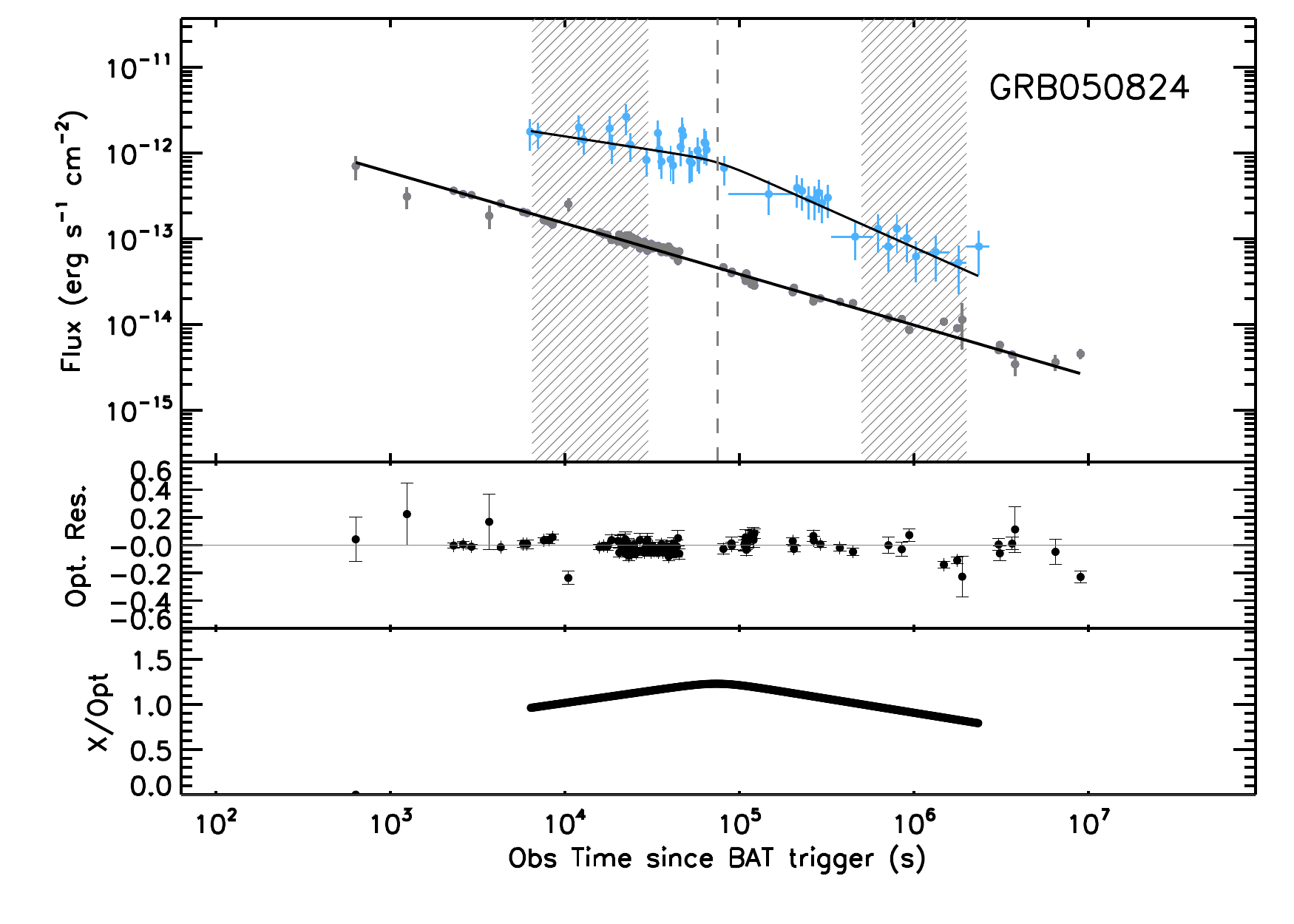}\\
\includegraphics[width=0.45 \hsize,clip]{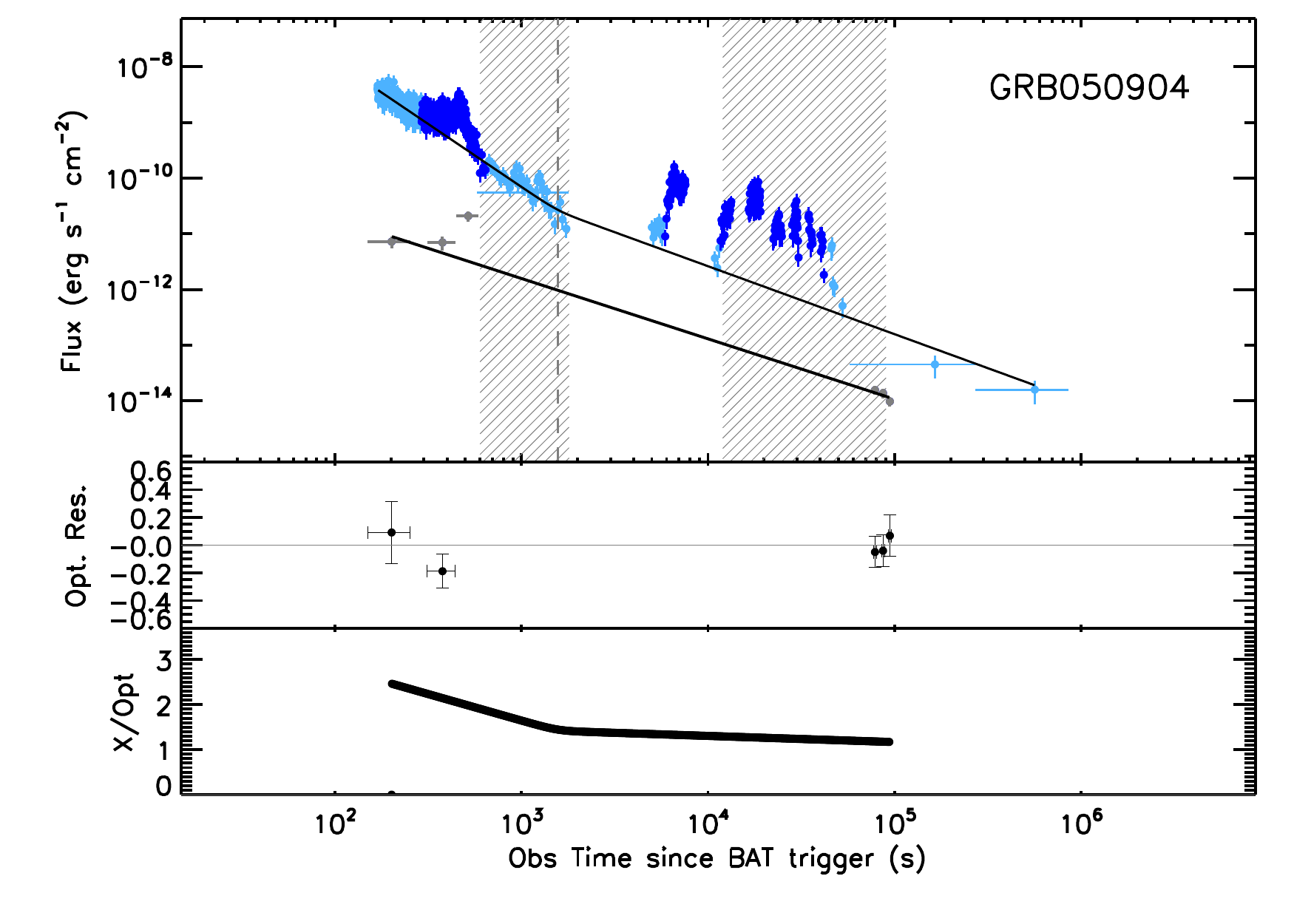}
\includegraphics[width=0.45 \hsize,clip]{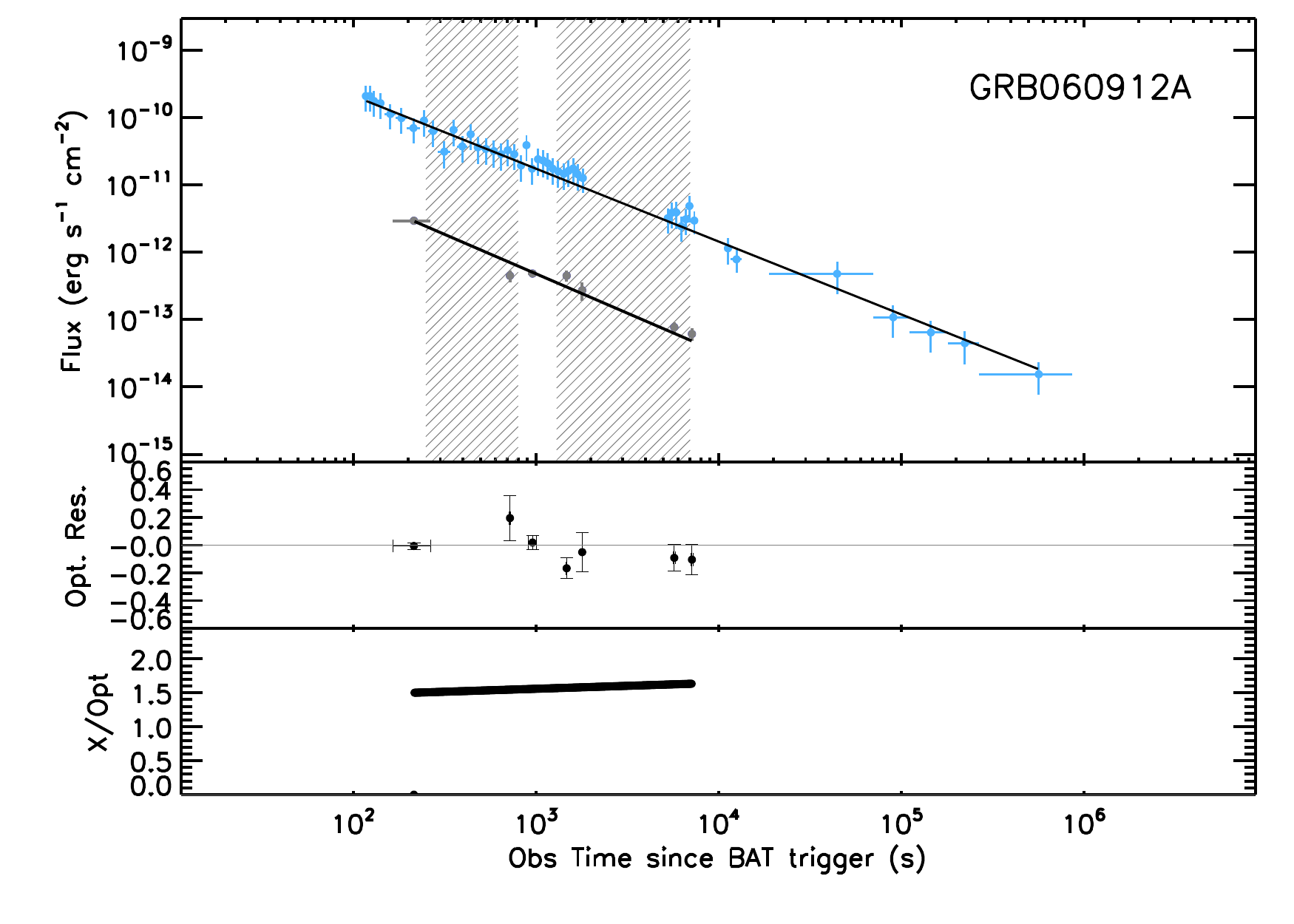}\\
\includegraphics[width=0.45 \hsize,clip]{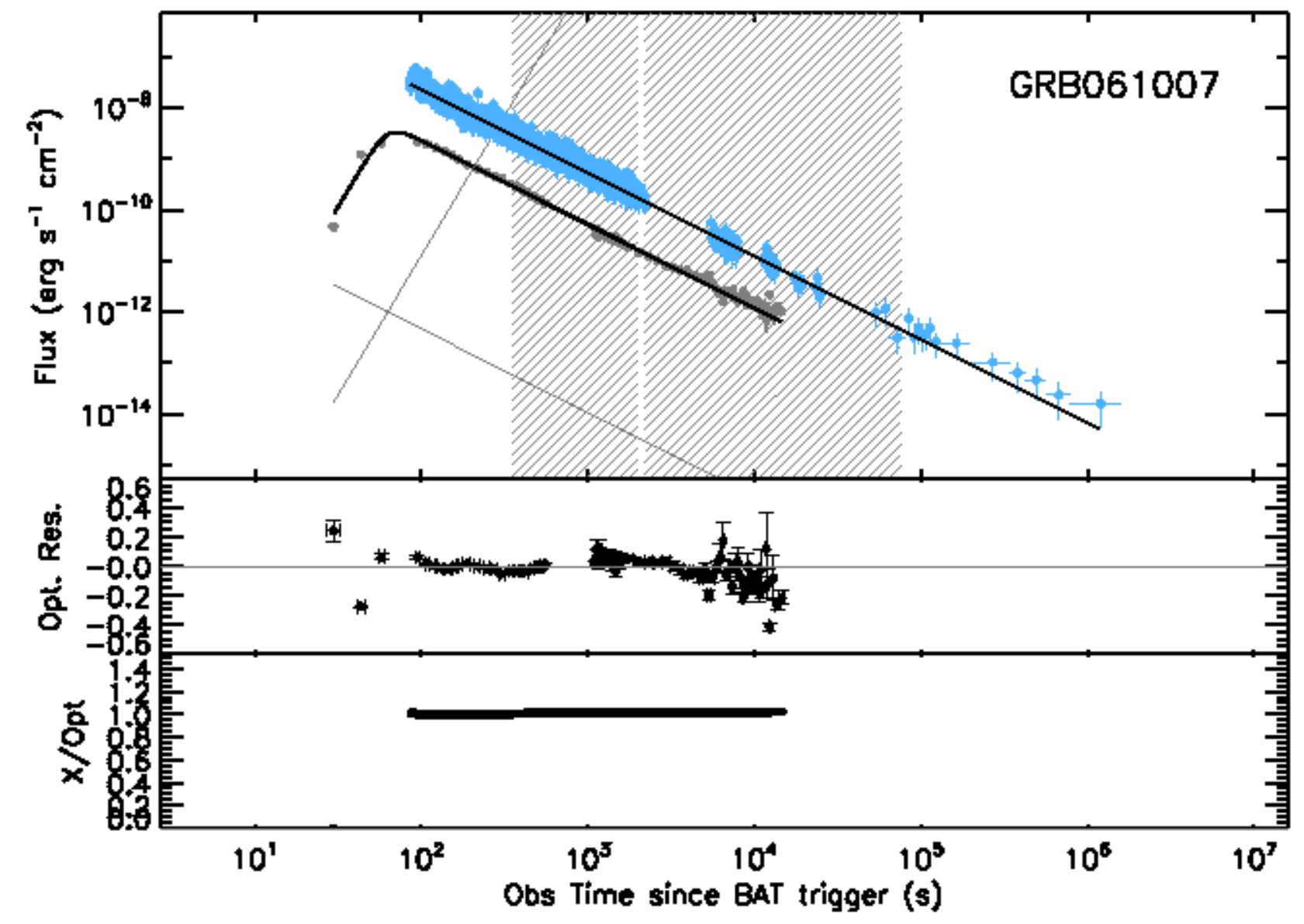}
\includegraphics[width=0.45 \hsize,clip]{FIGURE/LC/080310_OP1X-eps-converted-to.pdf}\\
\includegraphics[width=0.45 \hsize,clip]{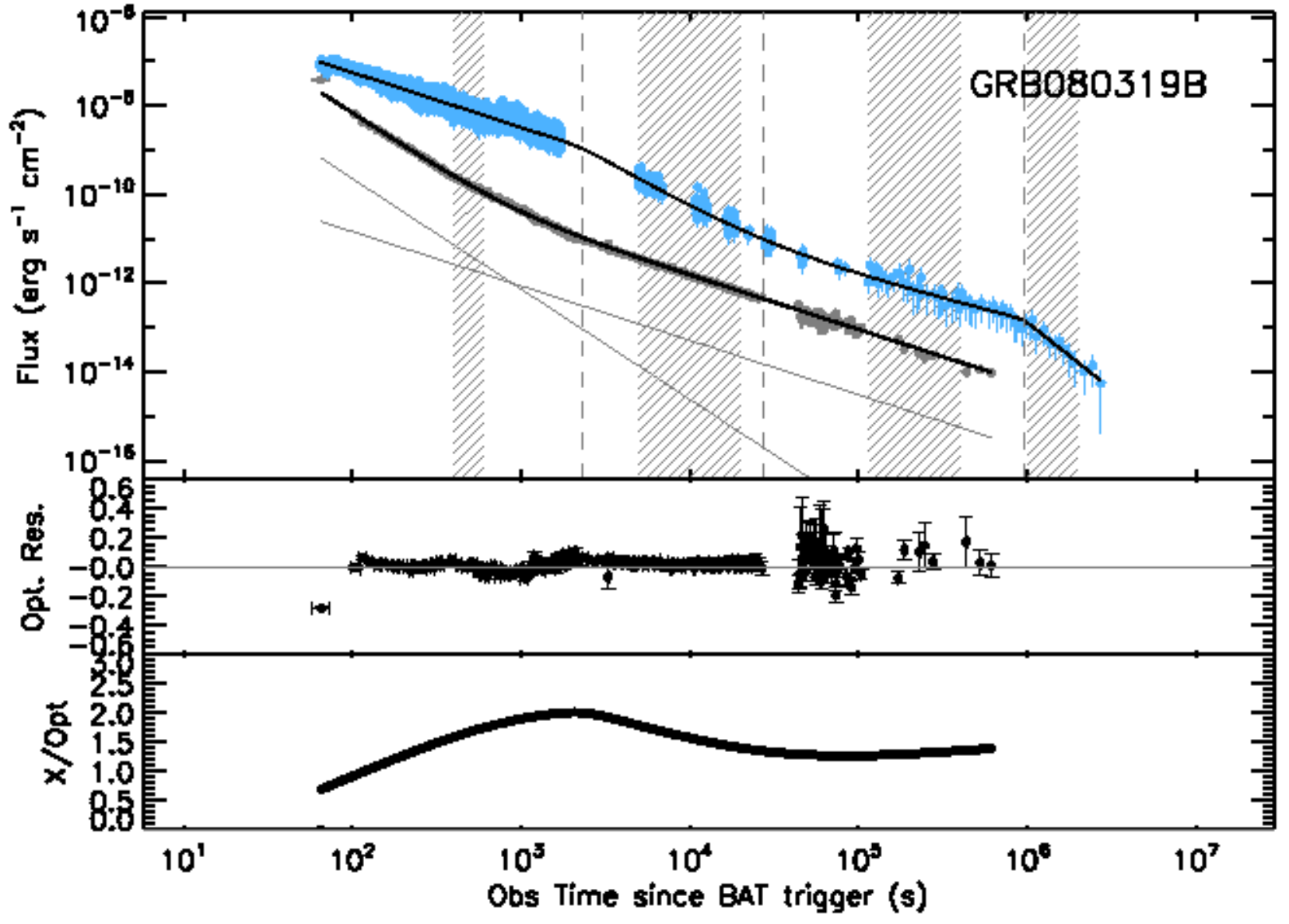}
\includegraphics[width=0.45 \hsize,clip]{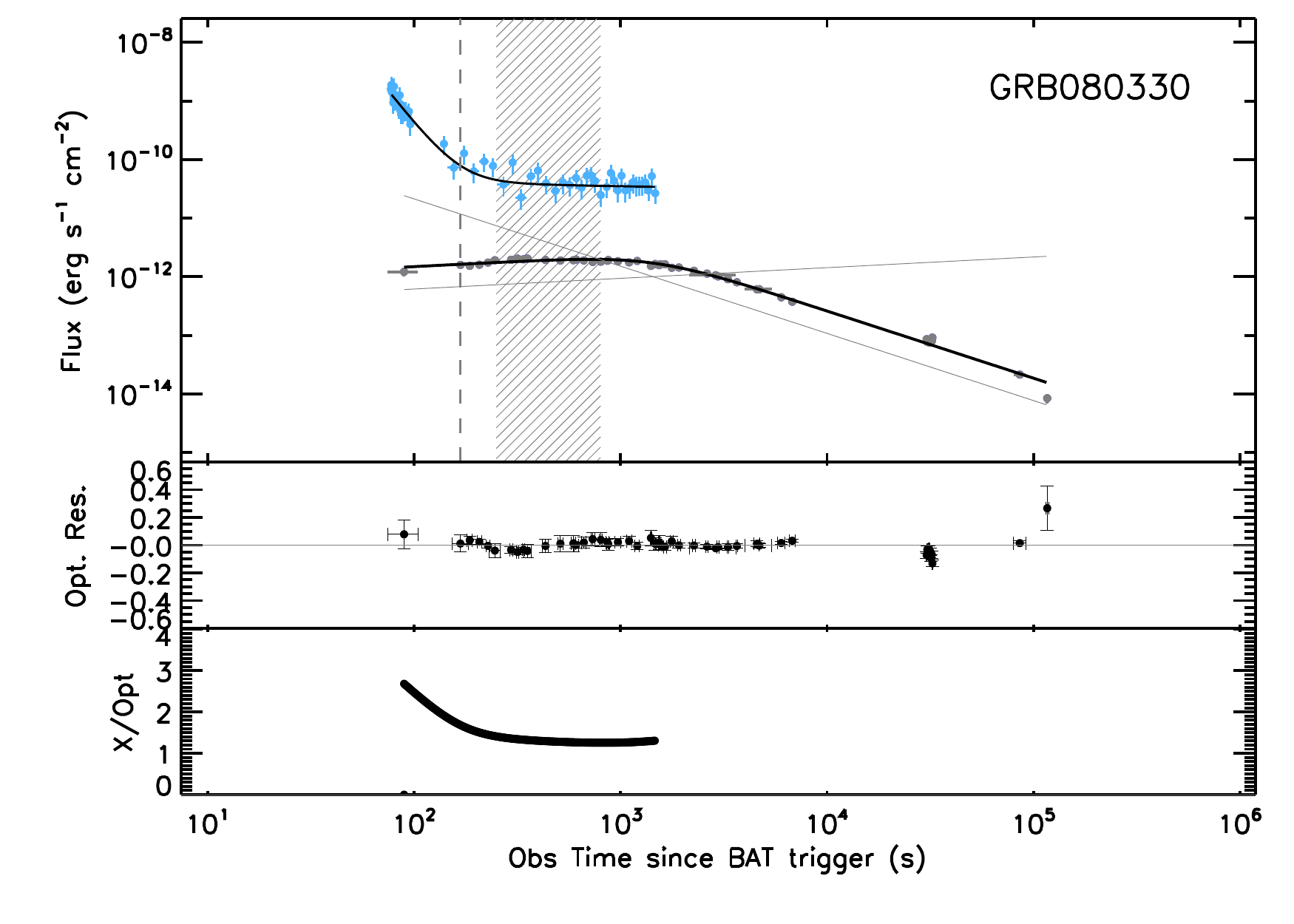}
\caption{\small{Comparison between optical and X-ray LCs. \textit{Top}. \textit{Colored points}: X-ray data. Dark color represents the excesses and light colors the continuum as calculated in M13. Group A: blue/lightblue. Group B: red/orange. Group C: purple/magenta. \textit{Gray points}: optical data. Black solid line: fit of the data. \textit{Gray solid line}: components of the fit function used to fit the optical data. \textit{Middle}. Ratio between the optical data and their fit function. The points have different colors when the optical data come from different filters. \textit{Bottom.} Ratio between the X-ray flux and the optical flux. \textit{Hashed gray boxes}: SED time intervals.}}\label{confronto1} 
\end{figure}

\begin{figure}
\includegraphics[width=0.45 \hsize,clip]{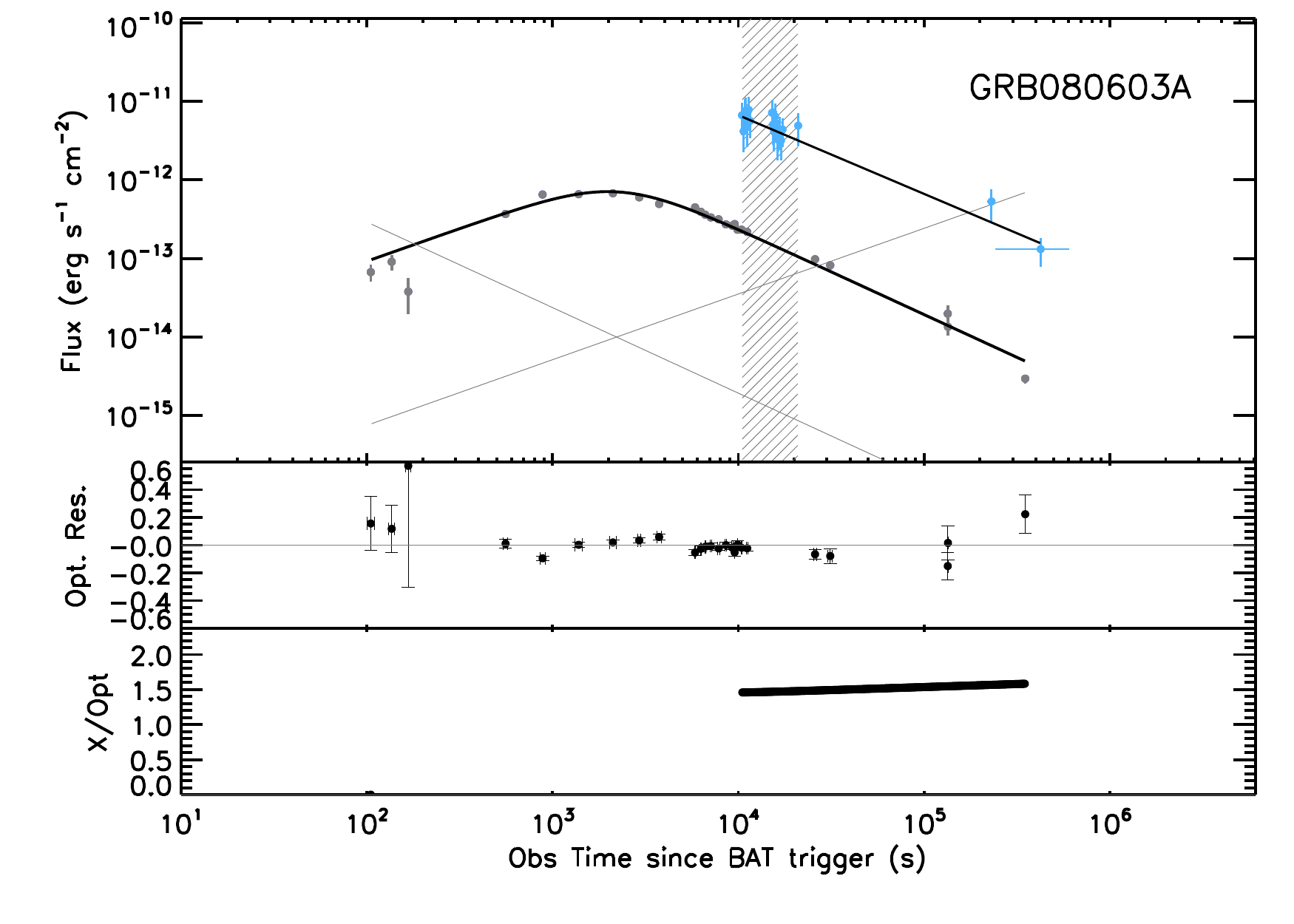}
\includegraphics[width=0.45 \hsize,clip]{FIGURE/LC/080607_OP1X-eps-converted-to.pdf}\\
\includegraphics[width=0.45 \hsize,clip]{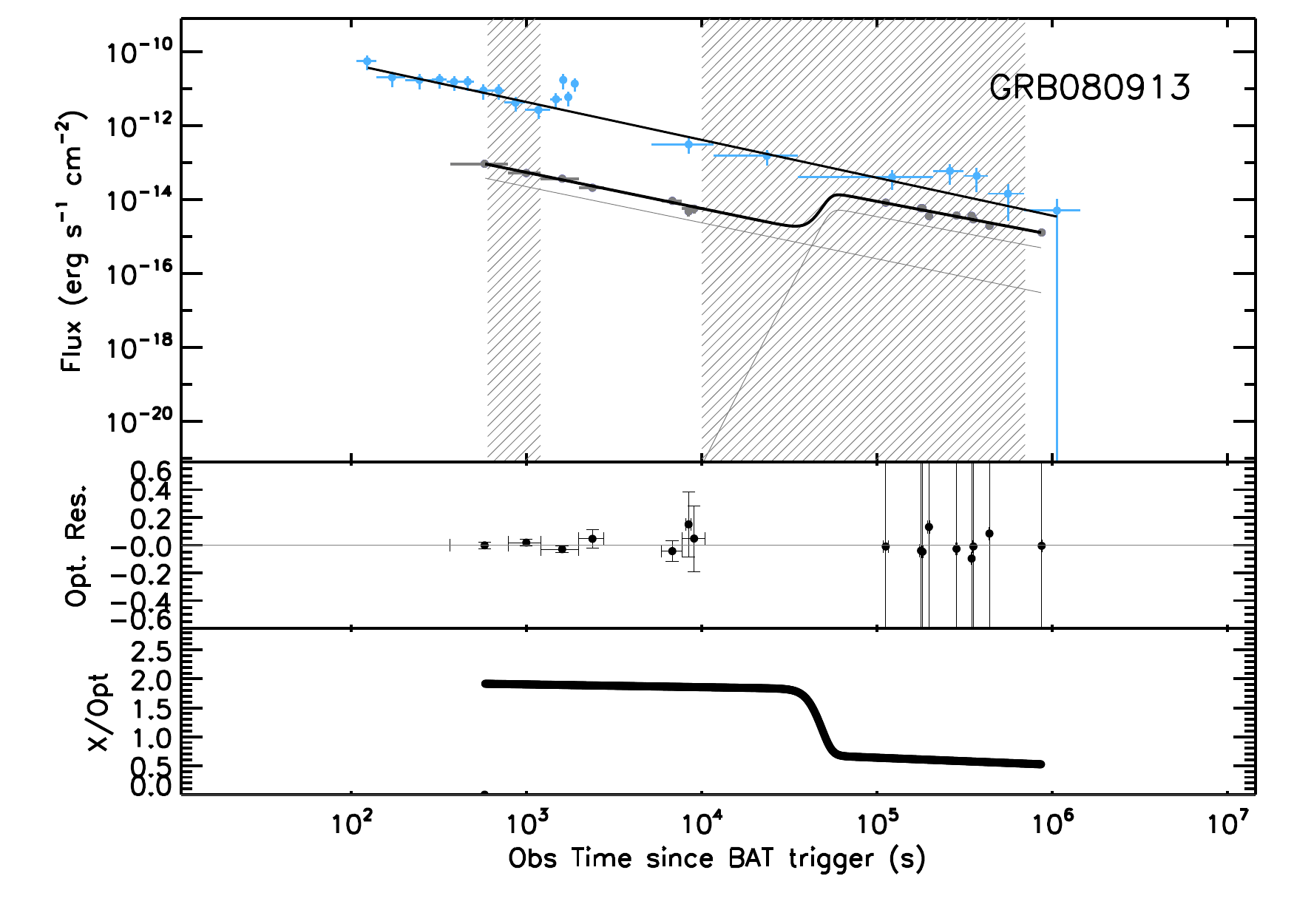}
\includegraphics[width=0.45 \hsize,clip]{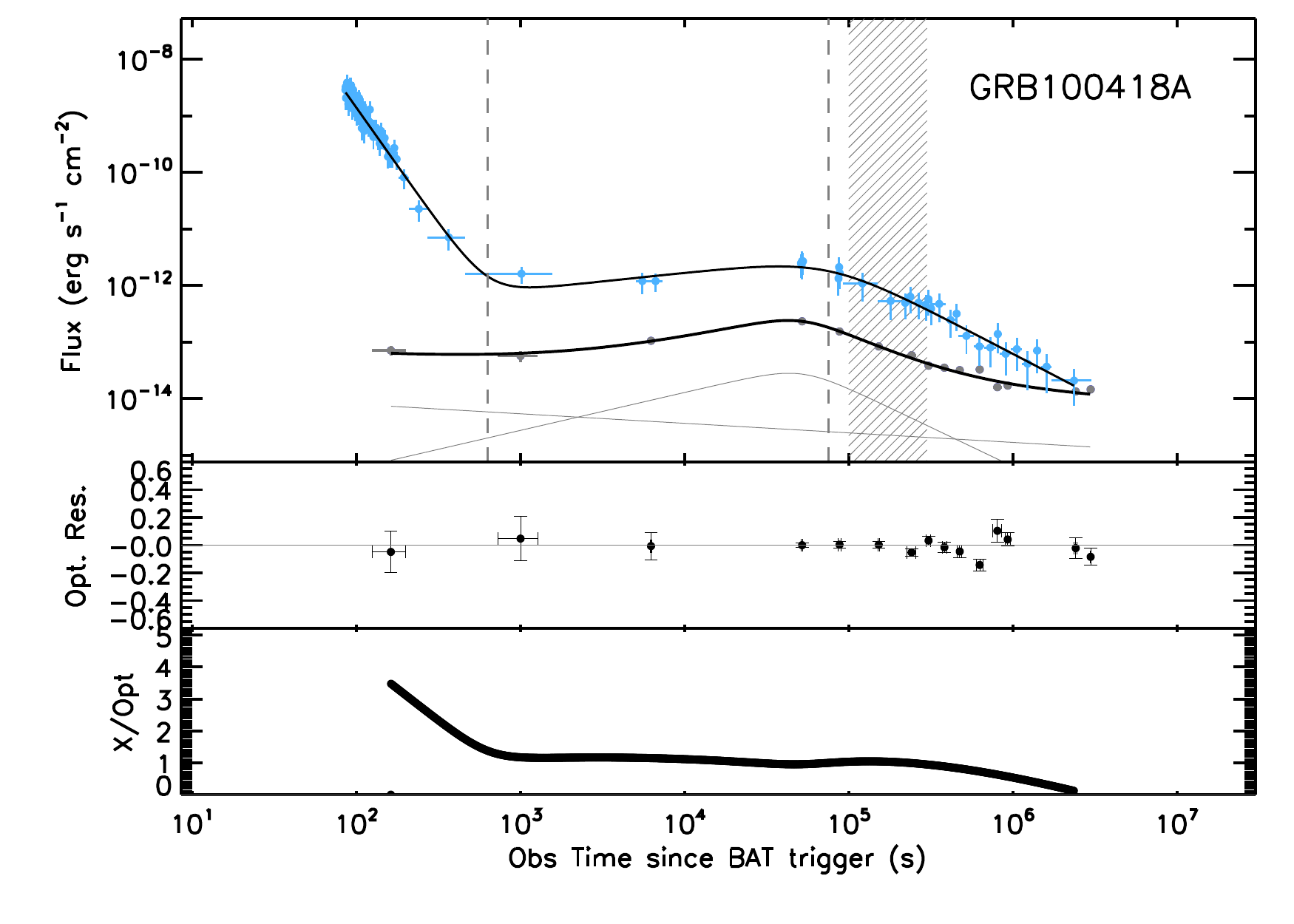}\\
\includegraphics[width=0.45 \hsize,clip]{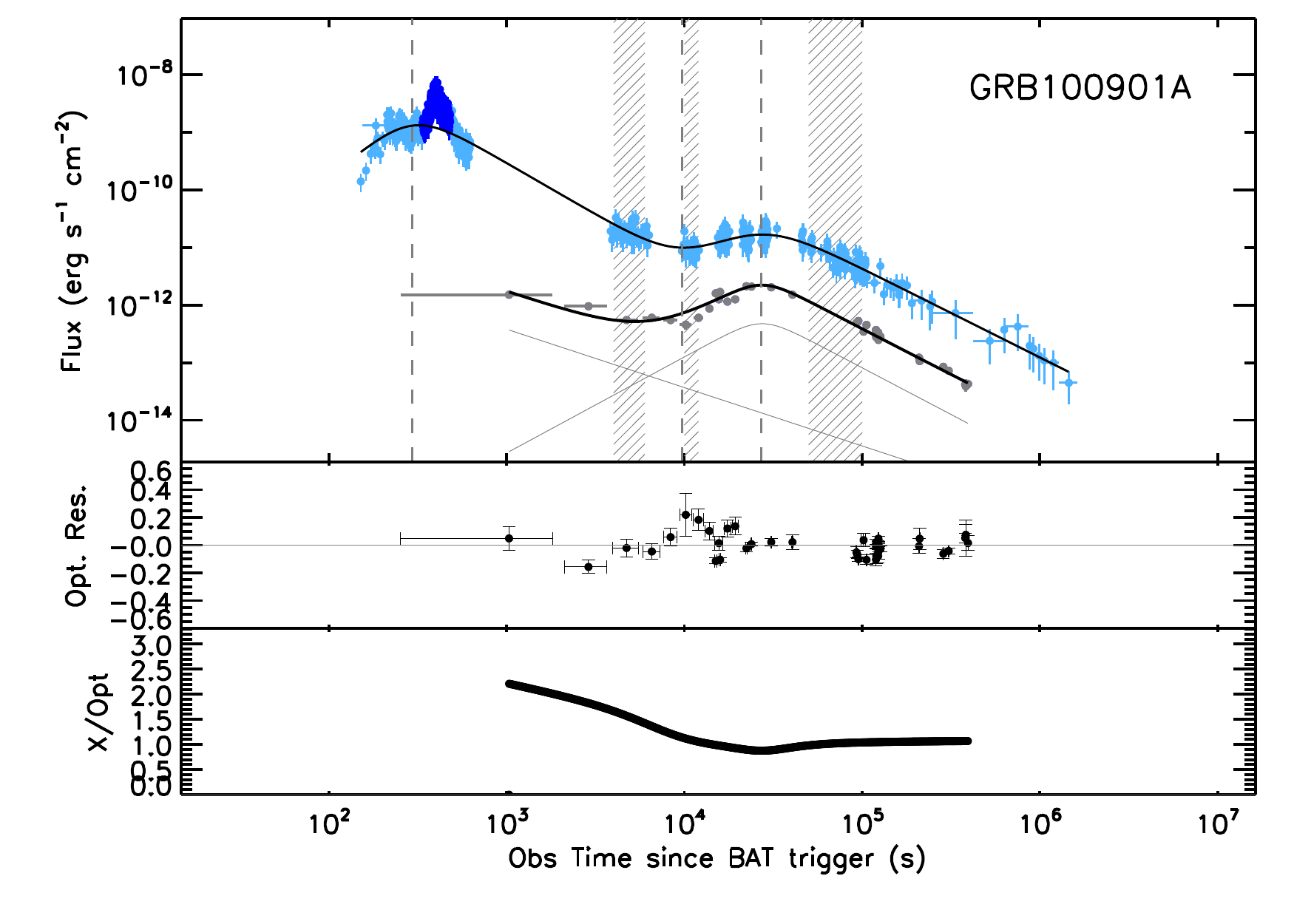}
\includegraphics[width=0.45 \hsize,clip]{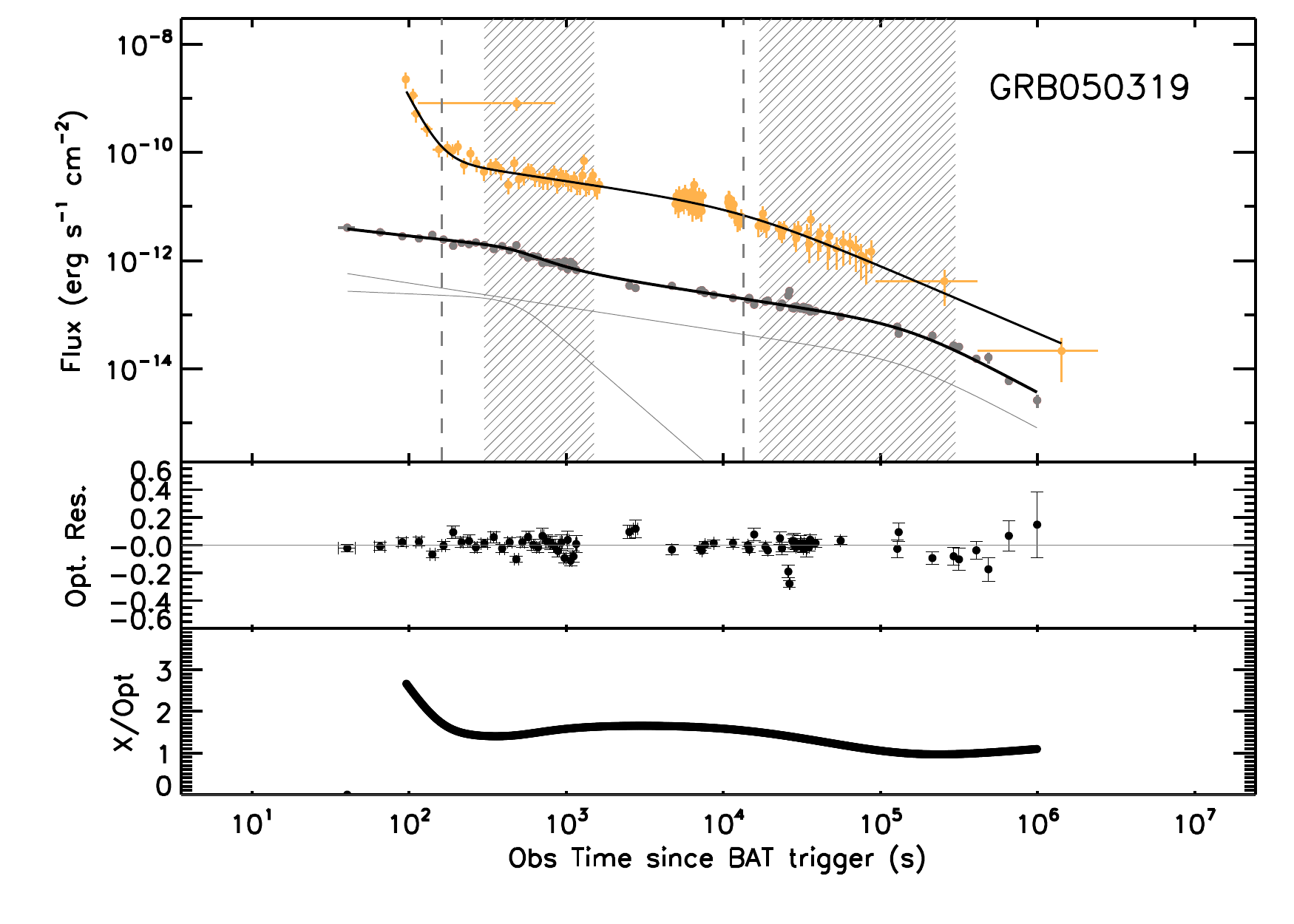}\\
\includegraphics[width=0.45 \hsize,clip]{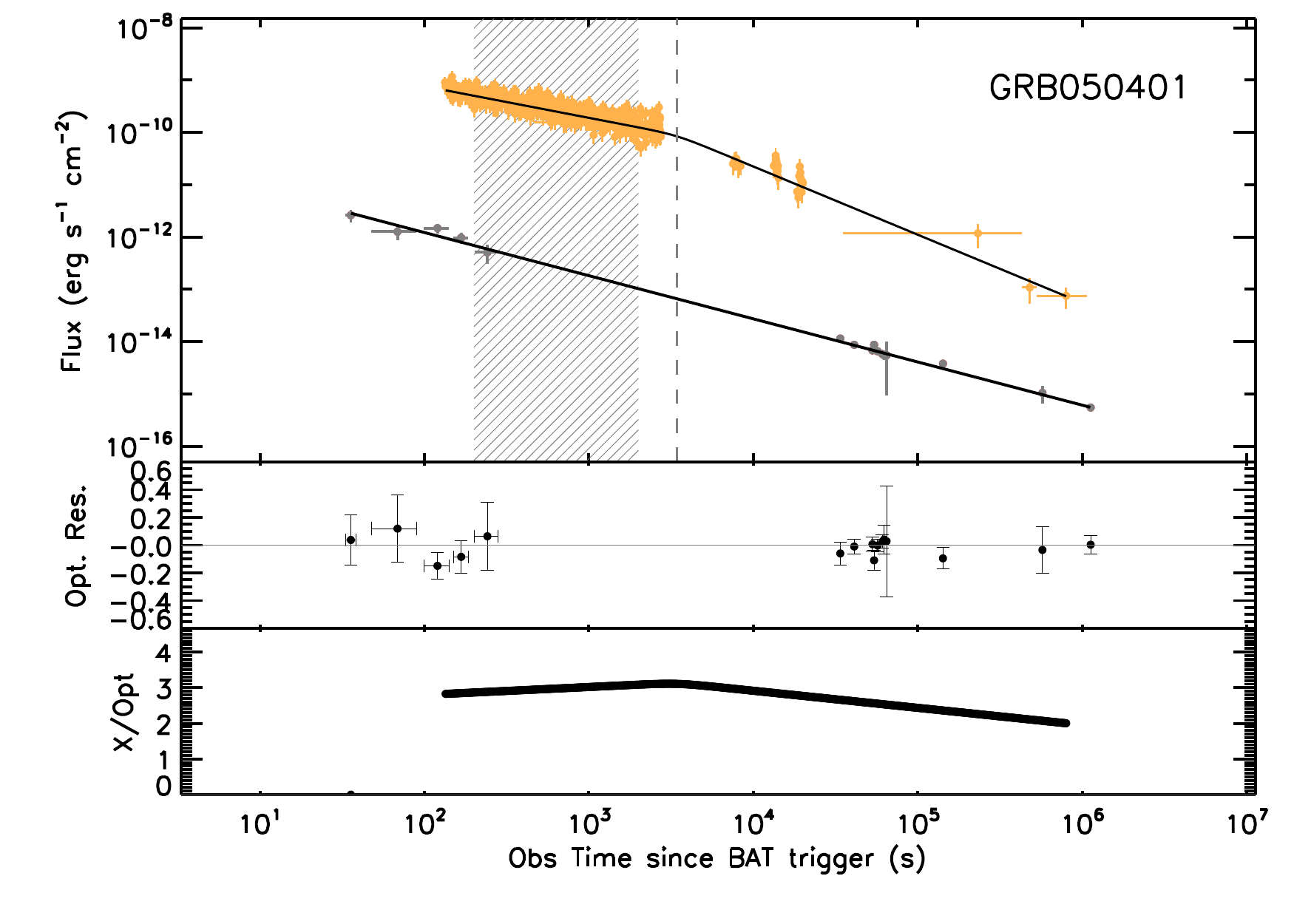}
\includegraphics[width=0.45 \hsize,clip]{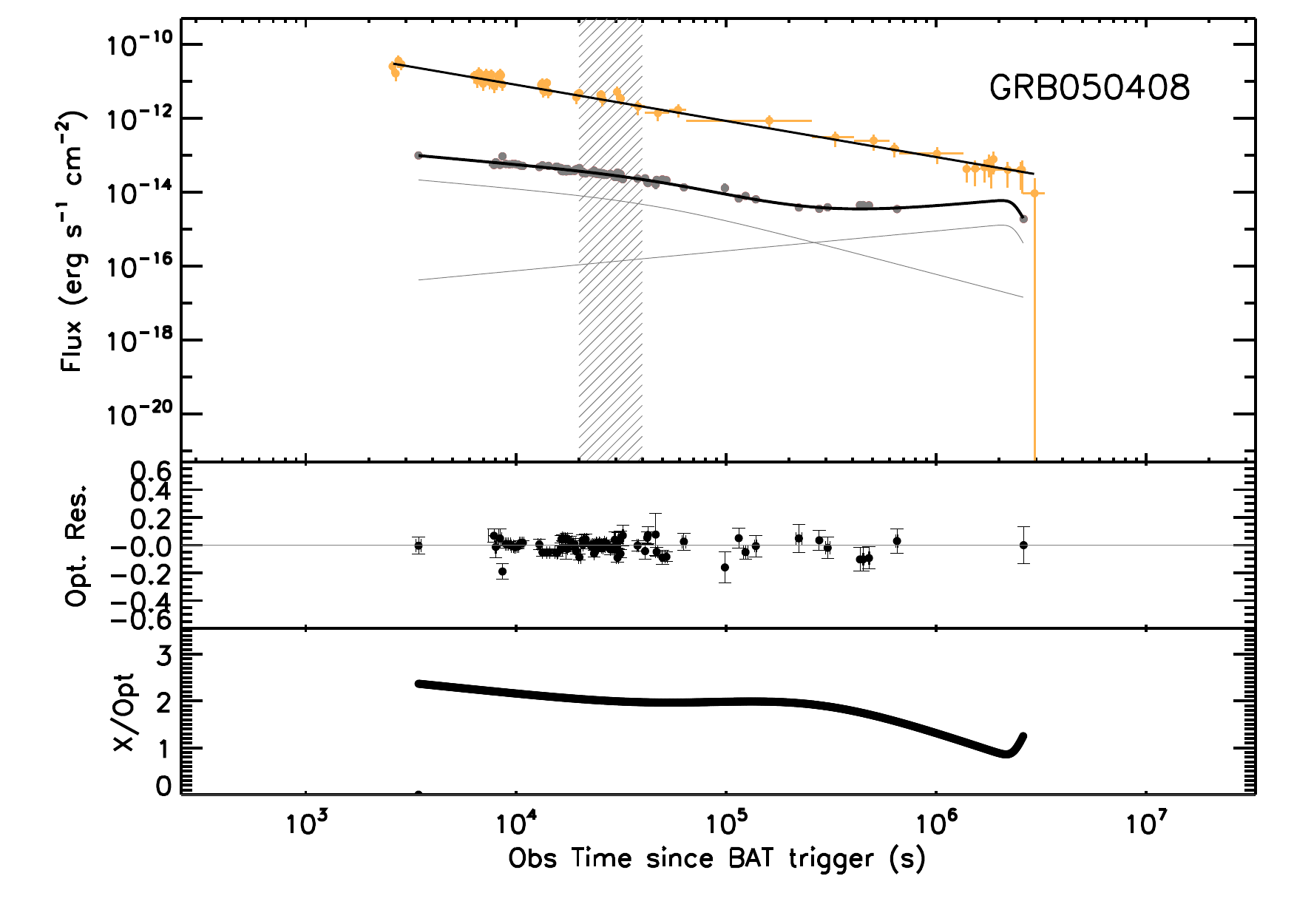}
\caption{\small{Comparison between optical and X-ray LCs. color-coding as in Figure~\ref{confronto1}.}}\label{confronto2} 
\end{figure}

\begin{figure}
\includegraphics[width=0.45 \hsize,clip]{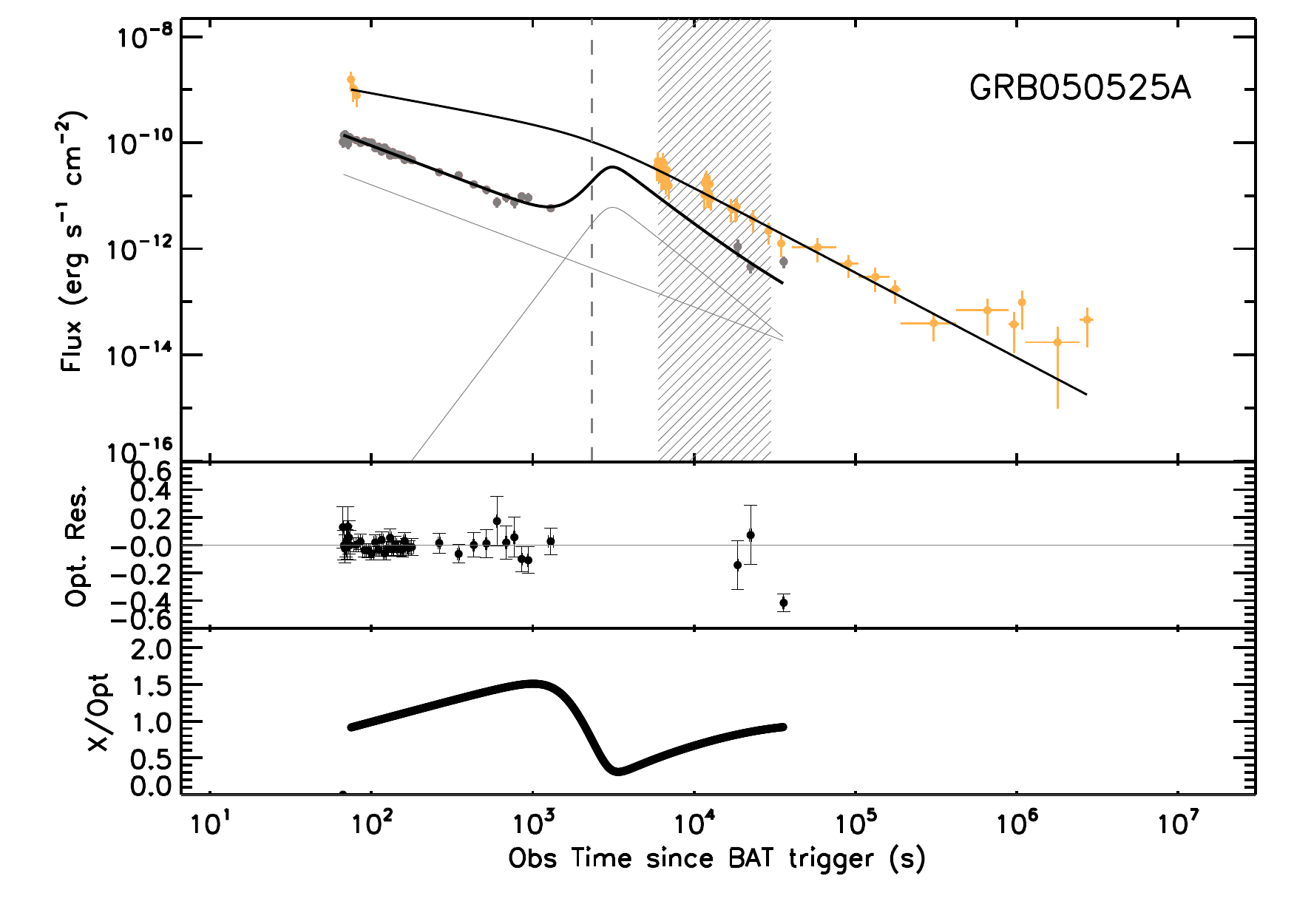}
\includegraphics[width=0.45 \hsize,clip]{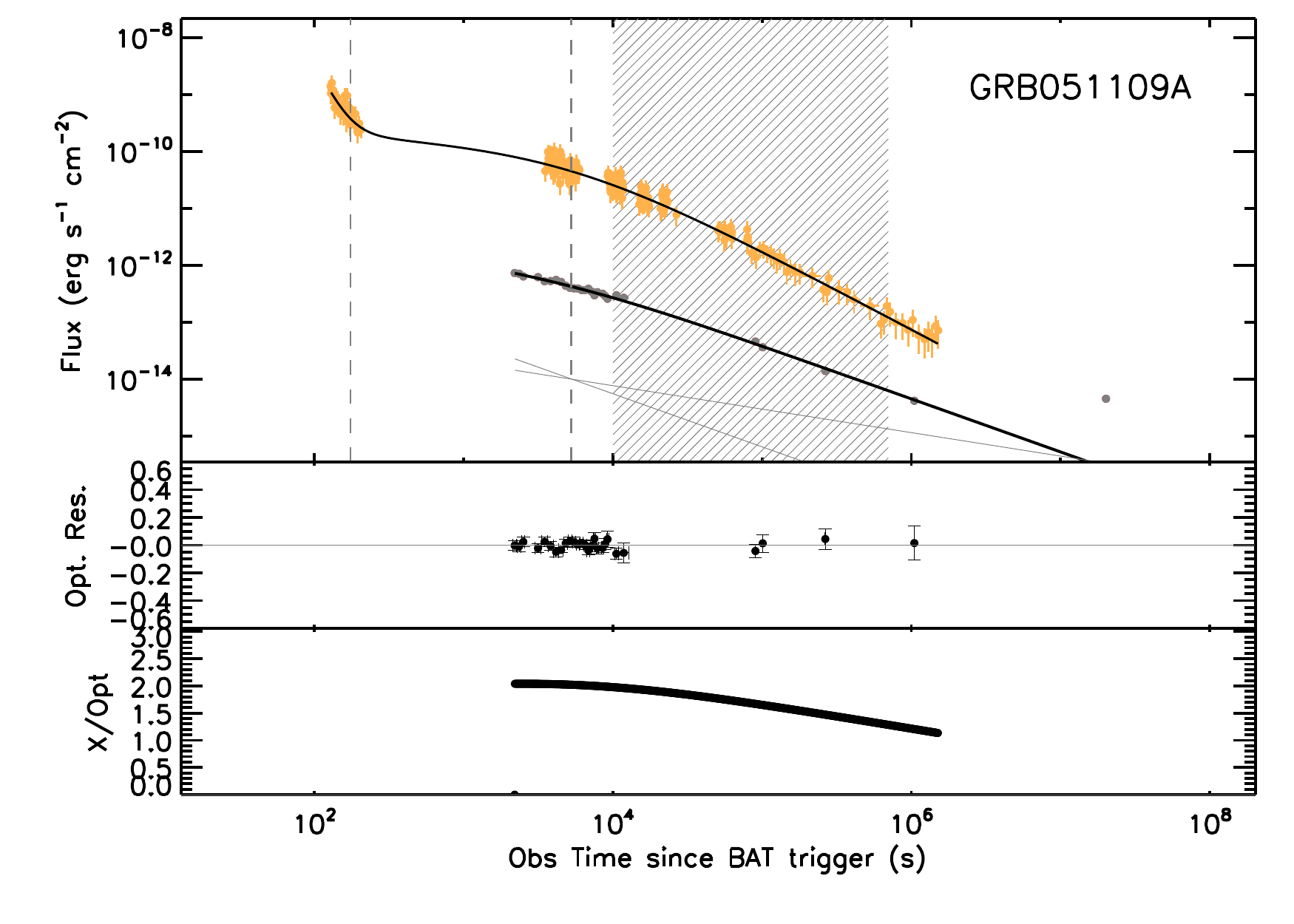}\\
\includegraphics[width=0.45 \hsize,clip]{FIGURE/LC/060124_OP1X-eps-converted-to.pdf}
\includegraphics[width=0.45 \hsize,clip]{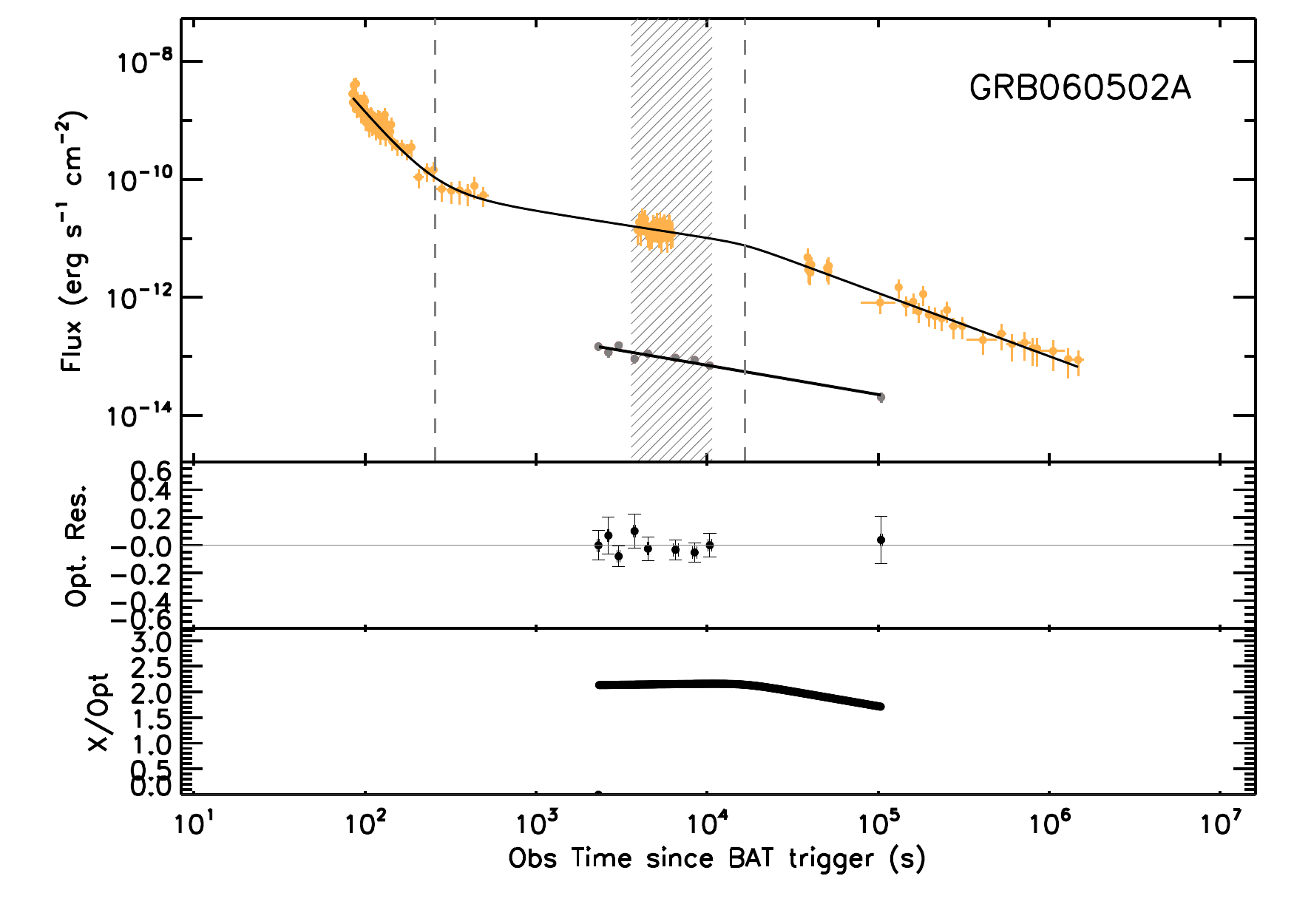}\\
\includegraphics[width=0.45 \hsize,clip]{FIGURE/LC/060512_OP1X-eps-converted-to.pdf}
\includegraphics[width=0.45 \hsize,clip]{FIGURE/LC/060526_OP1X-eps-converted-to.pdf}\\
\includegraphics[width=0.45 \hsize,clip]{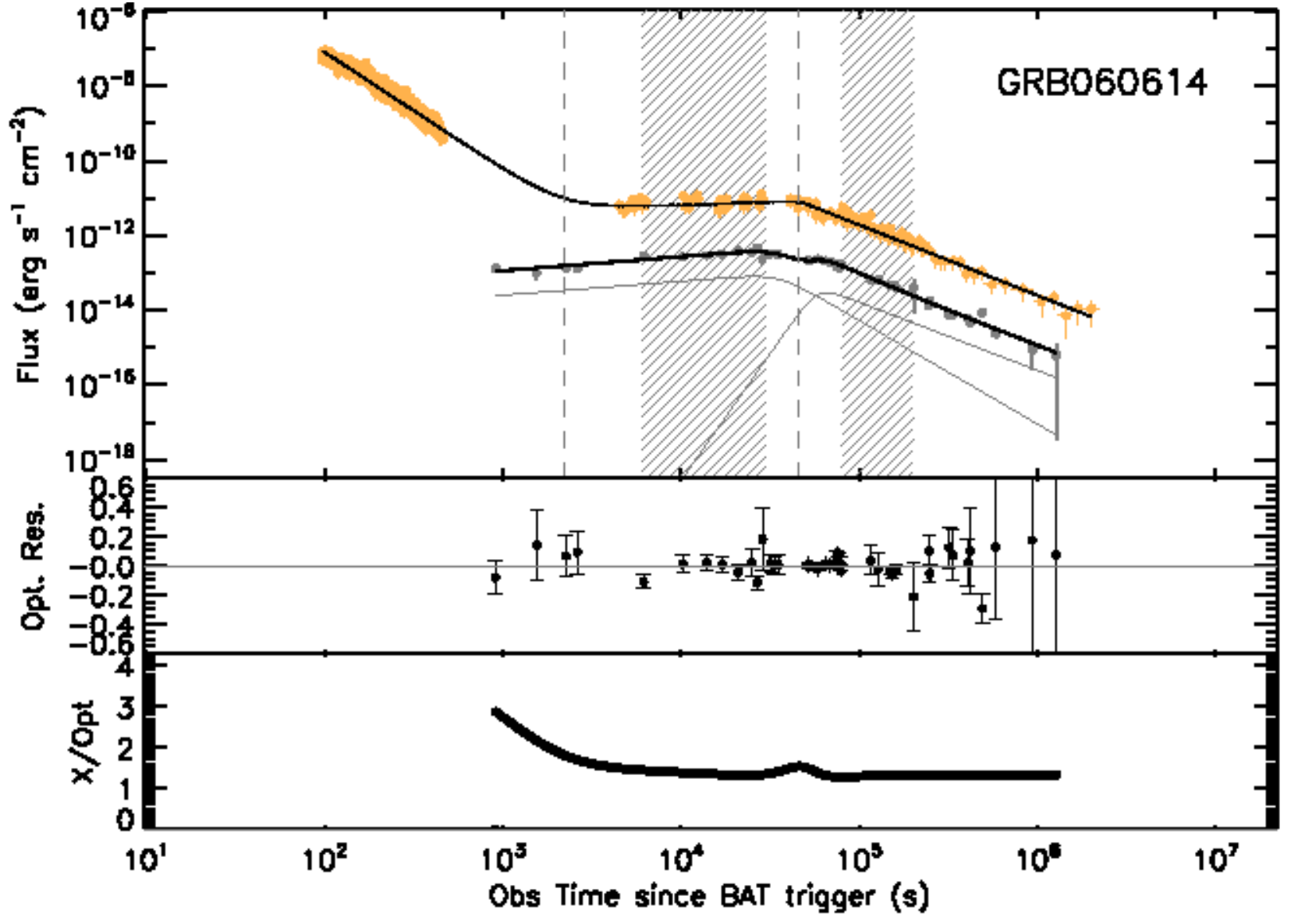}
\includegraphics[width=0.45 \hsize,clip]{FIGURE/LC/060729_OP1X-eps-converted-to.pdf}
\caption{\small{Comparison between optical and X-ray LCs. color-coding as in Figure~\ref{confronto1}.}}\label{confronto3} 
\end{figure}

\begin{figure}
\includegraphics[width=0.45 \hsize,clip]{FIGURE/LC/060904B_OP1X-eps-converted-to.pdf}
\includegraphics[width=0.45 \hsize,clip]{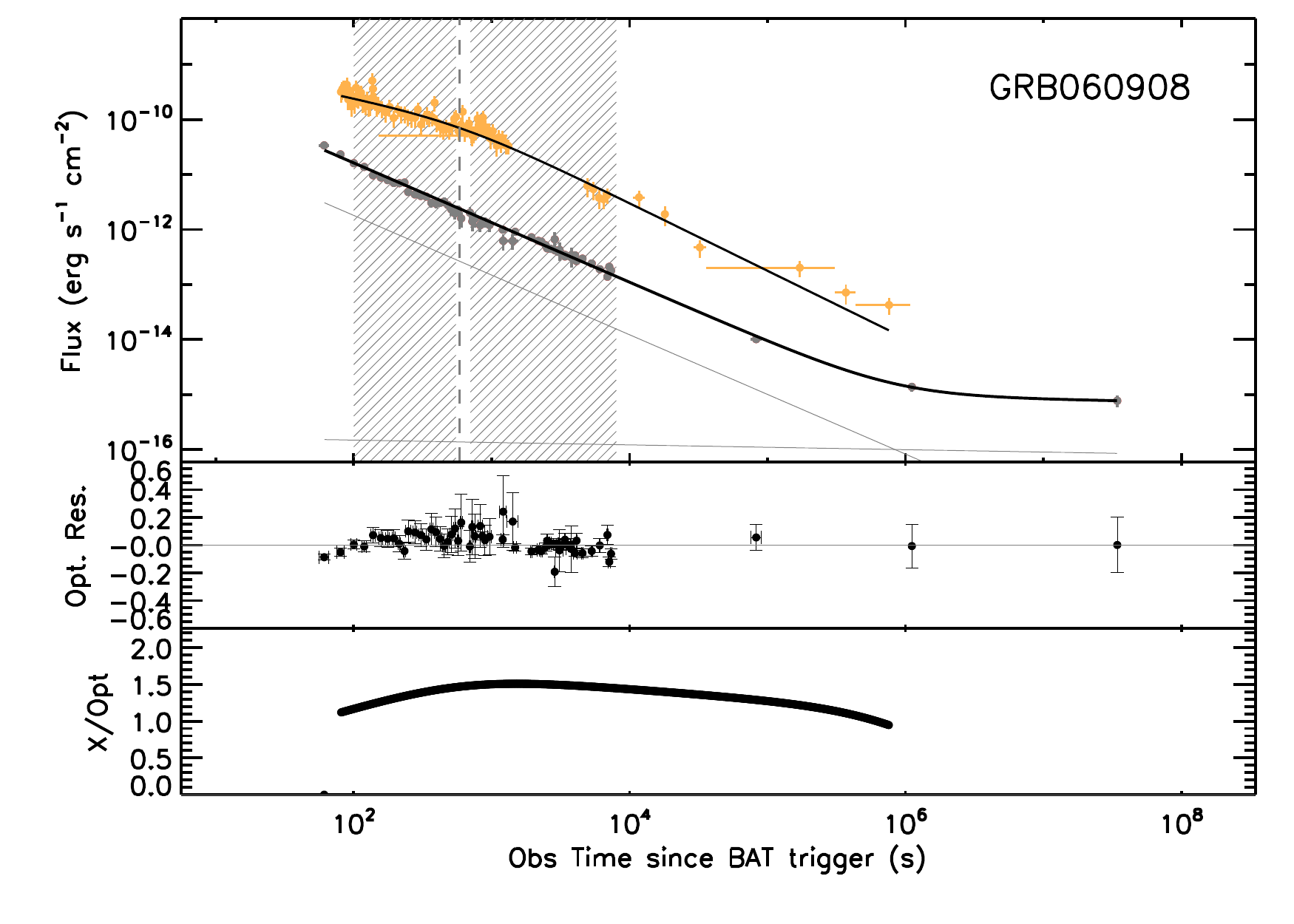}\\
\includegraphics[width=0.45 \hsize,clip]{FIGURE/LC/061121_OP1X-eps-converted-to.pdf}
\includegraphics[width=0.45 \hsize,clip]{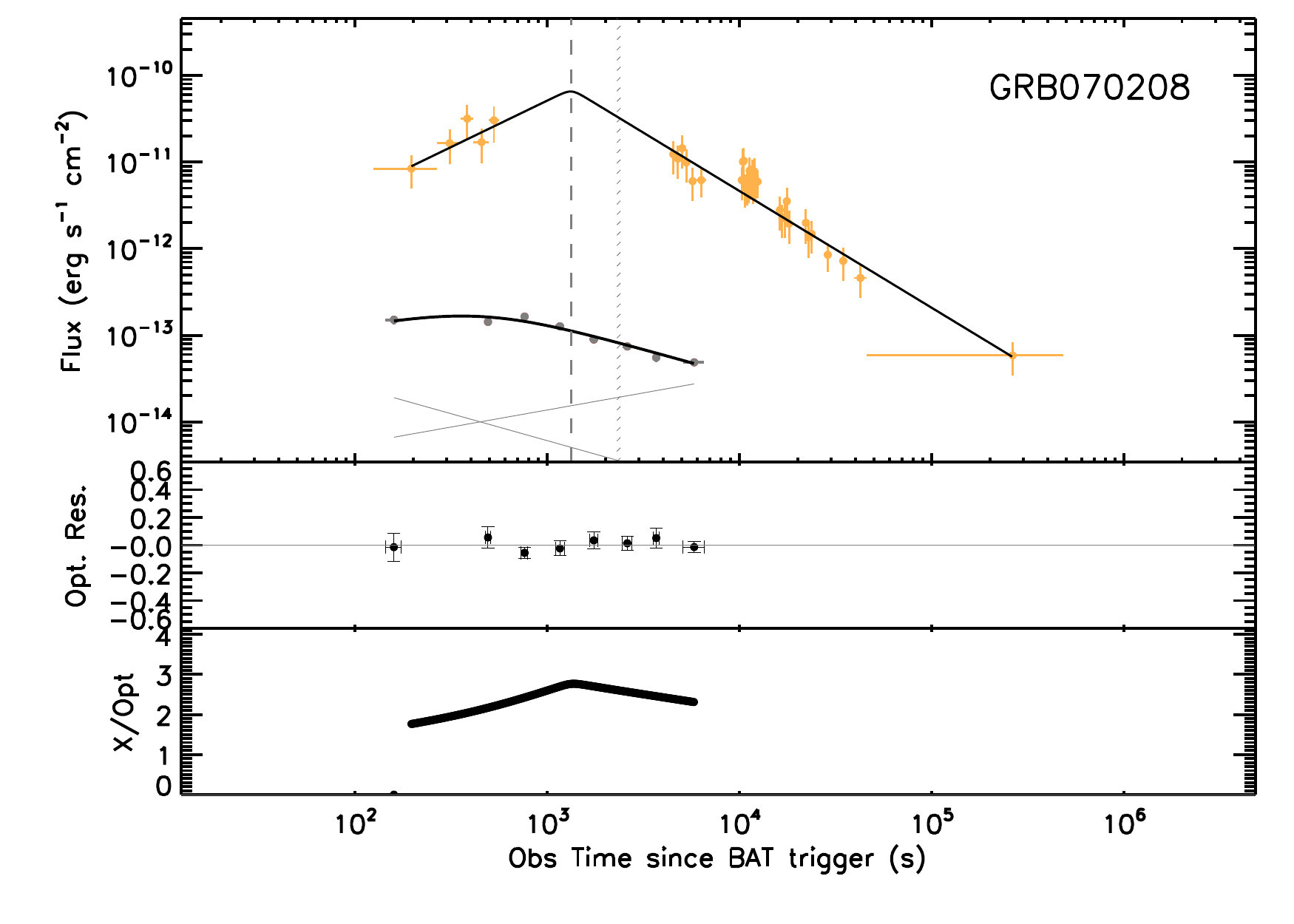}\\
\includegraphics[width=0.45 \hsize,clip]{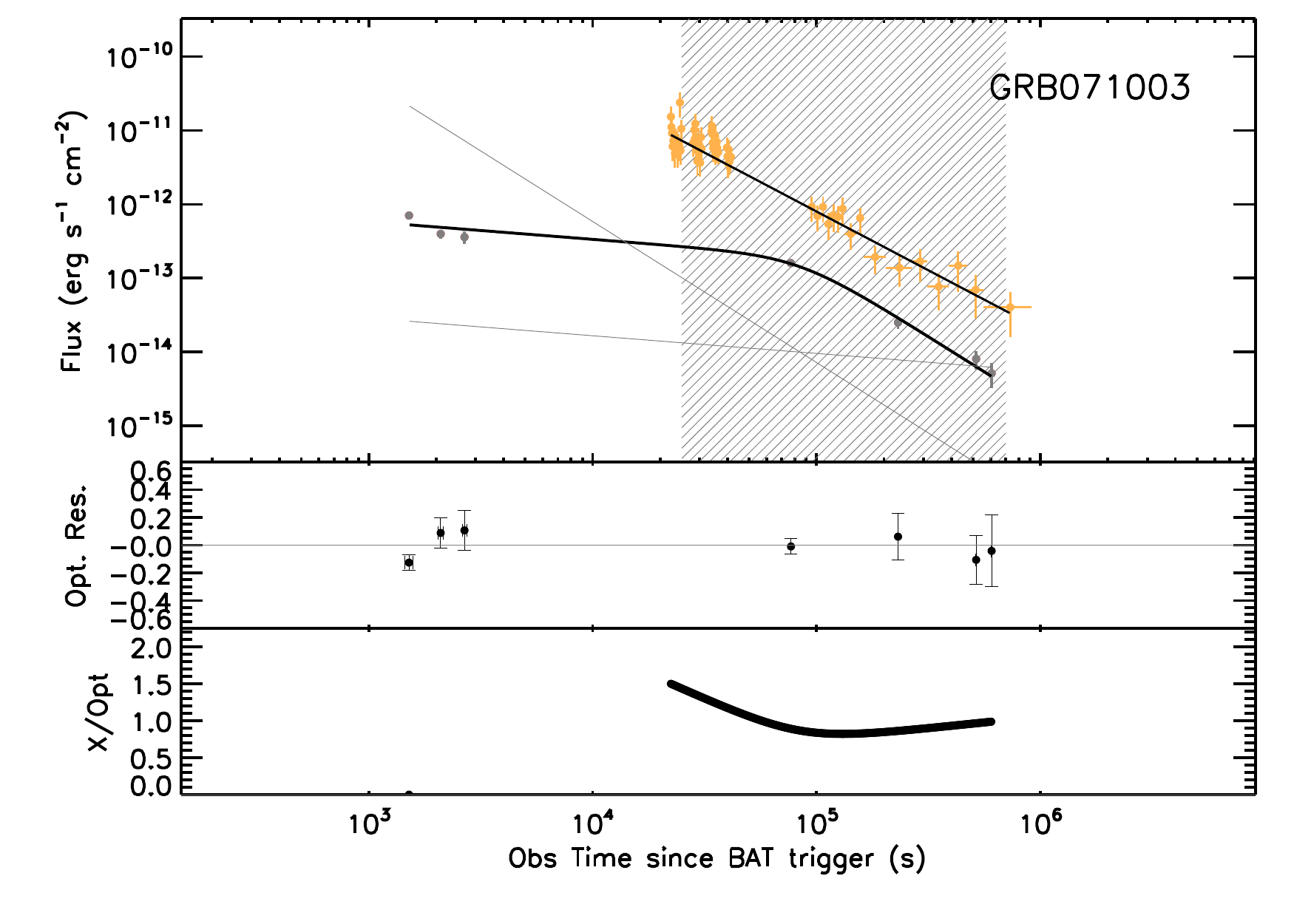}
\includegraphics[width=0.45 \hsize,clip]{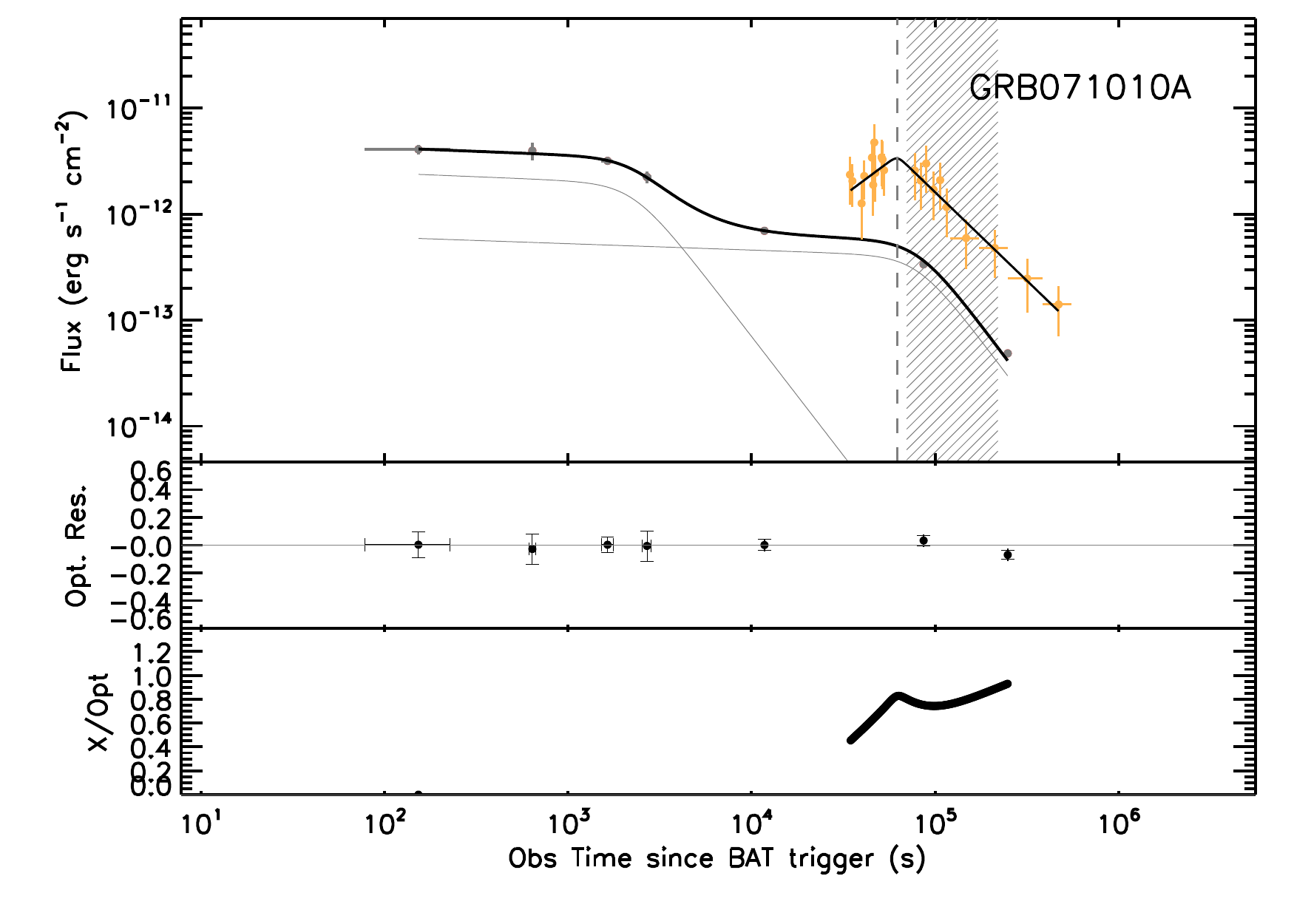}\\
\includegraphics[width=0.45 \hsize,clip]{FIGURE/LC/071031_OP1X-eps-converted-to.pdf}
\includegraphics[width=0.45 \hsize,clip]{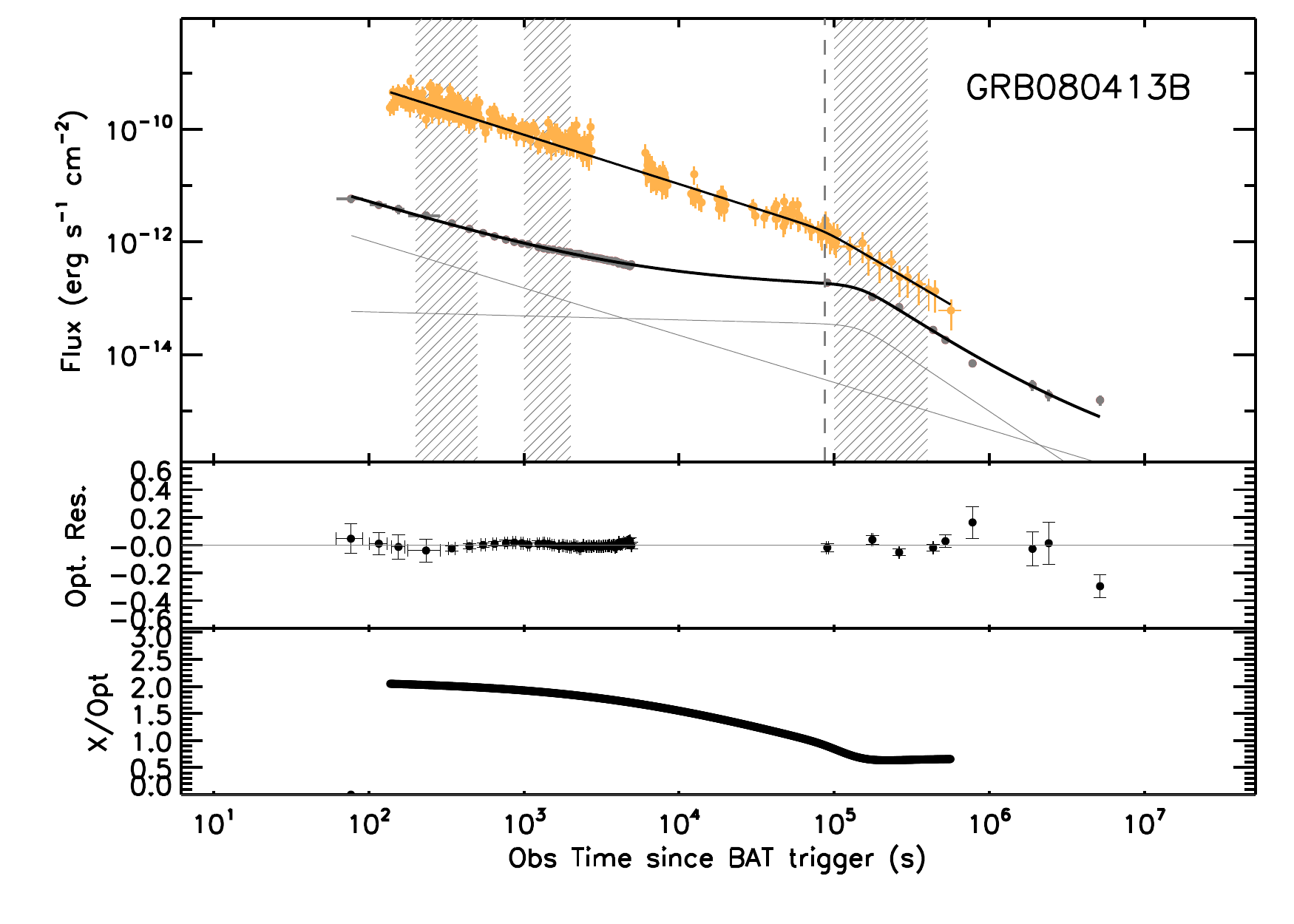}
\caption{\small{Comparison between optical and X-ray LCs. color-coding as in Figure~\ref{confronto1}.}}\label{confronto4} 
\end{figure}

\begin{figure}
\includegraphics[width=0.45 \hsize,clip]{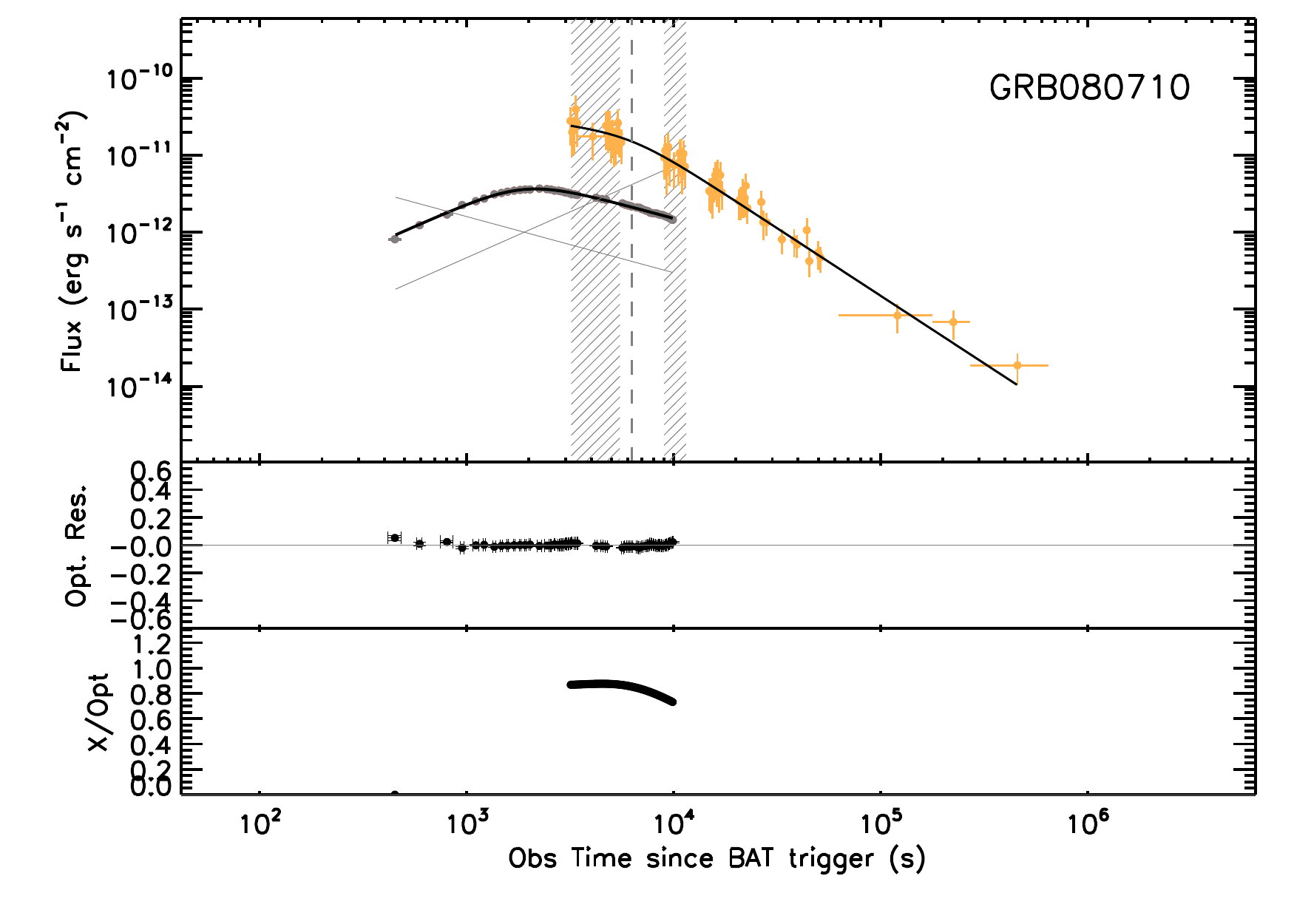}
\includegraphics[width=0.45 \hsize,clip]{FIGURE/LC/081008_OP1X-eps-converted-to.pdf}\\
\includegraphics[width=0.45 \hsize,clip]{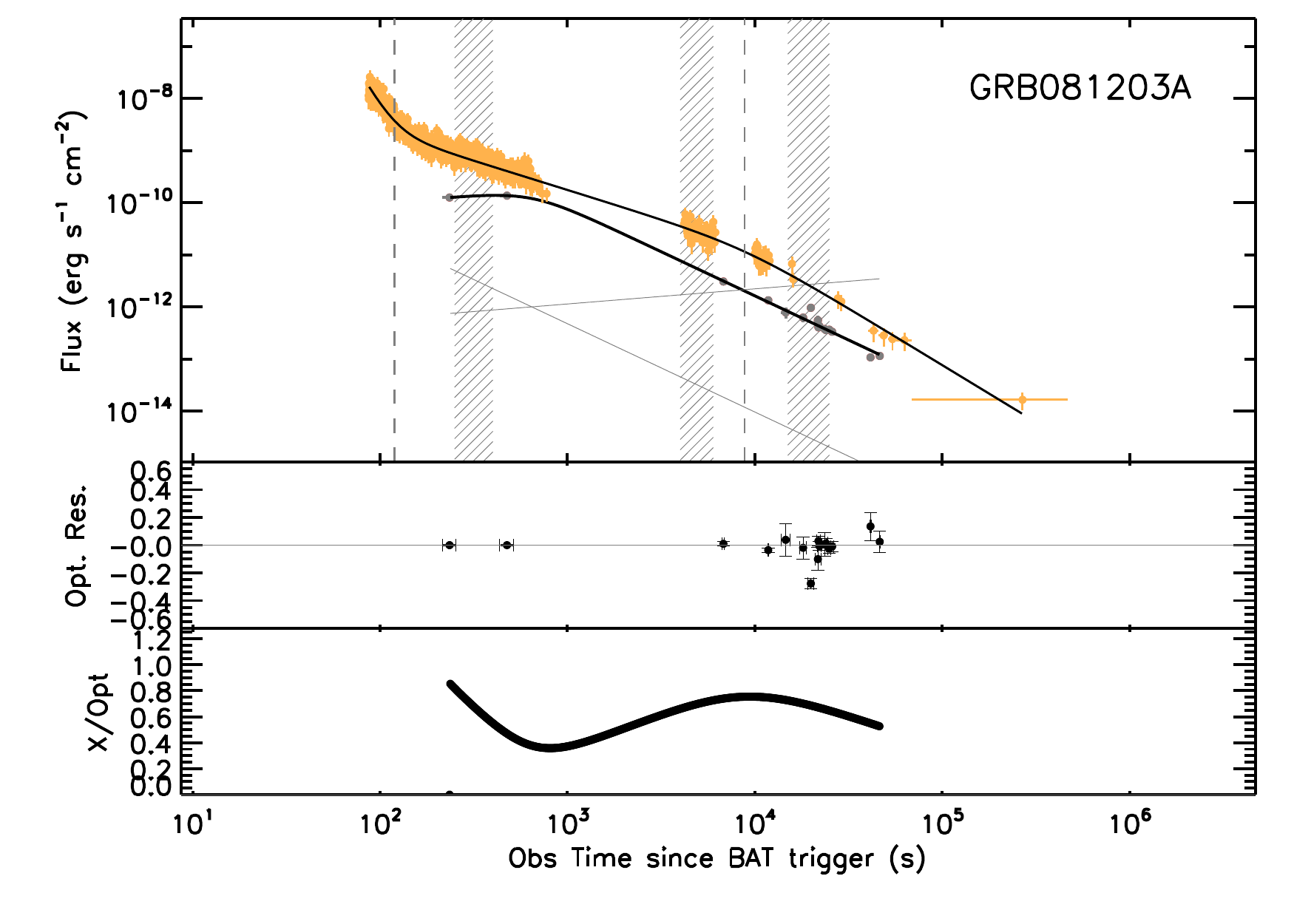}
\includegraphics[width=0.45 \hsize,clip]{FIGURE/LC/080928_OP1X-eps-converted-to.pdf}\\
\includegraphics[width=0.45 \hsize,clip]{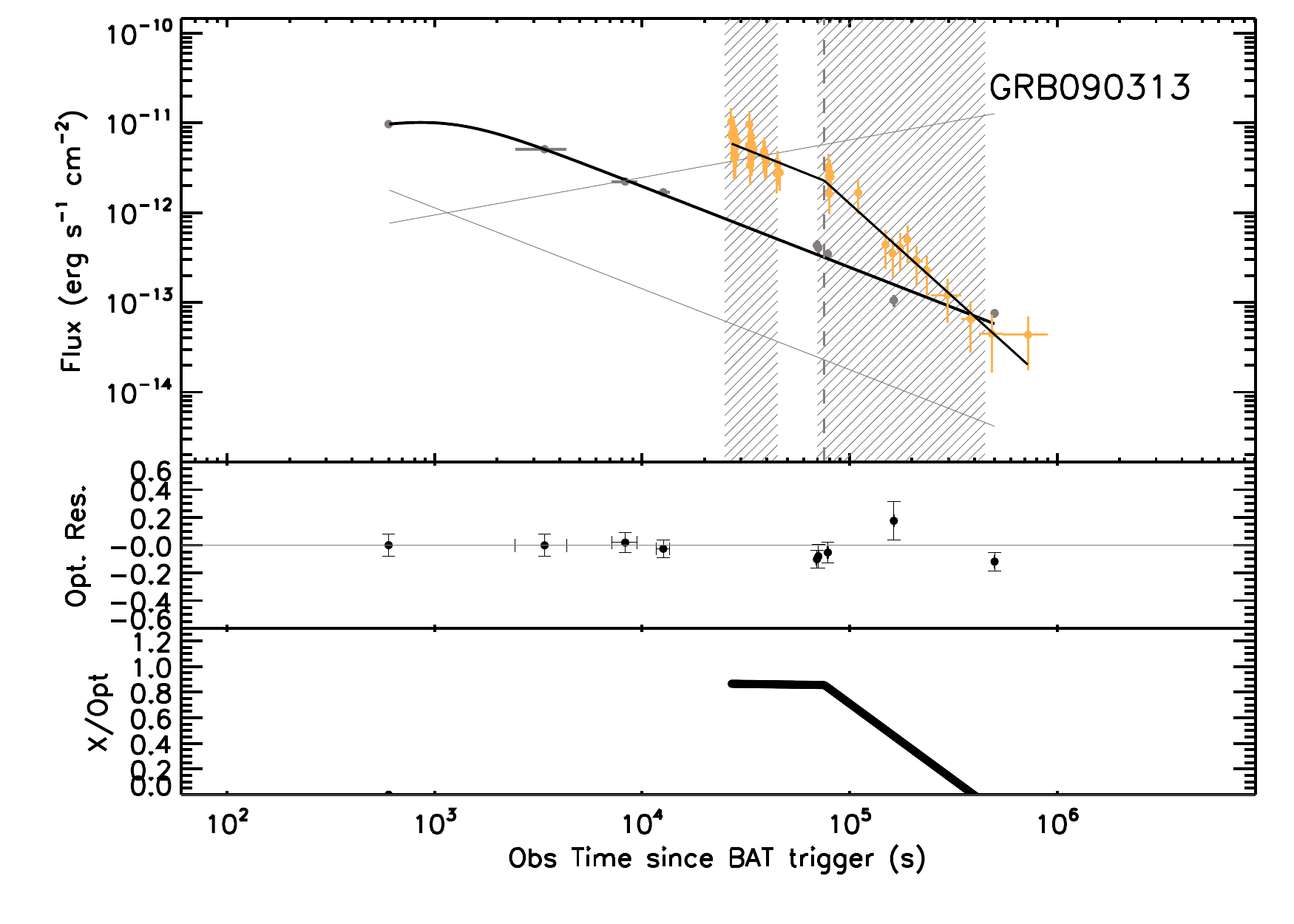}
\includegraphics[width=0.45 \hsize,clip]{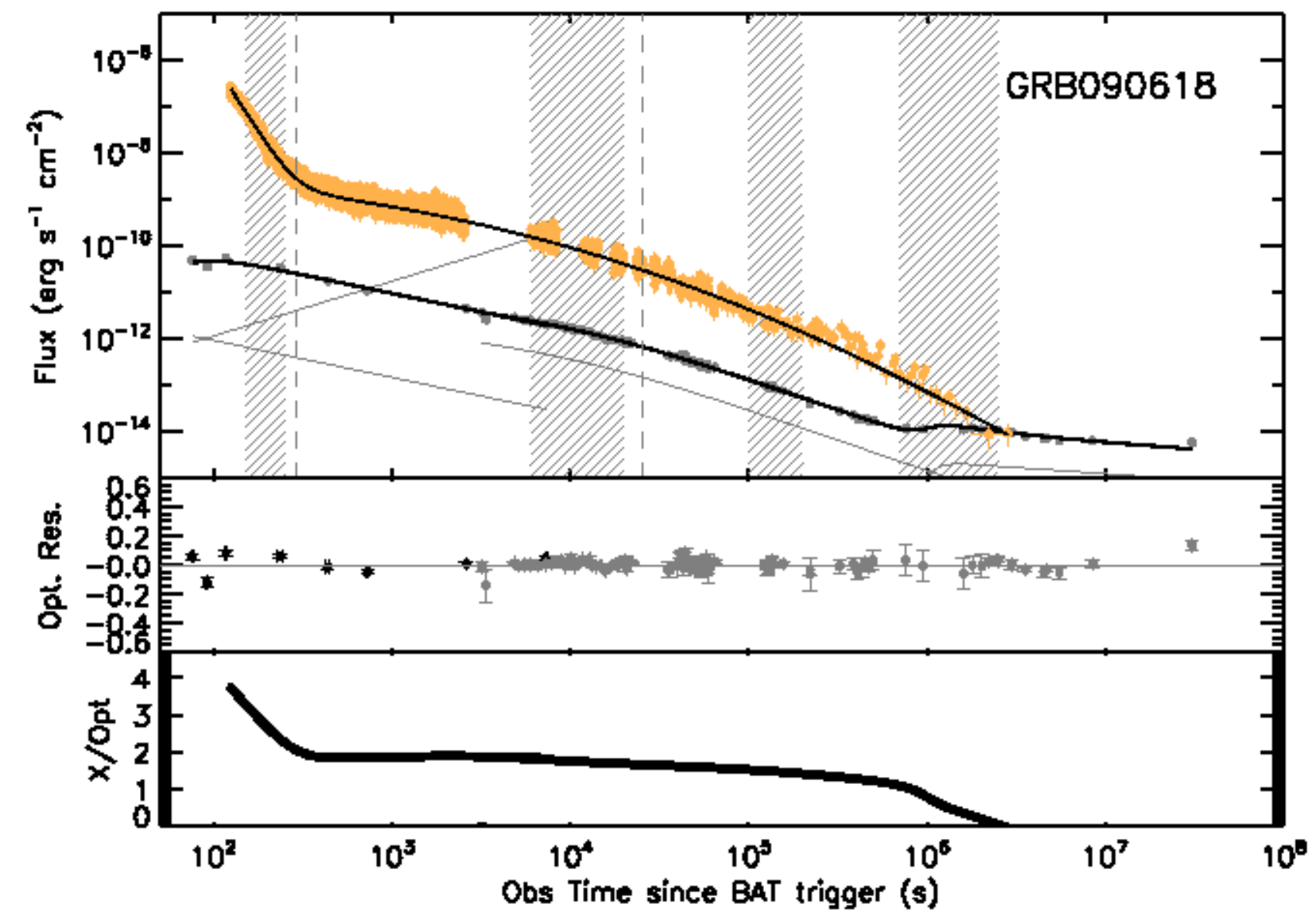}\\
\includegraphics[width=0.45 \hsize,clip]{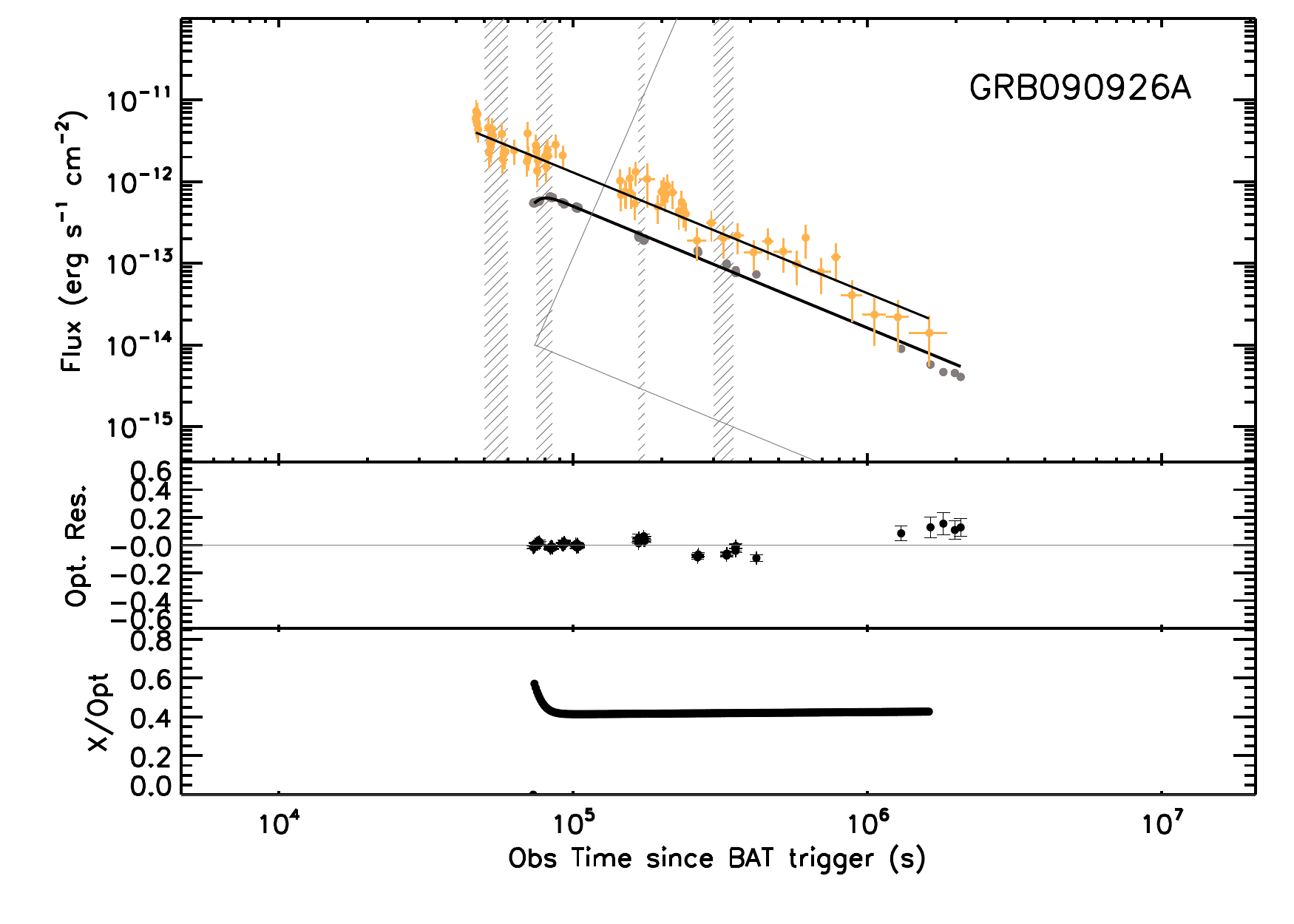}
\includegraphics[width=0.45 \hsize,clip]{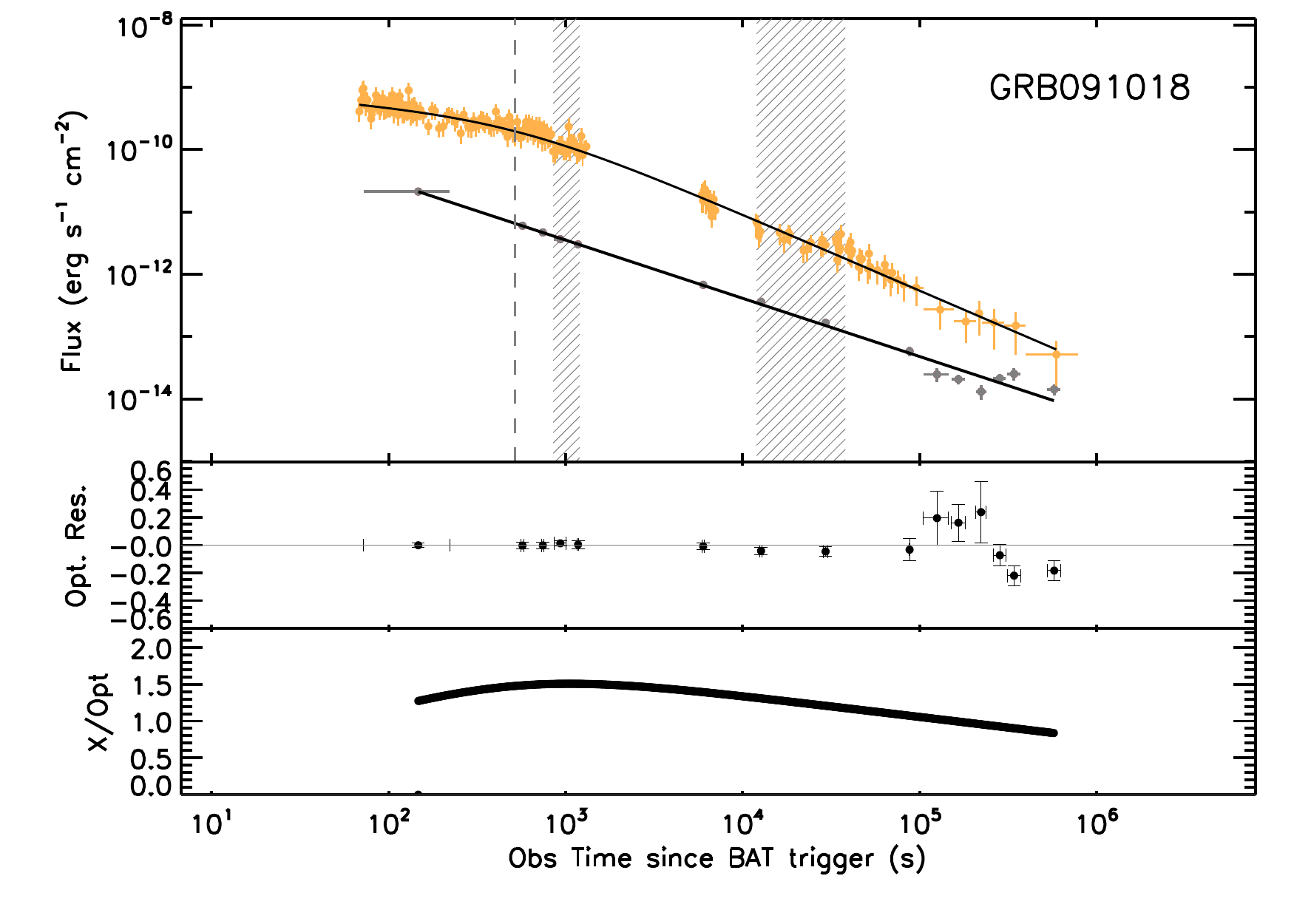}
\caption{\small{Comparison between optical and X-ray LCs. color-coding as in Figure~\ref{confronto1}.}}\label{confronto5} 
\end{figure}

\begin{figure}
\includegraphics[width=0.45 \hsize,clip]{FIGURE/LC/050730_OP1X-eps-converted-to.pdf}
\includegraphics[width=0.45 \hsize,clip]{FIGURE/LC/050820A_OP1X-eps-converted-to.pdf}\\
\includegraphics[width=0.45 \hsize,clip]{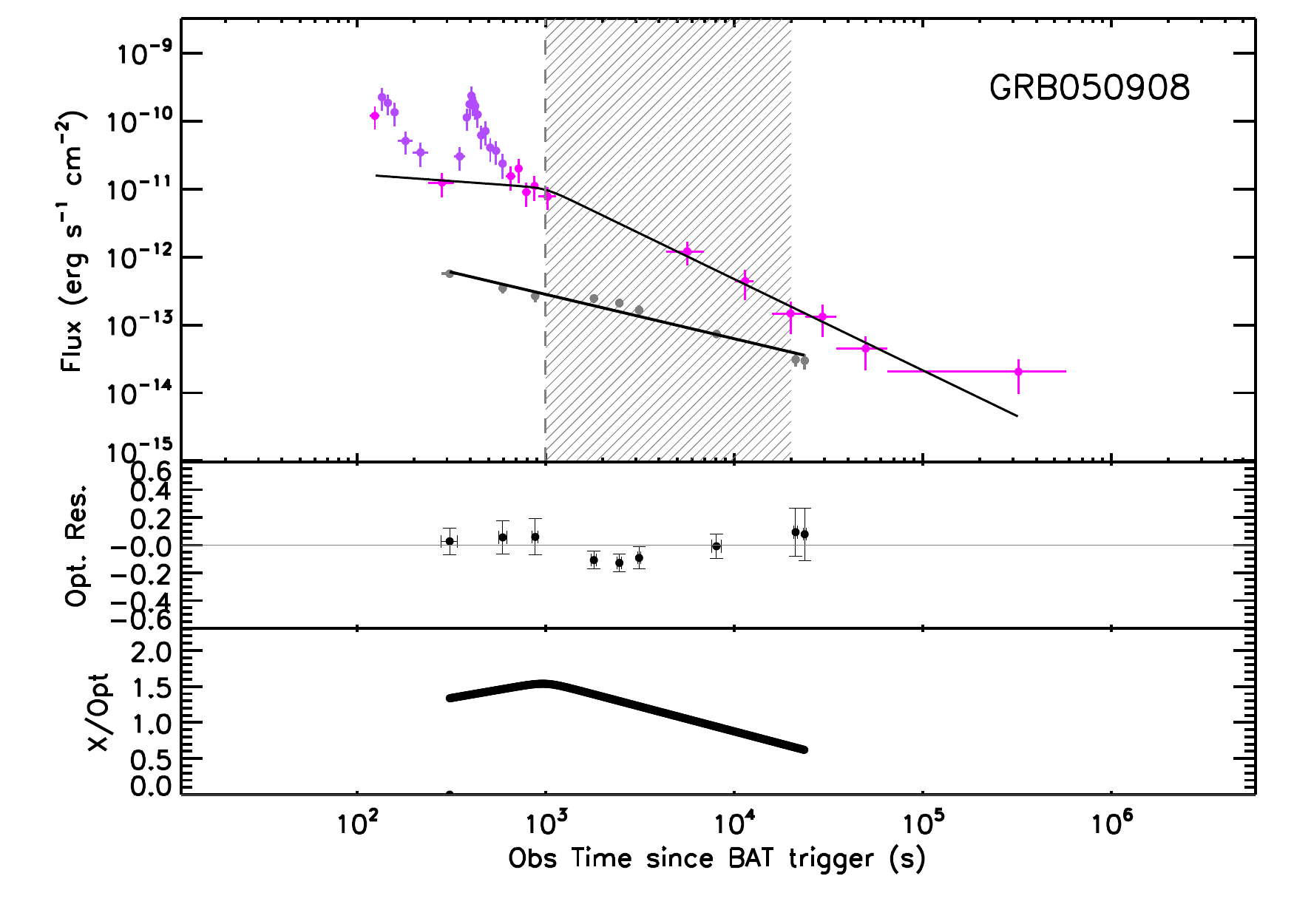}
\includegraphics[width=0.45 \hsize,clip]{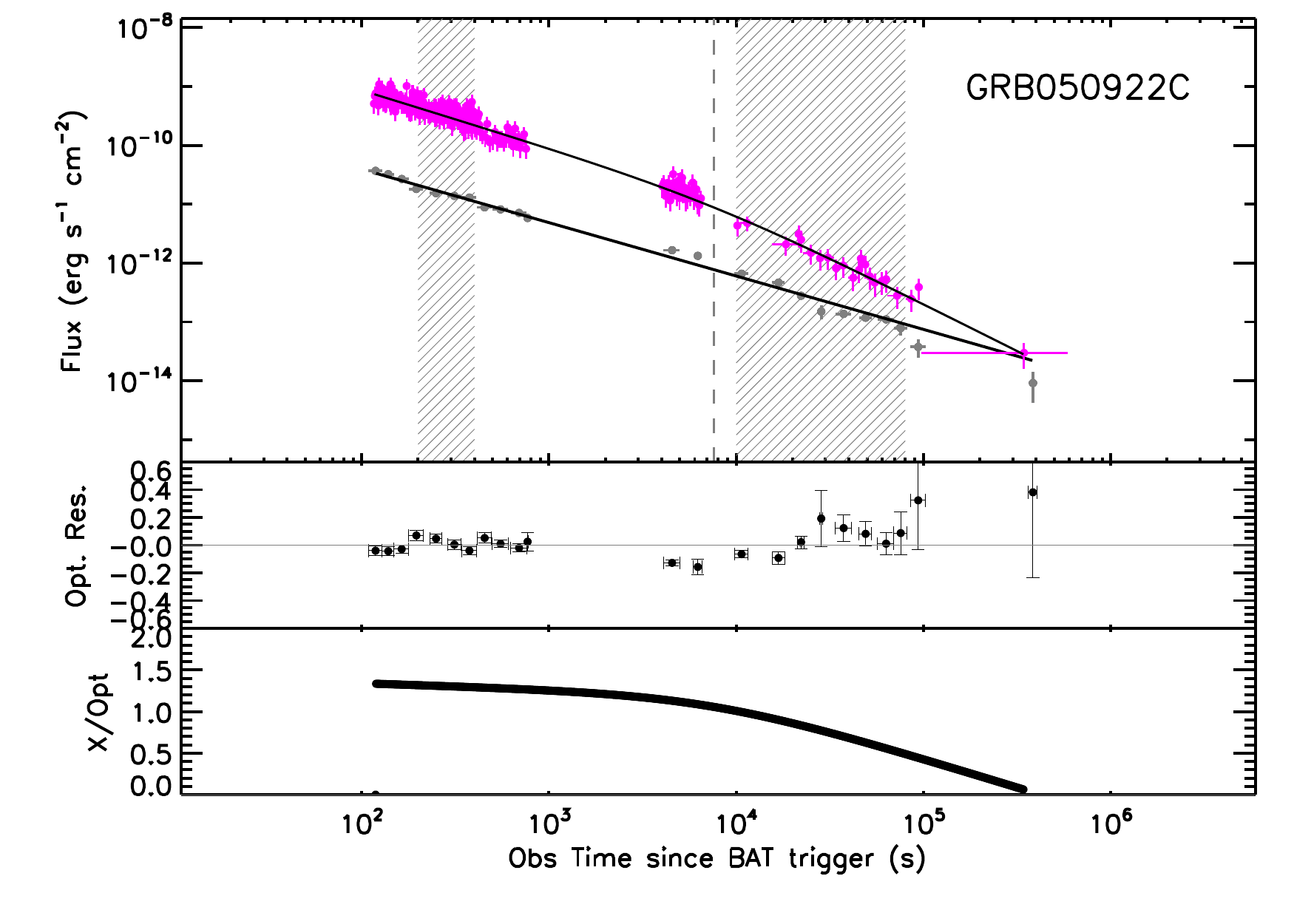}\\
\includegraphics[width=0.45 \hsize,clip]{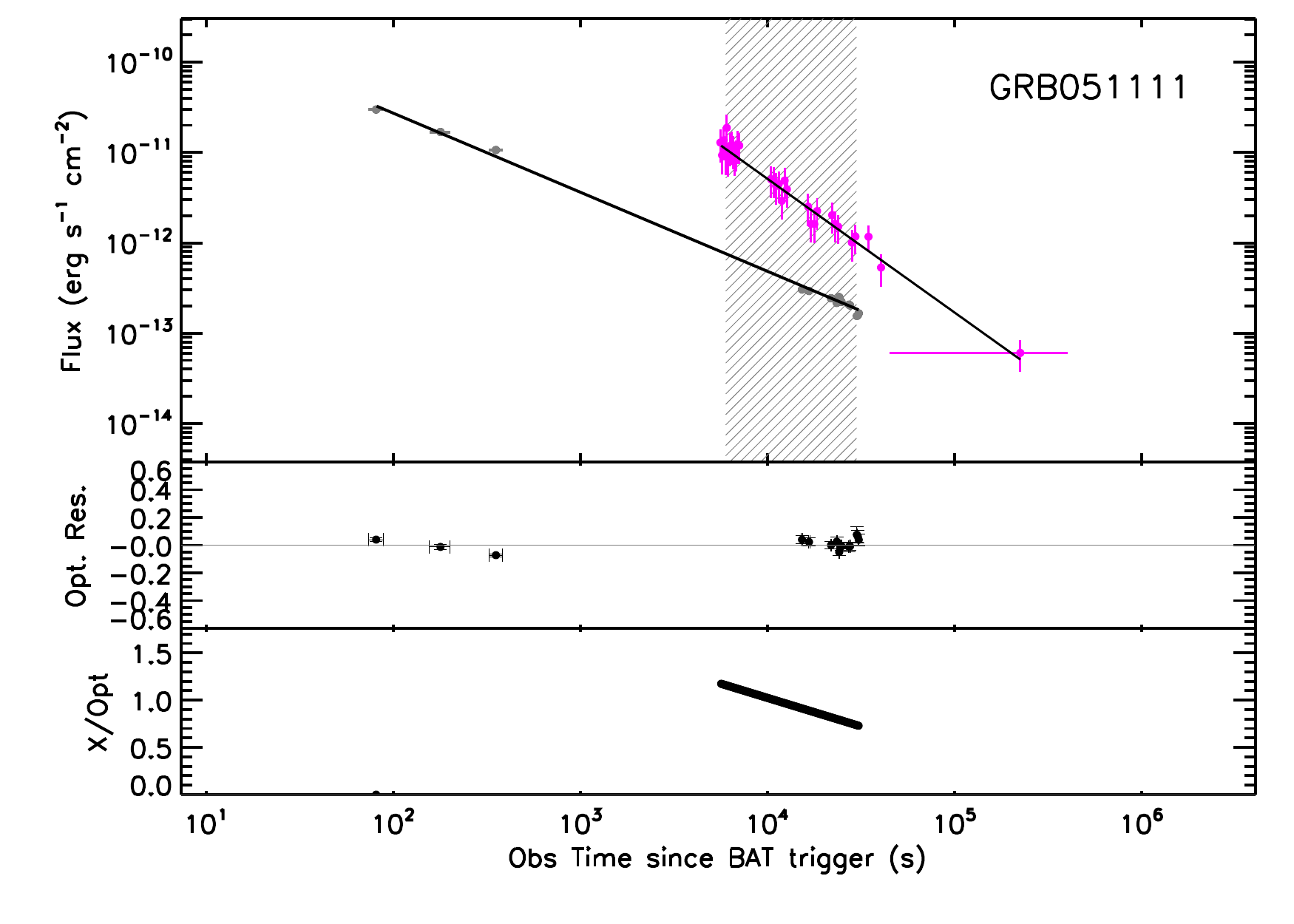}
\includegraphics[width=0.45 \hsize,clip]{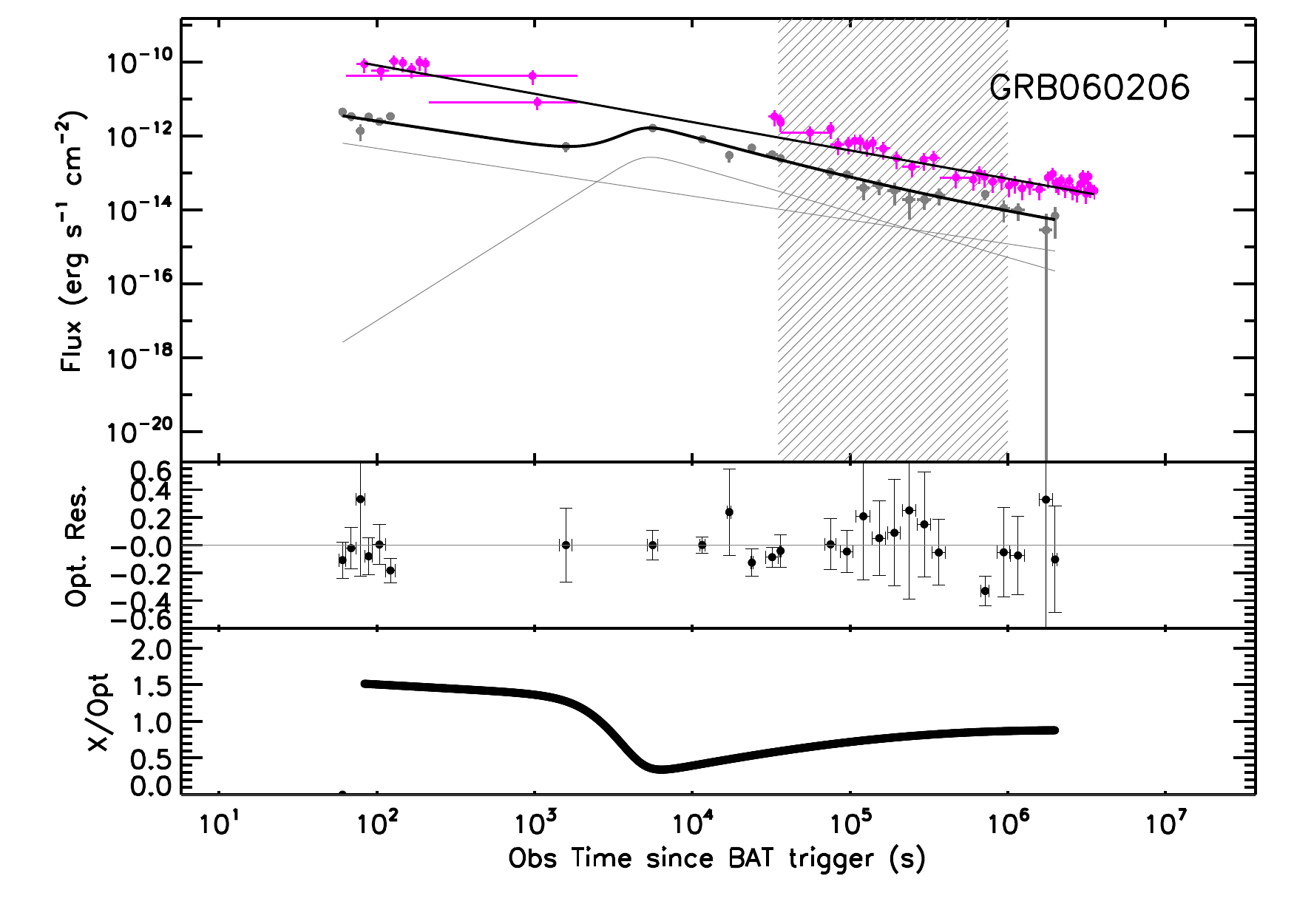}\\
\includegraphics[width=0.45 \hsize,clip]{FIGURE/LC/060210_OP1X-eps-converted-to.pdf}
\includegraphics[width=0.45 \hsize,clip]{FIGURE/LC/060418_OP1X-eps-converted-to.pdf}
\caption{\small{Comparison between optical and X-ray LCs. color-coding as in Figure~\ref{confronto1}.}}\label{confronto6} 
\end{figure}

\begin{figure}
\includegraphics[width=0.45 \hsize,clip]{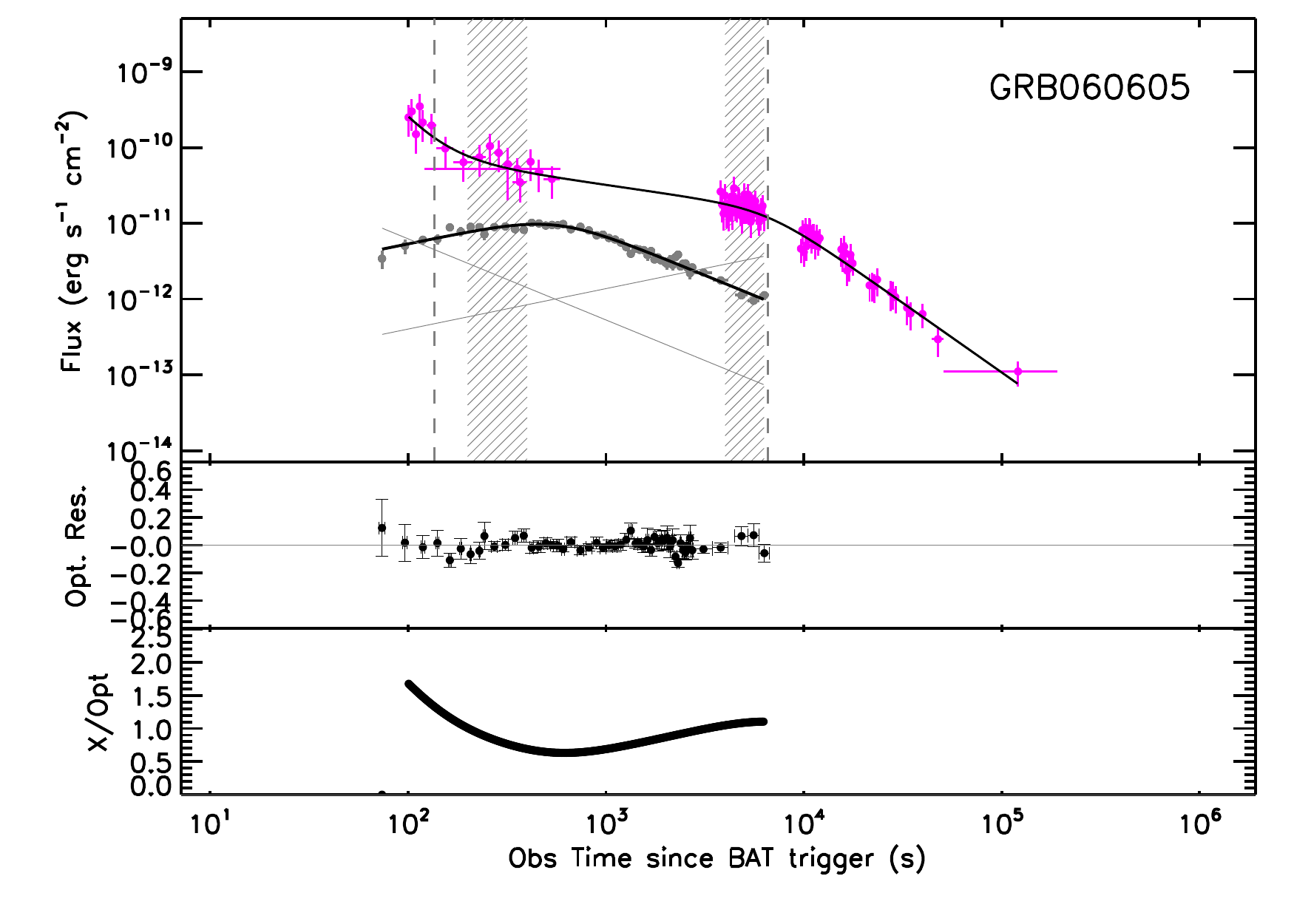}
\includegraphics[width=0.45 \hsize,clip]{FIGURE/LC/060607A_OP1X-eps-converted-to.pdf}\\
\includegraphics[width=0.45 \hsize,clip]{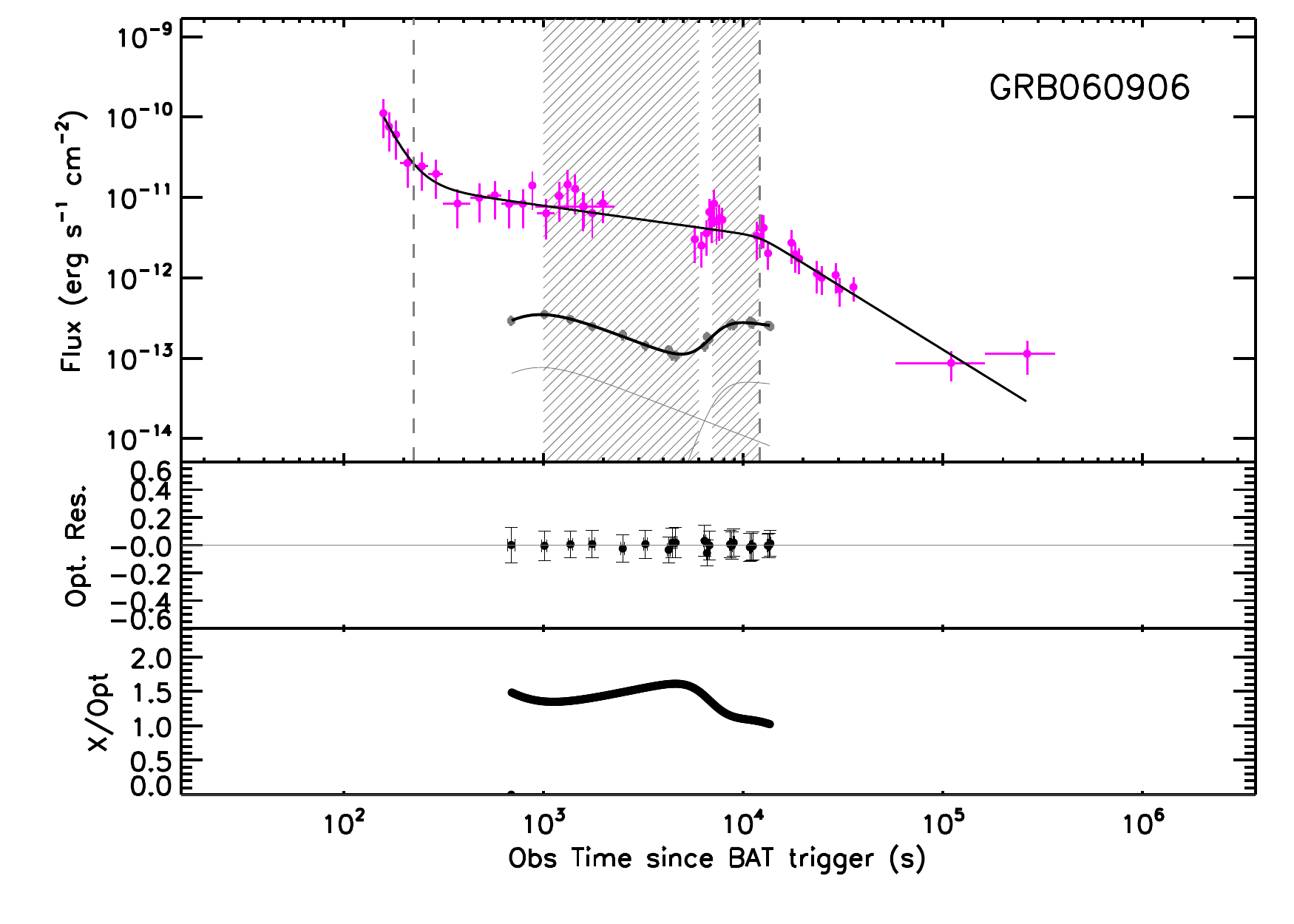}
\includegraphics[width=0.45 \hsize,clip]{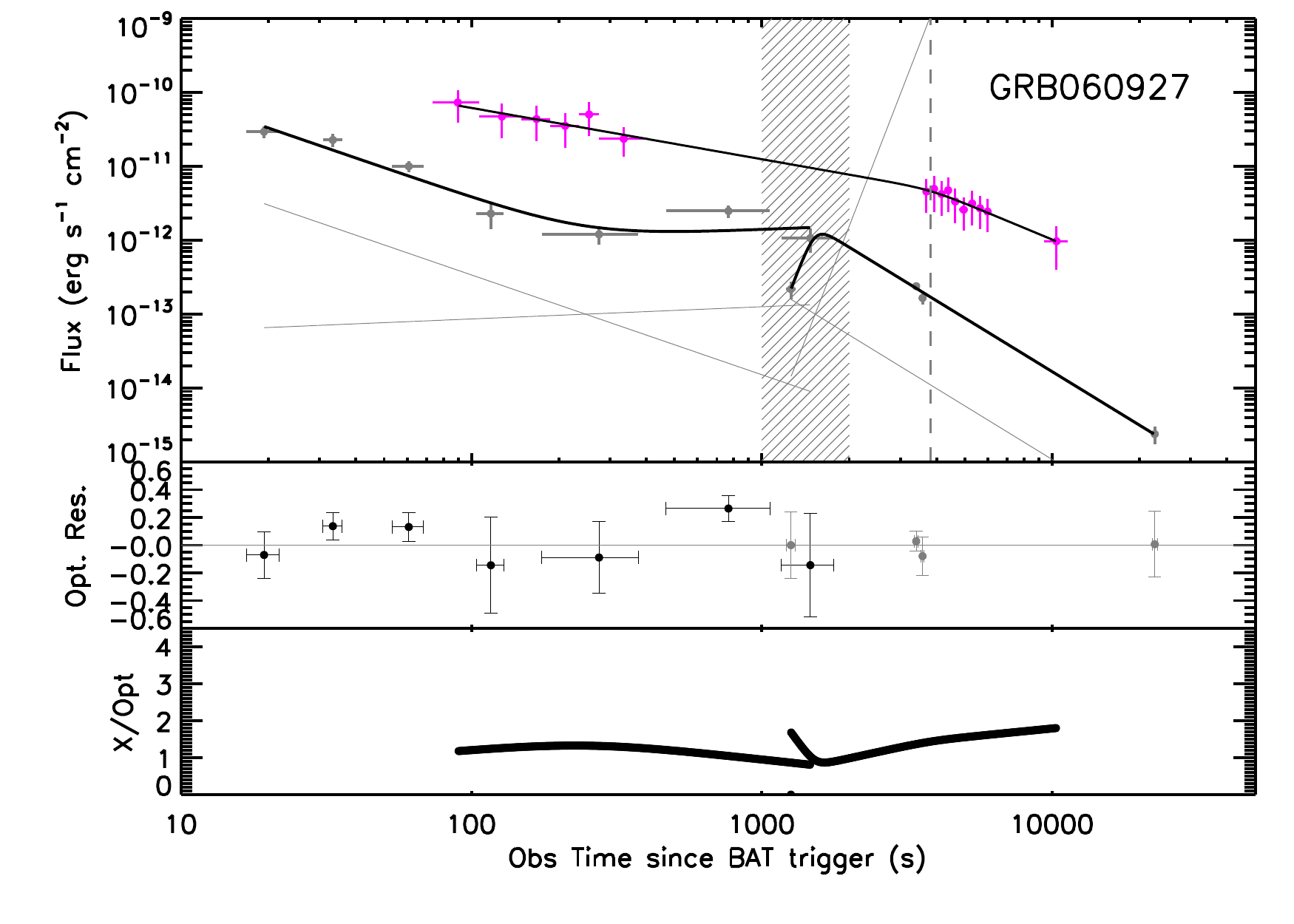}\\
\includegraphics[width=0.45 \hsize,clip]{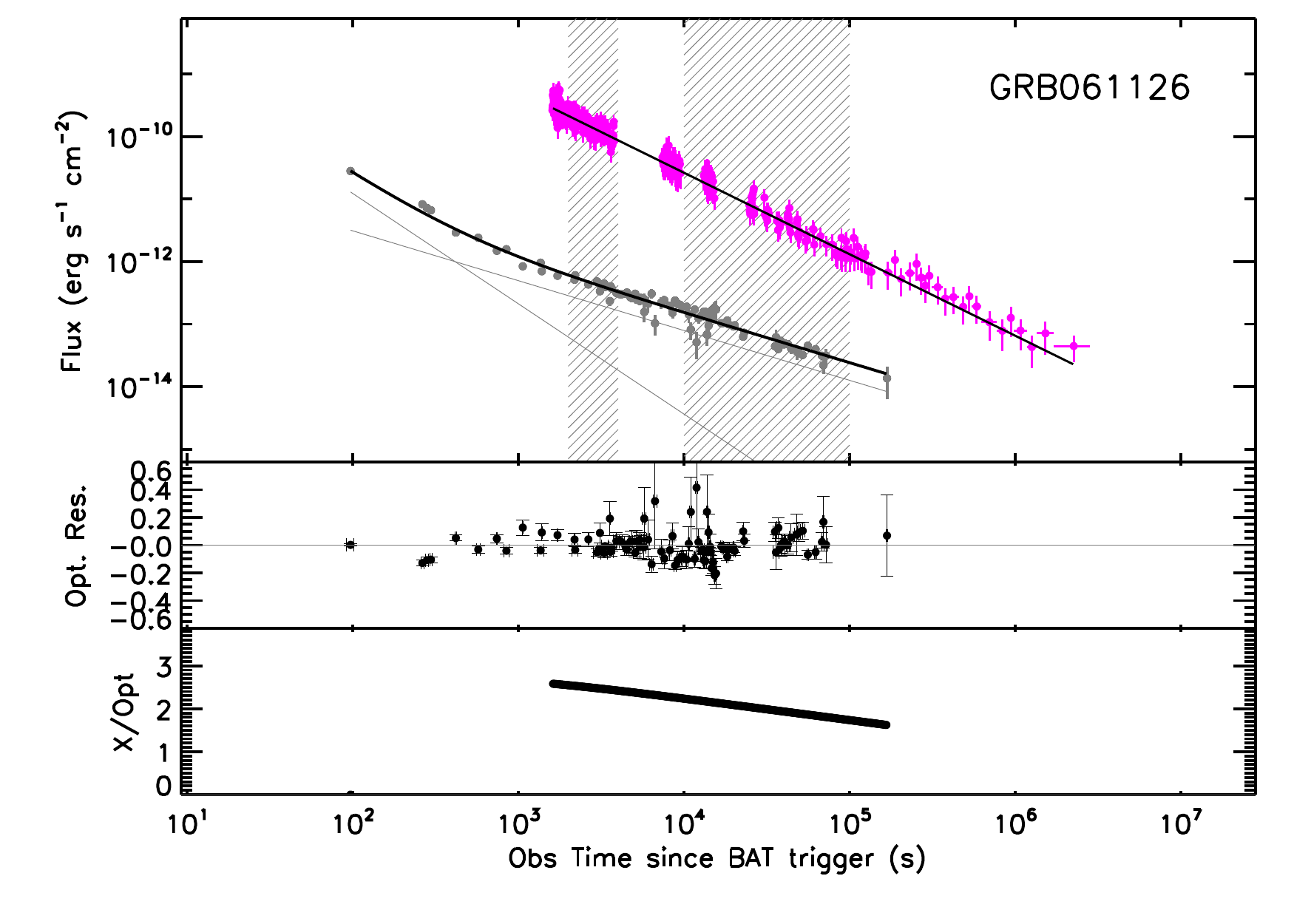}
\includegraphics[width=0.45 \hsize,clip]{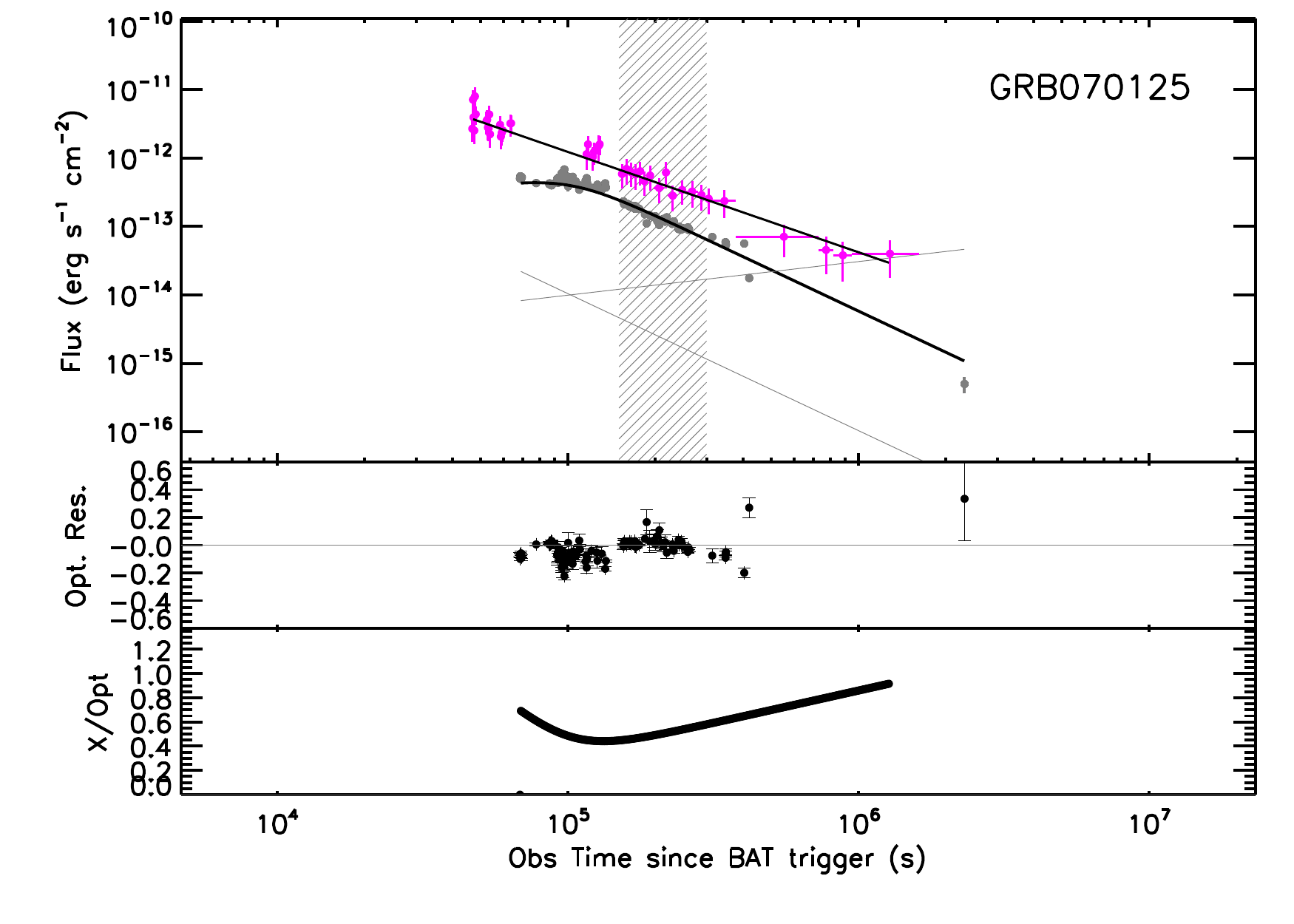}\\
\includegraphics[width=0.45 \hsize,clip]{FIGURE/LC/070318_OP1X-eps-converted-to.pdf}
\includegraphics[width=0.45 \hsize,clip]{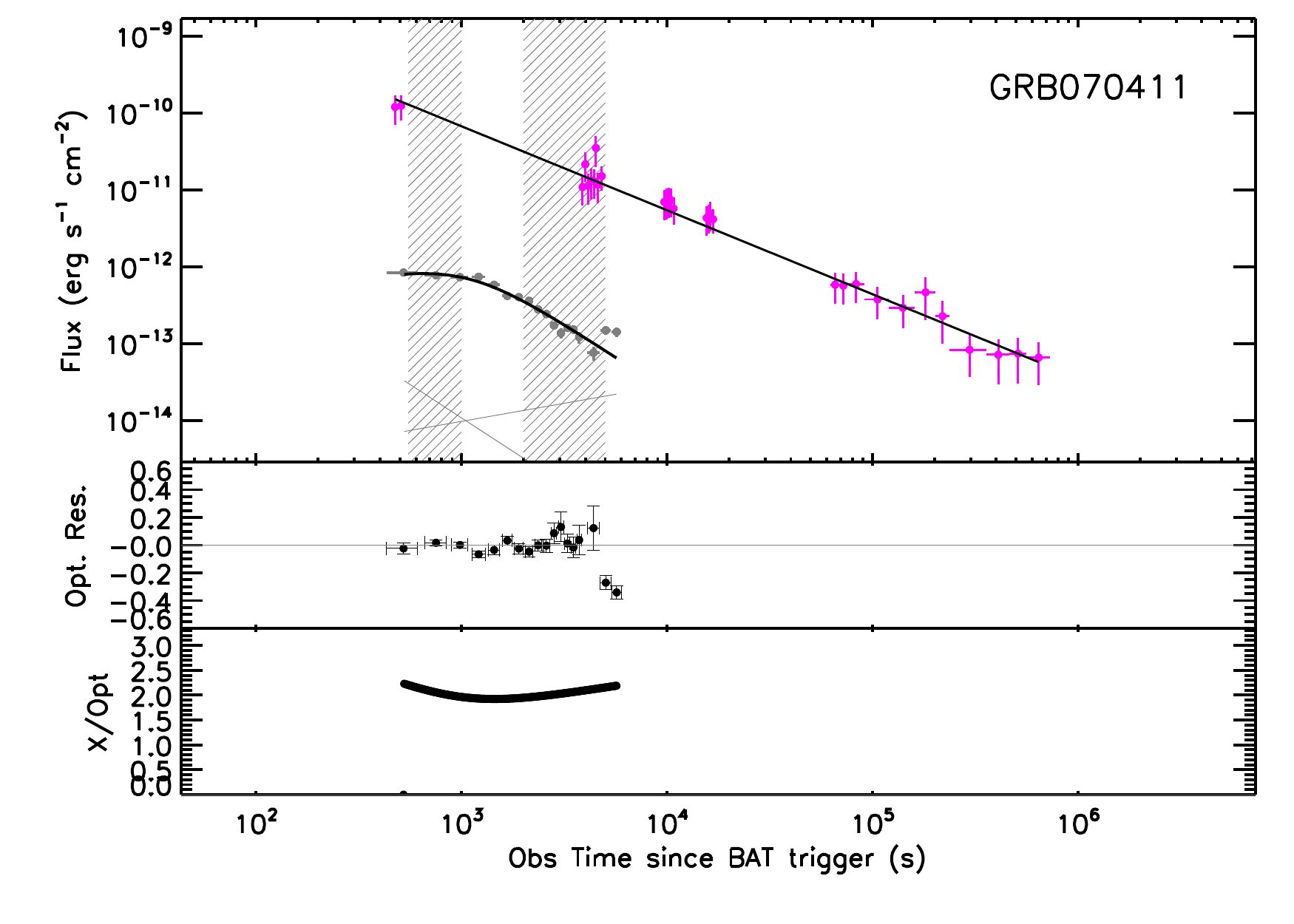}
\caption{\small{Comparison between optical and X-ray LCs. color-coding as in Figure~\ref{confronto1}.}}\label{confronto7} 
\end{figure}

\begin{figure}
\includegraphics[width=0.45 \hsize,clip]{FIGURE/LC/070419A_OP1X-eps-converted-to.pdf}
\includegraphics[width=0.45 \hsize,clip]{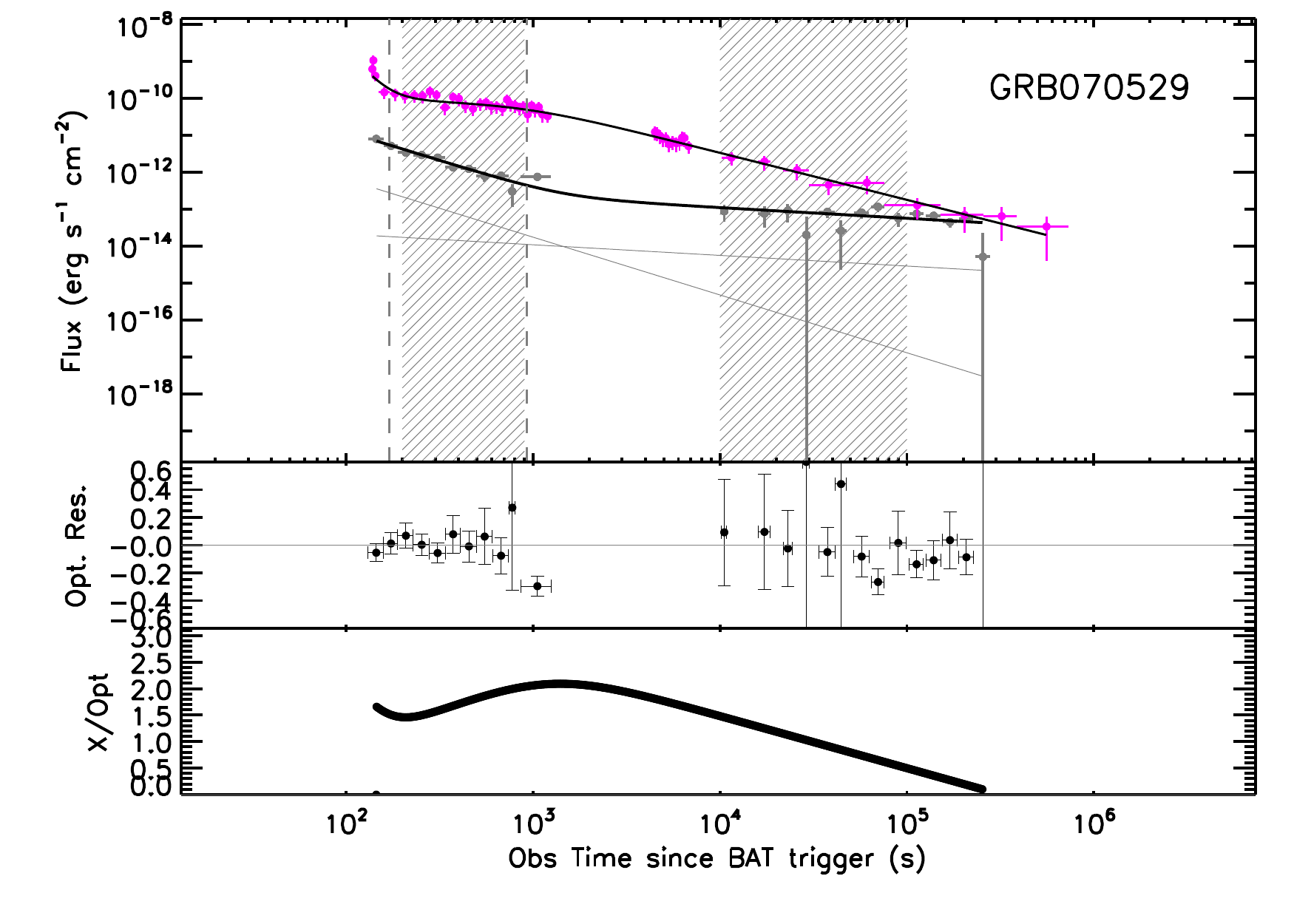}\\
\includegraphics[width=0.45 \hsize,clip]{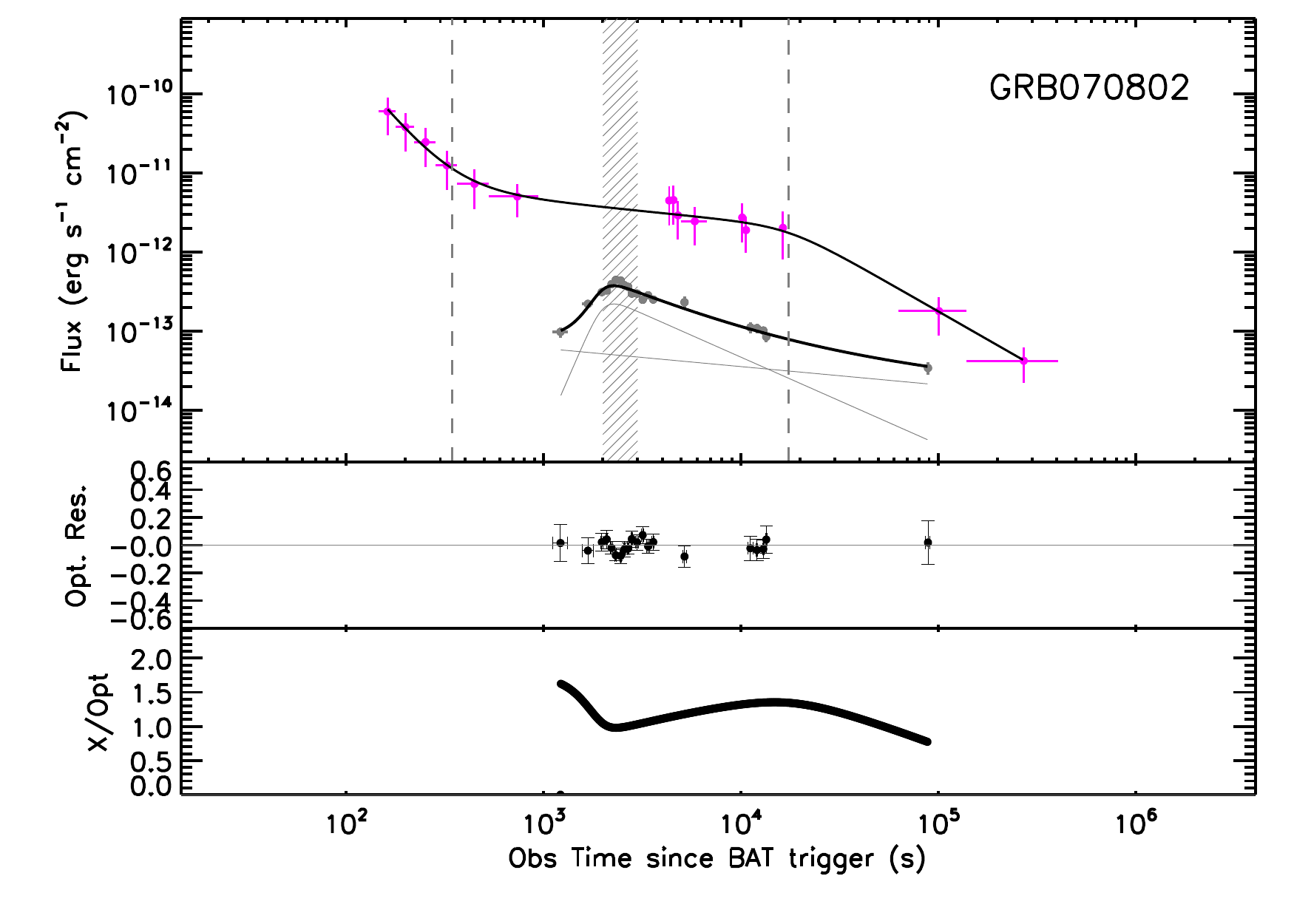}
\includegraphics[width=0.45 \hsize,clip]{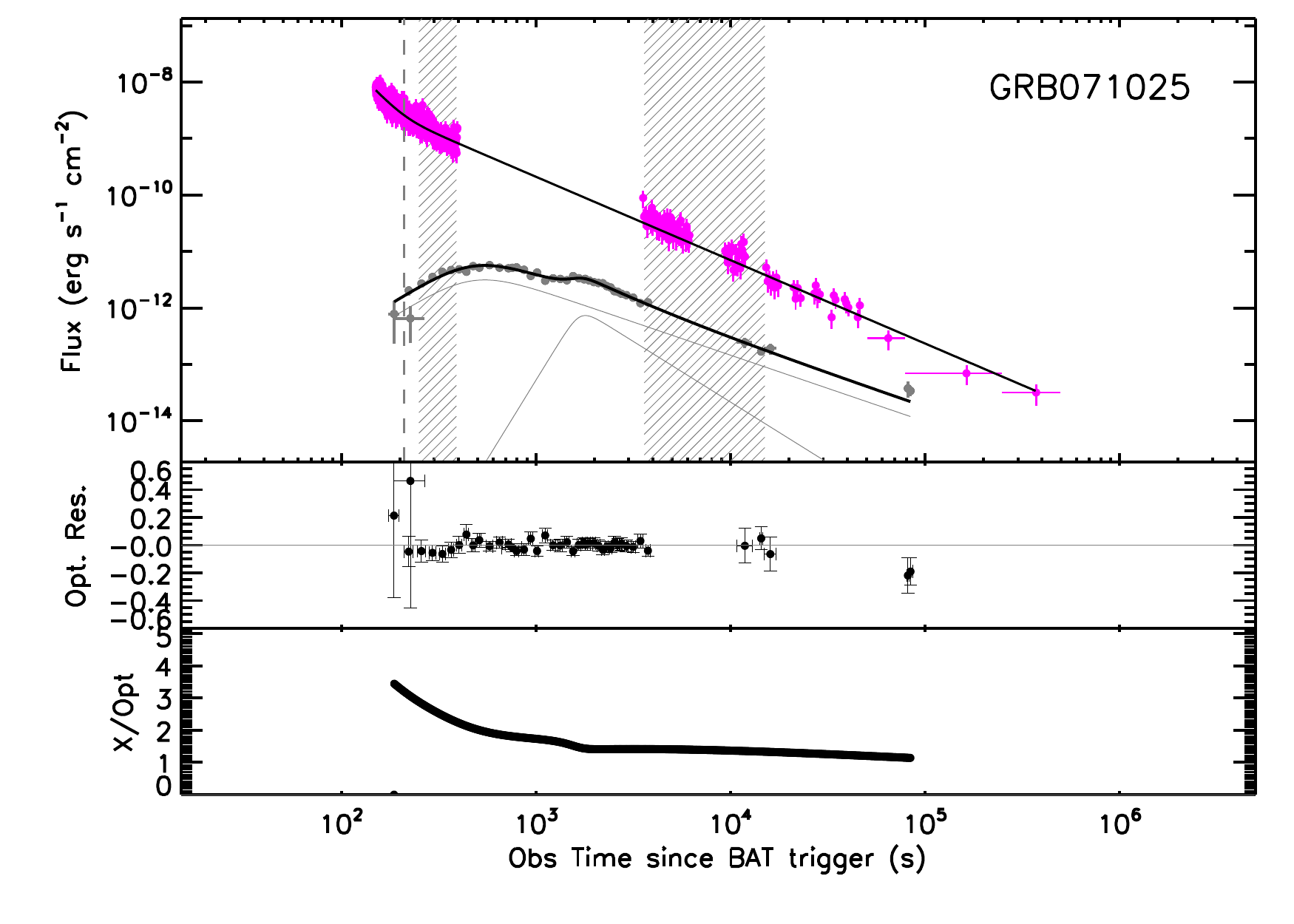}\\
\includegraphics[width=0.45 \hsize,clip]{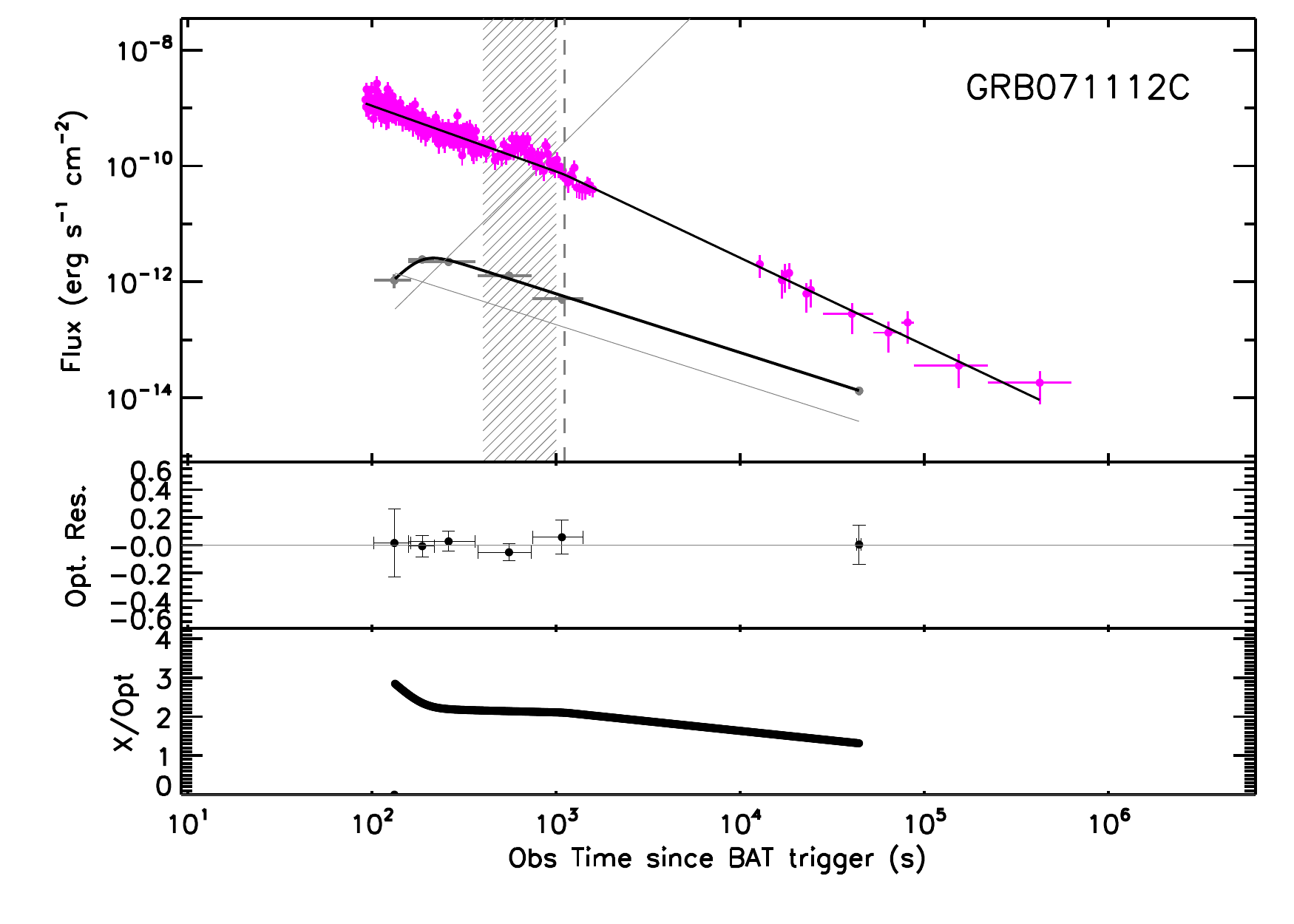}
\includegraphics[width=0.45 \hsize,clip]{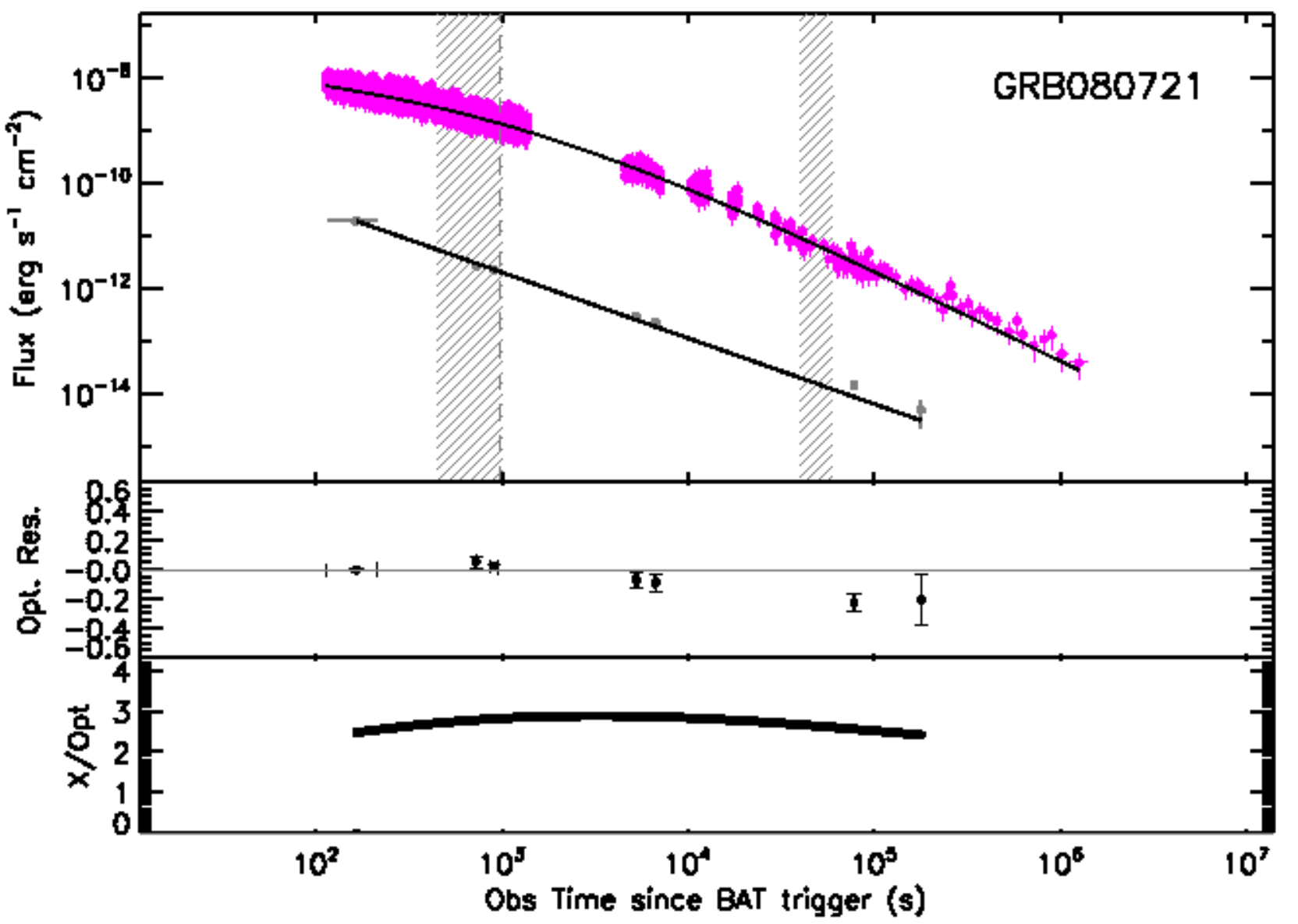}\\
\includegraphics[width=0.45 \hsize,clip]{FIGURE/LC/080810_OP1X-eps-converted-to.pdf}
\includegraphics[width=0.45 \hsize,clip]{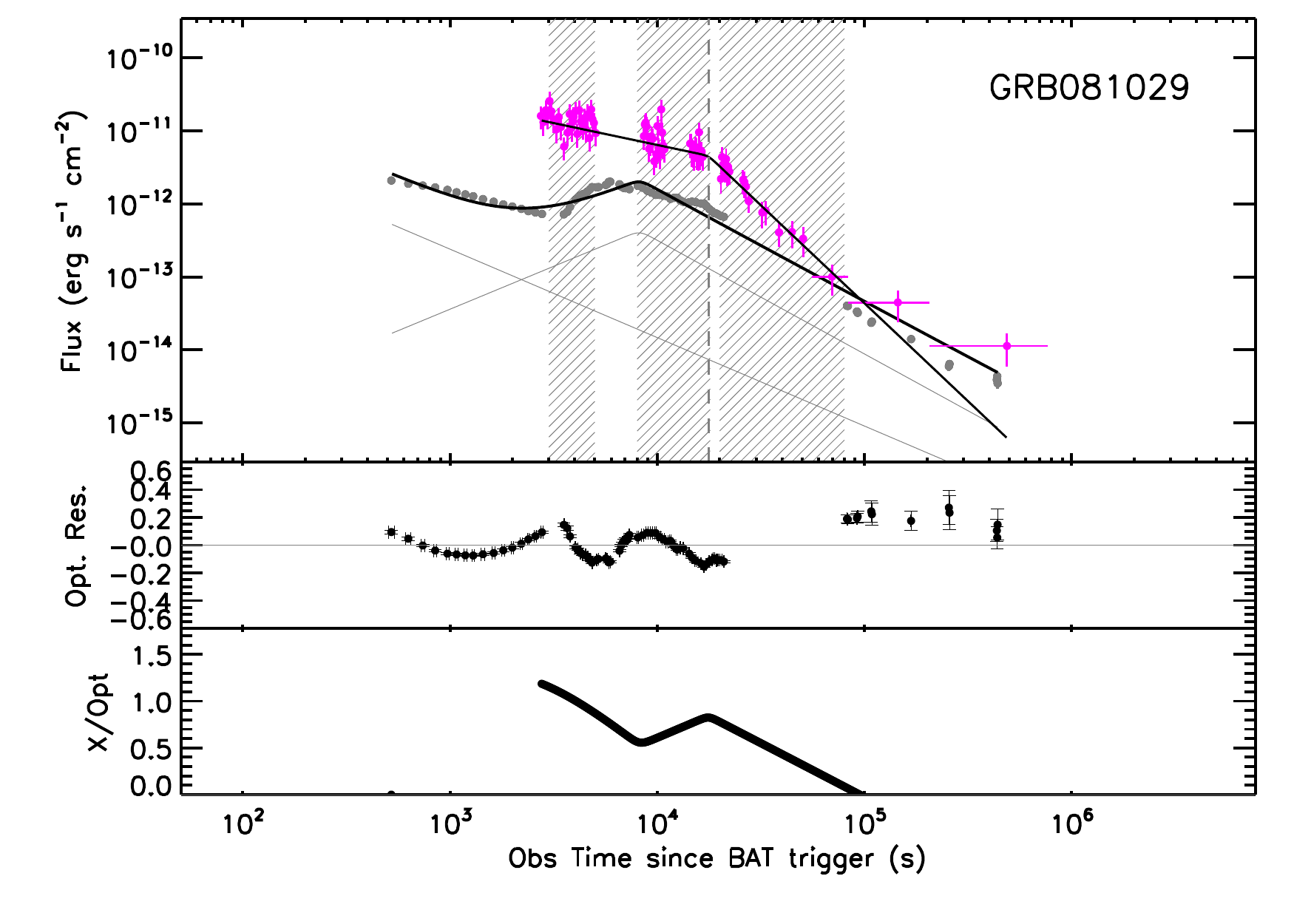}
\caption{\small{Comparison between optical and X-ray LCs. color-coding as in Figure~\ref{confronto1}.}}\label{confronto8} 
\end{figure}

\begin{figure}
\includegraphics[width=0.45 \hsize,clip]{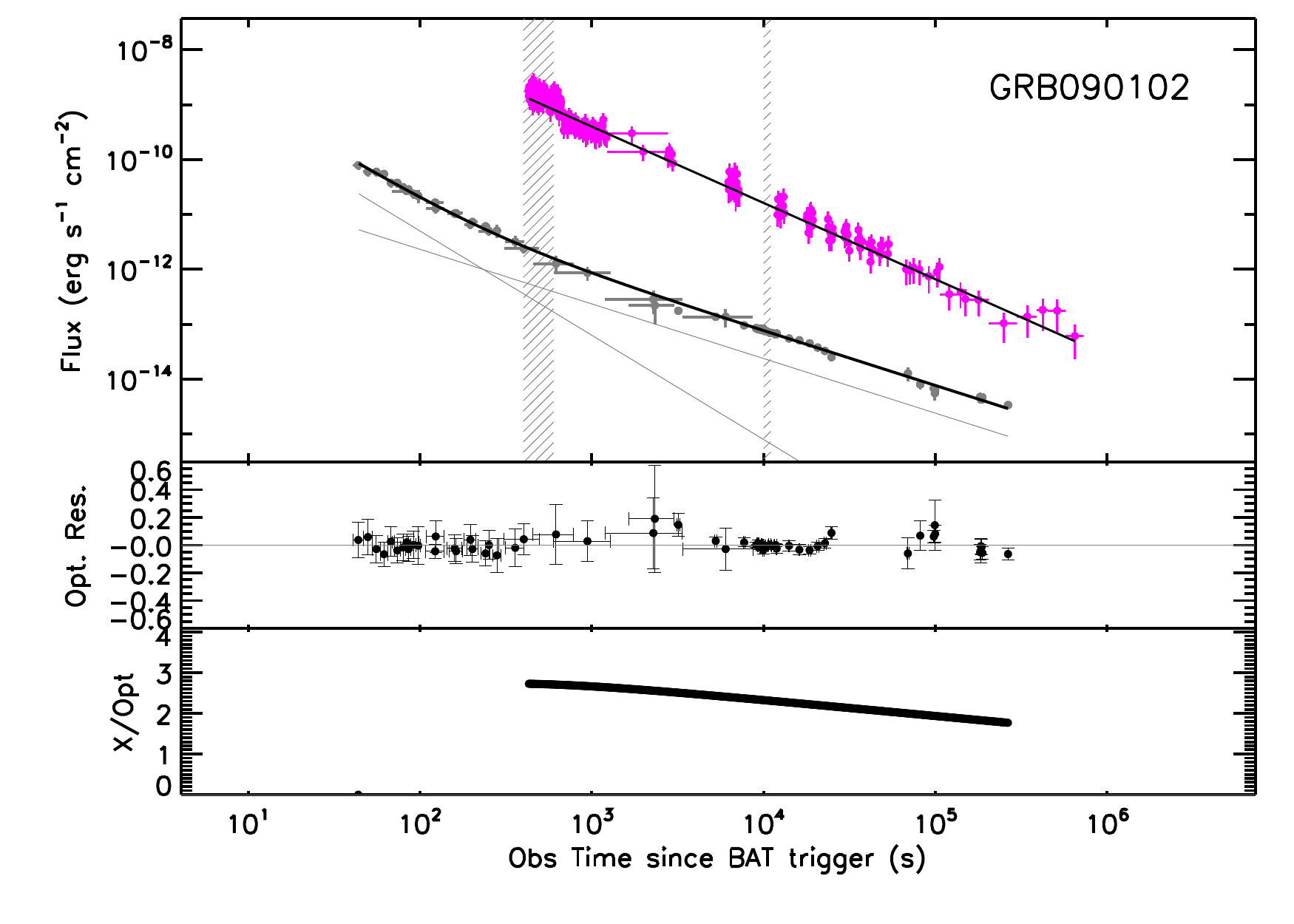}
\includegraphics[width=0.45 \hsize,clip]{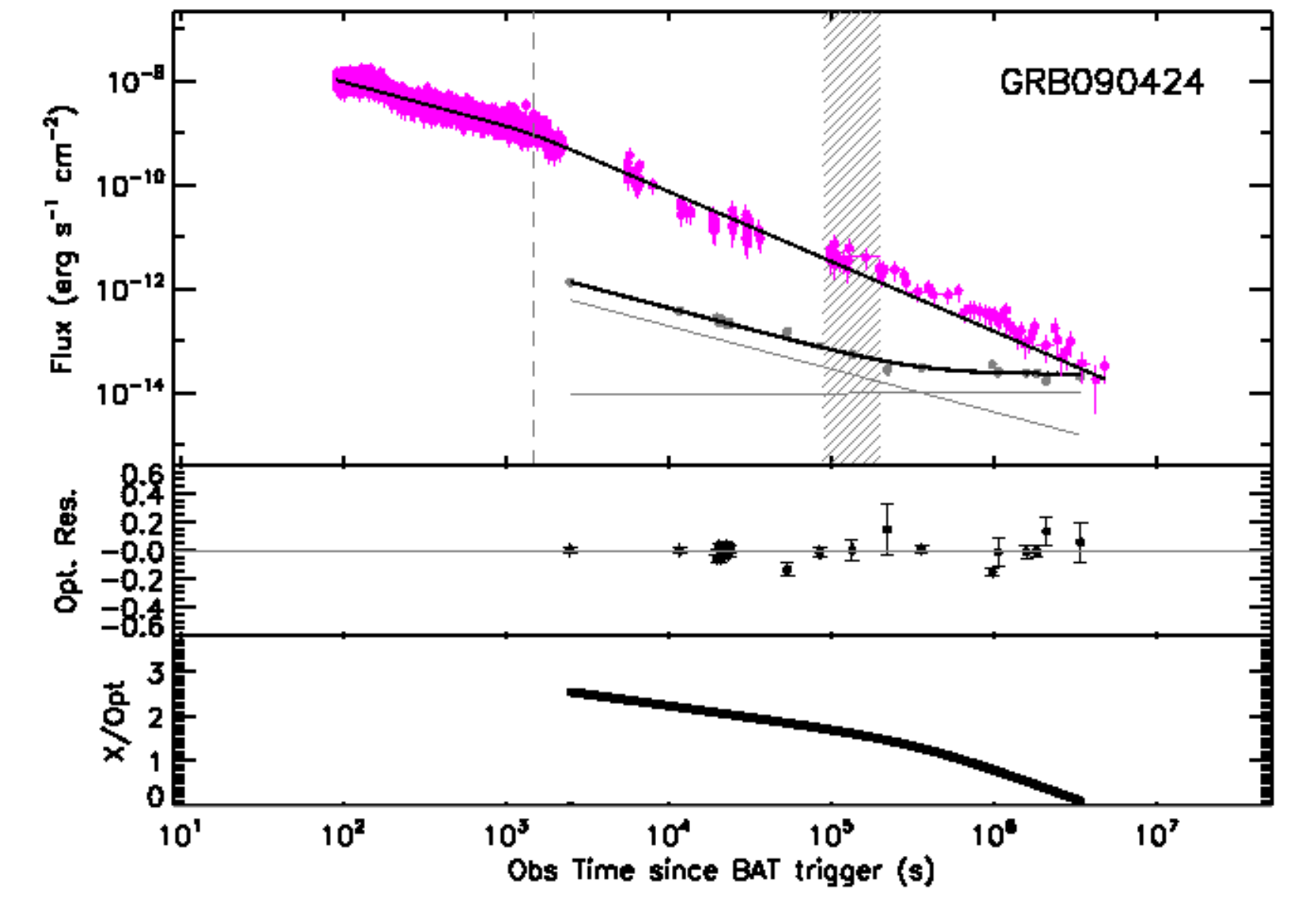}\\
\includegraphics[width=0.45 \hsize,clip]{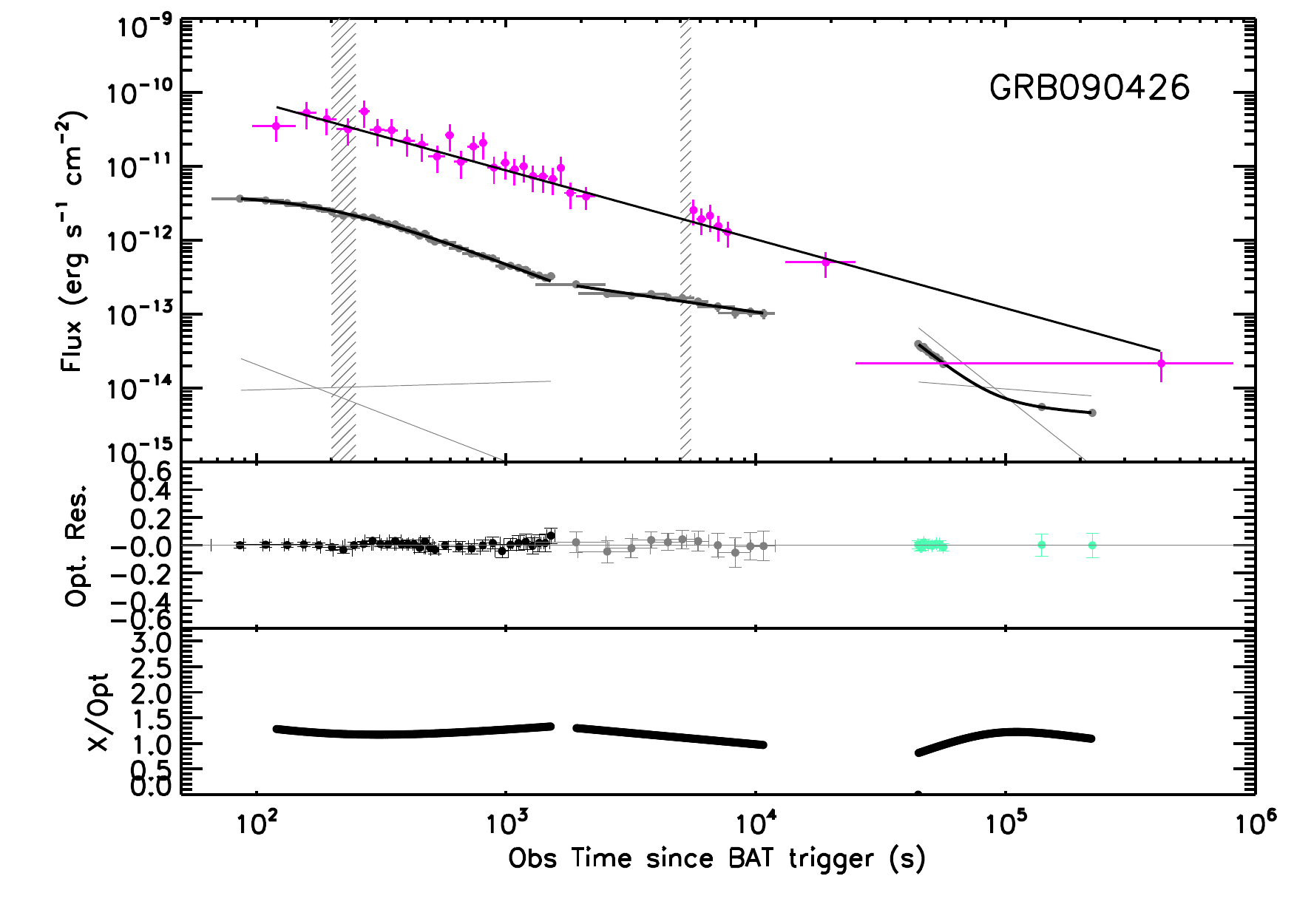}
\includegraphics[width=0.45 \hsize,clip]{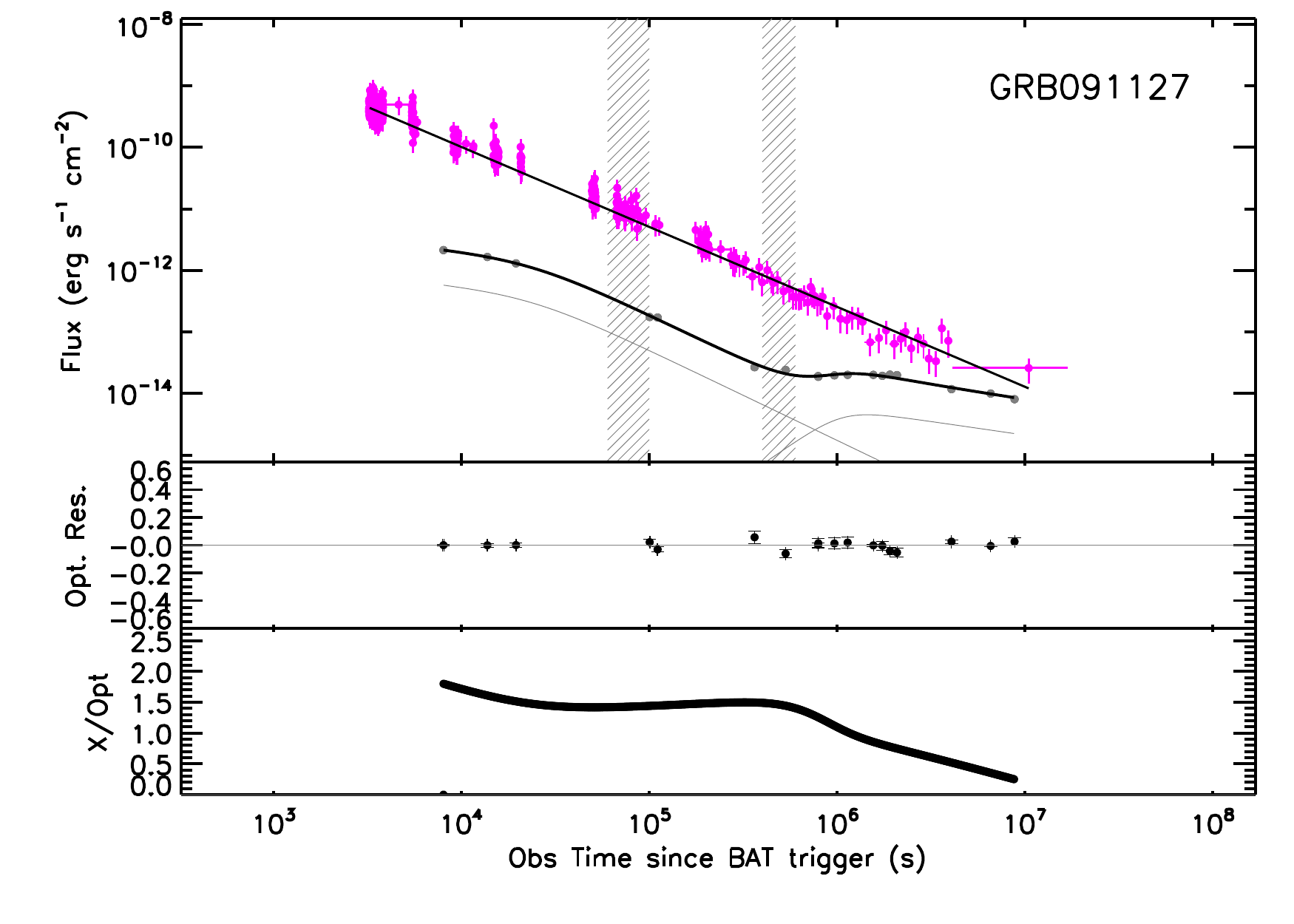}
\caption{\small{Comparison between optical and X-ray LCs. color-coding as in Figure~\ref{confronto1}.}}\label{confronto9} 
\end{figure}

\begin{figure}
\includegraphics[width=0.3 \hsize,clip]{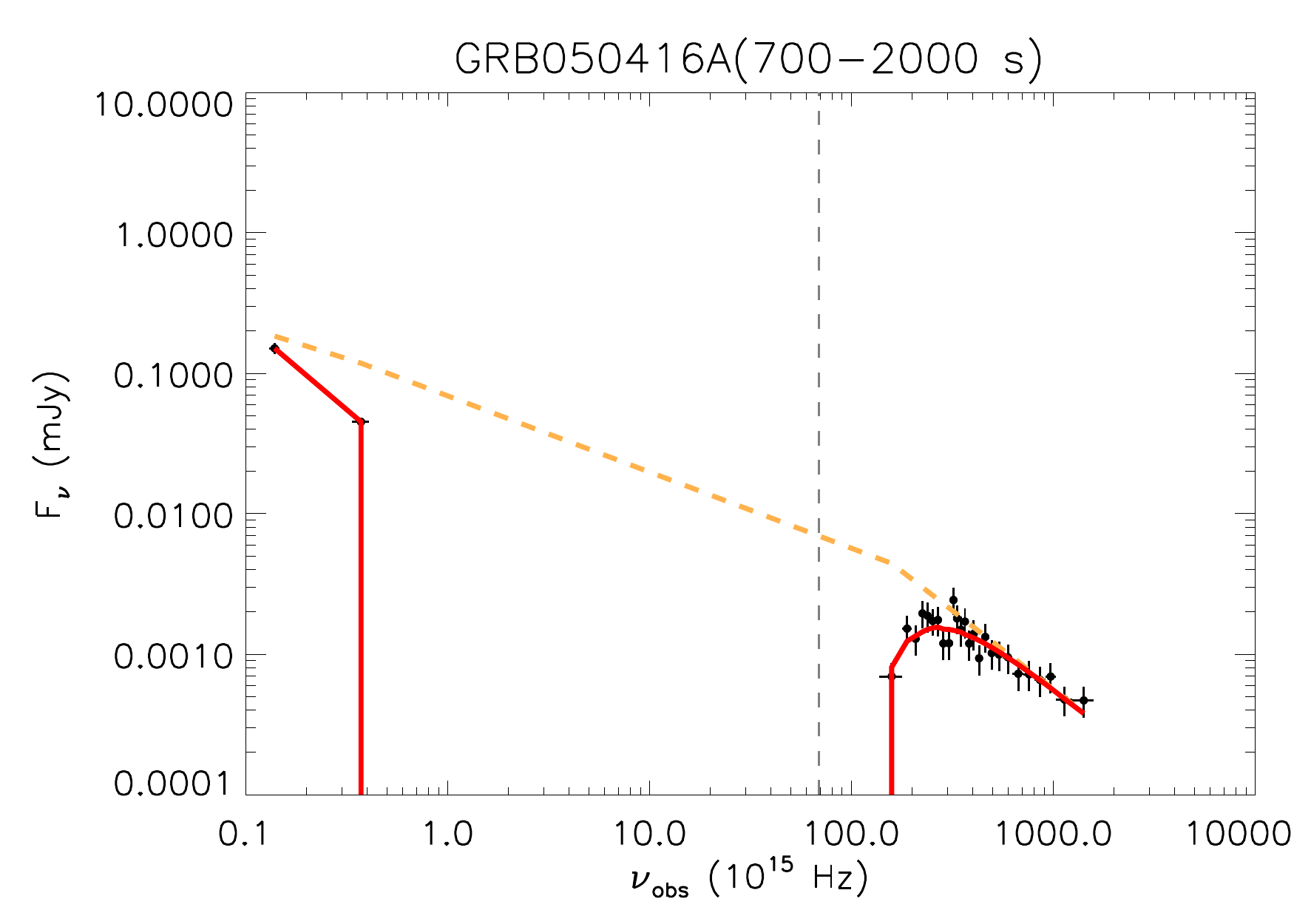}
\includegraphics[width=0.3 \hsize,clip]{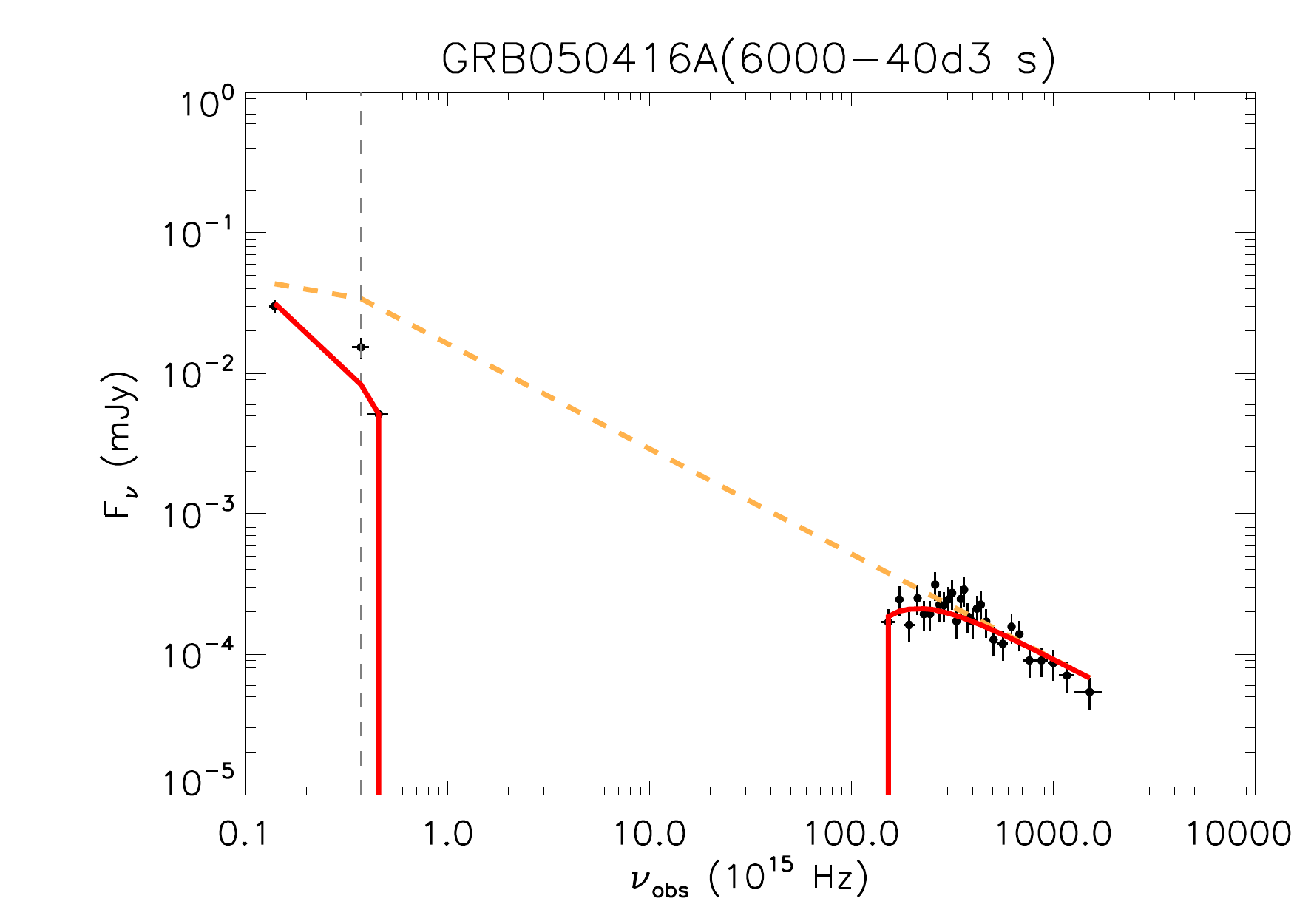}
\includegraphics[width=0.3 \hsize,clip]{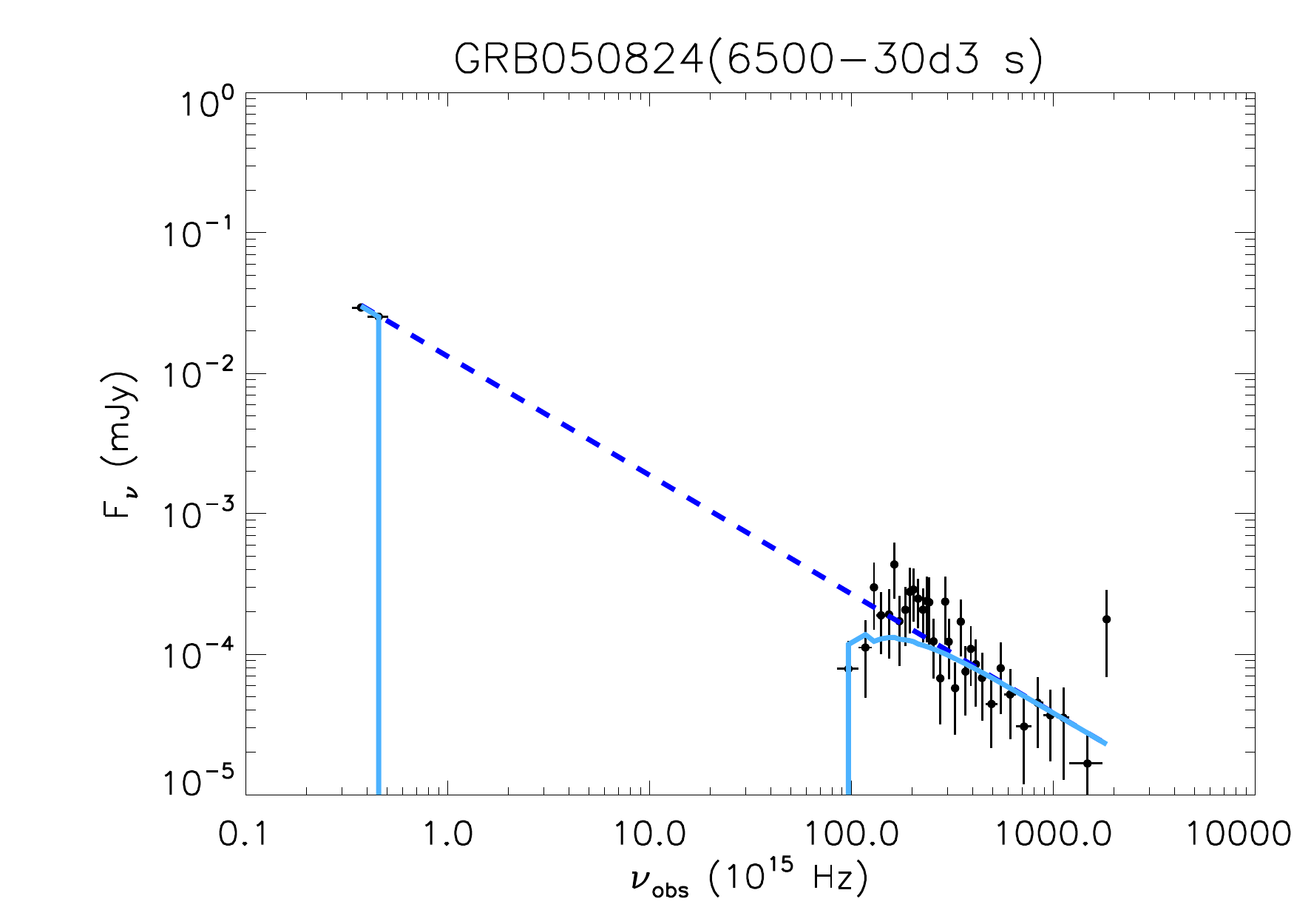}\\
\includegraphics[width=0.3 \hsize,clip]{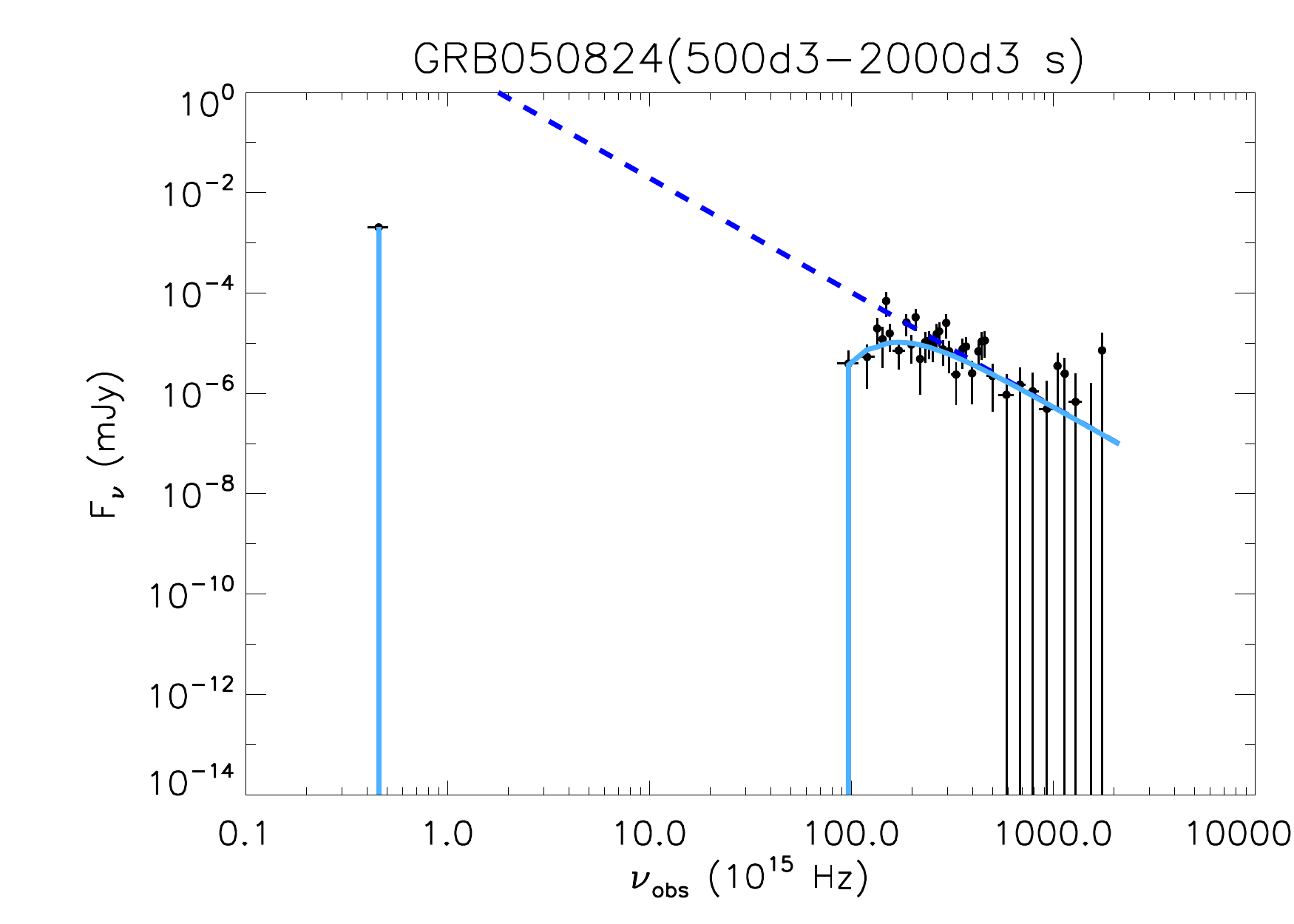}
\includegraphics[width=0.3 \hsize,clip]{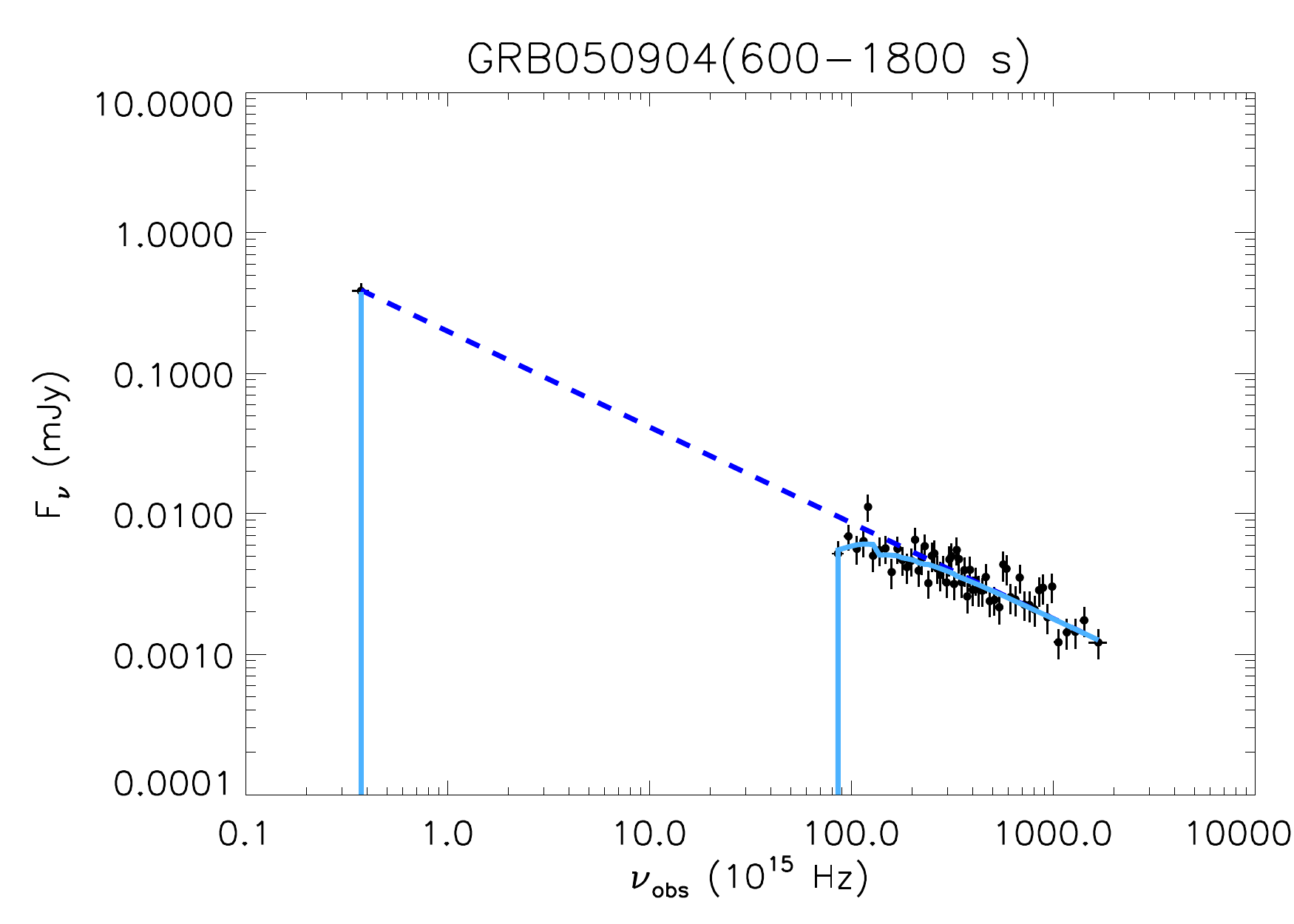}
\includegraphics[width=0.3 \hsize,clip]{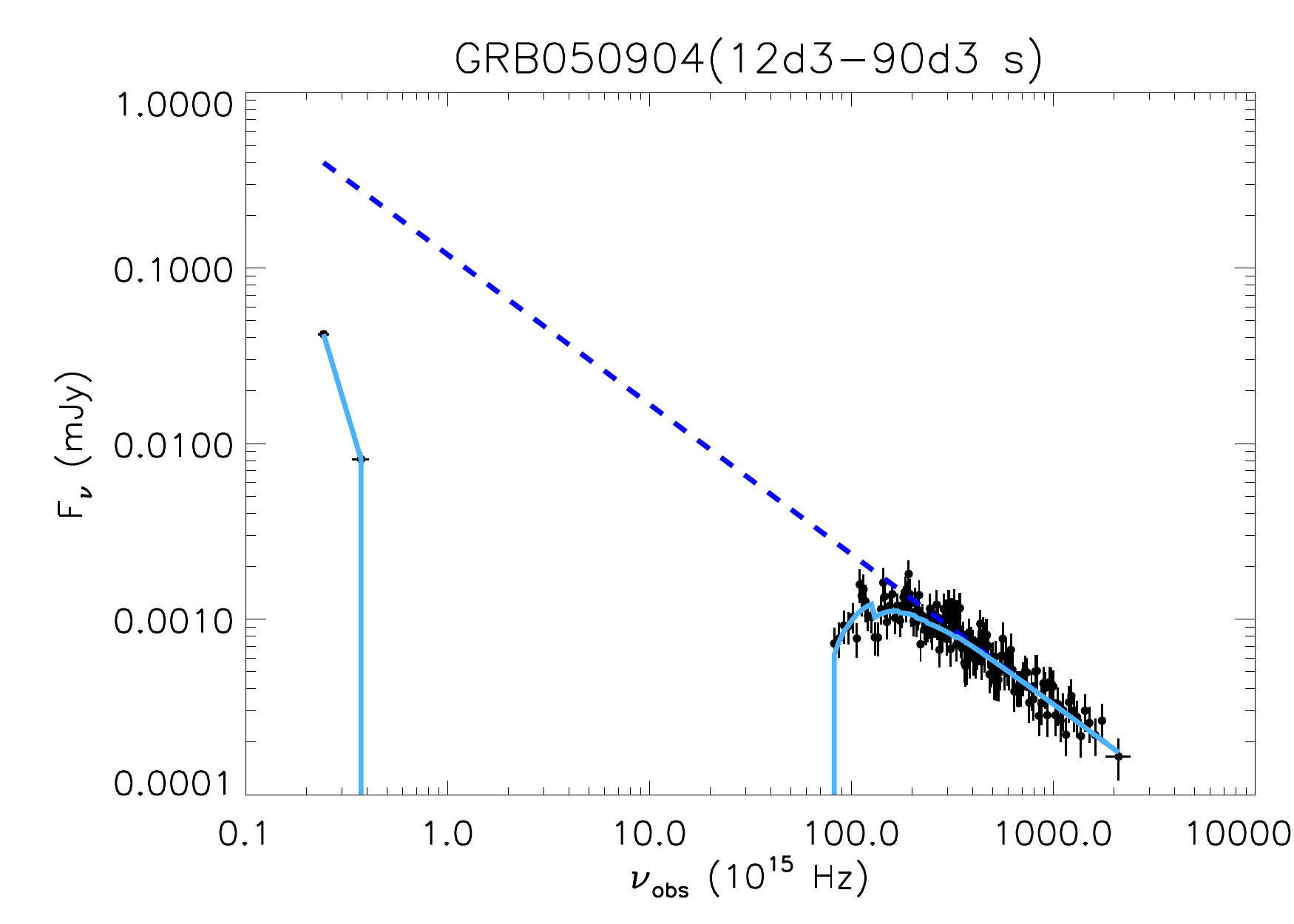}\\
\includegraphics[width=0.3 \hsize,clip]{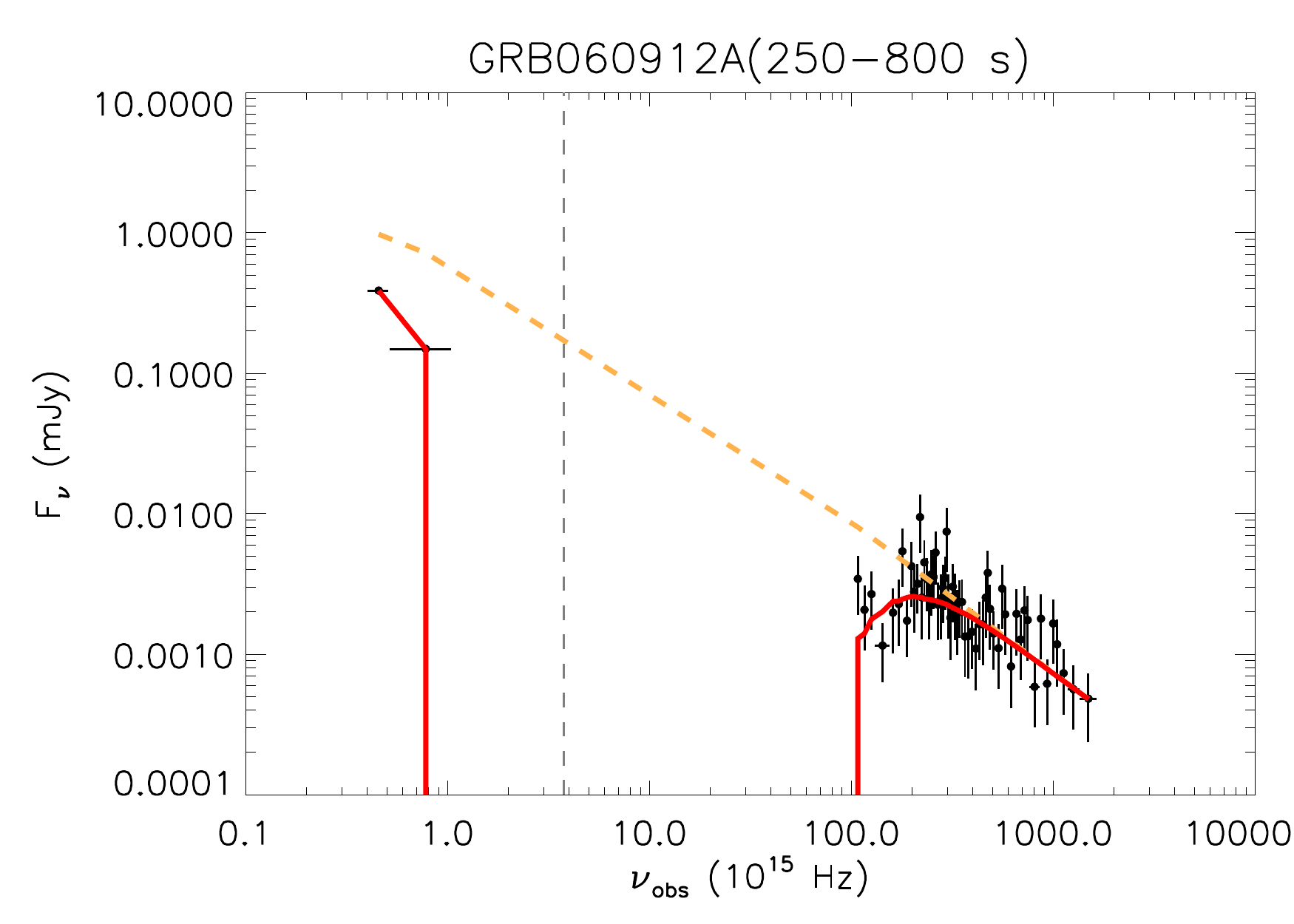}
\includegraphics[width=0.3 \hsize,clip]{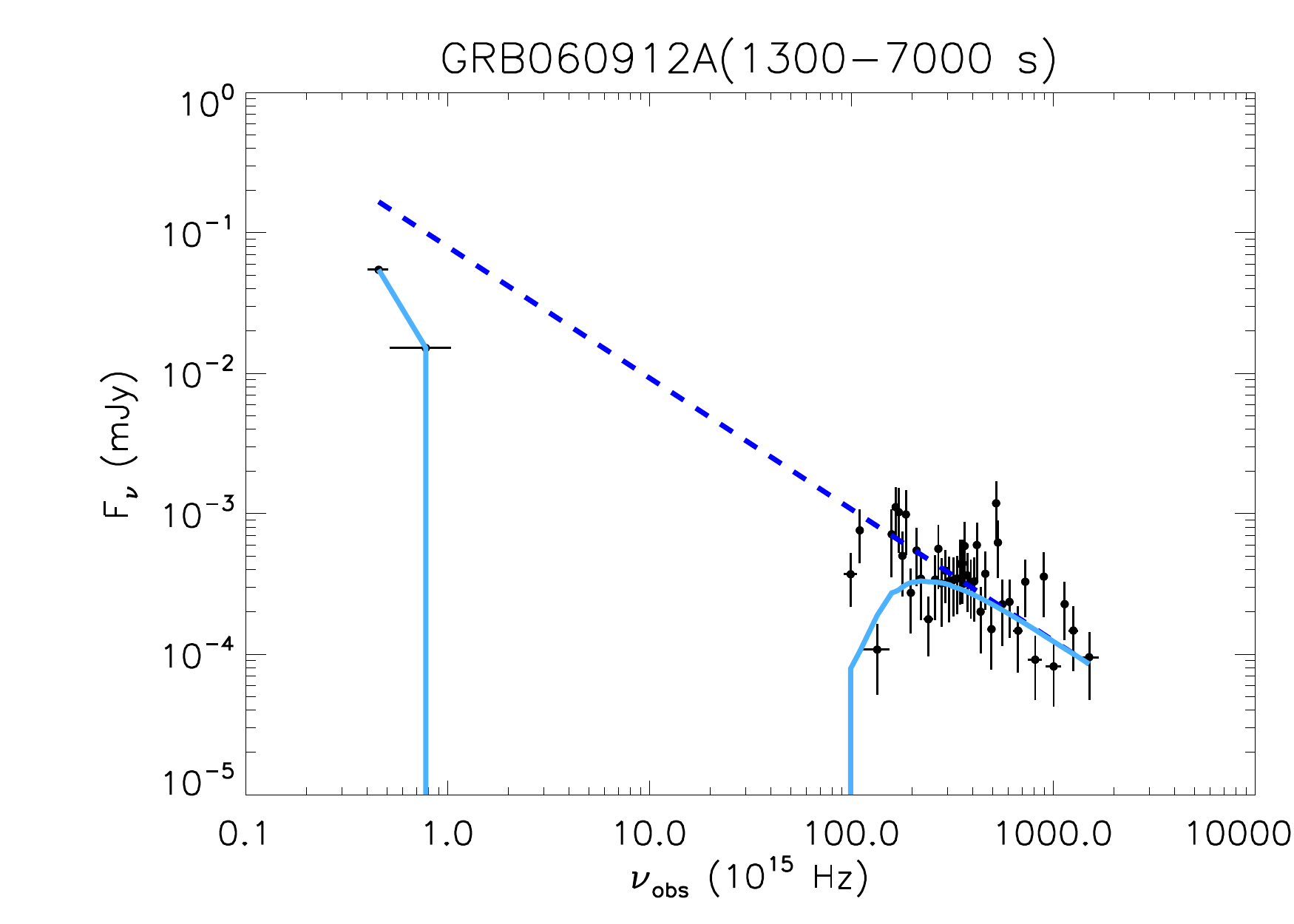}
\includegraphics[width=0.3 \hsize,clip]{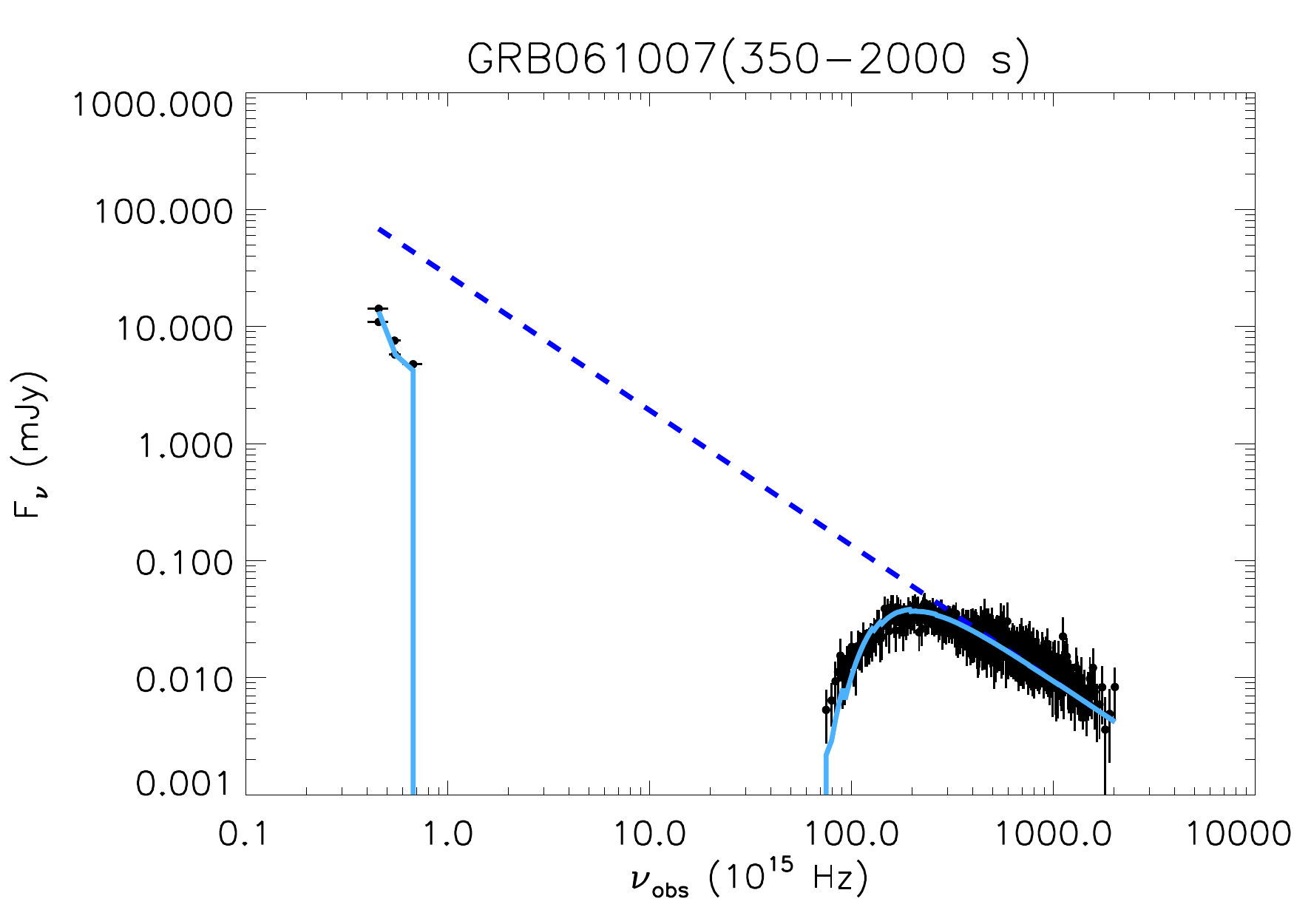}\\
\includegraphics[width=0.3 \hsize,clip]{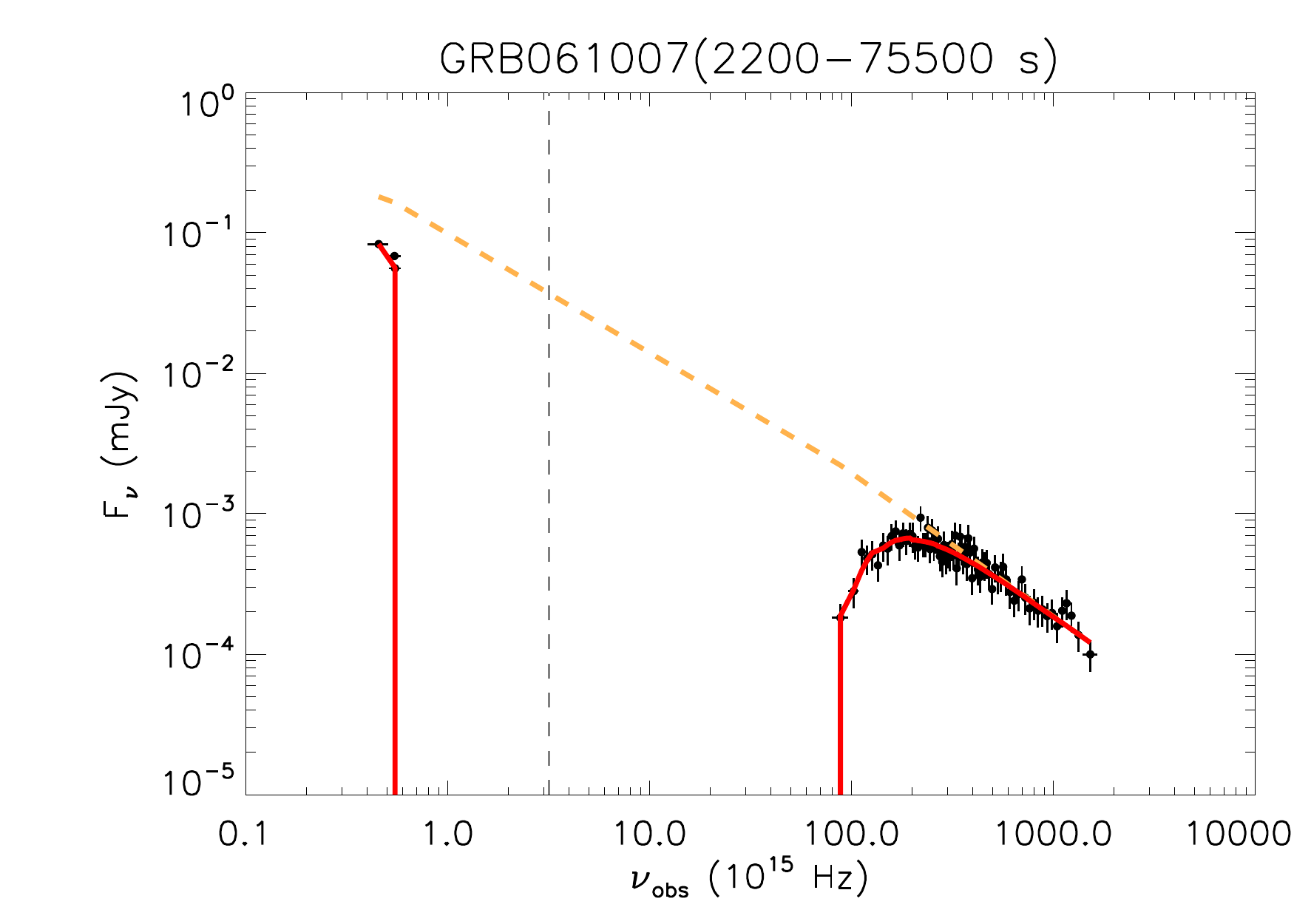}
\includegraphics[width=0.3 \hsize,clip]{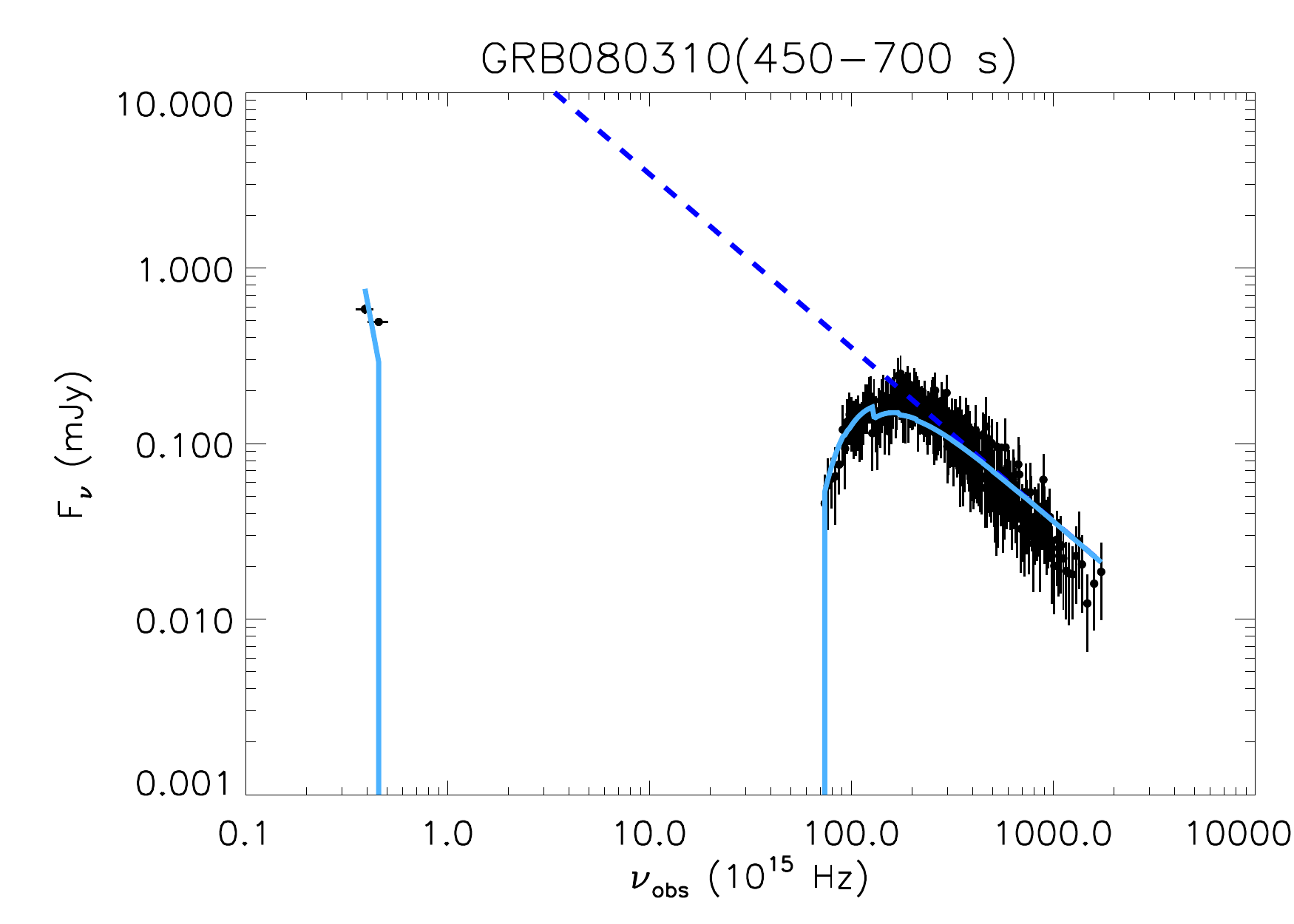}
\includegraphics[width=0.3 \hsize,clip]{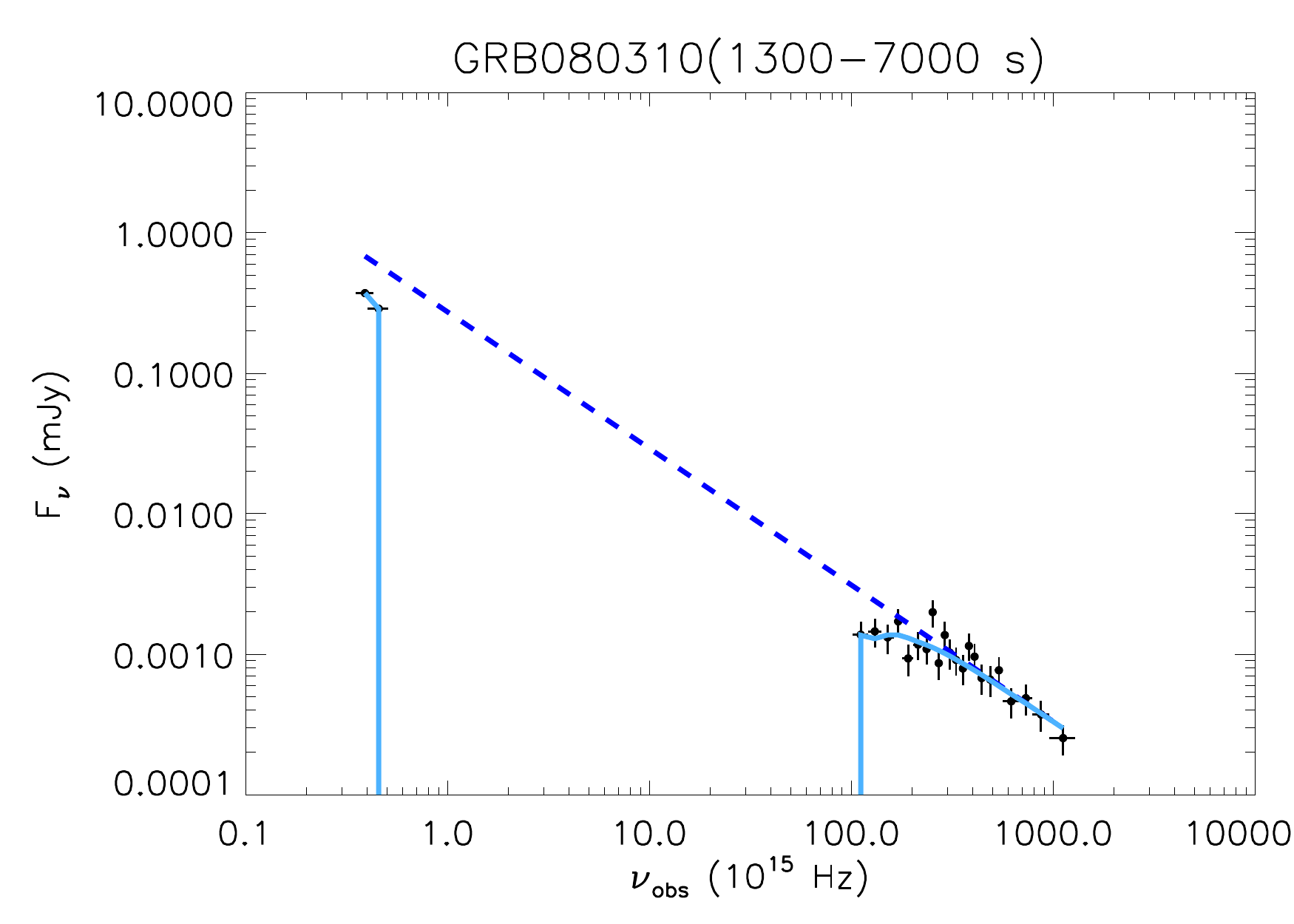}\\
\includegraphics[width=0.3 \hsize,clip]{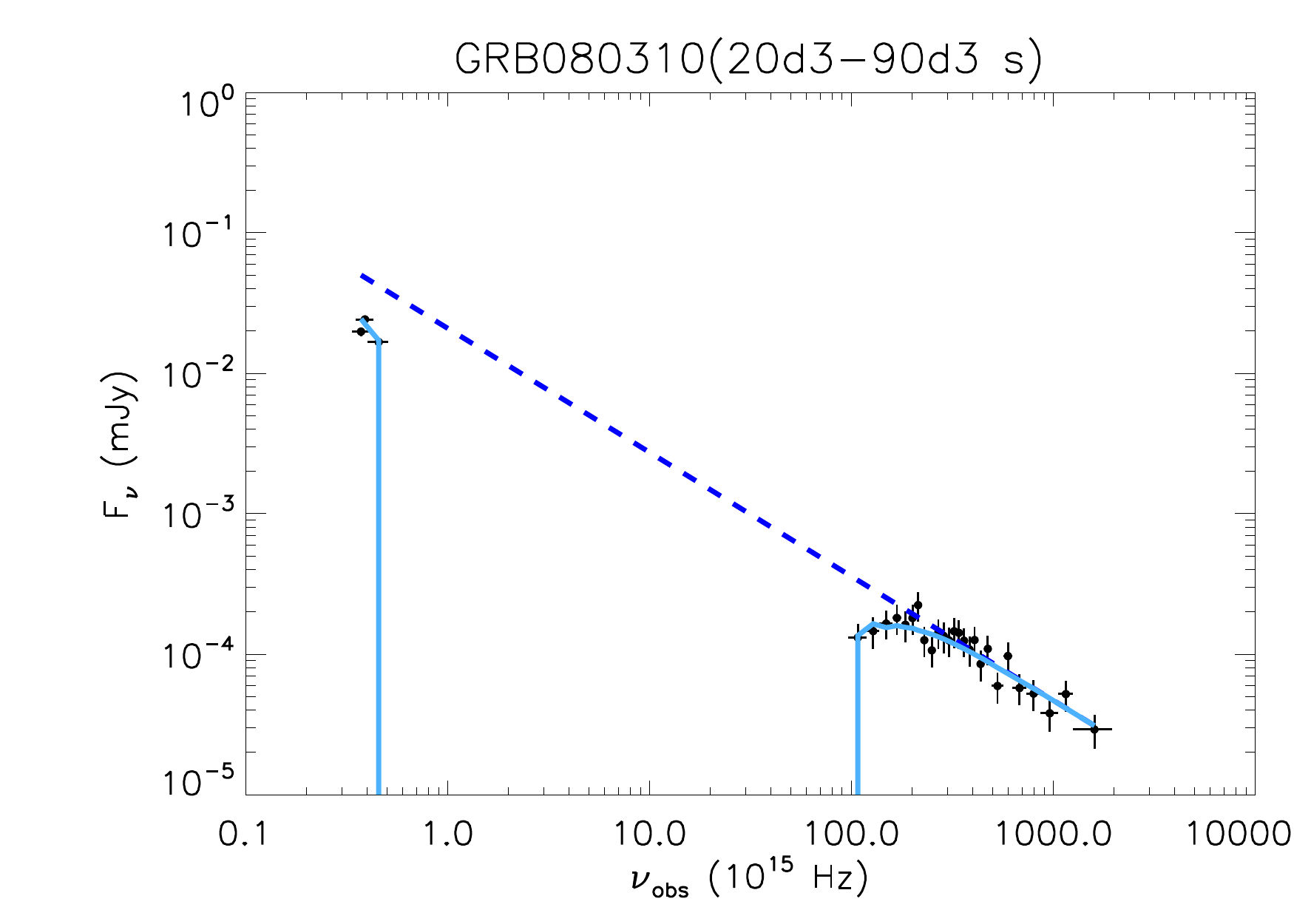}
\includegraphics[width=0.3 \hsize,clip]{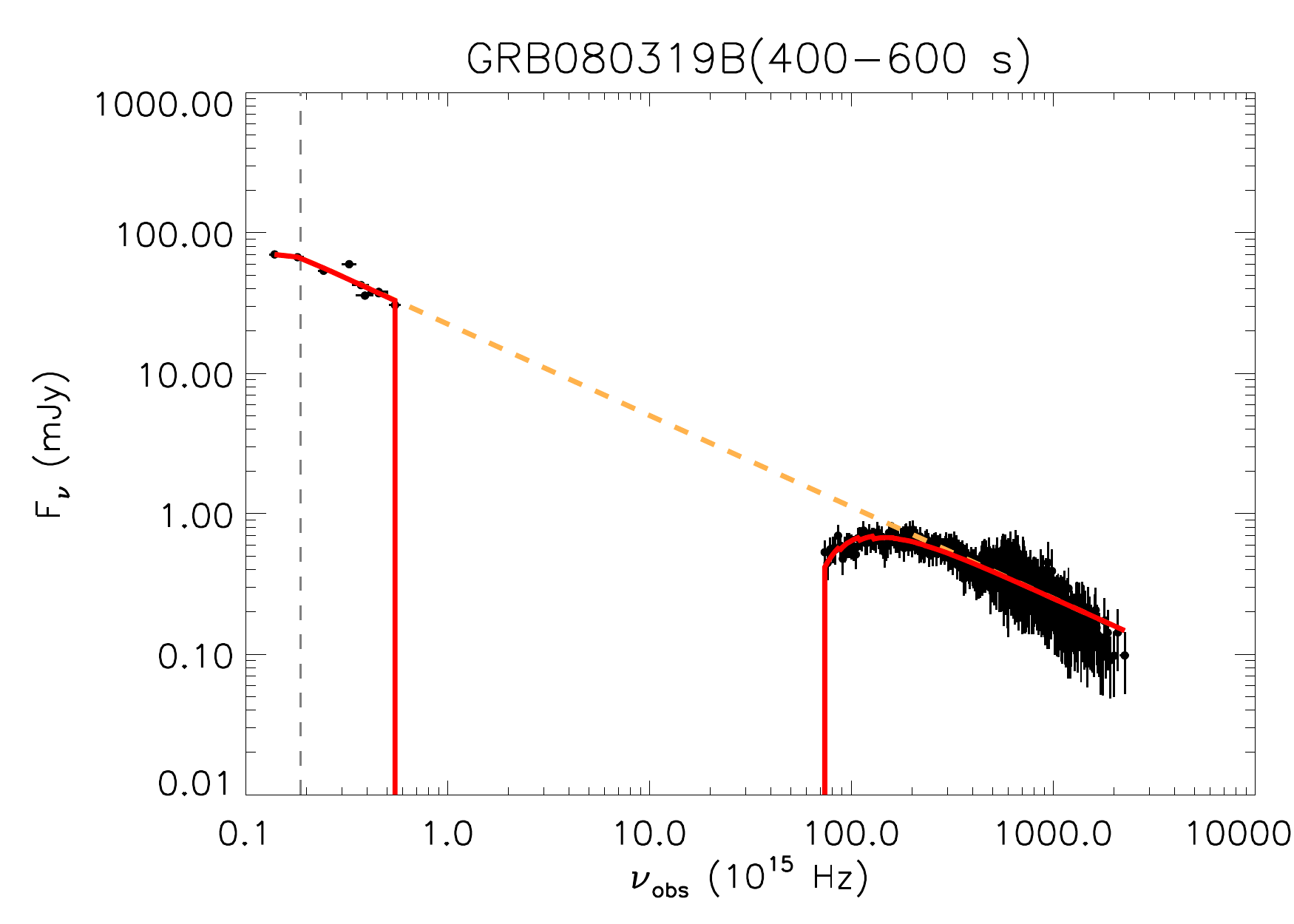}
\includegraphics[width=0.3\hsize,clip]{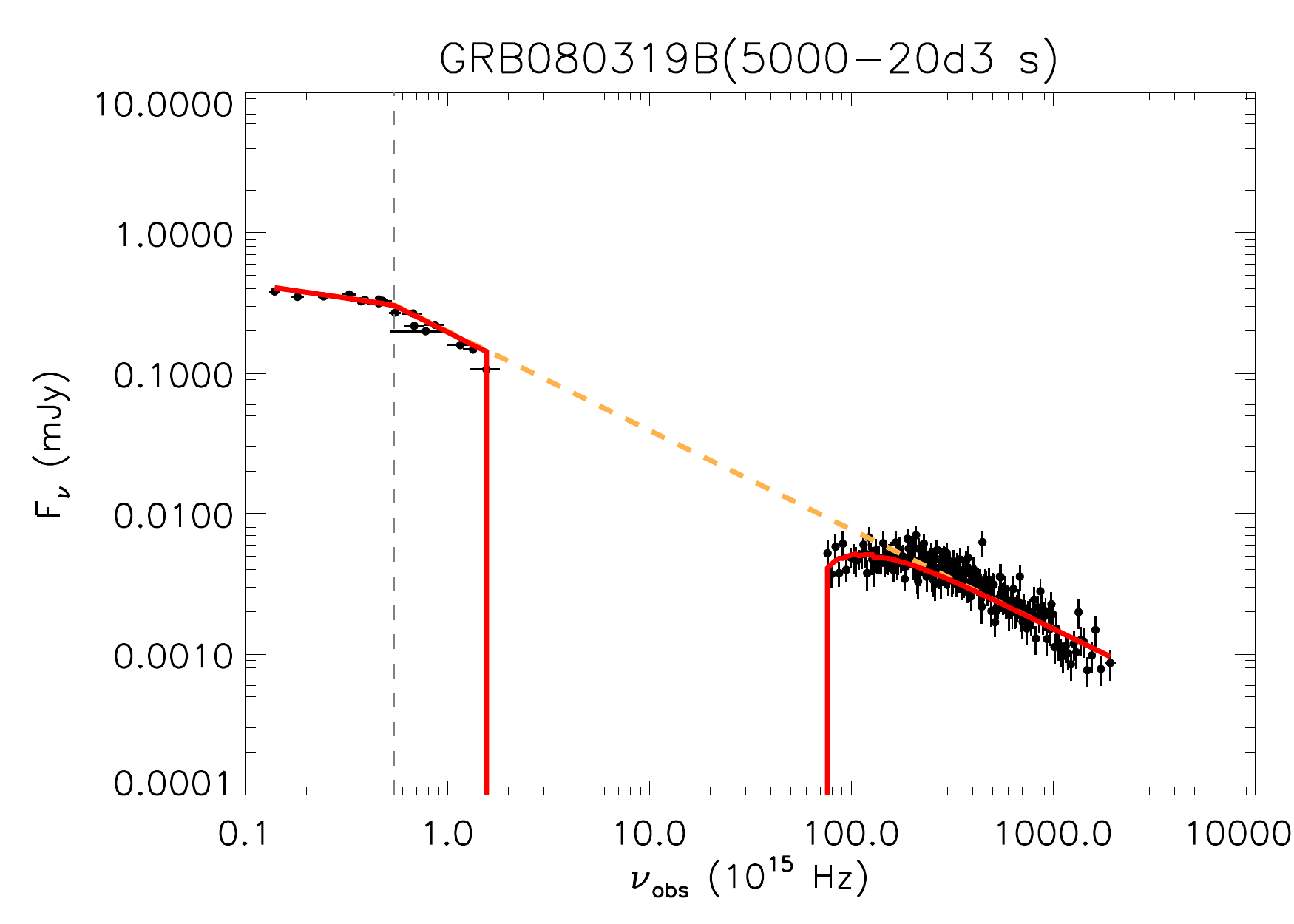}\\
\includegraphics[width=0.3 \hsize,clip]{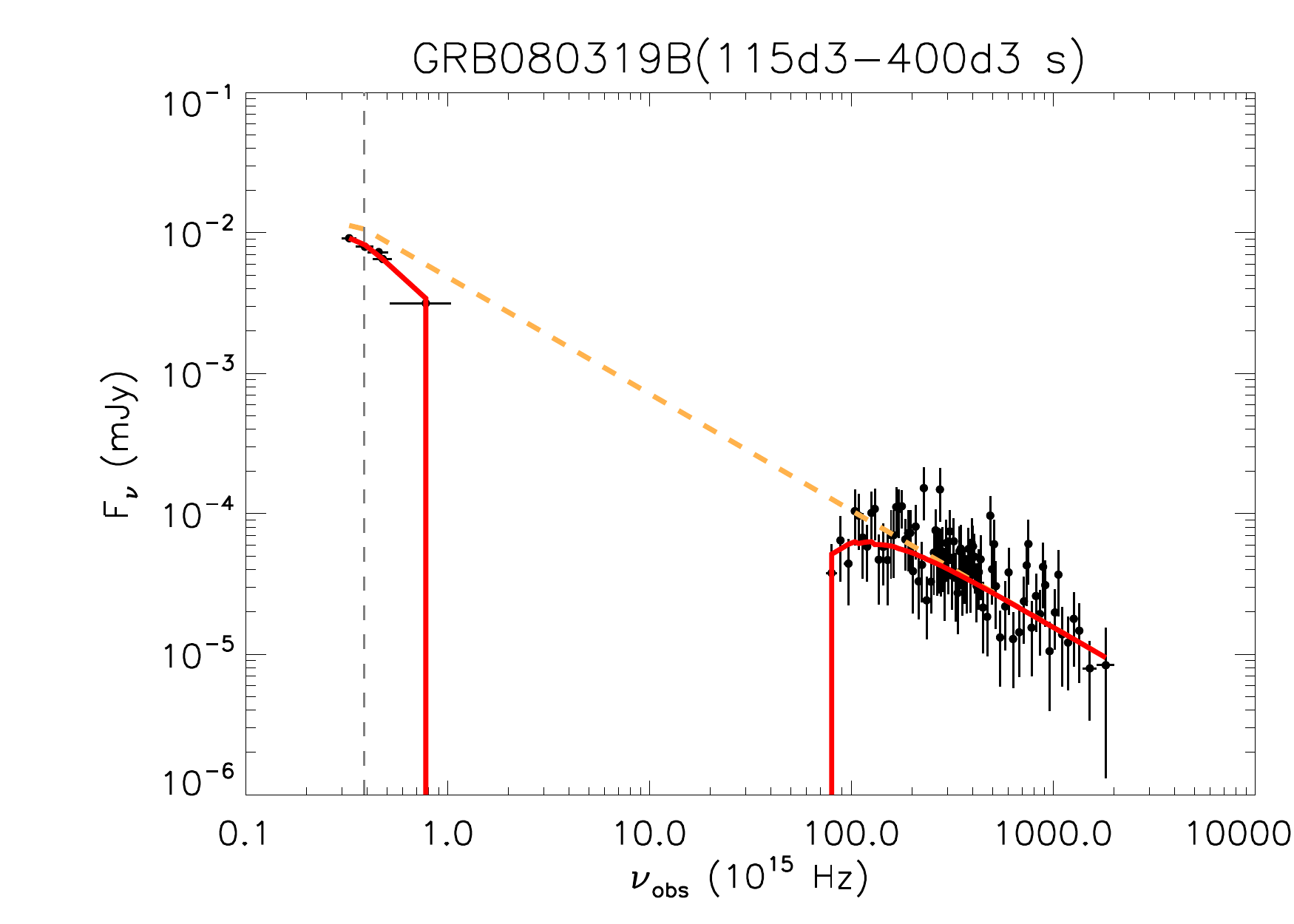}
\includegraphics[width=0.3 \hsize,clip]{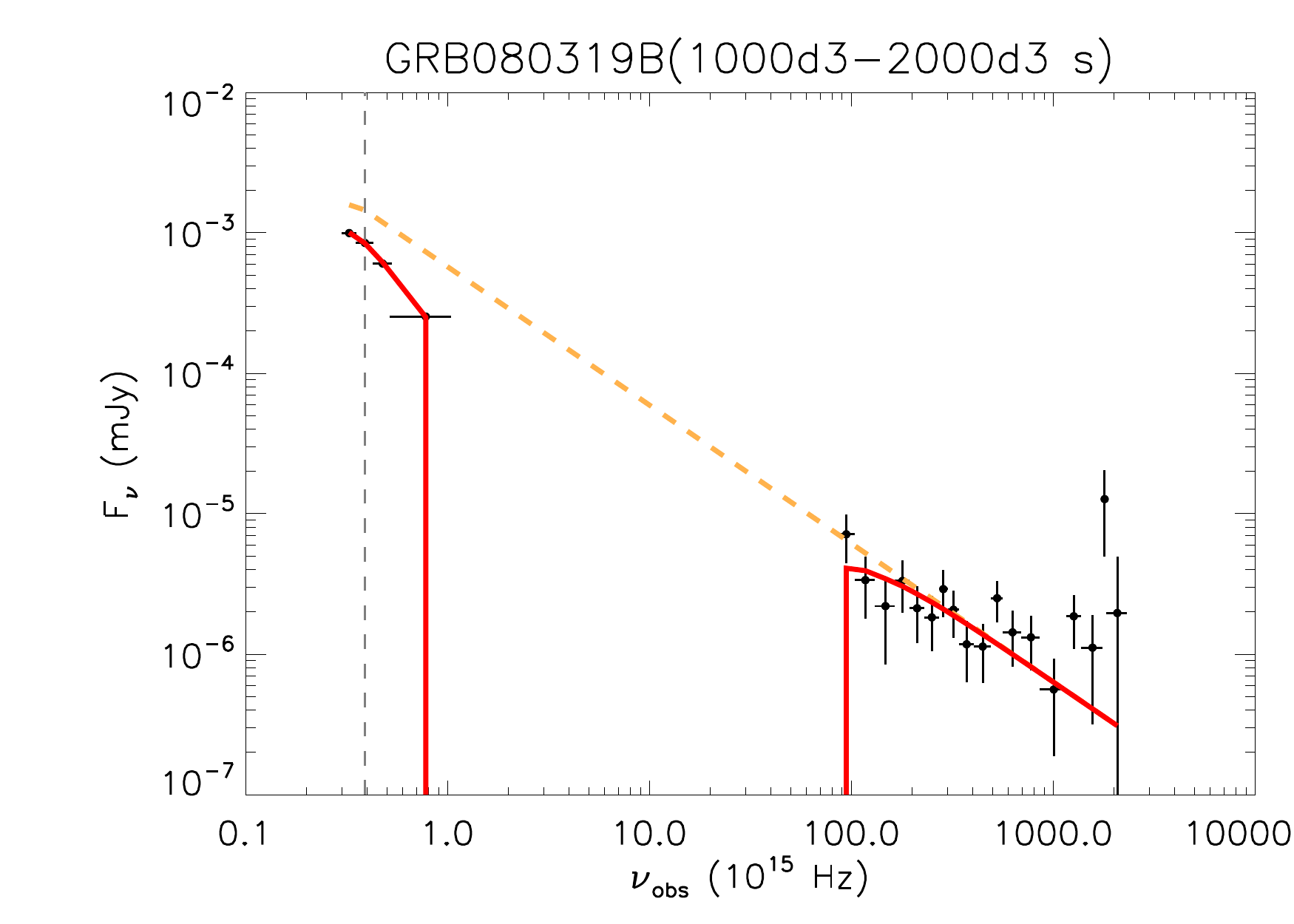}
\includegraphics[width=0.3 \hsize,clip]{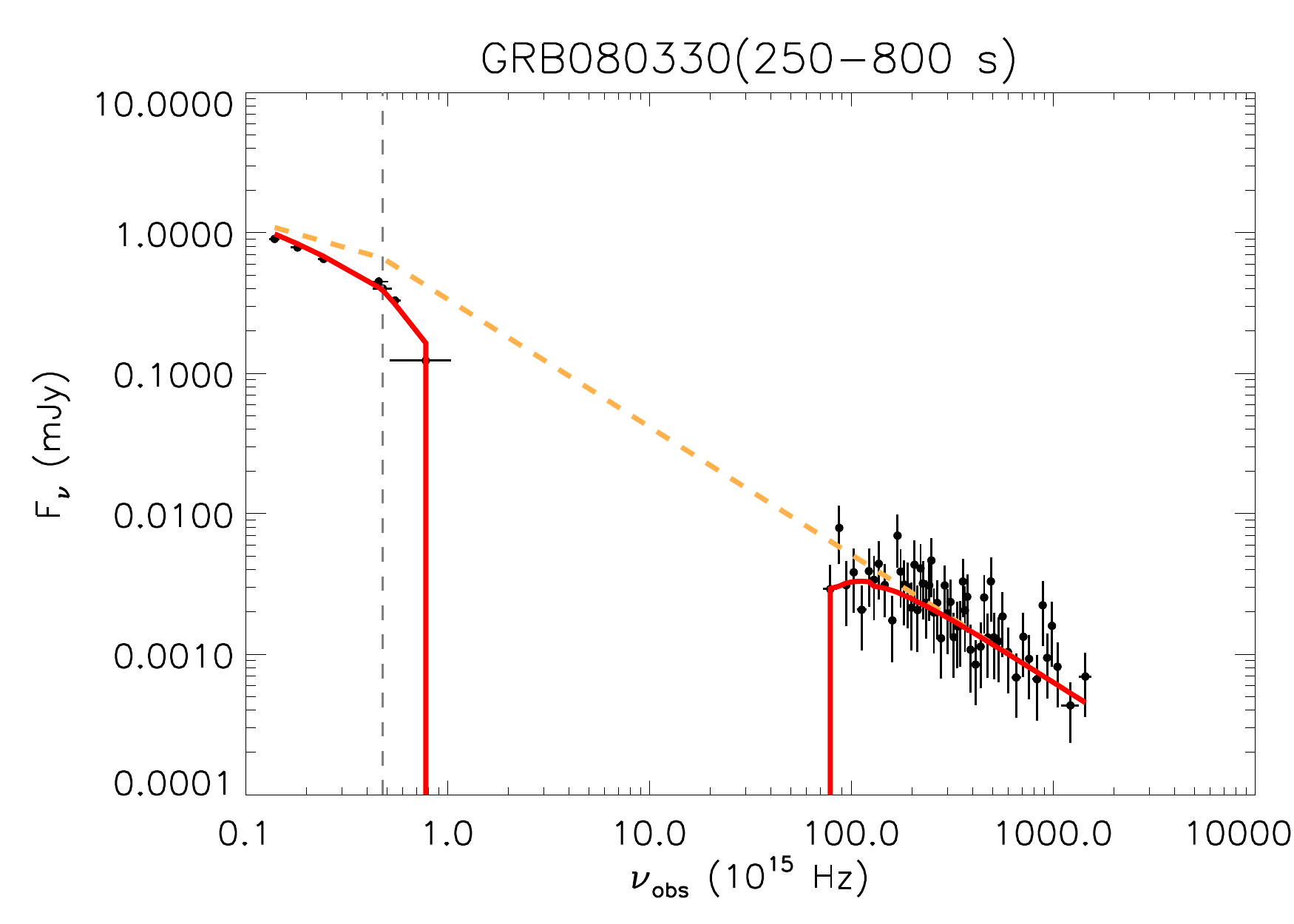}
\caption{\small{Optical/X-ray SEDs for GRBs belonging to Group A. \textit{Solid line}: the fitting function. \textit{Dotted line}: power-law (blue) or broken power-law (orange) fitting function. \textit{Light blue/blue lines} stand for th power-law fitting functions. \textit{Red/orange lines} correspond to the fitting function with the broken power-law. The distinction between the two different laws follows Table~\ref{ftest}.}}\label{sed1} 
\end{figure}

\begin{figure}
\includegraphics[width=0.3 \hsize,clip]{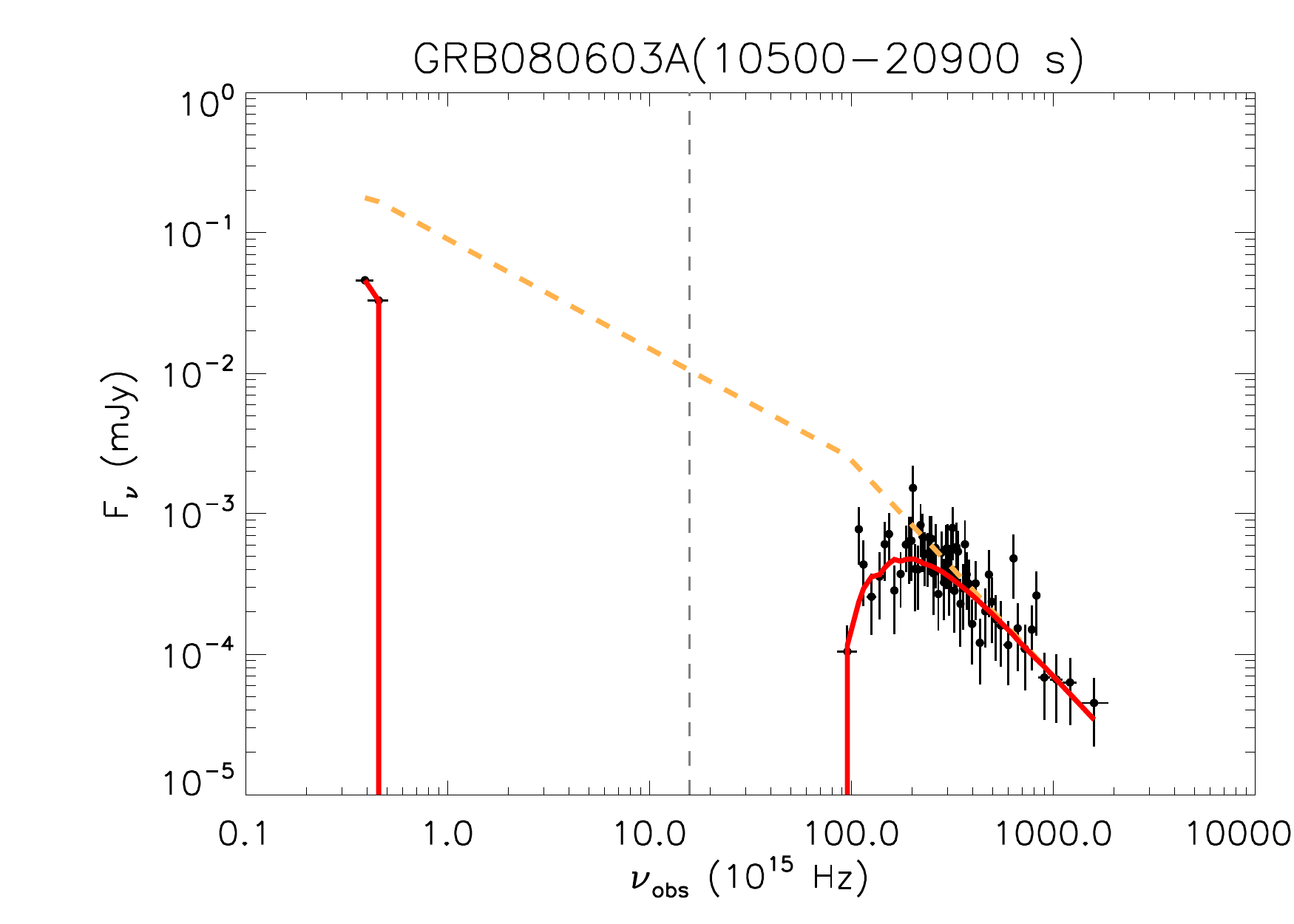}
\includegraphics[width=0.3 \hsize,clip]{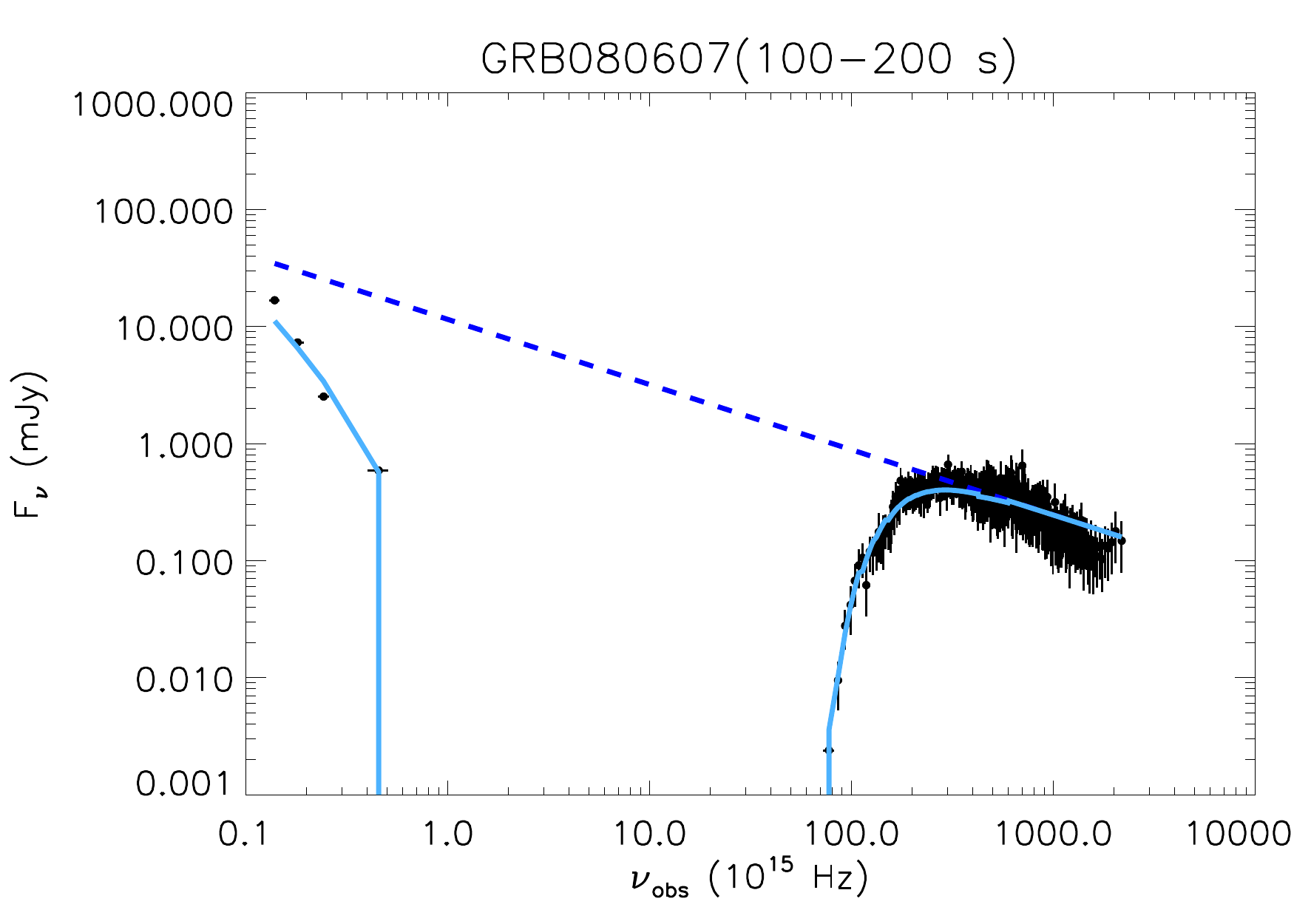}
\includegraphics[width=0.3\hsize,clip]{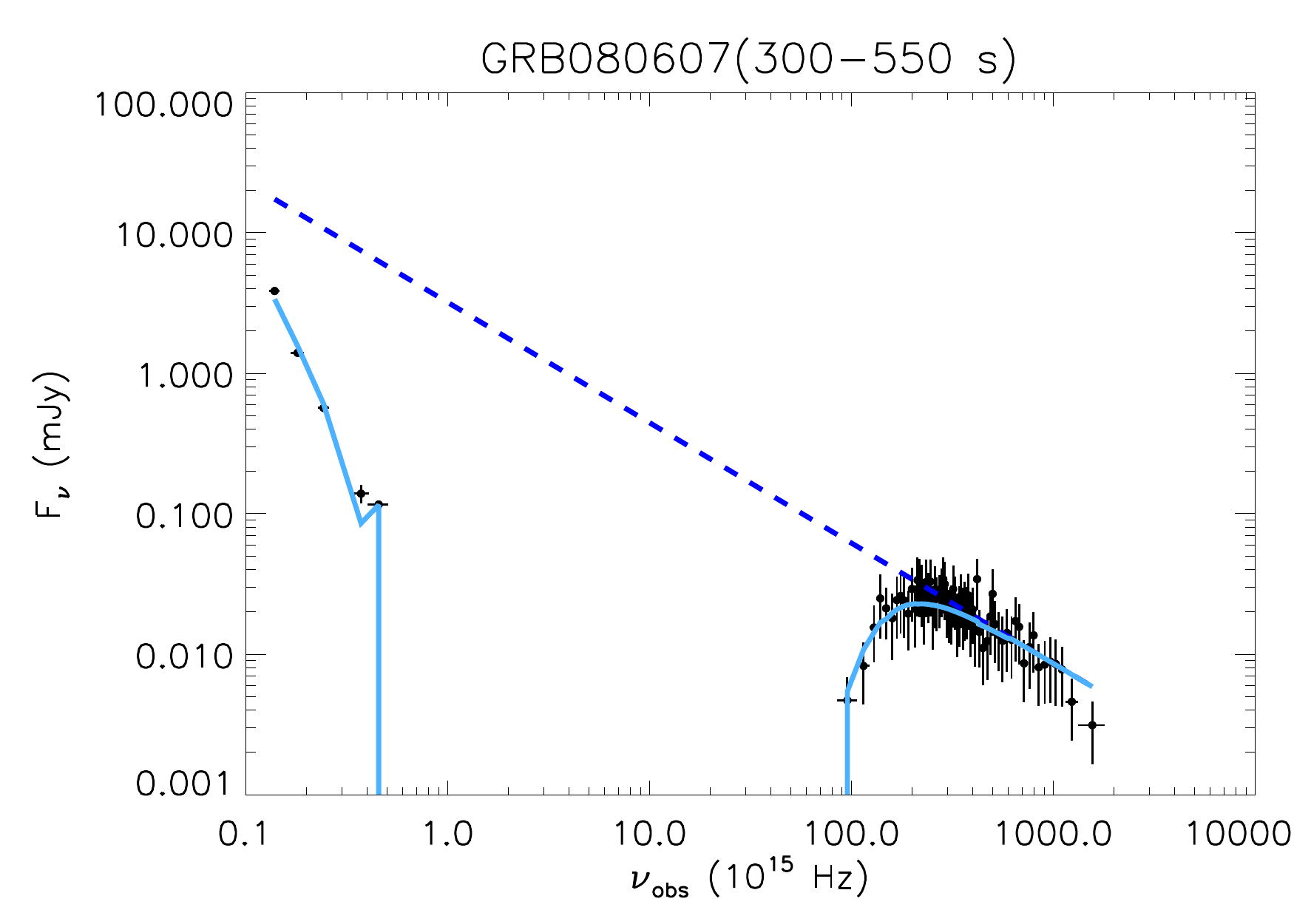}\\
\includegraphics[width=0.3 \hsize,clip]{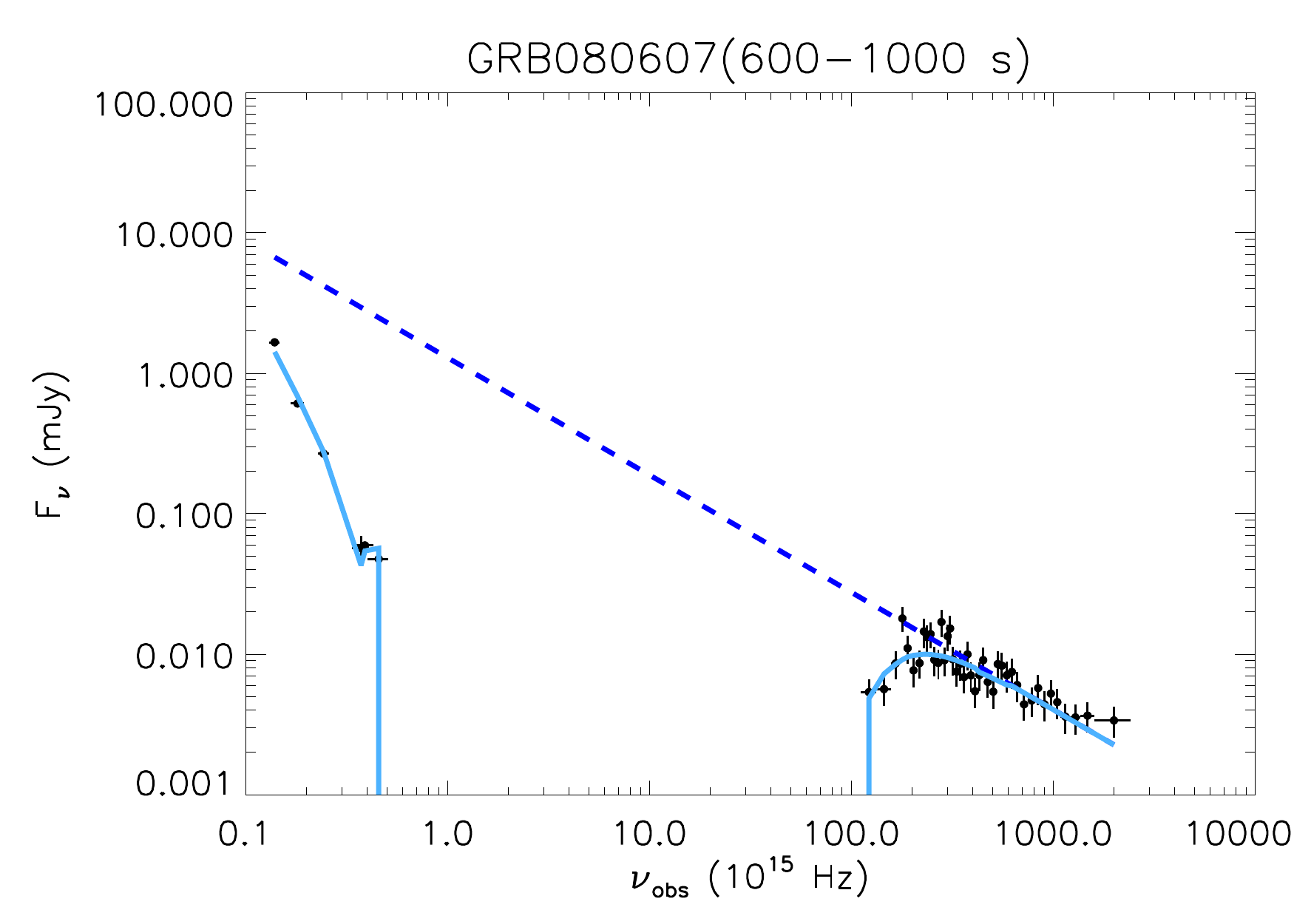}
\includegraphics[width=0.3 \hsize,clip]{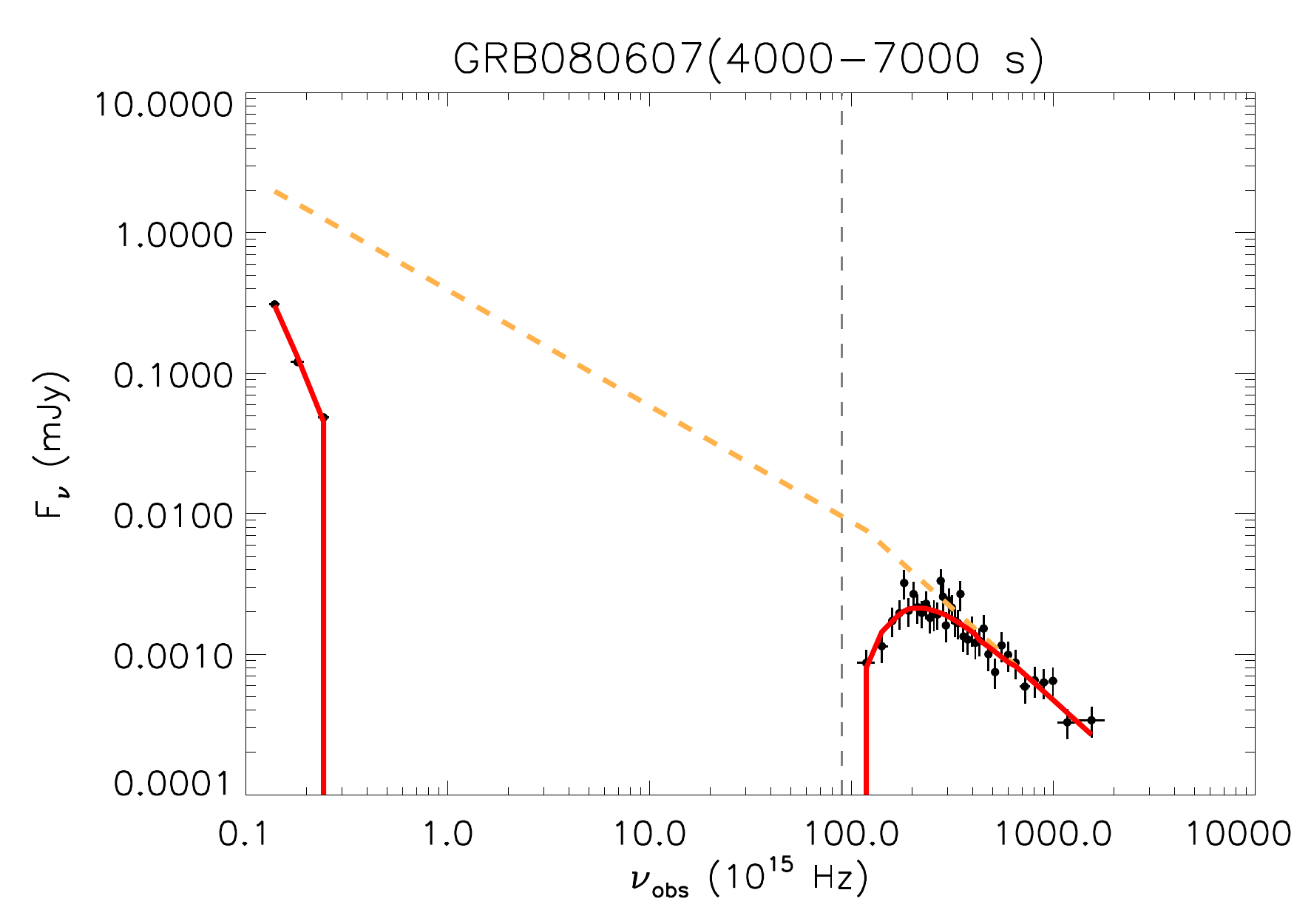}
\includegraphics[width=0.3\hsize,clip]{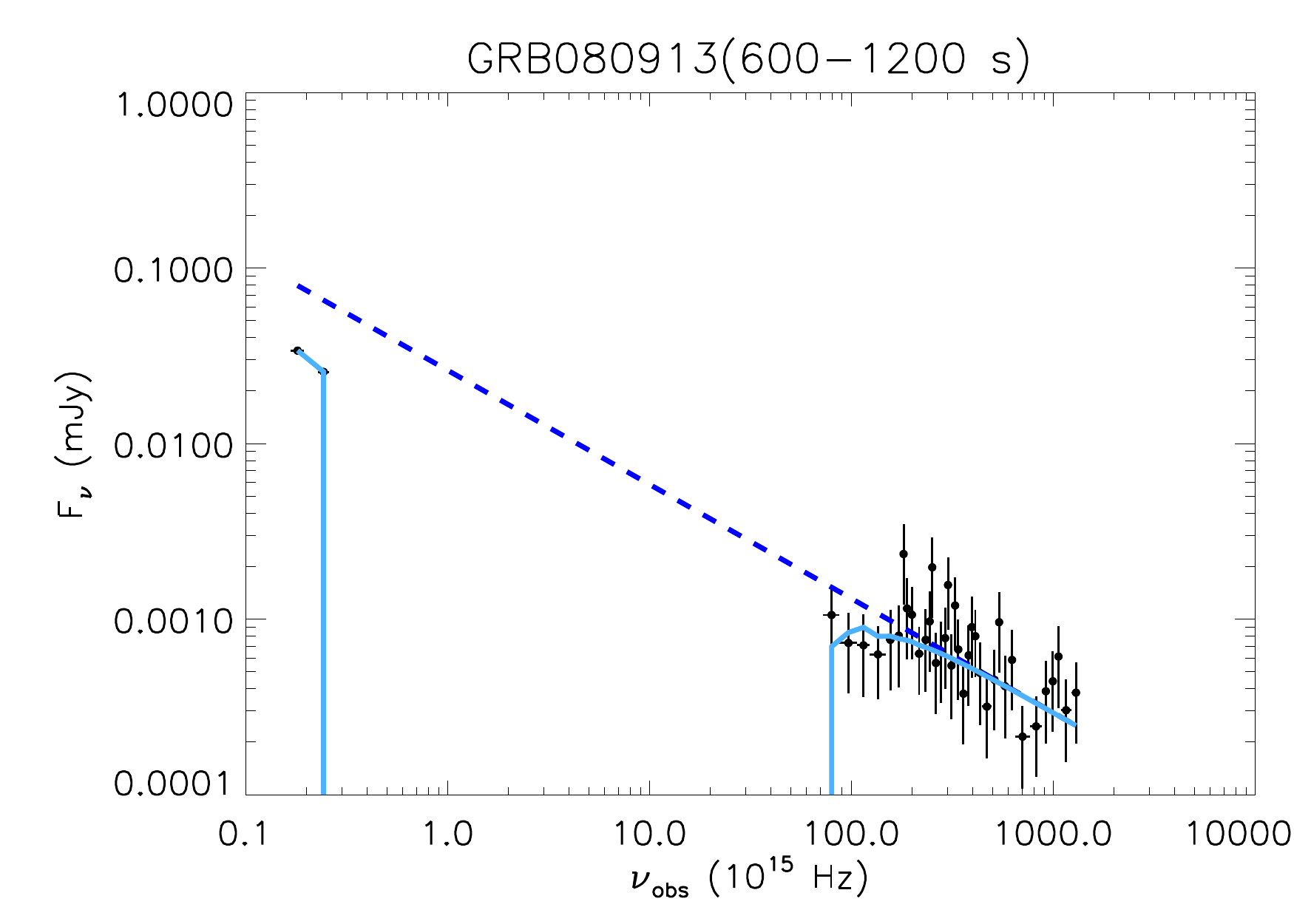}\\
\includegraphics[width=0.3 \hsize,clip]{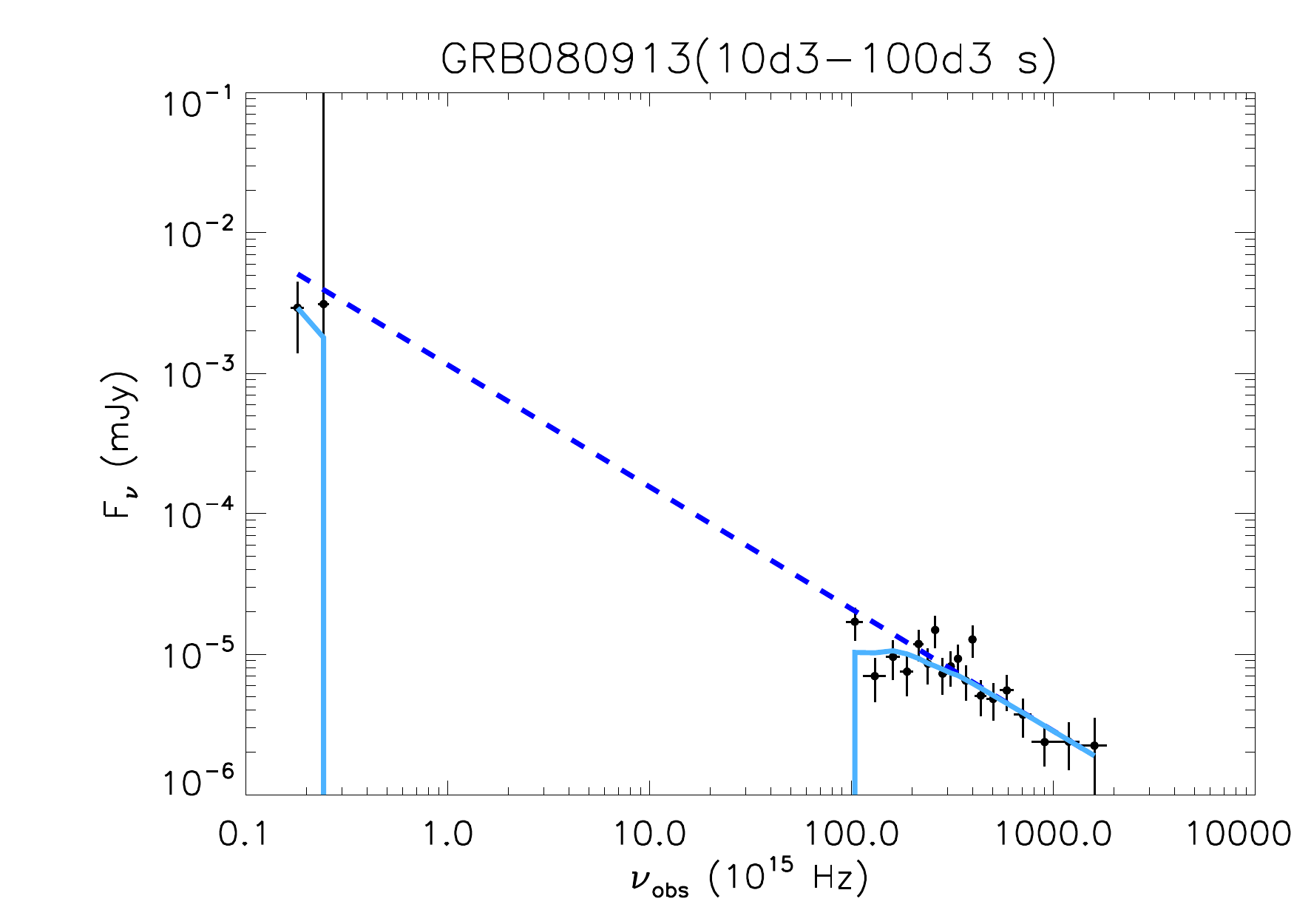}
\includegraphics[width=0.3 \hsize,clip]{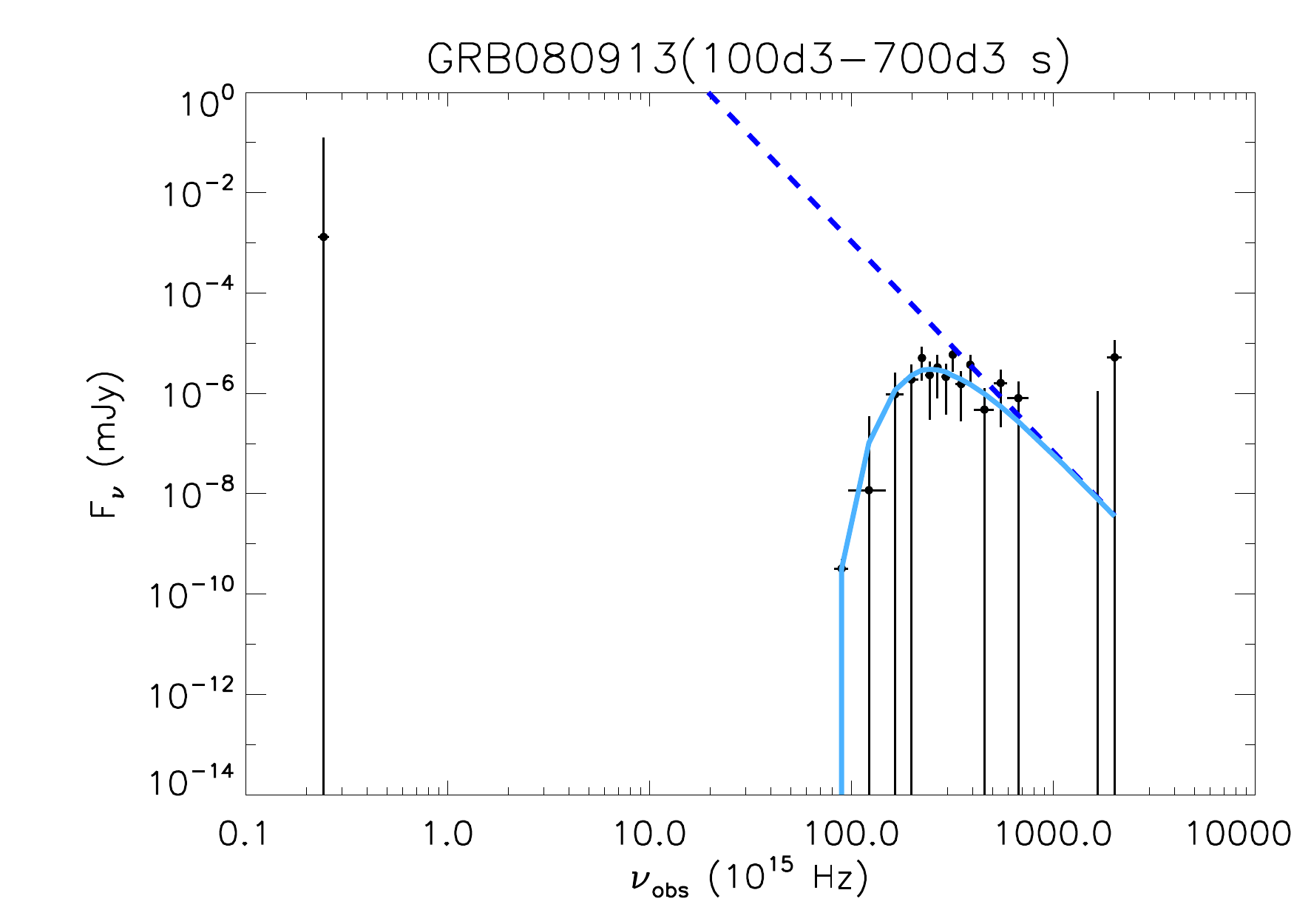}
\includegraphics[width=0.3\hsize,clip]{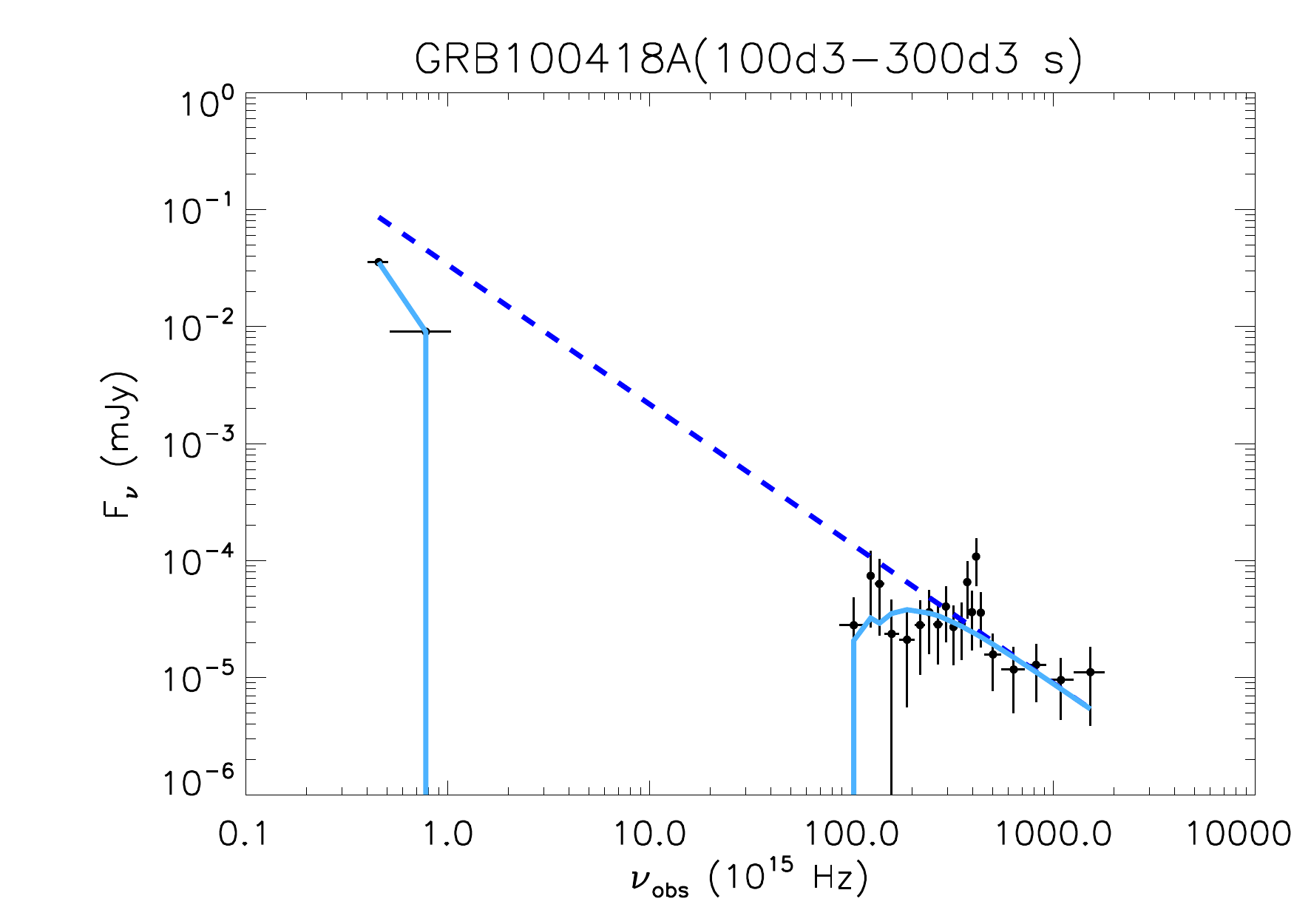}\\
\includegraphics[width=0.3 \hsize,clip]{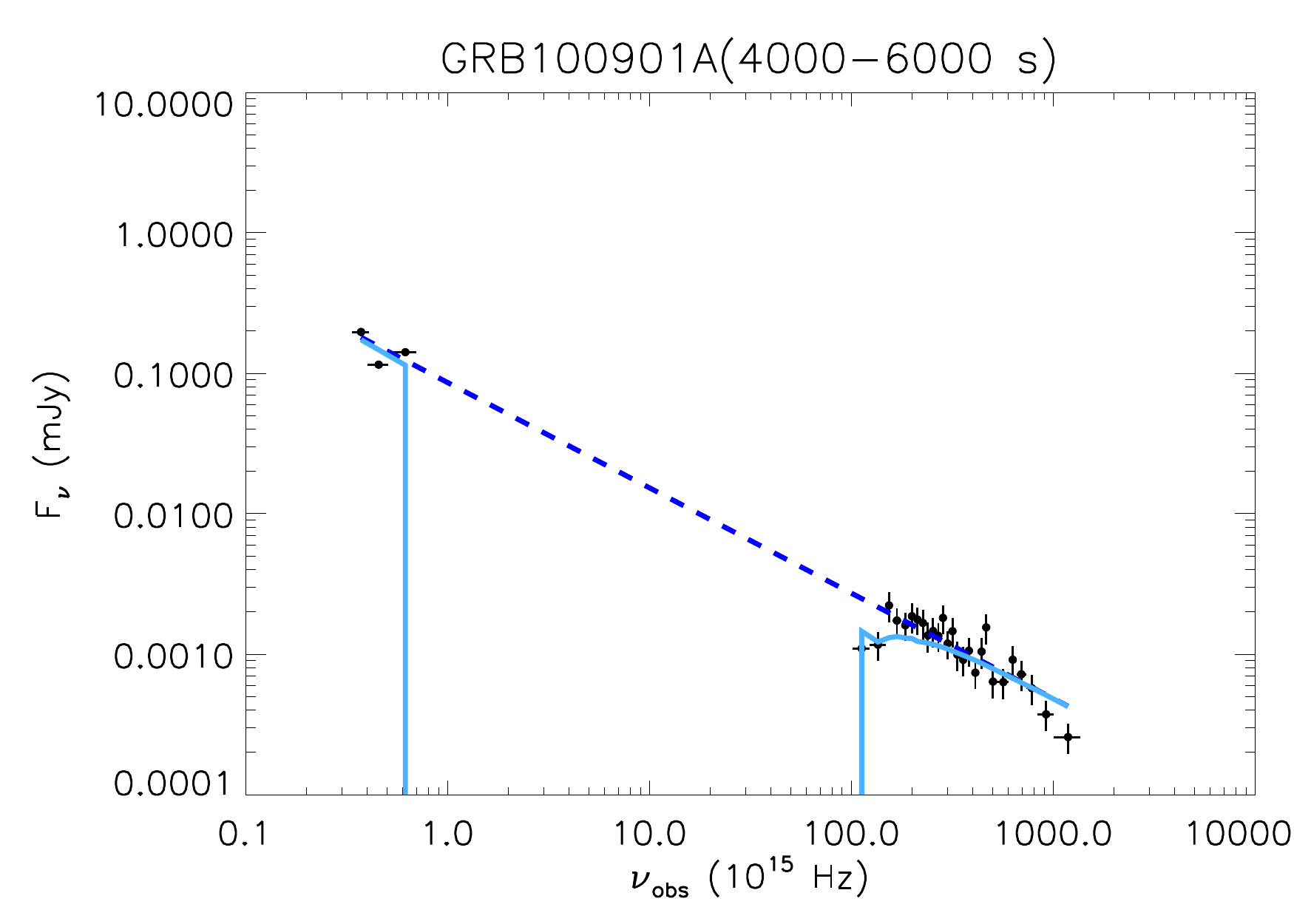}
\includegraphics[width=0.3 \hsize,clip]{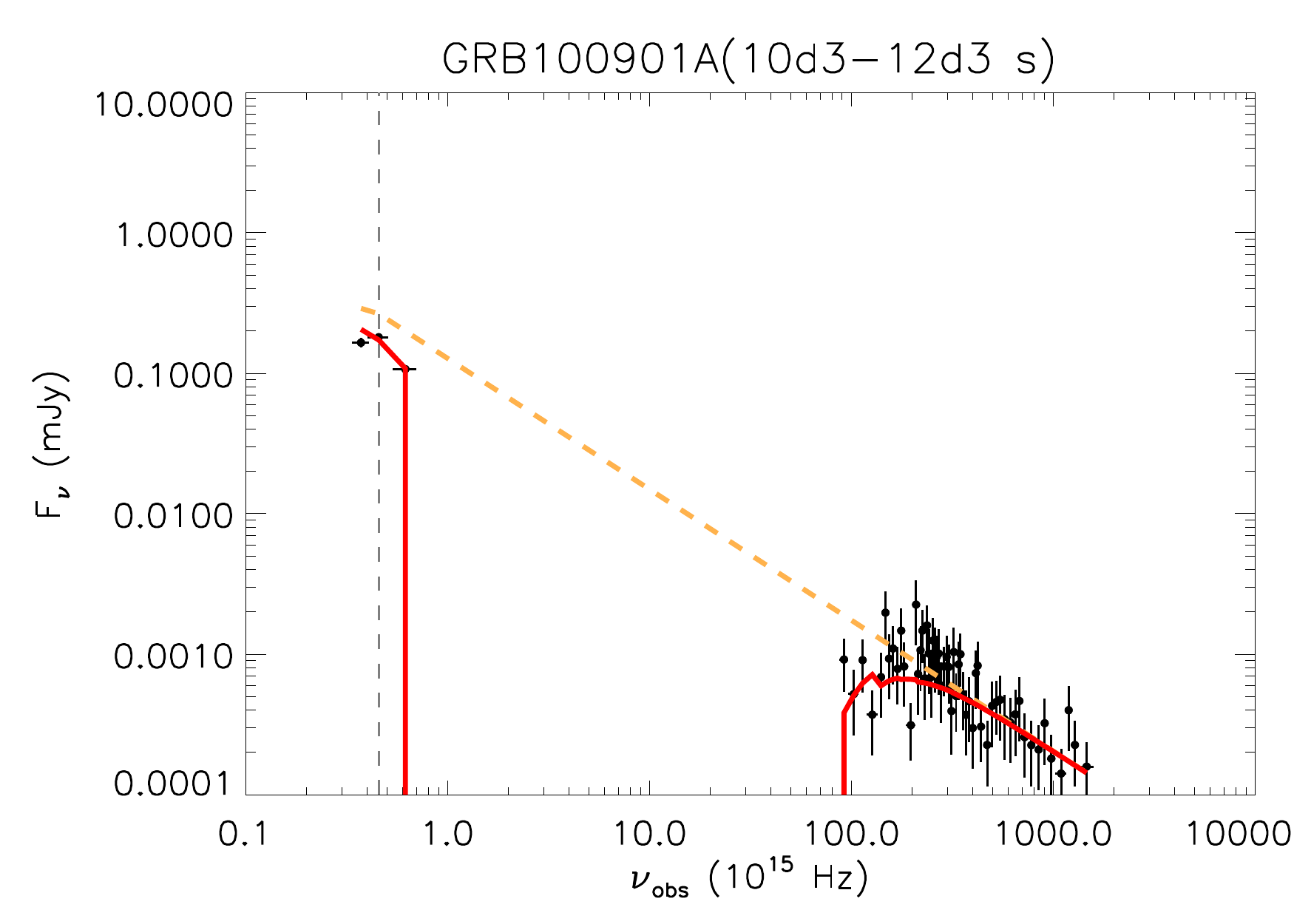}
\includegraphics[width=0.3\hsize,clip]{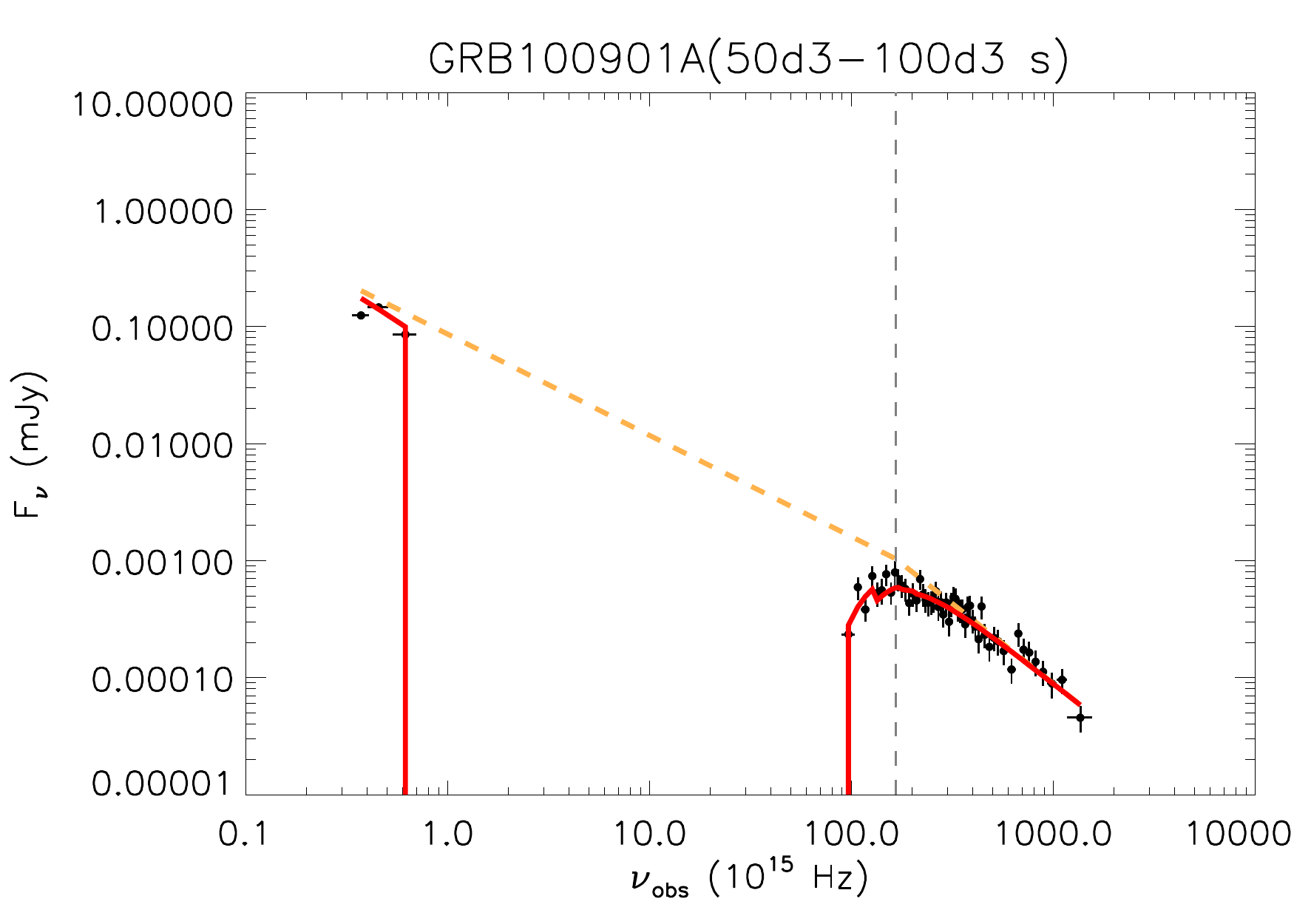}
\caption{\small{Optical/X-ray SEDs for GRBs belonging to Group A. color-coding as in Figure~\ref{sed1}.}}\label{sed2} 
\end{figure}

\begin{figure}
\includegraphics[width=0.3 \hsize,clip]{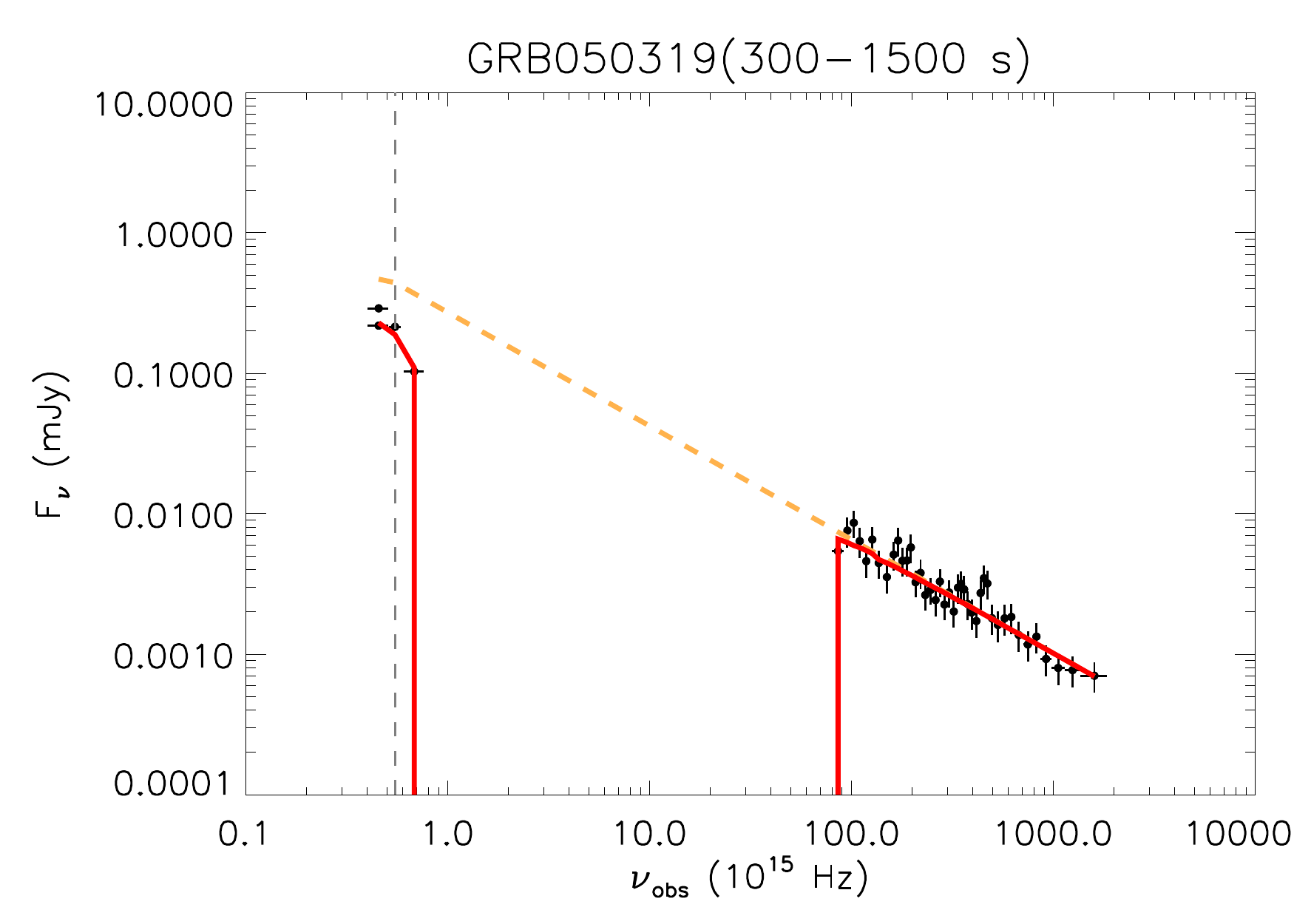}
\includegraphics[width=0.3 \hsize,clip]{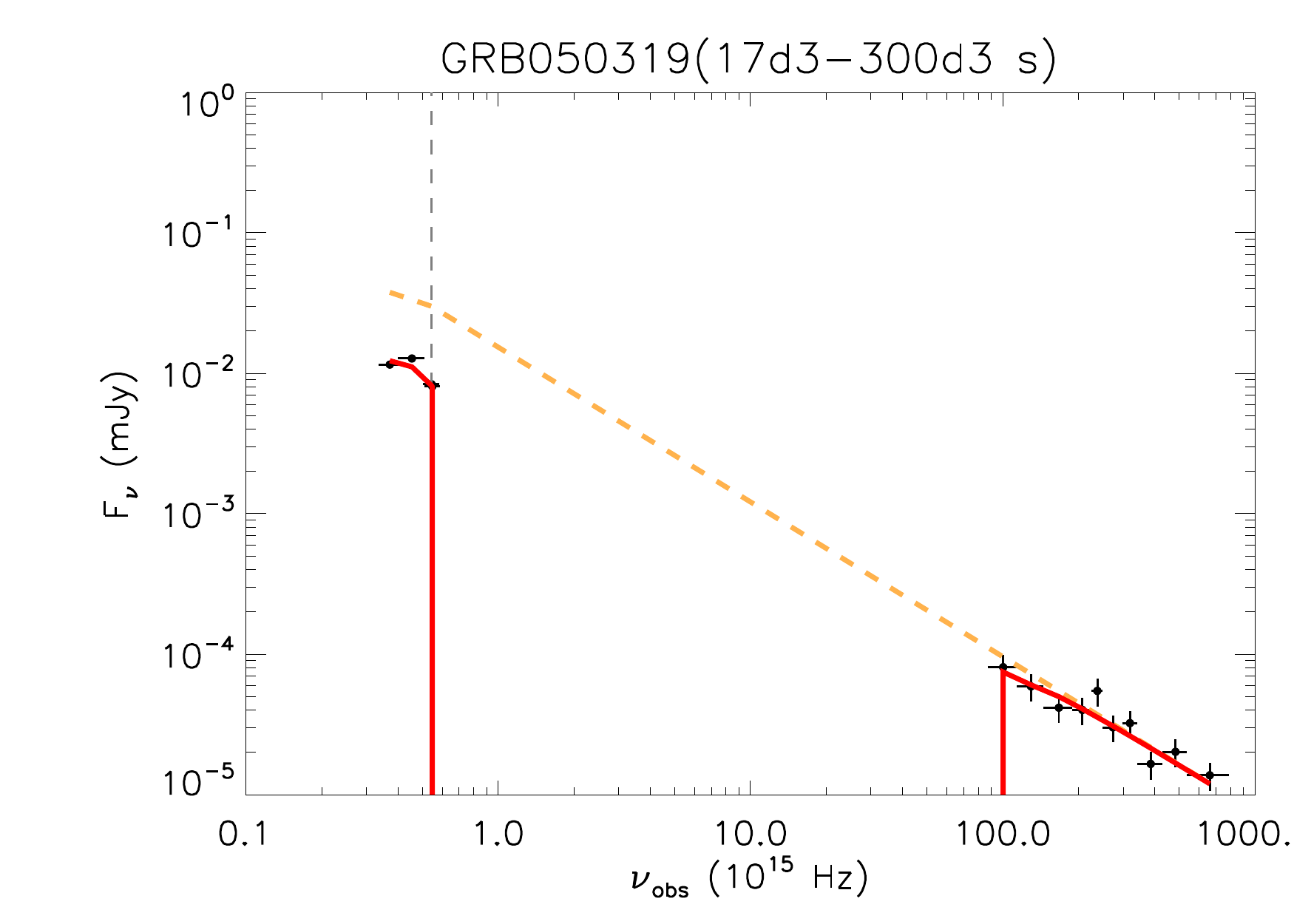}
\includegraphics[width=0.3 \hsize,clip]{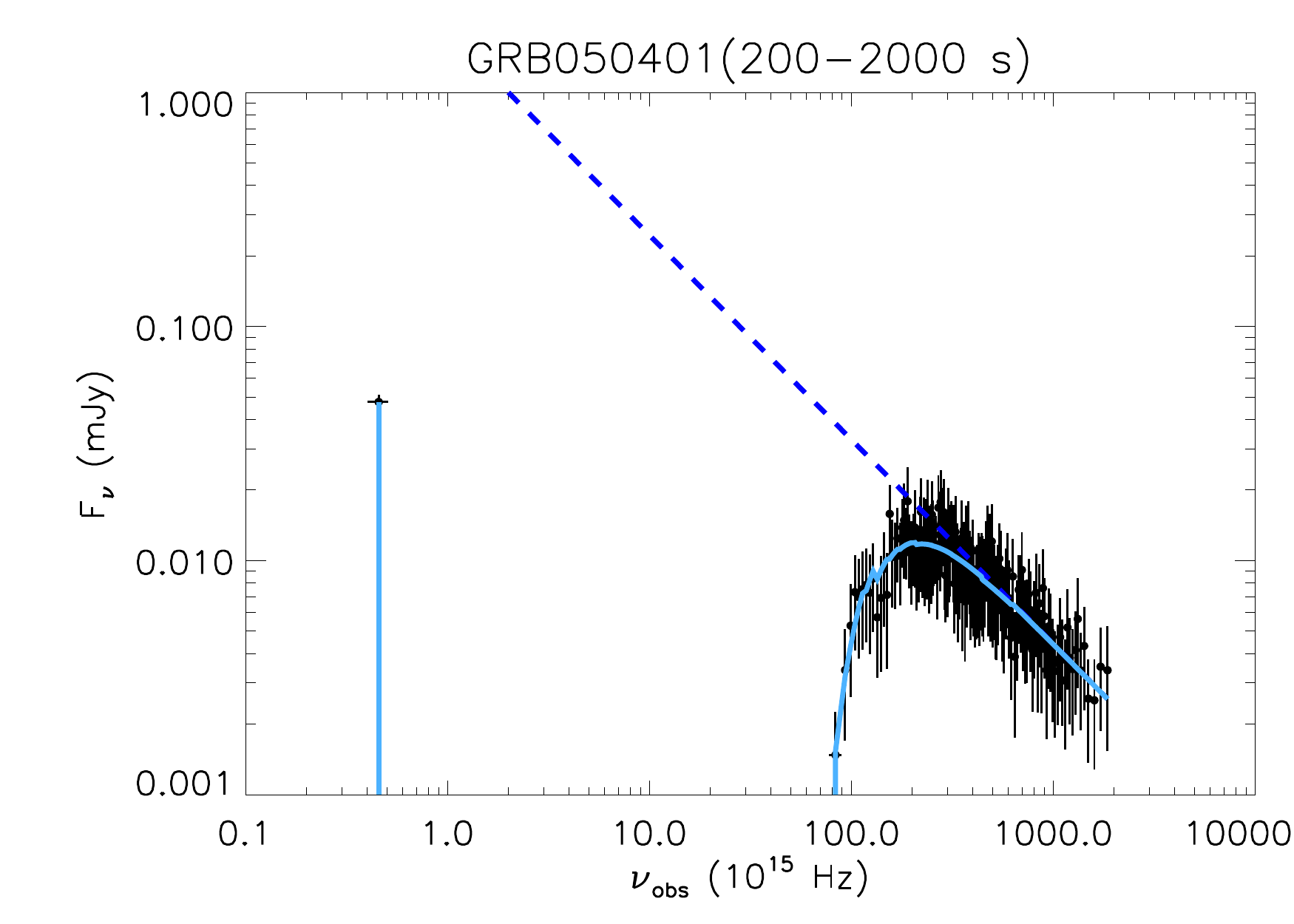}\\
\includegraphics[width=0.3 \hsize,clip]{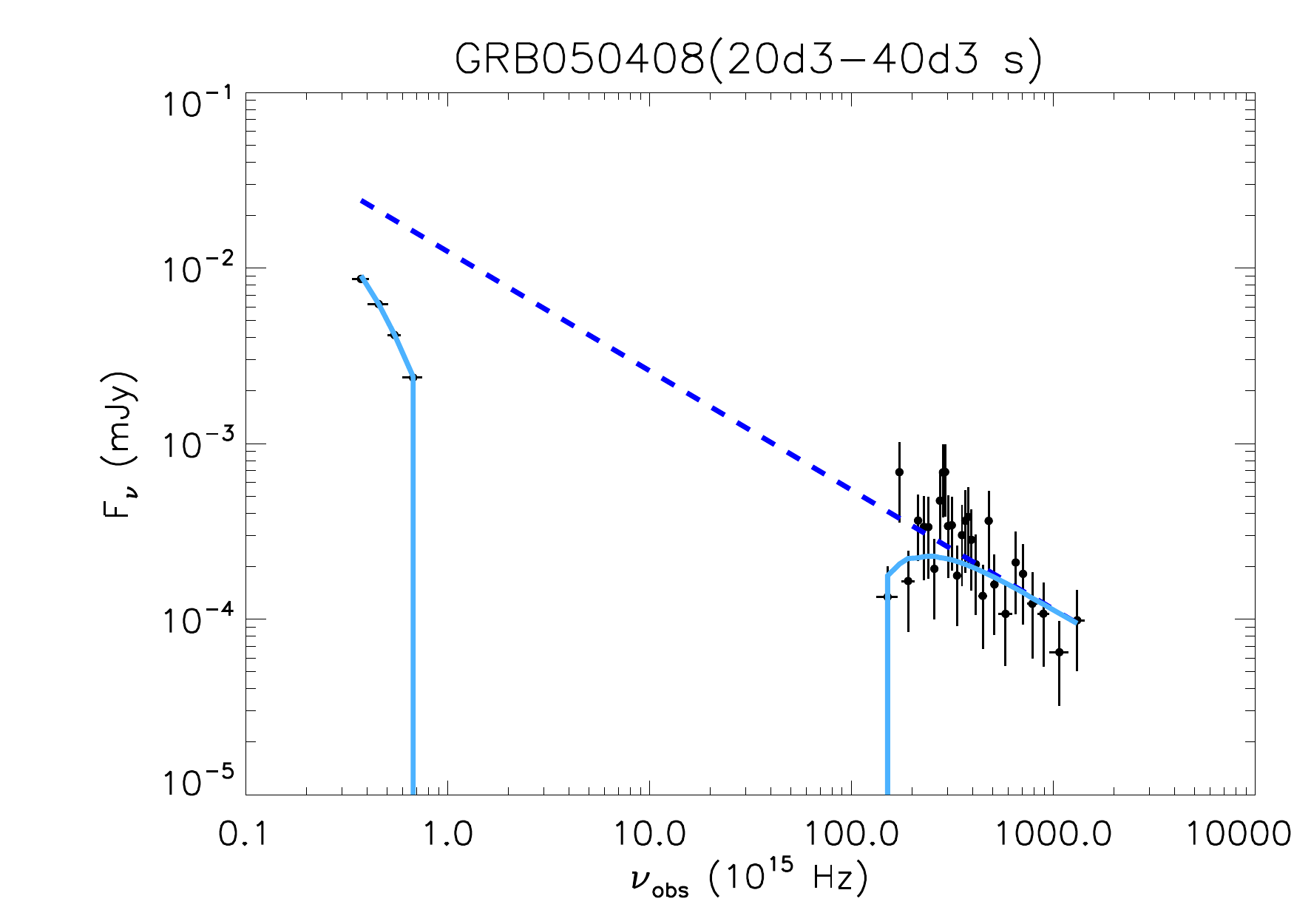}
\includegraphics[width=0.3 \hsize,clip]{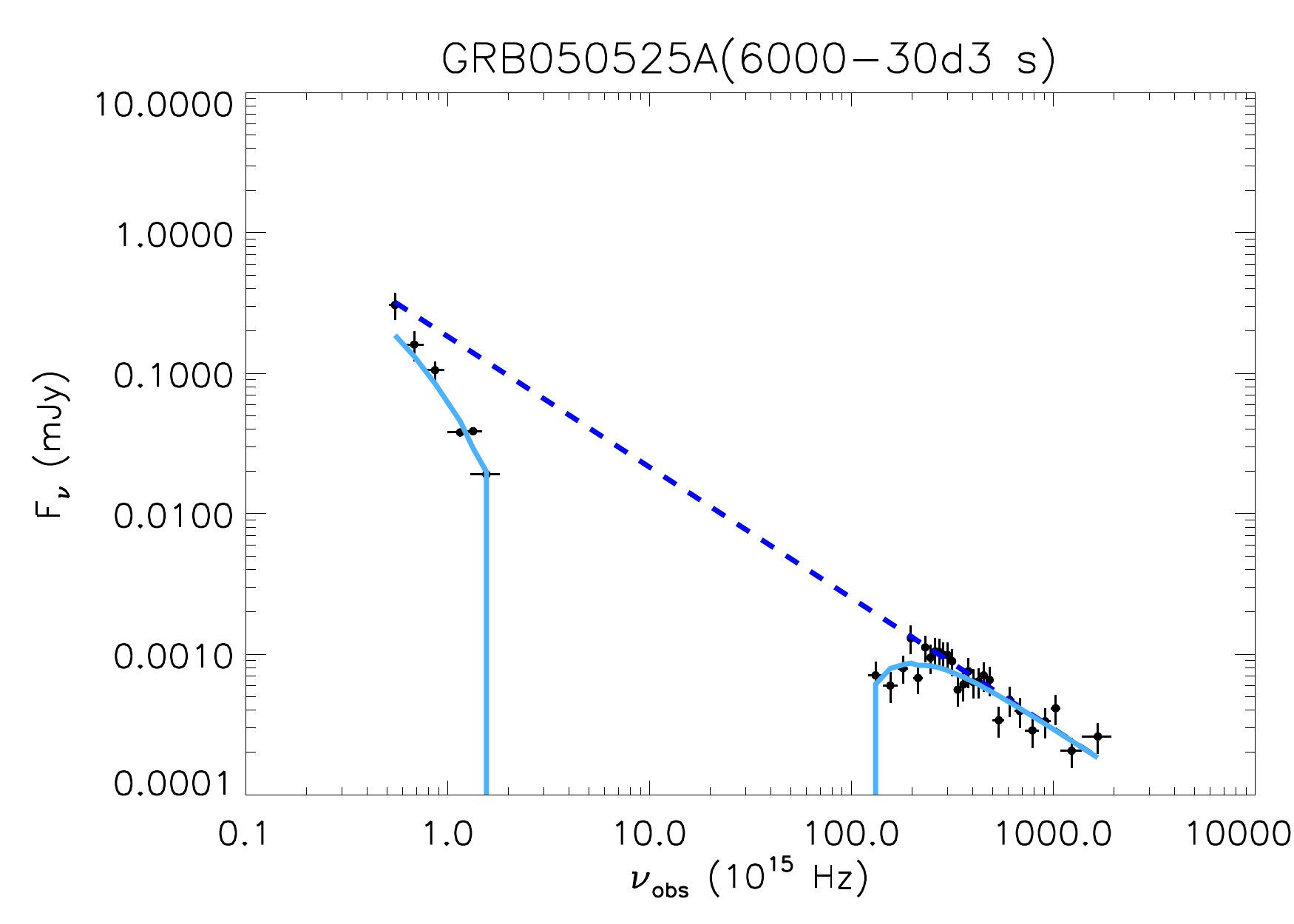}
\includegraphics[width=0.3 \hsize,clip]{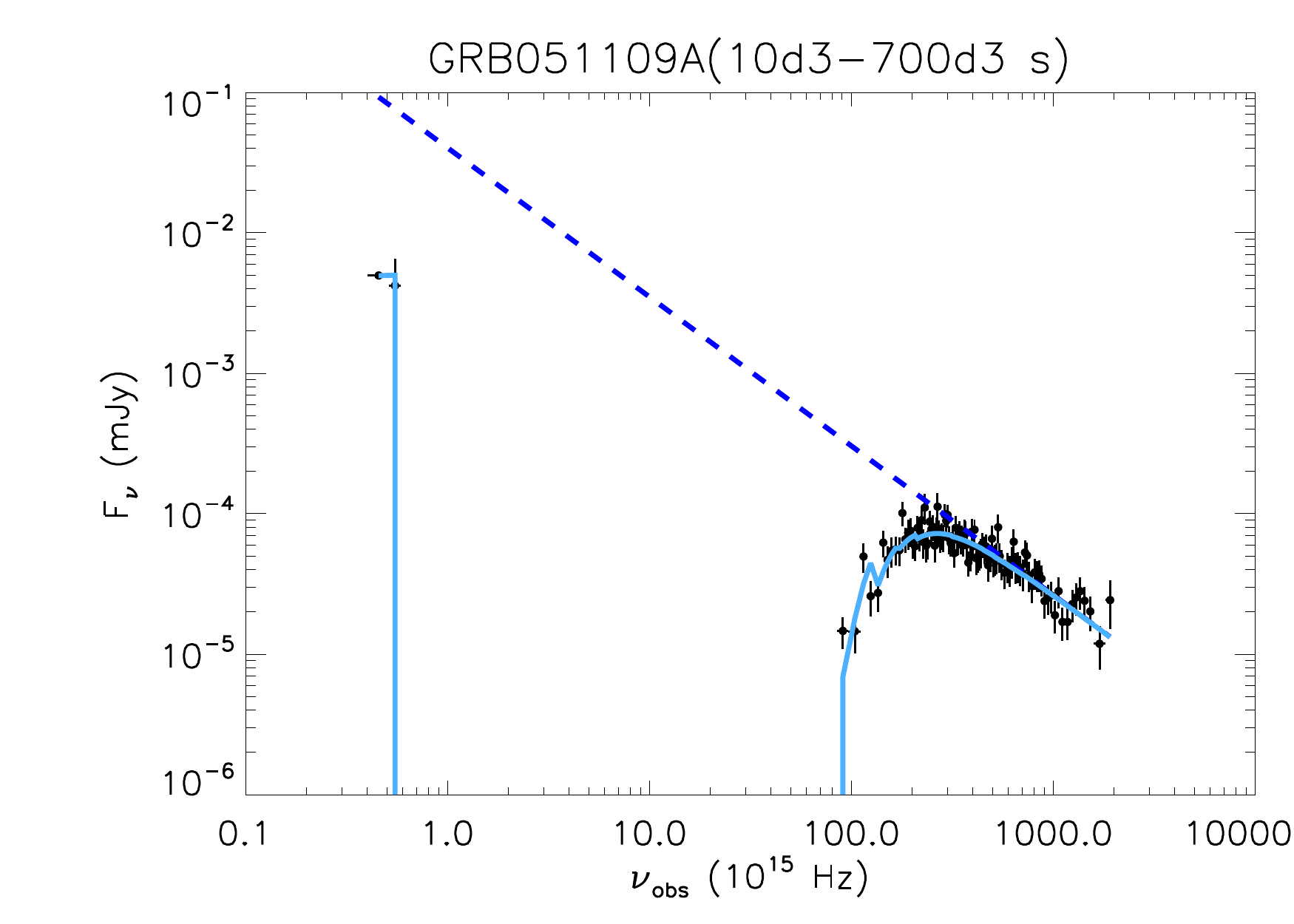}\\
\includegraphics[width=0.3 \hsize,clip]{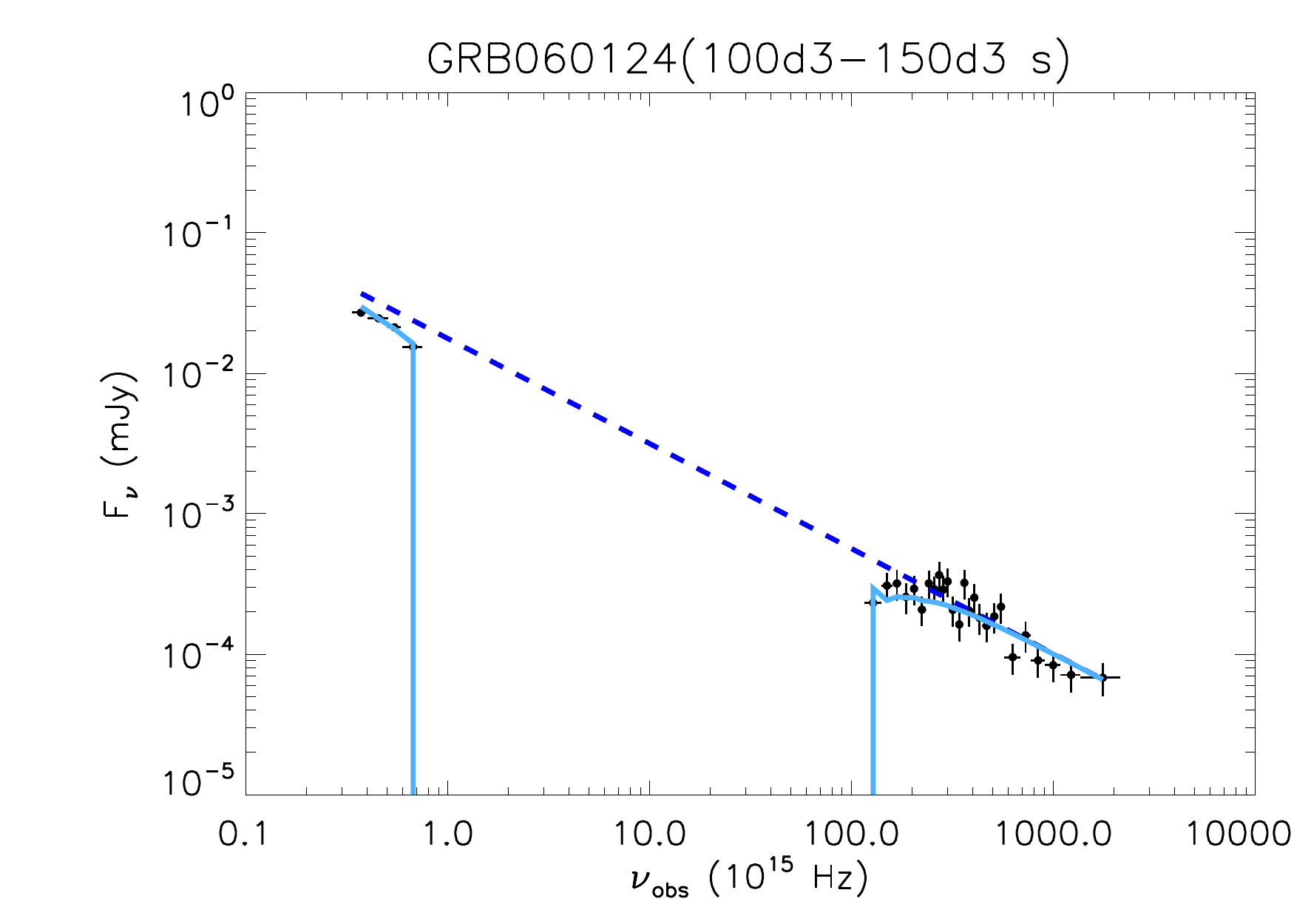}
\includegraphics[width=0.3 \hsize,clip]{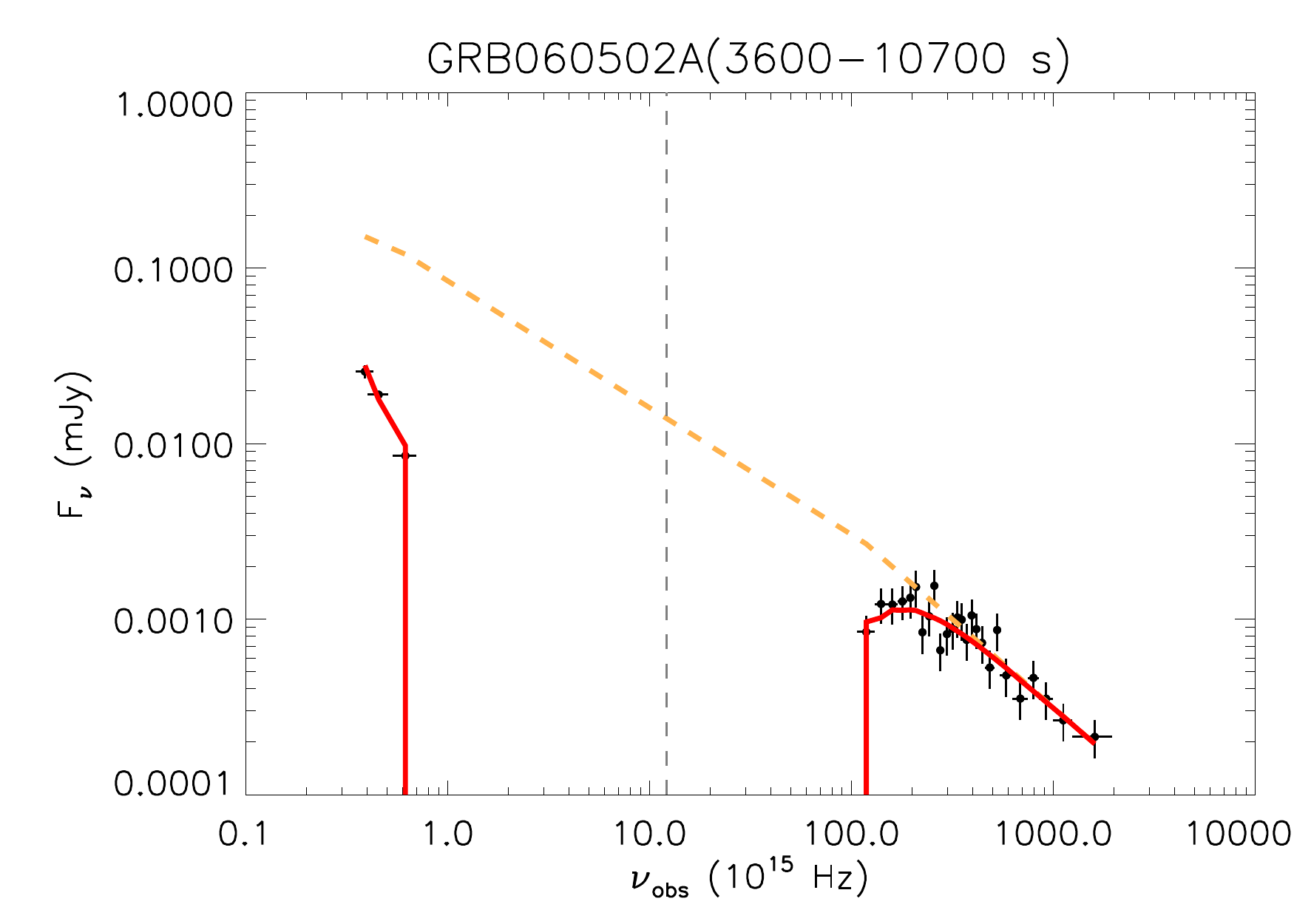}
\includegraphics[width=0.3\hsize,clip]{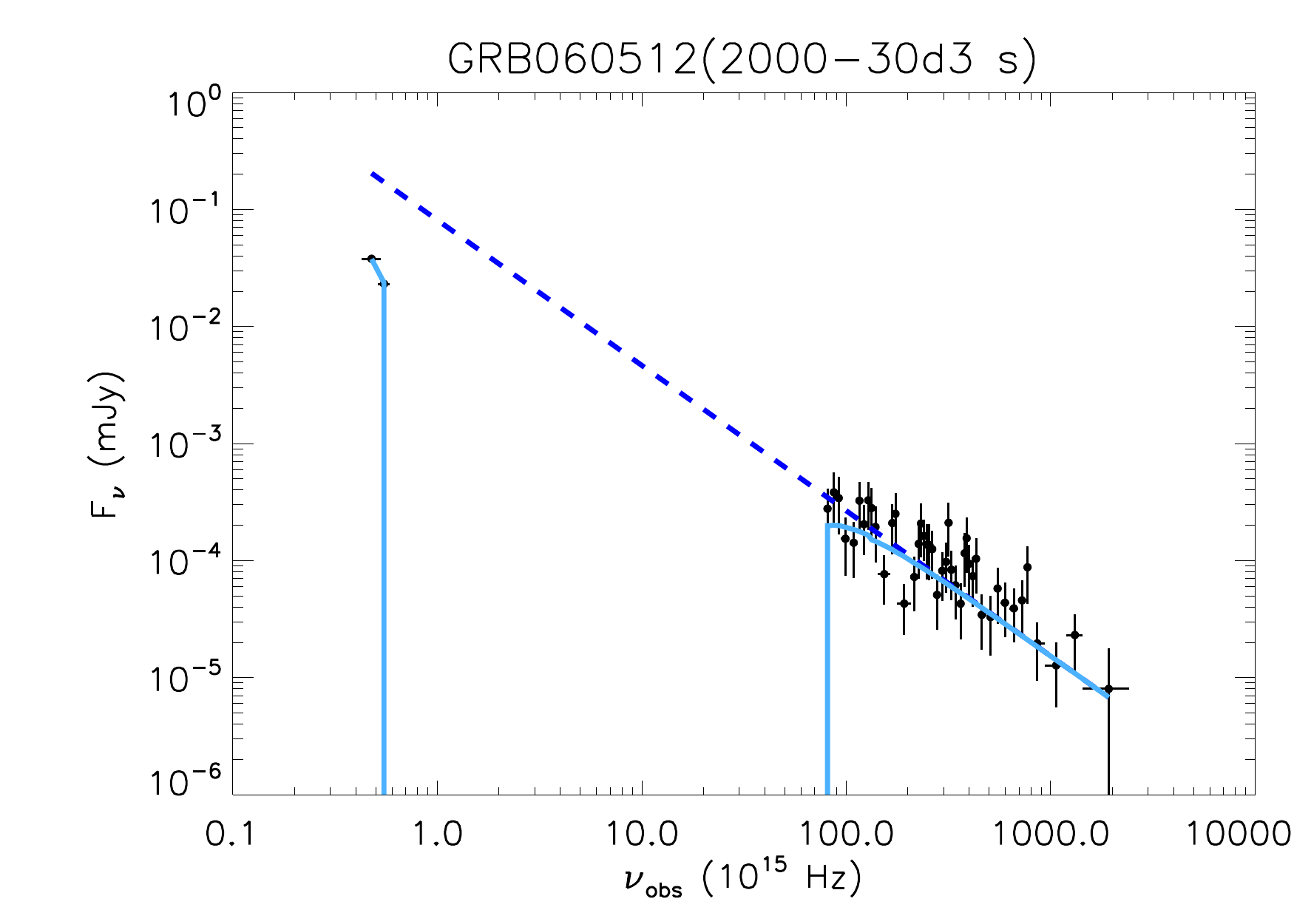}\\
\includegraphics[width=0.3 \hsize,clip]{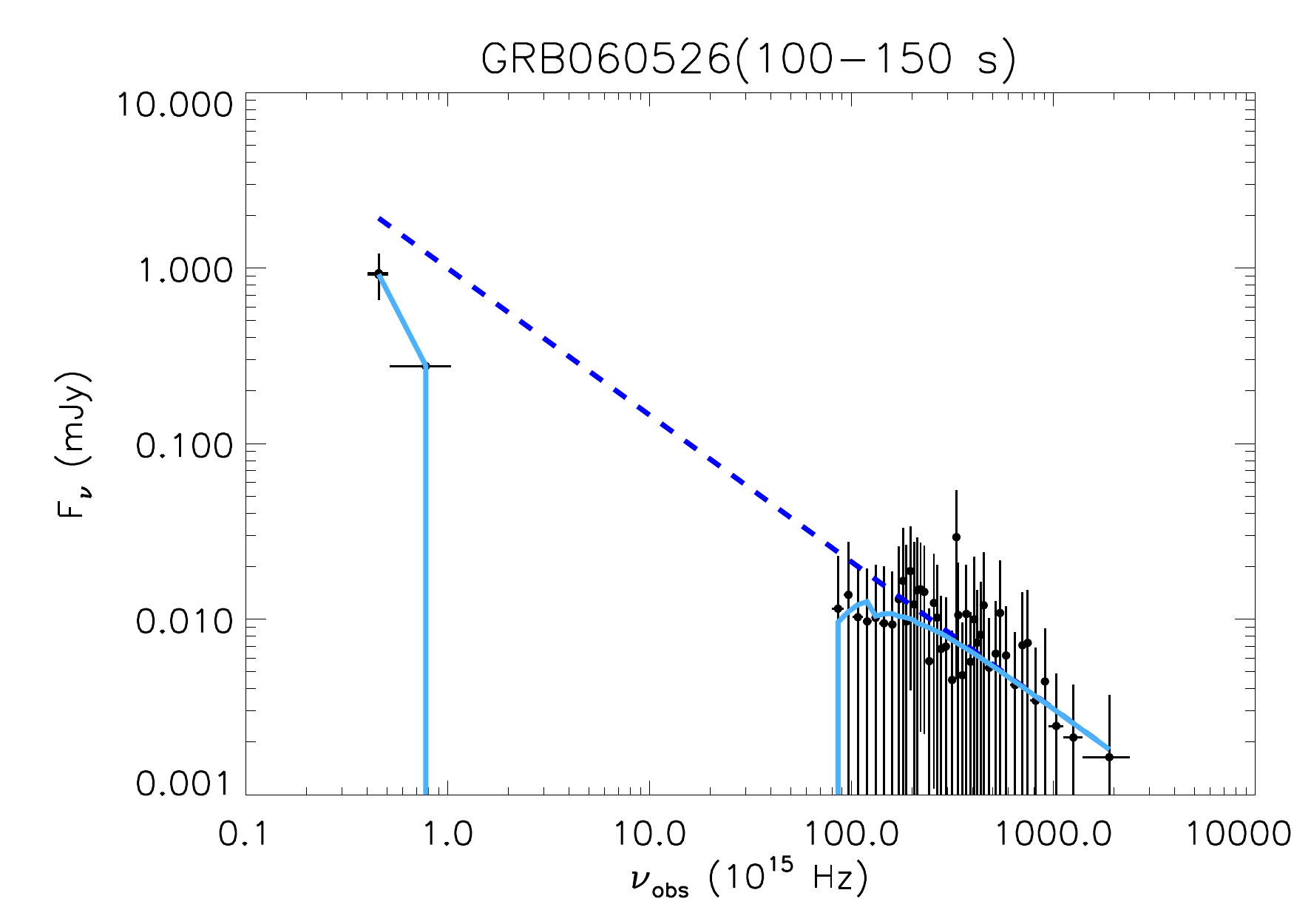}
\includegraphics[width=0.3 \hsize,clip]{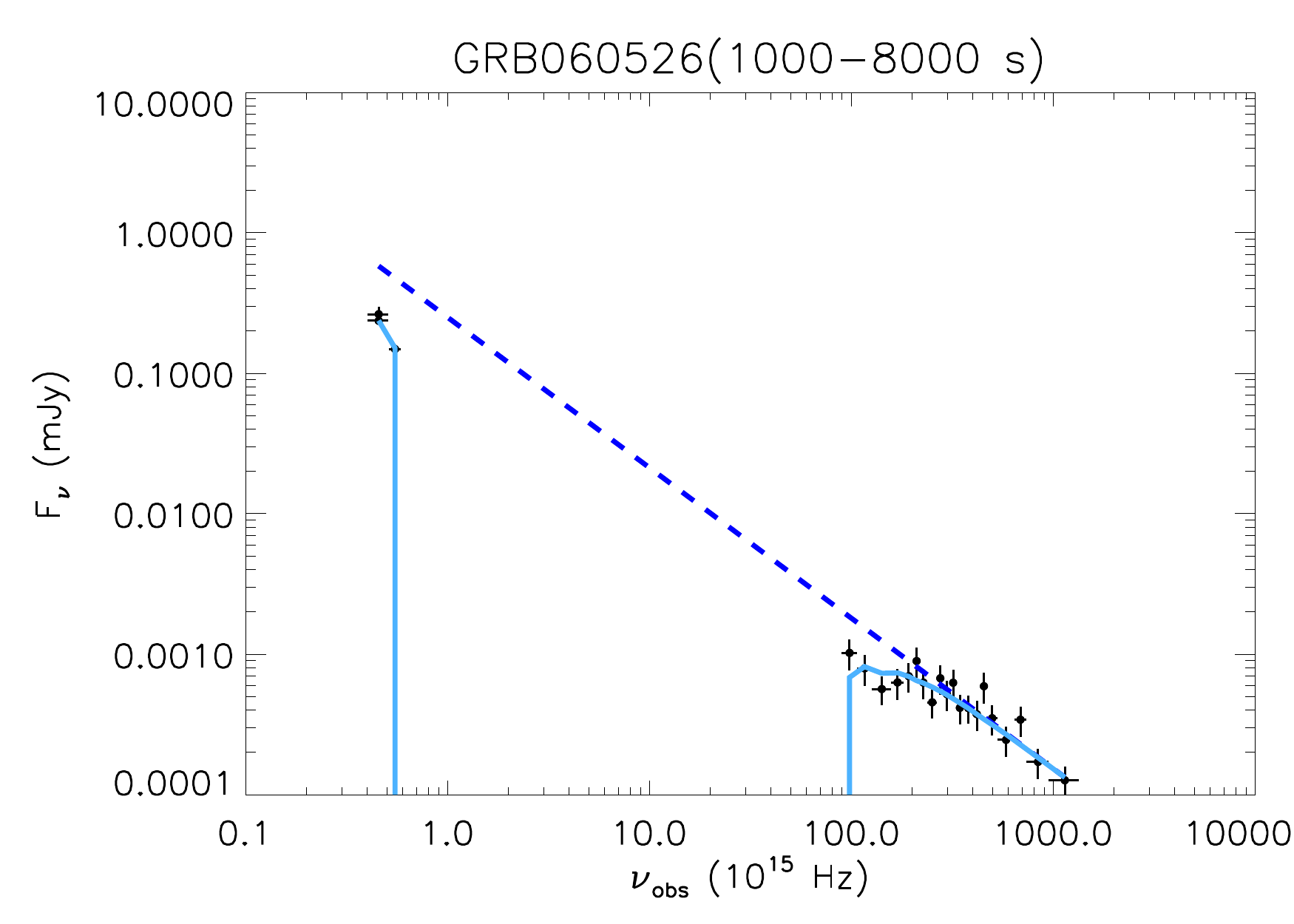}
\includegraphics[width=0.3 \hsize,clip]{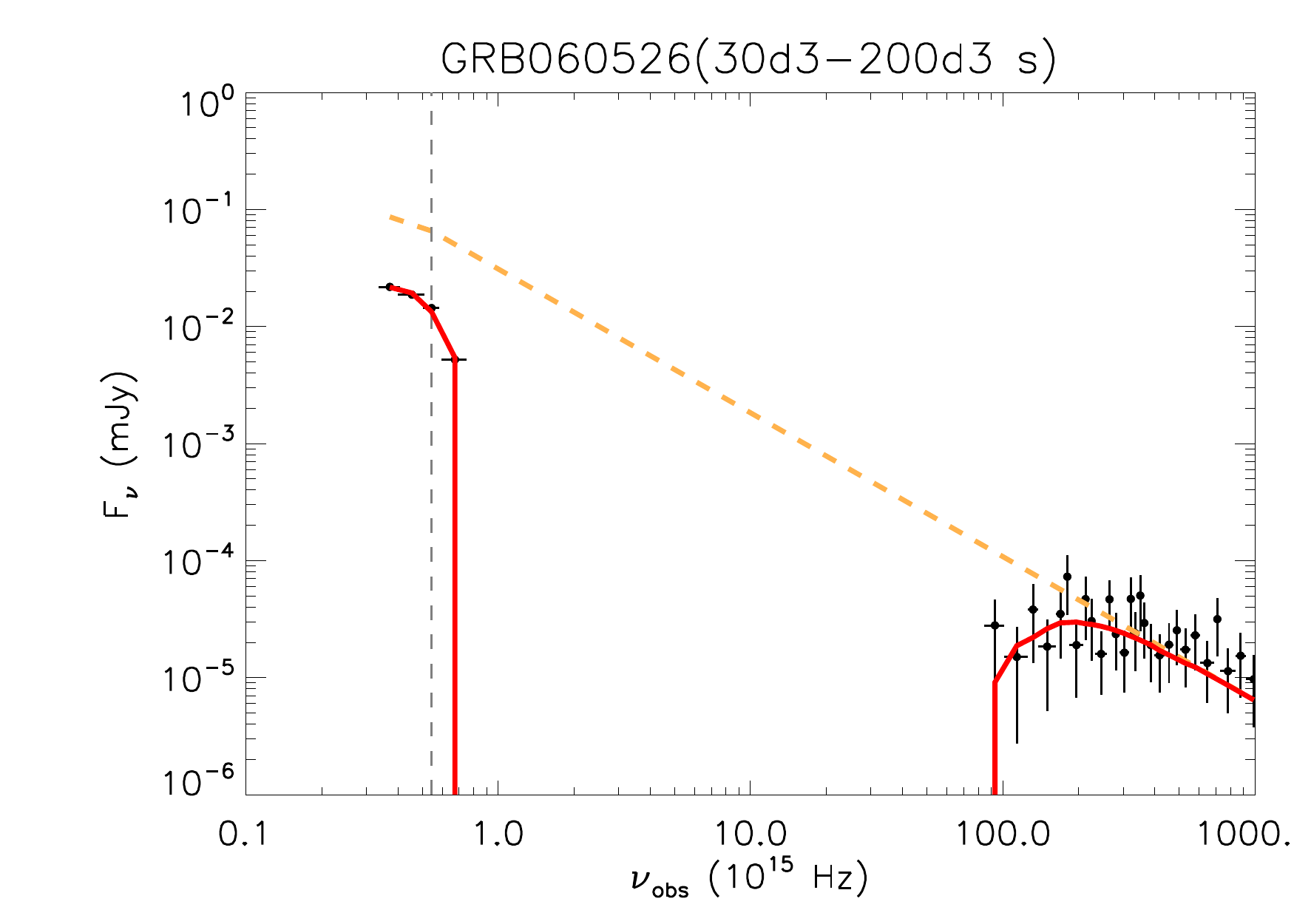}\\
\includegraphics[width=0.3 \hsize,clip]{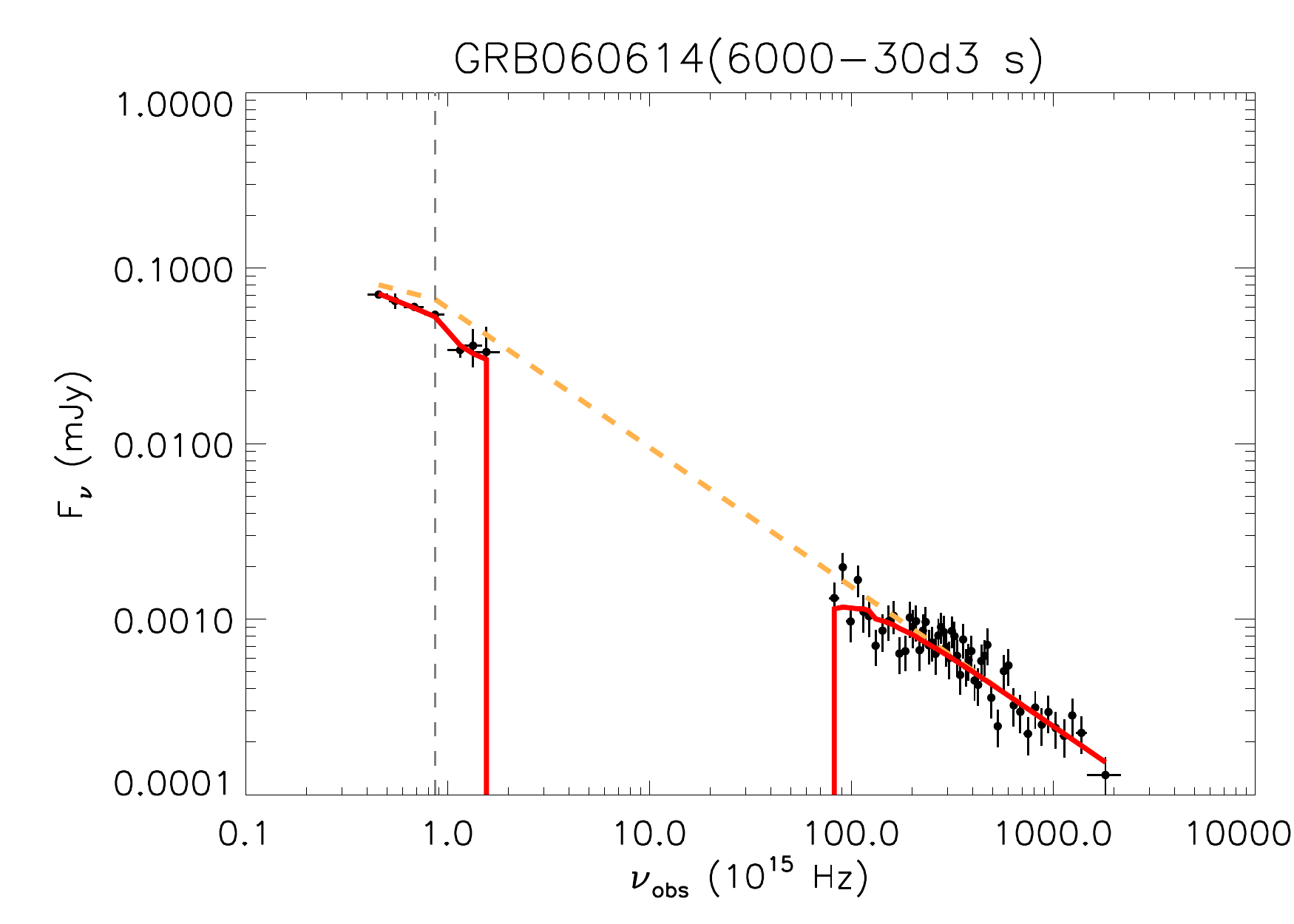}
\includegraphics[width=0.3 \hsize,clip]{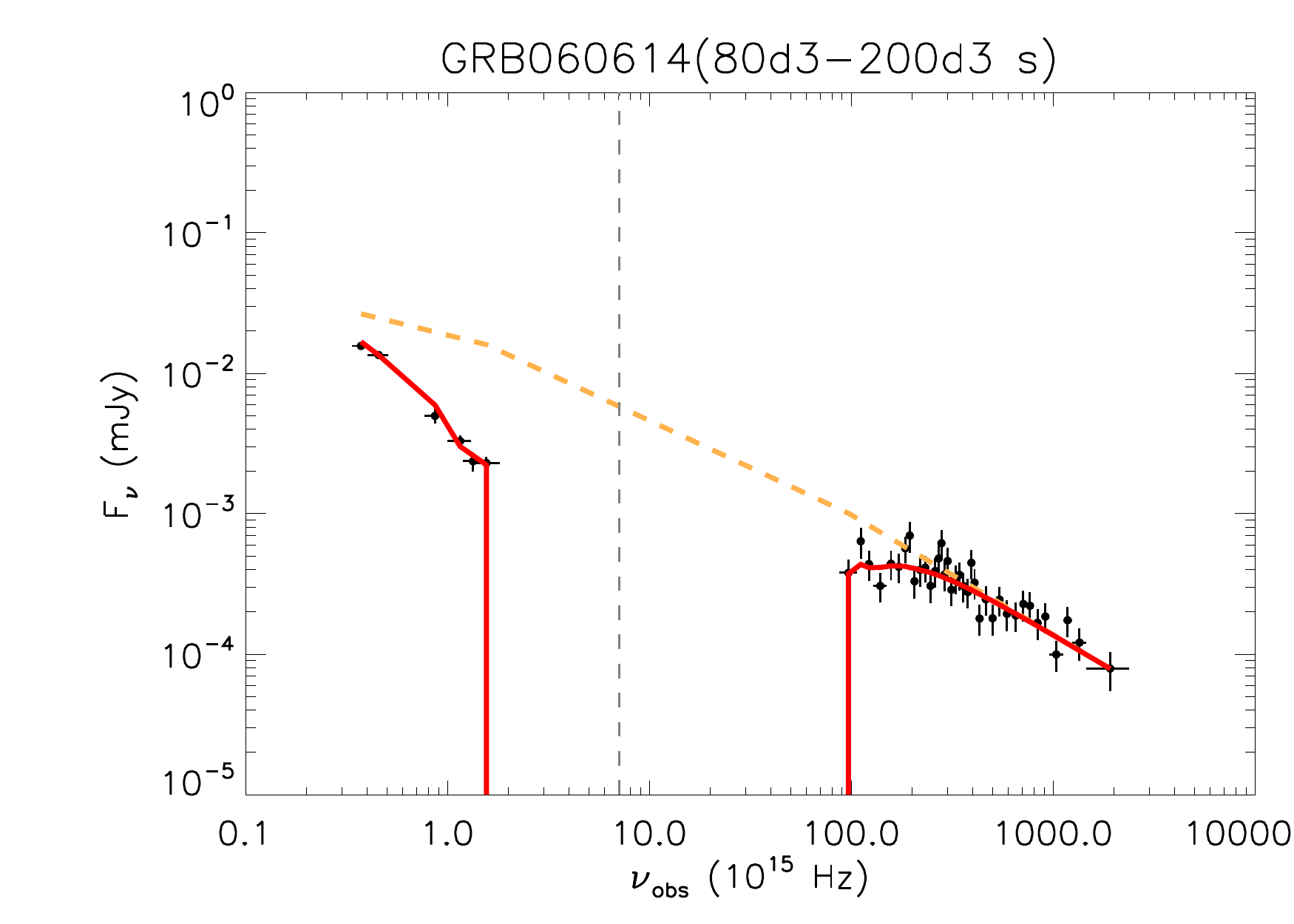}
\includegraphics[width=0.3 \hsize,clip]{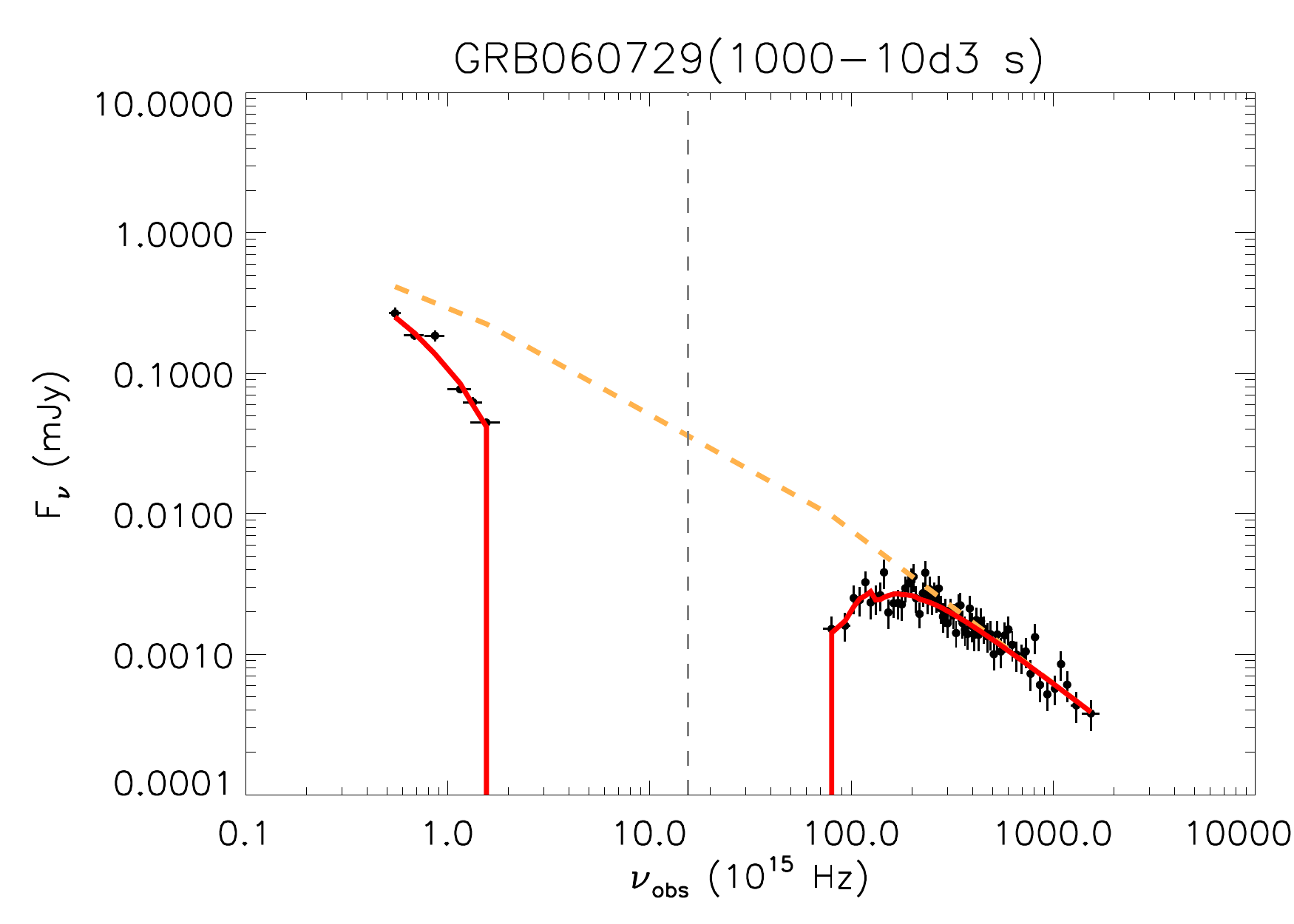}\\
\includegraphics[width=0.3 \hsize,clip]{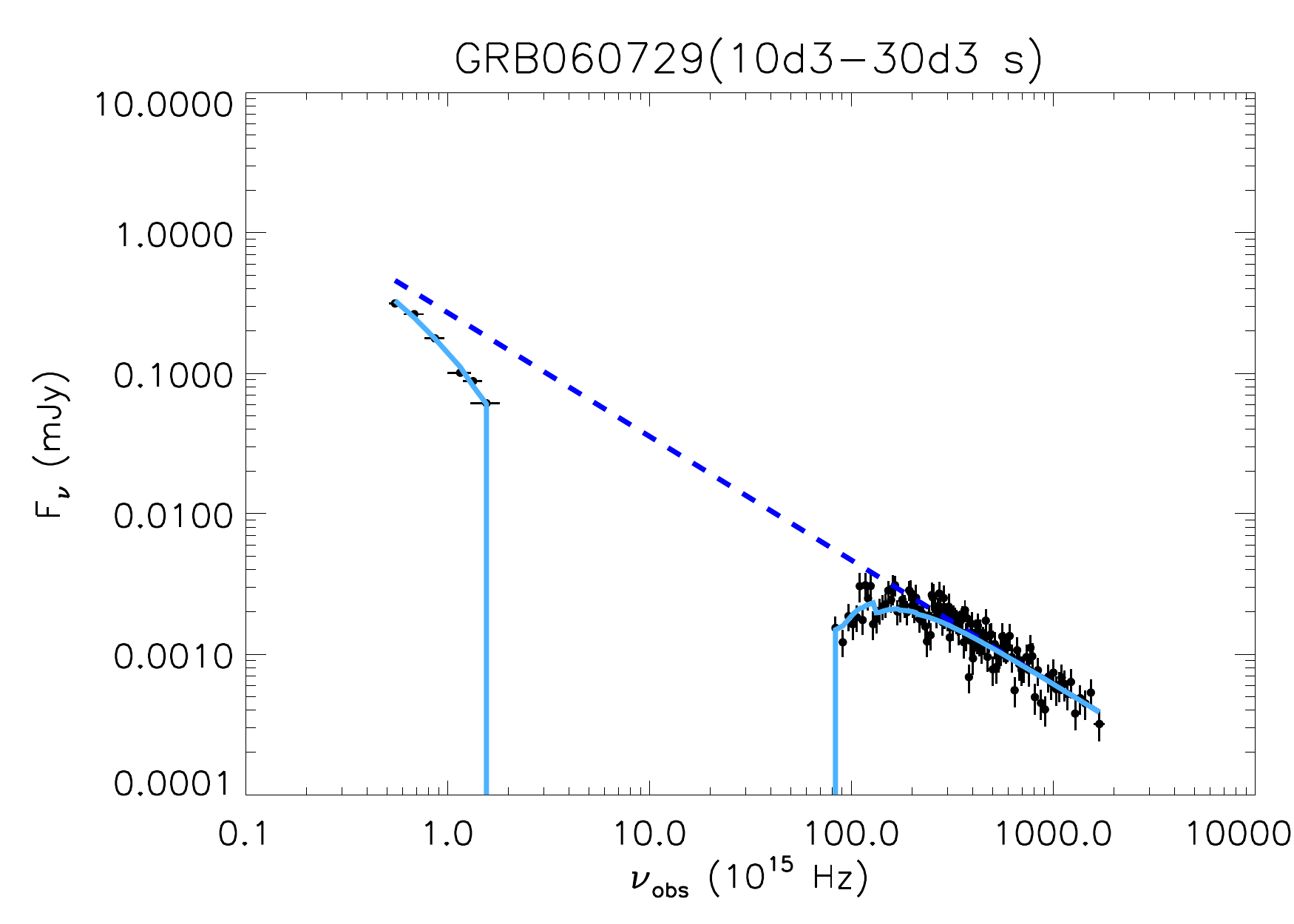}
\includegraphics[width=0.3 \hsize,clip]{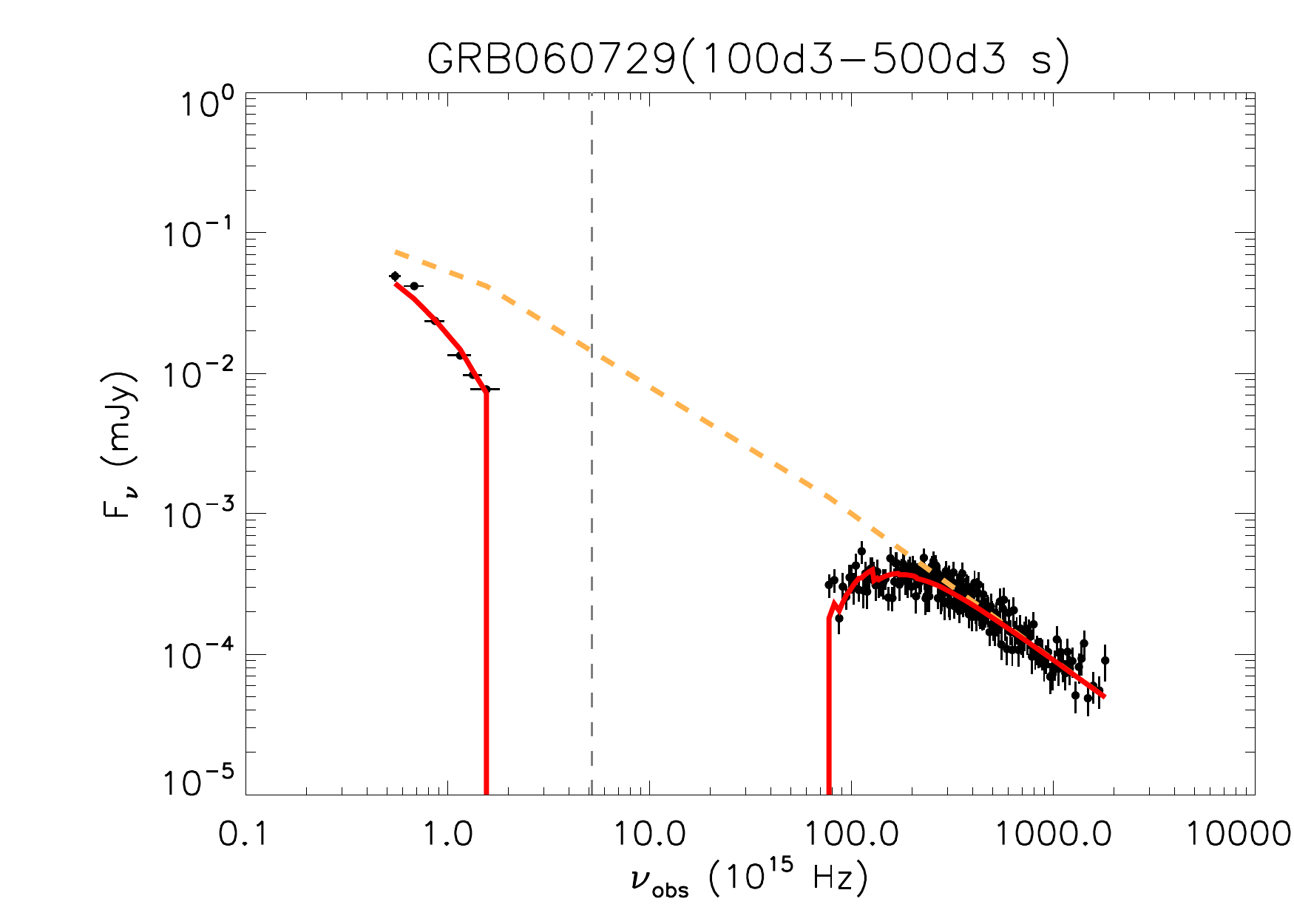}
\includegraphics[width=0.3 \hsize,clip]{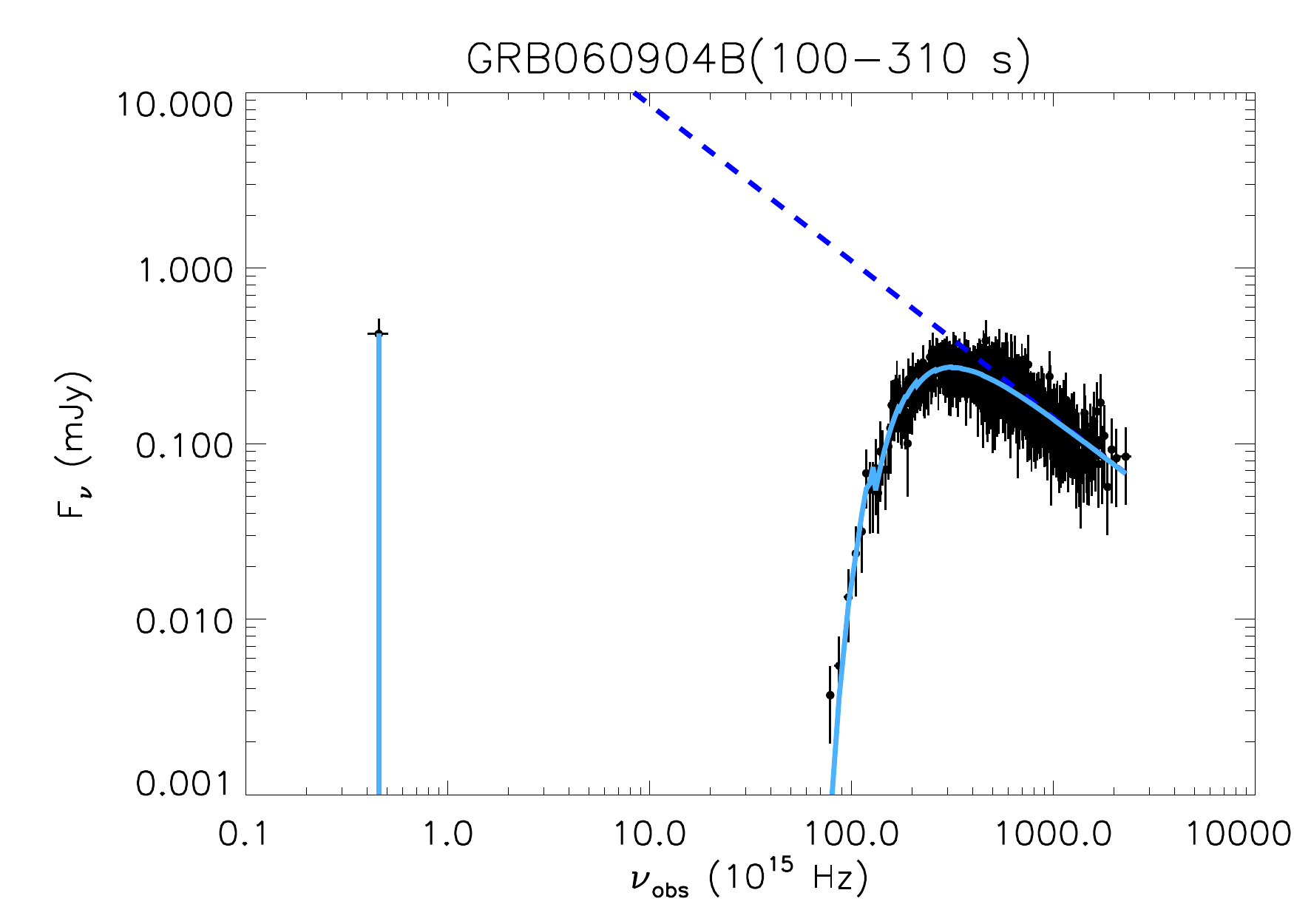}
\caption{\small{Optical/X-ray SEDs for GRBs belonging to Group B. color-coding as in Figure~\ref{sed1}.}}\label{sed3} 
\end{figure}

\begin{figure}
\includegraphics[width=0.3 \hsize,clip]{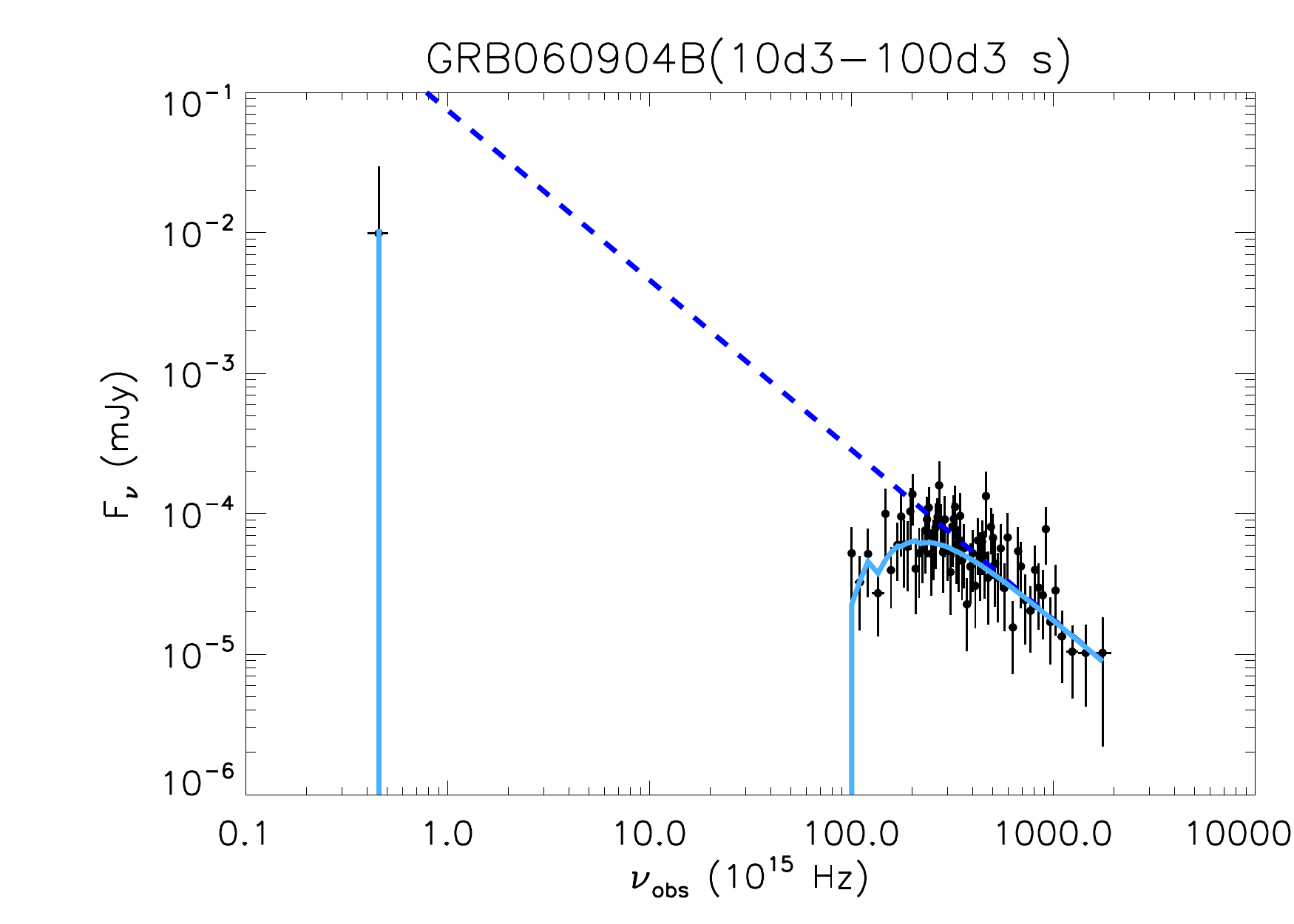}
\includegraphics[width=0.3 \hsize,clip]{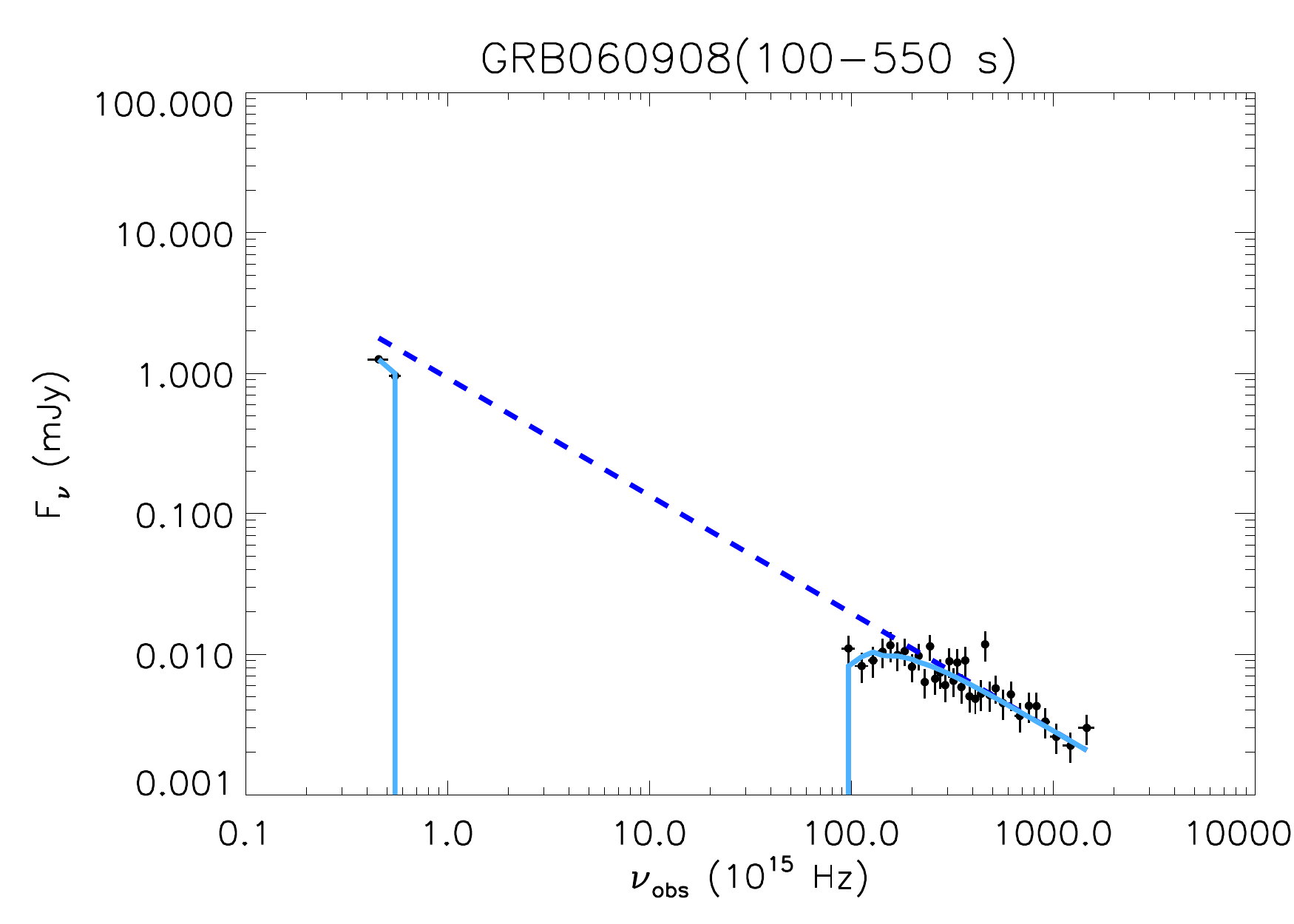}
\includegraphics[width=0.3\hsize,clip]{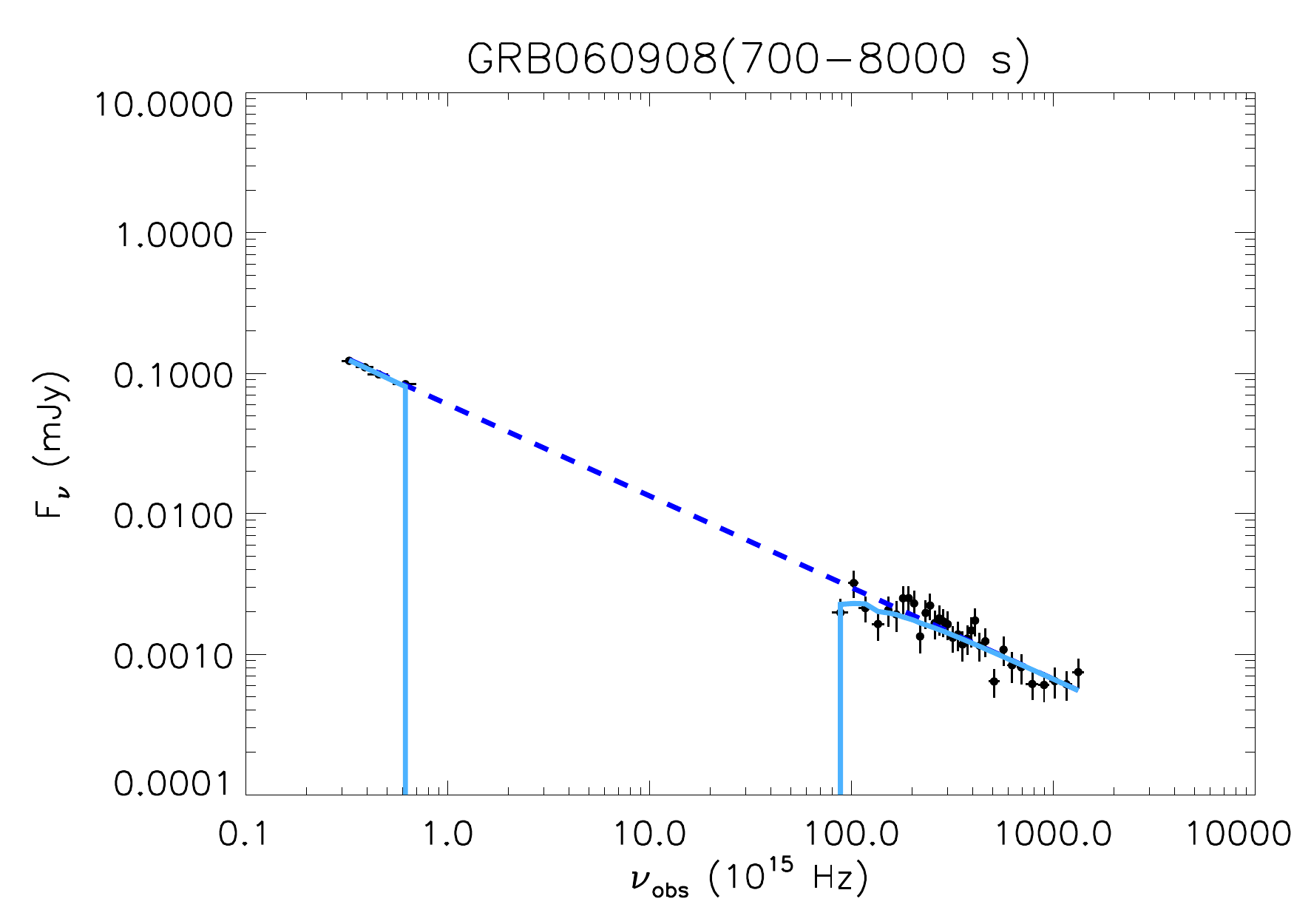}\\
\includegraphics[width=0.3 \hsize,clip]{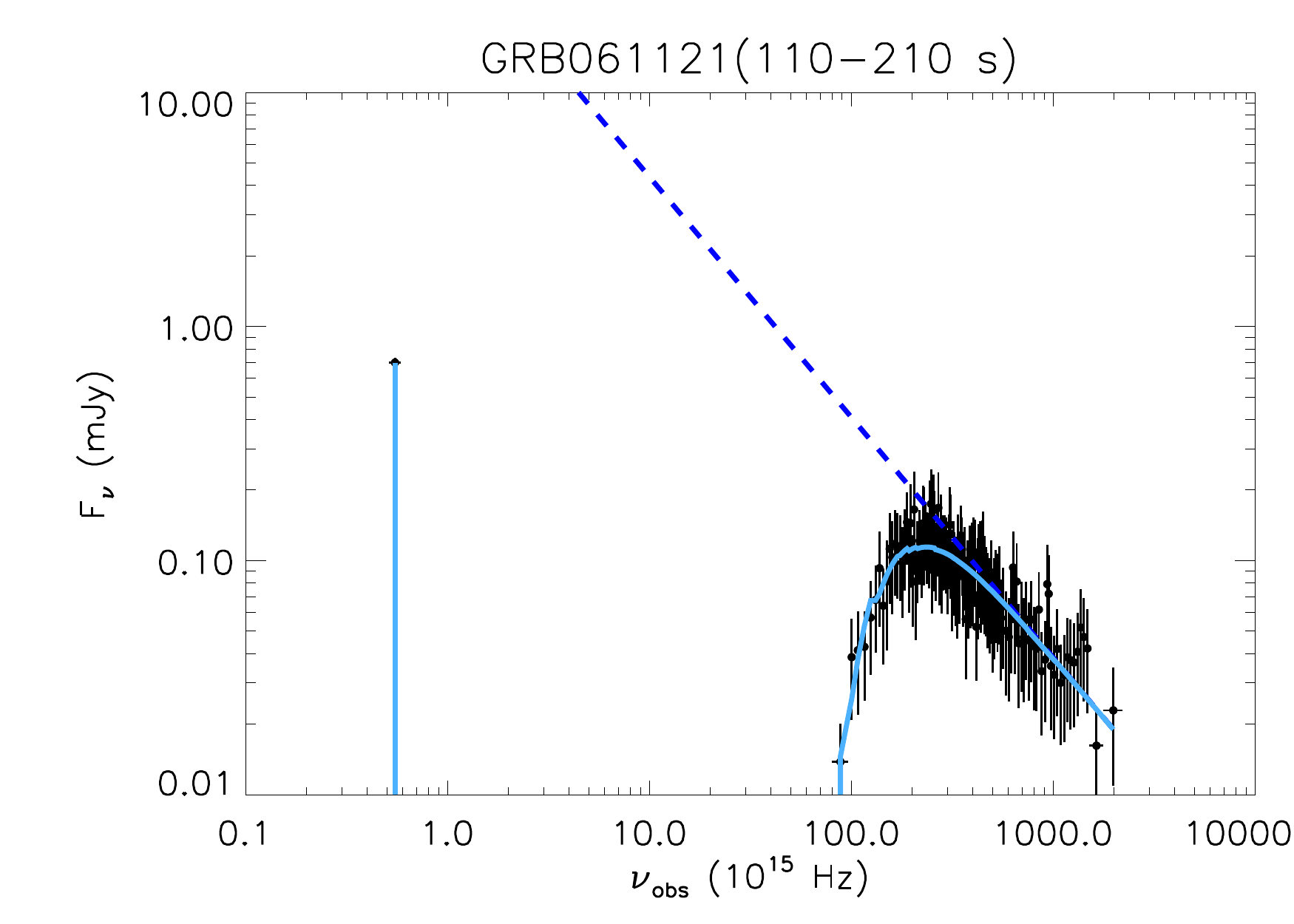}
\includegraphics[width=0.3 \hsize,clip]{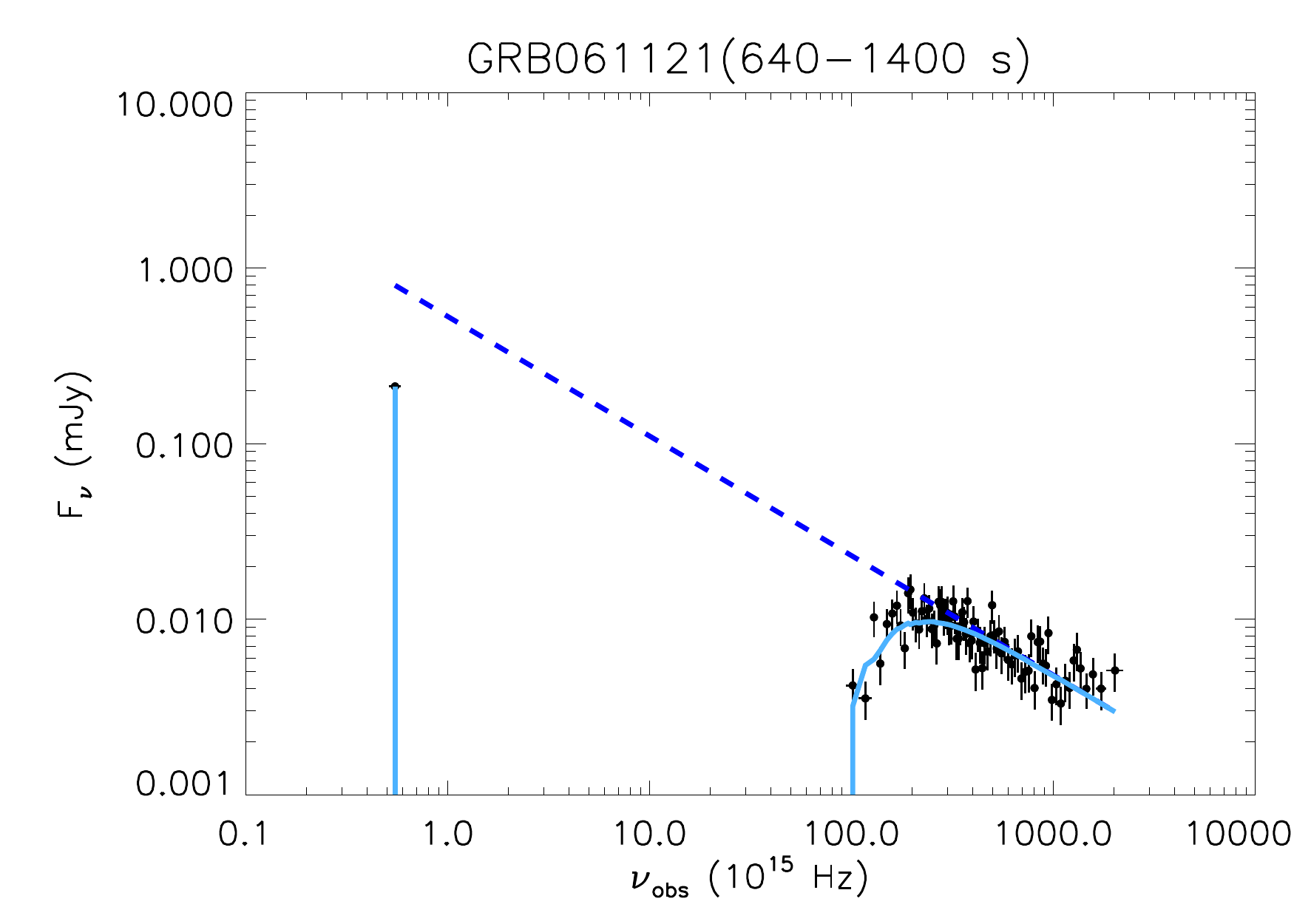}
\includegraphics[width=0.3 \hsize,clip]{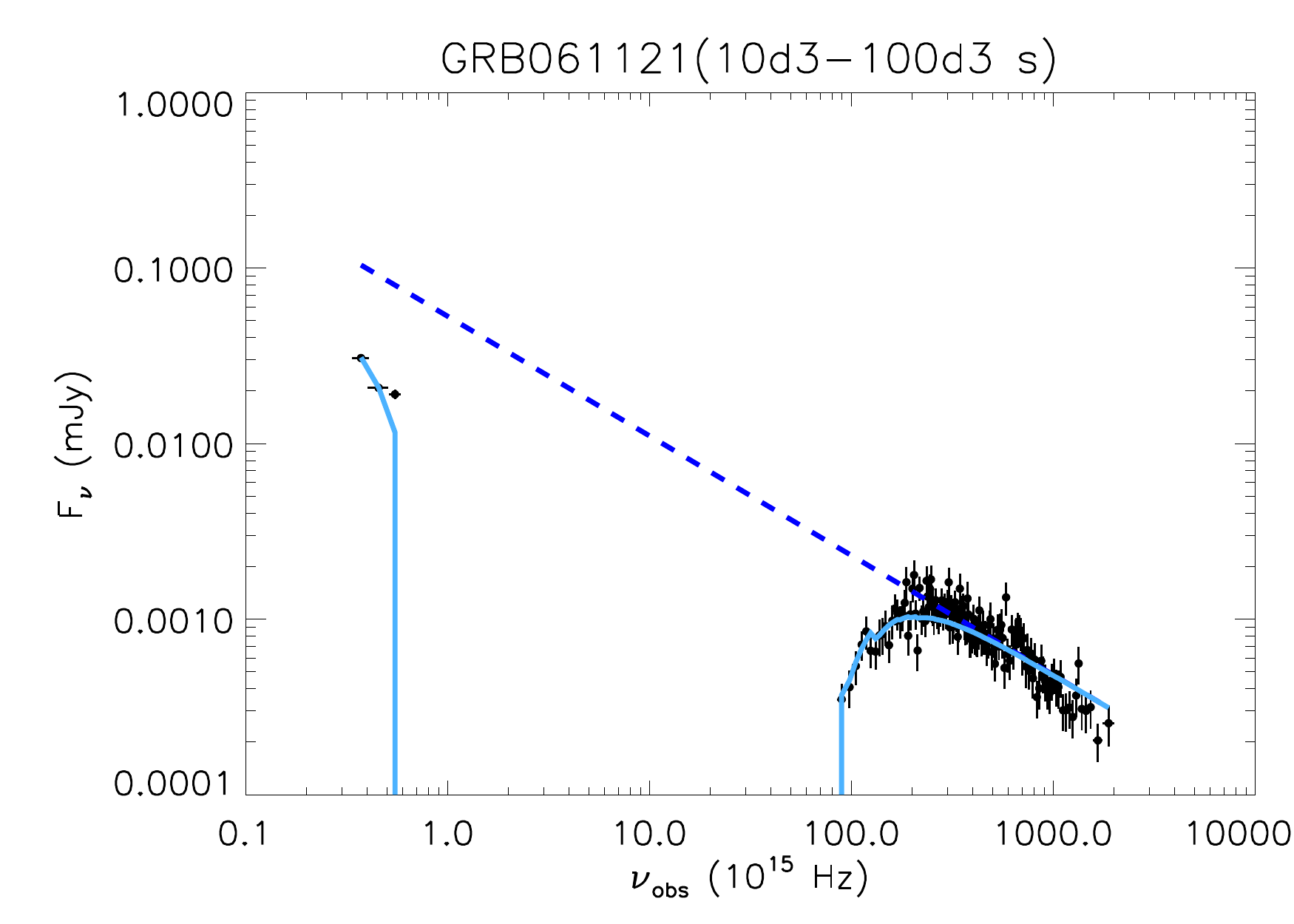}\\
\includegraphics[width=0.3 \hsize,clip]{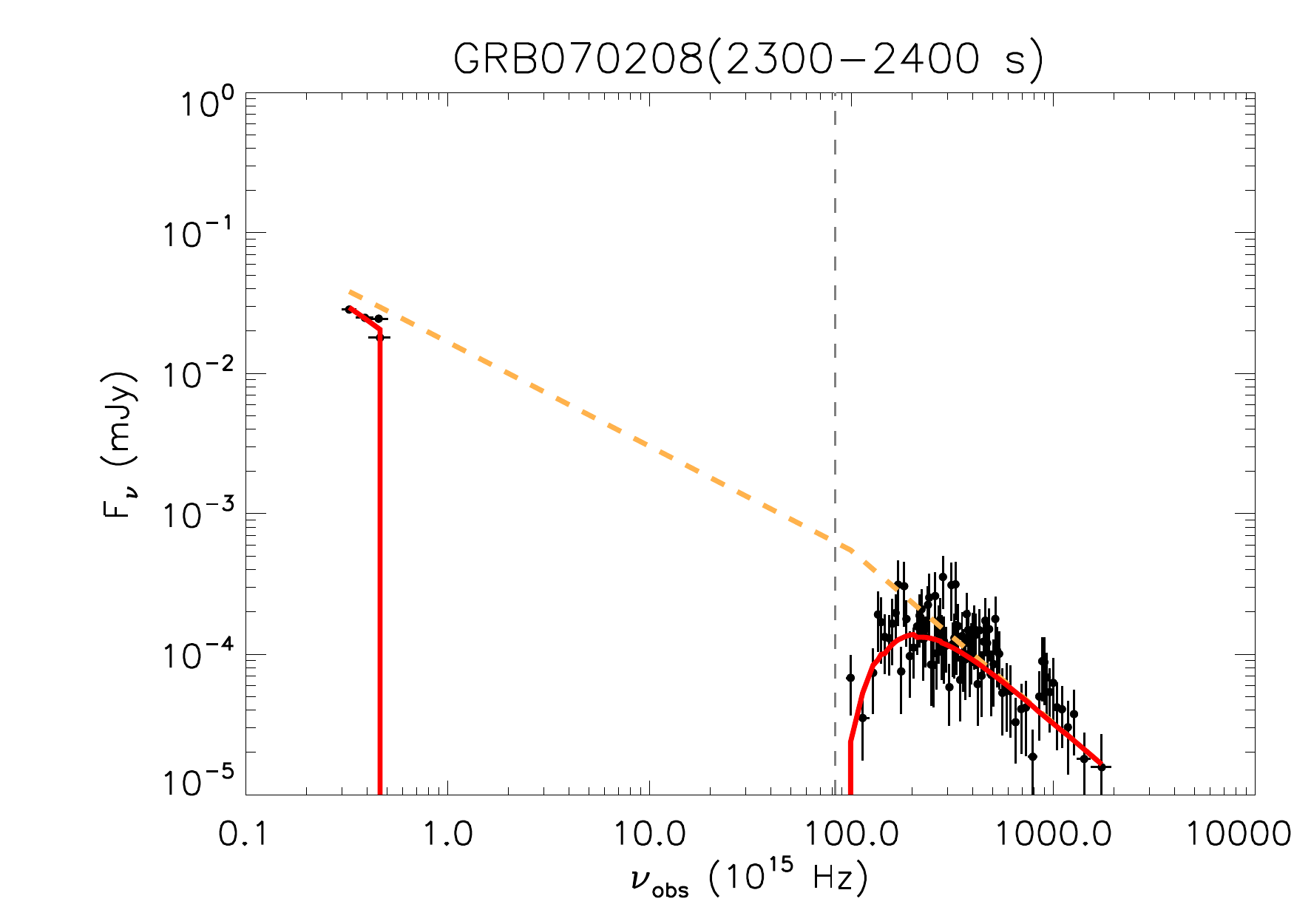}
\includegraphics[width=0.3 \hsize,clip]{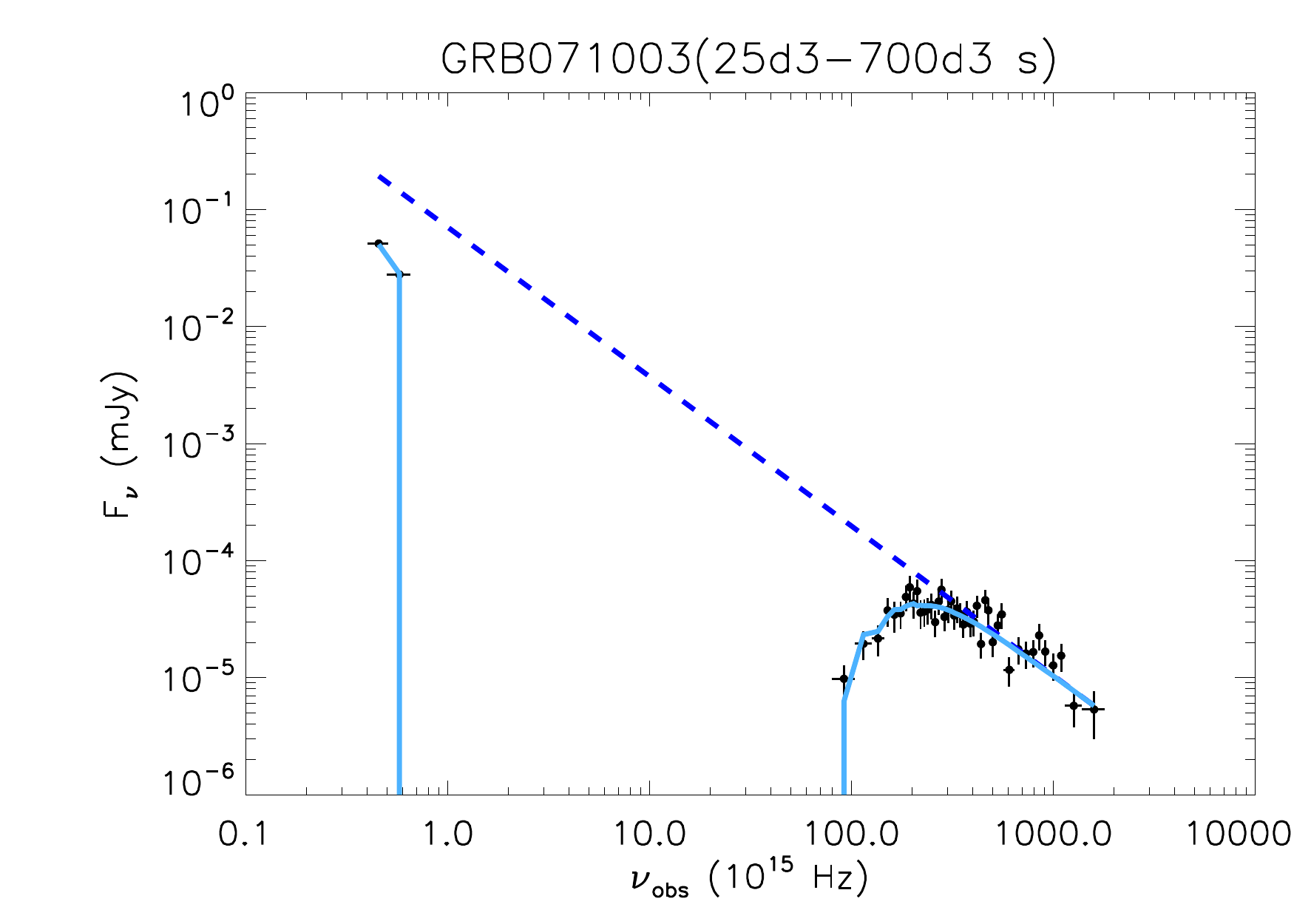}
\includegraphics[width=0.3 \hsize,clip]{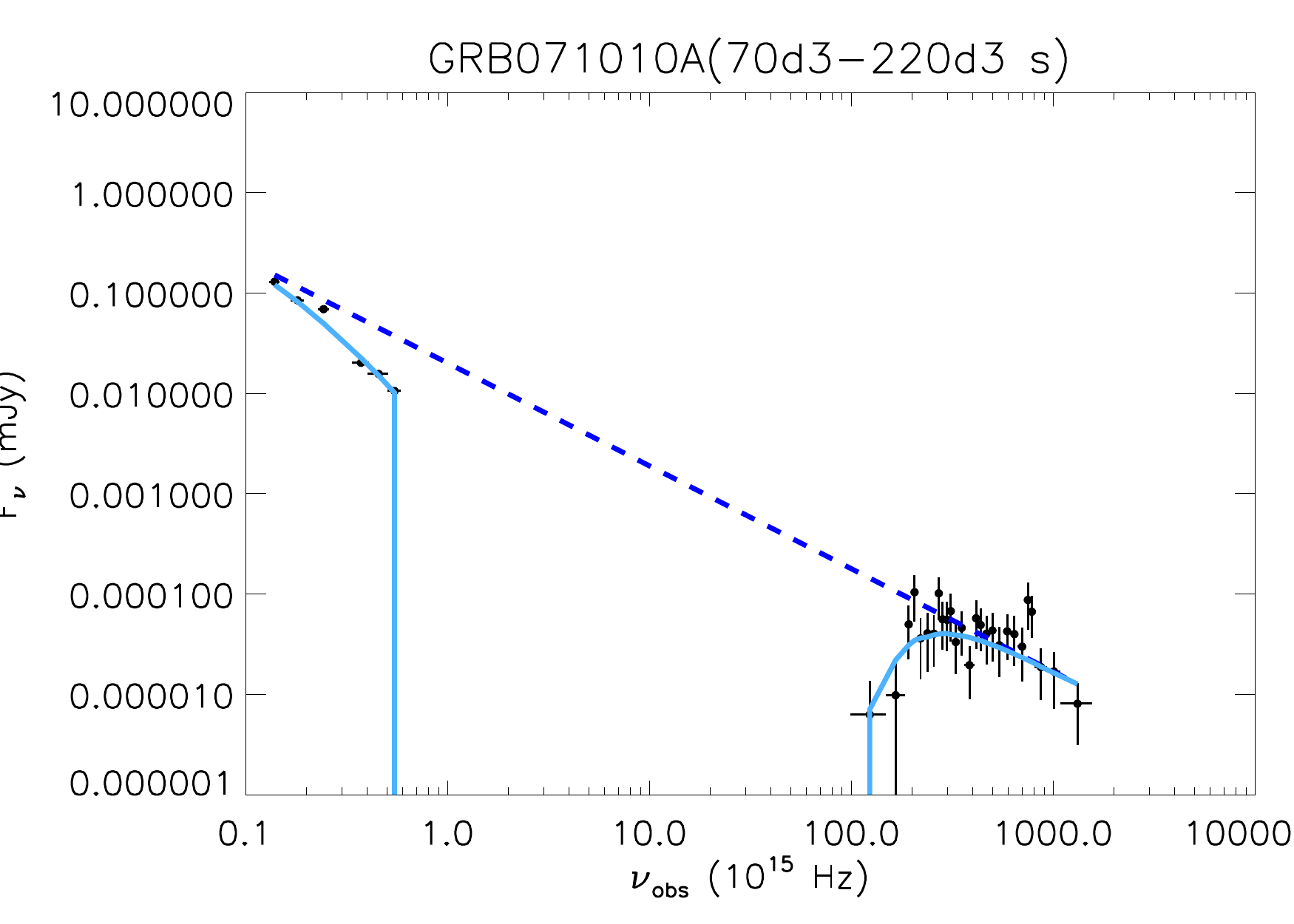}\\
\includegraphics[width=0.3 \hsize,clip]{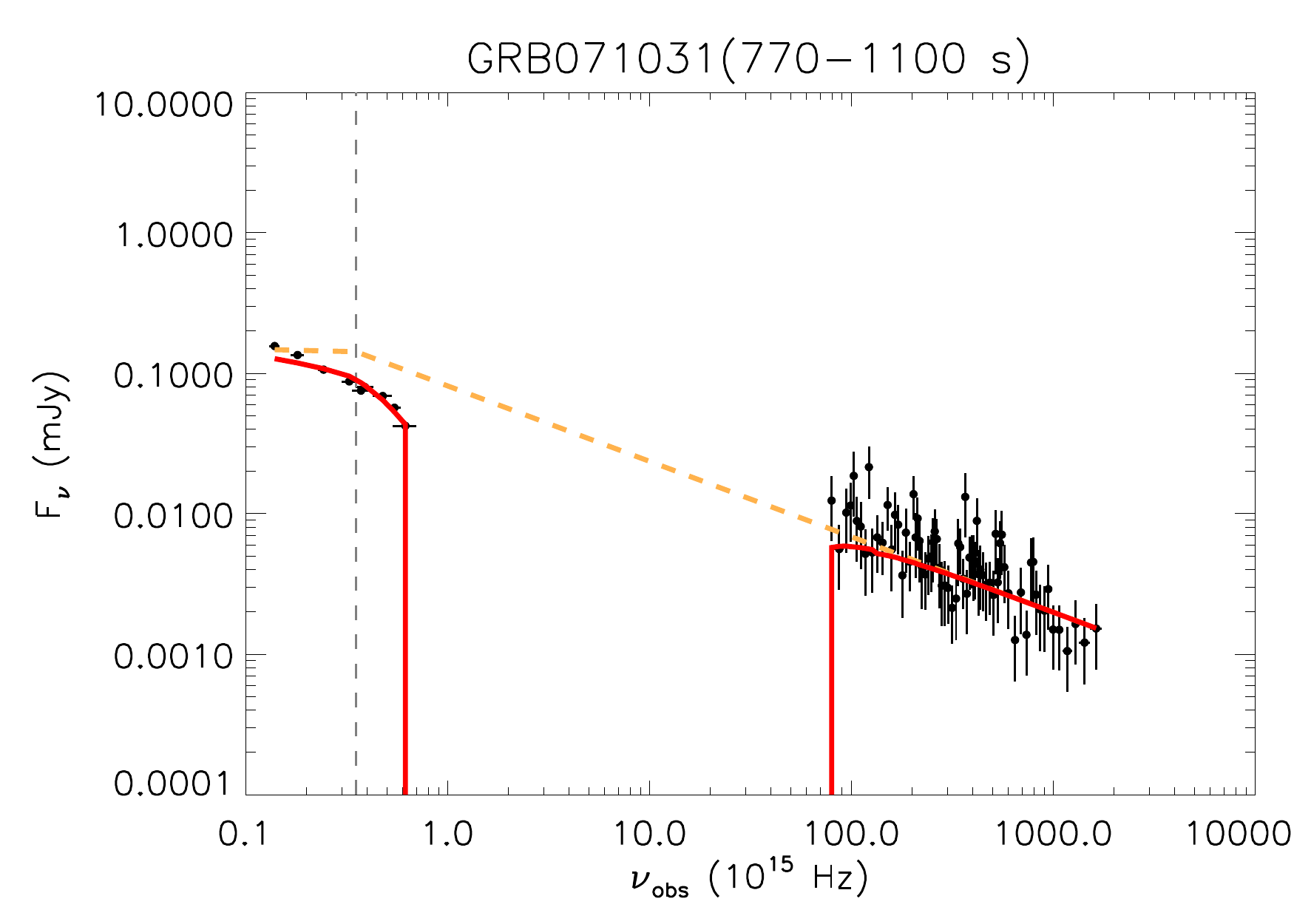}
\includegraphics[width=0.3\hsize,clip]{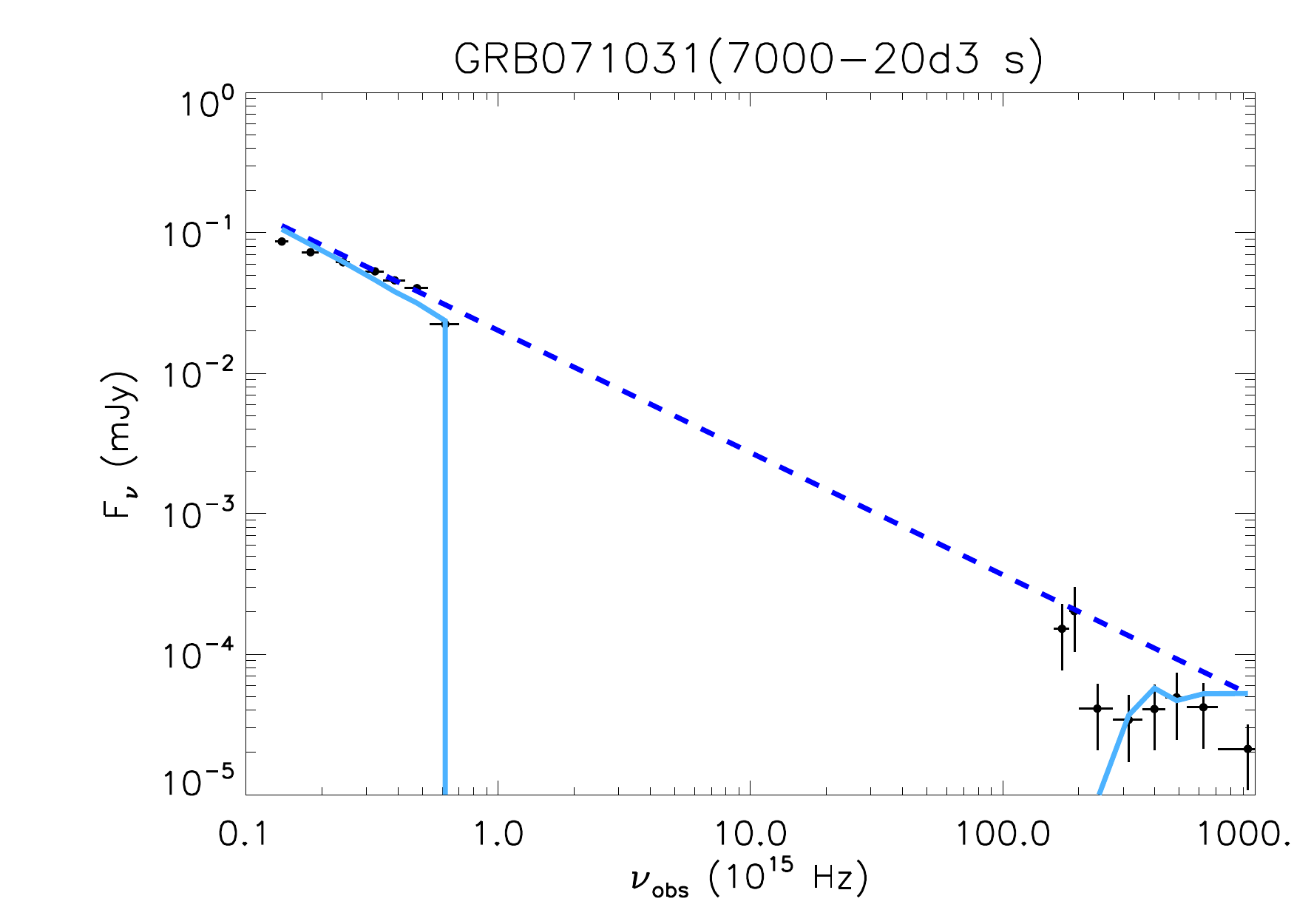}
\includegraphics[width=0.3 \hsize,clip]{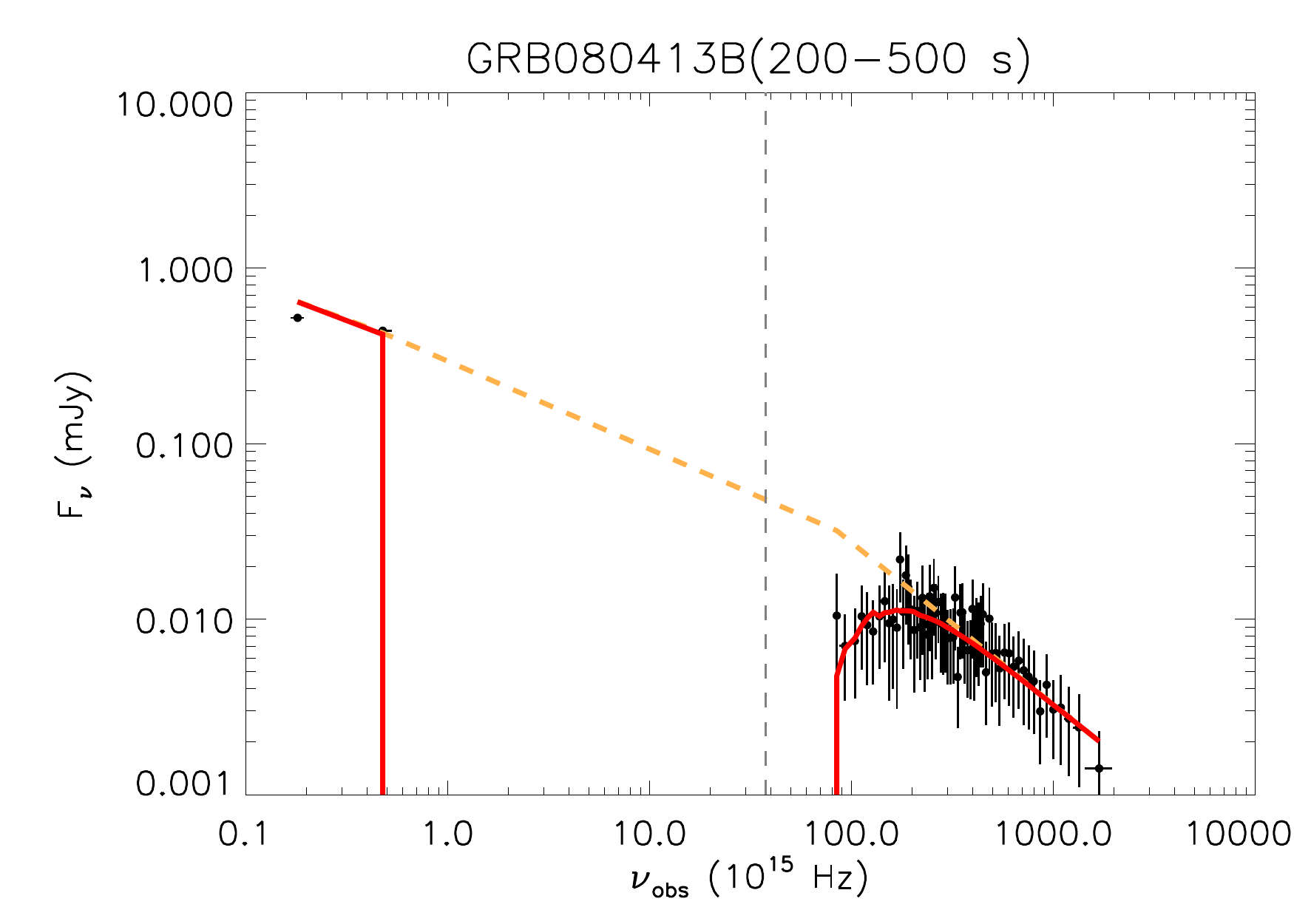}\\
\includegraphics[width=0.3 \hsize,clip]{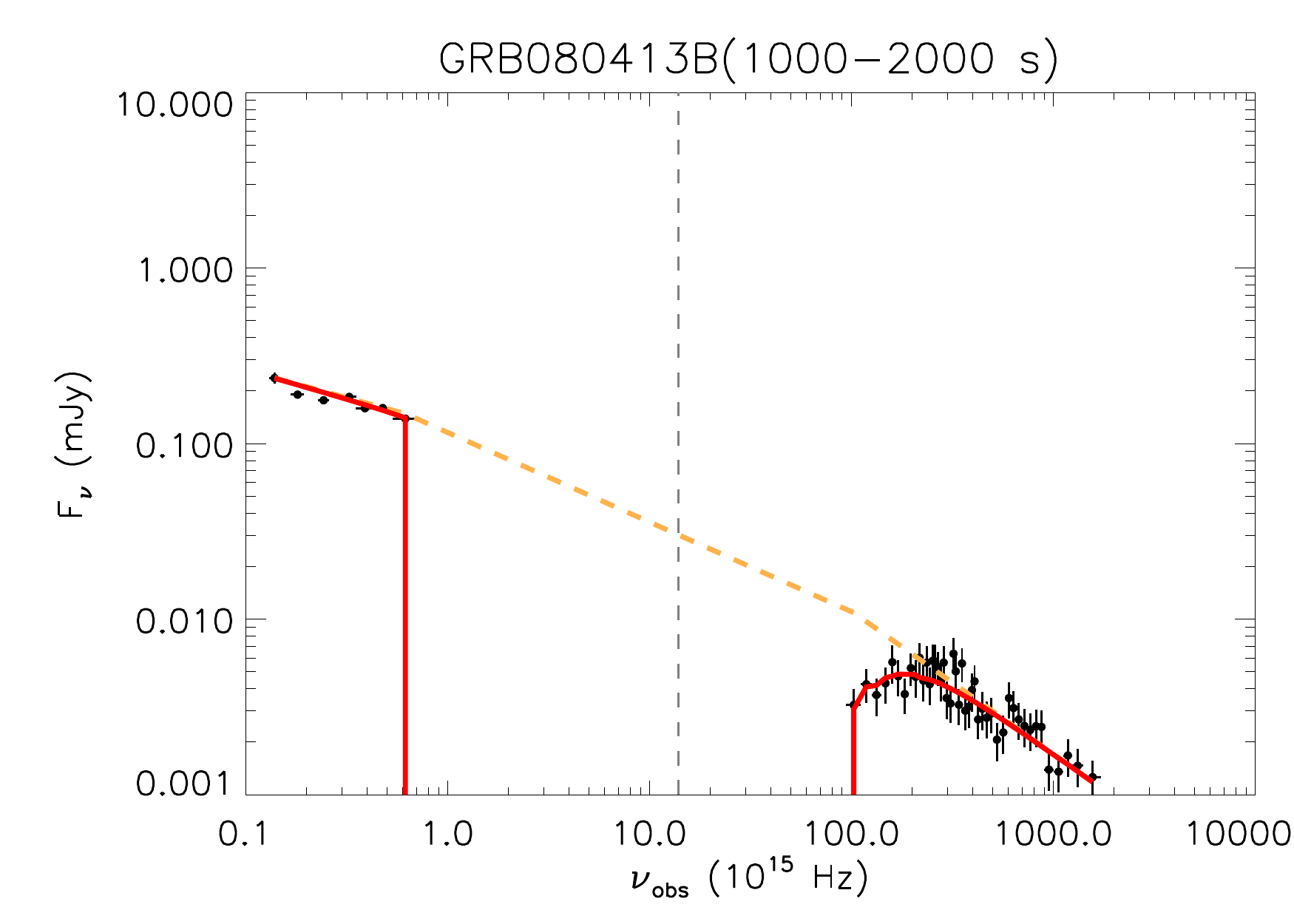}
\includegraphics[width=0.3 \hsize,clip]{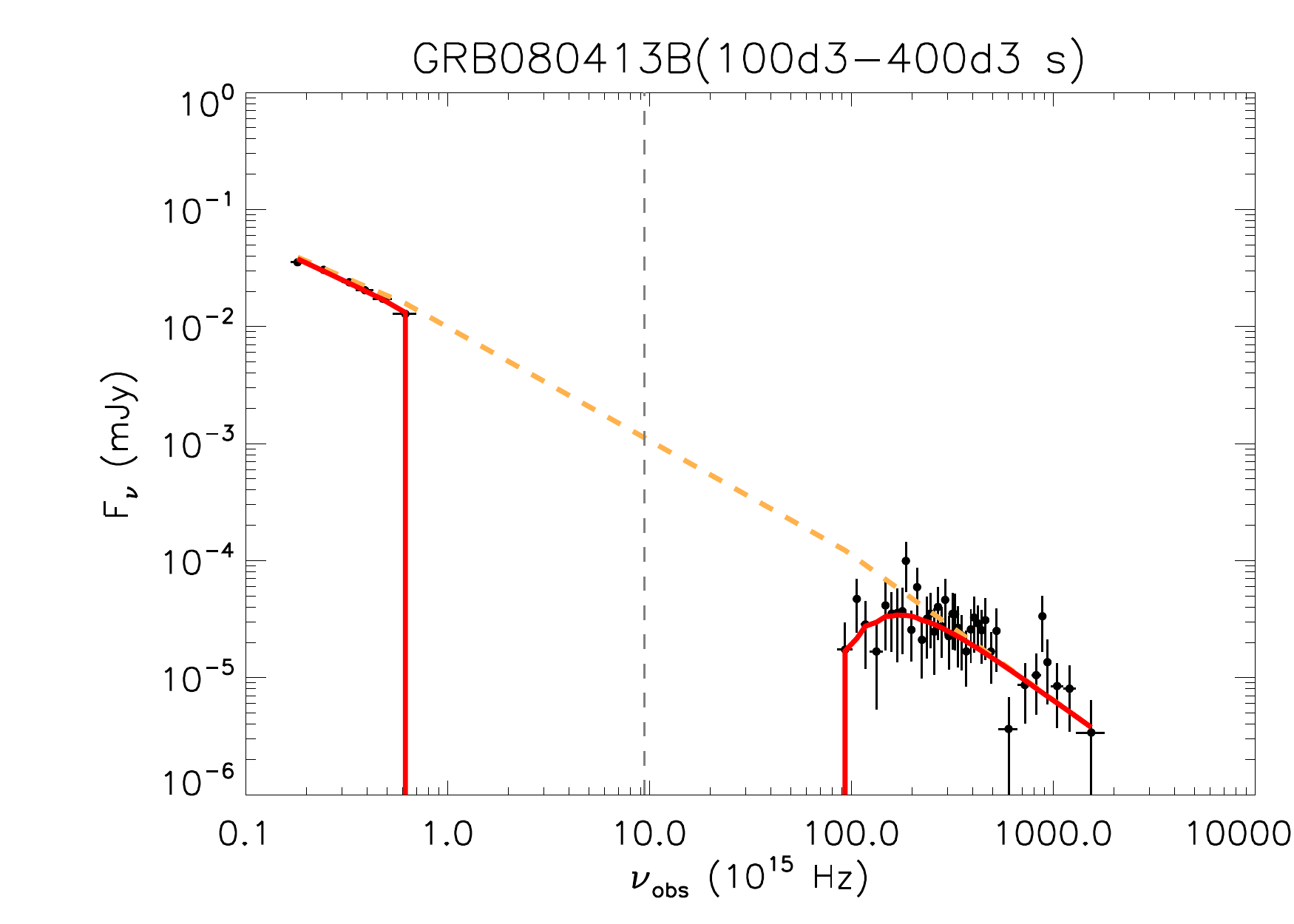}
\includegraphics[width=0.3 \hsize,clip]{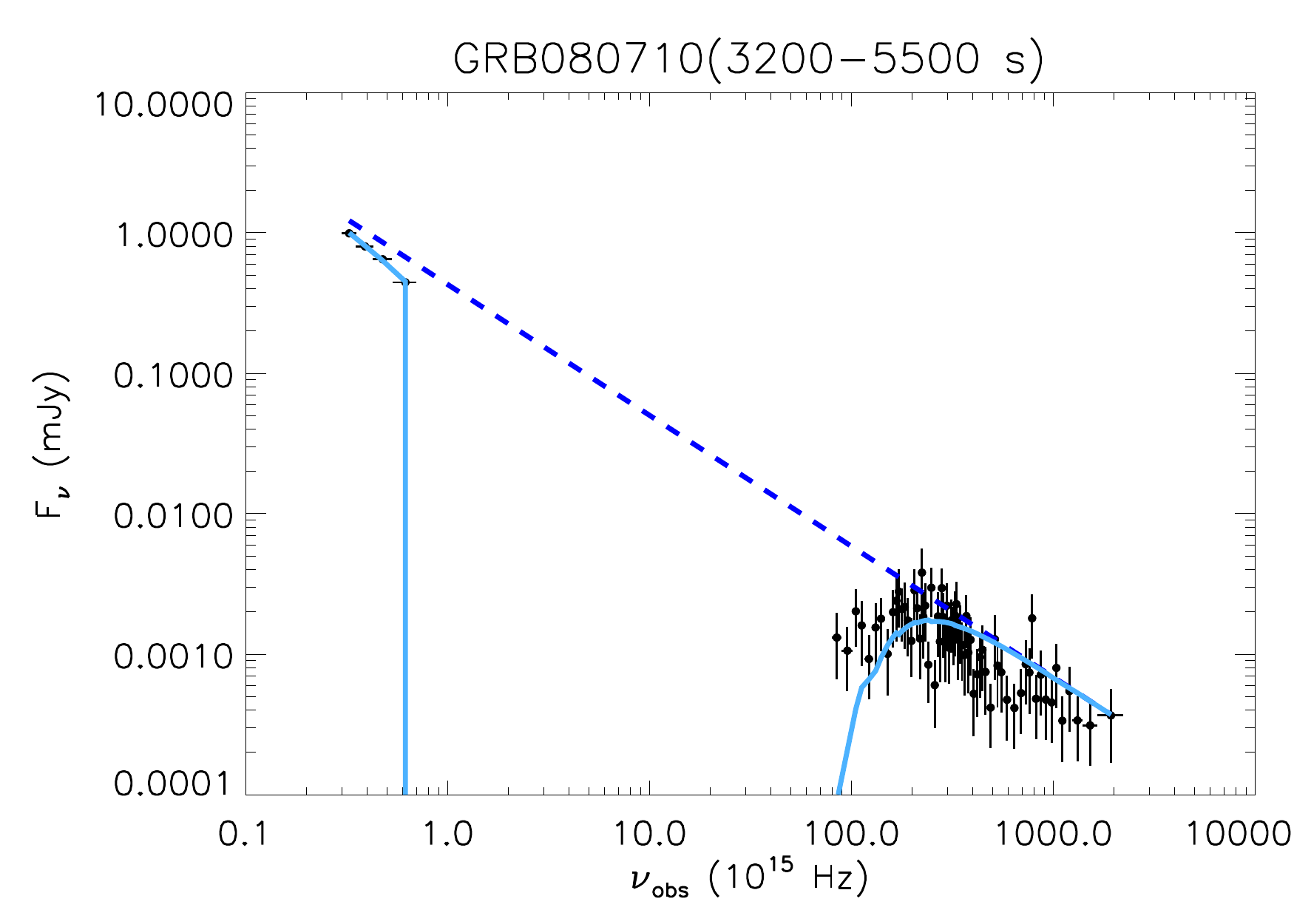}\\
\includegraphics[width=0.3 \hsize,clip]{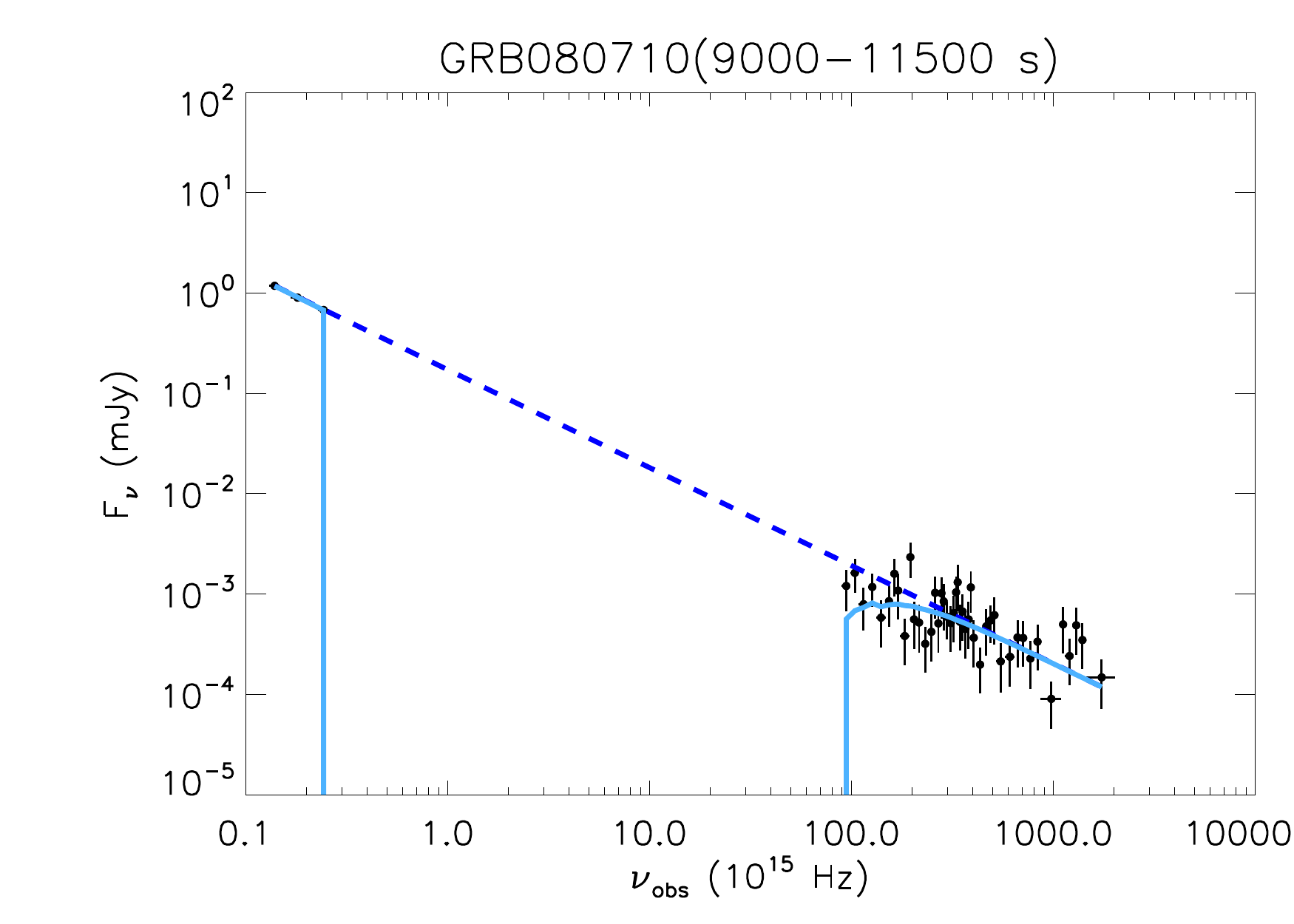}
\includegraphics[width=0.3\hsize,clip]{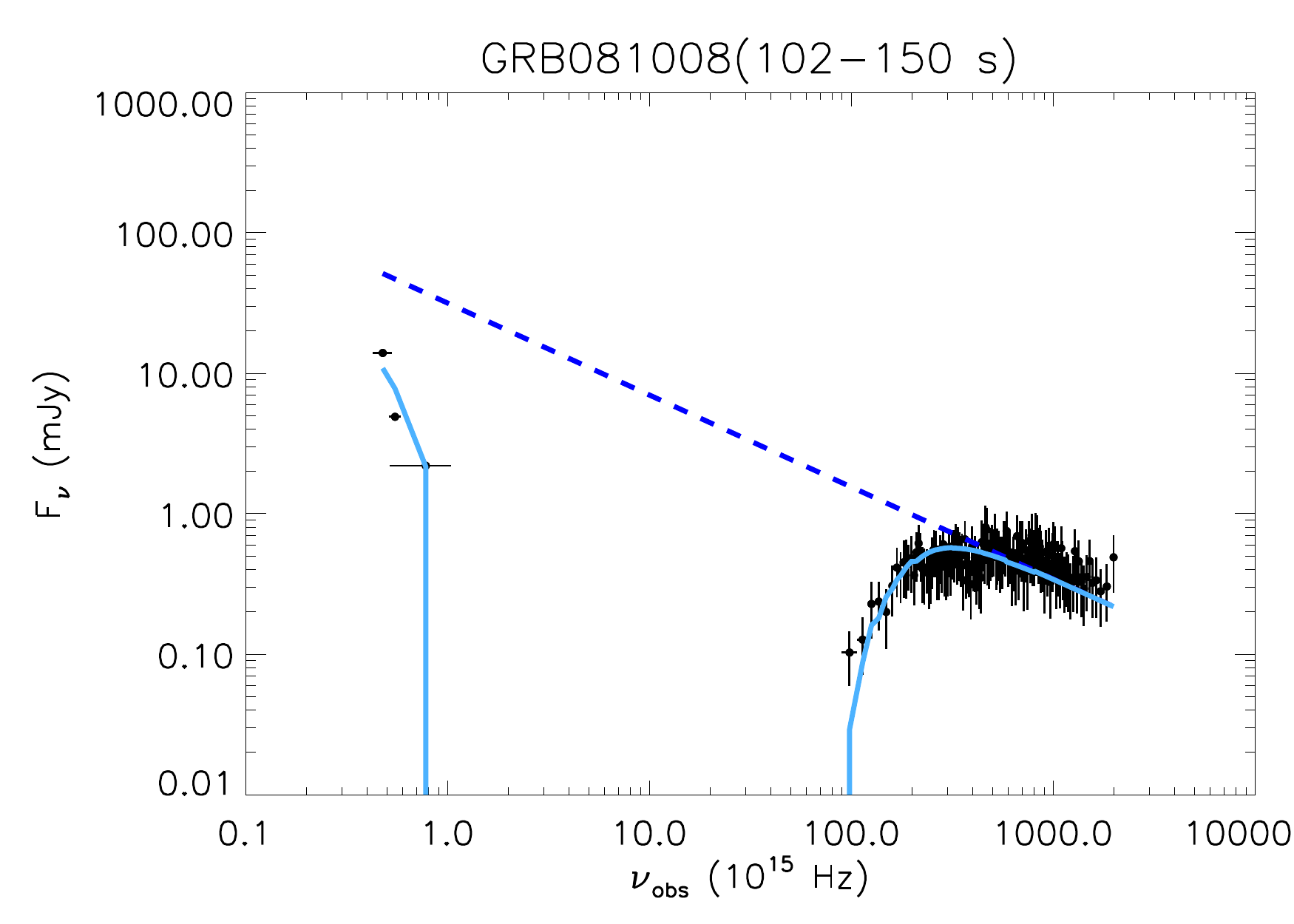}
\includegraphics[width=0.3 \hsize,clip]{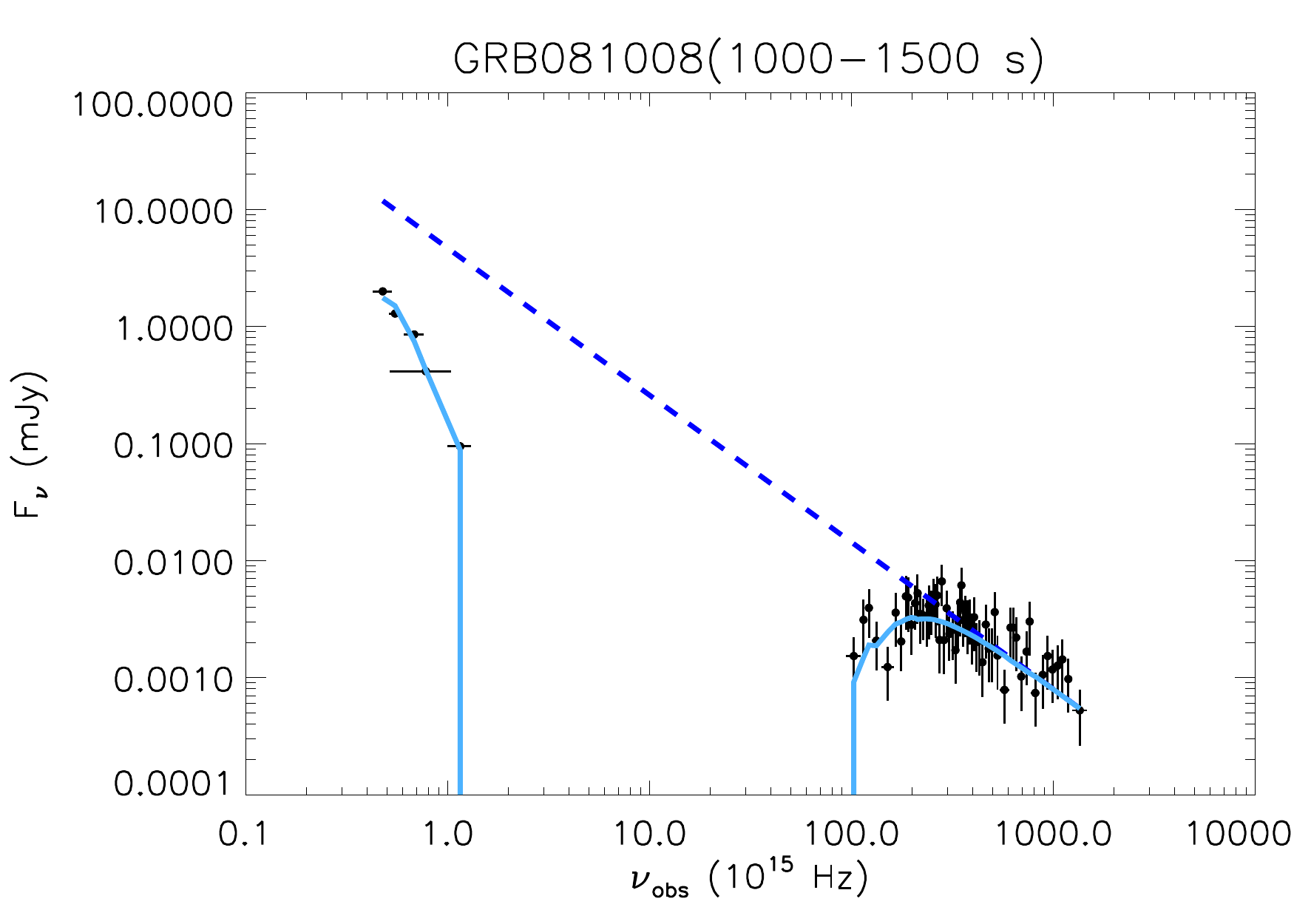}
\caption{\small{Optical/X-ray SEDs for GRBs belonging to Group B.  color-coding as in Figure~\ref{sed1}.}}\label{sed4} 
\end{figure}

\begin{figure}
\includegraphics[width=0.3\hsize,clip]{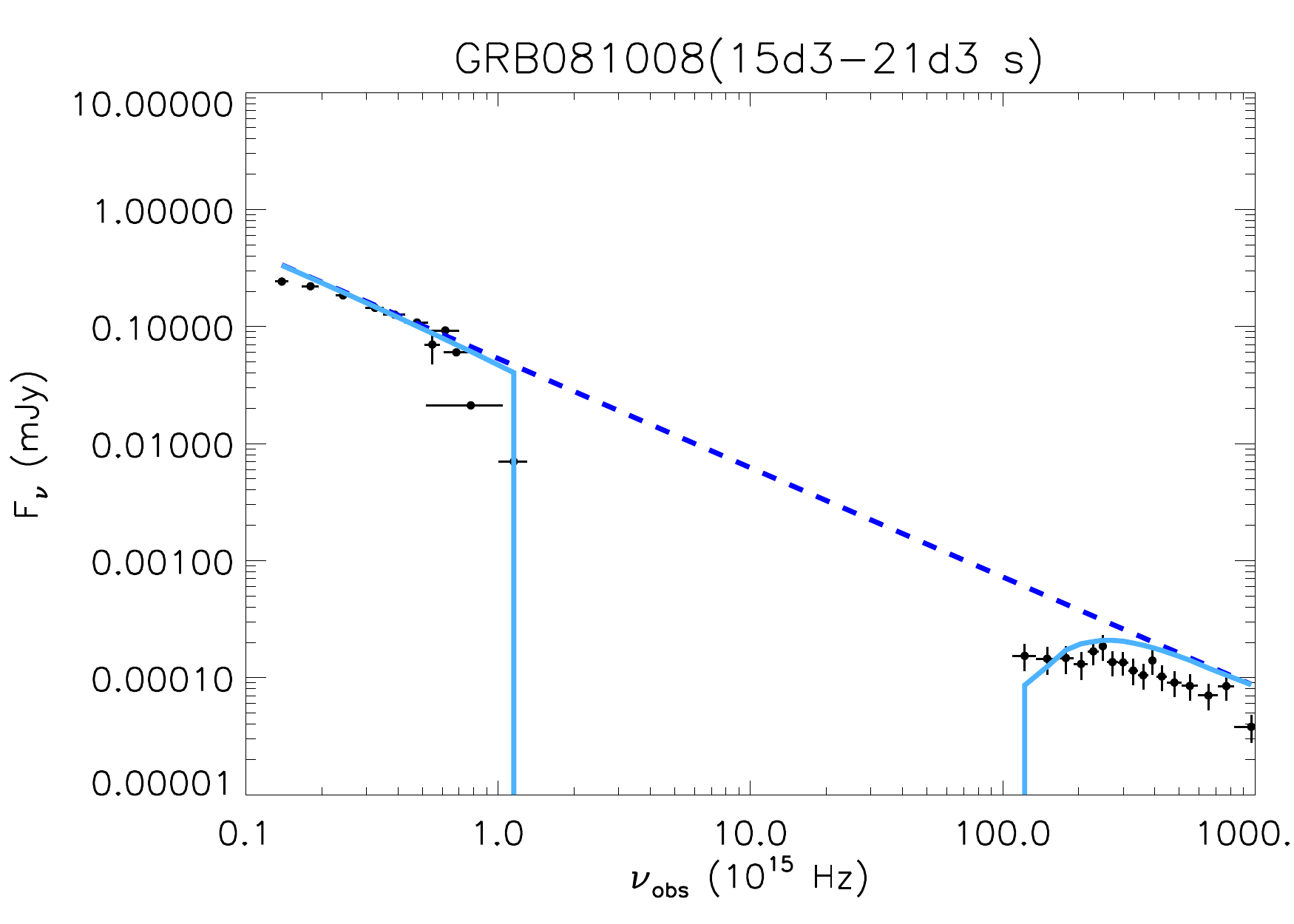}
\includegraphics[width=0.3 \hsize,clip]{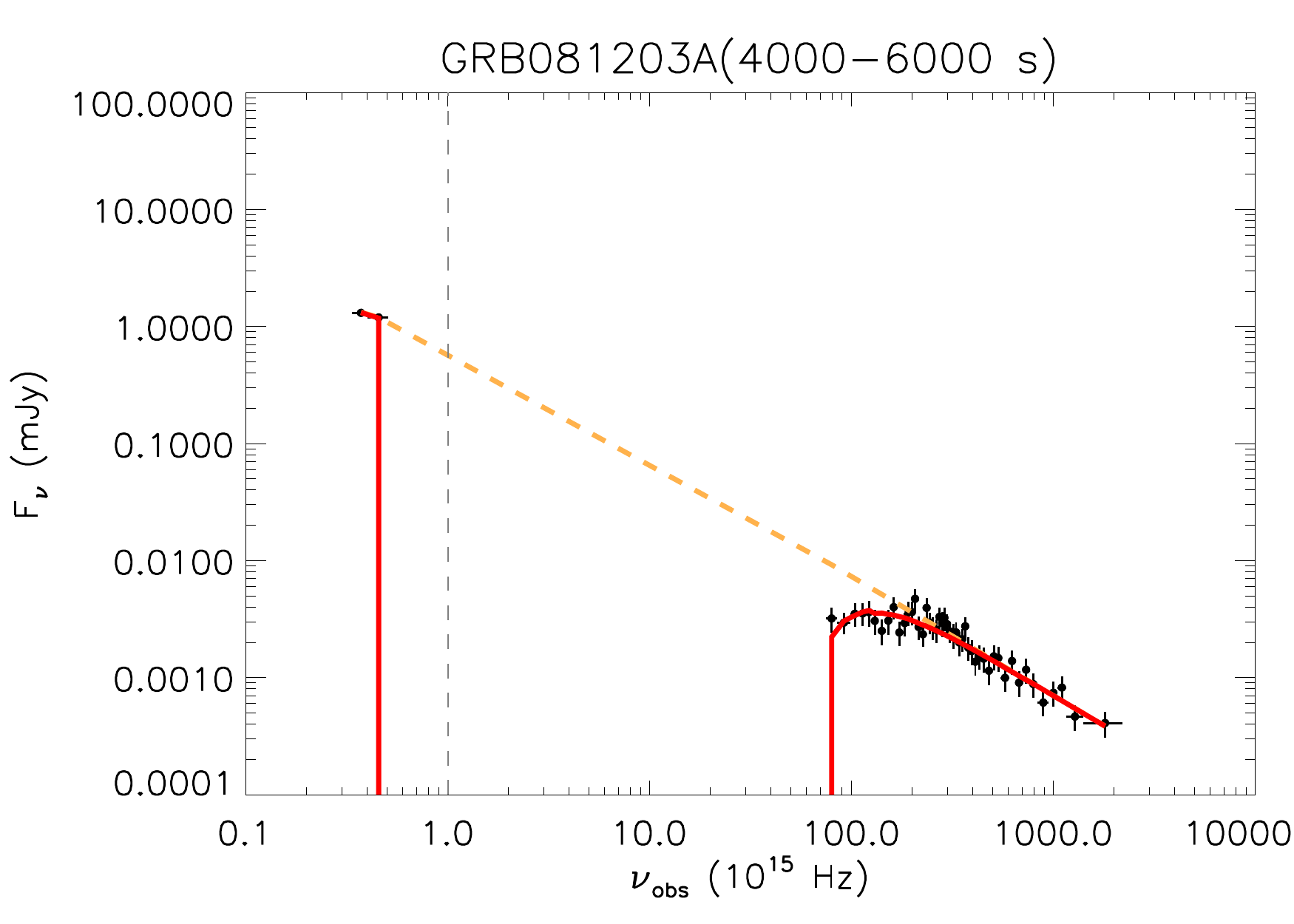}
\includegraphics[width=0.3 \hsize,clip]{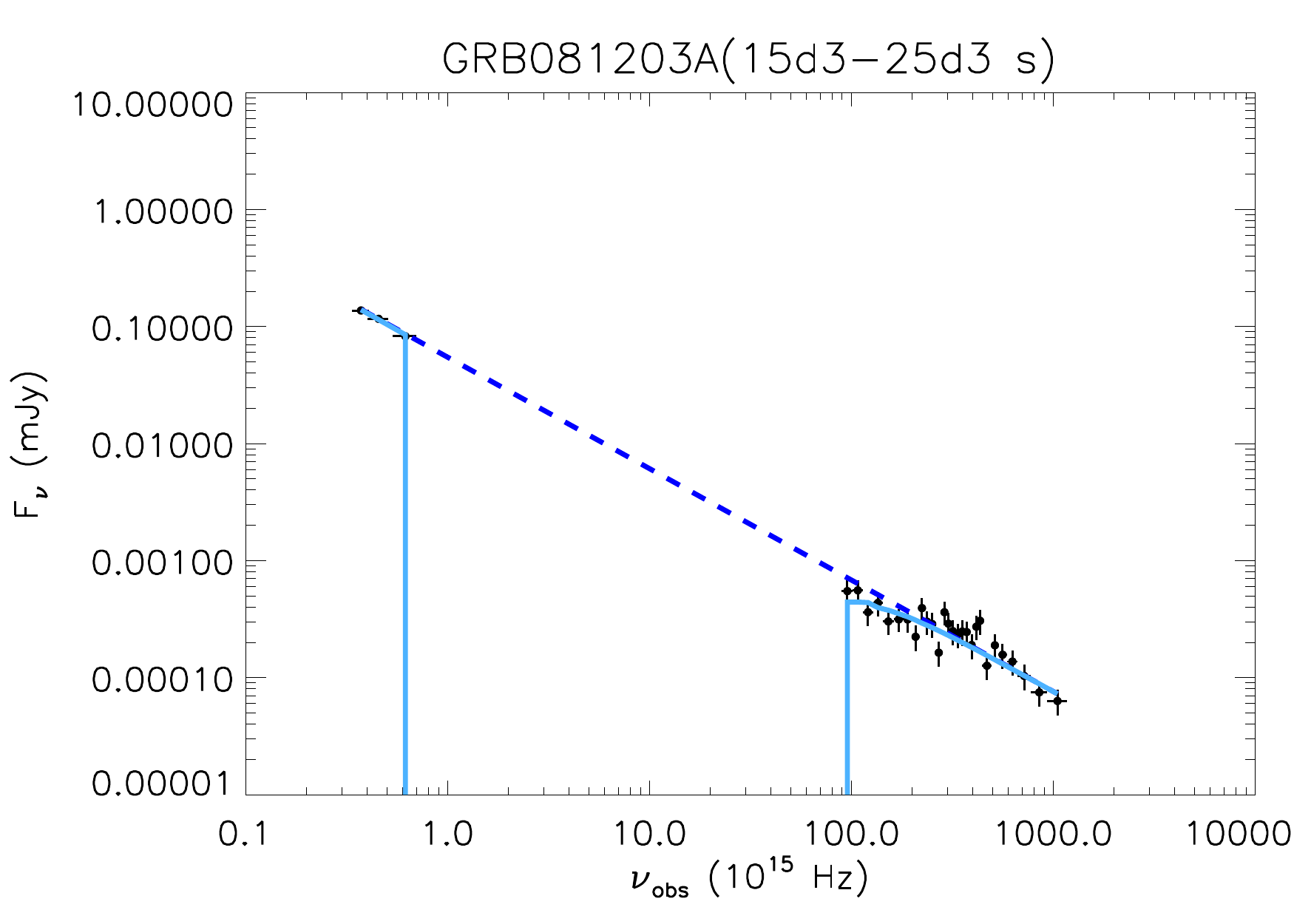}\\
\includegraphics[width=0.3\hsize,clip]{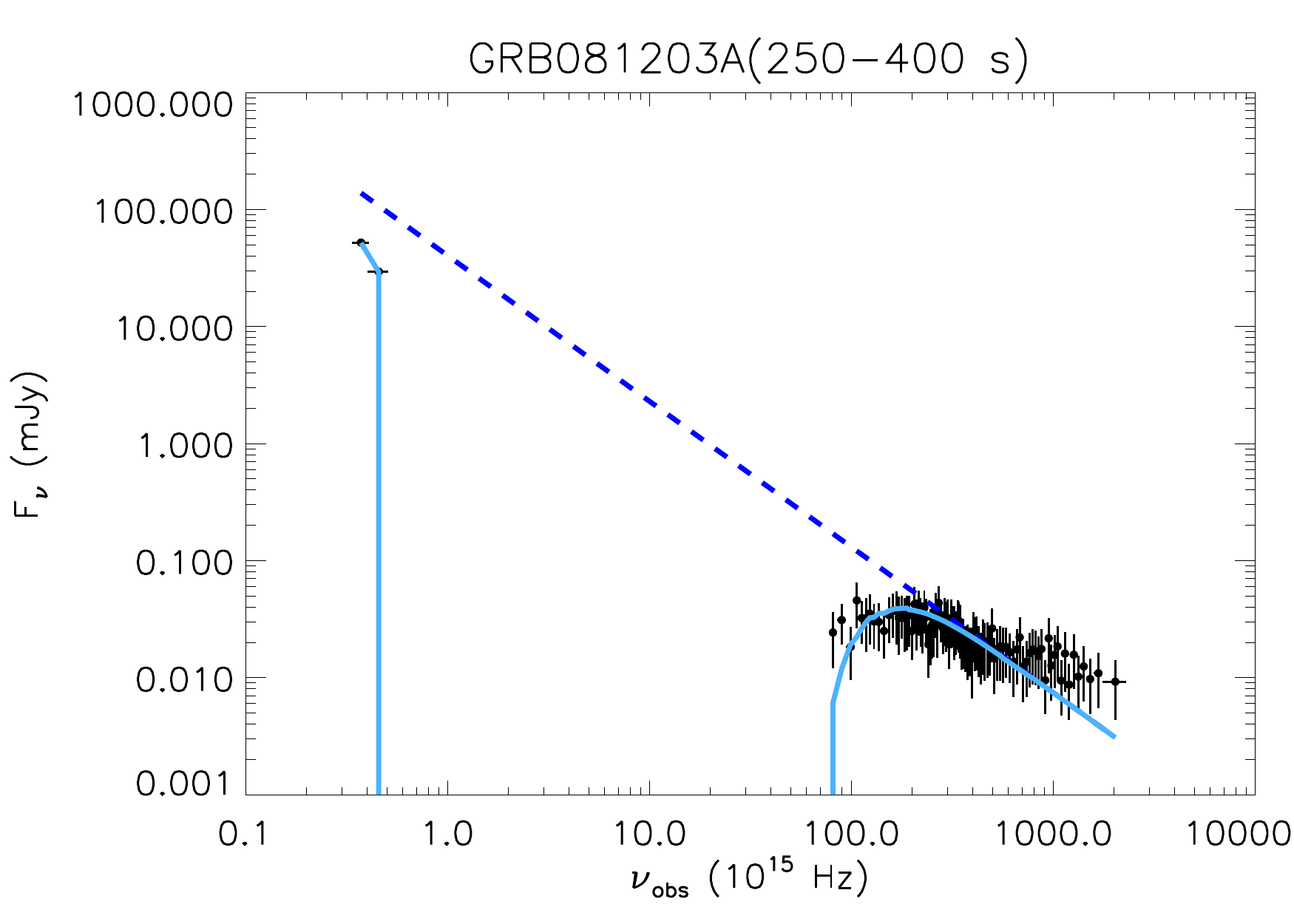}
\includegraphics[width=0.3 \hsize,clip]{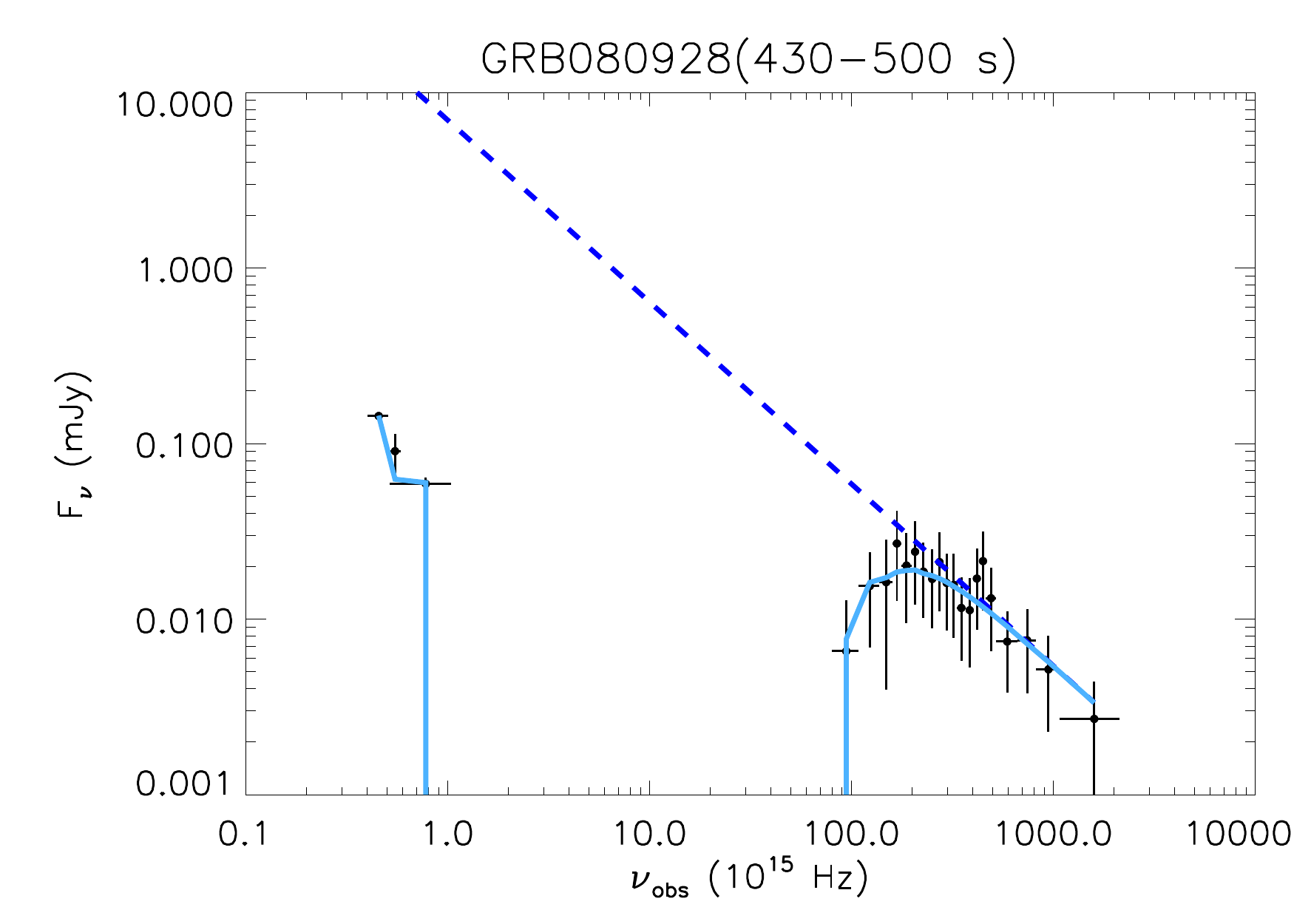}
\includegraphics[width=0.3 \hsize,clip]{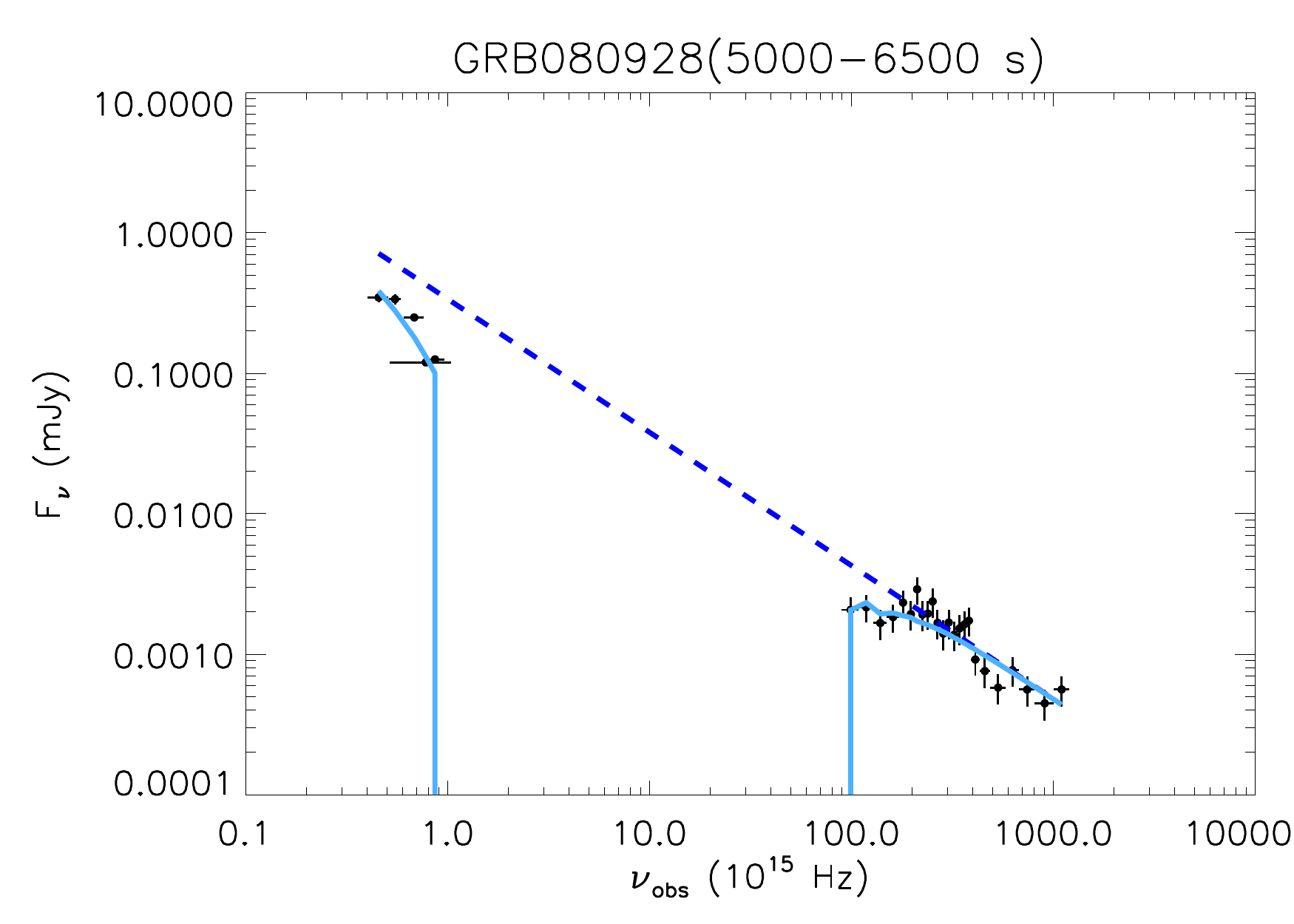}\\
\includegraphics[width=0.3 \hsize,clip]{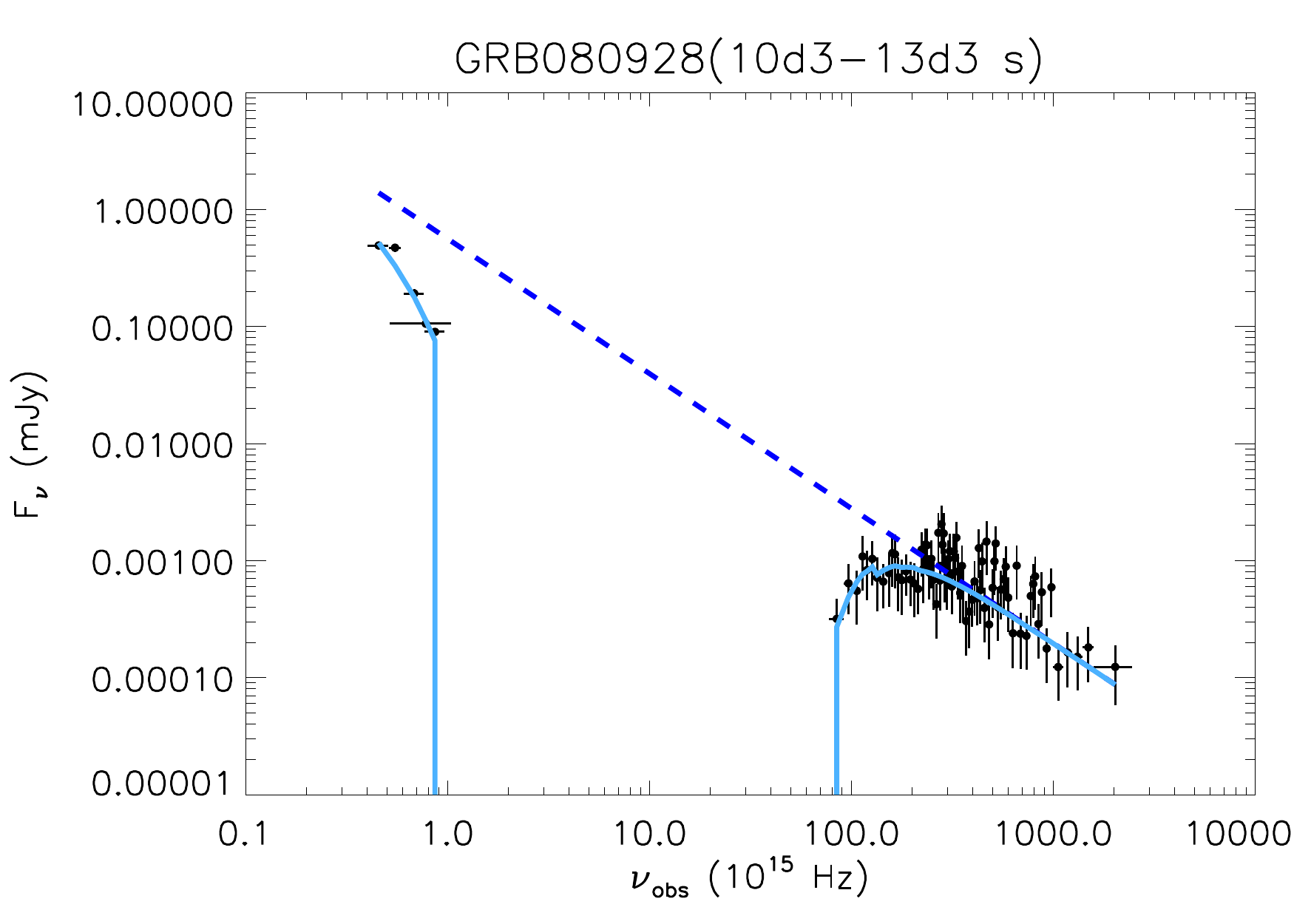}
\includegraphics[width=0.3 \hsize,clip]{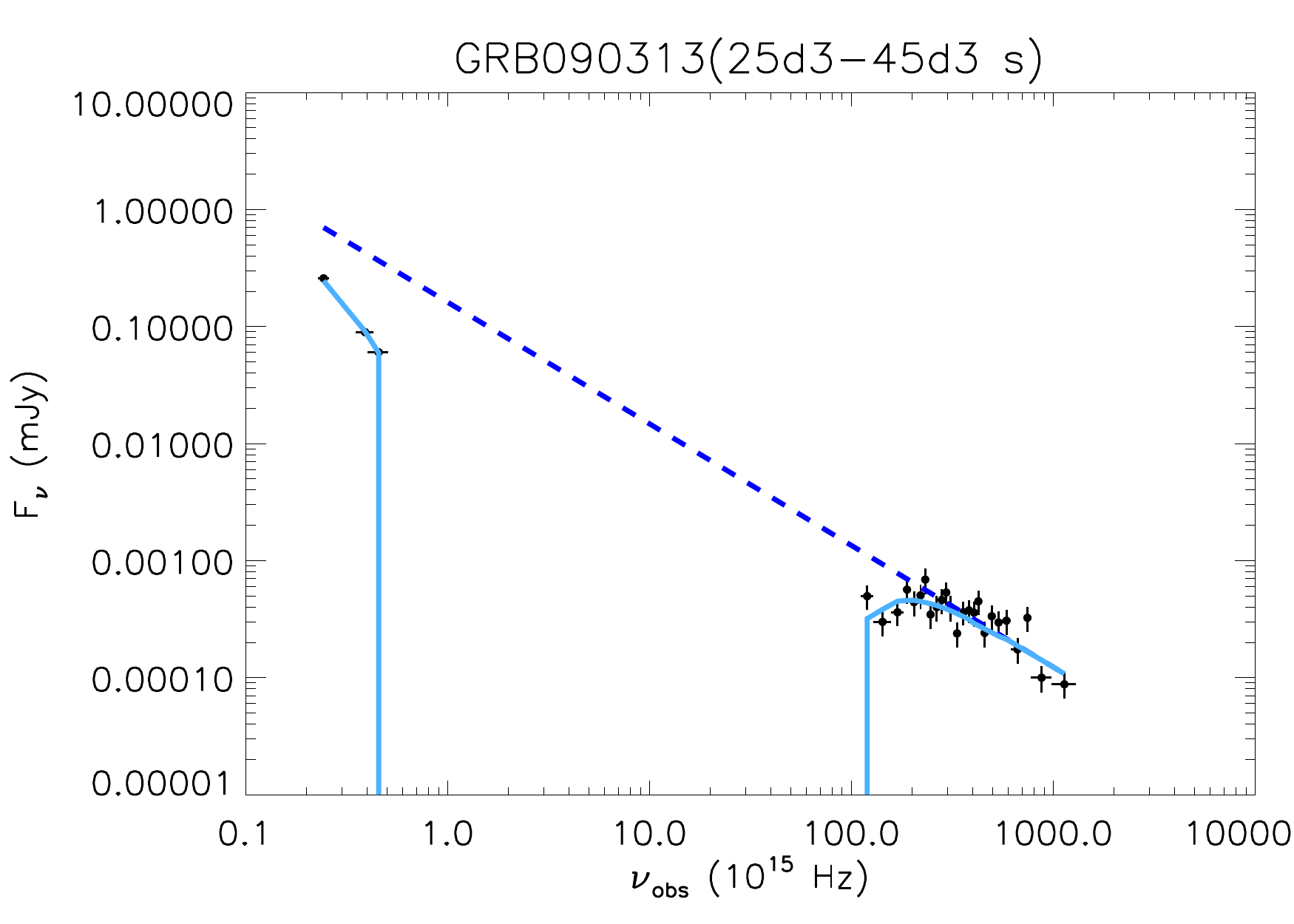}
\includegraphics[width=0.3\hsize,clip]{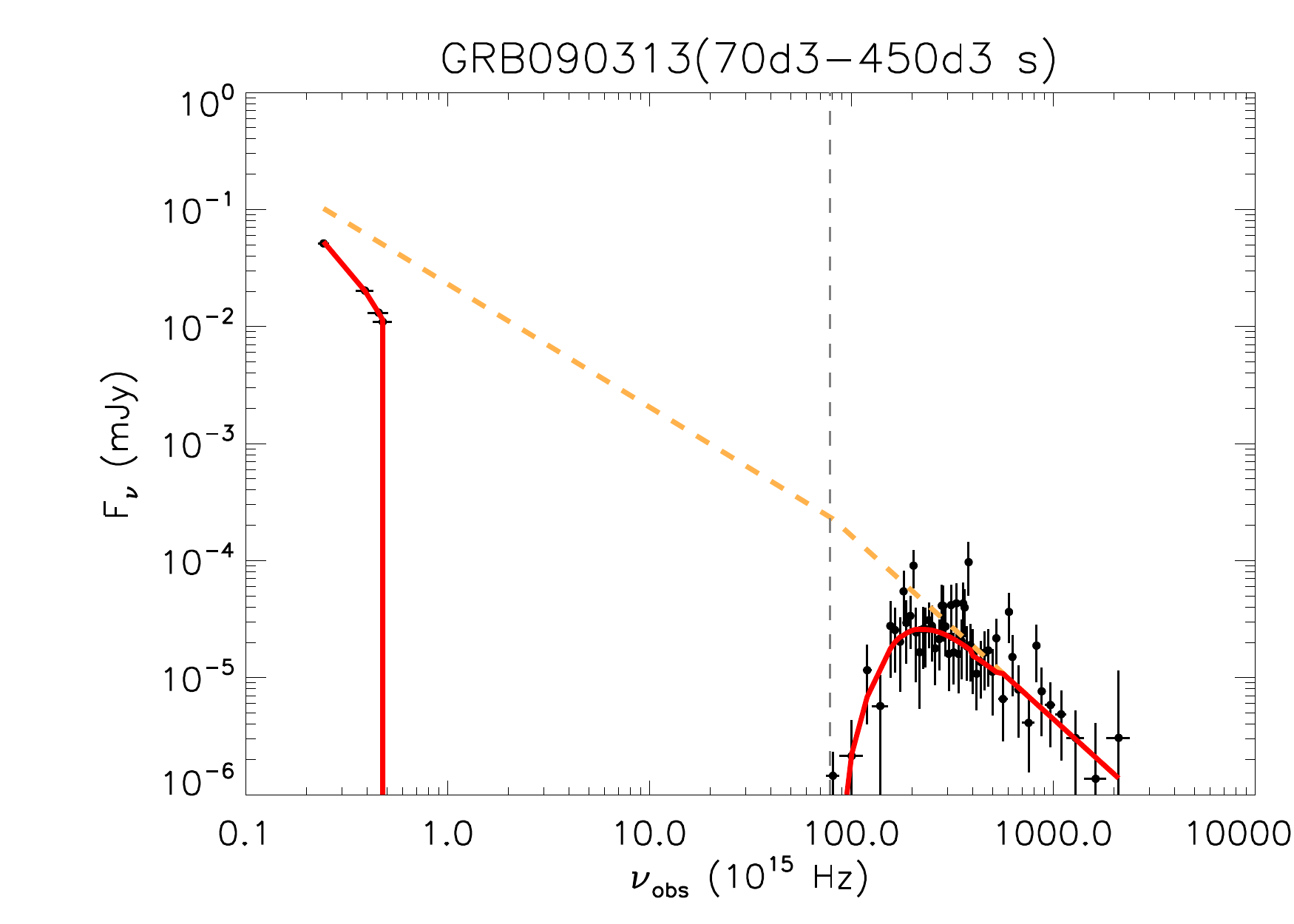}\\
\includegraphics[width=0.3 \hsize,clip]{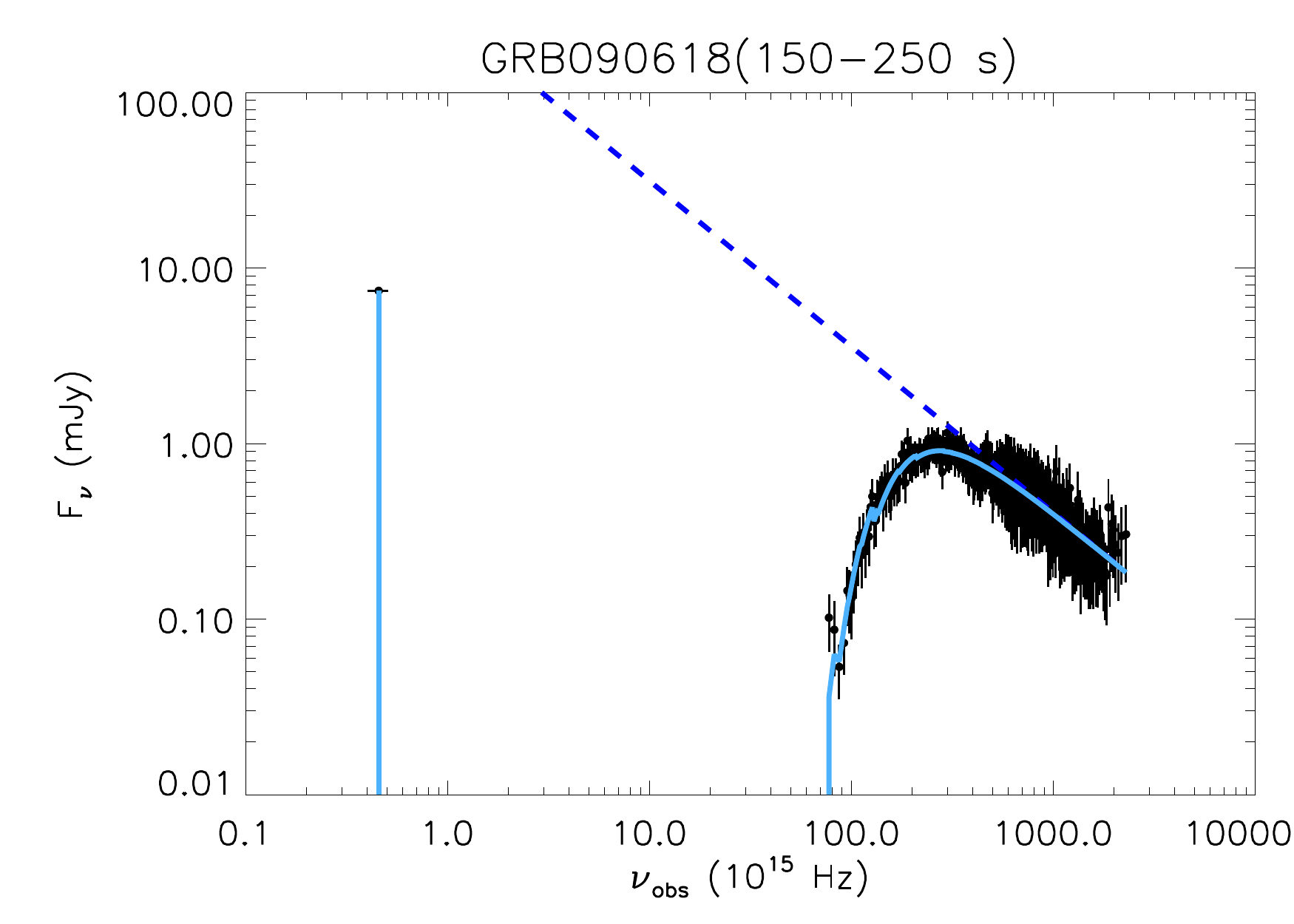}
\includegraphics[width=0.3 \hsize,clip]{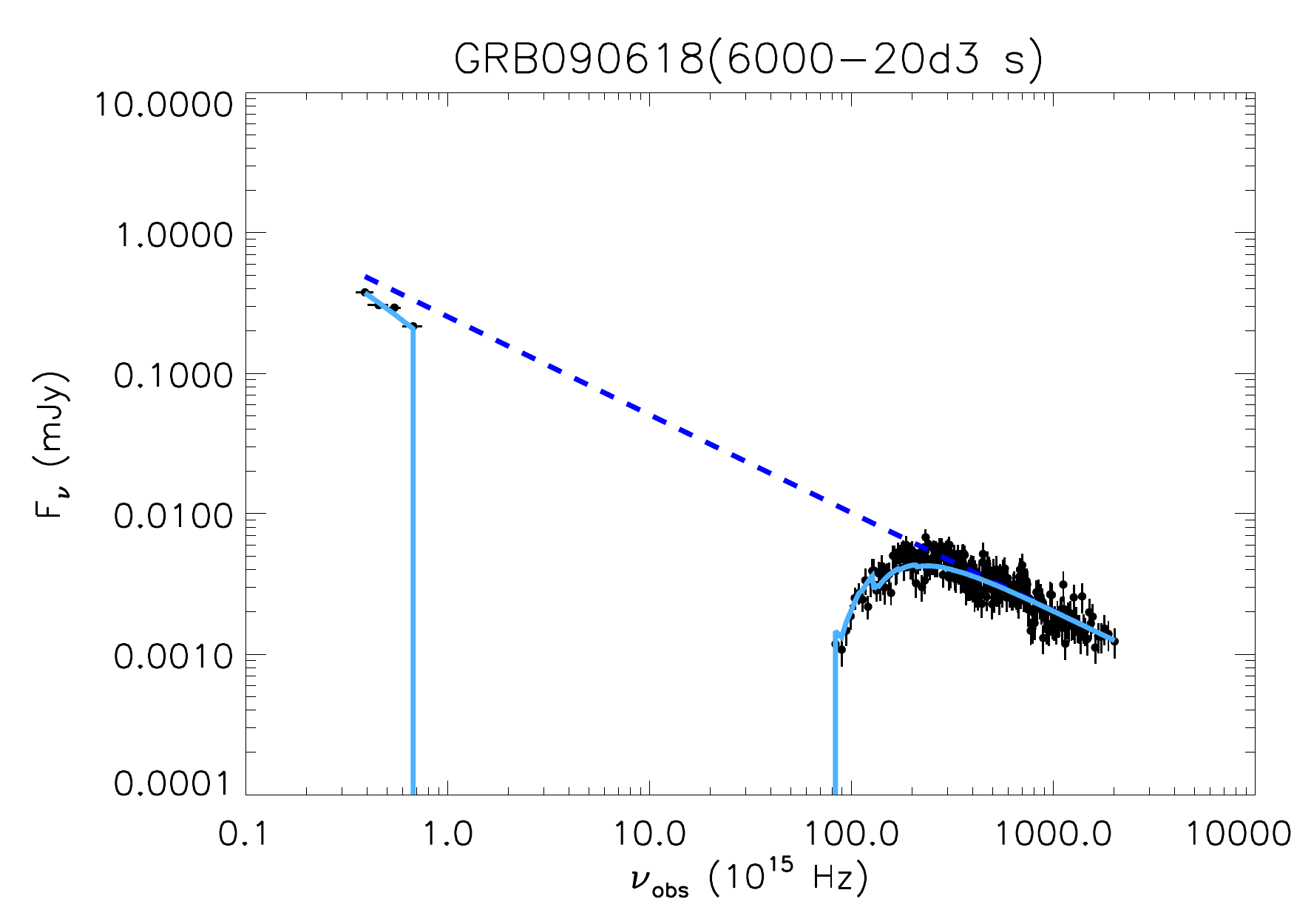}
\includegraphics[width=0.3 \hsize,clip]{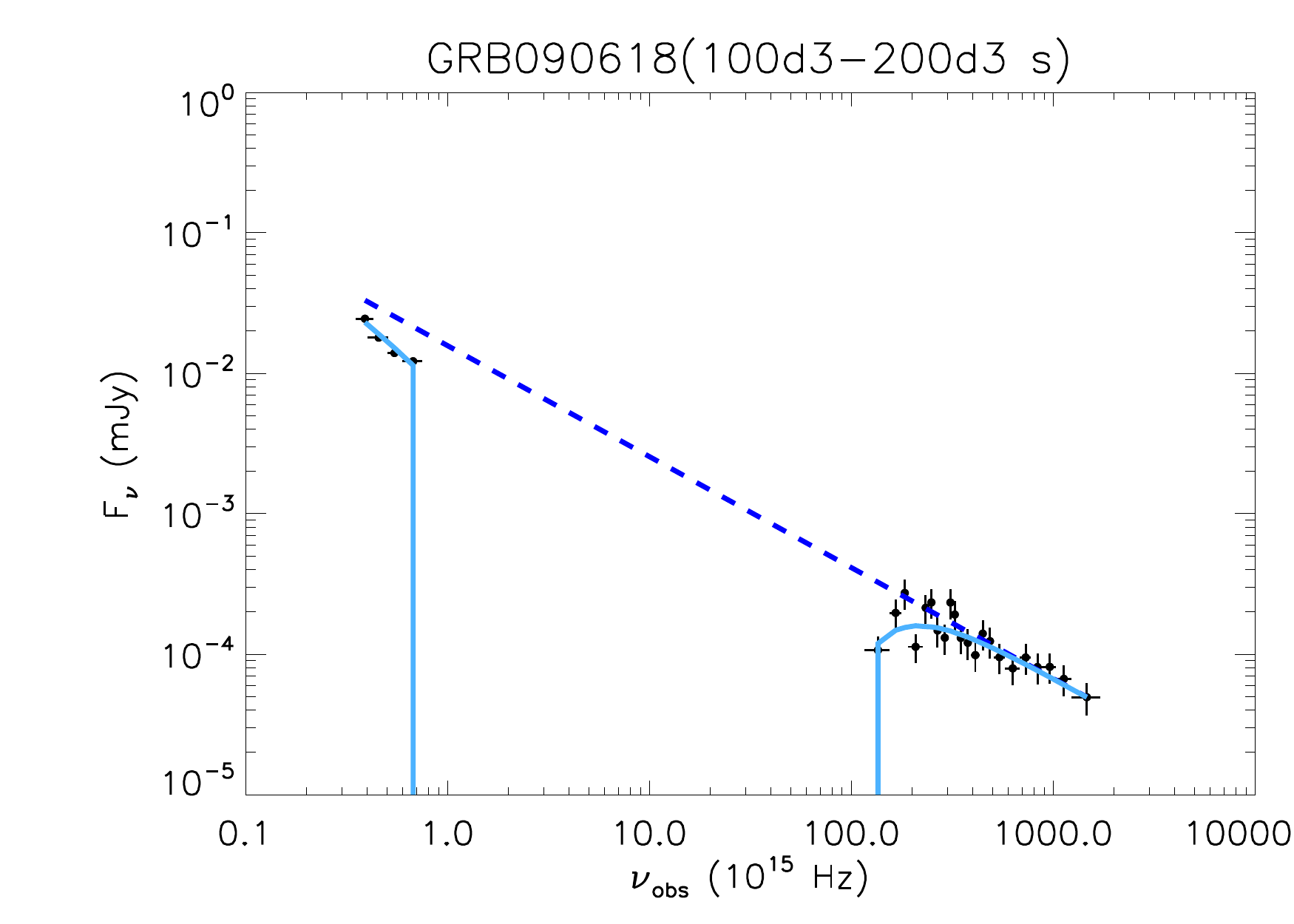}\\
\includegraphics[width=0.3\hsize,clip]{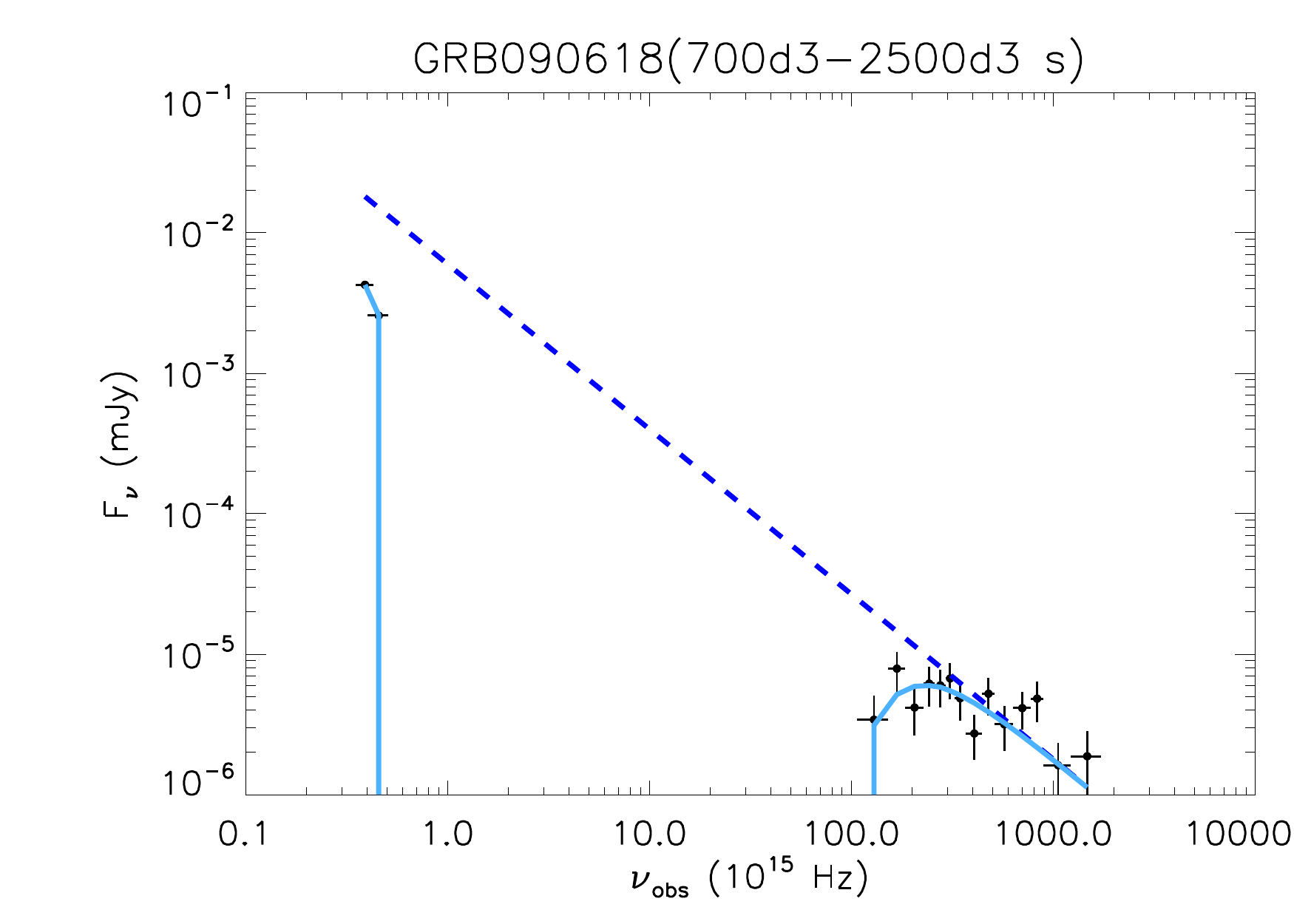}
\includegraphics[width=0.3 \hsize,clip]{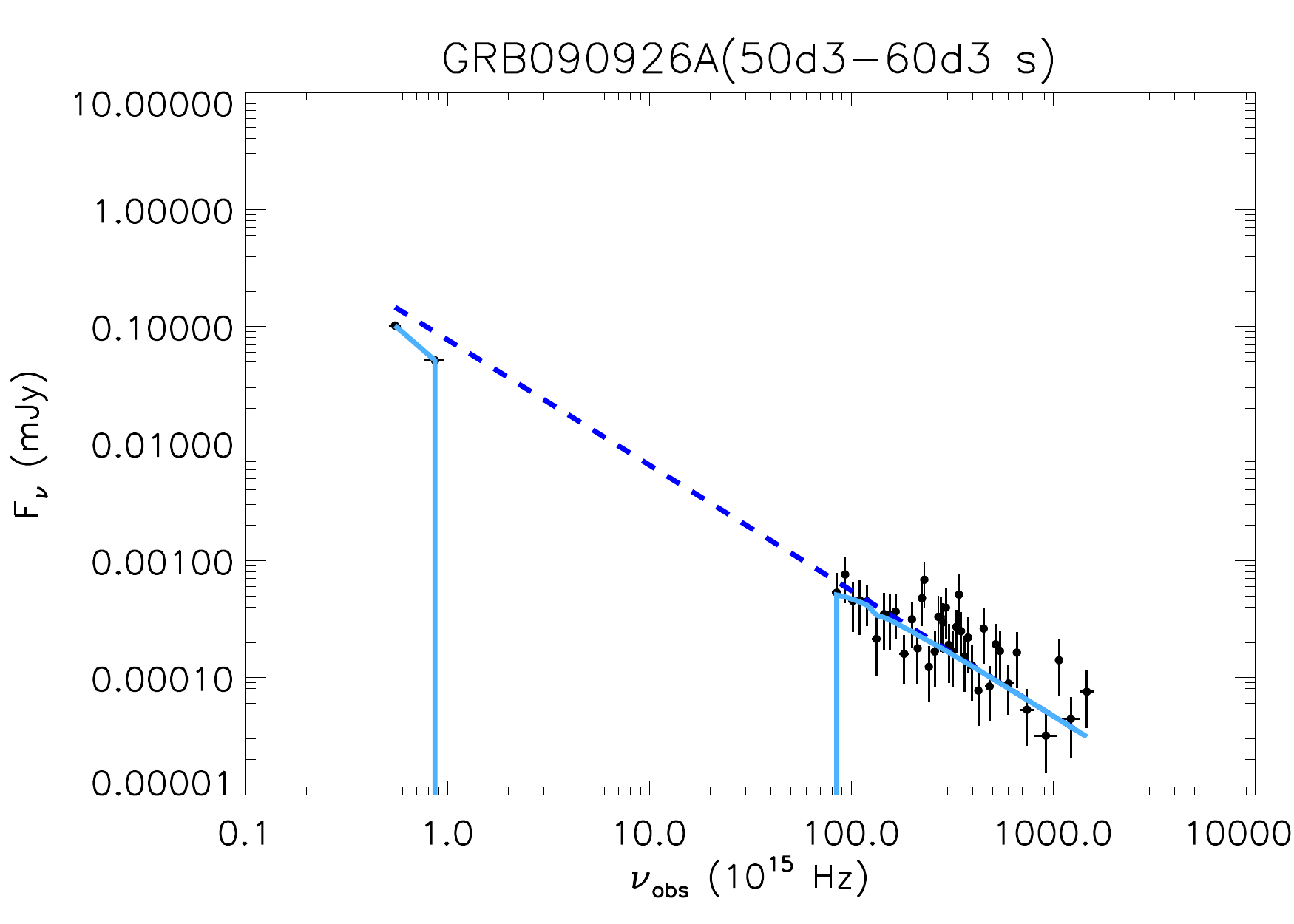}
\includegraphics[width=0.3 \hsize,clip]{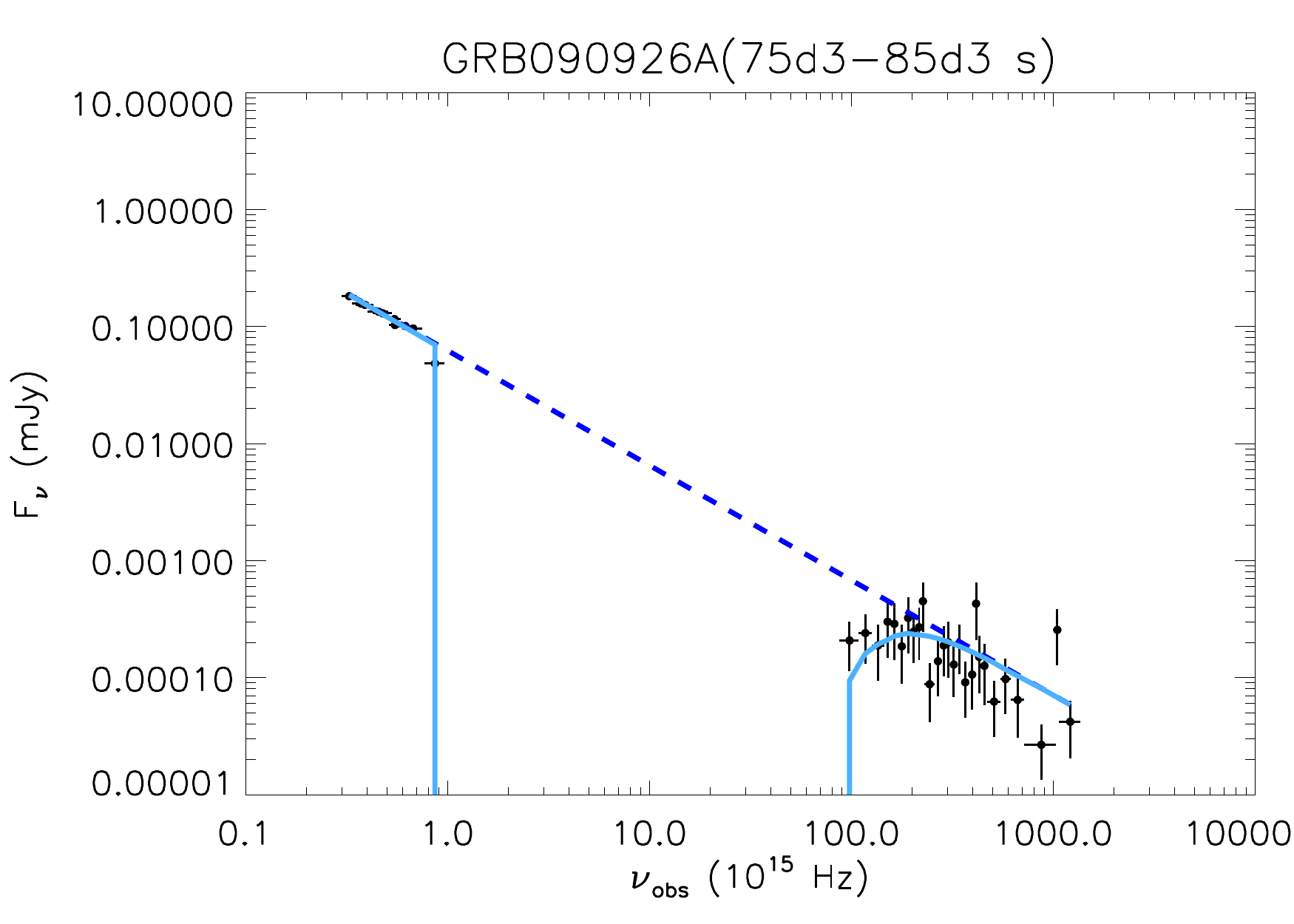}\\
\includegraphics[width=0.3 \hsize,clip]{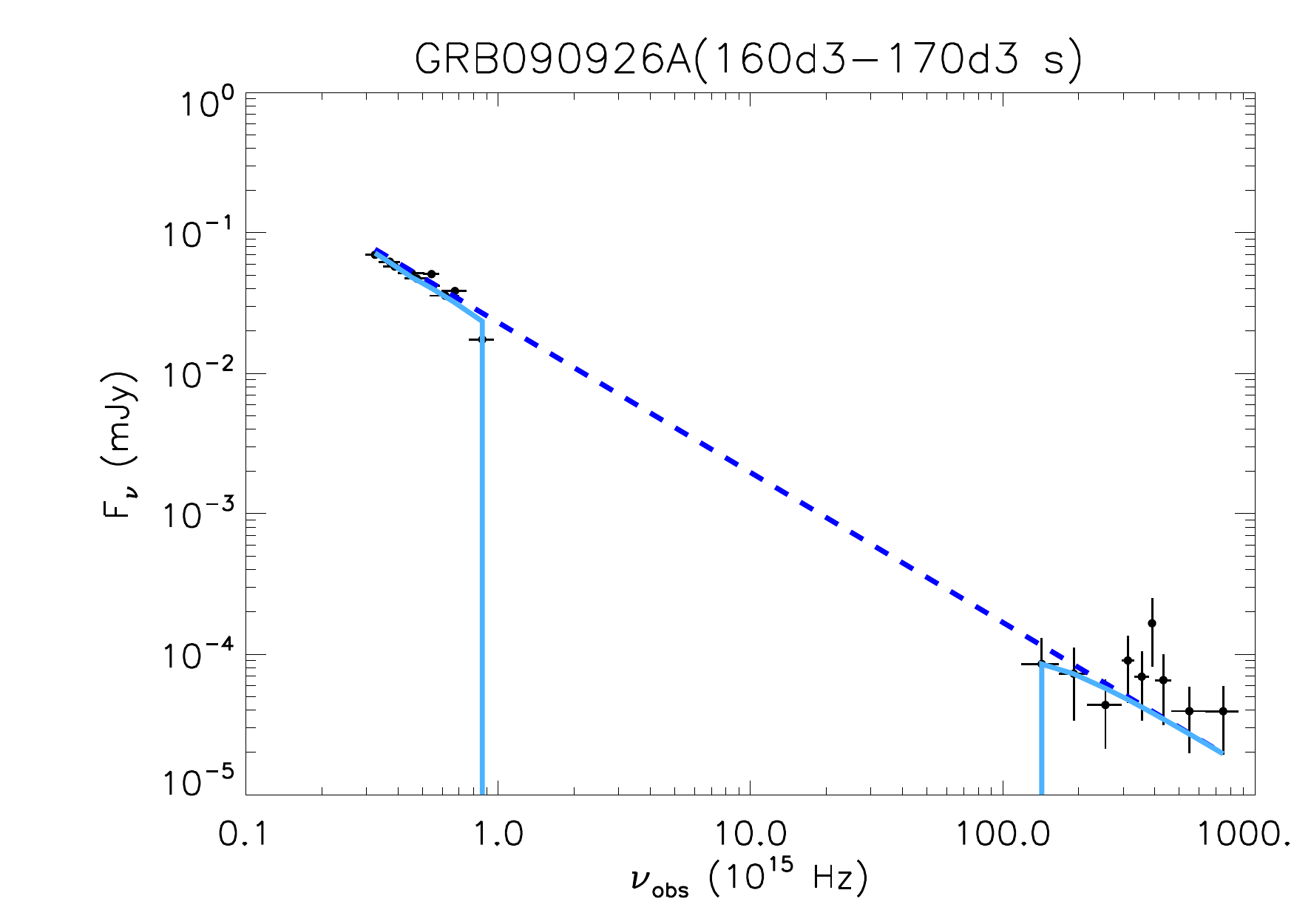}
\includegraphics[width=0.3 \hsize,clip]{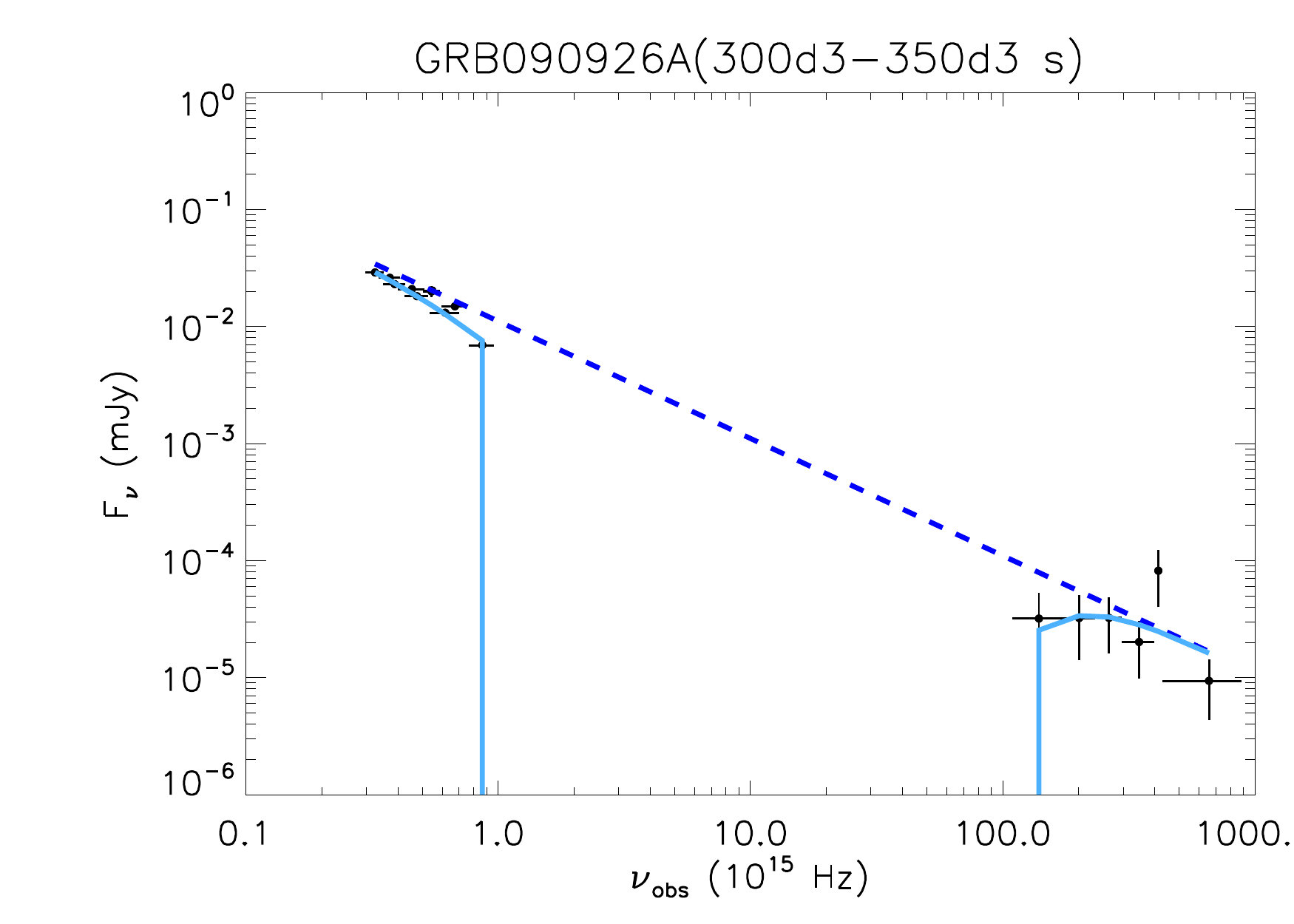}
\includegraphics[width=0.3 \hsize,clip]{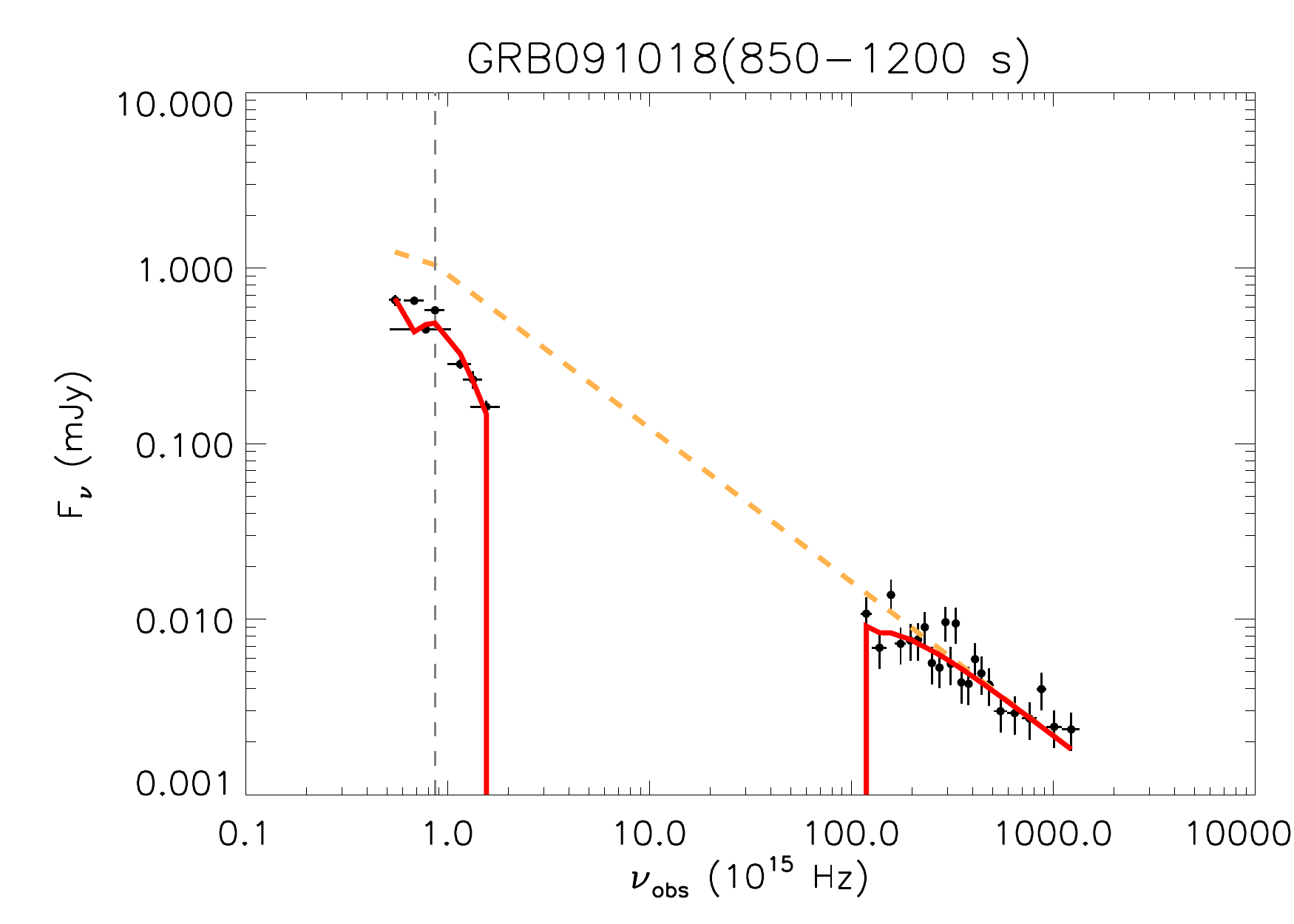}
\caption{\small{Optical/X-ray SEDs for GRBs belonging to Group B. color-coding as in Figure~\ref{sed1}.}}\label{sed5b} 
\end{figure}

\begin{figure}
\includegraphics[width=0.3\hsize,clip]{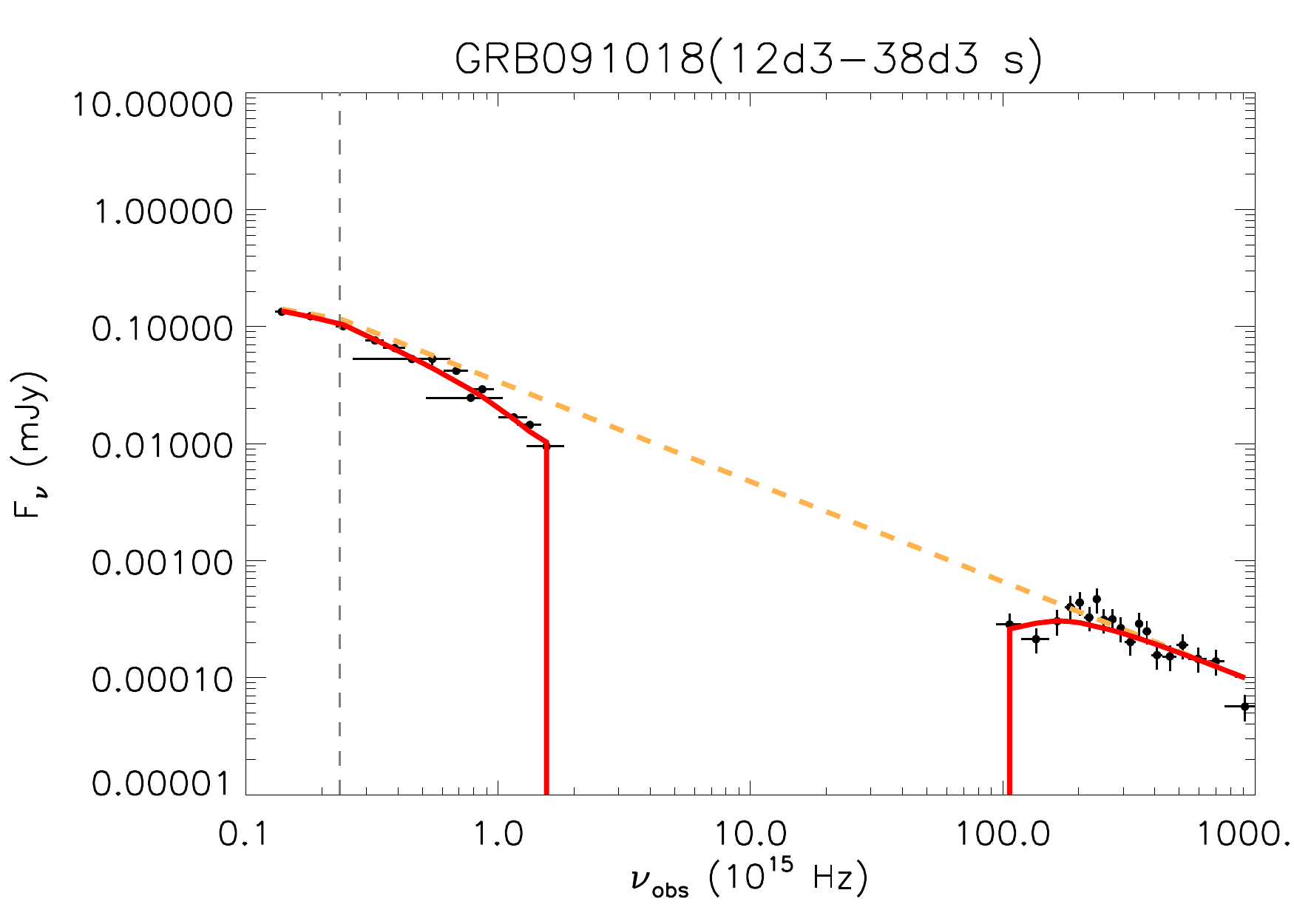}
\caption{\small{Optical/X-ray SEDs for GRBs belonging to Group B. color-coding as in Figure~\ref{sed1}.}}\label{sed5} 
\end{figure}

\begin{figure}
\includegraphics[width=0.3 \hsize,clip]{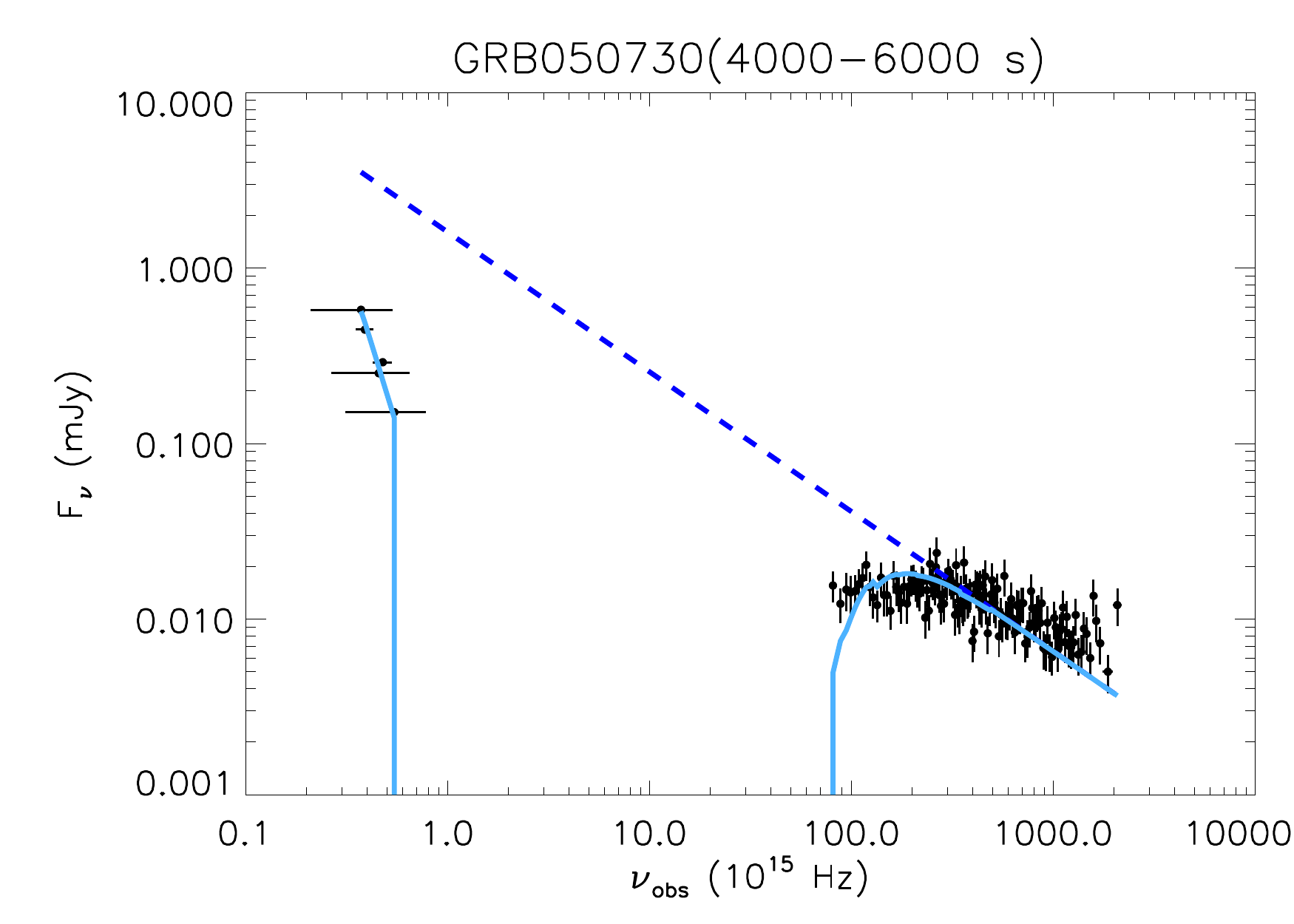}
\includegraphics[width=0.3 \hsize,clip]{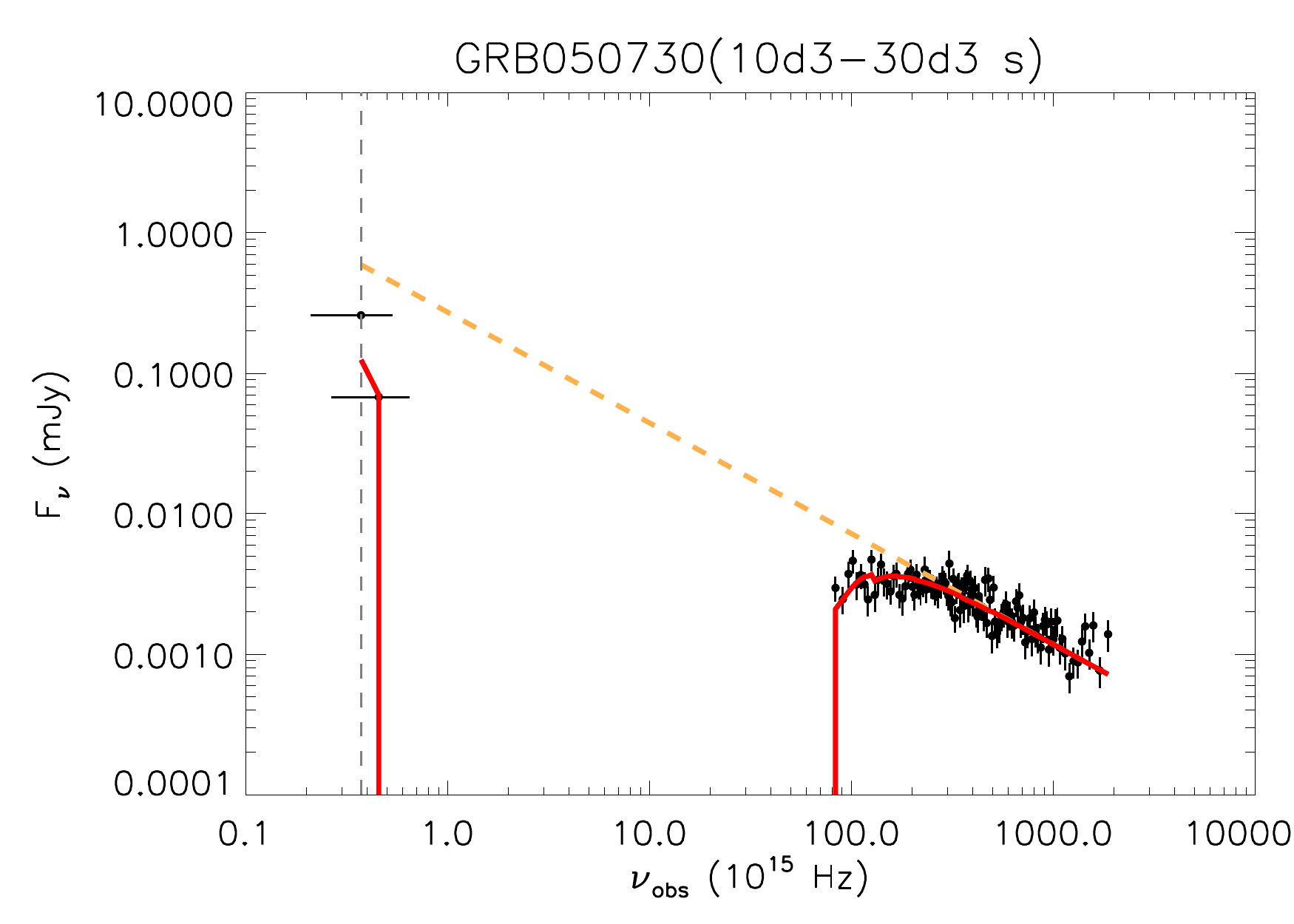}
\includegraphics[width=0.3 \hsize,clip]{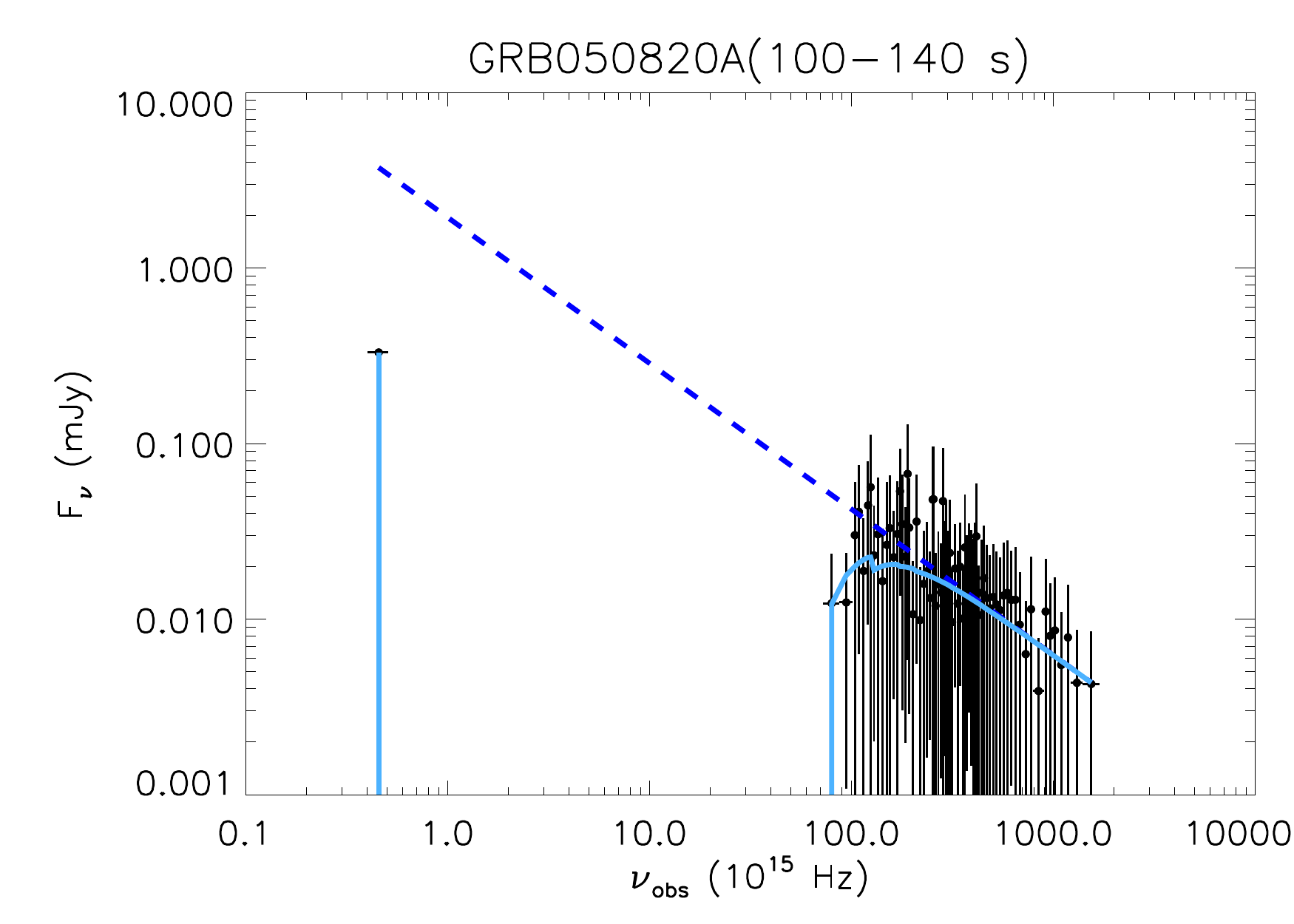}\\
\includegraphics[width=0.3 \hsize,clip]{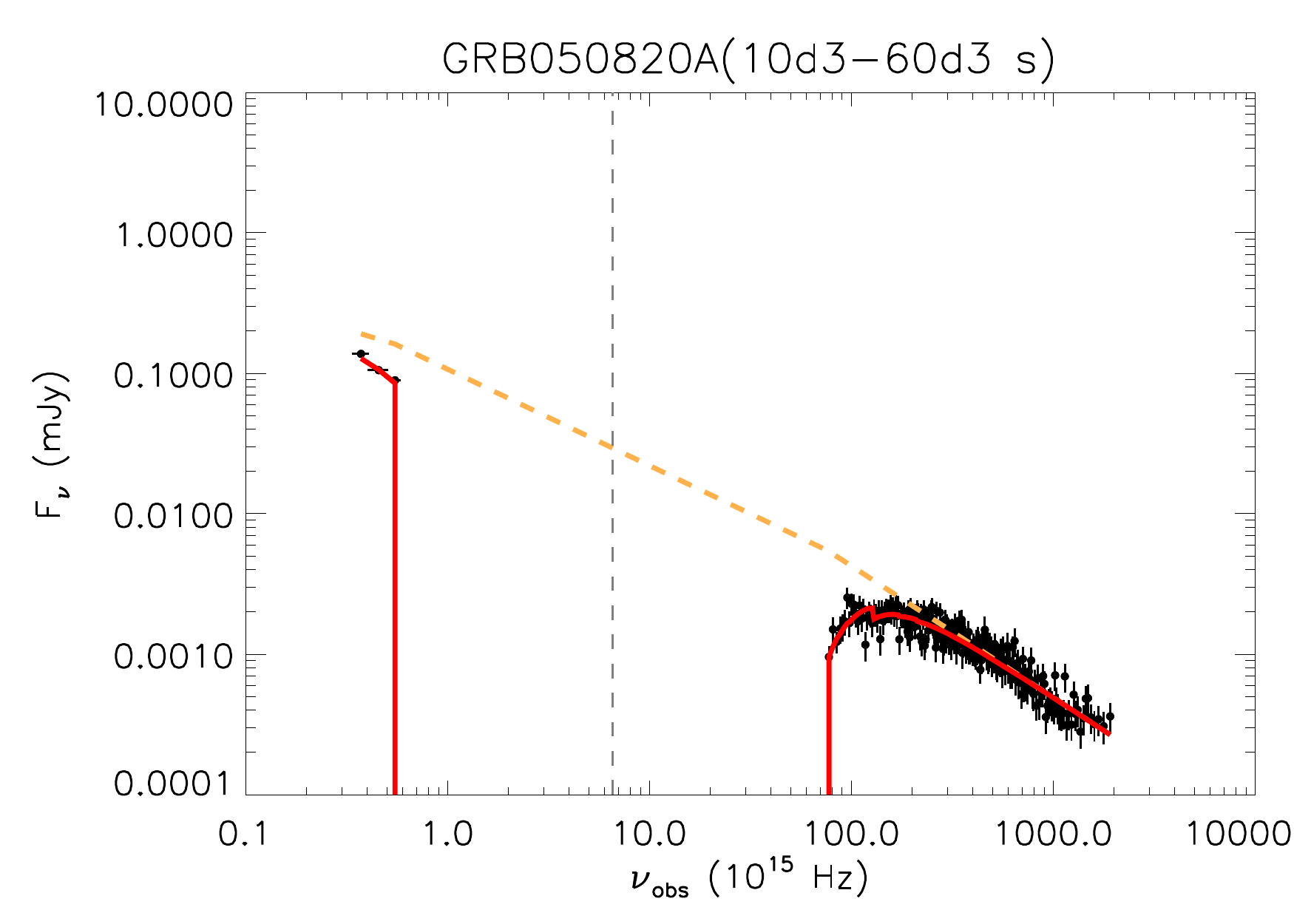}
\includegraphics[width=0.3\hsize,clip]{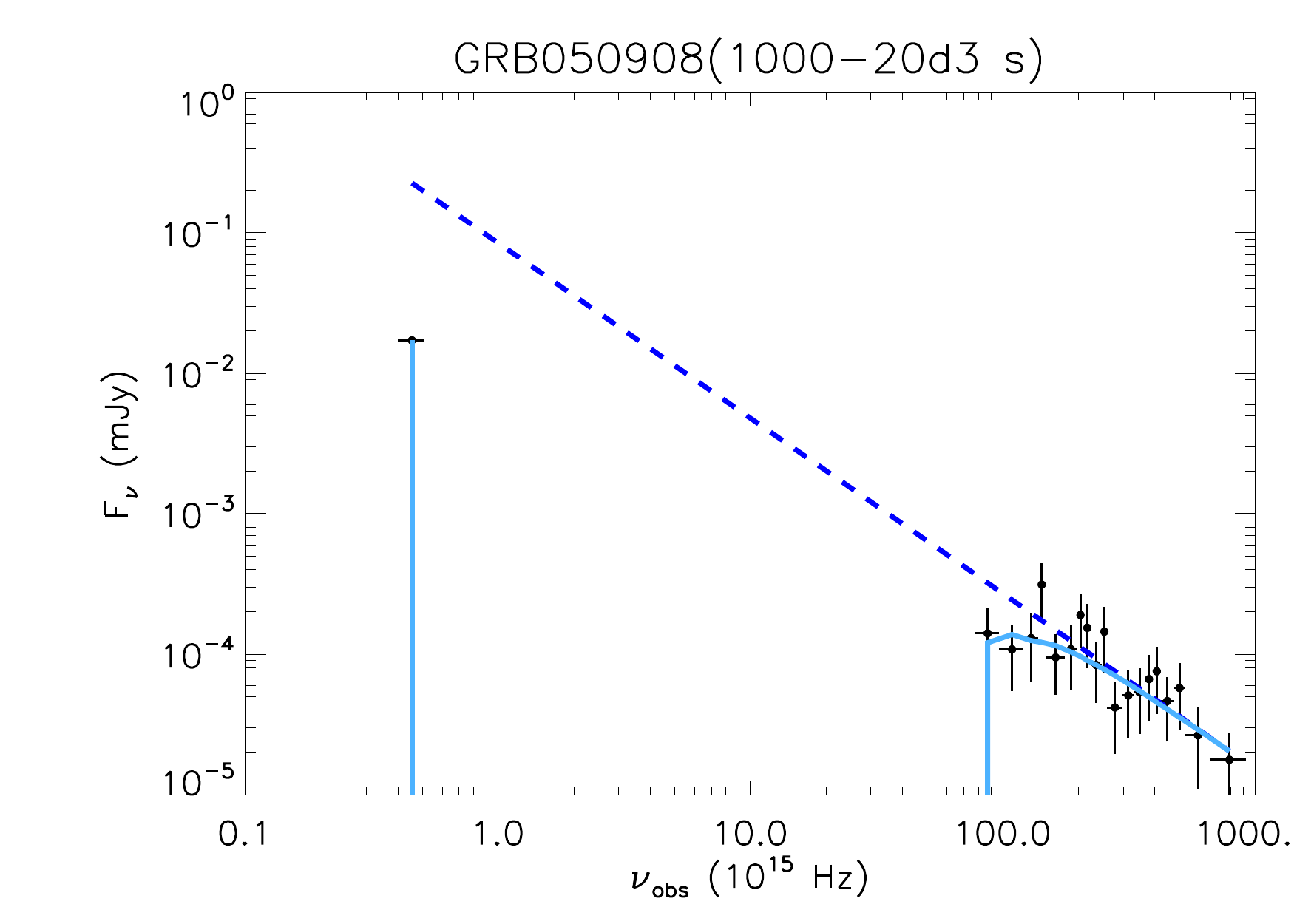}
\includegraphics[width=0.3 \hsize,clip]{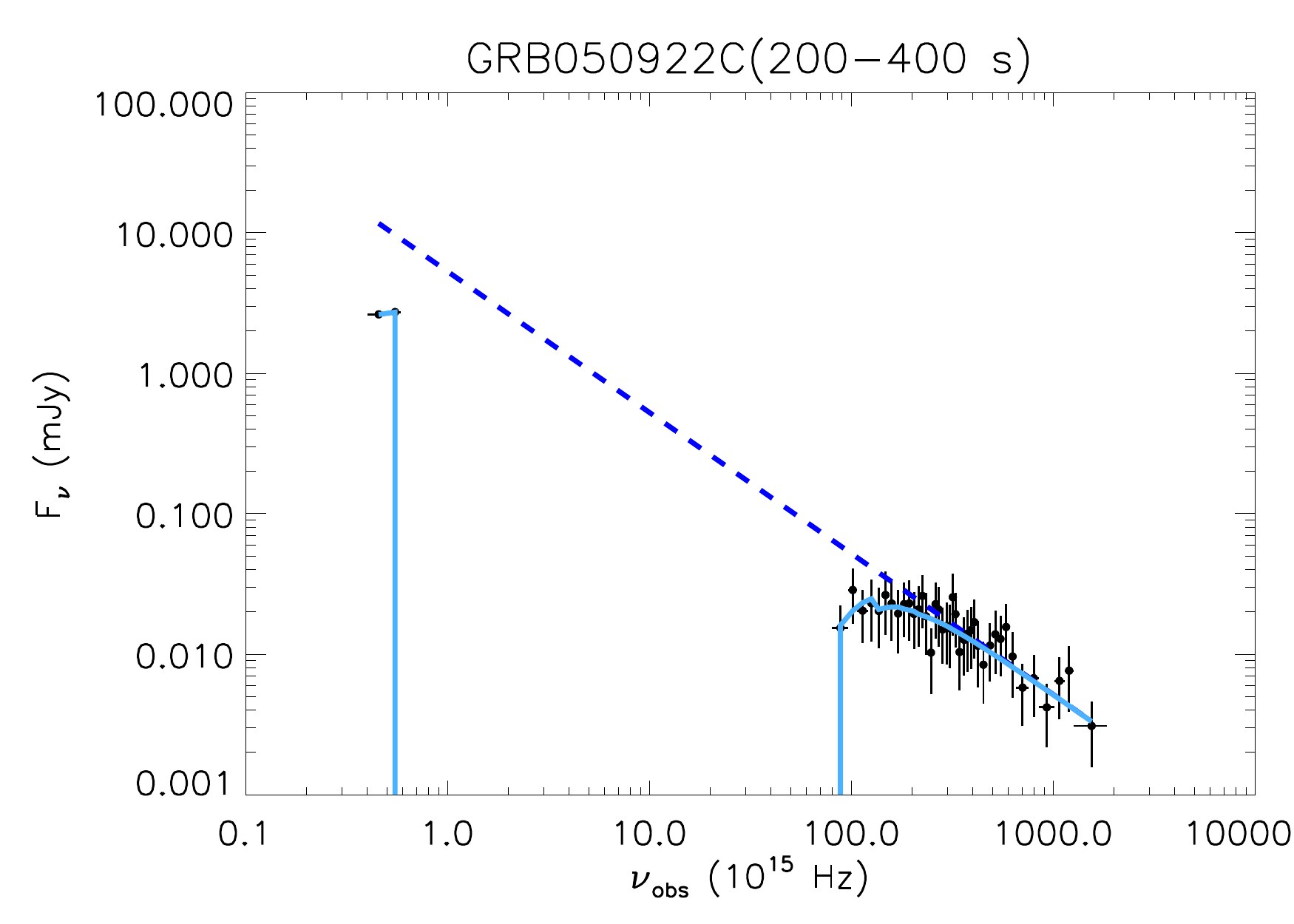}\\
\includegraphics[width=0.3 \hsize,clip]{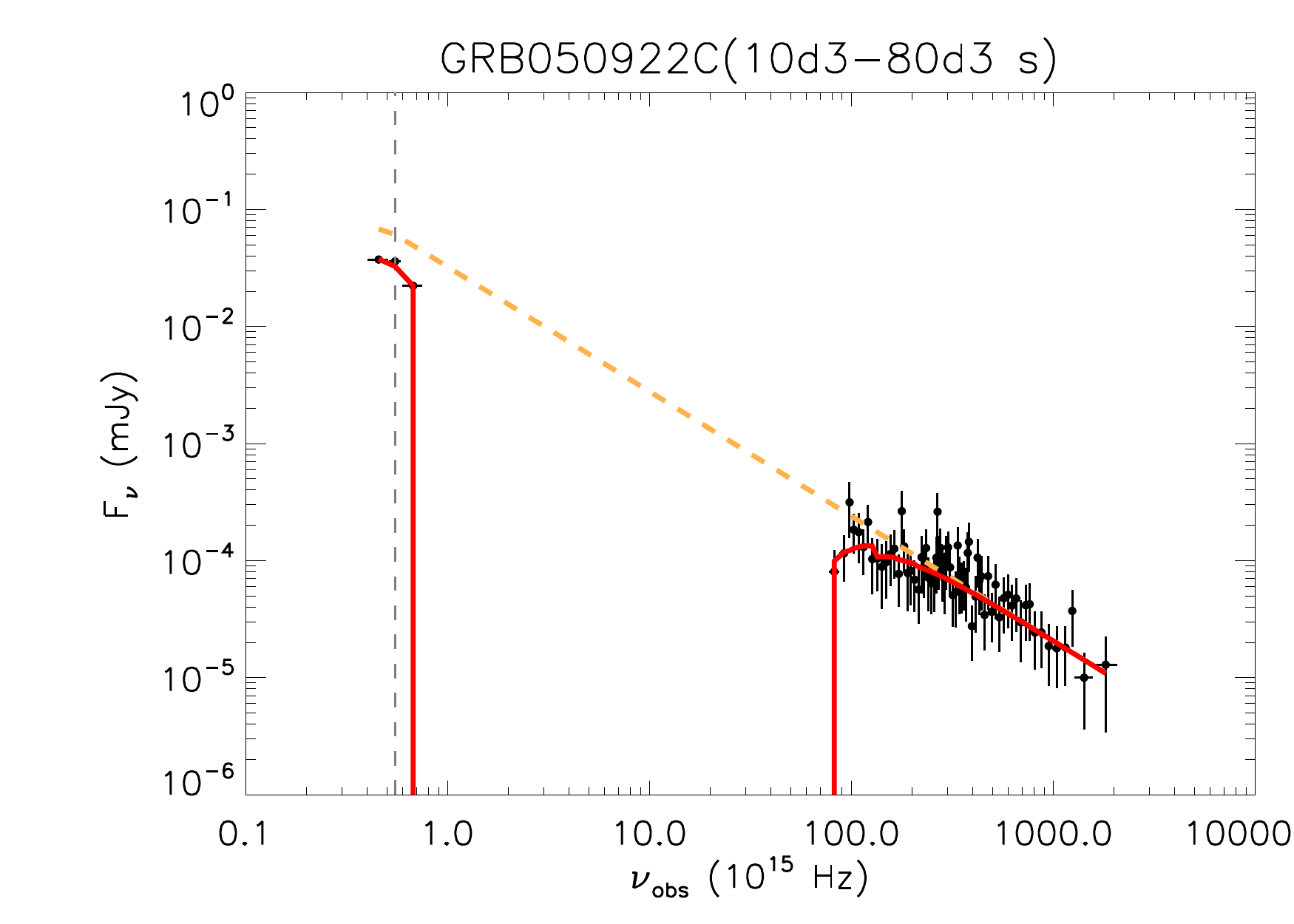}
\includegraphics[width=0.3\hsize,clip]{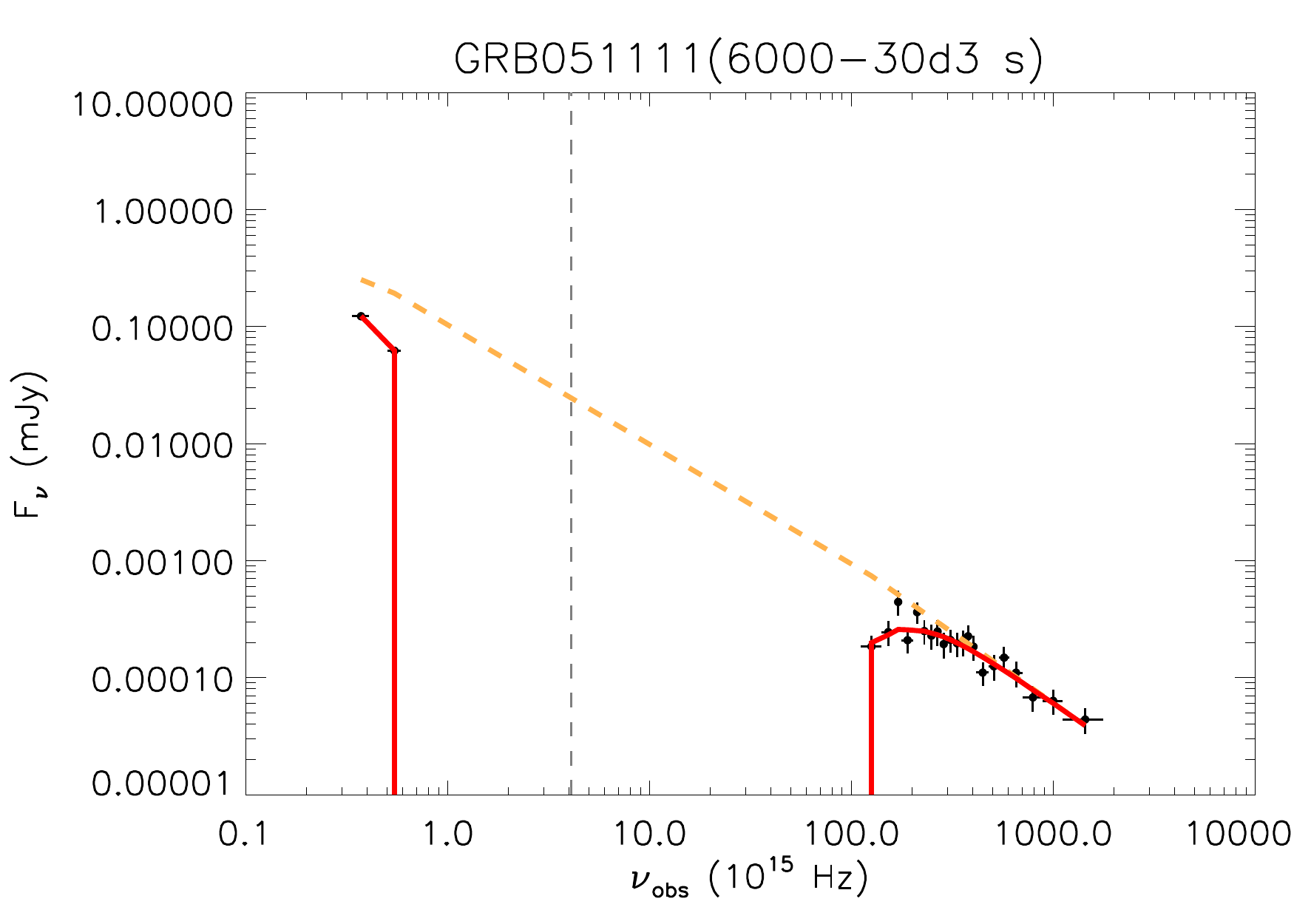}
\includegraphics[width=0.3 \hsize,clip]{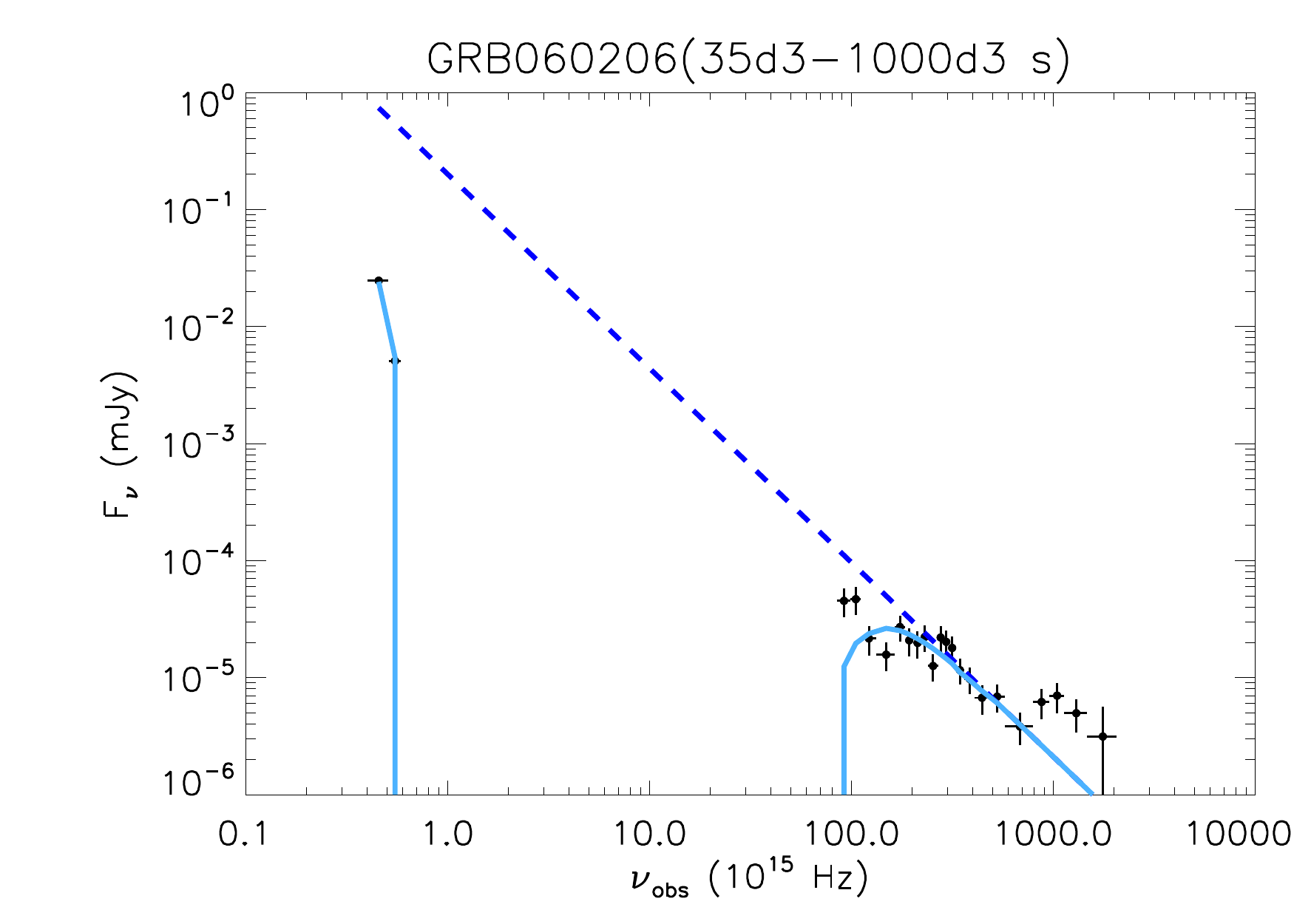}\\
\includegraphics[width=0.3 \hsize,clip]{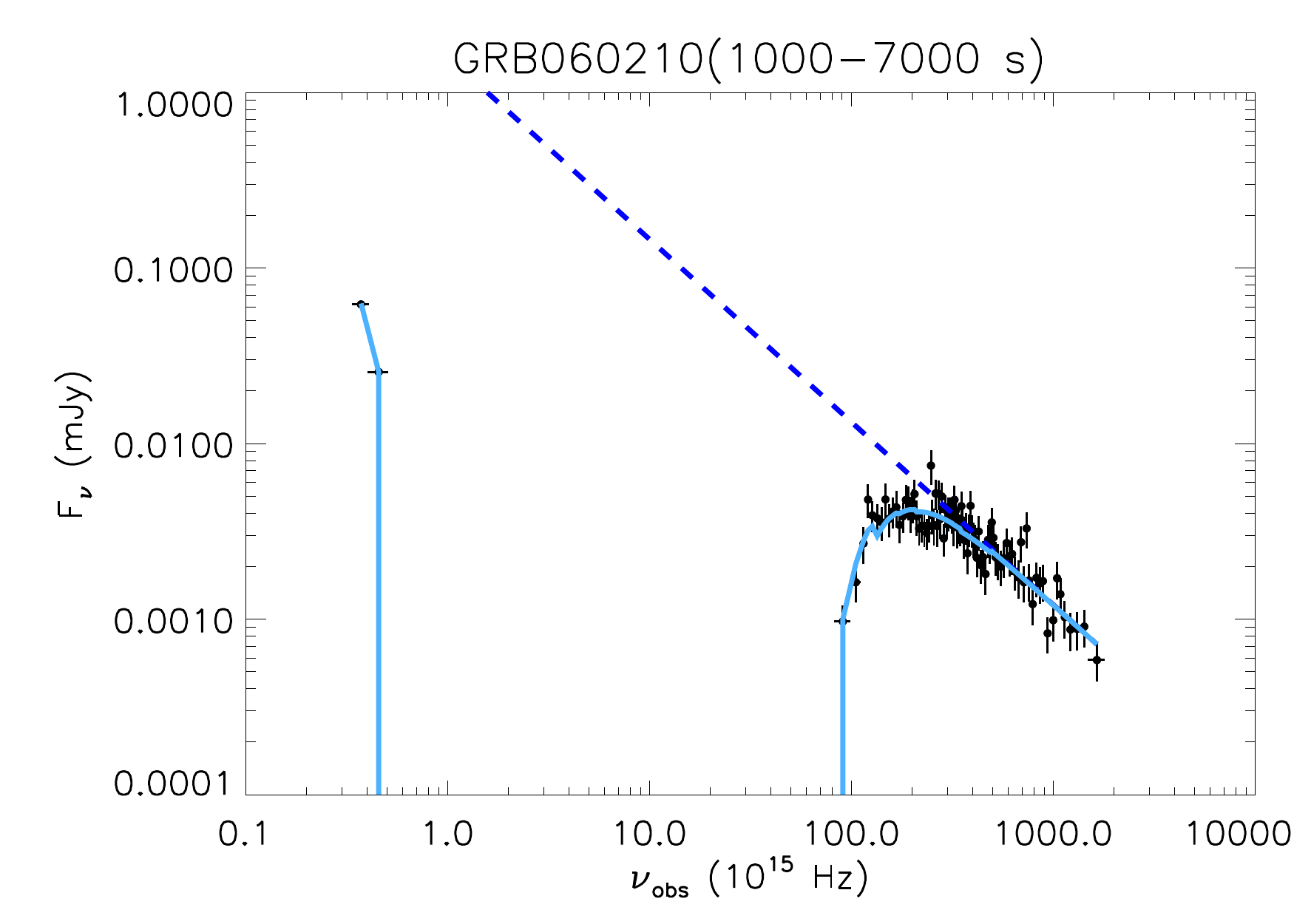}
\includegraphics[width=0.3 \hsize,clip]{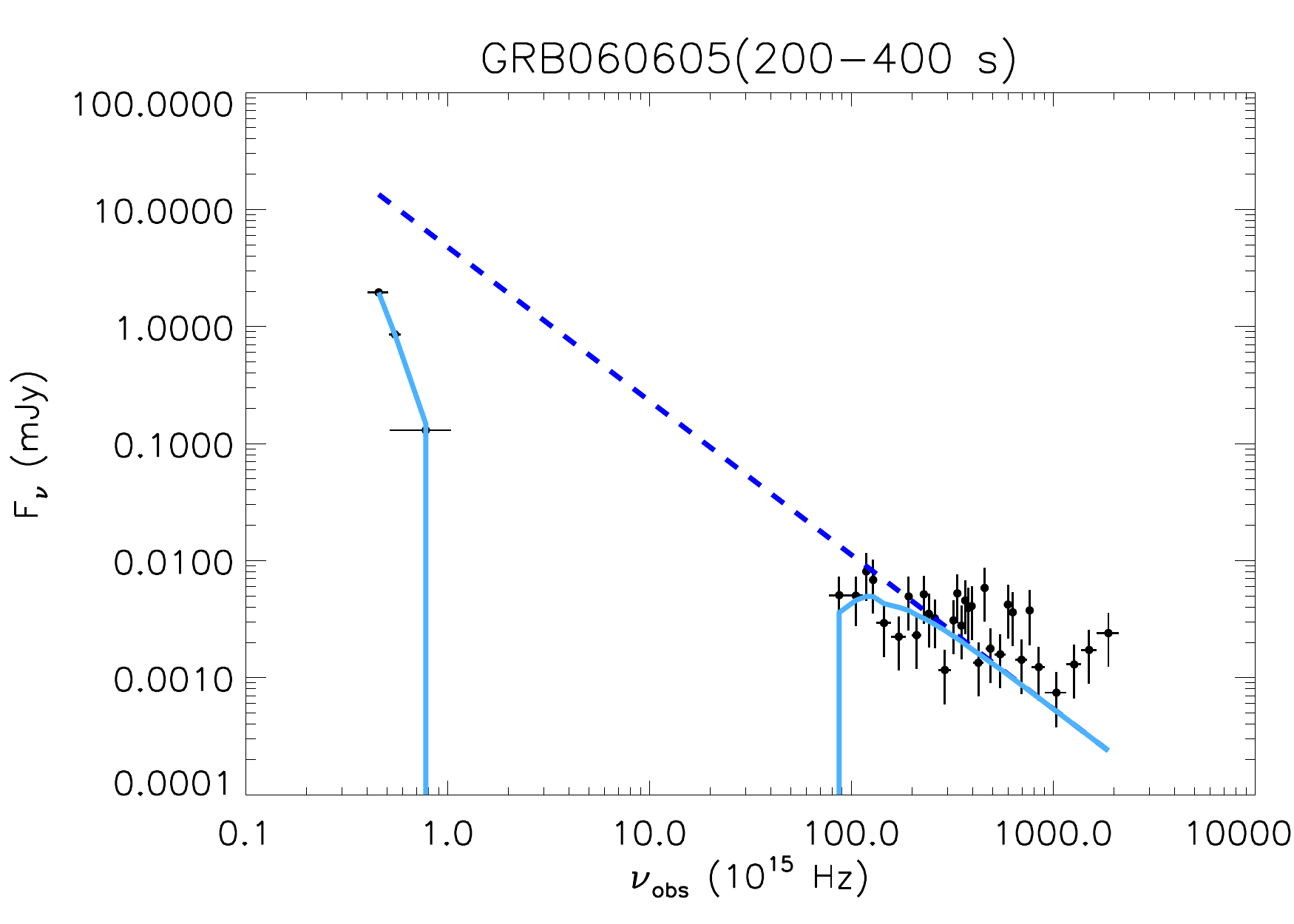}
\includegraphics[width=0.3 \hsize,clip]{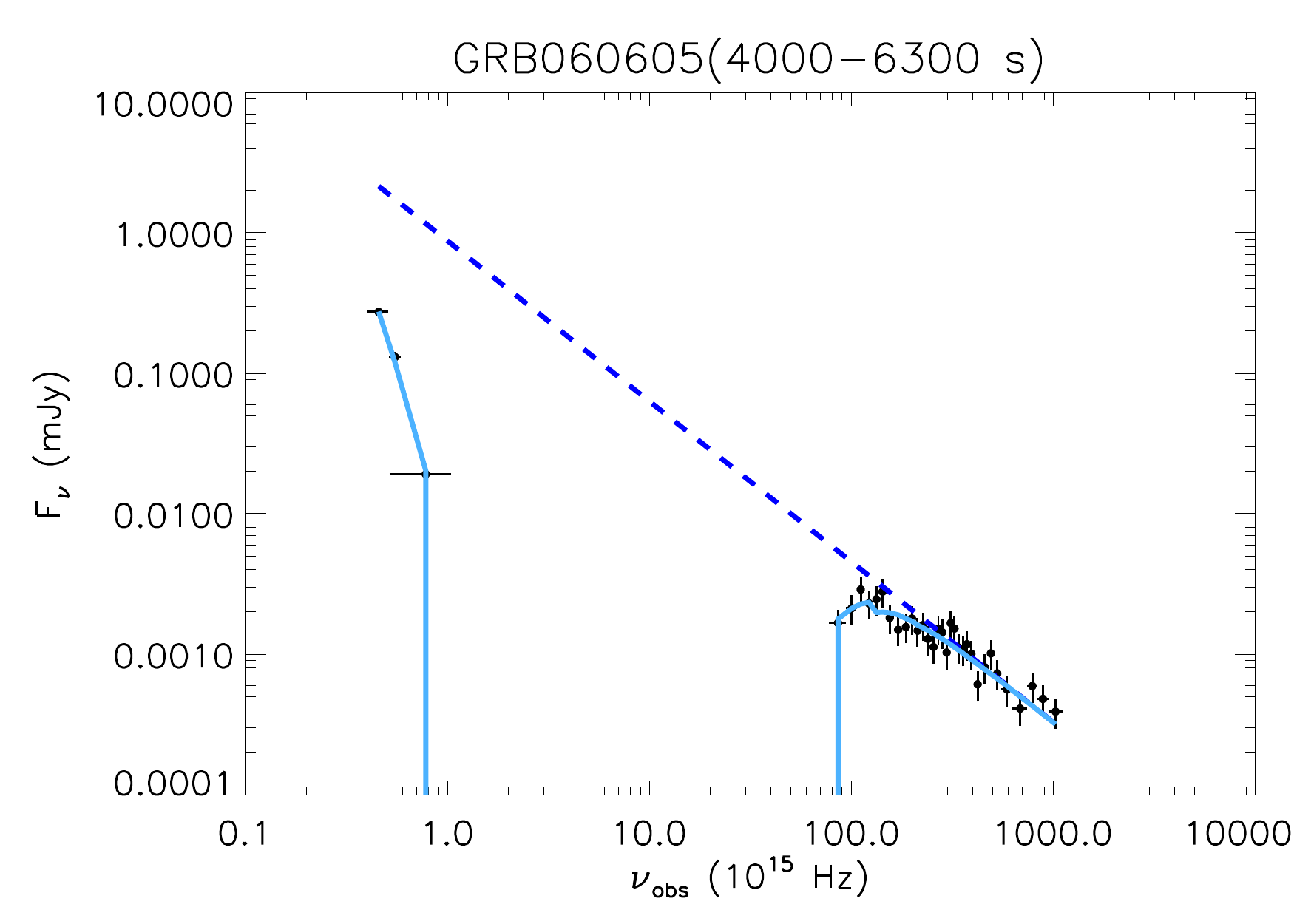}\\
\includegraphics[width=0.3 \hsize,clip]{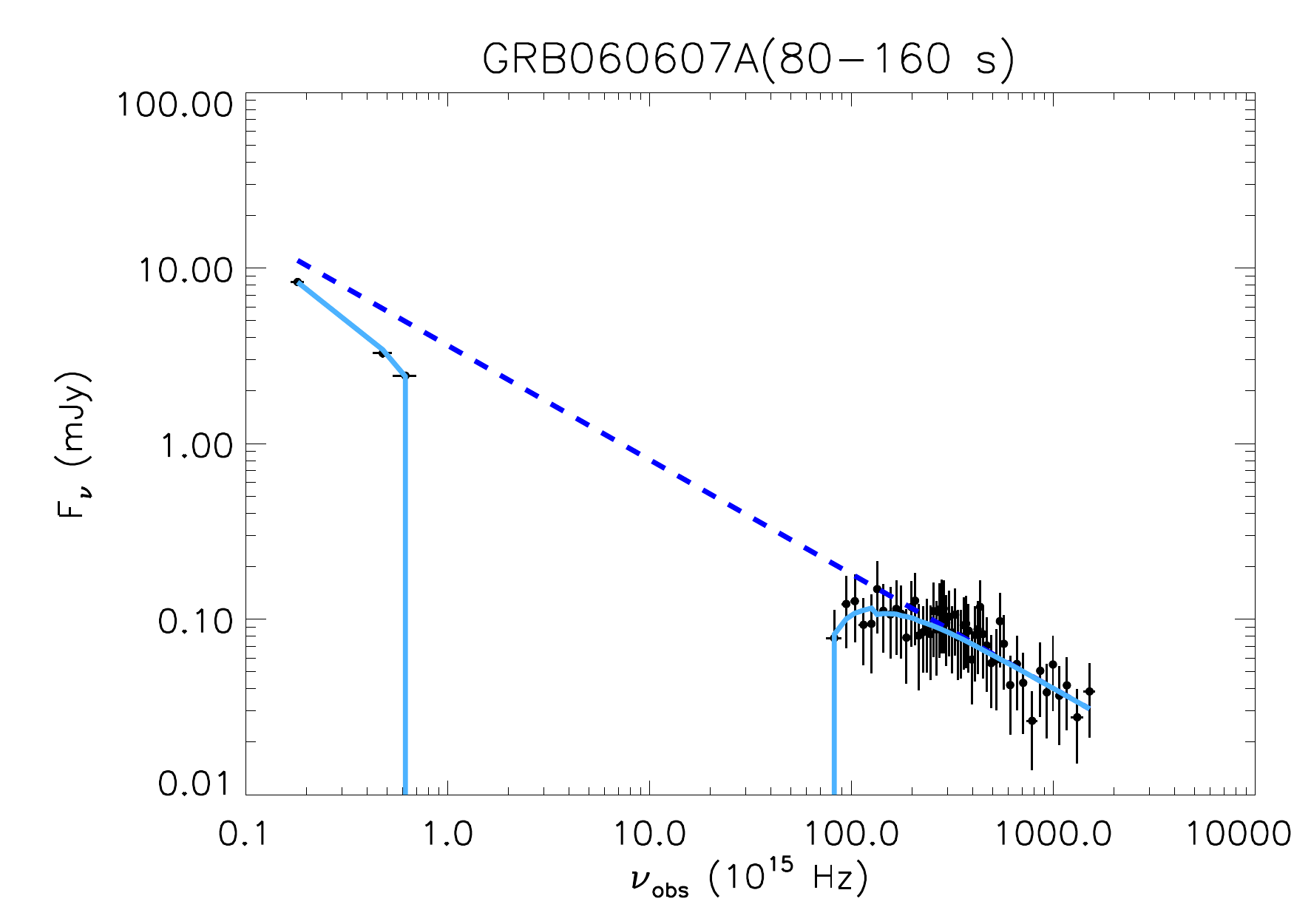}
\includegraphics[width=0.3 \hsize,clip]{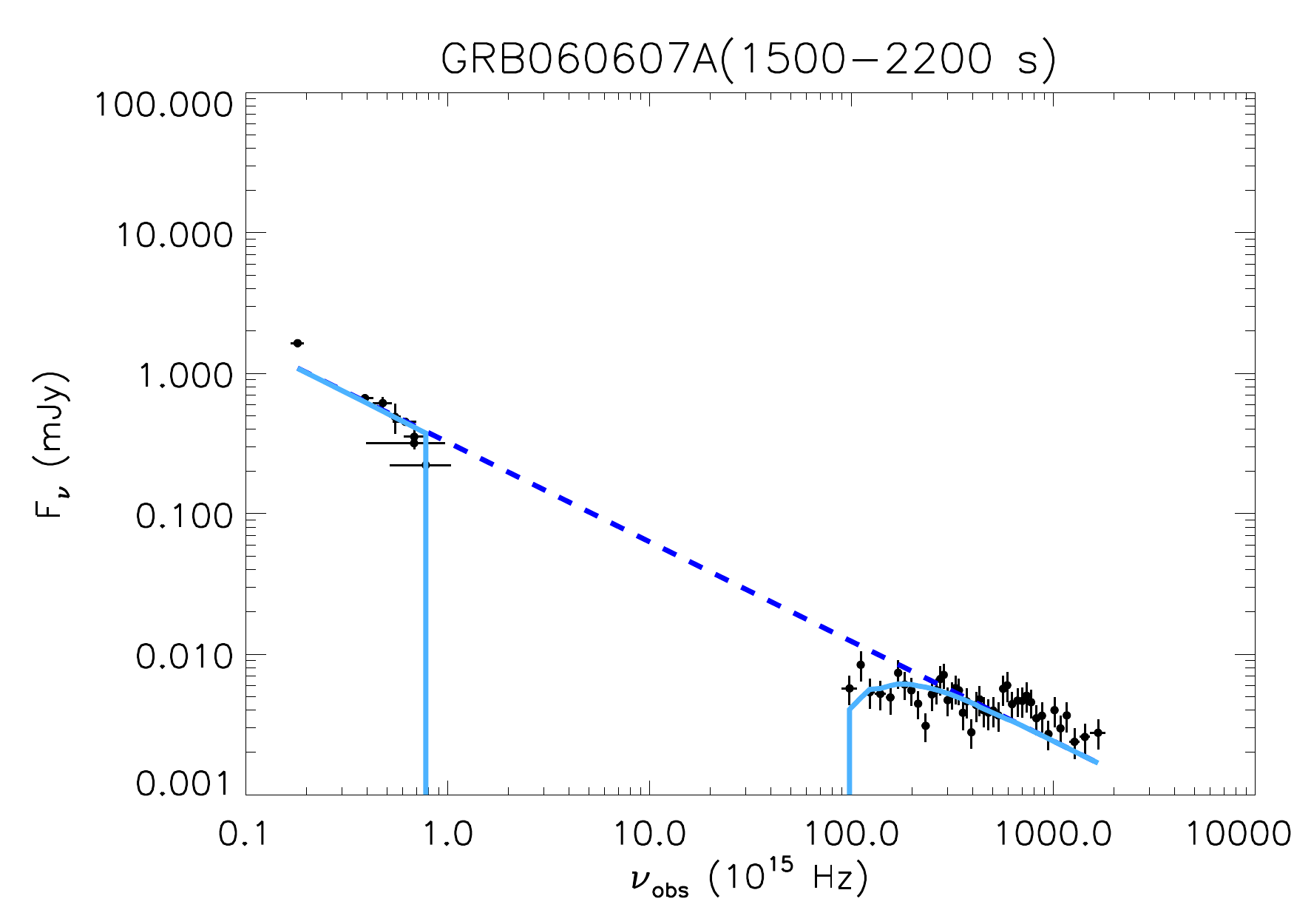}
\includegraphics[width=0.3 \hsize,clip]{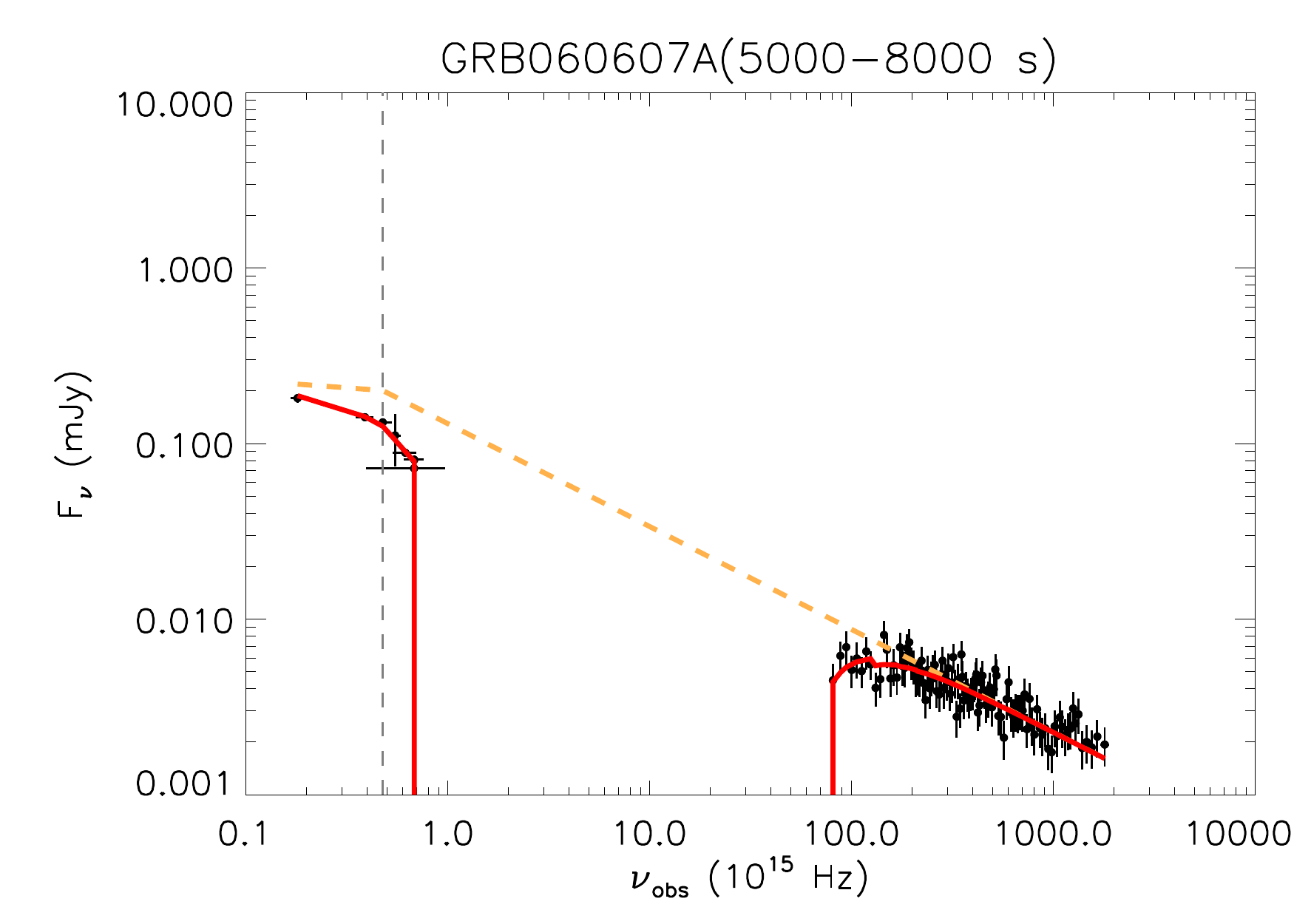}\\
\includegraphics[width=0.3 \hsize,clip]{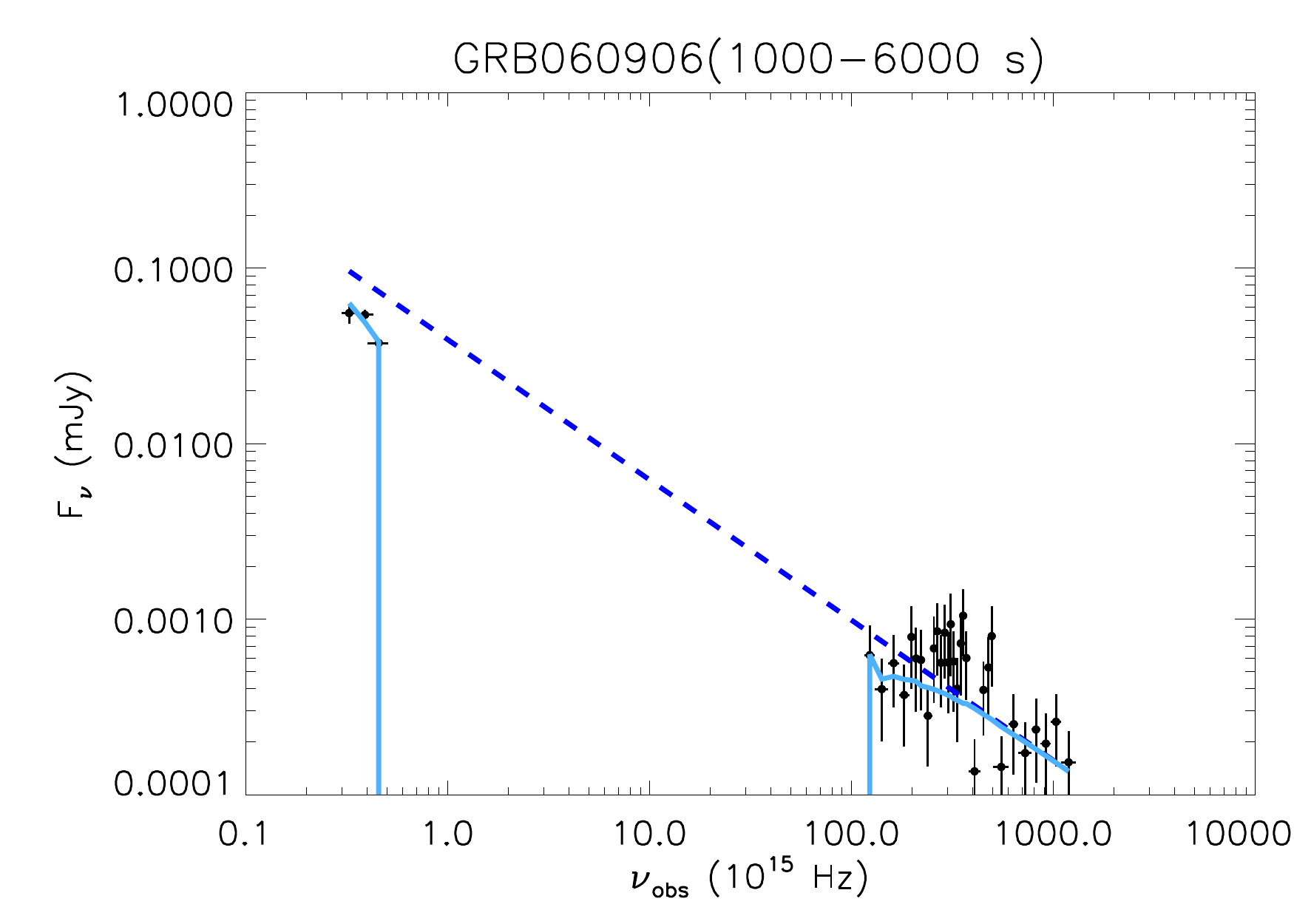}
\includegraphics[width=0.3 \hsize,clip]{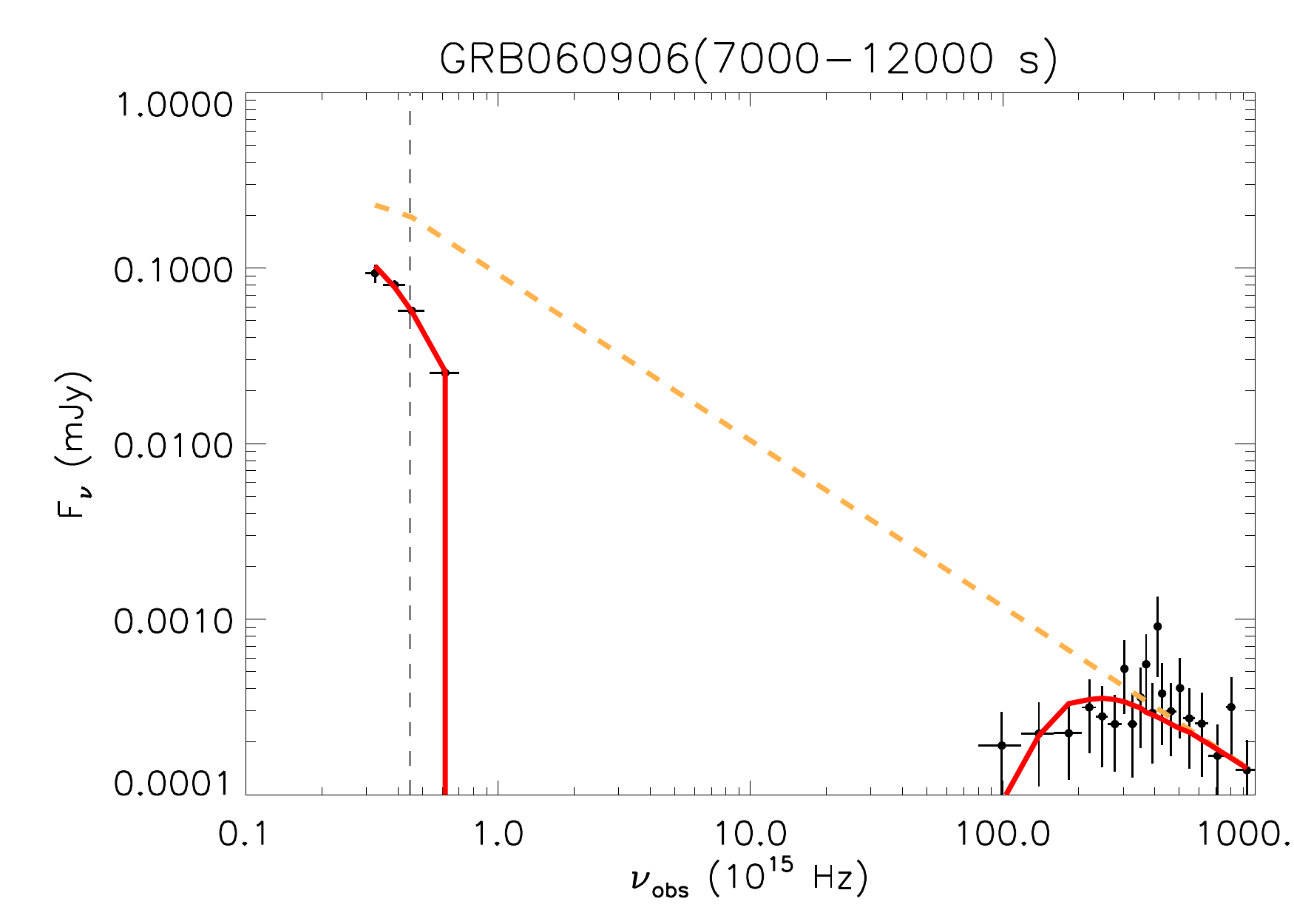}
\includegraphics[width=0.3\hsize,clip]{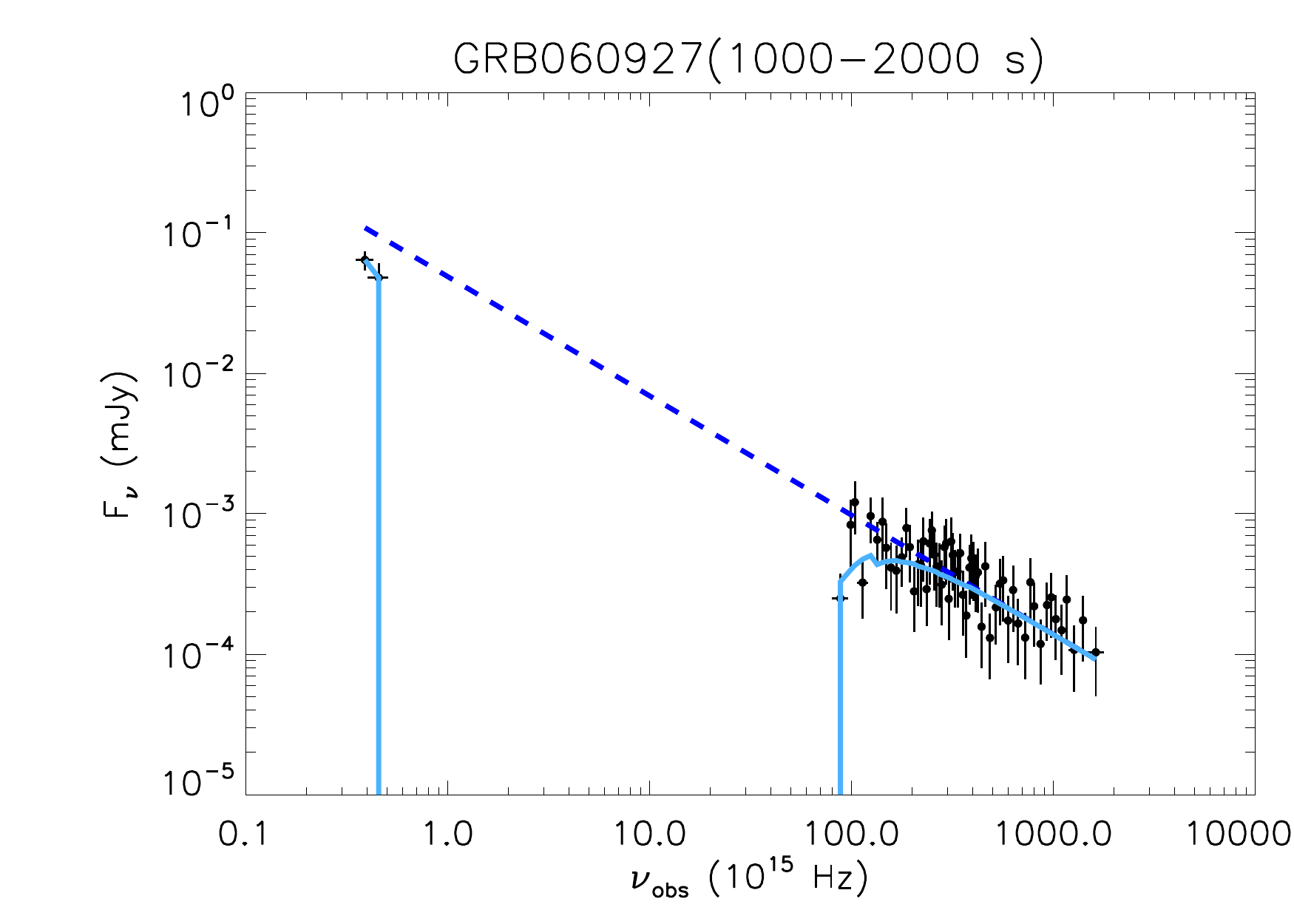}
\caption{\small{Optical/X-ray SEDs for GRBs belonging to Group C. color-coding as in Figure~\ref{sed1}.}}\label{sed6} 
\end{figure}

\begin{figure}
\includegraphics[width=0.3 \hsize,clip]{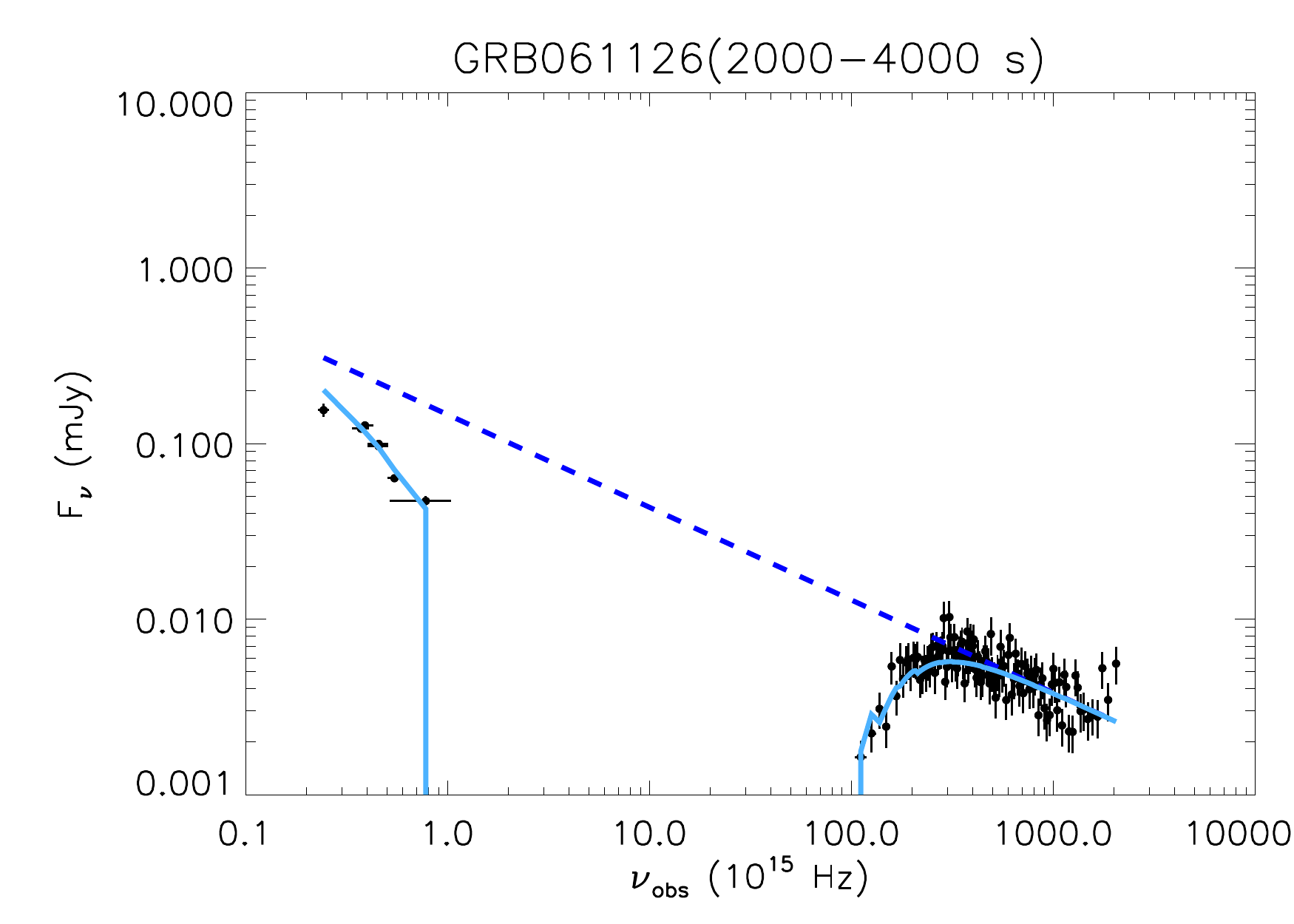}
\includegraphics[width=0.3\hsize,clip]{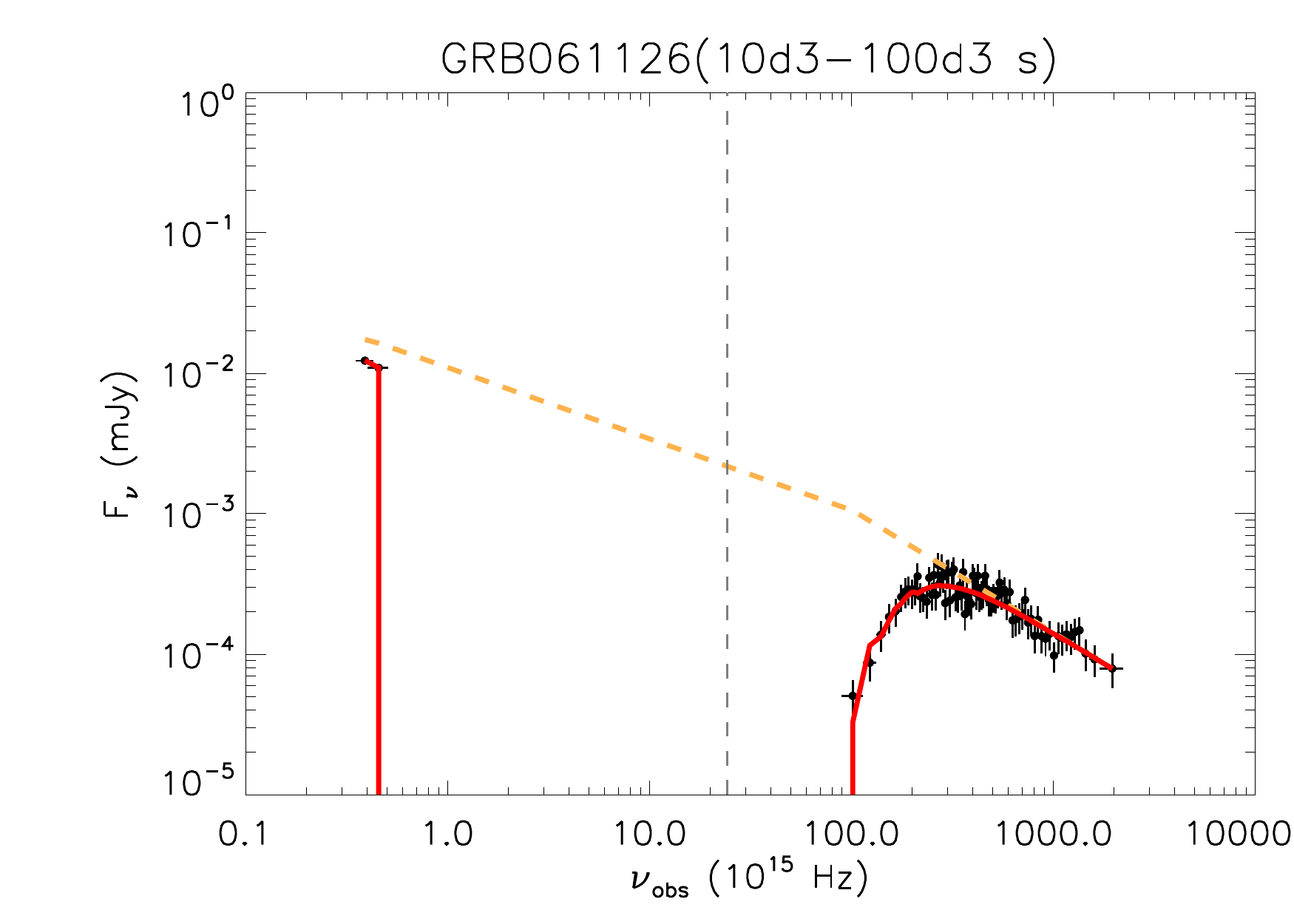}
\includegraphics[width=0.3\hsize,clip]{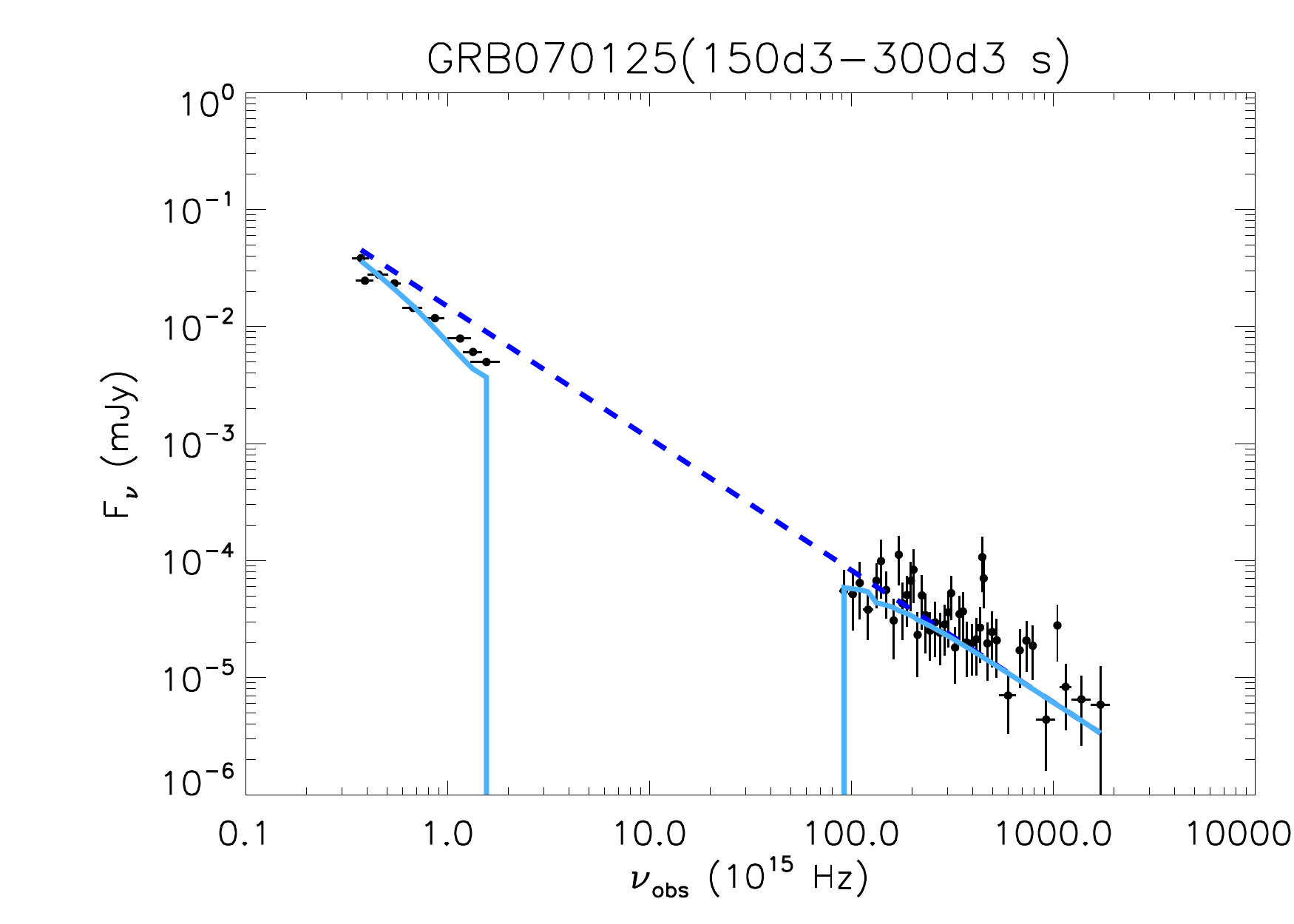}\\
\includegraphics[width=0.3 \hsize,clip]{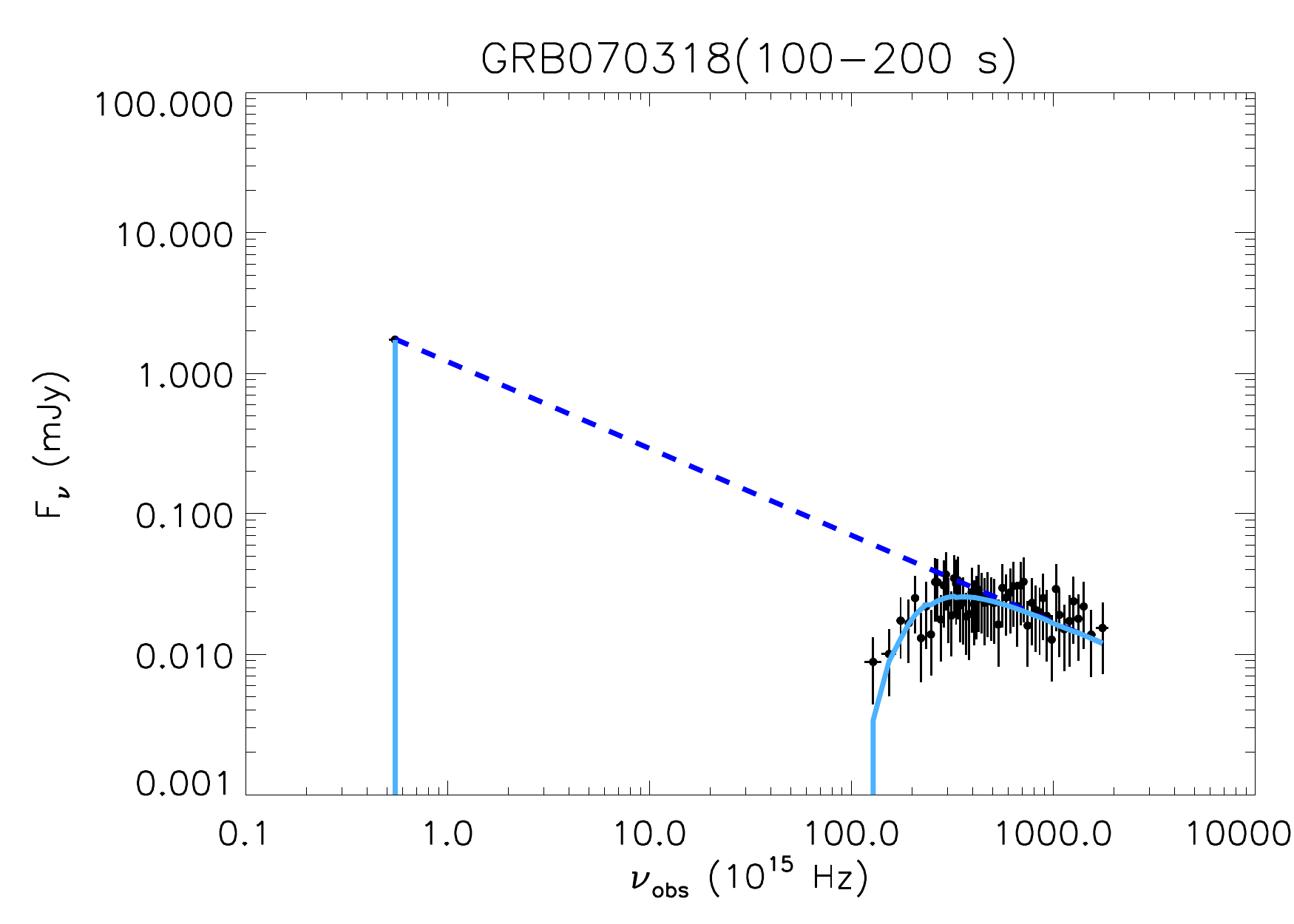}
\includegraphics[width=0.3 \hsize,clip]{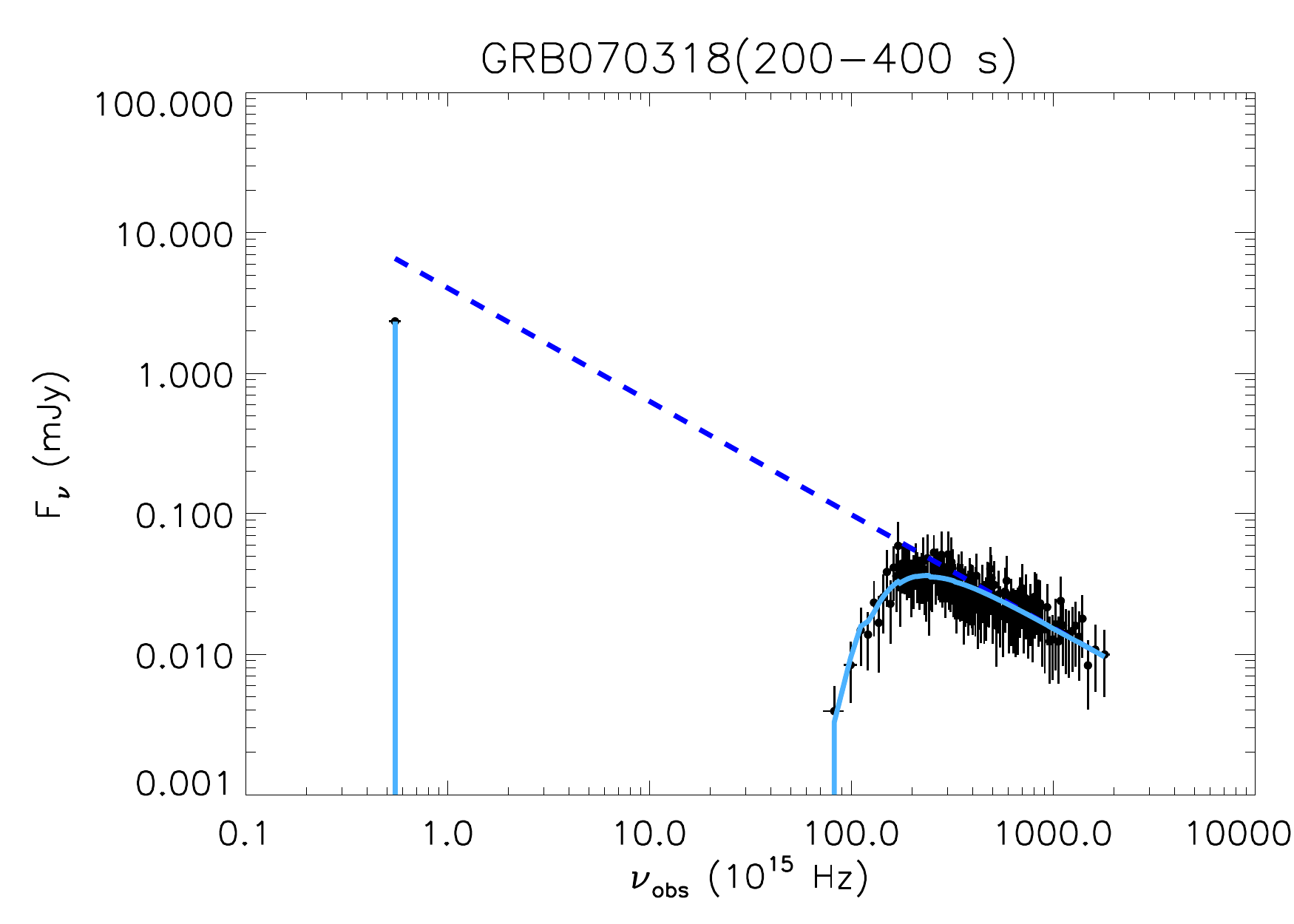}
\includegraphics[width=0.3 \hsize,clip]{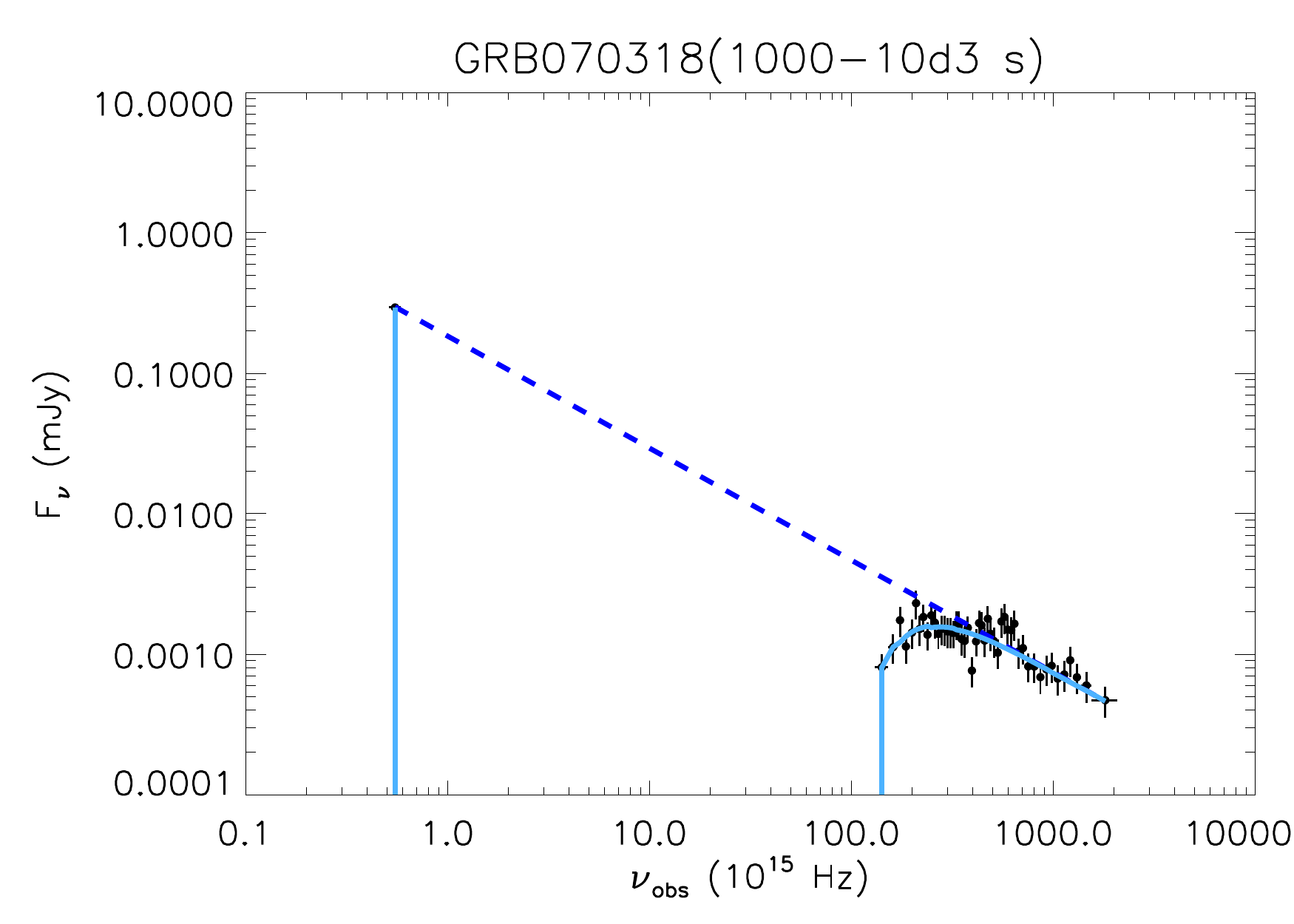}\\
\includegraphics[width=0.3 \hsize,clip]{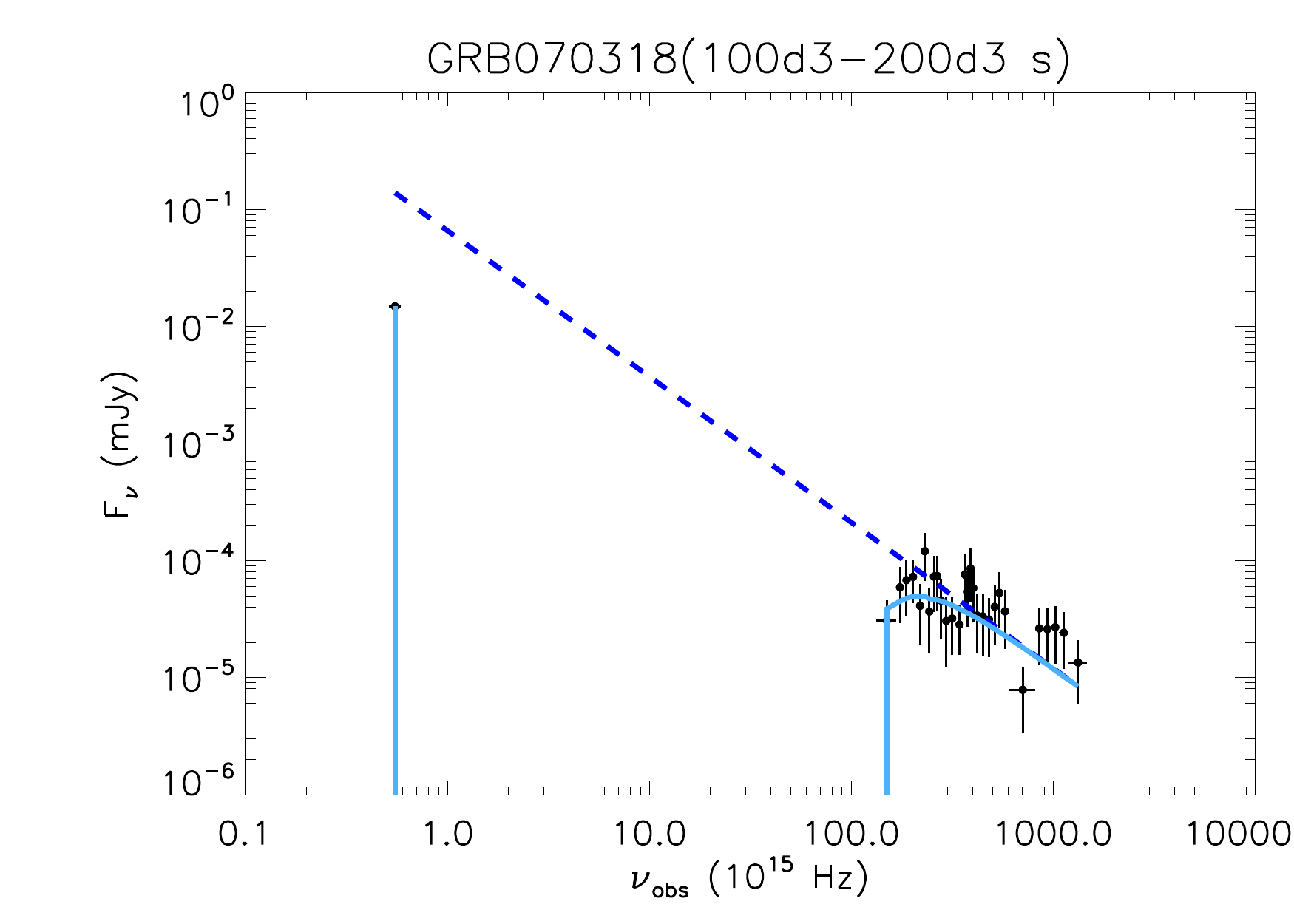}
\includegraphics[width=0.3 \hsize,clip]{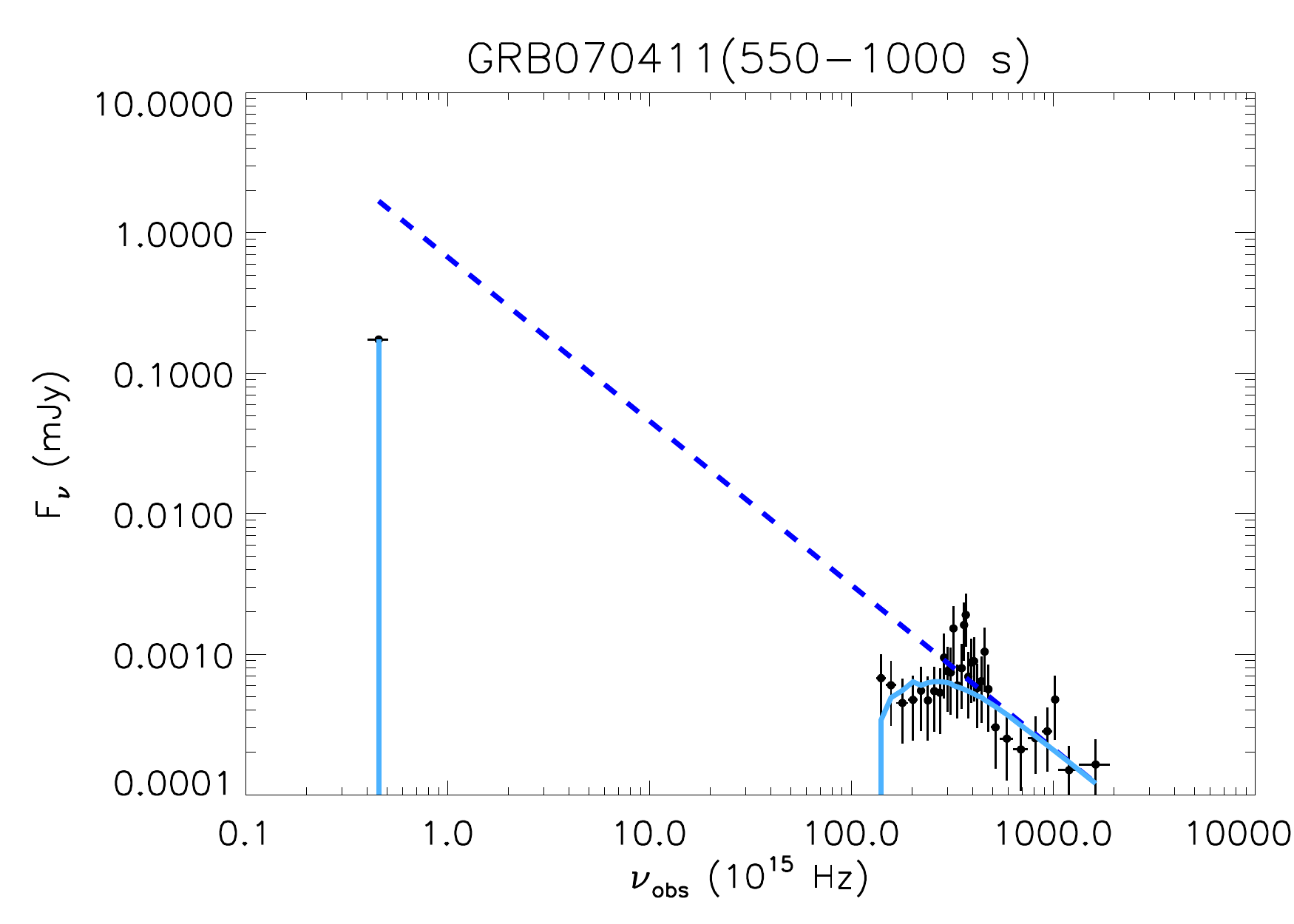}
\includegraphics[width=0.3 \hsize,clip]{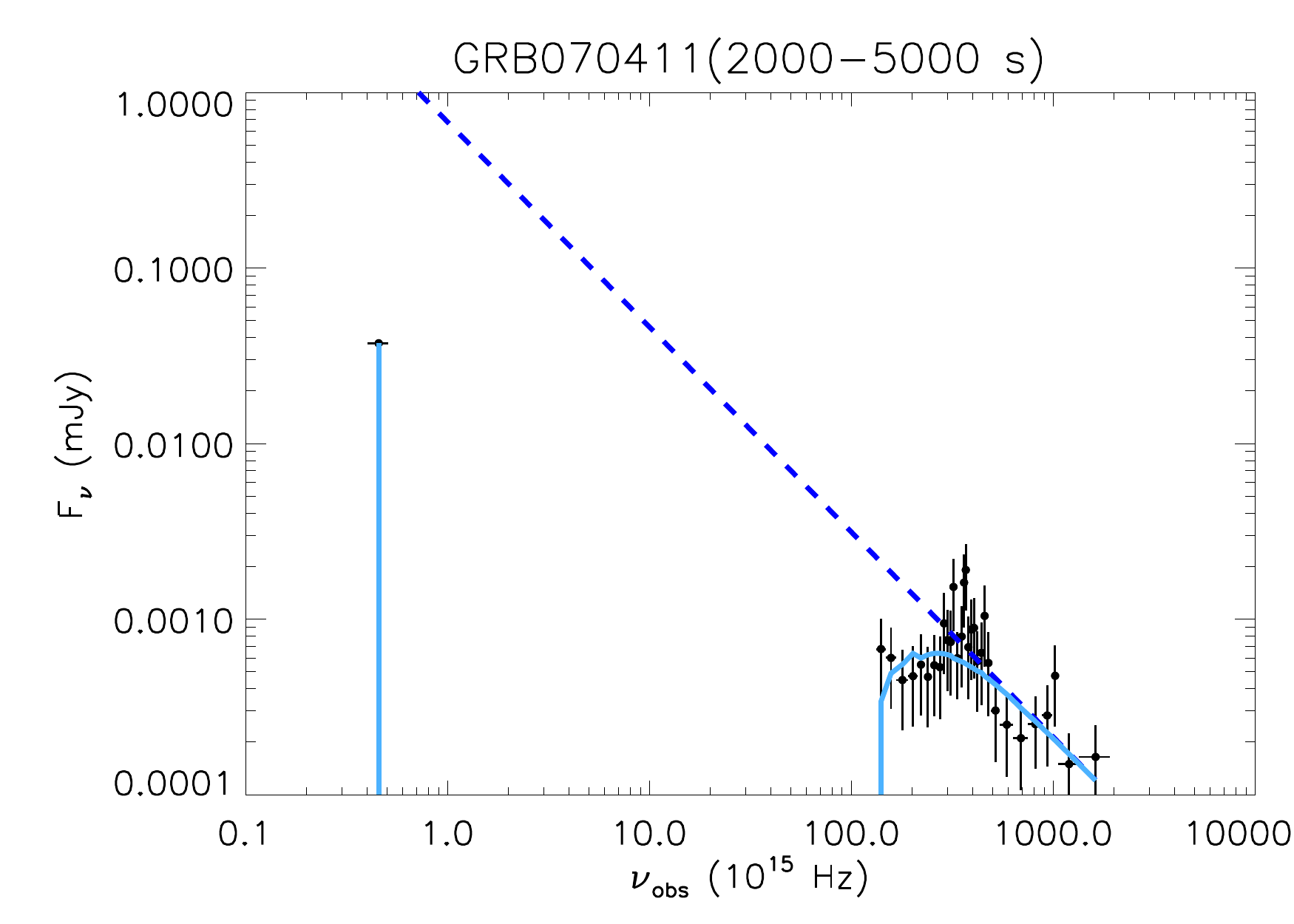}\\
\includegraphics[width=0.3 \hsize,clip]{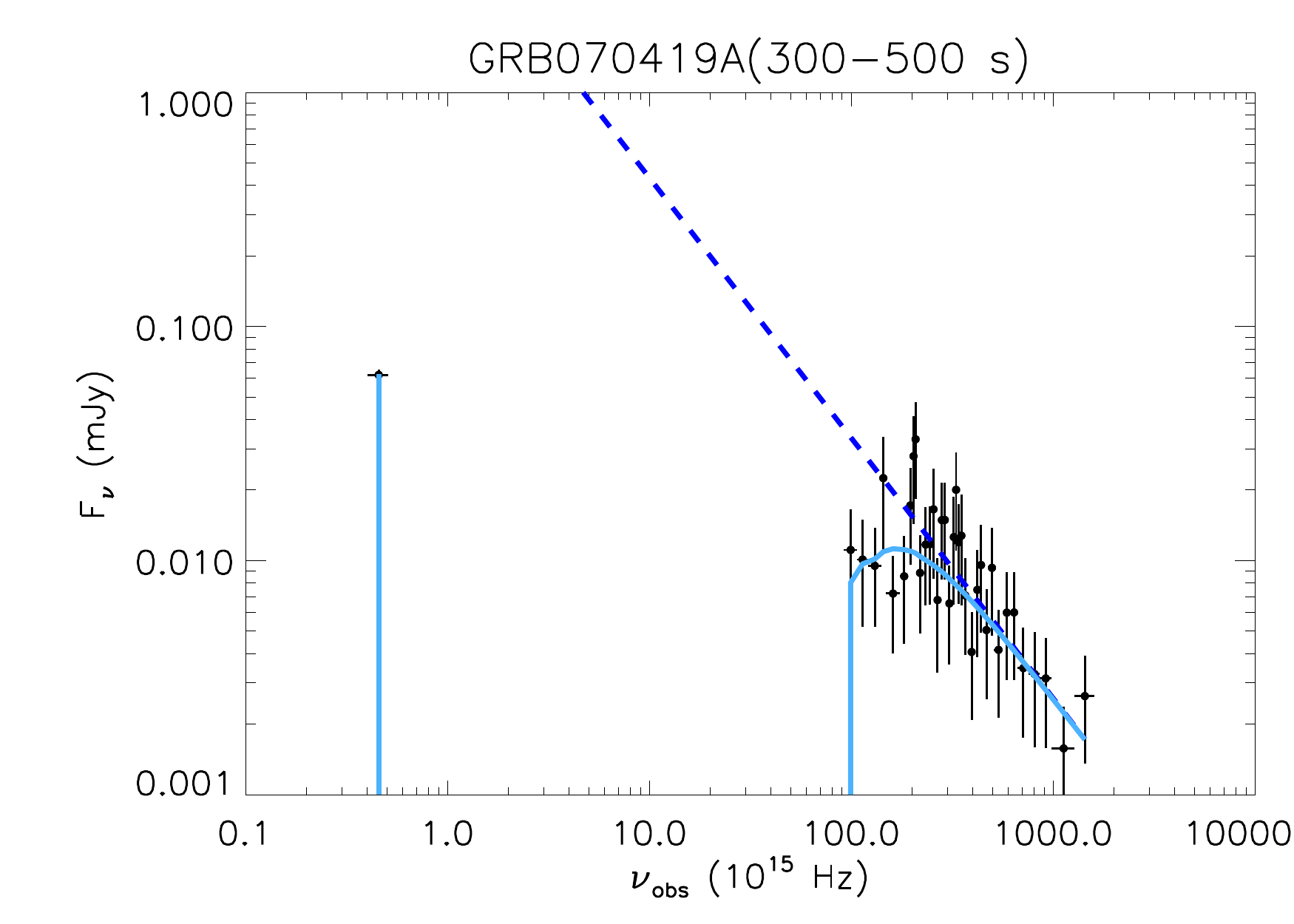}
\includegraphics[width=0.3 \hsize,clip]{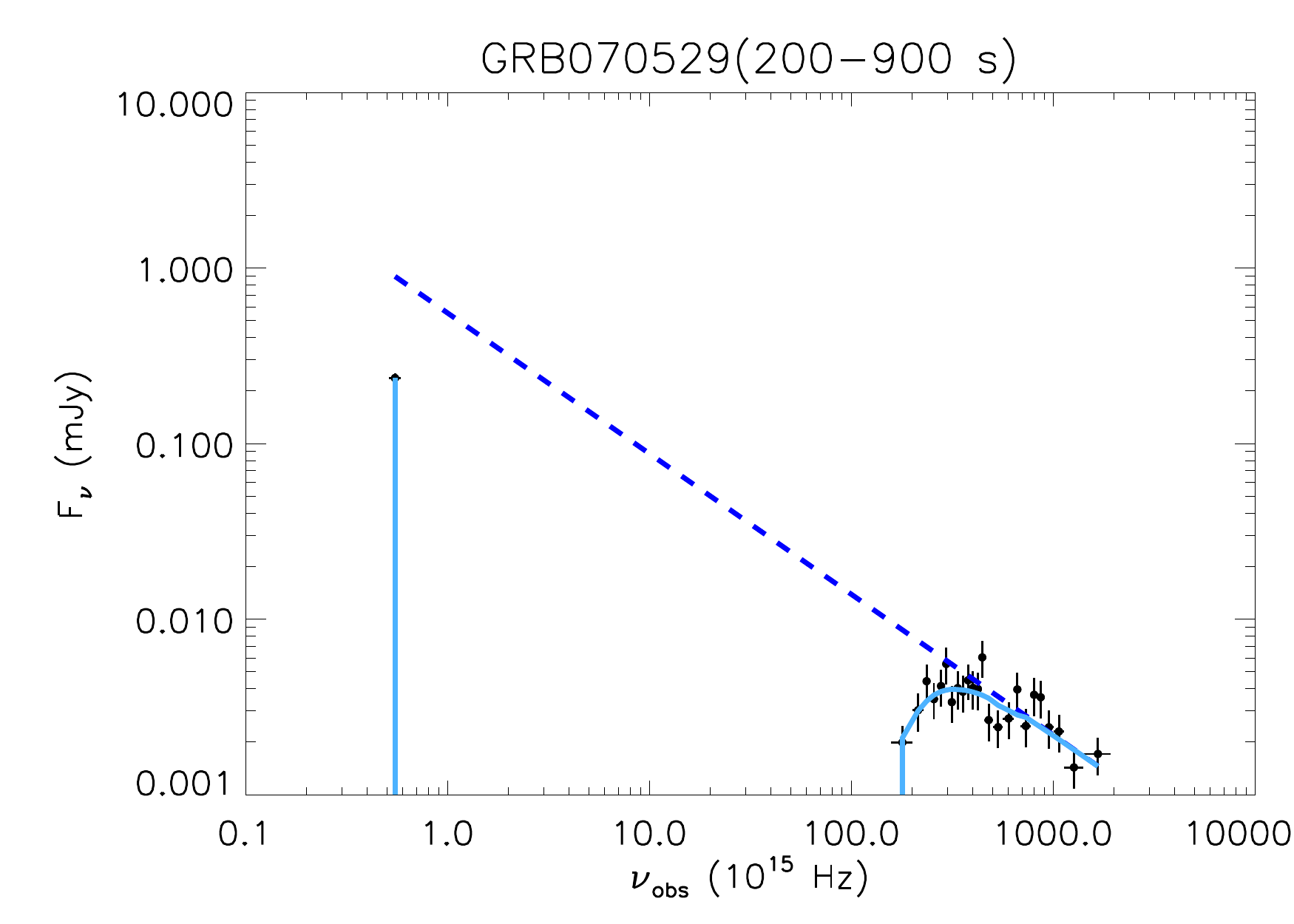}
\includegraphics[width=0.3 \hsize,clip]{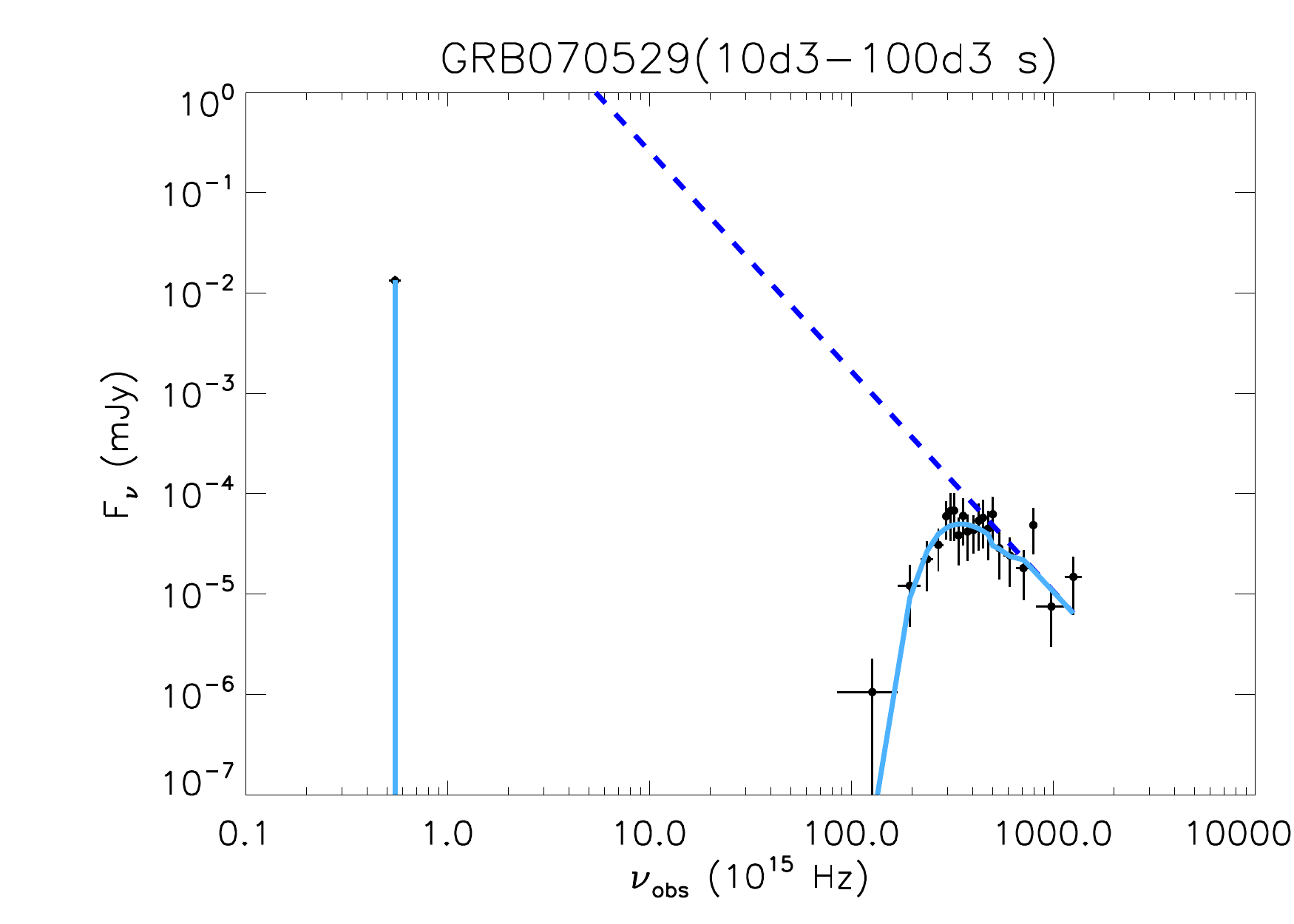}\\
\includegraphics[width=0.3 \hsize,clip]{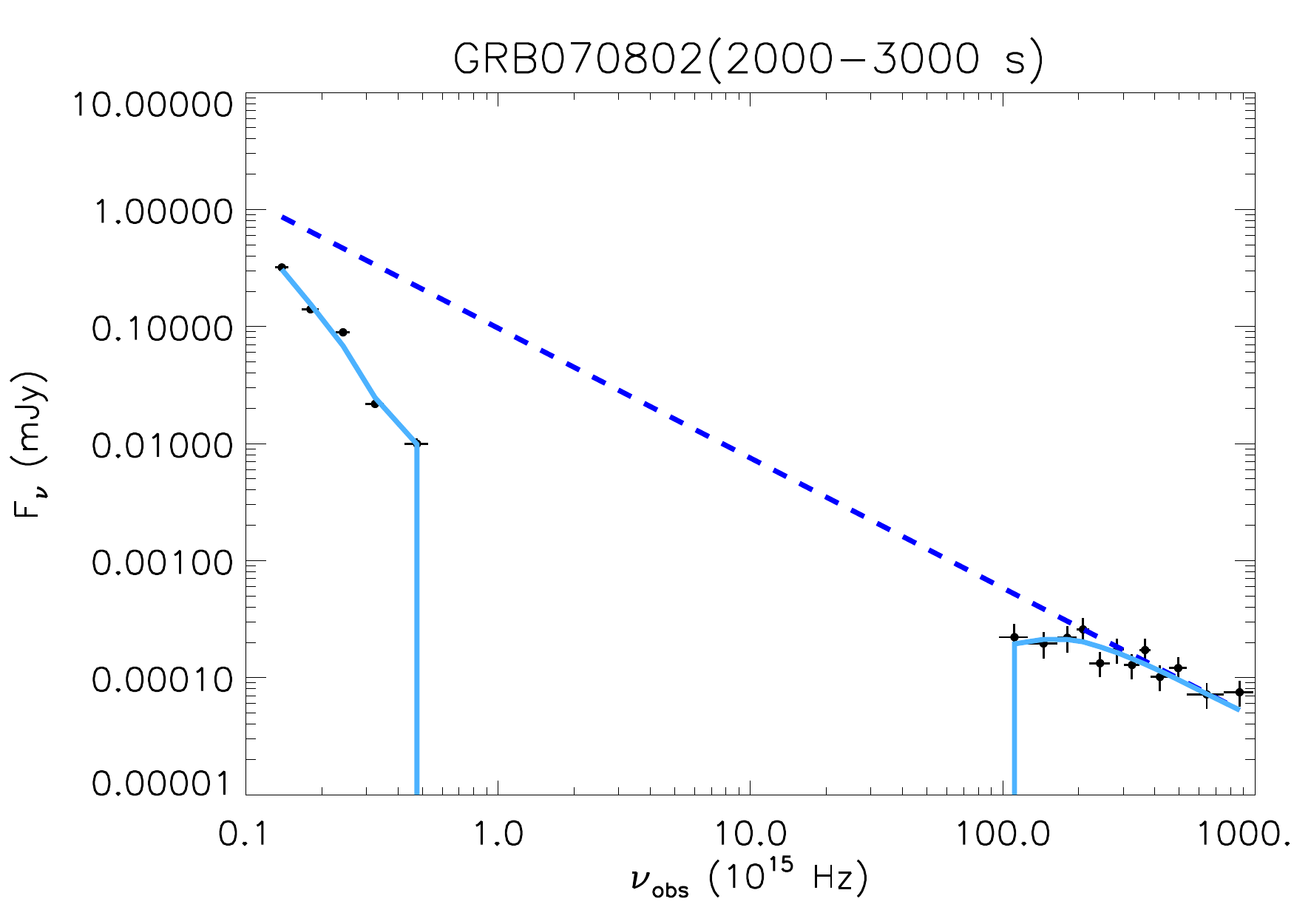}
\includegraphics[width=0.3 \hsize,clip]{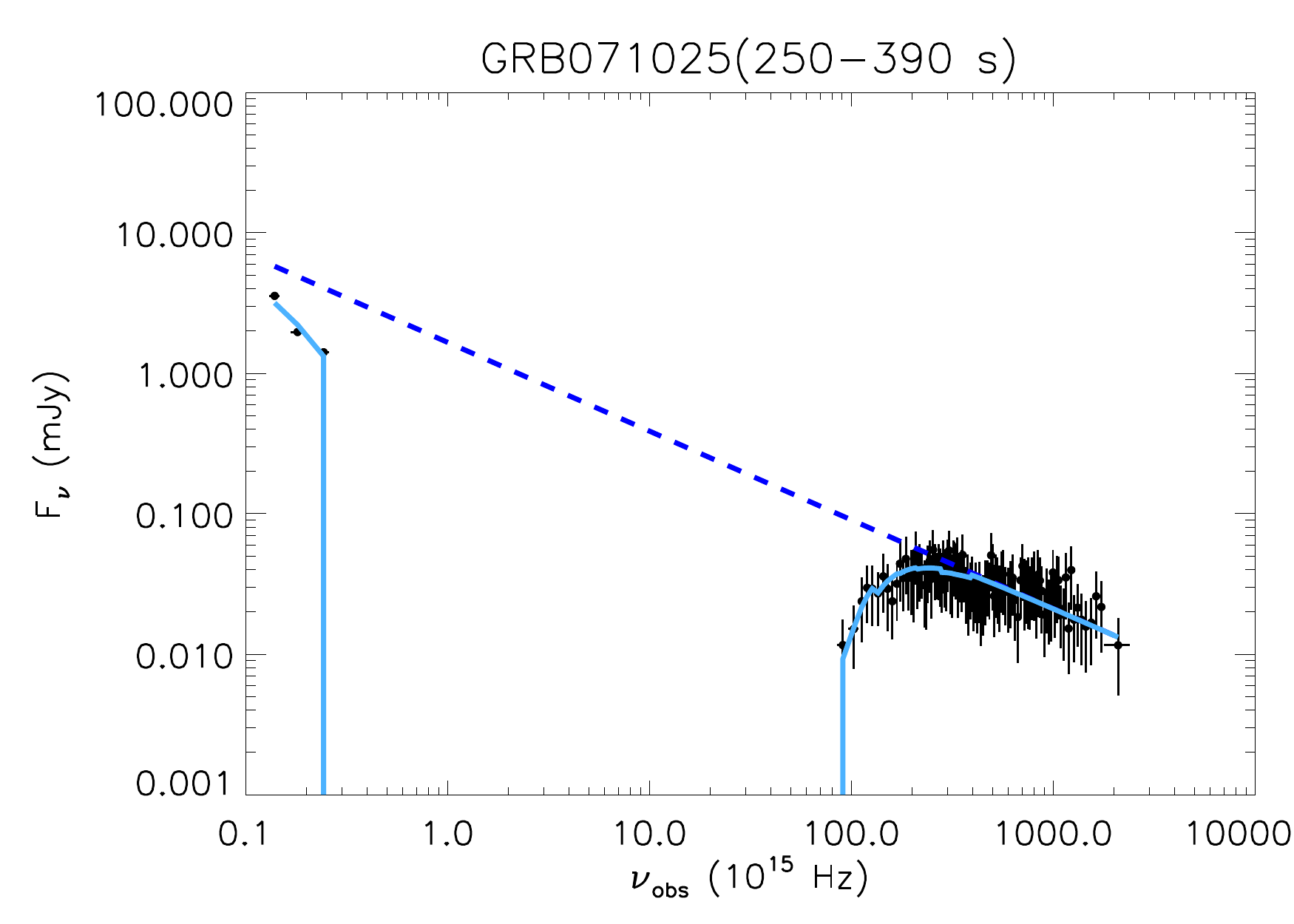}
\includegraphics[width=0.3 \hsize,clip]{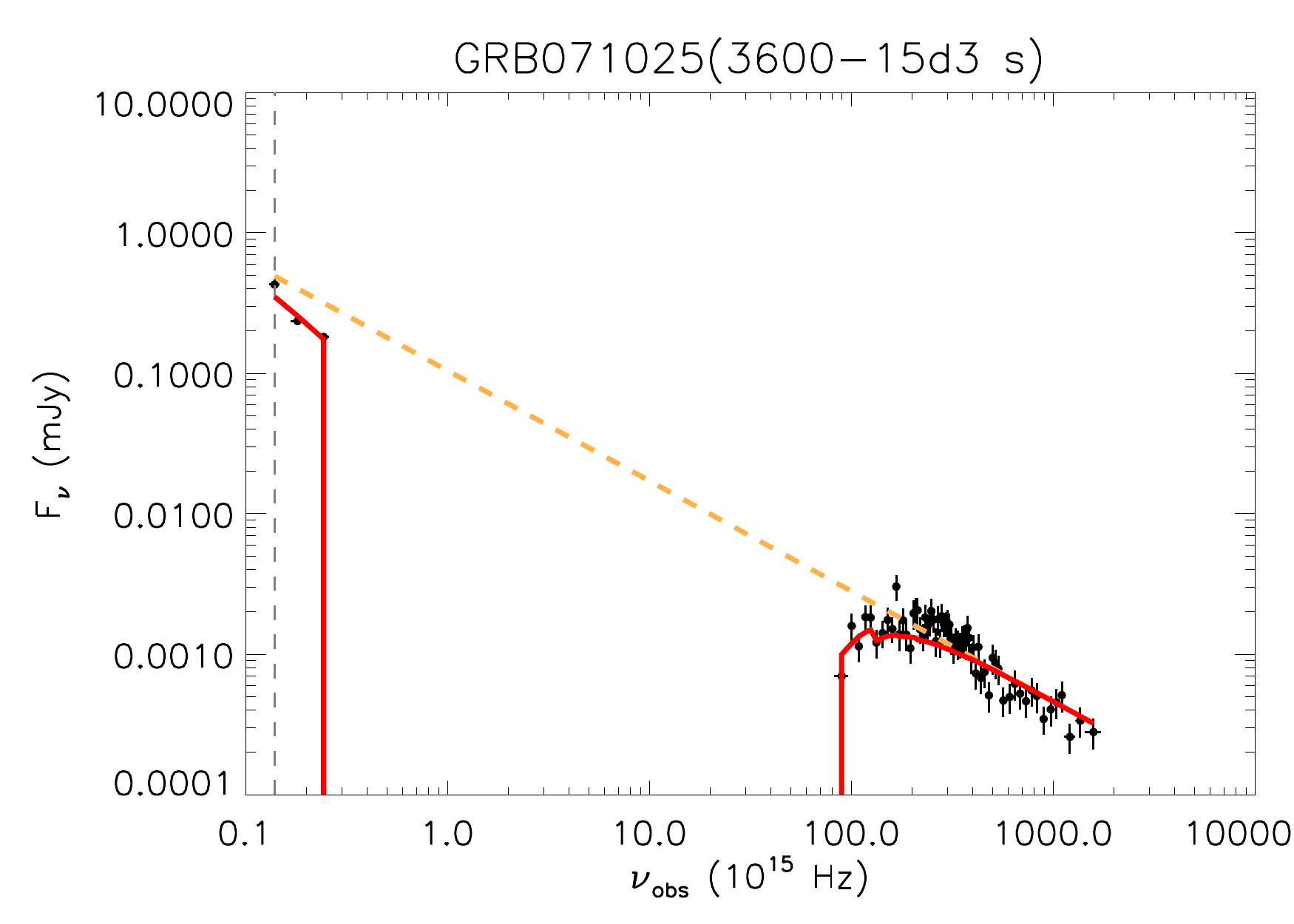}\\
\includegraphics[width=0.3 \hsize,clip]{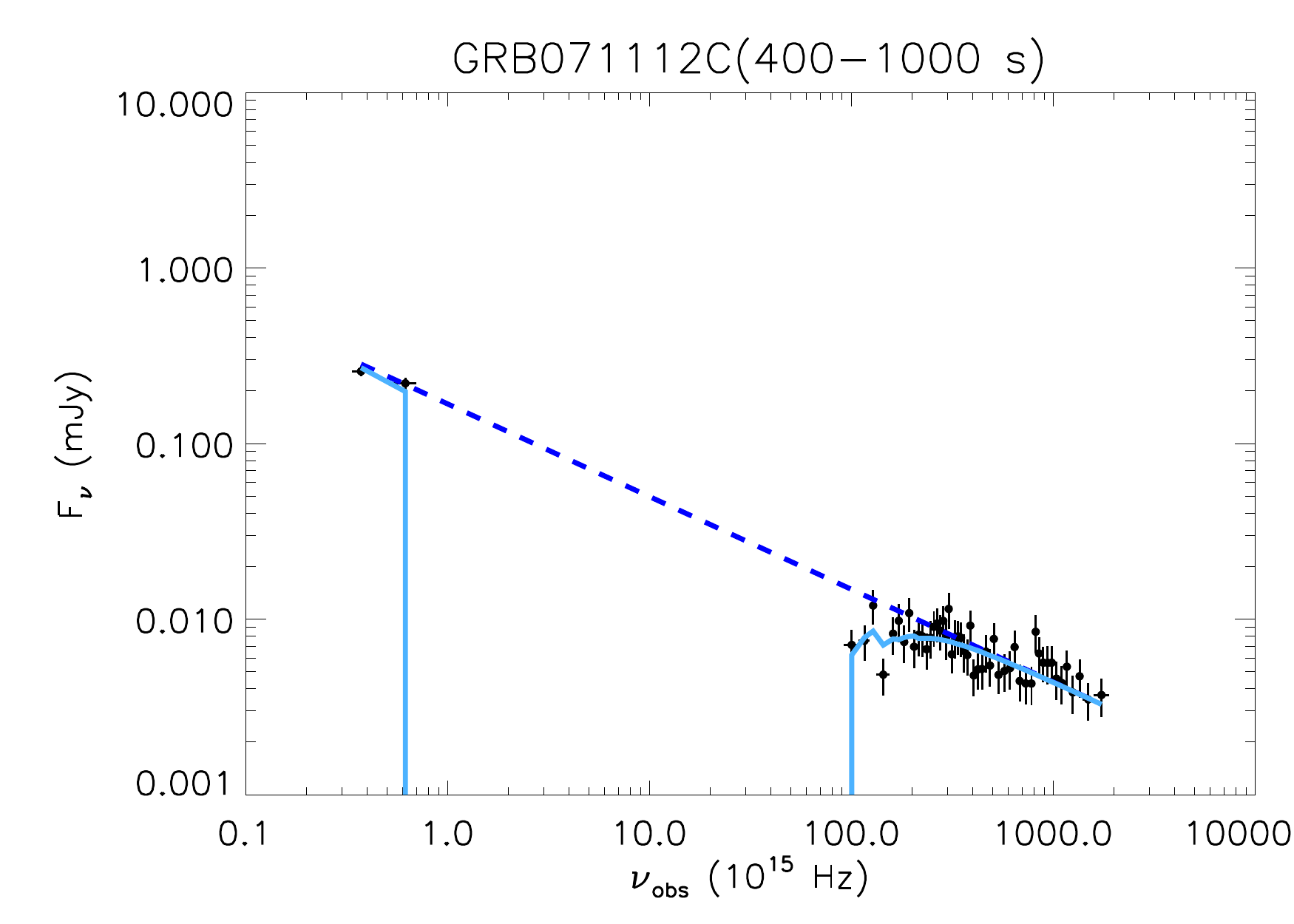}
\includegraphics[width=0.3 \hsize,clip]{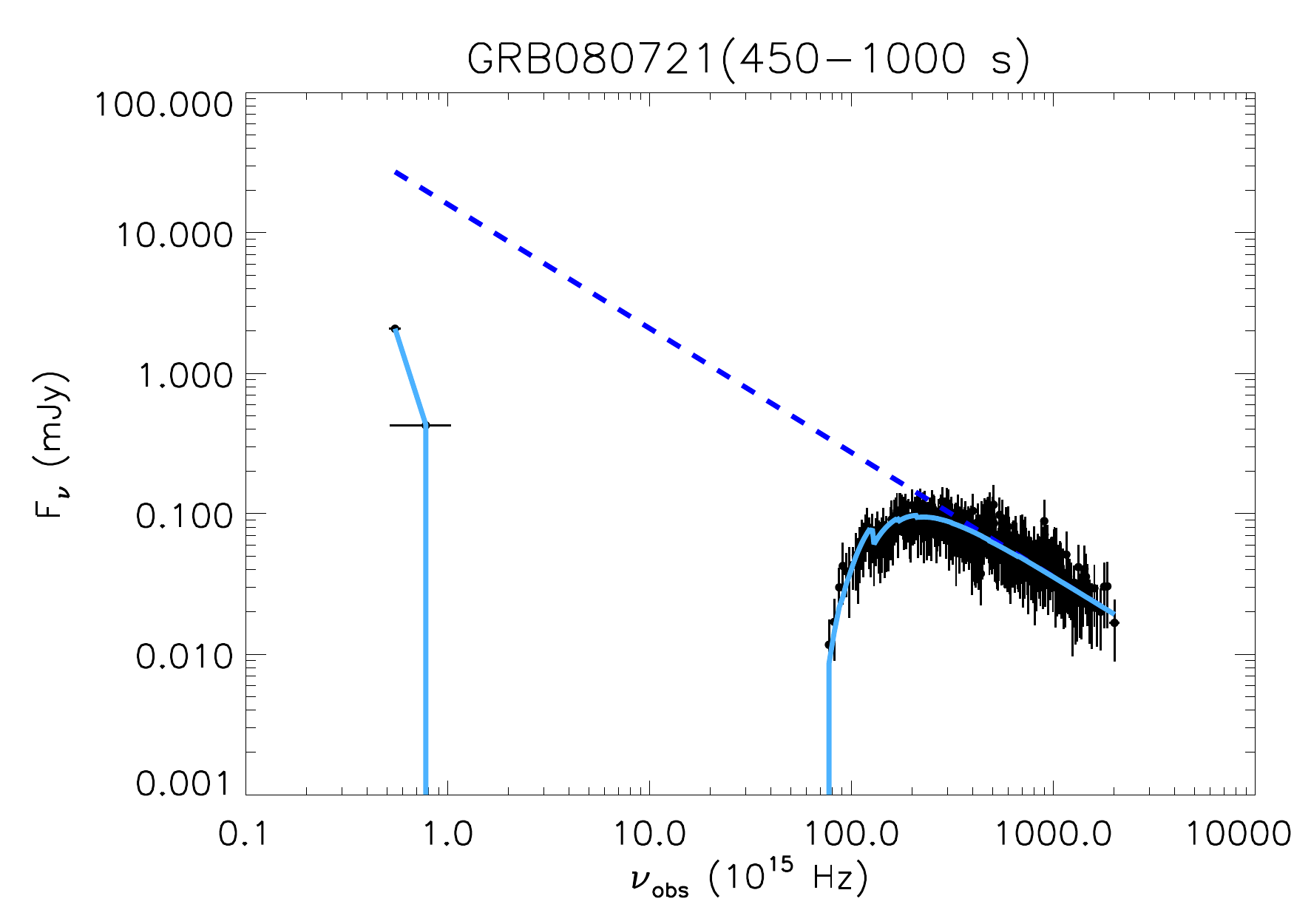}
\includegraphics[width=0.3 \hsize,clip]{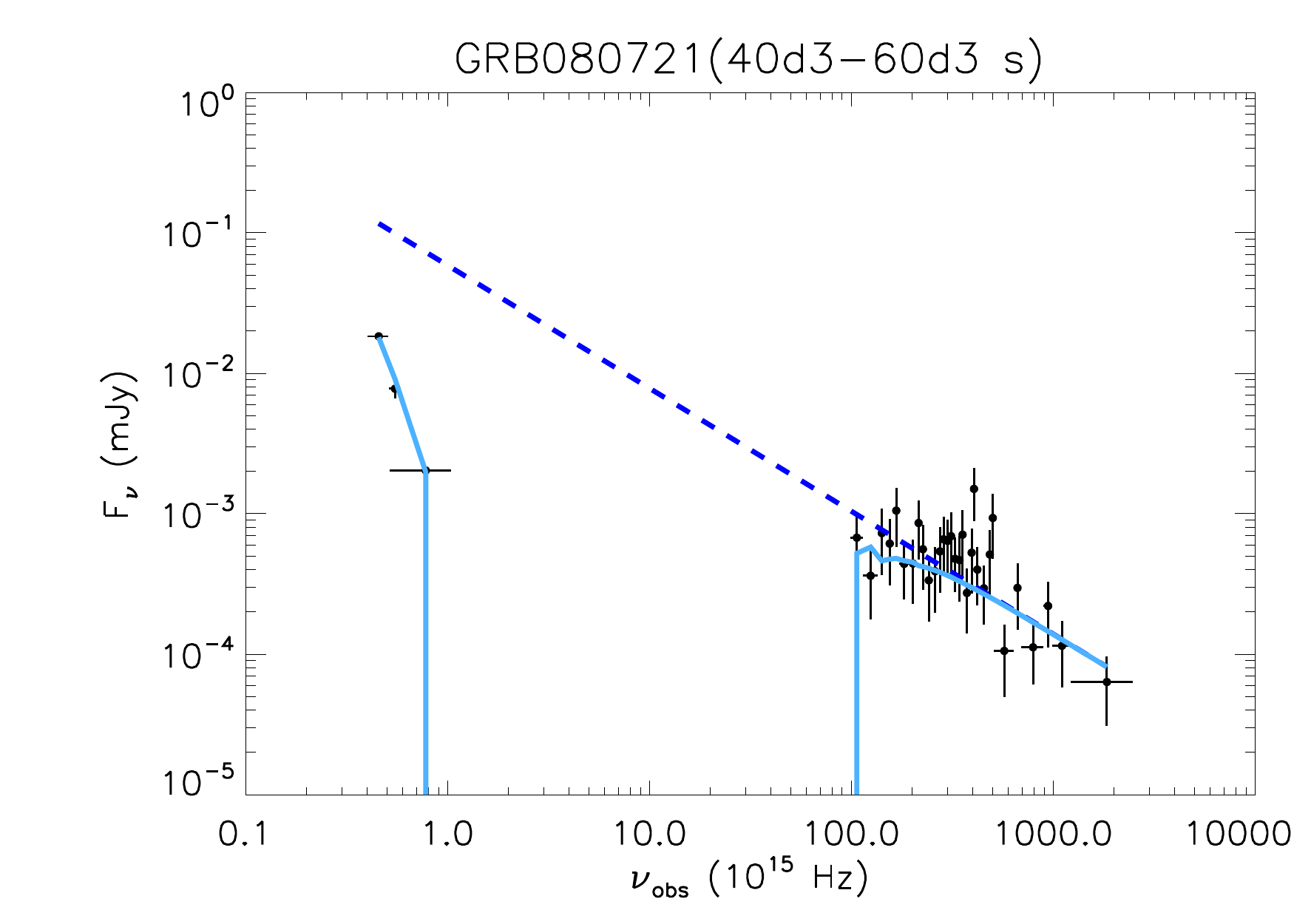}
\caption{\small{Optical/X-ray SEDs for GRBs belonging to Group C. color-coding as in Figure~\ref{sed1}.}}\label{sed7} 
\end{figure}

\begin{figure}
\includegraphics[width=0.3 \hsize,clip]{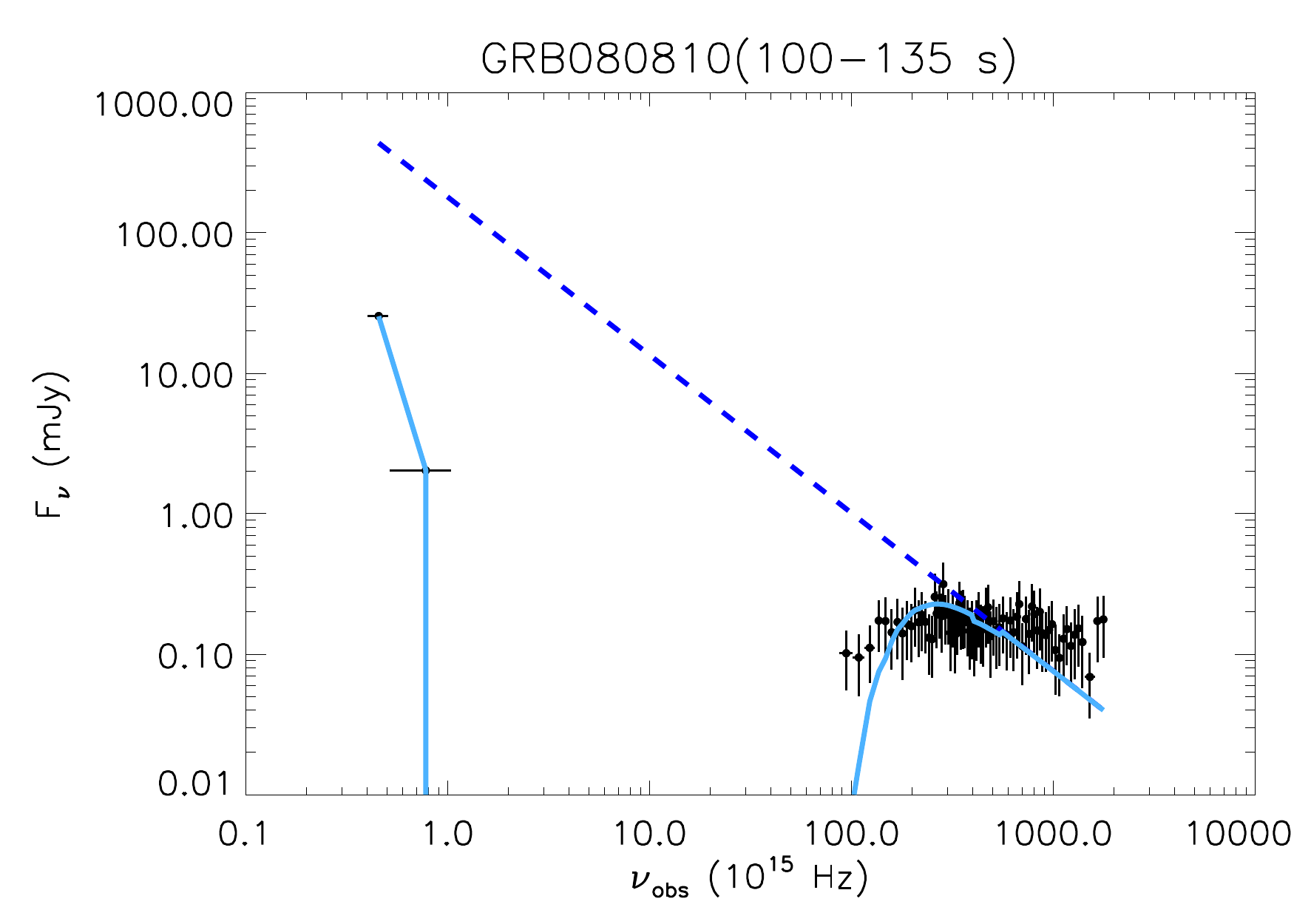}
\includegraphics[width=0.3 \hsize,clip]{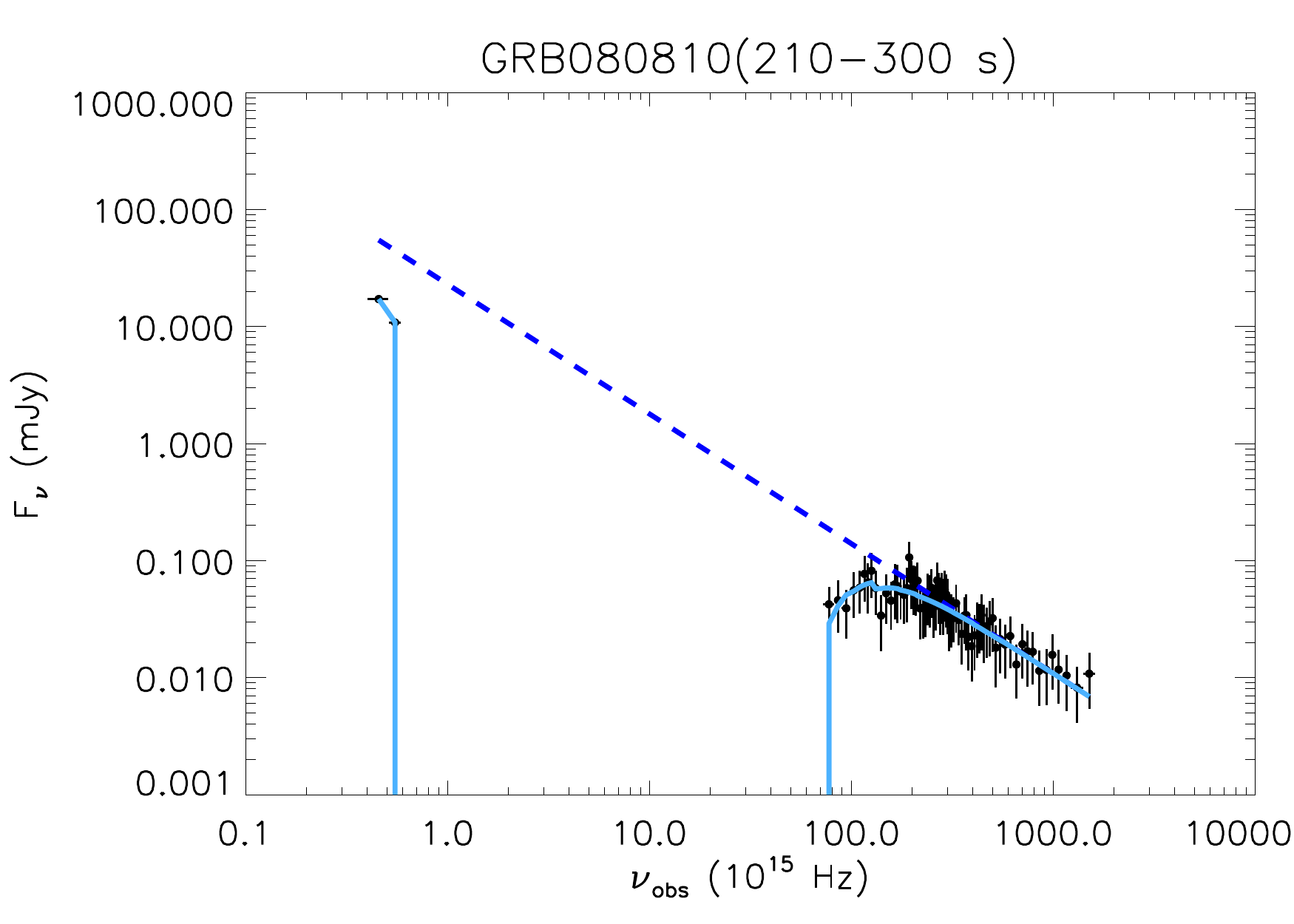}
\includegraphics[width=0.3 \hsize,clip]{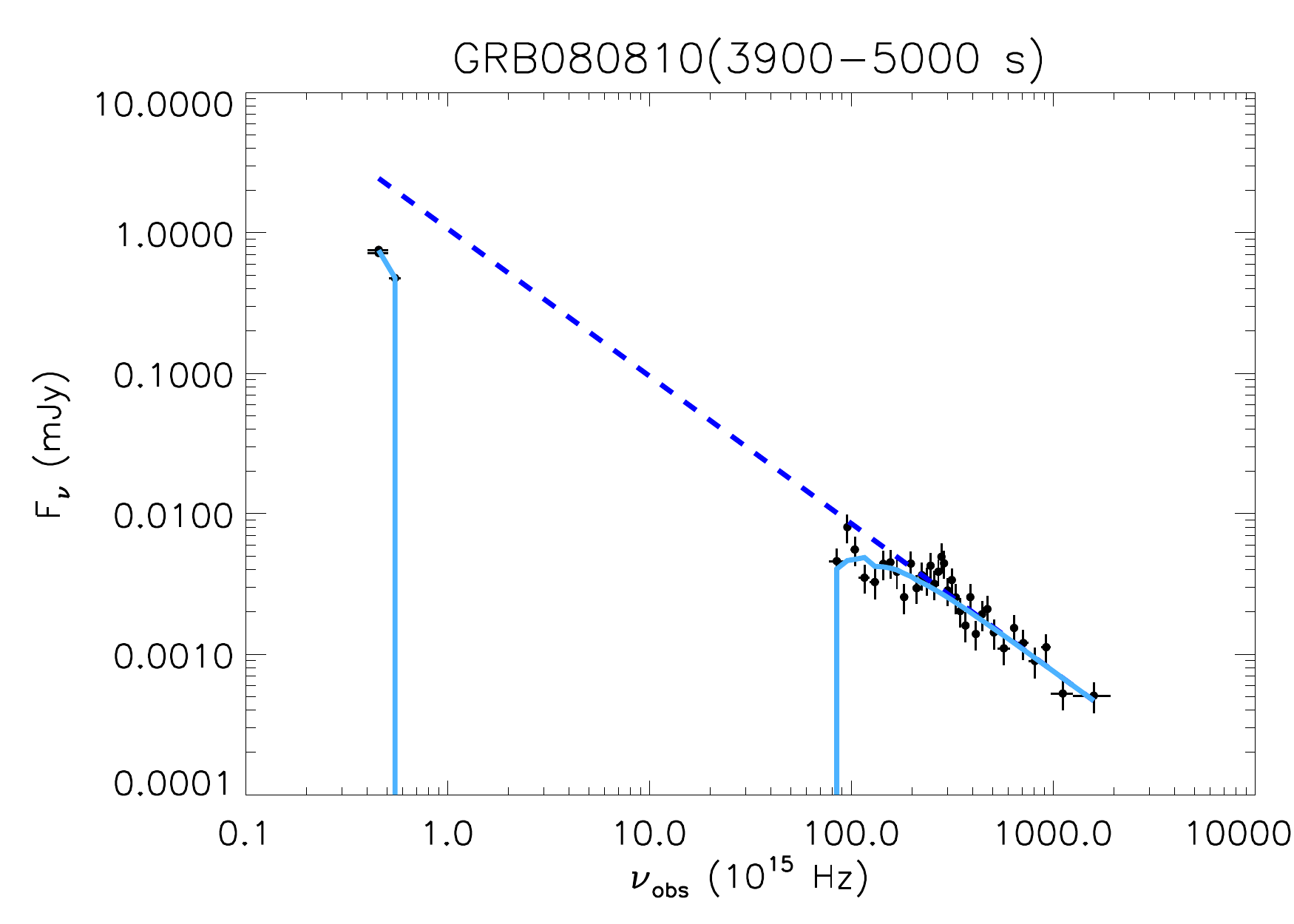}\\
\includegraphics[width=0.3 \hsize,clip]{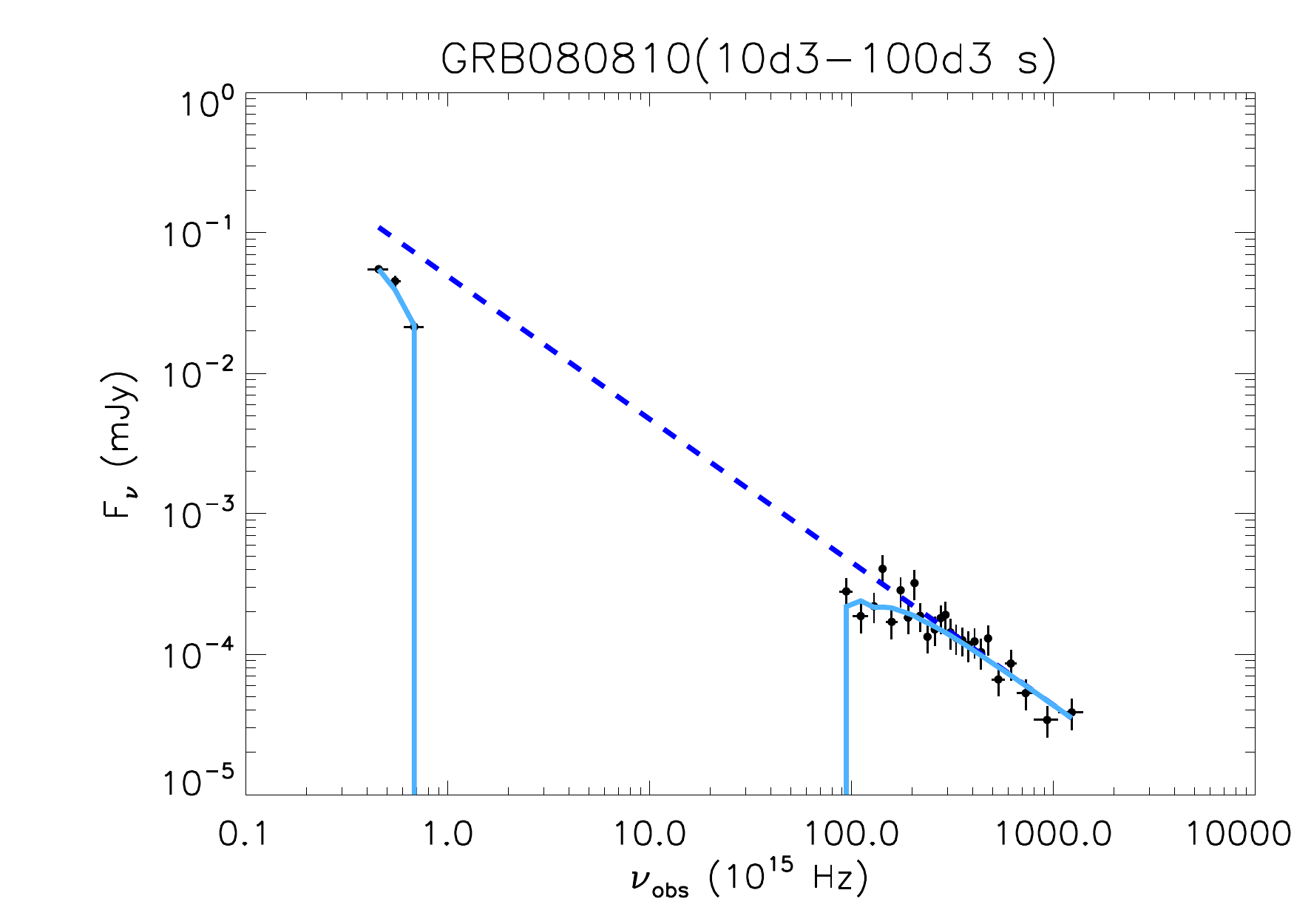}
\includegraphics[width=0.3 \hsize,clip]{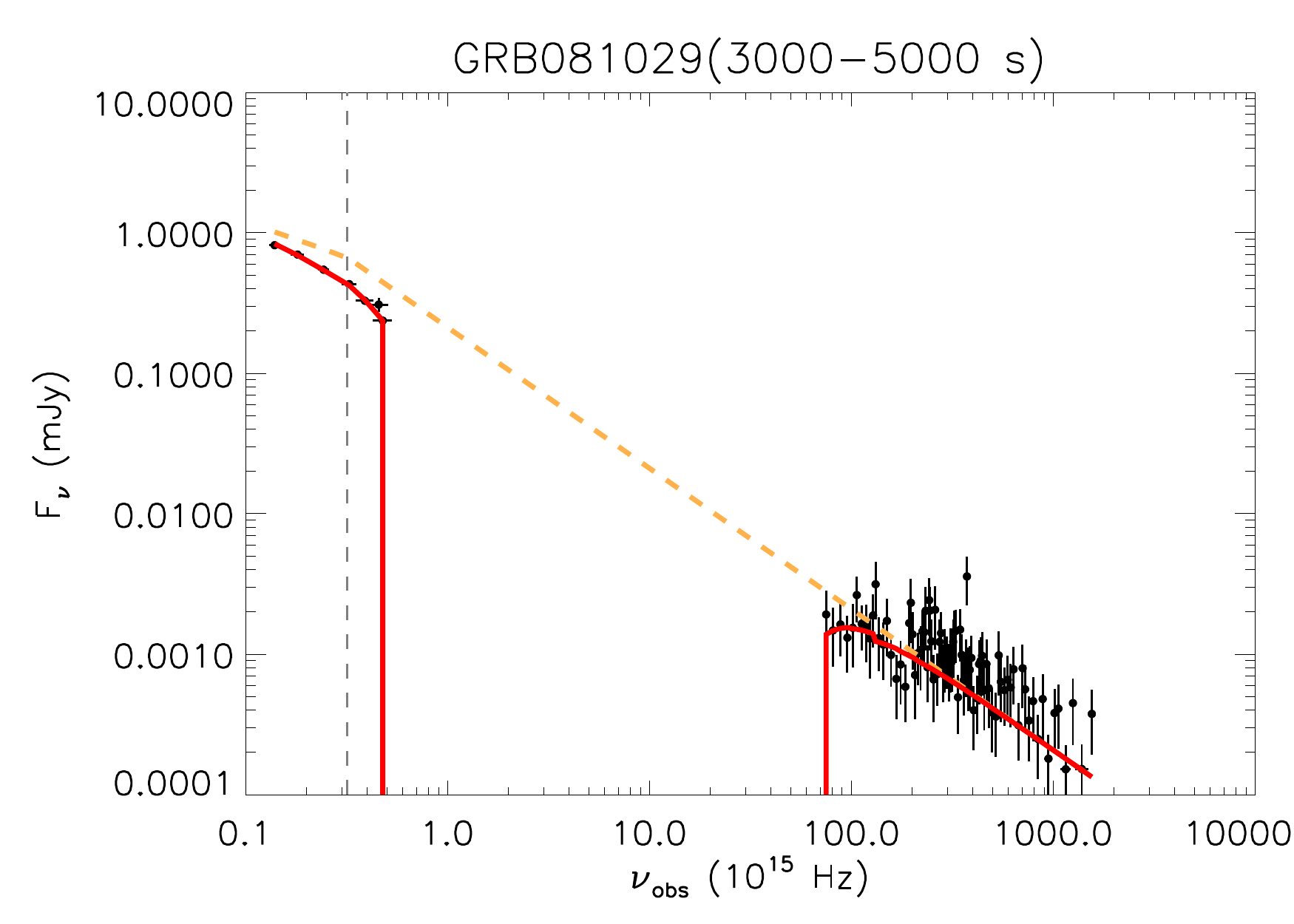}
\includegraphics[width=0.3 \hsize,clip]{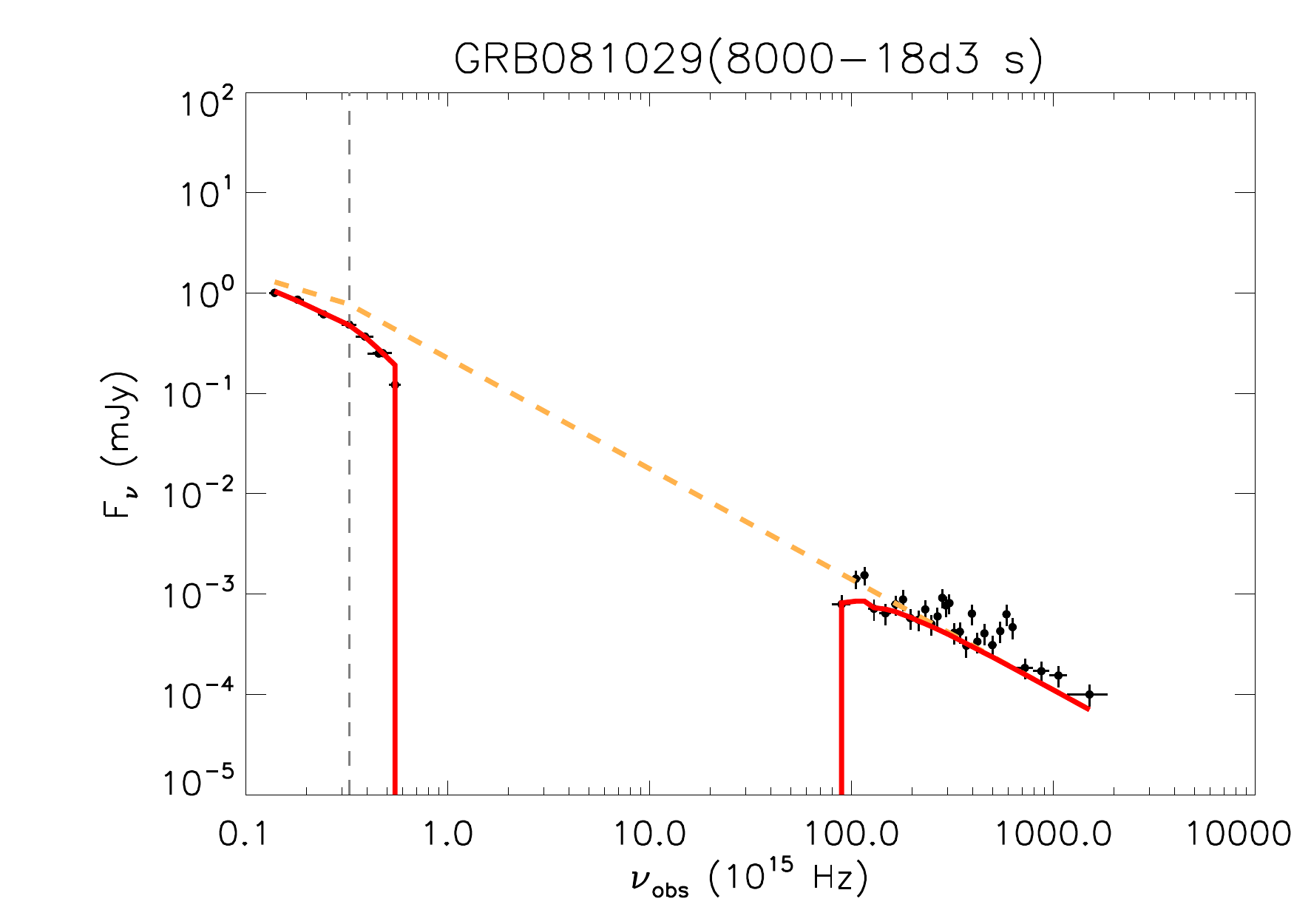}\\
\includegraphics[width=0.3 \hsize,clip]{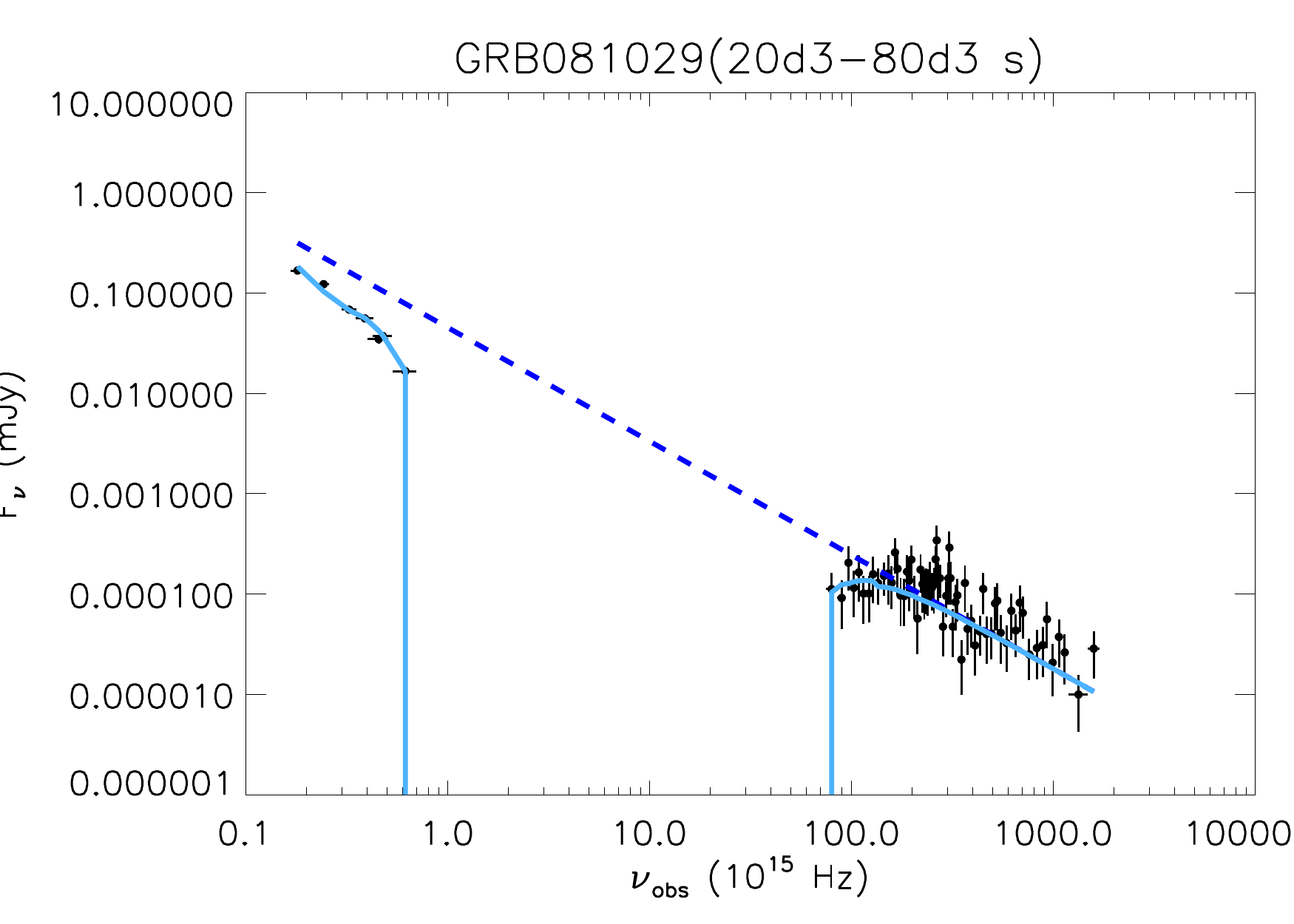}
\includegraphics[width=0.3 \hsize,clip]{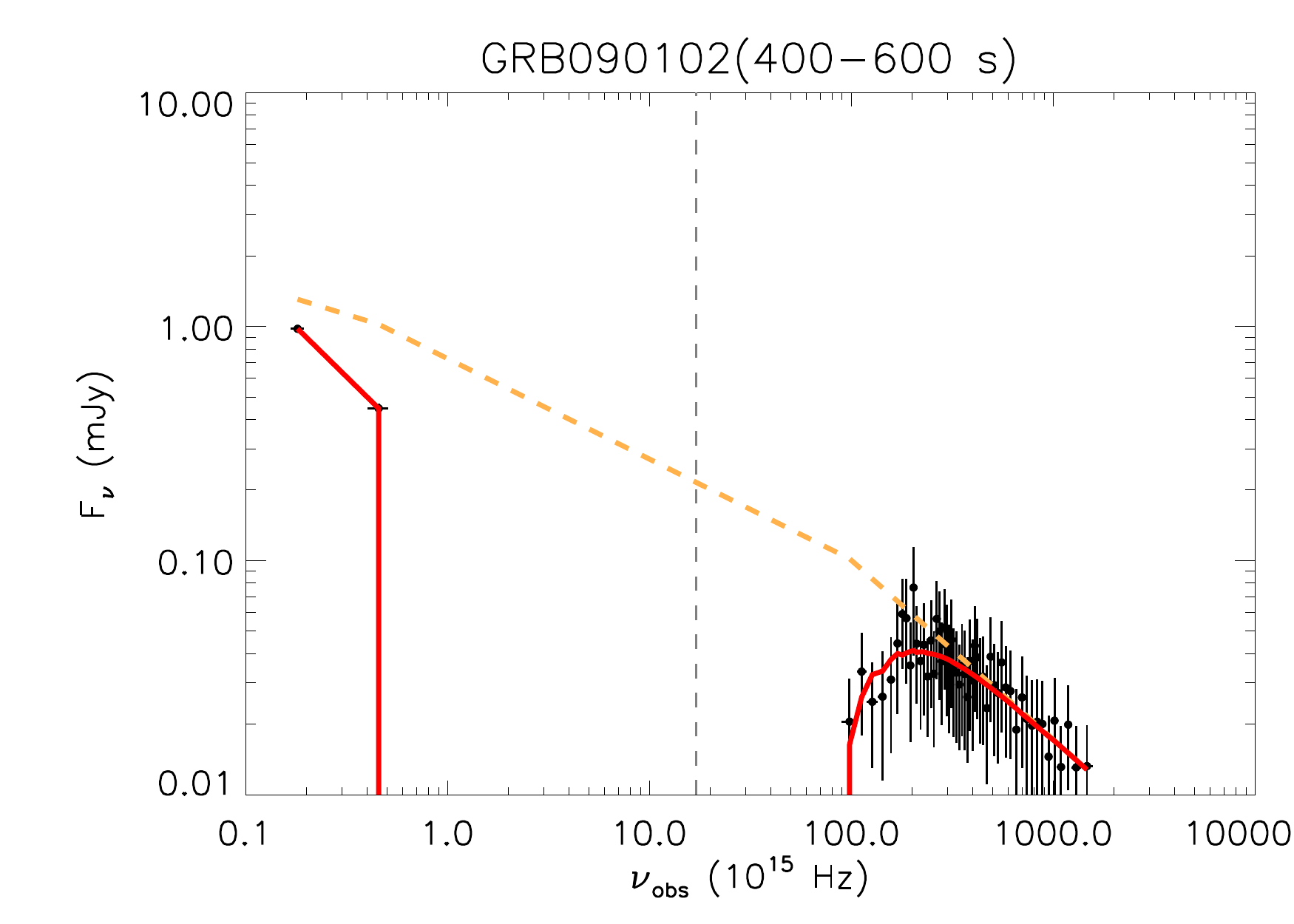}
\includegraphics[width=0.3 \hsize,clip]{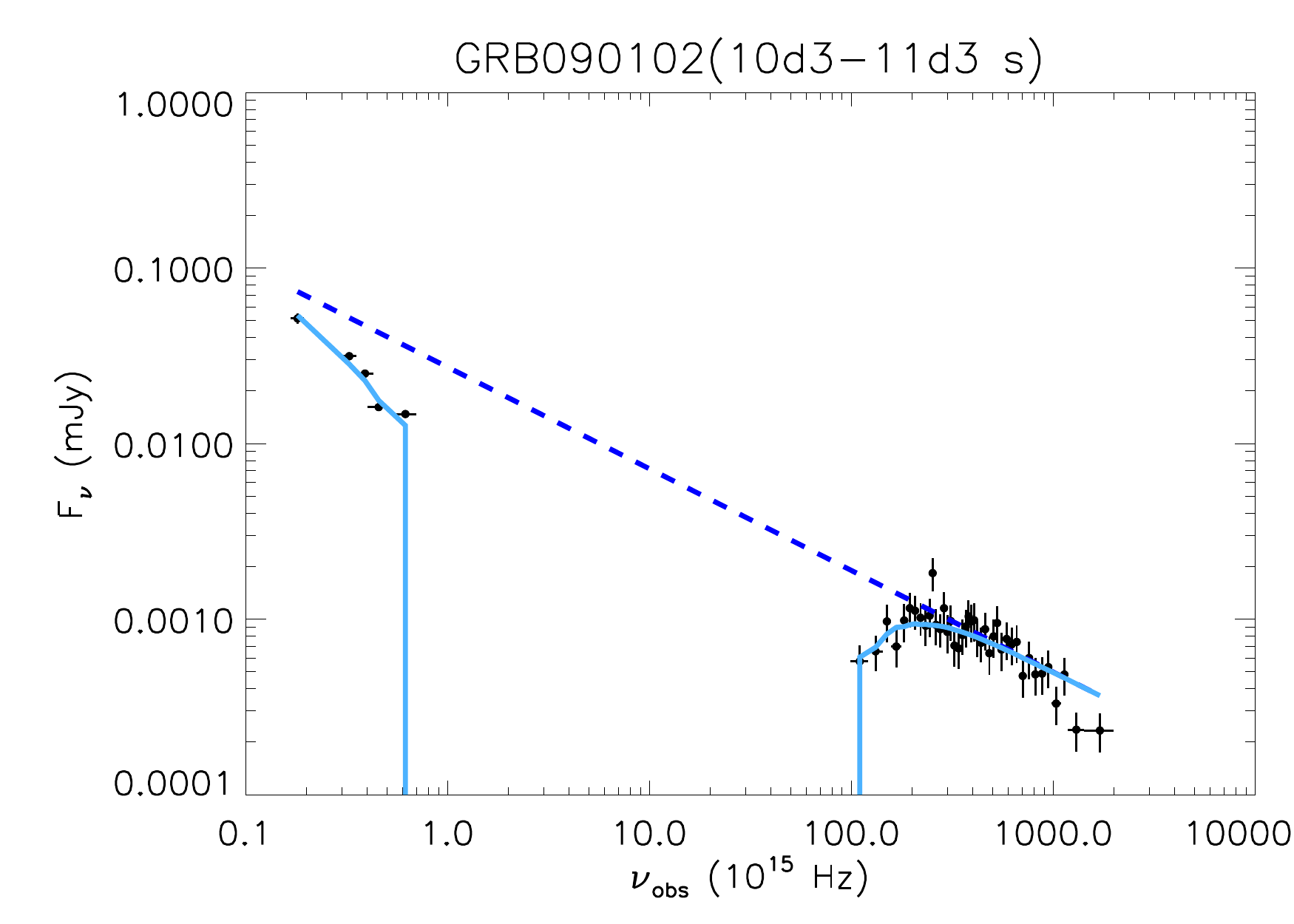}\\
\includegraphics[width=0.3 \hsize,clip]{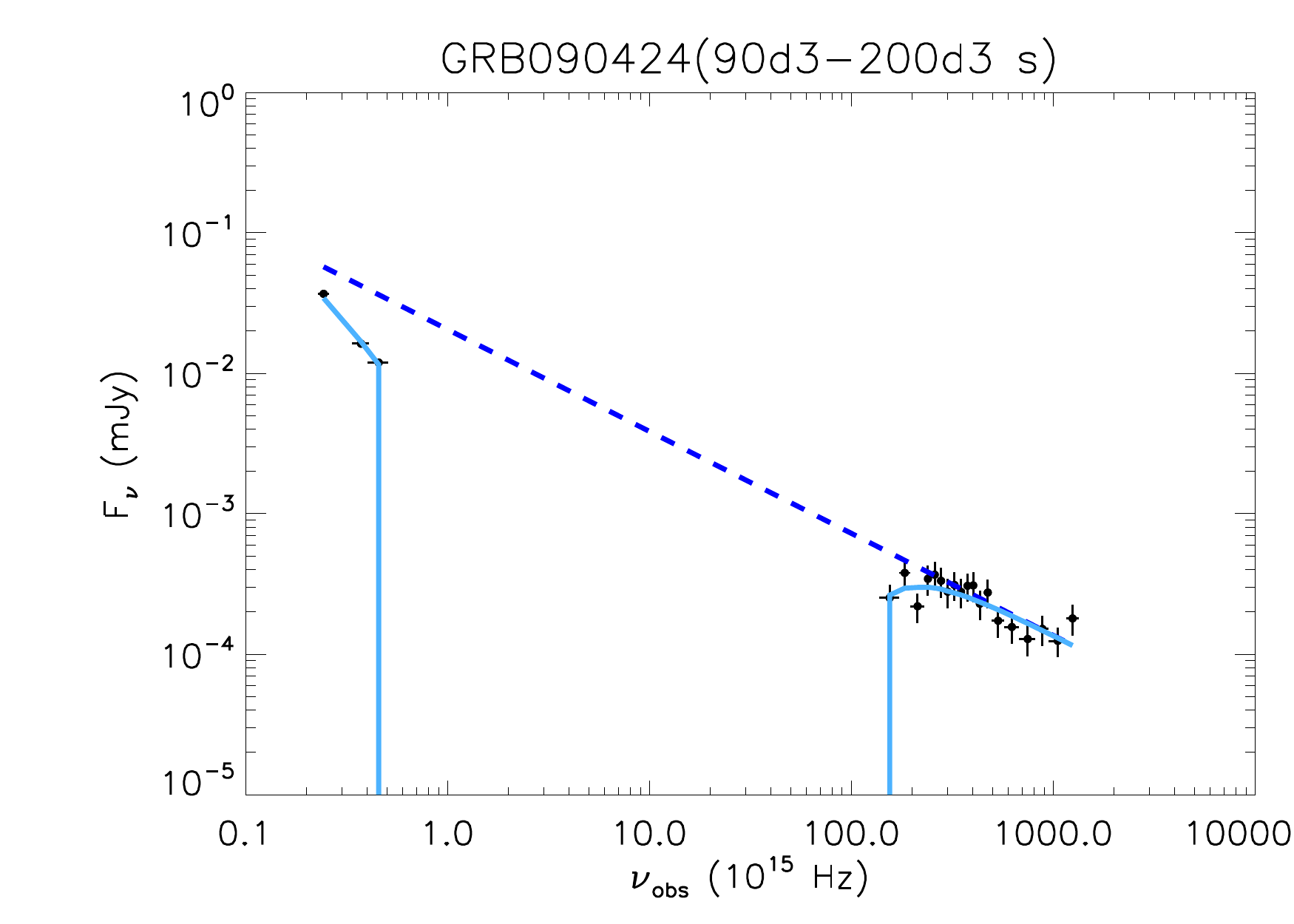}
\includegraphics[width=0.3 \hsize,clip]{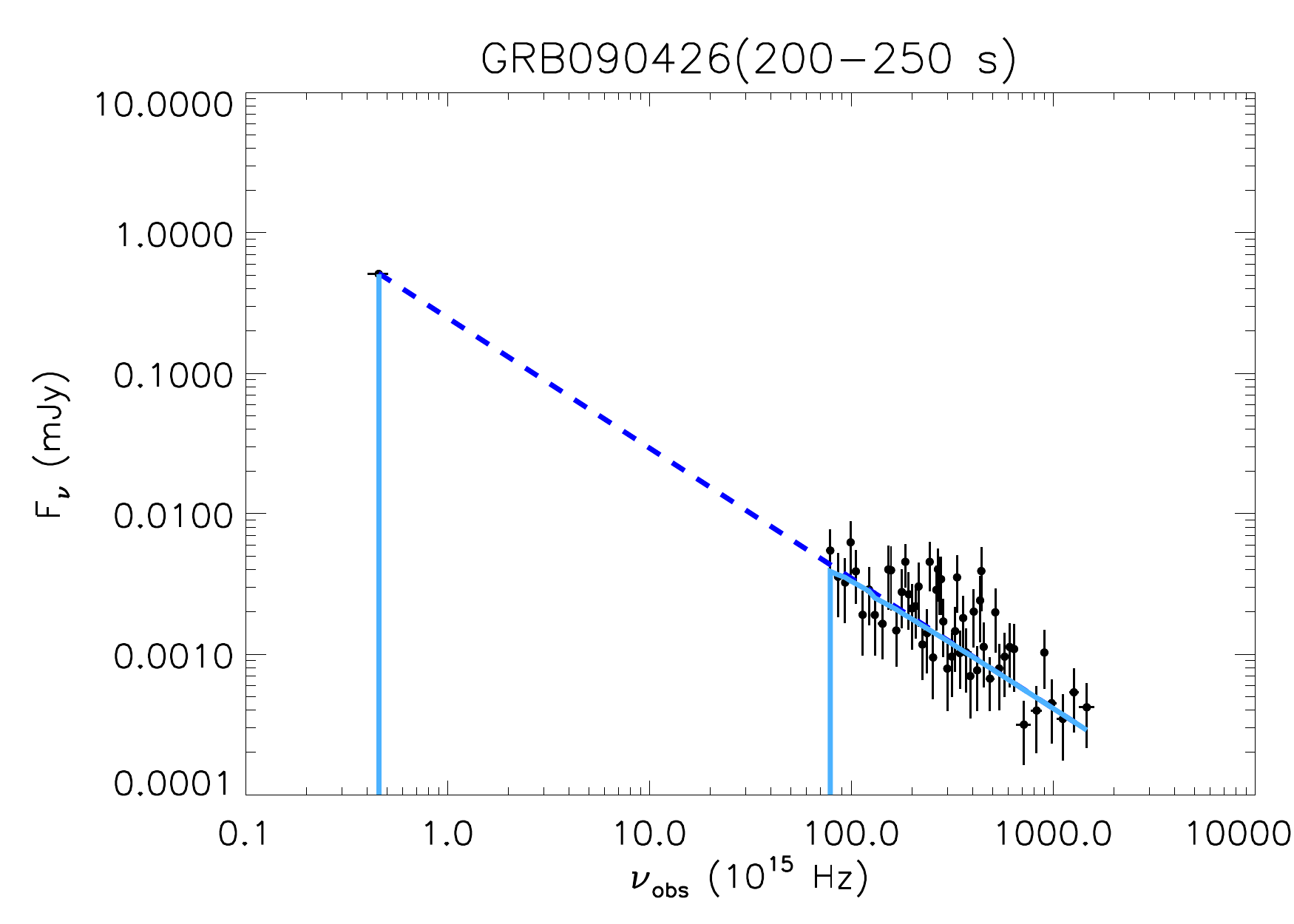}
\includegraphics[width=0.3 \hsize,clip]{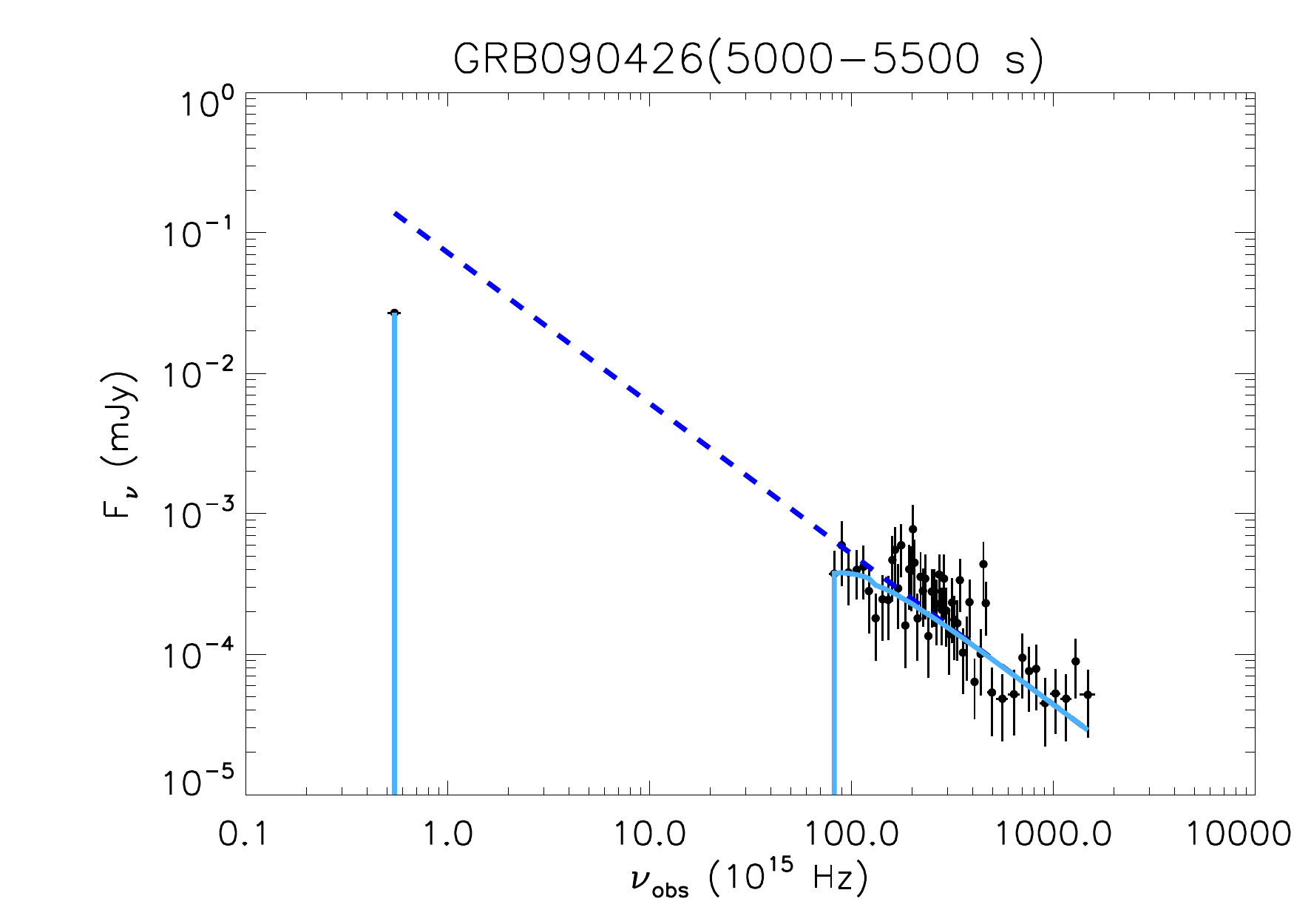}\\
\includegraphics[width=0.3 \hsize,clip]{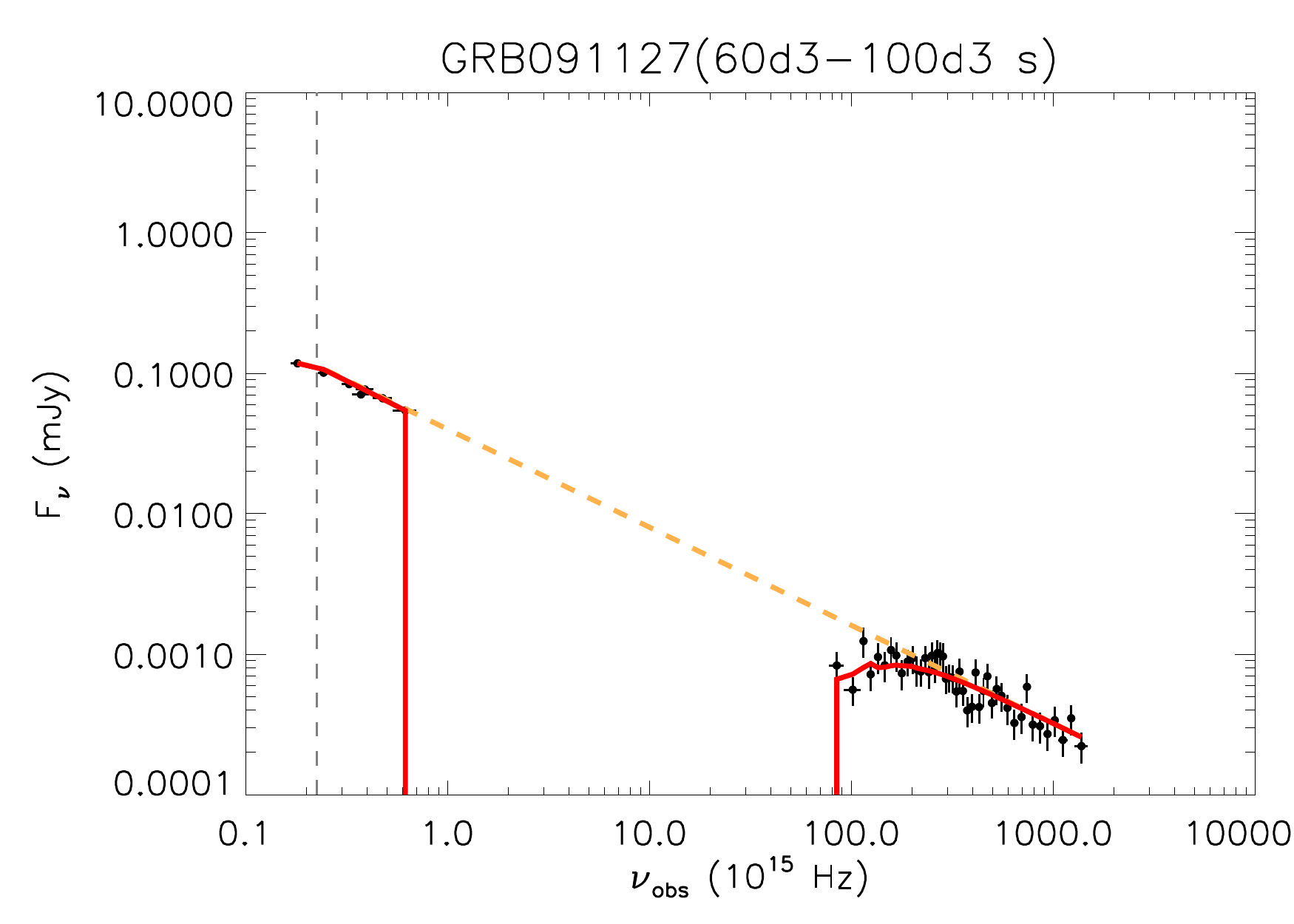}
\includegraphics[width=0.3 \hsize,clip]{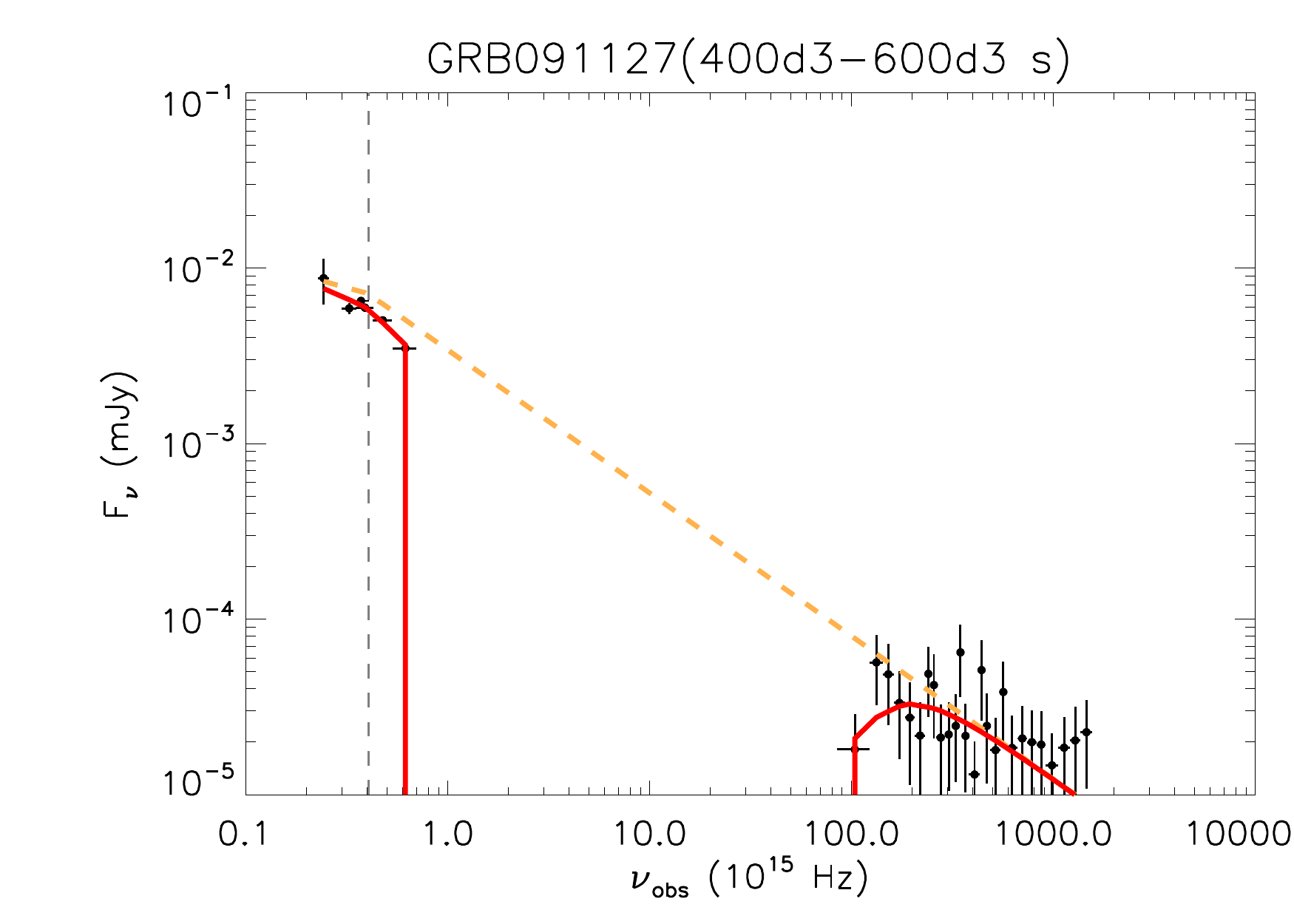}
\caption{\small{Optical/X-ray SEDs for GRBs belonging to Group C.  color-coding as in Figure~\ref{sed1}.}}\label{sed8} 
\end{figure}

We provide here:
\begin{itemize}
\item the examples of the online tables of CDS: the optical LC fit parameters (Tables~\ref{table1_a}); the parameters of the optical/X-ray SEDs fitted with a single power-law (Table~\ref{table2_a}) and with a broken power-law (Table~\ref{table3_a}); the X-ray spectrum fit parameters (Table~\ref{table4_a}); redshifts and luminosity distances of the GRBs in our sample (Table~\ref{table5_a}); the optical data used for the SEDs (Table~\ref{table6_a});
\item the result of the F-test over the optical/X-ray SEDs to choose the better fit function (Table~\ref{ftest});
\item the factors to convert magnitudes into flux densities (Table~\ref{conversion_factors});
\item the sample of the 165 GRBs with known redshift from which we started the data selection, with references to papers and GCNs whith optical data (Table~\ref{elenco});
\item plots of the optical and X-ray LCs for the GRBs in our sample (Figures~\ref{confronto1}-\ref{confronto9});
\item plots of the optical/X-ray SEDs (Figures~\ref{sed1}-\ref{sed8}).
\end{itemize}

\newpage

%
%%----------------------------------------------------------------

\begin{table*}
\caption{\small{Optical LC fit parameters. \textbf{GRB}: GRB name. \textbf{Fil}: filter name. \textbf{Fit}: fitting function. 0: $f(t) = N_1t^{-\alpha_1}$; 1: $f(t) = N_1((t/t_{br,1})^{-\alpha_1/s_1}+(t/t_{br,2})^{-\alpha_2/s_1})^{s_1}$; 2: $f(t) = N_1((t/t_{br,1})^{-\alpha_1/s_1}+(t/t_{br,2})^{-\alpha_2/s_1})^{s_1}+N_2((t/t_{br,3})^{-\alpha_3/s_2}+(t/t_{br,4})^{-\alpha_4/s_2})^{s_2}$; 4: $f(t) = N_1((t/t_{br,1})^{-\alpha_1/s_1}+(t/t_{br,2})^{-\alpha_2/s_1})^{s_1}+N_2(t)^{-\alpha_3}$; 6: $f(t) = N_1((t/t_{br,1})^{-\alpha_1/s_1}+(t/t_{br,2})^{-\alpha_2/s_1})^{s_1}+N_2(t)^{-\alpha_3}$; 8: $f(t) = N_1((t/t_{br,1})^{-\alpha_1/s_1}+(t/t_{br,2})^{-\alpha_2/s_1})^{s_1}+N_2((t/t_{br,3})^{-\alpha_3/s_2}+(t/t_{br,4})^{-\alpha_4/s_2})^{s_2}+N_3((t/t_{br,5})^{-\alpha_5/s_3}+(t/t_{br,6})^{-\alpha_6/s_3})^{s_3}$; 9: $f(t) = N_1((t/t_{br,1})^{-\alpha_1/s_1}+(t/t_{br,2})^{-\alpha_2/s_1})^{s_1}+N_2((t/t_{br,3})^{-\alpha_3/s_2}+(t/t_{br,4})^{-\alpha_4/s_2})^{s_2}+N_3(t)^{-\alpha_5}$. \textbf{Ok}: '1' denotes the data used for the comparison with the X-ray data and plotted in Figures \ref{confronto1}-\ref{confronto4} (\textit{Online Material}). \textbf{Tmin (Tmax)}: initial (end)  time of the observations in rest frame (s). $\mathbf{t_{br}}$, $\mathbf{\sigma(t_{br})}$: trigger times with relative errors (s). $\mathbf{\alpha}$, $\mathbf{\sigma(\alpha)}$: slopes with relative errors. \textbf{s}: smoothness parameters. $\mathbf{N}$, $\mathbf{\sigma{N}}$: normalization with relative error (mJy). $\mathbf{\chi^2}$: chi-square. \textbf{DOF}: degree of freedom. \textbf{pval}: p\_value. A ``-9'' indicates that the LC does not show this phase and the value of that parameter is therefore absent. This table is available in its entirety in a machine-readable form at the CDS (table1c.dat, ReadMe). A portion is shown here for guidance.}}\label{table1_a}          
\centering         
\tiny{                 
\begin{tabular}{l l l l l l l l l l l l l l}      
\hline\hline             
GRB&Fil&Fit&ok&Tmin&Tmax&$t_{br,1}$&$\sigma (t_{br,1})$&$t_{br,2}$&$\sigma (t_{br,2})$&$t_{br,3}$&$\sigma (t_{br,3})$&$\alpha{1}$&$\sigma(\alpha_1)$\\   
\hline                      
    050319    &     b &0&  0      &    243.0    &    28491.0   &        -9.0    &       -9.0   &        -9.0 &          -9.0     &      -9.0  &         -9.0   &   7.040e-01  &    7.955e-02\\      
    050319    &     v &0&  0      &    259.1    &   767337.7   &        -9.0    &       -9.0   &        -9.0 &          -9.0     &      -9.0  &         -9.0   &   6.618e-01  &    2.600e-02\\     
    050319    &    BJ &0&  0      &   5203.9    &    57196.8   &        -9.0    &       -9.0   &        -9.0 &          -9.0     &      -9.0  &         -9.0   &   4.112e-01  &    4.117e-02\\      
    050319    &    VJ &0&  0      &   6944.0    &   400550.4   &        -9.0    &       -9.0   &        -9.0 &          -9.0     &      -9.0  &         -9.0   &   6.905e-01  &    1.928e-02\\      
    050319    &    RC &2&  1      &     40.3    &   994118.4   &    157243.4    &    19520.9   &       482.5 &          30.1     &      -9.0  &         -9.0   &   4.457e-01  &    1.429e-02\\            
\end{tabular}
\begin{tabular}{l l l l l l l l l l }    
\hline\hline 
$\alpha_2$&$\sigma(\alpha_2)$&$\alpha_3$&$\sigma(\alpha_3)$&$\alpha_4$&$\sigma(\alpha_4)$&$\alpha_5$&$\sigma(\alpha_5)$&$\alpha_6$&$\sigma(\alpha_6)$\\
\hline 
-9.000e+00   &  -9.000e+00    & -9.000e+00  &   -9.000e+00 &    -9.000e+00  &   -9.000e+00   &  -9.000e+00  &   -9.000e+00    & -9.000e+00  &   -9.000e+00 \\ 
-9.000e+00   &  -9.000e+00    & -9.000e+00  &   -9.000e+00 &    -9.000e+00  &   -9.000e+00   &  -9.000e+00  &   -9.000e+00    & -9.000e+00  &   -9.000e+00  \\ 
-9.000e+00   &  -9.000e+00    & -9.000e+00  &   -9.000e+00 &    -9.000e+00  &   -9.000e+00   &  -9.000e+00  &   -9.000e+00    & -9.000e+00  &   -9.000e+00  \\ 
-9.000e+00   &  -9.000e+00    & -9.000e+00  &   -9.000e+00 &    -9.000e+00  &   -9.000e+00   &  -9.000e+00  &   -9.000e+00    & -9.000e+00  &   -9.000e+00  \\ 
 1.559e+00   &   1.039e-01    &  2.520e+00  &    2.744e-01 &     1.240e-01  &    7.844e-02   &  -9.000e+00  &   -9.000e+00    & -9.000e+00  &   -9.000e+00  \\  
 \end{tabular}
 \begin{tabular}{l l l l l l l l l l l l l }      
\hline\hline 
s$_1$&s$_2$&$s_3$&$N_1$&$\sigma(N_1)$&$N_2$&$\sigma(N_2)$&$N_3$&$\sigma(N_3)$&$\chi^2$&DOF&p\_val\\
\hline 
-9.0 &-9.0 &-9.0   & 1.107515e-02 & 6.496899e-03 &  -9.000000e+00  & -9.000000e+00  & -9.000000e+00 &  -9.000000e+00 &     8.40  &    7.00 &     2.985e-01\\
-9.0 &-9.0 &-9.0   & 1.747728e-02& 3.708279e-03 &  -9.000000e+00  & -9.000000e+00  & -9.000000e+00 &  -9.000000e+00 &    37.76  &   19.00 &     6.367e-03\\
 -9.0 &-9.0 &-9.0   & 8.883478e-04&3.640030e-04 &  -9.000000e+00  & -9.000000e+00  & -9.000000e+00 &  -9.000000e+00 &    18.59  &   10.00 &     4.578e-02\\
-9.0 &-9.0 &-9.0   & 2.510829e-02& 4.822976e-03 &  -9.000000e+00  & -9.000000e+00  & -9.000000e+00 &  -9.000000e+00 &    85.86  &   18.00 &     7.473e-06\\
 -0.5 &-0.5 &-9.0   & 1.446221e-05&1.248363e-06 &   2.000459e-04  &  0.000000e+00  & -9.000000e+00 &  -9.000000e+00 &   203.49  &   64.00 &    -3.484e-07\\
\hline                                 
\end{tabular}
}
\end{table*}
%%%%%%%%%%%%%%%%%%%%%%%%%%%
\begin{table}
\caption{\small{Parameters of the optical/X-ray SEDs fitted with a single power-law. \textbf{GRB}: GRB name. \textbf{Host}: the chosen extinction law; Milky Way (MW), Large Magellanic Cloud (LMC), Small Magellanic Cloud (SMC). \textbf{Part}: X-ray LC part considered; S steep decay, P plateau, N normal decay. $\mathbf{\beta_{op,X}}$, $\mathbf{\sigma(\beta_{op,X})}$: spectral index with error. \textbf{NH}, $\mathbf{\sigma(NH)}$: hydrogen column density with relative error ($10^{22}\ \rm cm^{-2}$). \textbf{E(B-V)}, $\mathbf{
\sigma}$\textbf{(E(B-V))}: optical absorption with relative error (mag). \textbf{N}, $\mathbf{\sigma}$\textbf{(N)}: normalization with error (mJy). $\mathbf{\chi^2}$: chi-square. \textbf{DOF}: degree of freedom. \textbf{pval}: p\_value. $\mathbf{T_i}$, $\mathbf{T_f}$: initial (end) time of the interval of the SED in observer time (s). \textbf{ok}: 1 if the best fitting function is a power-law, otherwise 0. This table is available in its entirety in a machine-readable form at the CDS (table2c.dat, ReadMe). A portion is shown here for guidance.}}\label{table2_a}             
\centering               
\tiny{                 
\begin{tabular}{p{0.7cm}p{0.3cm}p{0.3cm}p{0.65cm}p{0.65cm}p{1.cm}p{0.5cm}p{0.9cm}p{1.4cm}p{0.6cm}p{0.6cm}p{1.cm}p{0.6cm}p{1.3cm}p{0.6cm}p{0.6cm}p{0.3cm}}      
\hline\hline 
GRB&Host&Part&$\beta_{op,X}$&$\sigma(\beta_{op,X})$&NH&$\sigma(NH)$&E(B-V)&$\sigma$(E(B-V))&N&$\sigma(N)$&$\chi^2$&DOF&pval&T$_i$&T$_f$&ok\\
\hline
    050319  &MW &P&          0.755         & 0.020&         -0.269        &  0.184 &         0.082       &   0.019&          0.215          &0.025&      9.976e+01        &42  &    1.340e-06    &300.&      1500.&0\\
    050319  &MW &N  &        0.945        &  0.026 &        -0.204      &    0.420  &        0.030      &    0.025  &        0.006         & 0.001 &     7.258e+01       & 10   &   3.190e-07 &17000.  &   300000. &0\\
    050401  &MW &P     &     0.873       &   0.075   &       1.840       &   0.322    &      0.667       &   0.082     &     1.830        &  0.851   &   3.331e+01       &146     & 1.000e+00 &  200. & 2000.&1\\
    050408 &SMC &N      &    0.679    &      0.016   &       0.632    &      0.337    &      0.247    &      0.007      &    0.012     &     0.000   &   2.273e+01     &   28     & 7.470e-01&20000. & 40000.&0\\
   050416A &LMC& P        &  0.627&          0.015    &      0.307 &         0.086     &     0.232 &         0.041        &  0.054  &        0.005    &  2.702e+01  &      24   &   3.040e-01& 700.&    2000.&0\\
\hline                                 
\end{tabular}
}
\end{table}

%%%%%%%%%%%%%%%%%%%%%%%%%%%
\begin{table*}
\caption{\small{Parameters of the optical/X-ray SEDs fitted with a broken power-law. \textbf{GRB}: GRB name. \textbf{Host}: galaxy chosen as model to fit the SED; Milky Way (MW), Large Magellanic Cloud (LMC), Small Magellanic Cloud (SMC). \textbf{Part}: X-ray LC part considered; ``S'' steep decay, ``P'' plateau, ``N'' normal decay. $\mathbf{\beta_{op}}$, $\mathbf{\sigma(\beta_{op})}$: optical spectral index with error. $\mathbf{\beta_{X}}$, $\mathbf{\sigma(\beta_{X})}$: X-ray spectral index with error. $\mathbf{\nu}$, $\mathbf{\sigma(\nu)}$: observer frame break frequency ($10^{15}$ Hz). \textbf{NH}, $\mathbf{\sigma(NH)}$: hydrogen column density with relative error ($10^{22}\ \rm cm^{-2}$). \textbf{E(B-V)}, $\mathbf{
\sigma}$\textbf{(E(B-V))}: optical absorption with relative error (mag). \textbf{N}, $\mathbf{\sigma}$\textbf{(N)}: normalization with error (mJy). $\mathbf{\chi^2}$: chi-square. \textbf{DOF}: degree of freedom. \textbf{pval}: p\_value. $\mathbf{T_i}$, $\mathbf{T_f}$: initial (end) time of the interval of the SED in observer time (s). \textbf{ok}: ``1'' if the best fitting function is a broken power-law, otherwise ``0''. This table is available in its entirety in a machine-readable form at the CDS (table3c.dat, ReadMe). A portion is shown here for guidance.}}\label{table3_a}             
\centering               
\tiny{                 
\begin{tabular}{lllllllllll}      
\hline\hline 
GRB&Host&Part&$\beta_{op}$&$\sigma(\beta_{op})$&$\beta_X$&$\sigma(\beta_X)$&$\nu_{br}$&$\sigma(\nu_{br)}$&NH&$\sigma$(NH)\\
\hline
 050319&  MW &P&             0.358             &0.000 &            0.858             &0.077&             1.271             &1.198&            -0.030             &0.283 \\
 050319 & MW &N&             0.607          &   0.000   &          1.107          &   0.045  &           0.545             &0.118  &           0.114            & 0.437 \\    
 050401 &SMC &P&             0.372       &      0.000     &        0.872       &      0.146    &         0.455          &  87.600   &          1.850           &  2.540 \\    
 050408 &LMC &N&             0.342    &         0.000        &     1.210    &         0.310       &     22.350      &      26.190    &         1.560       &      0.690 \\    
050416A &SMC& P&             0.619&             0.000         &    0.951&             0.156        &   339.900&           131.310  &           0.398&             0.097 \\         
\hline\hline 
E(B-V)&$\sigma$(E(B-V))&N&$\sigma(N)$&$\chi^2$&DOF&pval&T$_i$&T$_f$&ok&\\
\hline
   0.103        &     0.022           &  0.327&             0.048    &4.050e+01 &     40   & 4.467e-01&      300.000     &1500.000& 1&\\
   0.165           &  0.034        &     0.021   &          0.005    &4.050e+01    &   9   & 3.770e-05   & 17000.000   &300000.000& 1&\\
   0.462       &     10.500    &         2.700    &       263.000  &  2.050e+02   &  145&    7.460e-04 &     200.000 &    2000.000& 0&\\
   0.290         &    0.000   &          0.020        &     0.001   & 2.654e+01     & 27&   4.885e-01    &20000.000  &  40000.000& 1&\\
   0.321           &  0.044&             0.060           &  0.005&    1.890e+01      &23&    7.059e-01     & 700.000 &    2000.000& 1&\\
\hline                                 
\end{tabular}
}
\end{table*}
%%%%%%%%%%%%%%%%%%%%%%%%%%%%
\begin{table*}
\caption{\small{Parameters of the X-ray spectrum fit.  \textbf{GRB}: GRB name. \textbf{N}, $\mathbf{\sigma}$(N): normalization with error (keV). $\mathbf{\beta}$, $\mathbf{\sigma(\beta)}$: spectral index. \textbf{NH}, $\mathbf\sigma$\textbf{(NH)}: hydrogen column density with error ($10^{22}\ \rm cm^{-2}$). $\mathbf{T_i}$, $\mathbf{T_f}$: initial (end) time of the interval of the SED in observer time (s). This table is available in its entirety in a machine-readable form at the CDS (table4c.dat, ReadMe). A portion is shown here for guidance.}}\label{table4_a}             
\centering               
\tiny{                 
\begin{tabular}{lllllllll}      
\hline\hline 
GRB&N&$\sigma(N)$&$\beta$&$\sigma(\beta)$&NH&$\sigma$(NH)&T$_i$&T$_f$\\
\hline
050319         &    0.365           &  0.170 &            0.857            & 0.077 &           -0.005            & 0.284 &          300.000          &1500.000\\
  050319         &    0.006       &      0.009  &           0.941         &    0.246  &          -0.310         &    0.823  &       17000.000     &   300000.000\\
  050401           &  1.719     &        0.790     &        0.864       &      0.075     &        2.000       &      0.334     &      200.000      &    2000.000\\
  050408            & 0.345   &          0.686        &     1.200     &        0.314        &     1.620    &        0.692         &20000.000   &      40000.000\\
 050416A             &1.159&             1.082         &    1.104  &           0.146         &    0.791 &            0.168         &  700.000 &         2000.000\\
\hline                                 
\end{tabular}
}
\end{table*}
%%%%%%%%%%%%%%%%%%%%%%%%%%%%
\begin{table*}
\caption{\small{Information on the GRBs in our sample. \textbf{GRB}: GRB name. \textbf{z}: redshift. \textbf{DL}: luminosity distance (Gpc). Redshifts and luminosity distances are taken from M13. This table is available in its entirety in a machine-readable form at the CDS (table5c.dat, ReadMe). A portion is shown here for guidance.}}\label{table5_a}             
\centering               
\tiny{                 
\begin{tabular}{lllllllll}      
\hline\hline 
GRB&z&DL\\
\hline
    050319  &   3.240         &28.369\\
        050401    & 2.900         &24.806\\
    050408     &1.236       &   8.644\\
   050416A    & 0.653   &       3.913\\
   050525A     &0.606&          3.567\\
\hline                                 
\end{tabular}
}
\end{table*}
%%%%%%%%%%%%%%%%%%%%%%%%%%%%
\begin{table*}
\caption{\small{Optical data used in the SED. \textbf{GRB}: GRB name. \textbf{Fil}: filter. $\lambda_c$: central wavelength of the filter ($\AA$). \textbf{FWHM}: full width half maximum of the filter ($\AA$). \textbf{F}, $\mathbf{\sigma}$\textbf{(F)}: flux with error (mJy). This table is available in its entirety in a machine-readable form at the CDS (table6c.dat, ReadMe). A portion is shown here for guidance.}}\label{table6_a}             
\centering               
\tiny{                 
\begin{tabular}{llllll}      
\hline\hline 
GRB&Filter&$\lambda_c$&FWHM&F&$\sigma$(F)\\
\hline
  050319 &      b           &4392&            975      &1.029e-01 &     1.229e-02\\
    050319  &     v          & 5468 &           769      &2.144e-01  &    1.297e-02\\
    050319     & RC    &       6588 &          1568   &   2.184e-01&      3.237e-03\\
    050319      &CR &          6588    &       1568&      2.901e-01   &   8.830e-03\\
    050319       &v &          5468          &  769&      8.072e-03      &8.283e-04\\
\hline                                 
\end{tabular}
}
\end{table*}
%%%%%%%%%%%%%%%%%%%%%%%%%%%%
\begin{table}
\caption{\small{Conversion factors used to convert magnitude into flux density (Jy). Reference: (1) http://www.ipac.caltech.edu/2mass/releases/allsky/faq.html\#jansky; (2) UVOT calibration files; (3) \cite{1995PASP..107..945F}; (4) http://svo.cab.inta-csic.es/; (5) \cite{1983ApJ...264..337S}.}}
\centering
\tiny{
\begin{tabular}{lccccc}
&&&&&\\
\hline
\hline
Filter    &$\lambda_c$ ($\AA$) & FWHM ($\AA$)&   F$_0$ (Jy)& Reference\\
\hline
J         & 12350  &1620  &1594  & (1)\\
H        & 16620  &2510  &1024        & (1)\\ 
KS      & 21590  & 2620  &666.7 & (1)\\
\hline
v	  &5468&   769&   3647   &  (2)\\             
b         &4392&   975&   4058   &  (2)\\       
u         & 3465&   785&   1423   &  (2)\\     
uvw1  & 2600&   693&   901&  (2)\\      
uvm2  & 2246&   493&	 767&  (2)\\  
uvw2   &1928&   657&	 716&  (2)\\    
white   &3850&   2600&  1732&  (2)\\     
\hline
U$_J$   &     3652&  526  & 1890  &(3)\\        
B$_J$    &    4448 &  1008&  4020 &(3)\\      
V$_J$    &    5505 &  827   &3590 &(3)\\       
R$_C$   &     6588 &  1568 &3020 &(3)\\      
I$_C$     &   8060 &  1542  &2380&(3)\\    
\hline
u'    &    3585 &  556   &1568.5& (4)\\                
g'     &   4858	&  1297	& 3965.9& (4)\\                 
r'	&  6290&	  1358	 &3162.0&(4)\\                
i'	&  7706&	  1547	 &2602.0& (4)\\               
z'	&  9222&	  1530	& 2244.7&   (4)\\               
\hline
g$_{Gunn}$ &	  5200 &     1400&  3730 &(5)\\                 
r$_{Gunn}$&	  6700&	  1400&	 4490  &(5)\\                
i$_{Gunn}$&	  7900&	  1600&	 4760. &(5)\\                
z$_{Gunn}$&	  9100&	  1300&	 4810 &(5)\\              
\hline
U$_{Bessel}$&	  3605.1  &3031&  1803.1  &(4)\\      
B$_{Bessel}$&	  4400.0  &3701&	 4000.0  &(4)\\      	            
V$_{Bessel}$&        5512.1  &4733&	 3579.8  &(4)\\      	
R$_{Bessel}$&	  6585.9  &5504&	 2971.4  &(4)\\      	
I$_{Bessel}$&	  8059.9  &7042  &2405.3  &(4)\\  
\hline    
\end{tabular}
}
\label{conversion_factors}
\end{table}
%%%%%%%%%%%%%%%%%%%%%%%%%%%%%%%%%%%%%
\begin{longtab}
\begin{landscape}
\tiny{
\begin{longtable}{lllllllllllllll}
\caption{ Result of the F-test over the optical/X-ray SEDs. \textit{GRB}: GRB name. $h_{PL}$: extinction curve type used to fit the data with a power-law. $h_{BR}$: host galaxy type used to fit the data with a broken power-law. \textit{Part}: segment of the X-ray LC considered: S=steep decay, P=plateau, N=normal decay. $t_{start}$ and $t_{end}$: initial and final time of the SED. $\chi^{2}_{PL}$, DOF$_{PL}$, $p-val_{PL}$: $\chi^{2}$, degree of freedom and $p$-value for the power-law fit function. $\chi^{2}_{BR}$, DOF$_{BR}$, $p-val_{BR}$: $\chi^{2}$, degree of freedom and $p$-value for the broken power-law fit function. \textit{PROB}: the F-test probability computed for each SED to comapre the single and broken power-law. \textit{PL} = power-law, \textit{BR} = broken power-law: $\surd$ indicates the best-fit model. $\star$ indicates the cases for which we decided that the simple power-law fits better the SED than the broken power-law fit, regardless of the F-test.}\label{ftest}\\
\hline\hline
GRB    &h$_{PL}$&h$_{BR}$&Part&$t_{start}$&$t_{end}$&$\chi^{2}_{PL}$&DOF$_{PL}$&p-val$_{PL}$&$\chi^{2}_{BR}$&DOF$_{BR}$&p-val$_{BR}$ &PROB&PL&BR\\
\hline
\endfirsthead
\caption{Continued.}\\
\hline\hline
GRB    &h$_{PL}$&h$_{BR}$&Part&$t_{start}$&$t_{end}$&$\chi^{2}_{PL}$&DOF$_{PL}$&p-val$_{PL}$&$\chi^{2}_{BR}$&DOF$_{BR}$&p-val$_{BR}$ &PROB&PL&BR\\
\hline
\endhead
\hline
\endfoot
050319 & MW & MW & P & 300 & 1500 &      99.76 &      42. &    1.3e-06 &      40.50 &      40. &       0.45 &   7.01e-20 & - & $\surd$ \\
050319 & MW & MW & N & 17e+03 & 300e+03 &      72.58 &      10. &    3.2e-07 &      40.50 &       9. &    3.8e-05 &    0.00342 & - & $\surd$ \\
050401 & MW & SMC & P & 200 & 2000 &      33.31 &     146. &        1.0 &     205.00 &     145. &    0.00075 &      -9.00 & $\surd$ & - \\
050408 & SMC & SMC & N & 20e+03 & 40e+03 &      22.73 &      28. &       0.75 &      20.00 &      37. &       0.83 &       0.00 & - & $\surd$ \\
050416A & LMC & SMC & P & 700 & 2000 &      27.02 &      24. &       0.30 &      18.90 &      23. &       0.71 &   2.98e-07 & - & $\surd$ \\
050416A & SMC & MW & N & 6000 & 40e+03 &      29.54 &      27. &       0.34 &      22.40 &      26. &       0.66 &   3.51e-07 & - & $\surd$ \\
050525A & SMC & SMC & N & 6000 & 30e+03 &      49.56 &      29. &      0.010 &      51.70 &      28. &     0.0041 &      -9.00 & $\surd$ & - \\
050730 & SMC & SMC & P & 4000 & 6000 &     759.74 &     140. &   -8.6e-07 &   16200.00 &     139. &    -0.0036 &      -9.00 & $\surd$ & - \\
\hline
\end{longtable}
\tablefoot{\tablefoottext{a}{We where unable to perform the fit with a broken power-law function.}}
}
\end{landscape}
\end{longtab}
%%%%%%%%%%%%%%%%%%%%%%
\begin{longtab}
\begin{landscape}
\begin{longtable}{l p{0.3cm} p{0.6cm} p{0.6cm} p{8.4cm} p{8.4cm}}
\caption{\small{165 GRBs with redshift from the beginning of Swift observations in December 2004 until December 2011. \textit{GRB}: GRB name. (S) stands for short GRB. Boldface: radio sample from \cite{2012ApJ...746..156C}. (1) Supernova. \textit{Opt.}: $\times$ means that there are not sufficient data to analyze the optical LC; $\surd$ indicates GRBs in our sample (69). \textit{References Optical Data} and \textit{Telescopes and Filters}: citations and names of telescopes in parenthesis mean that we did not use these data.}}\label{elenco}\\
\hline\hline
GRB &Opt.& XLC & z & References Optical Data & Telescopes\tablefootmark{a} and Filters \\ 
\hline
\endfirsthead
\caption{Continued.}\\
\hline\hline
GRB &Opt.& XLC & z & References Optical Data & Telescopes\tablefootmark{a} and Filters \\ 
\hline
\endhead
\hline
\endfoot
050126   &$\times$&ICN &1.290 &     & \\
050315   &$\times$&IICN &1.949 &     & \\
050318   & $\times$&0UN &1.440 & \cite{still05}.& \scriptsize{UVOT (u,v,b).} \\
050319   &$\surd$&IICN &3.240 & \cite{wozniak05}; \cite{quimby06}; \cite{mason06}; \cite{kamble07}; \cite{huang07}; \cite{george06};  \cite{2010ApJ...720.1513K}.  &\scriptsize{RAPTOR(CR); ROTSE(CR); UVOT(v,b); SNT(B,V,R,I); Kiso and Lulin(B,V,R,I); Mt.Abu(J); MAO(R,B); RTT150(R); NOT(B,R,I,K$_S$); TNG(V,R,I).}             \\    
050401   &$\surd$&ICN &2.900	 & \cite{watson06}; \cite{rykoff05} \cite{depasquale06}; \cite{kamble09}. &\scriptsize{VLT(R,K); Subaru(R); NOT(R); D1.5(R); TNG(R); MAO(R); CTIO4m(H); UKIRT(J,H,K); ROTSE(CR) SSO(R); SNT(R); MAO(R); Loiano(R); TNG(R); SNT(V,R,I).}\\
050408   & $\surd$&0UN &1.236	 & \cite{foley06}; \cite{deugarte07}; \cite{2010ApJ...720.1513K}. &\scriptsize{UVOT(v); Keck(V,I); Magellan(R,I); CTIO1.3m(I,J,V); PAIRITEL(J,H,K); REM(V,R,I,J); PROMPT(V); Kuiper(V,R,I); WIDGET(R); ART(R); Kiso(R); MITSuME(R); SNT(R); TLS(R); RTT150(R); SAO1m(R); Mercator(R); MAO(R,I); ARC(z'); D1.5(U,B,V); KPNO4m(U); BTA(B,V,R); CAHA3.5m(B); RTT150(R); TNG(U,B,V,R,I).} \\
050416A &$\surd$&ICN &0.654	 & \cite{soderberg07}; \cite{holland07}; \cite{2010ApJ...720.1513K}.&\scriptsize{P60(IRz); P200(K); ATT2.3m(R); HST(AB); UVOT(u,v,b,uvw1,uvm2,uvw2); NOT(K); D1.5(RVBI); MAO(R,B); SMARTS(I,J).}\\
050505   &$\times$&IUN &4.270	 & \cite{hurkett06}.&\scriptsize{BOOTES(R); TAROT(I); AAVSO(R); SARA(R); Keck(I,g); UKIRT(K); FTN(R,B,i').} \\
\hline
\hline
\end{longtable}
\tablefoottext{a}{List of the telescopes acronyms:
\begin{itemize}
\item[-] \textbf{AAVSO}: American Association of Variable Star Observers (International High Energy Network).
\item[-] \textbf{AEOS}: 3.6-m US Air Force Advanced Electro-Optical System telescope.
\item[-] \textbf{ARC}: Astrophysical Research Consortium 3.5-m telescope at Apache Point Observatory.
\item[-] \textbf{ARIES}: Aryabhatta Research Institute of Observational Sciences (SNT and DFOT).
\item[-] \textbf{ART}: Automated Response Telescope.
\item[-] \textbf{Asiago}: 1.82-m telescope at Cima Ekar (Asiago). 
\item[-] \textbf{ATT2.3m}: 2.3-m Advanced Tecnology Telescope at SSO.
\item[-] \textbf{AZT-8}: AZT-8 70	-cm Telescope at Kharkiv observatory .
\item[-] \textbf{AZT-11}: 1.25-cm telescope at CrAO.
\item[-] \textbf{AZT-33IK}: Russian IR telescope in Sayanskaya observatory.
\item[-] \textbf{Banon}: T50-Banon telescope of the Observatoire de Chante-Perdrix at Banon (France).
\item[-] \textbf{BOAO}: 15.5-cm telescope at Bohyunsan Optical Astronomy Observatory.
\item[-] \textbf{Bok}: 2.3-m Bok Telescope at Kitt Peak.
\item[-] \textbf{BOOTES}: Burst Observer and Optical Transient Exploring System.
\item[-] \textbf{BTA}: Big Telescope Alt-Azimuth at SAO.
\item[-] \textbf{Brno}: 40-cm Newtonian telescope.
\item[-] \textbf{BYUWMO}: 0.91m telescope at West Mountain Observatory (Brigham Young University).
\item[-] \textbf{CAHA}: telescope at Calar Alto Observatory: \textbf{CAHA3.5m}, \textbf{CAHA2.2m}, \textbf{CAHA1.3m}.
\item[-] \textbf{CrAO}: Crimean Astrophysical Observatory 2.6-m telescope; it corresponds to \textbf{Shajin}.
\item[-] \textbf{Crni}: \v{C}rni Vrh Observatory.
\item[-] \textbf{CTIO}: Cerro Tololo Inter-American Observatories. \textbf{CTIO4m}: 4-m Victor M. Blanco telescope. \textbf{CTIO1.3m}: 1.3-m (ex-2MASS) telescope.
\item[-] \textbf{D1.5}: Danish 1.54-m telescope.
\item[-] \textbf{DAO}: 16'' Dolomiti Astronomical Observatory telescope.
\item[-] \textbf{DFOT}: Devasthal Fast Optical telescope at ARIES.
\item[-] \textbf{EST}: 1-m EOS telescope.
\item[-] \textbf{FTN}: Faulkes Telescope North. 
\item[-] \textbf{FTS}: Faulkes Telescope South. 
\item[-] \textbf{GAO}: 150-cm telescope at Gamma Astronomical Observatory.
\item[-] \textbf{Gemini-N}: Gemini North telescope.
\item[-] \textbf{Gemini-S}: Gemini South telescope.
\item[-] \textbf{GMG}: 2.4m telescope at Gao-Mei-Gu. 
\item[-] \textbf{GORT}: 14" Glast Optical Robotic Telescope at Hume observatory in California. 
\item[-] \textbf{GRAS-04}: Global-Rent-A-Scope telescope at Mayhill (AAVSO).
\item[-] \textbf{GRAS-005}: Global-Rent-A-Scope telescope at New Mexico.
\item[-] \textbf{GRAS-17}: Global-Rent-A-Scope CDK17 17'' telescope.
\item[-] \textbf{GROND}: Gamma-Ray Burst Optical/Near-Infrared detector.
\item[-] \textbf{Hankasalmi}: 0.4-m telescope at Hankasalmi Observatory.
\item[-] \textbf{HET}: 9.2-m Hobby-Eberly Telescope.
\item[-] \textbf{HST}: Hubble Space Telescope.
\item[-] \textbf{HCT}: 2-m Himalayan Chandra Telescope. 
\item[-] \textbf{IAC80}: Instituto de Astrof\'isica de Canarias 0.8-m telescope.
\item[-] \textbf{IGO}: IUCAA Girawali 2-m Optical telescope.
\item[-] \textbf{IUCAA}: Inter-University Center for Astronomy and Astrophysics. 
\item[-] \textbf{INT}:  Isaac Newton Telescope.
\item[-] \textbf{IRSF}: IRSF 1.4-m telescope at SAAO (South Africa).
\item[-] \textbf{ISON-NM}: 0.45-m telescope at ISON-NM observatory.
\item[-] \textbf{KAIT}: Katz Automatic Imaging Telescope. 
\item[-] \textbf{Kanata}: Kanata 1.5-m telescope.
\item[-] \textbf{Keck}: Keck telescopes.
\item[-] \textbf{Kiso}: Kiso Observatory (105-cm Schmidt telescope).
\item[-] \textbf{Konkoly}: 60/90-cm telescope a Kolnkoly observatory.
\item[-] \textbf{KPNO4m}: Mayall 4-m telescope at Kitt Peak National Observatory. 
\item[-] \textbf{Kuiper}: Kuiper telescope.
\item[-] \textbf{LBT}: Large Binocular Telescope.
\item[-] \textbf{Lemmon1m}: Mt. Lemmon 1-m telescope.
\item[-] \textbf{Lick}: 3-m Shane telescope at Lick Observatory.
\item[-] \textbf{Lightbukets}: 0.61-m Lightbukets rental telescope in Rodeo (NM).
\item[-] \textbf{Loiano}: 1.52-m Cassini telescope at Loiano Astronomical Observatory.
\item[-] \textbf{LOTIS}: Super-LOTIS (Livermore Optical Transient Imaging System) 
\item[-] \textbf{LT}: Liverpool Telescope.
\item[-] \textbf{Lulin}: Lulin Observatory.
\item[-] \textbf{Magellan1}: Magellan 1-Baade Telescope.
\item[-] \textbf{MAGNUM}: Multicolor Active Galactic Nuclei Monitoring 2-m telescope on Haleakala.
\item[-] \textbf{MAO}: 1.5-m telescope at Maidanak Astronomical Observatory.
\item[-] \textbf{MASTER}: Mobile Astronomy System of Telescope Robots.
\item[-] \textbf{MDM}: Michigan-Dartmouth-MIT observatory. \textbf{MDM2.4m}: 2.4-m Hiltern telescope. \textbf{MDM1.3m}: Mc Graw-Hill telescope.
\item[-] \textbf{Meade-35}: telescope at Terskol.
\item[-] \textbf{Mercator}: 1.2-m Mercator telescope.
\item[-] \textbf{miniTAO}: 1-m miniTAO telescope at the University of Tokyo Atacama Observatory.
\item[-] \textbf{MIRO}: 1.2-m telescope of Mt. Abu InfraRed Observatory (India); it corresponds to \textbf{Mt.Abu}.
\item[-] \textbf{MITSuME}: Multicolor Imaging Telescopes for Survey and Monstrous Explosions.
\item[-] \textbf{Murikabushi}: 105-cm Murikabushi telescope (Japan).
\item[-] \textbf{MMT}: 6.5-m telescope at the Wipple Observatory on Mount Opkins in Arizona.
\item[-] \textbf{Mt.Abu}: 1.2 telescope of Physical Research Laboratory at Gurushikar (Mt. Abu/India). it corresponds to \textbf{MIRO}.
\item[-] \textbf{Nickel}: Lick 1-m Nickel telescope.
\item[-] \textbf{NMSU}: NMSU 1m telescope at the Apache Point Observatory.
\item[-] \textbf{NOT}: Noth Optical Telescope.
\item[-] \textbf{NTT}: New Technology Telescope.
\item[-] \textbf{OAUV}: 0.4-m OAUV telescope at the Observatorio Astronomico de Aras.
\item[-] \textbf{OHP}: T80 telescope of the Observatoire the Haute-Provence.
\item[-] \textbf{Ondrejov}: 500/1975-mm Newtonian telescope in Ondrejov. 
\item[-] \textbf{OSN}: 1.5-m telescope at Observatorio de Sierra Nevada. 
\item[-] \textbf{P60}: Palomar 60-Inch telescope.
\item[-] \textbf{P200}: Palomar 200-Inch telescope.
\item[-] \textbf{PAIRITEL}: Peters Authomated InfraRed Imaging TELescope.
\item[-] \textbf{PI}: Pi of the Sky.
\item[-] \textbf{PROMPT}: Panchromatic Robotic Optical Monitoring and Polarimetry Telescopes.
\item[-] \textbf{RAPTOR}: RAPid Telescope for Optical Response.
\item[-] \textbf{REM}: Rapid Eye Mount.
\item[-] \textbf{ROTSE}: Robotic Optical Transient Search Experiment.
\item[-] \textbf{RTT150}: Russian-Turkish 1.5-m Telescope.
\item[-] \textbf{SAO}: Special Astrophysical Observatory (1-m telescope Zeiss and 6-m BTA).
\item[-] \textbf{SARA}: Southeastern Association for Research in Astronomy. \textbf{SARA0.5m} and \textbf{SARA0.6m} telescopes.
\item[-] \textbf{Skinakas}: 1.3-m telescope at Skinakas Observatory.
\item[-] \textbf{SMARTS}: Small and Moderate Aperture Research Telescope System (1.3-m telescope); it corresponds to \textbf{CTIO1.3m}. 
\item[-] \textbf{SNT}: Sunpurnanand Telescope at ARIES.
\item[-] \textbf{SR-22}: 0.22-m telescope.  
\item[-] \textbf{SOAR}: 4.1-m SOuthern Astrophysical Research telescope.
\item[-] \textbf{SRO}: Sonoita Research Observatory 35-cm telescope.
\item[-] \textbf{SSO}: Siding Spring Observatory. \textbf{SSO40Inch} and \textbf{ATT2.3m}.
\item[-] \textbf{Stardome}: Stardome 0.4-m telescope in Auckland (New Zealand).
\item[-] \textbf{Subaru}: Subaru telescope.
\item[-] \textbf{T100}: 1-m telescope at TUBITAK National Observatory (Turkey).
\item[-] \textbf{T1M}: T1M telescope at Observatoire du Pic du Midi.
\item[-] \textbf{TAOS (A,B,C,D)}: Taiwanese-American Occultation Survey Telescopes. 
\item[-] \textbf{TAROT}:  T\'elescopes \`a  Action Rapide pour les Objects Transitoires.
\item[-] \textbf{Terskol}: Zeiss-600 of Mt. (Pik) Terskol observatory.
\item[-] \textbf{THO}: Taurus Hill Observatory.
\item[-] \textbf{TLS}: Th\"uringer Landessternwarte Tauteburg. \textbf{TLS2m} and \textbf{TLS1.34m}.
\item[-] \textbf{TNG}: Telescopio Nazionale Galileo.
\item[-] \textbf{TNT}: 0.8-m Tsinghua-NOAC Telescope.
\item[-] \textbf{TORTORA}: camera mounted on REM telescope.
\item[-] \textbf{UKIRT}: United Kingdom Infra-Red Telescope.
\item[-] \textbf{UVOT}: Ultra-Violet/Optical Telescope on board \textit{Swift} satellite.
\item[-] \textbf{VLT}: Very Large Telescope.
\item[-] \textbf{Watcher}: Watcher telescope.
\item[-] \textbf{WHT}: 4.2-m William Herchel Telescope.
\item[-] \textbf{WIDGET}: WIDe-field camera for Gamma-ray bursts Early Timing.
\item[-] \textbf{WIRO}: 2.3m Wyoming Infrared Observatory telescope in Wyoming (USA).
\item[-] \textbf{XLT}: 0.35-m C14 XLT telescope at Taurus Hill Observatory.
\item[-] \textbf{Z1000}: 1-m Zeiss-telescope at CrAO.   
\item[-] \textbf{Z2000}: telescope at Mt. Terskol.
\item[-] \textbf{Zadko}: 1-m Zadko telescope.
\item[-] \textbf{ZTE}: 1.25-m ZTE telescope at the Crimean station of the Sternberg Astronomical Institute.
\end{itemize}}
\end{landscape}
\end{longtab}
%----------------------------------------------------------------
\newpage
%%%%%%%%%%%%%%%%%%%%%%%%%%%%%%%%%%%%
\clearpage
%%%%%%%%%%%%%%%%%%%%%%%%%%%%%%%%%%%%
%%%%%%%%%%%%%%%%%%%%%%%%%%%%%%%%%%%%
%%%%%%%%%%%%%%%%%%%%%%%%%%%%%%%%%%%%
%%%%%%%%%%%%%%%%%%%%%%%%%%%%%%%%%%%%
%%%%%%%%%%%%%%%%%%%%%%%%%%%%%%%%%%%%
\clearpage
%%%%%%%%%%%%%%%%%%%%%%%%%%%%%%%%%%%%
%%%%%%%%%%%%%%%%%%%%%%%%%%%%%%%%%%%%
%%----------------------------------------------------------------
\newpage
%%%%%%%%%%%%%%%%%%%%%%%%%%%%%%%%%%%%%
\clearpage
%%%%%%%%%%%%%%%%%%%%%%%%%%%%%%%%%%%%%
%%%%%%%%%%%%%%%%%%%%%%%%%%%%%%%%%%%%%
%%%%%%%%%%%%%%%%%%%%%%%%%%%%%%%%%%%%%
%%%%%%%%%%%%%%%%%%%%%%%%%%%%%%%%
%%%%%%%%%%%%%%%%%%%%%%%%%%%%%%%%
\clearpage
%%%%%%%%%%%%%%%%%%%%%%%%%%%%%%%%%%%%%
%%%%%%%%%%%%%%%%%%%%%%%%%%%%%%%%%%%%%
\clearpage
%----------------------------------------------------------------
%\end{appendix}

\end{document}